\documentclass[11pt]{article}

\usepackage{epsfig,lscape}
\usepackage{amssymb,amsfonts,amsmath,amsthm}
\usepackage{rotating}
\usepackage{bbm}

\usepackage{threeparttable}
\usepackage{booktabs}
\usepackage{subcaption}

\usepackage[table]{xcolor} 
\usepackage{float}
\usepackage{hyperref}

\usepackage[left=1in,right=1in,top=.7in,bottom=.75in]{geometry}

\usepackage[medium,compact]{titlesec}

\titlespacing*{\section} {0pt}{2ex}{1ex}
\titlespacing*{\subsection} {0pt}{2ex}{1ex}
\titlespacing*{\subsubsection} {0pt}{2ex}{1ex}

\def\me{\mathrm e}

\def\dif{\mathrm d}

\def\E{\mathrm E}
\def\P{\mathrm P}
\def\var{\mathrm{var}}
\def\cov{\mathrm{cov}}
\def\N{\mathrm{N}}

\def\diag{\mathrm{diag}}

\def\T{ {\mathrm{\scriptscriptstyle T}} }

\def\bbR{\mathbb R}

\def\argmin{\mathrm{argmin}}

\newenvironment{prf}
{\noindent \textbf{Proof.}}{\hfill $\Box$ \vspace{.1in}}

\newtheorem{thm}{Theorem}
\newtheorem{lem}{Lemma}
\newtheorem{pro}{Proposition}
\newtheorem{cor}{Corollary}
\newtheorem{ass}{Assumption}

\theoremstyle{definition}

\theoremstyle{definition}
\newtheorem{rem}{Remark}

\begin{document}

\begin{titlepage}

\begin{center}
{\Large High-dimensional model-assisted inference for treatment effects with multi-valued treatments}

\vspace{.1in} Wenfu Xu\footnotemark[1] and Zhiqiang Tan\footnotemark[1]

\today
\end{center}

\footnotetext[1]{{\small Wenfu Xu is Assistant Professor, College of Economics and Management, China Jiliang University, Hangzhou 310018, China (E-mail:wf.xu@cjlu.edu.cn), and 
Zhiqiang Tan is Professor, Department of Statistics, Rutgers University, Piscataway, NJ 08854, USA (E-mail:ztan@stat.rutgers.edu). 
This work was completed when Wenfu Xu was visiting Rutgers University.
}}

\paragraph{Abstract}
Consider estimation of average treatment effects with multi-valued treatments using
augmented inverse probability weighted (IPW) estimators, depending on outcome regression and propensity score models
in high-dimensional settings.
These regression models are often fitted by regularized likelihood-based estimation, while ignoring how the fitted functions are used in the subsequent
inference about the treatment parameters.
Such separate estimation can be associated with known difficulties in existing methods.
We develop regularized calibrated estimation for fitting propensity score and outcome regression models,
where sparsity-including penalties are employed to facilitate variable selection but the loss functions are carefully chosen
such that valid confidence intervals can be obtained under possible model misspecification.
Unlike in the case of binary treatments, the usual augmented IPW estimator is generalized 
by allowing different copies of coefficient estimators in outcome regression to ensure just-identification.
For propensity score estimation, the new loss function and estimating functions are directly tied to achieving covariate balance between weighted treatment groups.
We develop practical numerical algorithms for computing the regularized calibrated estimators with group Lasso by innovatively exploiting Fisher scoring,
and provide rigorous high-dimensional analysis for the resulting augmented IPW estimators under suitable sparsity conditions,
while tackling technical issues absent or overlooked in previous analyses.
We present simulation studies and an empirical application to estimate the effects of maternal smoking on birth weights.
The proposed methods are implemented in the R package \texttt{mRCAL}.

\vspace{.1in}

\end{titlepage}

\section{Introduction}\label{sec:Introduction}

Estimation of average treatment effects (ATEs) has been extensively studied in the potential-outcome framework for causal inference (Neyman 1923; Rubin 1974).
As a distinctive feature of the problem, doubly robust (DR) methods are available to achieve consistent estimation
if either an outcome regression (OR) model is correctly specified for the outcome $Y$ given treatment $T$ and covariates $X$
or a propensity score (PS) model is correctly specified
for the treatment $T$ given covariates $X$ (e.g., Tan 2007).
All such methods can be viewed to involve two stages of estimation.
First, OR and PS models are built and fitted as regression models.
Then the fitted functions are substituted
into a DR point estimator, notably the augmented inverse probability weighted (IPW) estimator (Robins et al.~1994).
Conventionally, the two stages are treated separately:
regression models in the first stage are built and fitted by maximum likelihood or variants, while ignoring how
the fitted functions are used in the second stage.
Such separate estimation can be associated with known difficulties
in the existing methods. For example, augmented IPW estimation may perform poorly if the PS model appears
to be nearly correct and fitted by maximum likelihood (Kang and Schafer 2007). There is another difficulty more recently
recognized in the high-dimensional settings (Tan 2020b).
If the OR and PS models are fitted by regularized maximum likelihood, then
DR point estimation can be achieved by the augmented IPW estimator, but
valid confidence intervals are obtained under suitable sparsity conditions only when both the OR
and PS models are correctly specified (or with negligible specification biases) (Chernozhukov et al. 2018).


In this article, we develop new methods and theory for fitting OR and PS models while using augmented IPW estimation to
draw inferences about ATEs with multi-valued treatments in high-dimensional settings,
where the number of regressors $p$ is close to or greater than the sample size $n$. Estimation of
average treatment effects on the treated (ATTs) are also handled in a unified manner.
Compared with existing methods, the aforementioned two stages of estimation are integrated in our approach, called regularized calibrated estimation.
We employ regularized estimation for fitting the OR and PS models, where sparsity-inducing penalties such as
the Lasso or group Lasso penalty are
used to facilitate variable selection (Tibshirani 1996; Yuan and Lin 2006).
However, we carefully choose the loss functions (or equivalently estimating functions determined from the gradients),
depending on augmented IPW estimation in the second stage, such that valid confidence intervals
can be obtained for the treatment parameters under suitable sparsity conditions in high-dimensional settings, while accommodating possible model misspecification.
In fact, our methods not only lead to DR point estimation but also provide DR confidence intervals for ATEs and ATTs, which
are valid if either a multi-class logistic PS model or a linear OR model is correctly specified.
Our methods also provide model-assisted confidence intervals for the treatment parameters,
which are valid if a multi-class logistic PS model is correctly specified, but a generalized linear OR model may be misspecified.
Another advantage of our approach is that
the new loss function and estimating functions for PS estimation are directly tied to evaluation of covariate balance
after inverse probability weighting with a multi-valued treatment.
By analysis of Bregman divergences, minimization of the expected loss function can be shown to control relative errors
of propensity scores more effectively than maximum likelihood with possible model misspecification. See Remark 1 and Appendix A.

Our work builds on the recent success of using regularized calibrated estimation for treatment effect estimation with binary treatments
and instruments (Tan 2020ab; Sun and Tan 2021), but needs to tackle various analytical and computational complications due to
multi-valued treatments. In the case of binary treatments, there are two sets of calibration equations associated with the augmented IPW estimator.
One set is just-identifying for the regression coefficients in the PS model, and the other is just-identifying for those in the OR model.
These calibration equations can be directly converted into two corresponding loss functions for regularized estimation (Tan 2020b).
For a multi-valued treatment,
calibration equations derived from the usual augmented IPW estimator are no longer just-identifying for
the coefficients in the PS model and separately in the OR model.
This complication is related to the fact that a multi-class logistic PS model involves $(K-1)p$ free coefficients,
but an OR model involves only $p$ coefficients within a fixed treatment group, where $K$ is the number of treatment groups.
To make progress, we exploit a natural relationship between the two types of expectations in ATEs and ATTs:
\begin{align}
\E (Y^{(t)} ) = \E (R^{(t)} Y) + \sum_{k\not=t} \E (Y^{(t)} | T=k) \P (T=k), \label{eq:decomp}
\end{align}
where $Y^{(t)}$ denotes the potential outcome for treatment $t$.
Then we define a new augmented IPW estimator for $\E ( Y^{(t)} )$,
by decomposing the usual augmented IPW estimator for $\E ( Y^{(t)} )$
in terms of $K-1$ augmented IPW estimators of $\E (Y^{(t)} | T=k)$  as in (\ref{eq:decomp}),
while allowing different copies of the coefficient vector
in the same OR model for $Y$ given $T=t$ and $X$ for $k\not=t$.
We show that this approach leads to calibration equations which are just-identifying for the PS coefficient vectors
and separately for the $K-1$ copies of the OR coefficient vector
and can be converted into convex loss functions for regularized estimation.
In particular, the new loss function for PS estimation
properly extends the calibration loss in Tan (2020a) to multi-class logistic regression (see Remark 4).

We develop practical numerical algorithms for computing the regularized calibrated estimators with group Lasso penalties (Yuan and Lin 2006).
Our algorithms use quadratic approximation (Friedman et al.~2010) and the majorization-minimiation (MM) technique (Wu and Lange 2010),
but innovatively exploit Fisher scoring (McCullagh and Nelder 1989) to construct closed-form updates based on the block coordinate descent (Simon et al.~2013).
Furthermore, we provide rigorous high-dimensional analysis of the group-Lasso regularized calibrated estimators and the resulting augmented IPW estimators
for the treatment parameters with possible model misspecification.
Our analysis establishes that doubly robust or model-assisted Wald confidence intervals can be obtained for the treatment parameters
as described earlier under the sparsity condition,
$(|S_\gamma| + |S_{\alpha_t}|)(K-1)^{1/2}\{(K - 1) + \log(p)\} = o(n^{1/2})$,
where $|S_\gamma|$ and $|S_{\alpha_t}|$ denote the non-sparsity sizes in PS and OR estimation.
Compared with previous high-dimensional analyses,
we carefully tackle several technical issues, including the interdependency between the gradient vectors of
the loss functions in both PS and OR estimation and the data-dependency of an estimated weight in OR estimation (see Remarks 7--9).
These two issues are absent in previous analysis of multi-task linear regression with group Lasso, where tasks are independent of each other (Lounici et al.~2011).
The interdependency of gradient vectors appears to be overlooked in analysis of
regularized likelihood estimation with group Lasso for multi-class logistic regression in Farrell (2015).

A theoretical limitation of the proposed method is that the confidence intervals are not doubly robust for the treatment parameters
unless a linear OR model is used. This limitation can be traced to the fact that PS estimation and evaluation of covariate balance
are designed to be applicable in our method without access to any outcome data, which is a desirable property for the PS methodology
to avoid bias from outcome modeling (Rubin 2001).
Doubly robust confidence intervals can be developed, by adapting the approach of Ghosh and Tan (2021),
but PS and OR estimation would then be coupled to each other. See Tan (2020b, Section 3.5) for related discussion.


\vspace{.05in}
{\bf Related work.} There is a vast and growing literature on estimation of average treatment effects.
We discuss directly related work to ours, in addition to the earlier discussion.

Theory and methods for ATE estimation have been extensively studied in low-dimensional settings with binary or multi-valued treatments.
The case of a multi-valued treatment is often treated as a direct extension from the case of a binary treatment (e.g., Cattaneo 2010; Tan 2010).
For DR estimation, it is common to fit a generalized linear OR model and a multi-class logistic PS model
by maximum likelihood or quasi-likelihood, from which the fitted values can be combined, sometimes with additional adjustment,
for augmented IPW estimation.
For PS estimation with a multi-valued treatment, Imai and Ratkovic (2014, Section 4.1) briefly discussed a set of covariate balancing equations, which
appear related to our calibration equations (\ref{eq:EE2-gamma}) or equivalently (\ref{eq:EE2-gamma-b}) later.
However, there exist important differences: the balancing equations are
based on contrasts of IPW averages of covariates between successive treatment groups
(0 vs 1, 1 vs 2, etc.), instead of contrasts of probability ratio weighted and simple averages between one treatment group and the remaining ones.
In general, the set of balance equations cannot be converted into a loss function (much less a convex loss) and need to be solved
as nonlinear equations or combined with score equations from maximum likelihood using the generalized method of moments.
Adaption of such equations seems difficult numerically and theoretically in high-dimensional settings.

As an alternative to maximum likelihood estimation,
calibrated estimation and related methods have also been studied in low-dimensional settings for fitting PS models in causal inference with binary treatments
or fitting response probability models in survey sampling and missing-data problems,
where the non-missingness probability represents a propensity score
(e.g., Folsom 1991; Tan 2010; Graham et al.~2012;
Hainmueller 2012; Imai and Ratkovic 2014; Kim and Haziza 2014; Vermeulen and Vansteelandt 2015).
In the binary case, calibrated estimation requires that the IPW averages of covariates in the treated subsample are equal to the simple averages in
the overall sample. These equations can also be seen to match the probability ratio weighted averages of covariates in the treated subsample with
the simple averages in the untreated subsample (Tan 2020a, Section 7.4), thereby aligned with calibration equations (\ref{eq:EE2-gamma-b})
in our multi-valued extension.

For DR estimation of ATEs with binary treatments, Kim and Haziza (2014) and Vermeulen and Vansteelandt (2015)
proposed estimating equations which are equivalent to the calibration equations in Tan (2020b).
In low-dimensional settings, one of the benefits is to enable computationally simpler variance estimation for augmented IPW estimators,
compared with likelihood-based estimation in fitting PS and OR models.
In contrast, calibration equations are exploited by Tan (2020b) to develop regularized calibrated estimation for fitting PS and OR models
such that doubly robust or model-assisted confidence intervals can be obtained for ATEs in high-dimensional settings.
Regularized calibrated estimation has also been proposed by Sun and Tan (2021)
for estimating local average treatment effects (LATEs) with instrumental variables,
and extended by Ghosh and Tan (2021) to general semiparametric estimation based on DR estimating functions.
The calibration equations in the latter work are assumed to be just-identifying for two sets of nuisance parameters to be estimated.
But this assumption is not satisfied in ATE estimation with a multi-valued treatment, which is a major challenge addressed in the current work.

In high-dimensional settings, DR estimating functions have also been used to derive valid confidence intervals
under suitable sparsity conditions in Belloni et al.~(2014) with binary treatments
and Farrell (2015) with multi-valued treatments for ATEs, and in Chernozhukov et al.~(2018) for more general treatment parameters.
As mentioned earlier, regularized likelihood-based estimation is employed in these methods, and the confidence intervals
are only shown to be valid for the treatment parameters when all working models for the nuisance parameters are correctly specified
(or with negligible biases).
To alleviate this limitation, there has been considerable research in high-dimensional causal inference.
Examples related to regularized calibrated estimation include
Avagyan and Vansteelandt (2017), Smucler et al.~(2019), Bradic et al.~(2019), and Ning et al.~(2020) among others.
However, these methods mainly deal with ATE estimation with binary treatments,
and would not be applicable to the case of multi-valued treatments.

\section{Setup and existing methods} \label{sec:setup}

Suppose that $\{(Y_i,T_i,X_i ):i=1, \ldots, n\}$ are independent and identically distributed observations of $(Y, T, X)$,
where $Y$ is an observed outcome, $T$ is an observed treatment,  and $X$ is a $d\times 1$ vector of covariates.
The treatment $T$ is assumed to take $K\,(\ge 2)$ possible values, denoted as $\mathcal T = \{0,1,\ldots, K-1\}$, where $0$ denotes the null treatment.
Let $Y^{(t)}$ be the potential outcome that would be observed under treatment $t$ (Neyman 1923; Rubin 1974). We make the consistency assumption that $Y=Y^{(t)}$ if $T=t$.
The average treatment effect (ATE) for treatment $t$ versus $k$ is defined as
$\E(Y^{(t)}-Y^{(k)} )= \mu_t - \mu_k$, where $\mu_t=\E(Y^{(t)} )$ for $t\in\mathcal T$. 
The average treatment effect in the $k$th treated group (ATT) is defined as $\E (Y^{(t)} - Y^{(k)} | T= k )$ for $t\not= k \in \mathcal T$.
We mainly discuss estimation of $\{\mu_t: t\in\mathcal T\}$ and ATEs until Section~\ref{sec:ATT} on ATT estimation.

A fundamental difficulty in estimating ATEs is that for each subject $i$, only one potential outcome is observed, $Y^{(t)}_i$ if $T_i=t$, and the others are missing.
Nevertheless, the means $\{\mu_t: t\in\mathcal T\}$ can be identified from observed data under two assumptions: \vspace{-.1in}
\begin{itemize} \addtolength{\itemsep}{-.1in}
\item Unconfoundedness: $R^{(t)}$ and $Y^{(t)}$ are conditionally independent given $X$ for $t \in \mathcal T$ (Rubin 1976), where $R^{(t)} = 1\{T=t\}$,
equal to 1 if $T=t$ or 0 otherwise;
\item Overlap: $\pi^*(t,X) >0$ almost surely for $t \in \mathcal T$, where $\pi^*(t,X)= \P(T=t | X)$ is called the propensity score (Rosenbaum and Rubin 1983).
\end{itemize} \vspace{-.1in}
%
%
Under the foregoing assumptions, ATE estimation from sample data customarily involves two stages.
First, regression models are built and fitted for the outcome regression (OR) function $m^*(t,X)=E(Y| T=t,X)$ or the propensity score (PS) $\pi^*(t,X)= \P(T=t | X)$.
In the second stage, the fitted functions are substituted into various estimators for $\mu_t$ and ATEs.
To facilitate discussion in Section~\ref{sec:method}, we describe regression models with pre-specified regression terms and regularized likelihood estimation commonly used in these models.
We also introduce augmented IPW estimation for $\mu_t$, which is used in the proposed method and existing ones.

Consider an outcome regression model
\begin{align}
\E(Y | T=t, X) & = m(t, X;\alpha_t)= \psi\{\alpha_t^\T g(X)\}, \quad t\in\mathcal T, \label{eq:ORmodel}
\end{align}
where $\psi(\cdot)$ is an inverse link function, 
$g(X)= (g_0(X), g_1(X), \ldots, g_q(X))^\T$ is a $(q+1)\times 1$ vector of {\it known} functions of covariates (for example, main effects or interactions) with $g_0(X)\equiv 1$,
and $\alpha_t = (\alpha_{jt}: j=0,1, \ldots, q )^\T$ is a $(q+1)\times 1$ vector of unknown coefficients for $t \in\mathcal T$.
For concreteness, model (\ref{eq:ORmodel}) is specified to allow separate coefficient vectors $\alpha_t$ associated with different treatments $t \in \mathcal T$.
For a generalized linear model with a canonical link (McCullagh and Nelder 1989),
the average negative log-(quasi)-likelihood function in $\alpha_t$ is
$\ell_{\text{ML}} (\alpha_t ) = \tilde \E ( R^{(t)}  [ - Y \alpha_t^\T g(X) +  \Psi \{\alpha_t^\T g(X)\}  ] ) $,
where  $\Psi (u) = \int_0^u \psi(u^\prime) \,\dif u^\prime$.  
Throughout, $\tilde \E (\cdot)$ denotes the sample average, for example,
$\tilde \E \{  R^{(t)}  Y g(X) \} =  n^{-1} \sum_{i=1}^n  R^{(t)}_i Y_i g(X_i)$.
In high-dimensional settings (i.e., large $q$ relative to $n$), a regularized maximum likelihood (RML) estimator, $\hat\alpha_{t,\text{RMLs}}$, for $\alpha_t$
can be defined by separately minimizing the likelihood loss with a Lasso penalty (Tibshirani 1996):
\begin{align}
\ell_{\text{RMLs}} (\alpha_t ) =\ell_{\text{ML}} (\alpha_t ) + \lambda_t \| \alpha_{1:q,t} \|_1 , \quad t \in \mathcal T,  \label{eq:RML-alpha-obj1}
\end{align}
where $\|\cdot\|_1$ denotes the $L_1$ norm, $\alpha_{1:q,t}$ is $\alpha_t$ excluding the intercept $\alpha_{0t}$, and $\lambda_t \ge 0$ is a tuning parameter.
Alternatively, a group Lasso penalty (Yuan and Lin 2006; Farrell 2015) can be used to define regularized estimators $\{\hat\alpha_{t,\text{RMLg}}: t\in \mathcal T\} $ jointly as a minimizer to
\begin{align}
\ell_{\text{RMLg}} (\alpha_0,\ldots,\alpha_{K-1} ) = \sum_{t \in \mathcal T} \ell_{\text{ML}} (\alpha_t ) + \lambda \sum_{j=1}^q \| \alpha_{j\cdot} \|_2,  \label{eq:RML-alpha-obj2}
\end{align}
where $\| \cdot\|_2$ denotes the $L_2$ norm,  $\alpha_{j \cdot} = ( \alpha_{jt} : t\in\mathcal T)^\T$ consists of $K$ coefficients associated with
the covariate term $g_j(X)$, and $\lambda \ge 0$ is a tuning parameter.

Next, consider a multi-class logistic model for the propensity score:
\begin{align}
\P(T=t |  X) & = \pi(t,X;\gamma) = \frac{\exp\{\gamma_t^\T f(X)\}}{\sum_{k\in\mathcal T} \exp\{\gamma_k^\T f(X)\}} , \quad t\in\mathcal T,  \label{eq:PSmodel}
\end{align}
where $f(X) = (f_0(X), f_1(X), \ldots, f_p(X))^\T$ is a $(p+1)\times 1$ vector of {\it known} functions of covariates with $f_0(X)\equiv 1$, 
$\gamma_k =(\gamma_{jk}: j=0,1,\ldots,p)^\T$ is a $(p+1)\times 1$ vector of unknown coefficients for $k \in \mathcal T$,
and $\gamma = (\gamma_0, \ldots, \gamma_{K-1})$ is a $(p+1)\times K$ matrix.
The average negative log-likelihood function is
$\ell_{\text{ML}} (\gamma ) = \tilde \E [ - \sum_{k\in\mathcal T} R^{(k)} \gamma_k^\T f(X) +  \log \sum_{k \in\mathcal T} \exp\{\gamma_k^\T f(X)\} ] $.
In high-dimensional settings (i.e., large $p$ relative to $n$),
a regularized likelihood estimator $\hat\gamma_{\text{RML}} = (\hat\gamma_{k,\text{RML}}: k\in \mathcal T) $
can be defined by minimizing the likelihood loss with a group Lasso penalty (Simon et al.~2013; Farrell 2015):
\begin{align}
\ell_{\text{RML}} (\gamma ) =  \ell_{\text{ML}} (\gamma ) + \lambda \sum_{j=1}^p \| \gamma_{j\cdot} \|_2,  \label{eq:RML-gamma-obj}
\end{align}
where the intercepts are not penalized, $\gamma_{j\cdot} = ( \gamma_{jk} : k\in\mathcal T)^\T$ consists of $K$ coefficients associated with the covariate term $f_j(X)$,
and $\lambda \ge 0$ is a tuning parameter.
For non-penalized estimation ($\lambda=0$), two constraints are commonly used to ensure identification.
One is a one-to-zero constraint, for example, $\gamma_0\equiv 0$.
The other is the sum-to-zero constraint $\sum_{k\in\mathcal T}\gamma_k\equiv 0$.
For penalized estimation, there is a slight difference between use of the two constraints.
If the one-to-zero constraint $\gamma_0\equiv 0$ is used,
then $\gamma$ can be reduced to a $(p+1)\times (K-1)$ matrix $(\gamma_k: k\in\mathcal T\backslash\{0\})$ and $\gamma_{j\cdot}$ to
a $(K-1)\times 1$ vector $( \gamma_{jk} : k\in\mathcal T\backslash\{0\})^\T$.
If the sum-to-zero constraint is used, then only the intercepts need to be explicitly constrained, $\sum_{k\in\mathcal T} \gamma_{0k} \equiv 0$,  because minimization of (\ref{eq:RML-gamma-obj}) with $\lambda>0$ automatically implies that
$\sum_{k\in \mathcal T} \hat \gamma_{jk,\text{RML}} =0$ holds for $j=1,\ldots,p$.
See the Supplement Section \ref{sec:sum-to-zero-relationship} for a corrected proof of the sum-to-zero relationship originally discussed in Simon et al.~(2013).


Various estimators of $\mu_t$ can be employed, using fitted values from OR model (\ref{eq:ORmodel}) or PS model (\ref{eq:PSmodel}) or both (Tan 2007, 2010a).
In particular, there are doubly robust (DR) estimators depending on both OR and PS models in the augmented IPW form (Robins et al.~1994)
\begin{align}
\hat\mu_t(\hat m, \hat \pi ) =
\tilde \E \left\{ \varphi_t( Y, T, X; \hat\alpha_t, \hat\gamma) \right\},  \label{eq:ATE-aipw}
\end{align}
where $\hat m(t,X) = m(t,X;\hat\alpha_t)$ and $\hat \pi(t,X) = \pi(t,X; \hat\gamma)$ for some estimators $\hat\alpha_t$ and $\hat\gamma$, and
\begin{align*}
\varphi_t( Y, T, X; \alpha_t, \gamma) = \frac{ R^{(t)} Y }{\pi(t,X; \gamma) } - \left\{\frac{ R^{(t)} }{\pi(t,X; \gamma)} -1 \right\} m(t,X; \alpha_t)  .
\end{align*}
In high-dimensional settings, Farrell (2015) studied the estimator $\hat\mu_t(\hat m_{\text{RMLg}}, \hat \pi_{\text{RML}} )$,
using the group Lasso penalized estimators $\hat \alpha_{t,\text{RMLg}}$ and $\hat \gamma_{\text{RML}}$ described above,
where $\hat m_{\text{RMLg}} (t,X) = m(t,X; \hat\alpha_{t,\text{RMLg}})$ and $\hat\pi_{\text{RML}} = \pi(t,X ; \hat\gamma_{\text{RML}})$.
Two types of results are obtained, each under suitable sparsity conditions. First, $\hat\mu_t(\hat m_{\text{RMLg}}, \hat \pi_{\text{RML}} )$ is
shown to be pointwise doubly robust, i.e., remain consistent if either model (\ref{eq:ORmodel}) or model (\ref{eq:PSmodel}) is correctly specified.
Second, $\hat\mu_t(\hat m_{\text{RMLg}}, \hat \pi_{\text{RML}} )$ is shown to admit an $n^{-1/2}$ asymptotic expansion which leads
to valid Wald confidence intervals for $\mu_t$ if both models (\ref{eq:ORmodel}) and (\ref{eq:PSmodel}) are correctly specified.
See Remark \ref{rem:Farrell} for further discussion.

\section{Proposed method} \label{sec:method}

We develop regularized calibrated estimation (RCAL) to tackle ATE estimation  with multi-valued treatments in high-dimensional settings, where a large number of regression terms are allowed in
OR and PS models. Our approach exploits an interesting generalization of the augmented IPW estimator $\hat\mu_t(\hat m, \hat \pi )$ in (\ref{eq:ATE-aipw}),
and derives a novel set of regularized calibrated estimators  when fitting OR model (\ref{eq:ORmodel}) and PS model (\ref{eq:PSmodel}),
such that valid Wald confidence intervals can be obtained without requiring both models (\ref{eq:ORmodel}) and (\ref{eq:PSmodel}) are correctly specified.

\subsection{Regularized calibrated estimation} \label{sec:RCAL}

For technical convenience, consider the OR model (\ref{eq:ORmodel}) and PS model (\ref{eq:PSmodel}) with the same regressor vectors used (hence $q=p$).
Otherwise, models (\ref{eq:ORmodel}) and (\ref{eq:PSmodel}) can be enlarged by taking the union of the regressors.
Then the OR model can be stated as
\begin{align}
\E(Y | T=t, X) & = m(t, X;\alpha_t)= \psi\{\alpha_t^\T f(X)\}, \quad t\in\mathcal T, \label{eq:ORmodel2}
\end{align}
where $f(X)$ is the same vector of regressors as in model (\ref{eq:PSmodel}), including main effects and interactions of the covariate vector $X$.
This choice seems inconsequential, because different subsets of $f(X)$ with nonzero coefficients are allowed in models (\ref{eq:PSmodel}) and (\ref{eq:ORmodel2}).

As mentioned in Section~\ref{sec:setup}, ATE estimation can be viewed as two-stage semi-parametric estimation, where
OR and PS models are fitted to obtain $(\hat m, \hat\pi)$, and then the augmented IPW estimator $\hat\mu_t(\hat m, \hat \pi )$ is used to estimate $\mu_t$.
Our approach involves two main elements in the first-stage estimation.
First, we employ regularized estimation with sparsity-inducing penalties such as the Lasso or
group Lasso penalty (Tibshirani 1996; Yuan and Lin 2006), to deal with the large number of regressors under sparsity assumptions.
Second, we carefully choose the loss functions for regularized estimation,
such that the resulting estimator for $\mu_t$ admits an asymptotic expansion about $\mu_t$ in the usual order $O_p(n^{-1/2})$,
and hence valid Wald confidence intervals can be obtained for $\mu_t$ and ATEs, while allowing for model misspecification.
Moreover, the new loss function for PS estimation can be directly tied to evaluation of covariate balance.
Unless otherwise stated, we discuss estimation of $\mu_t$ for a fixed treatment $t$ and denote as $k\in\mathcal T$ a generic treatment.

\vspace{.1in}
\textbf{Calibration equations.}\;
We derive calibration equations for $\hat\alpha_t$ and $\hat\gamma$ in the augmented IPW estimator $\hat\mu_t(\hat m, \hat \pi )$ with possible
model misspecification, by similar reasoning as in Tan (2020b) and Ghosh and Tan (2020).
Suppose that $\hat\alpha_t$ converges in probability to a limit (or target) value $\bar\alpha_t$ and $\hat\gamma$ converges in probability to a limit (or target) value $\bar \gamma$ as $n\to\infty$,
such that\vspace{-.1in}
\begin{itemize}\addtolength{\itemsep}{-.1in}
\item $\bar\alpha_t$ coincides with the true value $\alpha_t^*$ in model (\ref{eq:ORmodel2}) with
$m^*(t,X) = m(t,X; \alpha_t^*)$ if model (\ref{eq:ORmodel2}) is correctly specified, but otherwise $m(t,X;\bar\alpha_t)$ may differ from $m^*(t,X)$.

\item $\bar\gamma$ coincides with the true value $\gamma^*$ in model (\ref{eq:PSmodel}) with
$\pi^*(k,X) = \pi(k,X; \gamma^*)$ if model (\ref{eq:PSmodel}) is correctly specified, but otherwise $\pi(k,X;\bar\gamma)$ may differ from $\pi^*(k,X)$ for $k\in\mathcal T$.
\end{itemize}\vspace{-.1in}
For convenience, we fix $\gamma_t \equiv 0$ in $\gamma$ in the following discussion.
Then a Taylor expansion of $\hat\mu_t(\hat m, \hat \pi )= \tilde \E \{ \varphi_t( Y, T, X; \hat \alpha_t, \hat \gamma) \}$ yields
\begin{align}
\hat\mu_t(\hat m, \hat \pi ) & = \hat \mu_t (\bar m, \bar\pi)
+ (\hat\alpha_t - \bar\alpha_t)^\T \times (\partial /\partial \alpha_t) \tilde \E\{ \varphi_t (Y,T,X;\alpha_t,\gamma)\} \big|_{(\alpha_t,\gamma)=(\bar\alpha_t,\bar\gamma)} \nonumber \\
& \quad + \sum_{k \not= t}(\hat\gamma_k - \bar\gamma_k)^\T \times (\partial /\partial \gamma_k) \tilde \E\{ \varphi_t (Y,T,X;\alpha_t,\gamma)\} \big|_{(\alpha_t,\gamma)=(\bar\alpha_t,\bar\gamma)} + o_p(n^{-1/2}), \label{eq:general-expan}
\end{align}
where $\bar m(t,X) = m(t,X;\bar\alpha_t)$, $\bar \pi(t,X) = \pi(t,X; \bar\gamma)$, and the remainder is taken to be $o_p(n^{-1/2})$ under suitable conditions.
For calibrated estimation, a basic idea is that if the derivatives with respect to $\alpha_t$ and $\gamma_k$, $k\not=t$, in (\ref{eq:general-expan}) have means 0,
referred to as calibration equation:
\begin{align}
& 0 = (\partial /\partial \alpha_t) \, \E\{ \varphi_t (Y,T,X;\alpha_t,\gamma)\} \big|_{(\alpha_t,\gamma)=(\bar\alpha_t,\bar\gamma)} , \label{eq:derv-alpha} \\
& 0 = (\partial /\partial \gamma_k) \, \E\{ \varphi_t (Y,T,X;\alpha_t,\gamma)\} \big|_{(\alpha_t,\gamma)=(\bar\alpha_t,\bar\gamma)} , \quad k \not= t, \label{eq:derv-gamma}
\end{align}
then it can be shown that the second and third terms on the right-hand side of (\ref{eq:general-expan}) reduces to $o_p(n^{-1/2})$ and hence $\hat\mu_t(\hat m, \hat \pi )$
admits the asymptotic expansion
\begin{align}
\hat\mu_t(\hat m, \hat \pi ) & = \hat \mu_t (\bar m, \bar \pi) + o_p(n^{-1/2}), \label{eq:desired-expan}
\end{align}
under certain sparsity conditions.
The expansion (\ref{eq:desired-expan}) appears similar to the expansion
\begin{align}
\hat\mu_t(\hat m, \hat \pi )  = \hat \mu_t ( m^*, \pi^*) + o_p(n^{-1/2}) , \label{eq:existing-expan}
\end{align}
which is satisfied for the regularized likelihood estimators $(\hat m, \hat\pi)= (\hat m_{\text{RMLg}}, \hat \pi_{\text{RML}} )$ under suitable conditions
if both models (\ref{eq:PSmodel}) and (\ref{eq:ORmodel2}) are correctly specified (Farrell 2015).
However, an important difference between (\ref{eq:desired-expan}) and (\ref{eq:existing-expan})
is that the expansion (\ref{eq:desired-expan}) is expected to hold even when models (\ref{eq:PSmodel}) and (\ref{eq:ORmodel2}) may be misspecified,
provided that the estimators $\hat\alpha_t$ and $\hat\gamma_k$, $k\not=t$, are constructed with the limit values satisfying calibration equations (\ref{eq:derv-alpha})--(\ref{eq:derv-gamma}).
If model (\ref{eq:PSmodel}) or (\ref{eq:ORmodel2}) is correctly specified, then
$\bar \pi(t,X) = \pi^*(t,X)$ or $\bar m(t,X) = m^*(t,X)$ respectively and $\hat \mu_t (\bar m, \bar \pi)$ has mean equal to $\mu_t$
by the double robustness of augmented IPW estimation.
In either case, the expansion (\ref{eq:desired-expan}) implies that the estimator
$\hat\mu_t(\hat m, \hat \pi )$ admits $\varphi_t (Y,T,X; \bar \alpha_t, \bar\gamma) - \mu_t$ as the influence function,
and hence valid Wald confidence intervals for $\mu_t$ can be obtained in the usual manner.

\vspace{.1in}
\textbf{Sequential calibration estimation.}\;
While the preceding discussion outlines basic reasoning for our approach, there are nontrivial complications which we need to address.
For simplicity, assume that OR model (\ref{eq:ORmodel2}) is linear with $m(t,X;\alpha_t) = \alpha_t^\T f(X)$. Then
calibration equations (\ref{eq:derv-alpha}) and (\ref{eq:derv-gamma}), with $\gamma_t\equiv 0$, can be directly shown to yield
\begin{align}
&  0 = \E \big[ \{ 1- R^{(t)} / \pi(t,X; \bar\gamma) \} f(X) \big] ,  \label{eq:EE-gamma}  \\
&  0 = \E \big[ R^{(t)} \{ \pi(k,X; \bar\gamma) / \pi(t,X; \bar\gamma)\} \{ Y - m(t,X; \bar \alpha_t) \} f(X) \big]  , \quad k \not= t.  \label{eq:EE-alpha}
\end{align}
There are $(K-1)p$ free coefficients in $\gamma$, but only $p$ equations in (\ref{eq:EE-gamma}),
whereas there are $p$ free coefficients in $\alpha_t$, but $(K-1)p$ equations in (\ref{eq:EE-alpha}).
For multi-valued treatments ($K\ge 3$),
equations (\ref{eq:EE-gamma})--(\ref{eq:EE-alpha}) are under-identifying in $\gamma_k$, $k\not=t$, but over-identifying in $\alpha_t$.
In addition, solving nonlinear equations such as sample versions of (\ref{eq:EE-gamma})--(\ref{eq:EE-alpha}), even if theoretically just-identifying,
may suffer the issue of no solution or multiple solutions (Small et al. 2000).

To tackle the foregoing issues, we introduce $\hat\alpha_t^{(k)}$, $k\not= t$, as separate versions of $\hat\alpha_t$ and
generalize the augmented IPW estimator $\hat\mu_t(\hat m, \hat \pi )$ in (\ref{eq:ATE-aipw}) as
\begin{align}
\hat\mu_t(\hat m^\#, \hat \pi ) &= \tilde \E \left\{ \varphi_t( Y, T, X; \hat\alpha_t^\#, \hat\gamma) \right\} , \label{eq:ATE-aipw2}
\end{align}
where $\hat m^\# (t,X) = \{ m(t,X;\hat\alpha_t^{(k)}): k \not= t\}$, $\hat\alpha_t^\# = (\hat\alpha_t^{(k)}: k \not=t)$, and
\begin{align}
\varphi_t( Y, T, X; \hat\alpha_t^\#, \hat\gamma)  & =  R^{(t)} Y +  \sum_{k\not=t} \varphi_t^{(k)} (Y,T,X; \hat\alpha_t^{(k)},\hat\gamma)  ,  \label{eq:phi2} \\
\varphi_t^{(k)} (Y,T,X; \alpha_t^{(k)},\gamma) & = R^{(t)} Y  \frac{ \pi(k,X; \gamma) }{ \pi(t,X; \gamma) } -
\left\{ R^{(t)} \frac{ \pi(k,X; \gamma) }{ \pi(t,X; \gamma) } - R^{(k)} \right\} m(t,X; \alpha_t^{(k)}) . \label{eq:phi3}
\end{align}
Because different estimators $\hat\alpha_t^{(k)}$, $ k\not= t$, are allowed, $\hat\mu_t(\hat m^\#, \hat \pi )$ in (\ref{eq:ATE-aipw2}) depends on
$\pi(k,X; \hat\gamma)$ for all $k\in\mathcal T$, not just $\pi(t,X; \hat\gamma)$ as in (\ref{eq:ATE-aipw}).
If $\hat\alpha_t^{(k)}$, $ k\not= t$, are all defined to be the same as $\hat \alpha_t$, then (\ref{eq:ATE-aipw2}) reduces to (\ref{eq:ATE-aipw}).
Interestingly, $\tilde \E ( R^{(t)} Y )$ is directly an unbiased estimator of $\E ( R^{(t)} Y ) = E(Y|T=t) \P(T=t)$, and
$ \tilde \E \{\varphi_t^{(k)} (Y,T,X; \hat\alpha_t^{(k)},\hat\gamma) \} $ can be identified as an augmented IPW estimator
of the expectation $\E (R^{(k)}  Y^{(t)} ) = \E (Y^{(t)} | T=k) \P (T=k)$ for $k \not= t$.
Hence (\ref{eq:ATE-aipw2}) corresponds to a natural decomposition of the mean $\mu_t =\E (Y^{(t)} )$ in ATEs into the means of potential outcomes involved in ATTs,
as stated in equation (\ref{eq:decomp}). See Section~\ref{sec:ATT} for further discussion.

Next, we apply similar reasoning as (\ref{eq:general-expan})--(\ref{eq:derv-gamma}) to the estimator $\hat\mu_t(\hat m^\#, \hat \pi )$.
With $\gamma_t \equiv 0$, a Taylor expansion of $\hat\mu_t(\hat m^\#, \hat \pi )$ yields
\begin{align}
\hat\mu_t(\hat m^\#, \hat \pi ) & = \hat \mu_t (\bar m^\#, \bar \pi)
+ \sum_{k\neq t}(\hat\alpha_t^{(k)} - \bar\alpha_t^{(k)})^\T \times (\partial /\partial \alpha_t^{(k)}) \tilde \E\{ \varphi_t^{(k)} (Y,T,X;\alpha_t^{(k)},\gamma)\} \big|_{(\alpha_t^{(k)},\gamma)=(\bar\alpha_t^{(k)},\bar\gamma)}  \nonumber \\
& \quad + \sum_{k\neq t}(\hat\gamma_k - \bar\gamma_k)^\T \times (\partial /\partial \gamma_k) \tilde \E\{ \varphi_t^{(k)} (Y,T,X;\alpha_t^{(k)},\gamma)\} \big|_{(\alpha_t^{(k)},\gamma)=(\bar\alpha_t^{(k)},\bar\gamma)} + o_p(n^{-1/2})  . \label{eq:general-expan2}
\end{align}
where $\bar m^\#(t,X) = \{ m(t,X; \bar\alpha_t^{(k)}): k\not=t\}$ and $\bar \alpha_t^{(k)}$ is the limit value of  $\hat\alpha_t^{(k)}$, $k \not= t$.
Setting the expectations of the derivative terms above to 0 leads to the calibration equations:
\begin{align}
&  0 = \E \big[ \{ R^{(k)} - R^{(t)} \me^{(\bar \gamma_k - \bar \gamma_t)^\T f(X)} \} f(X) \big] ,  \quad k \not=t, \label{eq:EE2-gamma}  \\
&  0 = \E \big[ R^{(t)} \me^{(\bar \gamma_k - \bar \gamma_t)^\T f(X) } \{ Y - m(t,X; \bar \alpha_t^{(k)}) \} f(X) \big] , \quad k \not= t,  \label{eq:EE2-alpha}
\end{align}
For symmetry, $\bar \gamma_k$ is replaced by $\bar\gamma_k -\bar\gamma_t$ for $k\in\mathcal T$ including $k=t$ to remove the constraint $\gamma_t \equiv 0$.
Remarkably, equations (\ref{eq:EE2-gamma}) are just-identifying in $\{\gamma_k-\gamma_t: k\not=t\}$,
and, given $\bar\gamma$, equations (\ref{eq:EE2-alpha}) are just-identifying in $\{ \alpha_t^{(k)}: k\not=t\}$.
In fact, (\ref{eq:EE2-gamma}) can be shown to be the stationary condition for minimizing the expected loss
$\E \{ \ell_{\text{CAL}}(\gamma) \}$, where
\begin{align}
\ell_{\text{CAL}}(\gamma) =  \tilde \E \Big[ \sum_{k\not=t} \left\{  R^{(t)} \me^{(\gamma_k - \gamma_t)^\T f(X)} - R^{(k)} (\gamma_k - \gamma_t)^\T f(X) \right\} \Big], \label{eq:PSloss}
\end{align}
is a convex loss in $\gamma$, different from the likelihood loss $\ell_{\text{ML}}(\gamma)$.
Moreover, (\ref{eq:EE2-alpha}) can be shown to be the stationary condition for minimizing the expected loss
$\E \{ \ell_{\text{WL}}(\alpha_t^{(k)} ; \bar \gamma) \}$, where
\begin{align}
\ell_{\text{WL}}(\alpha_t^{(k)}; \gamma) =  \tilde \E  \big[ R^{(t)} \me^{(\gamma_k - \gamma_t)^\T f(X)}  \{ Y - m(t,X; \alpha_t^{(k)}) \}^2 \big] /2, \quad k \not= t, \label{eq:ORloss}
\end{align}
is a weighted least squares loss for regression of $Y$ on $X$ in treatment group $t$,
with the weight $ \me^{(\gamma_k - \gamma_t)^\T f(X)} $ depending on $\gamma$.
Different $\ell_{\text{WL}}(\alpha_t^{(k)}; \gamma)$ for $k\not=t$ is  associated with different weights.
The losses $\ell_{\text{CAL}}(\gamma)$ and $\ell_{\text{WL}}(\alpha_t^{(k)}; \gamma)$ are called the calibration (CAL) loss in $\gamma$ and $\alpha_t^{(k)}$ respectively, and
$\ell_{\text{WL}}(\alpha_t^{(k)}; \gamma)$ is also called the weighted least squares or likelihood (WL) loss.

For a nonlinear OR model (\ref{eq:ORmodel2}),
calibration equations obtained from the derivative terms with respect to $\gamma$ in (\ref{eq:general-expan2}) are the same as (\ref{eq:EE2-alpha}),
whereas those from the derivative terms with respect to $\alpha_t^{(k)}$ are of a more complicated form than (\ref{eq:EE2-gamma}):
\begin{align}
&  0 = \E \big[ \{ R^{(k)} - R^{(t)} \me^{(\bar \gamma_k - \bar \gamma_t)^\T f(X)} \} \psi_2(\bar \alpha_t^{(k)}f(X)) f(X) \big] ,  \quad k \not=t, \label{eq:EE3-gamma}
\end{align}
where $\psi_2(\cdot)$ denotes the derivative of $\psi(\cdot)$.
Compared with (\ref{eq:EE2-gamma}) for a linear OR model, equations (\ref{eq:EE3-gamma}) involve both $\gamma$ and $\alpha_t^{(k)}$, $k\not= t$,
and hence the sample versions of (\ref{eq:EE2-alpha}) and (\ref{eq:EE3-gamma}) cannot be solved sequentially or inverted to define loss functions sequentially as above.
To circumvent this complication, we employ sequential calibration as follows: retain the loss function $\ell_{\text{CAL}}(\gamma) $ in $\gamma$
and then invert the calibration equation (\ref{eq:EE2-alpha}) to obtain the loss function in $\alpha_t^{(k)}$:
\begin{align}
\ell_{\text{WL}}(\alpha_t^{(k)}; {\gamma}) = \tilde{\E}\left(R^{(t)}\me^{(\gamma_k -\gamma_t)^\T f(X) } [-Y f^\T(X)\alpha_t^{(k)} + \Psi \{f^\T (X)\alpha_t^{(k)} \} ]\right),
\quad k\not=t, \label{eq:GORloss}
\end{align}
where  $\Psi (u) = \int_0^u \psi(u^\prime) \,\dif u^\prime$.
The loss (\ref{eq:GORloss}) is a weighted version of the likelihood loss $\ell_{\text{ML}}(\alpha_t)$ in Section~\ref{sec:setup}.
The weighted least squares loss (\ref{eq:ORloss}) is recovered in the special case of a linear OR model, $\psi(u) = u$ and $\Psi(u) = u^2/2$.
This approach has two main advantages. First, as discussed below, the loss functions can be used sequentially for regularized estimation in a computationally convenient manner.
Second, the calibration loss $\ell_{\text{CAL}}(\gamma)$ can be desirable for PS estimation, independently of outcome regression,
in terms of both the informative form of calibration equations (\ref{eq:EE2-gamma}) and a strong relationship between  minimization of the expected calibration loss and
reduction of relative errors in propensity scores, which are discussed in Remark \ref{rem:balance} and Appendix A.

\begin{rem}[PS calibration equations] \label{rem:balance}
Although derived for achieving desirable asymptotic expansions,
the calibration equations for PS estimation are directly related to covariate balance after inverse probability weighting.
In fact, the calibration equations (\ref{eq:EE2-gamma})  can be rewritten as
\begin{align}
&  0 = \E \left[ \left\{ R^{(k)} - R^{(t)} \frac{\pi(k,X; \bar\gamma)}{ \pi(t,X; \bar\gamma) } \right\}f(X) \right] ,  \quad k \not=t, \label{eq:EE2-gamma-b}
\end{align}
which indicates that the weighted mean of $f(X)$ in the $t$th treated group is matched with the simple mean of $f(X)$ in the $k$th treated group.
The weight used in (\ref{eq:EE2-gamma-b}) is the probability ratio $\pi(k,X; \bar\gamma)/\pi(t,X; \bar\gamma)$ related to ATT estimation.
Summing the two sides of (\ref{eq:EE2-gamma-b}) over $k \not= t$ yields
$0= \E [\{ 1-R^{(t)} / \pi(t,X; \bar\gamma) \}f(X)  ]$, that is,
calibration equations (\ref{eq:EE-gamma}) derived from the usual augmented IPW estimator $\hat\mu_t(\hat m,\hat\pi)$.
Equation (\ref{eq:EE-gamma}) indicates that the weighted mean of $f(X)$  in the $t$th treated group is matched with the mean of $f(X)$ in the population,
with the weight being the inverse probability $1/\pi(t,X; \bar\gamma)$, typically found in ATE estimation.
However, equations (\ref{eq:EE2-gamma-b}) over all choices $k\not=t$ are just-identifying in $\gamma$
whereas equations (\ref{eq:EE-gamma}) are not, unless in the case of binary treatments where the equations are equivalent (Tan 2020a, Section 7.4).
\end{rem}

\textbf{Lasso regularized estimation.}\;
In high-dimensional settings, we combine the calibration losses in (\ref{eq:PSloss}) and (\ref{eq:ORloss}) or (\ref{eq:GORloss}) with Lasso-type penalties
to define regularized calibrated estimators $\hat\gamma$ and $\hat\alpha_t^{(k)}$, $k\not=t$, which are then substituted into the estimator (\ref{eq:ATE-aipw2}) for $\mu_t$.
To exploit group sparsity in coefficients associated with different covariate terms, we incorporate group Lasso penalties with the calibration losses.
First, define the estimator $\hat{\gamma}_{\text{RCAL}}$ as a minimizer of
\begin{align}
\ell_{\text{RCAL}}(\gamma) = \ell_{\text{CAL}}(\gamma) + \lambda_1\sum_{j=1}^p \lVert \gamma_{j\cdot} \rVert_2, \label{eq:PSobj}
\end{align}
where $\gamma_{j\cdot} = (\gamma_{jk}: k\in \mathcal T)^\T$ is the transpose of the row vector in $\gamma$ associated with $f_j(X)$, and $\lambda_1 \ge 0$ is a tuning parameter.
Similarly to $\hat\gamma_{\text{RML}}$ discussed earlier, either a one-to-zero constraint or the sum-to-zero constraint can be used for identification.
For simplicity, we fix the one-to-zero constraint $\gamma_t\equiv 0$
and use the notation $\gamma = (\gamma_k: k\in\mathcal T\backslash\{t\})$ as a $(p+1)\times (K-1)$ matrix and $\gamma_{j\cdot} = (\gamma_{jk}: k\in\mathcal T\backslash\{t\})^\T$ as a $(K-1)\times 1$ vector, whenever needed from the context.

Instead of the generic constraint $\gamma_0 \equiv 0$, our choice $\gamma_t\equiv 0$ is aligned with the fact that
the calibration loss $\ell_{\text{CAL}}(\gamma)$ in (\ref{eq:PSloss}) depends on the treatment $t$ for which $\mu_t$ is estimated,
whereas the likelihood loss $\ell_{\text{ML}}(\gamma)$ is invariant regardless of which treatment $t$ is considered for estimation of $\mu_t$
(see Remark \ref{rem:separate-PS}).
By the constraint $\gamma_t\equiv 0$, the loss function $\ell_{\text{CAL}}(\gamma)$
becomes separable in $(\gamma_k: k\in\mathcal T\backslash\{t\})$,
which then leads to both a simple quadratic approximation (\ref{eq:taylor-gamma}) to $\ell_{\text{CAL}}(\gamma)$ for numerical implementation
and a simple compatibility condition  in Assumption \ref{ass1}(iii)  for theoretical analysis.
We defer to Supplement Section \ref{sec:sum-to-zero} the material about use of the sum-to-zero constraint and comparison with the one-to-zero constraint.


Next, we form a combined loss $\ell_{\text{WL}}(\alpha^\#_t; \hat{\gamma}_{\text{RCAL}}) = \sum_{k\neq t}\ell_{\text{WL}}(\alpha_t^{(k)}; \hat{\gamma}_{\text{RCAL}})$,
and define the estimator
$\hat{\alpha}_{t,\text{RWL}}^\# = (\hat\alpha_{t,\text{RWL}}^{(k)}: k \not= t)$ as a minimizer of
\begin{align}
\ell_{\text{RWL}}(\alpha_t^\#; \hat \gamma_{\text{RCAL}}) = \ell_{\text{WL}}(\alpha^\#_t; \hat{\gamma}_{\text{RCAL}}) + \lambda_2\sum_{j=1}^p \lVert \alpha_{jt}^\# \rVert_2, \label{eq:ORobj}
\end{align}
where $\alpha_t^\# = (\alpha_t^{(k)}: k \not=t)$ is a $(p+1)\times (K-1)$ matrix, $\alpha_{jt}^\# = (\alpha_{jt}^{(k)}: k\neq t)^\T$ is the transpose of the row vector in $\alpha_t^\#$
associated with $f_j(X)$,
and $\lambda_2 \ge 0$ is a tuning parameter.
See Appendix B for implications from the Karush--Kuhn--Tucker conditions for minimization of (\ref{eq:PSobj}) and (\ref{eq:ORobj}).

\vspace{.1in}
\textbf{Wald confidence intervals.}\;
From regularized calibrated estimation, the resulting estimator (\ref{eq:ATE-aipw2}) for $\mu_t$ is $\hat \mu_t ( \hat m^\#_{\text{RWL}}, \hat \pi_{\text{RCAL}})$, where
$\hat m^\#_{\text{RWL}} (t,X) = \{\hat m(t,X;\hat\alpha^{(k)}_{t,\text{RWL}} ): k \not= t\}$ and $\hat\pi_{\text{RCAL}}(k,X) = \pi(k,X; \hat\gamma_{\text{RCAL}})$, $k\in\mathcal T$.
In Section~\ref{sec:theory-aipw},  we show that if PS model (\ref{eq:PSmodel}) is correctly specified but OR model (\ref{eq:ORmodel2}) may be misspecified,
then $\hat \mu_t ( \hat m^\#_{\text{RWL}}, \hat \pi_{\text{RCAL}})$ admits an asymptotic expansion similar to (\ref{eq:desired-expan}) under suitable sparsity conditions:
\begin{align}
\hat \mu_t ( \hat m^\#_{\text{RWL}}, \hat \pi_{\text{RCAL}})  = \hat \mu_t ( \bar m^\#_{\text{WL}}, \bar \pi_{\text{CAL}}) + o_p(n^{-1/2}), \label{eq:desired-expan2}
\end{align}
where $\bar m^\#_{\text{WL}} (t,X) = \{ m(t,X; \bar\alpha_{t,\text{WL}}^{(k)}): k\not=t\}$, $\bar \pi_{\text{CAL}}(k,X) = \pi(k,X; \bar \gamma_{\text{CAL}})$, $k\in\mathcal T$,
and $\bar \alpha_{t,\text{WL}}^{(k)}$ and $ \bar \gamma_{\text{CAL}}$ are the limit values of  $\hat \alpha_{t,\text{RWL}}^{(k)}$, $k\not=t$, and $\hat \gamma_{\text{RCAL}}$ respectively.
Then an asymptotic $(1-c)$ confidence interval for $\mu_t$ can be obtained as
$ \hat \mu_t ( \hat m^\#_{\text{RWL}}, \hat \pi_{\text{RCAL}})  \pm z_{c/2}\sqrt{\hat V_t / n}$,
where  $z_{c/2}$ is the $(1 - c/2)$ quantile of $\N(0, 1)$ and, with $\varphi_t (Y, T, X; \alpha_t^\#, \gamma)$ defined in (\ref{eq:phi2}),
\begin{align}
\hat V_t = \tilde{\E} \left[\{ \varphi_t (Y, T, X; \hat{\alpha}_{t,\text{RWL}}^\#, \hat{\gamma}_{\text{RCAL}}) -  \hat \mu_t ( \hat m^\#_{\text{RWL}}, \hat \pi_{\text{RCAL}}) \}^2\right] .\label{eq:estimatedV}
\end{align}
For a linear OR model, the asymptotic expansion (\ref{eq:desired-expan2}) can be established, with possible misspecification of
both models (\ref{eq:ORmodel2}) and (\ref{eq:PSmodel}). In this case, the confidence intervals for $\mu_t$ are doubly robust,
being also valid when model (\ref{eq:ORmodel2}) is correctly specified, but model (\ref{eq:PSmodel}) may be misspecified.

For a nonlinear OR model, our approach does not in theory lead to doubly robust confidence intervals for $\mu_t$,
because calibration equations (\ref{eq:EE2-alpha}) and (\ref{eq:EE3-gamma}) are not fully taken into account as discussed above.
Alternatively, doubly robust confidence intervals can be investigated,
while exploiting the calibration equations (\ref{eq:EE2-alpha}) and (\ref{eq:EE3-gamma}) based on  $\hat\mu_t(\hat m^\#, \hat \pi)$ proposed here.
Such methods tend to involve more complex theory and implementation, for example, sample splitting and cross fitting in Smucler et al. (2019) and
iterations of regularized calibrated estimation in Ghosh and Tan (2020).
Hence our approach is expected to remain useful in applications.

\begin{rem}[Separate PS estimation] \label{rem:separate-PS}
A subtle feature of our method is that for fitting the PS model (\ref{eq:PSmodel}), the loss $\ell_{\text{CAL}}(\gamma)$ and the estimator $\hat\gamma_{\text{RCAL}}$ depend on the treatment $t$
for which the mean $\mu_t$ is estimated. Such dependency on $t$ is suppressed in the notation for simplicity, but needs to be noticed throughout.
Hence when estimating different means $\mu_t$ for $t \in \mathcal T$, separate estimators of $\gamma$ are required in our method, whereas a single estimator of $\gamma$ is used in existing methods
as described in Section~\ref{sec:setup}.
This scheme of separate estimation of propensity scores is inherent to our method, and may be advantageous in allowing treatment-specific approximations in the presence of model misspecification.
See Tan (2020b), Section 3.5, for related discussion.
\end{rem}

\begin{rem}[Separate OR estimation] \label{rem:separate-OR}
There are also important differences between our method and existing methods in handling outcome regression.
Farrell (2015) employed the penalized objective function (\ref{eq:RML-alpha-obj2}), by combining individual loss functions in
separate treatment groups and adding a group Lasso penalty,
which encourages a small subset of important regressors, $f_j(X)$ with nonzero $\alpha_{j\cdot}$, for outcome regression over different treatment groups.
In contrast, our penalized objection function (\ref{eq:ORobj}) combines individual loss functions in the fixed treatment group $t$ but depending on
separate versions of  $\alpha_t$ through different weights in outcome regression. A group Lasso penalty is then introduced to encourage
a small subset of important regressors, $f_j(X)$ with nonzero $\alpha_{jt}^\#$, for outcome regression with different weights.
Hence for estimating the mean $\mu_t$, our method exploits separate versions of weighted outcome regression within treatment group $t$,
instead of combining outcome regression across all treatment groups as in Farrell (2015).

\end{rem}

\begin{rem}[Binary treatments] \label{rem:Tan}
We discuss how the proposed method generalizes that in Tan (2020ab) with binary treatments.
Take $\mathcal T=\{0,1\}$ and fix $t=1$.
First, the calibration loss $\ell_{\text{CAL}}(\gamma)$ in (\ref{eq:PSloss}) becomes
$\ell_{\text{CAL}}(\gamma) = \tilde{\E} [R^{(1)} \me^{{(\gamma_0 - \gamma_1)}^\T f(X)} - R^{(0)}(\gamma_0 - \gamma_1)^\T f(X)]$. 
By minimization of the penalized loss (\ref{eq:PSobj}) with the constraint $\gamma_t = \gamma_1 \equiv 0$,
the estimator $\hat\gamma_{\text{RCAL}} = (\hat{\gamma}_{0, \text{RCAL}} , 0)$ is defined
such that
$\hat{\gamma}_{0, \text{RCAL}}
 = \argmin_{\gamma_0} \{ \ell_{\text{CAL}}(\gamma_0,0) + \lambda_1 \lVert \gamma_{1:p,0} \rVert_1 \}$
or equivalently
$-\hat{\gamma}_{0, \text{RCAL}}= \argmin_{\gamma_1} \{  \ell_{\text{CAL}}(0,\gamma_1) + \lambda_1 \lVert \gamma_{1:p,1} \rVert_1 \}$,
where $\gamma_{1:p,k}$ is $\gamma_k$ excluding the intercept $\gamma_{0k}$ for $k=0,1$.
The loss function  in $\gamma_1$ only,
\begin{align*}
\ell_{\text{CAL}}(0,\gamma_1) = \tilde{\E}\left[R^{(1)} \me^{- \gamma_1^\T f(X)} + R^{(0)} \gamma_1^\T f(X)\right]
\end{align*}
is precisely  the calibration loss for fitting PS models with binary treatments in Tan (2020ab).
Hence the same fitted propensity score $\pi(1, X; \hat\gamma_{\text{RCAL}}) = \{ 1+ \exp( \hat\gamma_{0,\text{RCAL}}^\T f(X) )\}^{-1} $
can be obtained from Tan (2020ab) with the same tuning parameter $\lambda_1$.
Second, the loss $\ell_{\text{WL}}(\alpha_1^{(0)}; {\gamma})$ in (\ref{eq:GORloss})
can be easily shown to coincide with the weighted likelihood loss for fitting OR models in Tan (2020b),
where $\alpha_1^{(0)}$ is the only version of $\alpha_1$ and $\me^{(\gamma_0-\gamma_1)^\T f(X)} = \{1-\pi(1,X;\gamma)\}/\pi(1,X;\gamma)$.
The group Lasso penalty in (\ref{eq:ORobj}) also reduces to the Lasso penalty.
Hence the same fitted function $m (1,X; \hat\alpha_{1,\text{RWL}}^{(0)})$ can be obtained from Tan (2020b) using the same tuning parameter $\lambda_2$.
Finally, the generalized  estimator $\hat\mu_t(\hat m^\#, \hat \pi )$ in (\ref{eq:ATE-aipw2}) can be directly shown to coincide with the original version
$\hat\mu_t(\hat m, \hat \pi )$ in (\ref{eq:ATE-aipw}) with binary treatments. Therefore,
our estimator $\hat \mu_t ( \hat m^\#_{\text{RWL}}, \hat \pi_{\text{RCAL}})$ for $\mu_t$ recovers that in Tan (2020b), using the same fitted PS and OR functions.
\end{rem}

\subsection{Computation} \label{sec:computation}

We propose Fisher scoring block coordinate descent algorithms for computing the estimators $\hat{\gamma}_{\text{RCAL}}$ and $\hat{\alpha}^\#_{t, \text{RWL}}$, that is,
minimizing the objective functions \eqref{eq:PSobj} for a fixed $\lambda_1$ and \eqref{eq:ORobj} for a fixed $\lambda_2$.
Compared with existing algorithms for group-penalized multi-response and multinomial regression (Simon et al.~2013), our algorithms for
both $\hat{\gamma}_{\text{RCAL}}$ and $\hat{\alpha}^\#_{t, \text{RWL}}$ are derived by innovatively incorporating
Fisher scoring before forming a majorizing quadratic approximation to the loss function used and
solving a multi-response group-Lasso least-square problem in each iteration.
Previously, Fisher scoring is used in
the iterative reweighted least squares algorithm for fitting generalized linear models with noncanonical links, such as probit regression (McCullagh and Nelder 1989).

\vspace{.1in}
\textbf{Algorithm for computing $\hat{\gamma}_{\text{\normalfont RCAL}}$.}\;
We fix $\gamma_t \equiv 0$ and
use the notation $\gamma = (\gamma_k: k\in\mathcal T\backslash\{t\})$ and $\gamma_{j\cdot} = (\gamma_{jk}: k\in\mathcal T\backslash\{t\})^\T$
as mentioned below (\ref{eq:PSobj}).
A second-order Taylor expansion of $\ell_{\text{CAL}}(\gamma)$ at the current estimate, denoted as $\tilde\gamma$, gives the quadratic approximation
\begin{align}
\ell_{\text{CAL,Q1}}(\gamma) & = \ell_{\text{CAL}}(\tilde{\gamma}) +  \tilde{\E} \left\{ g^\T(X; \tilde{\gamma}) (\gamma - \tilde{\gamma})^\T f(X)\right\} \nonumber \\
& \quad + \frac{1}{2} \tilde{\E}\left\{f^\T(X)(\gamma - \tilde{\gamma})\frac{R^{(t)}}{\pi(t, X; \tilde{\gamma})} H(X; \tilde{\gamma})(\gamma - \tilde{\gamma})^\T f(X)\right\}, \label{eq:taylor-gamma}
\end{align}
where
$g(X; \gamma) = (g_k: k\in\mathcal T\backslash\{t\})^\T$ with $g_k = R^{(t)}\pi(k, X; \gamma) / \pi(t, X; \gamma) - R^{(k)}$ for $k\neq t$, and $H(X; \gamma) = \diag(\pi(k, X; \gamma): k\in\mathcal T\backslash\{t\})$.
Incidentally, if the sum-to-zero constraint is used on $\gamma$, then a quadratic approximation
to $\ell_{\text{CAL}}(\gamma)$ can also be obtained in the form (\ref{eq:taylor-gamma}),
but with $H(X; \tilde\gamma)$ a $K\times K$ non-diagonal matrix.
See Supplement Section \ref{sec:algorithm} for a discussion about computation of $\hat{\gamma}_{\text{RCAL}}$ with the sum-to-zero constraint,
where a diagonal matrix dominating $H(X; \tilde\gamma)$ is used.

The quadratic function $\ell_{\text{CAL,Q1}}(\gamma)$ cannot be directly related to a weighted least-square loss as in Simon et al.~(2013):
the quadratic term in $\ell_{\text{CAL,Q1}}(\gamma) $ depends on only the observations in the $t$th treated group,
but the linear term depends on all the observations from the $K$ treatment groups.
By Fisher scoring, we replace $R^{(t)} / \pi(t, X; \tilde{\gamma})$ by its model expectation, which equals 1, and obtain
the quadratic approximation
\begin{align*}
\ell_{\text{CAL,Q2}}(\gamma) &= \ell_{\text{CAL}}(\tilde{\gamma})  +  \tilde{\E} \left\{  g^\T(X; \tilde{\gamma}) (\gamma - \tilde{\gamma})^\T f(X)\right\}
 + \frac{1}{2} \tilde{\E}\left\{f^\T(X)(\gamma - \tilde{\gamma}) H(X; \tilde{\gamma})(\gamma - \tilde{\gamma})^\T f(X)\right\} ,
\end{align*}
where both the quadratic and linear terms depend on all observations in the sample.
Then $\ell_{\text{CAL,Q2}}(\gamma)$ can be viewed as a weighted least-square loss for multi-response linear regression.

To facilitate block coordinate descent with closed-form updates, we employ the majorization-minimization (MM) technique (Wu and Lange 2010), similarly as in related algorithms (Simon et al. 2013).
For the diagonal matrix $H(X;\tilde\gamma)$, we use the simple majorization
 $H(X_i; \tilde{\gamma}) \preceq b_1 I$, where $b_1 = \max_{i=1}^n\max_{k\neq t}\pi(k, X; \tilde{\gamma})$ and $I$ is the $(K-1)\times (K-1)$ identity matrix.
By the quadratic lower bound principle (Bohning and Lindsay 1988), a majorizing function of $\ell_{\text{CAL,Q2}}(\gamma)$ is
\begin{align*}
\ell_{\text{CAL,Q3}}(\gamma) =\ell_{\text{CAL}}(\tilde{\gamma})  +  \tilde{\E} \left\{  g^\T(X; \tilde{\gamma}) (\gamma - \tilde{\gamma})^\T f(X)\right\}
+ \frac{b_1}{2} \tilde{\E} \left\{ \lVert  (\gamma - \tilde{\gamma})^\T f(X)   \rVert_2^2 \right\} .
\end{align*}
Combining $\ell_{\text{CAL,Q3}}(\gamma)$ with the group Lasso penalty as in (\ref{eq:PSobj}) leads to
\begin{align}
\frac{1}{2} \tilde{\E} \left\{ \lVert \tilde{\gamma}^\T f(X) -  g(X; \tilde{\gamma}) / b_1 -  \gamma^\T f(X) \rVert_2^2 \right\}
 +  \frac{\lambda_1}{b_1 } \sum_{j=1}^p \lVert \gamma_{j.} \rVert_2. \label{eq:PSq}
\end{align}
Minimization of (\ref{eq:PSq}) corresponds to a group-penalized least-square estimation
for multi-response linear regression with the same design matrix for each response.
Then the block coordinate update has the closed form
\begin{align}
\tilde {\gamma}_{j.} \leftarrow \frac{1}{\tilde{\E}\{f^2_j(X)\}}\left\{1 - \frac{\lambda_1/b_1}{\lVert \tilde{\E}\{Z_{(j)}f_j(X)\} \rVert_2}\right\}_{+}\tilde{\E}\{Z_{(j)}f_j(X)\},
\quad j = 1, \ldots, p, \label{eq:PSupdate}
\end{align}
where $Z_{(j)} = \tilde{\gamma}^\T f(X) - g(X; \tilde{\gamma}) / b_1 - \sum_{j^\prime \neq j}f_{j^\prime}(X)\tilde {\gamma}_{j^\prime \cdot}$ is the partial residual.

A complication from Fisher scoring, i.e., replacement of $\ell_{\text{CAL,Q1}}(\gamma)$ by $\ell_{\text{CAL,Q2}}(\gamma)$
is that the quadratic function $\ell_{\text{CAL,Q3}}(\gamma)$, even though a majoring function of $\ell_{\text{CAL,Q2}}(\gamma)$,
may not be a majorizing function of $\ell_{\text{CAL}}(\gamma)$. Hence minimization of (\ref{eq:PSq}) may not guarantee a decrease in the objective function \eqref{eq:PSobj},
as otherwise would be achieved by the MM technique.
To restore the descent property, we incorporate a backtracking line search  similarly as in Tan (2020a).

From the preceding discussion, we obtain the algorithm for computing $\hat{\gamma}_{\text{\normalfont RCAL}}$.

\vspace{0.1in}
\hrule
\vspace{0.1in}
{\it Algorithm} 1. Fisher scoring block descent algorithm for minimizing \eqref{eq:PSobj}.
\vspace{0.05in}
\hrule \vspace{-.09in}
\begin{itemize}\addtolength{\itemsep}{-.1in}
\item[1.] Set an initial value $\gamma^{(0)}$.
\item[2.] Repeat the following updates for $s = 1, 2, ...$ until convergence to obtain $\hat{\gamma}_{\text{RCAL}}$:\vspace{-.1in}
	\begin{itemize}\addtolength{\itemsep}{-.1in}
		\item[(a)] Compute $\gamma^{(s - 1/2)}$ as a minimizer of \eqref{eq:PSq} using block coordinate descent with update formula \eqref{eq:PSupdate}.
		\item[(b)] If $\ell_{\text{RCAL}}(\gamma^{(s - 1/2)}) < \ell_{\text{RCAL}}(\gamma^{(s - 1)})$, then set $\gamma^{(s)} = \gamma^{(s - 1/2)}$; otherwise set $\gamma^{(s)} = (1 - c)\gamma^{(s-1)} + c\gamma^{(s - 1/2)}$ for some $0 < c < 1$ through a backtracking line search, such that $\ell_{\text{RCAL}}(\gamma^{(s)}) < \ell_{\text{RCAL}}(\gamma^{(s - 1)})$.
	\end{itemize}\vspace{-.1in}
\end{itemize}\vspace{-.02in}
\hrule

\vspace{.2in}
\textbf{Algorithm for computing $\hat{\alpha}^\#_{t, \text{\normalfont RWL}}$.}\;
A second-order Taylor expansion of the loss function in (\ref{eq:ORobj}), $\ell_{\text{WL}}(\alpha^\#_t; \hat{\gamma}_{\text{RCAL}}) = \sum_{k\neq t}\ell_{\text{WL}}(\alpha_t^{(k)}; \hat{\gamma}_{\text{RCAL}})$,
at the current estimate, denoted as $\tilde{\alpha}^\#_t$, gives the quadratic approximation
\begin{align*}
\ell_{\text{WL,Q1}}(\alpha^\#_t; \hat{\gamma}_{\text{RCAL}}) & = \ell_{\text{WL}}(\tilde{\alpha}^\#_t; \hat{\gamma}_{\text{RCAL}}) +  \tilde{\E}\left\{  g^\T(X; \tilde{\alpha}^\#_t, \hat{\gamma}_{\text{RCAL}}) (\alpha^\#_t - \tilde{\alpha}^\#_t)^\T  f(X) \right\} \\
& + \frac{1}{2} \tilde{\E}\left\{f^\T(X)(\alpha^\#_t - \tilde{\alpha}^\#_t)\frac{R^{(t)}}{\pi(t, X; \hat{\gamma}_{\text{RCAL}})}H(X; \tilde{\alpha}^\#_t, \hat{\gamma}_{\text{RCAL}})(\alpha^\#_t - \tilde{\alpha}^\#_t)^\T f(X)\right\},
\end{align*}
where $g(X; \alpha^\#_t, \gamma) = (g_k: k\in \mathcal T\backslash\{t\})^\T$ with $g_k = R^{(t)}\{\pi(k, X; \gamma) / \pi(t, X; \gamma)\} [-Y + \psi\{f^T(X)\alpha_t^{(k)}\}]$ and
$H(X; \alpha^\#_t, \gamma) = \diag\{ \pi(k, X; \gamma) \psi_2\{f^\T(X)\alpha_t^{(k)}\}:  k\in \mathcal T\backslash \{t\} \}$.

In contrast with $\ell_{\text{CAL,Q1}}(\gamma)$ discussed earlier,
the quadratic function $\ell_{\text{WL,Q1}}(\alpha^\#_t; \hat{\gamma}_{\text{RCAL}})$ can be recast as a weighted least-squares loss in multi-response linear regression,
because both the quadratic and linear terms here depend on only the observations from $t$th treated group.
Nevertheless, Fisher scoring can be exploited to address another complication.
For fast implementation of block coordinate descent, it is desirable to find a constant $b_2$ such that the weight matrices $\{R^{(t)}_i / \pi(t, X_i; \hat{\gamma}_{\text{RCAL}})\} H(X_i; \tilde{\alpha}^\#_t,
\hat{\gamma}_{\text{RCAL}})$, $i=1,\ldots,n$, are dominated by $b_2 I$, where $I$ is the $(K-1)\times (K-1)$ identity matrix. Although these weight matrices are diagonal,
the $k$th entry on the diagonal of the $i$th weight matrix is a product of $R^{(t)}_i \psi_2\{f^\T(X_i)\alpha_t^{(k)}\}$ and $\pi(k, X_i; \hat{\gamma}_{\text{RCAL}}) / \pi(t, X_i; \hat{\gamma}_{\text{RCAL}})$,
which may be inflated for small $\pi(t, X_i; \hat{\gamma}_{\text{RCAL}})$. Taking the maximum of the diagonal entries of the weight matrices would lead to unnecessarily large $b_2$ and slow convergence.

By Fisher scoring, we replace $R^{(t)} / \pi(t, X; \hat{\gamma}_{\text{RCAL}})$ by its model expectation, which equals 1, and obtain the quadratic approximation
\begin{align*}
\ell_{\text{WL,Q2}}(\alpha^\#_t; \hat{\gamma}_{\text{RCAL}}) & = \ell_{\text{WL}}(\tilde{\alpha}^\#_t; \hat{\gamma}_{\text{RCAL}}) +  \tilde{\E}\left\{  g^\T(X; \tilde{\alpha}^\#_t, \hat{\gamma}_{\text{RCAL}}) (\alpha^\#_t - \tilde{\alpha}^\#_t)^\T  f(X) \right\} \\
& \quad + \frac{1}{2} \tilde{\E}\left\{f^\T(X)(\alpha^\#_t - \tilde{\alpha}^\#_t)H(X; \tilde{\alpha}^\#_t, \hat{\gamma}_{\text{RCAL}})(\alpha^\#_t - \tilde{\alpha}^\#_t)^\T f(X)\right\}.
\end{align*}
The weight matrices can be dominated as  $H(X_i; \tilde{\alpha}^\#_t, \hat{\gamma}_{\text{RCAL}}) \preceq  b_2I$  for $i = 1, \ldots, n$, where $b_2 = \max_{i = 1}^n \max_{k\neq t}\pi(k, X_i; \hat{\gamma}_{\text{RCAL}})\psi_2\{f^\T(X_i)\tilde{\alpha}_t^{(k)}\}$, which does not suffer the inflation due to large probability ratios.
By the quadratic lower bound principle (Bohning and Lindsay 1988), a majorizing function of $\ell_{\text{WL,Q2}}(\alpha^\#_t; \hat{\gamma}_{\text{RCAL}})$ is
\begin{align*}
\ell_{\text{WL,Q3}}(\alpha^\#_t; \hat{\gamma}_{\text{RCAL}}) &= \ell_{\text{WL}}(\tilde{\alpha}^\#_t; \hat{\gamma}_{\text{RCAL}}) + \tilde{\E}\left\{  g^\T(X; \tilde{\alpha}^\#_t, \hat{\gamma}_{\text{RCAL}}) (\alpha^\#_t - \tilde{\alpha}^\#_t)^\T  f(X) \right\} \\
& \quad + \frac{b_2}{2} \tilde{\E}\left\{ \lVert  (\alpha^\#_t - \tilde{\alpha}^\#_t)^\T f(X)   \rVert_2^2 \right\} .
\end{align*}
Combining $\ell_{\text{WL,Q3}}(\alpha^\#_t; \hat{\gamma}_{\text{RCAL}})$ with the group Lasso penalty as in (\ref{eq:ORobj}) yields
\begin{align}
\frac{1}{2} \tilde{\E} \left\{ \lVert \tilde{\alpha}^{\# \T}_t f(X) -  g(X; \tilde{\alpha}^\#_t, \hat{\gamma}_{\text{RCAL}}) / b_2 -  \alpha^{\# \T}_t f(X) \rVert_2^2 \right\}
 + \frac{\lambda_2}{b_2 } \sum_{j=1}^p \lVert \alpha^\#_{jt} \rVert_2. \label{eq:ORq}
\end{align}
Minimization of (\ref{eq:ORq}) corresponds to a group penalized multi-response linear regression with the same design matrix for each response.
Then the block coordinate update has the closed form
\begin{align}
\tilde {\alpha}^\#_{jt} \leftarrow \frac{1}{\tilde{\E}\{f^2_j(X)\}}\left\{1 - \frac{\lambda_2/b_2}{\lVert \tilde{\E}\{Z_{(j)}f_j(X)\} \rVert_2}\right\}_{+}\tilde{\E}\{Z_{(j)}f_j(X)\},
\quad j = 1, \ldots, p, \label{eq:ORupdate}
\end{align}
where $Z_{(j)} = \tilde{\alpha}^{\# \T}_t f(X) - g(X; \tilde{\alpha}^\#_t, \hat{\gamma}_{\text{RCAL}}) / b_2 - \sum_{j^\prime \neq j}f_{j^\prime}(X)\tilde {\alpha}^\#_{j^\prime t}$ is the partial residual.

Similarly as in Algorithm 1 for computing $\hat{\gamma}_{\text{RCAL}}$,
we incorporate a backtracking line search to maintain the descent of the objective function \eqref{eq:ORobj}, which may be violated due to the use of Fisher scoring.
Hence our algorithm for computing $\hat{\alpha}^\#_{t, \text{\normalfont RWL}}$ is as follows.

\vspace{0.1in}
\hrule
\vspace{0.1in}
{\it Algorithm} 2. Fisher scoring block  descent algorithm for minimizing \eqref{eq:ORobj}.
\vspace{0.05in}
\hrule \vspace{-.1in}
\begin{itemize}\addtolength{\itemsep}{-.1in}
\item[1.] Set an initial value $\alpha_t^{\# {(0)}}$.
\item[2.] Repeat the following updates for $s = 1, 2, ...$ until convergence to obtain $\hat{\alpha}^\#_{t, \text{RWL}}$:\vspace{-.1in}
	\begin{itemize}\addtolength{\itemsep}{-.1in}
		\item[(a)] Compute $\alpha_t^{\# {(s - 1/2)}}$ as the minimizer of \eqref{eq:ORq} using block coordinate descent with update formula \eqref{eq:ORupdate}.
		\item[(b)] If $\ell_{\text{RWL}}(\alpha_t^{\# {(s - 1/2)}}; \hat{\gamma}_{\text{RCAL}}) < \ell_{\text{RWL}}(\alpha_t^{\# {(s - 1)}}; \hat{\gamma}_{\text{RCAL}})$, then set $\alpha_t^{\# {(s)}} = \alpha_t^{\# {(s - 1/2)}}$; otherwise set $\alpha_t^{\# {(s)}} = (1 - c)\alpha_t^{\# {(s - 1)}} + c\alpha_t^{\# {(s - 1/2)}}$ for some $0 < c < 1$, through a backtracking line search, such that $\ell_{\text{RWL}}(\alpha_t^{\# {(s)}}; \hat{\gamma}_{\text{RCAL}}) < \ell_{\text{RWL}}(\alpha_t^{\# {(s - 1)}}; \hat{\gamma}_{\text{RCAL}})$.
	\end{itemize}\vspace{-.1in}
\end{itemize}\vspace{-.1in}
\hrule

\vspace{.2in}
\section{Theoretical properties} \label{sec:theory}

We study statistical properties of the regularized calibrated estimators $ ( \hat \alpha^\#_{t,\text{RWL}}, \hat \gamma_{\text{RCAL}})$
and the augmented IPW estimator $\hat \mu_t ( \hat m^\#_{\text{RWL}}, \hat \pi_{\text{RCAL}})$, for a fixed treatment $t$ in high-dimensional settings.
In particular, we establish the asymptotic expansion (\ref{eq:desired-expan2}) and consistency of the estimated variance $\hat V_t$,
which lead to valid Wald confidence intervals for $\mu_t$ as described in Section~\ref{sec:RCAL}.
For $k\in\mathcal T\backslash \{t\}$, confidence intervals for the ATE, $\mu_t - \mu_k$,
can be obtained by standard arguments from the asymptotic expansions of $\hat \mu_t ( \hat m^\#_{\text{RWL}}, \hat \pi_{\text{RCAL}})$
and the corresponding estimator for $\mu_k$, which requires a separate set of regularized calibrated estimators as explained in Remarks~\ref{rem:separate-PS}--\ref{rem:separate-OR}.

\subsection{Estimation of regression coefficients} \label{sec:theory-rcal}

We develop theoretical analysis of the regularized calibrated estimators $ ( \hat \alpha^\#_{t,\text{RWL}}, \hat \gamma_{\text{RCAL}})$
with multi-class logistic PS model (\ref{eq:PSmodel}) and OR model (\ref{eq:ORmodel2}) in high-dimensional settings.
Compared with existing high-dimensional theory (Buhlmann and van de Geer 2011), our analysis needs to deal with several technical complications including
the interdependency of multiple responses on each subject in the loss functions
$\ell_{\text{CAL}}(\gamma) $ and $\ell_{\text{RWL}}(\alpha^\#_t; \hat{\gamma}_{\text{RCAL}})$,
the dependency of the estimator $\hat \alpha^\#_{t,\text{RWL}}$ on $\hat \gamma_{\text{RCAL}}$,
and possible misspecification of both models (\ref{eq:PSmodel}) and (\ref{eq:ORmodel2}).

First, we study the regularized calibrated estimator $\hat \gamma_{\text{RCAL}}$.
The tuning parameter in the penalized objective function \eqref{eq:PSobj} is specified as $\lambda = A_1\tilde{\lambda}_1$, with a constant $A_1 > 1$, and
\begin{align*}
 \tilde{\lambda}_1  = 4\sqrt{2}B_0^2B_1 \sqrt{(K - 1) / n + \log \{(p+1)/\epsilon\} / n } ,
\end{align*}
where $(B_0, B_1)$ comes from Assumption \ref{ass1} below, and $0 < \epsilon < 1$ is a tail probability for the error bound, to be discussed below.

With possible misspecification of model (\ref{eq:PSmodel}), the limit (or target) value of $\hat \gamma_{\text{RCAL}}$, denoted as $\bar \gamma_{\text{CAL}}$,
can be identified as a minimizer of the expected calibration loss $\E \{  \ell_{\text{CAL}}(\gamma) \}$, subject to
the constraint $\bar{\gamma}_{t, \text{CAL}} \equiv 0$,  where $\ell_{\text{CAL}}(\gamma)$ is defined in (\ref{eq:PSloss}). The one-to-zero constraint on $\bar \gamma_{\text{CAL}}$ corresponds to our definition of $\hat \gamma_{\text{RCAL}}$ with the same constraint.
If model (\ref{eq:PSmodel}) is correctly specified, then $\bar \gamma_{\text{CAL}}$ coincides with the true value $\gamma^*$
such that $\pi(k,X; \gamma^*) = \pi^*(k,X)$ for $k\in\mathcal T$ subject to $\gamma^*_t \equiv 0$.
Otherwise, $\bar \pi_{\text{CAL}}(k,X) = \pi(k,X; \bar \gamma_{\text{CAL}})$ may differ from $\pi^*(k,X)$ for $k\in\mathcal T$.
For two matrices $\gamma =(\gamma_k: k\in\mathcal T\backslash\{t\})$ and $\gamma^\prime = (\gamma^\prime_k: k \in \mathcal T\backslash\{t\})$,
the group $L_1$ norm of $\gamma-\gamma^\prime$ is defined as $\| \gamma - \gamma^\prime \|_{2,1} = \sum_{j=0}^p \lVert \gamma_{j\cdot} - \gamma^\prime_{j\cdot} \rVert_2$,
where $\gamma_{j\cdot}$ or $\gamma^\prime_{j\cdot}$ is the transpose of the $(j+1)$th row vector in $\gamma$ or $\gamma^\prime$.
The Bregman divergence associated with the convex loss $\ell_{\text{CAL}}(\cdot)$ is
\begin{align*}
D_{\text{CAL}}(\gamma, \gamma^\prime) =  \ell_{\text{CAL}}(\gamma) - \ell_{\text{CAL}} (\gamma^\prime)
- \sum_{k\in\mathcal T\backslash\{t\}} (\gamma_k -\gamma^\prime_k)^\T (\partial /\partial \gamma^\prime_k)  \ell_{\text{CAL}} (\gamma^\prime) .
\end{align*}
The symmetrized Bregman divergence is easily shown to be
\begin{align*}
D^\dagger_{\text{CAL}}(\gamma, \gamma^\prime) & =  D_{\text{CAL}}(\gamma, \gamma^\prime) + D_{\text{CAL}}(\gamma^\prime, \gamma) \\
& = \sum_{k\neq t}\tilde{\E}\left[R^{(t)}\left\{ \me^{\gamma_k^\T f(X)} - \me^{ \gamma^{\prime^\T}_k f(X) }\right\}
(\gamma_k - \gamma^\prime_k)^\T f(X) \right].
\end{align*}
The following assumptions are required in our analysis of convergence of $\hat \gamma_{\text{RCAL}}$ to $\bar \gamma_{\text{CAL}}$.

\begin{ass} \label{ass1}
Suppose that the following conditions are satisfied:\vspace{-.1in}
\begin{itemize}\addtolength{\itemsep}{-.1in}
\item[(i)] $\sup_{j=1, \ldots, p} |f_j(X)| \leq B_0$ almost surely for a constant $B_0\geq 1$;

\item[(ii)] $\pi^{-1}(t, X; \bar{\gamma}_{\text{CAL}}) \leq B_1$ almost surely for a constant $B_1 > 1$;

\item[(iii)] The theoretical compatibility condition below holds with the subset $S_{\gamma} = \{0\}\cup\{j: \bar{\gamma}_{j\cdot,\text{CAL}} \neq 0, j = 1, \ldots, p\}$
and some constants $\nu_1 > 0$ and $\xi_1 > 1$;

\item[(iv)] (a) $(\xi_1 + 1)^2\nu_1^{-2}|S_\gamma|\tilde{\lambda}_1 \leq \eta_{1,1}$ for a constant $0 < \eta_{1,1} < 1$, and (b) $B_0(\xi_1 + 1)^2(A_1 - 1)(1 - \eta_{1,1})^{-1}\nu_1^{-2}|S_\gamma|\tilde{\lambda}_1\leq \eta_{1,2}$ for a constant $0 < \eta_{1,2} < 1$.
\vspace{-.1in}
\end{itemize}\vspace{-.1in}
\end{ass}

The theoretical compatibility condition in Assumption~\ref{ass1} (iii) is defined as follows: for any $(p+1)\times (K - 1)$ matrix $b = (b_k: k\in\mathcal T\backslash\{t\})$ satisfying
\begin{align}
\sum_{j\notin S_\gamma} \lVert b_{j\cdot} \rVert_2 \leq \xi_1\sum_{j\in S_\gamma} \lVert b_{j\cdot} \rVert_2, \label{eq:ps-cc2}
\end{align}
it holds that
\begin{align}
\nu_1^2\left(\sum_{j\in S_\gamma }\lVert b_{j\cdot} \rVert_2\right)^2 \leq |S_\gamma|\, \sum_{k\neq t}b_k^\T\E\left\{R^{(t)} \omega(k,X; \bar{\gamma}_{\text{CAL}})f(X)f^\T(X)\right\}b_k.
\label{eq:ps-cc1}
\end{align}
where $b_k = (b_{jk}: j = 0, 1, \ldots, p)^\T$ for $k \in \mathcal T\backslash\{t\}$, $b_{j\cdot} = (b_{jk}: k\in\mathcal T\backslash\{t\})^\T$ is the transpose of the $(j+1)$th row vector in $b$,
and $\omega(k,X; \gamma) = \pi(k, X; \gamma) / \pi(t, X;\gamma)$ with dependency on $t$ suppressed.
As seen from the Taylor expansion (\ref{eq:taylor-gamma}),
the right-hand side of (\ref{eq:ps-cc1}) can be expressed as $|S_\gamma| \text{vec}^\T(b)\Sigma_{\gamma}\text{vec}(b)$,
where $\text{vec}(b) = (b^\T_k: k\in\mathcal T\backslash\{t\})^\T$, $\Sigma_{\gamma} = \E[\diag\{R^{(t)}\omega(k, X; \bar{\gamma}_{\text{CAL}}): k\neq t\}\otimes f(X)f^\T(X)]$ is the Hessian matrix of $\E\{\ell_{\text{CAL}}(\gamma)\}$ at $\gamma = \bar{\gamma}_{\text{CAL}}$, and $\otimes$ denotes the Kronecker product.
Hence Assumption \ref{ass1}(iii) amounts to a compatibility condition on the matrix $\Sigma_\gamma$, similarly as in Buhlmann and van de Geer (2011) and Tan (2020ab).
See Remark~\ref{rem:Lounici} for further discussion.

\begin{rem}[On Assumption \ref{ass1}] \label{rem:comments-ass1}
Assumption~\ref{ass1}(i) may often be satisfied in practice, although relaxation can be made to allow sub-Gaussian regressors $f_j(X)$ with increasing technical complexity.
Assumption~\ref{ass1}(ii), which is also used in Farrell (2015), is used among others to bound the gradient of the loss $\ell_{\text{CAL}}(\gamma)$ at $\gamma = \bar{\gamma}_{\text{CAL}}$.
The compatibility condition in Assumption \ref{ass1}(iii) is discussed in Remark \ref{rem:Lounici}, together with related conditions.
Assumption~\ref{ass1}(iv) requires that $|S_\gamma| \tilde{\lambda}_1$ is sufficiently small, and is used to obtain the empirical compatibility condition
(Lemma \ref{lem:ps-empicc}) and perform localized analysis with a non-quadratic loss function (Lemma \ref{lem:ps-error}).
\end{rem}

The following result establishes the convergence of $\hat \gamma_{\text{RCAL}}$ to $\bar \gamma_{\text{CAL}}$
in the $L_{2,1}$ norm at rate $|S_\gamma|\{K/n + \log(p) / n\}^{1/2}$ and in the associated Bregman divergence at the rate $|S_\gamma|\{K/n + \log(p) / n\}$.
While the rate $\log(p) / n$ is familiar in high-dimensional analysis,
our analysis appears to, for the first time, allow a theoretical compatibility condition
and account for the interdependency between the gradient vectors in $\gamma_k$, $k\not=t$, of a loss function
for group-Lasso penalized estimation in multi-class logistic regression (\ref{eq:PSmodel}).
See Remark \ref{rem:Farrell} for a comparison with Farrell (2015).

\begin{thm}\label{thm:ps-error}
Suppose that Assumption~\ref{ass1} holds. Then we have probability at least $1 - 3\epsilon$,
\begin{align}
D_{\text{CAL}}^{\dagger}(\hat{\gamma}_{\text{RCAL}}, \bar{\gamma}_{\text{CAL}}) + (A_1 - 1)\tilde{\lambda}_1\lVert \hat{\gamma}_{\text{RCAL}} - \bar{\gamma}_{\text{CAL}} \rVert_{2, 1} \leq \xi_{1, 1}^2\nu_{1, 1}^{-2}|S_\gamma|\tilde{\lambda}_1^2, \label{eq:thm-ps-error}
\end{align}
where $\xi_{1, 1}  = (\xi_1 + 1)(A_1 - 1)$, and $\nu_{1, 1} = \nu_1(1 - \eta_{1, 1})^{1/2}(1 - \eta_{1, 2})^{1/2}$.
\end{thm}

Next, we study the regularized calibrated (or weighted likelihood) estimator  $\hat \alpha^\#_{t,\text{RWL}}$.
The tuning parameter in the objective function \eqref{eq:ORobj} is specified as $\lambda_2 = A_2 \tilde{\lambda}_2$ with a constant $A_2 > 1$ and
\begin{align*}
\tilde{\lambda}_2 = \max\left[\tilde{\lambda}_1, \sqrt{3}B_0(B_1 - 1)\sigma_0\sqrt{(K - 1) / n + \log\{(p+1)/\epsilon\} / n} \right],
\end{align*}
where $(B_0, B_1)$  are from Assumptions \ref{ass1}(i)-(ii), $\sigma_0$ is from Assumption \ref{ass2}(i), and $0 < \epsilon < 1$ is a tail probability for the error bound.

With possible misspecification of model (\ref{eq:ORmodel2}) as well as model (\ref{eq:PSmodel}), the limit (or target) value of
$\hat{\alpha}_{t,\text{RWL}}^\# = (\hat\alpha_{t,\text{RWL}}^{(k)}: k \not= t)$, denoted as $\bar{\alpha}_{t,\text{WL}}^\# = (\bar\alpha_{t,\text{WL}}^{(k)}: k \not= t)$,
can be identified as a minimizer of the expected loss $\E \{\ell_{\text{WL}}(\alpha^\#_t; \bar{\gamma}_{\text{CAL}}) \}$,
where $ \ell_{\text{WL}}(\alpha^\#_t; \gamma ) = \sum_{k\not= t}  \ell_{\text{WL}}(\alpha^{(k)}_t; \gamma)$.
If model (\ref{eq:ORmodel2}) is correctly specified, then
$ \bar{\alpha}_{t, \text{WL}}^{(k)}$ for each $k\not=t$ coincides with $\alpha^*_t$ such that $m(t,X ; \alpha^*_t) = m^*(t,X)$.
Otherwise, $\bar m_{\text{RWL}}^{(k)} (t, X) = m(t, X; \bar{\alpha}_{t, \text{WL}}^{(k)})$, $k\not=t$, may differ from $m^*(t,X)$.
For two matrices $\alpha^\#_t =(\alpha^{(k)}_t: k\not=t )$ and $\alpha^{\prime \#}_t =(\alpha^{\prime (k)}_t: k\not=t )$,
the group $L_1$ norm is defined as $\|\alpha^\#_t - \alpha^{\prime \#}_t \|_{2,1} = \sum_{j=0}^p \lVert \alpha_{jt}^\# - \alpha_{jt}^{\prime\#} \rVert_2$,
where $\alpha_{jt}^\#$ or $\alpha_{jt}^{\prime\#}$ is the transpose of $(j+1)$th row vector in $\alpha^\#_t$ or $\alpha^{\prime \#}_t$.
The Bregman divergence associated with the convex loss $\ell_{\text{WL}}(\cdot; \hat{\gamma}_{\text{RCAL}})$ is
\begin{align*}
D_{\text{WL}}(\alpha^\#_t, \alpha^{\prime \#}_t; \hat{\gamma}_{\text{RCAL}}) & = \ell_{\text{WL}}(\alpha^\#_t; \hat{\gamma}_{\text{RCAL}}) - \ell_{\text{WL}}(\alpha^{\prime \#}_t; \hat{\gamma}_{\text{RCAL}}) \\
& \quad - \sum_{k\neq t}(\alpha^{(k)}_t - \alpha^{\prime(k)}_t)^\T(\partial / \partial \alpha^{\prime(k)}_t)\ell_{\text{WL}}(\alpha^{\prime \#}_t; \hat{\gamma}_{\text{RCAL}}).
\end{align*}
The symmetrized Bregman divergence is easily shown to be
\begin{align*}
& D^\dagger_{\text{WL}}(\alpha^\#_t, \alpha^{\prime \#}_t; \hat{\gamma}_{\text{RCAL}}) = D_{\text{WL}}(\alpha^\#_t, \alpha^{\prime \#}_t; \hat{\gamma}_{\text{RCAL}}) + D_{\text{WL}}(\alpha^{\prime \#}_t, \alpha^\#_t; \hat{\gamma}_{\text{RCAL}}) \\
& \quad = \sum_{k\neq t}\tilde{\E}\left(R^{(t)}\omega(k,X; \hat{\gamma}_{\text{RCAL}})[\psi\{\alpha^{(k) \T}_tf(X)\} - \psi\{\alpha^{\prime (k) \T}_tf(X)\}]\{\alpha^{(k) \T}_tf(X) - \alpha^{\prime (k) \T}_tf(X)\}\right).
\end{align*}
The following assumptions are adapted from Tan (2020b).
The theoretical compatibility condition is the same as in Assumption~\ref{ass1}(iii), except with the sparsity subset $S_{\alpha_t}$
and possibly different constants $(\nu_2,\xi_2)$.
Assumption~\ref{ass2}(iv) is not needed here, but will be used in later results.

\begin{ass} \label{ass2}
Suppose that the following conditions are satisfied:\vspace{-.1in}
\begin{itemize}\addtolength{\itemsep}{-.1in}
\item[(i)] $Y^{(t)} - m(t, X; \bar{\alpha}^{(k)}_{t, \text{WL}})$ are uniformly sub-gaussian random variables with parameter $\sigma_0^2$ given $X$  for $k = 0, 1, \ldots, K-1~\text{and}~k\neq t$;
\item[(ii)] The theoretical compatibility condition in Assumption~\ref{ass1}(iii) holds with $S_\gamma$, $\nu_1$, and $\xi_1$ replaced by respectively $S_{\alpha_t} = \{0\}\cup\{j: \bar{\alpha}^\#_{jt, \text{WL}} \neq 0, j = 1, \ldots, p\}$ and some constants $\nu_2 > 0$ and $\xi_2 > 1$ in (\ref{eq:ps-cc2})--(\ref{eq:ps-cc1});
\item[(iii)] $\max_{k\in\mathcal T\backslash\{t\}} \psi_2\{\bar{\alpha}_{t, \text{WL}}^{(k)\T}f(X)\} \leq C_1$ almost surely for a constant $C_1 > 0$,
where $\psi_2(\cdot)$ denotes the derivative of $\psi(\cdot)$;
\item[(iv)] $\min_{k\in\mathcal T \backslash\{t\}} \psi_2\{\bar{\alpha}_{t, \text{WL}}^{(k)\T}f(X)\} \geq C_2$ almost surely for a constant $C_2 > 0$;
\item[(v)] For any $h, h^\prime\in\bbR$, $\psi_2(h) \leq \psi_2 (h^\prime) \me^{C_3|h - h^\prime|}$ holds for a constant $C_3\geq 0$;
\item[(vi)] (a) $(1 + \xi_2^2)\nu_2^{-2}|S_{\alpha_t}|\tilde{\lambda}_2 \leq \eta_2$ for a constant $0 < \eta_2 < 1$, (b) $\me^{\eta_{1, 3}}B_0C_3C_2^{-1}\nu_2^{-2}(\xi_2 + 1)^2(A_2 - 1)(1 - \eta_2)^{-1}|S_{\alpha_t}|\tilde{\lambda}_2 \leq \eta_{2, 1}$ for a constant $0 \leq \eta_{2, 1} < 1$, and (c) $\me^{3\eta_{1, 3}}B_0C_3C_2^{-1}\xi_{2, 3}^{-2}(A_2 - 1)^{-1}M_{1, 1}|S_\gamma|\tilde{\lambda}_1 \leq \eta_{2, 2}$ for a constant $0 \leq \eta_{2, 2} < 1$, where $(\eta_{1, 3}, \xi_{2, 3}, M_{1, 1})$ are as in Theorem \ref{thm:or-error-glm}.
\end{itemize}\vspace{-.1in}
\end{ass}

The following result establishes  the convergence of $\hat \alpha^\#_{t,\text{RWL}}$ to $\bar{\alpha}_{t,\text{WL}}^\#$
in the $L_{2,1}$ norm at rate $(|S_\gamma| + |S_{\alpha_t}| ) \{K/n + \log(p) / n\}^{1/2}$ and
in the associated Bregman divergence at the rate $(|S_\gamma| + |S_{\alpha_t}| )  \{K/n + \log(p) / n\}$.
Compared with related results (Farrell 2015), our analysis needs to handle
the interdependency between the gradient vectors of the weighted likelihood loss, corresponding to
outcome regression with different weights in the same treatment group $t$.
In addition, our error bound for $\hat \alpha^\#_{t,\text{RWL}}$ depends on the sparsity subsets of $\bar\gamma_{\text{CAL}}$ and $\bar\alpha^\#_{t, \text{WL}}$,
due to the construction of $\hat \alpha^\#_{t,\text{RWL}}$ depending on $\hat\gamma_{\text{RCAL}}$.
See Remarks \ref{rem:key-step}--\ref{rem:Farrell} for further discussion.

\begin{thm}\label{thm:or-error-glm}
Suppose that Assumption~\ref{ass1} holds and Assumption~\ref{ass2} except~\ref{ass2}(iii) hold. If $\log\{(K - 1) + \log(p+1)/\epsilon\} / n\leq 1$, then for $A_1 > (\xi_1 + 1) / (\xi_1 - 1)$ and $A_2 > (\xi_2 + 1) / (\xi_2 - 1)$, we have with probability at least $1 - 6\epsilon$,
\begin{align}
& D^{\dagger}_{\text{WL}}(\hat{\alpha}^\#_{t, \text{RWL}}, \bar{\alpha}^\#_{t, \text{WL}}; \bar{\gamma}_{\text{CAL}}) + \me^{\eta_{1, 3}}(A_2 - 1)\tilde{\lambda}_2\lVert \hat{\alpha}^\#_{t, \text{RWL}} - \bar{\alpha}^\#_{t, \text{WL}} \rVert_{2, 1} \nonumber \\
& \leq \me^{4\eta_{1, 3}}\xi_{2, 3}^{-2}\{M_{1, 1}|S_\gamma|\tilde{\lambda}_1^2\} + \me^{2\eta_{1, 3}}\xi_{2, 2}^2\{\nu_{2, 2}^{-2}|S_{\alpha_t}|\tilde{\lambda}_2^2\}, \label{eq:thm-or-error-glm}
\end{align}
where $\xi_{2, 2} = (\xi_2 + 1)(A_2 - 1)$, $\xi_{2, 3} = \xi_{2, 1}(1 - \eta_{2, 2})^{1/2}C_2^{1/2}$, $\xi_{2, 1} = 1 - 2A_2 / \{(\xi_2 + 1)(A_2 - 1)\}$, $\nu_{2, 2} = \nu_{2, 1}(1 - \eta_{2, 1})^{1/2}C_2^{1/2}$,  and $\nu_{2, 1} = \nu_2(1 - \eta_2)^{1/2}$,  depending only on $(A_2, \xi_2, \nu_2, \eta_2, \eta_{2, 1}, \eta_{2, 2})$, and
$M_{1, 1} = 3\sigma_0^2(A_1 - 1)^{-2}M_1^2\eta_1 + 2\sigma_0^2\me^{\eta_{1, 3}}M_1$, $\eta_{1, 3} = (A_1 - 1)^{-1}M_1\eta_1B_0$ and $M_1 = \xi_{1,1}^2\nu_{1, 1}^{-2}$, depending only on $(B_0, B_1, A_1, \xi_1, \nu_1)$ and $\sigma_0$, and $\eta_1$ is a constant such that $|S_\gamma|\tilde{\lambda}_1 \leq \eta_1$ under Assumption \ref{ass1}(iv), and $(\xi_{1, 1}, \nu_{1, 1})$ are as in Theorem \ref{thm:ps-error}.
\end{thm}

\begin{rem}[Linear outcome model] \label{rem:ORlm}
If linear OR model (\ref{eq:ORmodel2}) is used, the symmetrized Bregman divergence $D^\dagger_{\text{WL}}(\hat{\alpha}^\#_{t, \text{RWL}}, \bar{\alpha}^\#_{t, \text{WL}}; \bar{\gamma}_{\text{CAL}})$ becomes the weighted (in-sample) prediction error
\begin{align*}
Q_{\text{WL}}(\hat{\alpha}^\#_{t, \text{RWL}}, \bar{\alpha}^\#_{t, \text{WL}}; \bar{\gamma}_{\text{CAL}}) = \sum_{k\neq t}\tilde{\E}
\left[ R^{(t)}\omega(k,X; \bar{\gamma}_{\text{CAL}}) \{ \hat{\alpha}^{(k) \T}_{t, \text{RWL}} f(X) - \bar{\alpha}^{(k) \T}_{t, \text{WL}} f(X) \}^2 \right],
\end{align*}
In this case, Assumptions~\ref{ass2}(iii)--(v) hold with $C_1 = C_2 = 1$ and $C_3 = 0$ and Assumptions~\ref{ass2}(vi)(b) and~\ref{ass2}(vi)(c) hold with $ \eta_{2, 1} = \eta_{2, 2} = 0$. Therefore, under Assumptions~\ref{ass2}(i), (ii) and (vi)(a), we have from Theorem \ref{thm:or-error-glm} that
\begin{align}
& Q_{\text{WL}}(\hat{\alpha}^\#_{t, \text{RWL}}, \bar{\alpha}^\#_{t, \text{WL}}; \bar{\gamma}_{\text{CAL}}) + \me^{\eta_{1, 3}}(A_2 - 1)\tilde{\lambda}_2\lVert \hat{\alpha}^\#_{t, \text{RWL}} - \bar{\alpha}^\#_{t, \text{WL}} \rVert_{2, 1}  \nonumber \\
& \leq \me^{4\eta_{1, 3}}\xi_{2, 1}^{-2}\{M_{1, 1}|S_\gamma|\tilde{\lambda}_1^2\} + \me^{2\eta_{1, 3}}\xi_{2, 2}^2\{\nu_{2, 1}^{-2}|S_{\alpha_t}|\tilde{\lambda}_2^2\}, \label{eq:thm-or-error-lm}
\end{align}
where $(\xi_{2, 1}, \xi_{2, 2}, \nu_{2, 1}, M_{1, 1})$, only depending on $(A_2, \xi_2, \nu_2, \eta_2, \sigma_0, A_1, \xi_1, \nu_1, \eta_1, B_0, B_1)$, are the same as in Theorem \ref{thm:or-error-glm}.
\end{rem}

\begin{rem}[Data-dependent weights] \label{rem:key-step}
A key step in our proof is to upper-bound the product
\begin{align}
\sum_{k\neq t}(\hat{\alpha}^{(k)}_{t, \text{RWL}} - \bar{\alpha}^{(k)}_{t, \text{WL}})^\T\tilde{\E}[R^{(t)}\omega(k,X; \hat{\gamma}_{\text{RCAL}})\{Y - m(t, X; \bar{\alpha}^{(k)}_{t, \text{WL}})\}f(X)], \label{eq:product1}
\end{align}
which involves the estimated weight $\omega(k,X; \hat{\gamma}_{\text{RCAL}})$. If we replace $\hat{\gamma}_{\text{RCAL}}$ with $\bar{\gamma}_{\text{CAL}}$, then it is standard to use the following bound,
\begin{align}
& \sum_{k\neq t}(\hat{\alpha}^{(k)}_{t, \text{RWL}} - \bar{\alpha}^{(k)}_{t, \text{WL}})^\T\tilde{\E}[R^{(t)}\omega(k,X; \bar{\gamma}_{\text{CAL}})\{Y - m(t, X; \bar{\alpha}^{(k)}_{t, \text{WL}})\}f(X)] \label{eq:product2} \\
& = \sum_{j=0}^p(\hat{\alpha}^\#_{jt, \text{RWL}} - \bar{\alpha}^\#_{jt, \text{WL}})^\T\tilde{\E}[R^{(t)}\diag\{\omega(k,X; \bar{\gamma}_{\text{CAL}}): k\neq t\}\{Y - \bar{m}^\#_{t,\text{WL}} \}f_j(X)] \nonumber \\
& \leq \Vert \hat{\alpha}^\#_{t, \text{RWL}} - \bar{\alpha}^\#_{t, \text{WL}} \rVert_{2, 1}\times \max_{j=0,1,\ldots,p} \lVert \tilde{\E}[R^{(t)}\diag\{\omega(k,X; \bar{\gamma}_{\text{CAL}}): k\neq t\}\{Y - \bar{m}^\#_{t,\text{WL}} \}f_j(X)] \rVert_2, \nonumber
\end{align}
where $Y - \bar{m}^\#_{t,\text{WL}} = (Y - m(t, X; \bar{\alpha}^{(k)}_{t, \text{WL}}): k\neq t)^\T$. To handle the dependency on $\hat{\gamma}_{\text{RCAL}}$, we derive an upper bound on the difference between \eqref{eq:product1} and \eqref{eq:product2}
and a quadratic inequality in terms of
$Q_{\text{WL}}(\hat{\alpha}^\#_{t, \text{RWL}}, \bar{\alpha}^\#_{t, \text{WL}}; \bar{\gamma}_{\text{CAL}})$, which is then inverted to obtain the desired bound (\ref{eq:thm-or-error-glm}).
\end{rem}

To compare our results with related ones, we first summarize those in Lounici et al.~(2011) and Buhlmann and van de Geer (2011).
Consider fixed-design multi-task linear regression:
\begin{align}
Y^k_{i} = \alpha_k^{*\T} X^k_{i} + \epsilon^k_{i}, \quad i = 1, \ldots, n_0,\; k = 0, 1, \ldots, K - 1,
\label{eq:MT-model}
\end{align}
where $Y_i^k$ is the $i$th response, $X^k_i = ( X^k_{1i}, \dots, X^k_{pi})^\T$ is the $i$th fixed covariate vector,
and $\alpha^*_k = (\alpha^*_{1k}, \ldots, \alpha^*_{pk})^\T$ is the associated coefficient vector in task $k$,
and $\epsilon^k_{i}$ is $\N(0, \sigma^2)$ independently over $i = 1, \ldots, n_0$ and $k = 0, 1, \ldots, K-1$,
with $n_0$ the fixed sample size in each task.
Similarly to $\{\hat\alpha_{k,\text{RMLg}}: k\in \mathcal T\}$ in Section \ref{sec:setup}, the group-Lasso penalized estimators
$\{\hat\alpha_k: k =0,1,\ldots,K-1\}$ are defined as a minimizer to
$\sum_{k=0}^{K-1} \tilde{\E}_k \{ ( Y^k - \alpha_k^\T X^k)^2 \}/(2K) + \lambda \sum_{j=1}^p \| \alpha_{j\cdot} \|_2 $,
where $\alpha_{j\cdot}$ is the transpose of the $j$th row vector in the $p\times K$ matrix $\alpha=(\alpha_0,\ldots,\alpha_{K-1})$,
and $\tilde{\E}_k(\cdot)$ denotes the sample average over task $k$.
For example, $\tilde{\E}_k \{ ( Y^k - \alpha_k^\T X^k)^2 \} = n_0^{-1} \sum_{i=1}^{n_0}  ( Y_i^k - \alpha_k^\T X_i^k)^2 $.
The empirical compatibility condition assumed in Buhlmann and van de Geer (2011), which is weaker than
the restricted eigenvalue condition in Lounici et al. (2011), is as follows:
for any $p \times K$ matrix $b = (b_0, b_1, \ldots, b_{K-1})$ satisfying $\sum_{j\notin S^*} \lVert b_{j\cdot} \rVert_2 \leq 3\sum_{j\in S^*} \lVert b_{j\cdot} \rVert_2$, it holds that
\begin{align}
\nu^2_{\text{MT}} \left( \sum_{j\in S^*}\lVert b_{j\cdot} \rVert_2 \right)^2 \leq |S^*| \sum_{k=0}^{K-1}b_k^\T\tilde{\E}_k(X^k X^{k\,\T} )b_k,
\label{eq:MT-CC}
\end{align}
where
$S^* = \{j: \alpha^*_{j\cdot} \not=0\}$ is the sparsity subset for $\alpha^*=(\alpha^*_0, \ldots,\alpha^*_{K-1})$.
Then the following error bound is obtained with high probability (Buhlmann and van de Geer 2011, Theorem 8.4) :
\begin{align}
Q(\hat{\alpha}, \alpha^*)  + K^{-1/2} \dot{\lambda}\lVert \hat{\alpha} - \alpha^* \rVert_{2, 1} \leq O(1) \nu^{-2}_{\text{MT}} |S^*|\dot{\lambda}^2,
\label{eq:MT-error}
\end{align}
where $Q(\hat{\alpha}, \alpha^*) = K^{-1} \sum_{k=0}^{K-1} \tilde{\E}_k \{ (\hat{\alpha}_k X^k - \alpha_k^{*\T} X^k )^2 \}$
and $\dot{\lambda} = O(1) \{K/n + \log(p)/n\}^{1/2}$ with $n= n_0 K$.
We compare our results with related ones in the following remarks.

\begin{rem}[Comparison with multi-task linear regression] \label{rem:Lounici}
The results in Lounici et al.~(2011) and Buhlmann and van de Geer (2011)
can be transferred to an error bound on the group-RML estimator $\{\hat\alpha_{k,\text{RMLg}}: k\in \mathcal T\}$ in a linear OR model with regressor vector $f(X)$,
by taking the treatment group $\{i: T_i=k,i=1,\ldots,n\}$ as task $k$
and conditioning on $\{(T_i, X_i): i=1,\ldots,n\}$ such that all treatment groups are of the same size $n_0 = n/K$.
For the empirical compatibility condition, (\ref{eq:MT-CC}) can be rewritten in a similar form to (\ref{eq:ps-cc1}) as
\begin{align}
\tilde \nu_{\text{MT}}^2 \left( \sum_{j\in S^*}\lVert b_{j\cdot} \rVert_2 \right)^2 \leq |S^*| \sum_{k=0}^{K-1}b_k^\T\tilde{\E} (R^{(k)} f(X) f^\T(X) )b_k,
\label{eq:MT-CC2}
\end{align}
where $\tilde \nu_{\text{MT}} = \nu_{\text{MT}}/K^{1/2}$ plays the role of $\nu_1$ in (\ref{eq:ps-cc1}).
In this way, the compatibility condition in Buhlmann and van de Geer (2011) and
those in Assumptions~\ref{ass1}(iii) and~\ref{ass2}(ii) are comparable, while treating  $\omega(k,X; \bar{\gamma}_{\text{CAL}}) \approx 1$.
Then the error bound (\ref{eq:MT-error}) can be stated as
\begin{align}
Q(\hat\alpha_{\text{RMLg}}, \alpha^*)  + K^{-1/2} \dot{\lambda}\lVert \hat\alpha_{\text{RMLg}} - \alpha^* \rVert_{2, 1} \leq O(1) K^{-1} \tilde \nu_{\text{MT}}^{-2} |S^*|\dot{\lambda}^2, \label{eq:MT-error2}
\end{align}
where $Q(\hat\alpha_{\text{RMLg}}, \alpha^*) =  \sum_{k=0}^{K-1} \tilde{\E} [ R^{(k)}\{ \hat\alpha^\T_{k,\text{RMLg}} f(X) - \alpha_k^{*\T} f(X) \}^2 ]$.
Hence $Q(\hat\alpha_{\text{RMLg}}, \alpha^*)$ is of order $O(1) K^{-1} \{K/n + \log(p) / n\}$
and $\lVert \hat\alpha_{\text{RMLg}} - \alpha^* \rVert_{2, 1}$ is of order $O(1) K^{-1/2} \{K/n + \log(p) / n\}^{1/2}$,
which are smaller than the error bounds on $\hat \alpha^\#_{t,\text{RWL}}$ in Theorem \ref{thm:or-error-glm} by a factor of $K^{-1}$ and $K^{-1/2}$ respectively.
However, this comparison needs to be interpreted with caution, even after ignoring differences caused by the weight $\omega(k,X; \hat{\gamma}_{\text{RCAL}})$.
Our analysis involves a random design and a possibly misspecified OR model,
whereas the error bound (\ref{eq:MT-error2}) is derived in a fixed design (with fixed treatments and covariates) and a correctly specified linear model
(with $\alpha^*$ encoding true coefficients).
With possible model misspecification, a random design enables that the gradient of a loss function remains mean-zero unconditionally
when evaluated at the target parameter value,
but such a mean-zero property is lost in a fixed design.
On the other hand, the error bounds are improved in fixed-design multi-task linear regression (\ref{eq:MT-model}),
partly because the sup-$L_2$ norm of the gradient of the loss function 
at $\alpha^*$ can be more tightly controlled using the independence between different treatmeant subsamples,
which is further discussed in Appendix C.
\end{rem}

\begin{rem}[Comparison with Farrell 2015] \label{rem:Farrell}
We compare our results with those about the group-RML estimators in Farrell (2015).
For a correctly specified linear OR model, Farrell provided finite-sample analysis of the estimator $\{\hat\alpha_{k,\text{RMLg}}: k\in \mathcal T\}$,
using an empirical restricted eigenvalue condition as in Lounici et al.~(2011),
related to the empirical compatibility condition in Remark \ref{rem:Lounici}.
Such an empirical condition is further assumed to be satisfied in asymptotic analysis,
instead of being derived with high probability from a theoretical condition as in our analysis.
In addition, Farrell used Lemma 9.1 in Lounici et al.~(2011) to control the gradient norm of the least-square loss
by the independence between $\tilde E\{ R^{(k)} (Y - \alpha_k^{*\T} f(X)) f_j(X)\}$, $k\in\mathcal T$, from different treatment groups.
But the derivatives of our weighted least-square loss
$\tilde E\{ R^{(t)}\omega(k, X; \bar{\gamma}_{\text{CAL}}) (Y - \bar\alpha_{t, \text{WL}}^{(k)\T} f(X)) f_j(X)\}$, $k\not=t$,
are interdependent for each $j$,
being induced by different weights
in the same treatment group $t$. See Appendix C for further discussion about control of gradient norms.
For a correctly specified multi-class logistic PS model, Farrell's finite-sample analysis of the estimator $\hat\gamma_{\text{RML}}$
also used an empirical restricted eigenvalue condition, which is further assumed to hold in asymptotic analysis.
More importantly, Farrell in his Lemma B.1 also applied Lemma 9.1 in Lounici et al.~(2011) to control
the sup-$L_2$ norm of the gradient $\partial \ell_{\text{ML}}(\gamma^*) / \partial \gamma$ at the true value $\gamma^*$.
But this application appears to be flawed, because the derivatives
$\partial \ell_{\text{ML}}(\gamma^*) / \partial \gamma_{jk} = \tilde{\E}[\{\pi(k, X; \gamma^*) - R^{(k)}\}f_j(X)]$, $k = 1, \ldots, K-1$,
are interdependent for each $j$, violating the independence assumption in Lounici et al.'s Lemma 9.1.
In contrast, our analysis appropriately tackles a similar interdependency within the gradient of the calibration loss $\ell_{\text{CAL}}(\gamma)$
evaluated at $\bar\gamma_{\text{CAL}}$.
See Supplement Lemma \ref{lem:ps-score}, where the approach can also be applied to appropriately analyze $\hat\gamma_{\text{RML}}$.
\end{rem}

\subsection{Estimation of treatment means} \label{sec:theory-aipw}

With the preceding results on $ ( \hat \alpha^\#_{t,\text{RWL}}, \hat \gamma_{\text{RCAL}})$,
we turn to theoretical analysis of the augmented IPW estimator $\hat \mu_t ( \hat m^\#_{\text{RWL}}, \hat \pi_{\text{RCAL}})$ for the treatment mean $\mu_t$.
The convergence results in Section~\ref{sec:theory-rcal} are obtained with possible misspecification of both PS model (\ref{eq:PSmodel}) and OR model (\ref{eq:ORmodel2}).
However, statistical properties of $\hat\mu_t (\hat \alpha^\#_{t,\text{RWL}}, \hat \gamma_{\text{RCAL}})$ are model-dependent in various ways.
On one hand, $\hat\mu_t (\hat m^\#_{\text{RWL}}, \hat \pi_{\text{RCAL}})$
is expected be pointwise doubly robust similarly as the limit version
$\hat\mu_t (\bar m^\#_{\text{WL}}, \bar\pi_{\text{CAL}})$, i.e., remains consistent for $\mu_t$ if
either model (\ref{eq:PSmodel}) or (\ref{eq:ORmodel2}) is correctly specified.
On the other hand, to obtain valid Wald confidence intervals,
the deviation of $\hat\mu_t (\hat m^\#_{\text{RWL}}, \hat \pi_{\text{RCAL}})$ from
$\hat\mu_t (\bar m^\#_{\text{WL}}, \bar\pi_{\text{CAL}})$ can be shown to be of order $o_p (n^{-1/2})$ as stated in the asymptotic expansion (\ref{eq:desired-expan2}),
depending on whether linear or nonlinear OR model (\ref{eq:ORmodel2}) is used.

First, we assume that linear OR model (\ref{eq:ORmodel2}) is used together with PS model (\ref{eq:PSmodel}), and develop theoretical analysis which leads
to doubly robust Wald confidence intervals for $\mu_t$.

\begin{thm} \label{thm:mu-lm}
Suppose that Assumption~\ref{ass1} and Assumptions~\ref{ass2}(i)-(iii) hold. If $\log\{(K - 1) + \log(p+1)/\epsilon\} / n\leq 1$, then for $A_1 > (\xi_1 + 1) / (\xi_1 - 1)$ and $A_2 > (\xi_2 + 1) / (\xi_2 - 1)$, we have with probability at least $1 - 8\epsilon$,
\begin{align}
& |\hat{\mu}_t(\hat{m}^\#_{\text{RWL}}, \hat{\pi}_{\text{RCAL}}) - \bar{\mu}_t(\bar{m}^\#_{\text{WL}}, \bar{\pi}_{\text{CAL}})| \nonumber \\
& \leq M_{2,1}|S_\gamma|\tilde{\lambda}_1^2 + M_{2,2}|S_\gamma|\tilde{\lambda}_1\tilde{\lambda}_2 + M_{2,3}|S_{\alpha_t}|\tilde{\lambda}_1\tilde{\lambda}_2,
\label{eq:thm-mu-lm}
\end{align}
where $M_{2,1} = (\sqrt{2} + 1)\sigma_0\eta_{1, 4}\me^{\eta_{1, 3}} / 2 + M_{2,3} + \sqrt{2}\sigma_0M_1\me^{2\eta_{1, 3}} / 2$, $M_{2,2} = (A_1 - 1)^{-1}M_1$, $M_{2,3} = A_1(A_2 - 1)^{-1}M_2$, $\eta_{1, 4} = (A_1 - 1)^{-2}M_1^2\eta_1$, $M_2$ is a constant such that the right-hand side of \eqref{eq:thm-or-error-lm} in Remark \ref{rem:ORlm} is upper bounded by $\me^{\eta_{1, 3}}M_2(|S_\gamma|\tilde{\lambda}_1\tilde{\lambda}_2 + |S_{\alpha_t}|\tilde{\lambda}_2^2)$ and $(M_1, \eta_1)$ are as in Theorem \ref{thm:or-error-glm}.
\end{thm}

Theorem \ref{thm:mu-lm} shows that $\hat{\mu}_t(\hat{m}^\#_{\text{RWL}}, \hat{\pi}_{\text{RCAL}})$ is doubly robust for $\bar{\mu}_t(\bar{m}^\#_{\text{WL}}, \bar{\pi}_{\text{CAL}})$ provided $(|S_\gamma| + |S_{\alpha_t}|)\tilde{\lambda}_1^2 = o(1)$, that is, $(|S_\gamma| + |S_{\alpha_t}|)\{(K - 1) + \log(p+1)\} = o(n)$. In addition, Theorem \ref{thm:mu-lm} gives the $n^{-1/2}$ asymptotic expansion (\ref{eq:desired-expan2}) provided $n^{1/2}(|S_\gamma| + |S_{\alpha_t}|)\tilde{\lambda}_1^2 = o(1)$, that is $(|S_\gamma| + |S_{\alpha_t}|)\{(K - 1) + \log(p+1)\} = o(n^{1/2})$. To obtain valid confidence intervals for $\mu_t$ via the Slutsky theorem, the following result gives the consistency of the variance estimator $\hat{V}_t$ to $V_t$, where $\hat{V}_t$ is defined in \eqref{eq:estimatedV} and $V_t = \var\{\varphi_t(Y, T, X; \bar{\alpha}^\#_{t, \text{WL}}, \bar{\gamma}_{\text{CAL}})\}$ with $\varphi_t(Y, T, X; \alpha^\#_t, \gamma)$ defined in \eqref{eq:phi2}.

For notational simplicity, denote $\hat{\varphi}_t = \varphi_t(Y, T, X; \hat{\alpha}^\#_{t,\text{RWL}}, \hat{\gamma}_{\text{RCAL}})$ and $\hat{\varphi}_{tc} = \hat{\varphi}_t - \hat{\mu}_t(\hat{m}^\#_{\text{RWL}}, \hat{\pi}_{\text{RCAL}})$ such that $\hat{V}_t = \tilde{\E}(\hat{\varphi}^2_{tc})$. Similarity, denote $\bar{\varphi}_t = \varphi_t(Y, T, X; \bar{\alpha}^\#_{t,\text{WL}}, \bar{\gamma}_{\text{CAL}})$ and $\bar{\varphi}_{tc} = \bar{\varphi}_t - \bar{\mu}_t(\bar{m}^\#_{\text{WL}}, \bar{\pi}_{\text{CAL}})$ such that $V_t = \E(\bar{\varphi}^2_{tc})$.

\begin{thm} \label{thm:V-lm}
Under the conditions of Theorem \ref{thm:mu-lm}, if $\{(K - 1) + \log(p+1)/\epsilon\} / n \leq 1$, then we have with probability at least $1 - 8\epsilon$,
\begin{align}
|\tilde{\E}(\hat{\varphi}_{tc}^2 - \bar{\varphi}_{tc}^2)| \leq & 2M^{1/2}_{2, 4}\{\tilde{\E}(\bar{\varphi}_{tc}^2)\}^{1/2}(K - 1)^{1/2}(|S_\gamma|\tilde{\lambda}_1 + |S_{\alpha_t}|\tilde{\lambda}_2) \nonumber \\
& + M_{2, 4}(K - 1)(|S_\gamma|\tilde{\lambda}_1 + |S_{\alpha_t}|\tilde{\lambda}_2)^2,
\label{eq:thm-V-lm}
\end{align}
where $M_{2, 4}$ is a positive constant depending only on $(B_0, B_1, A_1, \xi_1, \nu_1, \eta_1)$ in Theorem \ref{thm:ps-error} and $(\sigma_0, A_2, \xi_2, \nu_2, \eta_2)$ in Theorem \ref{thm:or-error-glm}.
\end{thm}

Inequality \eqref{eq:thm-V-lm} shows that $\hat{V}_t$ is a consistent estimator of $V_t$, that is, $\hat{V} - V = o(1)$, provided $(|S_\gamma| + |S_{\alpha_t}|)(K-1)^{1/2}\tilde{\lambda}_1 = o(1)$, which means $(|S_\gamma| + |S_{\alpha_t}|)(K-1)^{1/2}\{(K - 1) + \log(p+1)\}^{1/2} = o(n^{1/2})$.
Combining Theorems \ref{thm:mu-lm} and \ref{thm:V-lm}, we have the following doubly robust Wald confidence intervals for $\mu_t$.
For simplicity,  the group Lasso tuning parameters are denoted as $\lambda_1 = A_1^{\dagger}[\{(K - 1) + \log(p+1)\}/n]^{1/2}$ for $\hat{\gamma}_{\text{RCAL}}$ and $\lambda_2 = A_2^{\dagger}[\{(K - 1) + \log(p+1)\}/n]^{1/2}$ for $\hat{\alpha}^\#_{t, \text{RWL}}$.

\begin{pro} \label{pro:double-mu-lm}
Suppose that Assumption~\ref{ass1} and Assumption~\ref{ass2}(i), \ref{ass2}(ii), and \ref{ass2}(iv)(a) hold, and $(|S_\gamma| + |S_{\alpha_t}|)(K-1)^{1/2}\{(K - 1) + \log(p+1)\} = o(n^{1/2})$. Then for $\hat{\gamma}_{\text{RCAL}}$ and $\hat{\alpha}^\#_{t, \text{RWL}}$ with sufficiently large constants $A_1^{\dagger}$ and $A_2^{\dagger}$, asymptotic expansion (\ref{eq:desired-expan2}) is valid. Moreover, if either PS model (\ref{eq:PSmodel}) or linear OR model (\ref{eq:ORmodel2}) is correctly specified, the following results hold:\vspace{-.1in}
\begin{itemize}\addtolength{\itemsep}{-.1in}
\item[(i)] $n^{1/2}\{\hat{\mu}_t(\hat{m}^\#_{\text{RWL}}, \hat{\pi}_{\text{RCAL}}) - \mu_t\}\overset{D}{\to}N(0, V_t)$, where $V_t =\var\{\varphi_t(Y, T, X; \bar{\alpha}^\#_{t, \text{WL}}, \bar{\gamma}_{\text{CAL}})\}$;
\item[(ii)] a consistent estimator of $V$ is
\begin{align*}
\hat{V}_t = \tilde{\E}\left[\{\varphi_t(Y, T, X; \hat{\alpha}^\#_{t, \text{RWL}}, \hat{\gamma}_{\text{RCAL}}) - \hat{\mu}_t(\hat{m}^\#_{\text{RWL}}, \hat{\pi}_{\text{RCAL}})\}^2\right];
\end{align*}
\item[(iii)] an asymptotic $(1 - c)$ confidence interval for $\mu_t$ is $\hat{\mu}_t(\hat{m}^\#_{\text{RWL}}, \hat{\pi}_{\text{RCAL}})\pm z_{c/2}\sqrt{\hat{V}_t / n}$, where $z_{c/2}$ is the $(1 - c/2)$ quantile of $N(0, 1)$.
\end{itemize}\vspace{-.1in}
That is, a doubly robust confidence interval for $\mu_t$ is obtained.
\end{pro}


Second, we assume that a generalized linear OR model (\ref{eq:ORmodel2}) is used together with PS model (\ref{eq:PSmodel}), and develop theoretical analysis which leads
to valid Wald confidence intervals for $\mu_t$ if model (\ref{eq:PSmodel}) is correctly specified.  Compared with the upper bound of $|\hat{\mu}_t(\hat{m}^\#_{\text{RWL}}, \hat{\pi}_{\text{RCAL}}) - \bar{\mu}_t(\bar{m}^\#_{\text{WL}}, \bar{\pi}_{\text{CAL}})|$ for linear OR model (\ref{eq:ORmodel2}) in Theorem \ref{thm:mu-lm}, the upper bound for generalized linear OR model (\ref{eq:ORmodel2}) depends on additional terms $\Lambda(r)$, which is defined as, for any $r\geq 0$,
\begin{align*}
\Lambda(r) = \sup_{j=0, 1, \ldots, p, \lVert \alpha^\#_{t} - \bar{\alpha}^\#_{t, \text{WL}} \rVert_{2, 1} \leq r} \left\lVert \E\left[\diag\{\psi_2\{\alpha^{(k)\T}_tf(X)\}: k\neq t\}g(X; \bar{\gamma}_{\text{CAL}})f_j(X)\right] \right\rVert_2
\end{align*}
with $g(X; \bar{\gamma}_{\text{CAL}}) = (g_k: k\neq t)^\T$ and $g_k =R^{(t)}\pi(k, X; \bar{\gamma}_{\text{CAL}}) / \pi(t, X; \bar{\gamma}_{\text{CAL}}) - R^{(k)}$.
By the definition of $\bar{\gamma}_{\text{CAL}}$, it holds that $E[g(X; \bar{\gamma}_{\text{CAL}})f_j(X)] = 0$ for $j = 0, 1, \ldots, p$ whether or not model \eqref{eq:PSmodel} is correctly specified. But $\Lambda(r)$ is in general either zero or positive respectively if outcome model \eqref{eq:ORmodel} is correctly specified or misspecified, except in the case of linear outcome model where $\Lambda(r)$ is automatically zero because $\psi_2(\cdot)$ is constant.

\begin{thm} \label{thm:mu-glm}
Suppose that Assumptions \ref{ass1} and \ref{ass2} hold. If $\{(K - 1) + \log(p+1) / \epsilon\} / n \leq 1$, then for $A_1 > (\xi_1 + 1) / (\xi_1 - 1)$ and $A_2 > (\xi_2 + 1) / (\xi_2 - 1)$, we have with probability at least $1 - 9\epsilon$
\begin{align}
& |\hat{\mu}_t(\hat{m}^\#_{\text{RWL}}, \hat{\pi}_{\text{RCAL}}) - \bar{\mu}_t(\bar{m}^\#_{\text{WL}}, \bar{\pi}_{\text{CAL}})| \nonumber \\
& \leq M_{3,1}|S_\gamma|\tilde{\lambda}_1^2 + M_{3,2}|S_\gamma|\tilde{\lambda}_1\tilde{\lambda}_2 + M_{3,3}|S_{\alpha_t}|\tilde{\lambda}_1\tilde{\lambda}_2 + \eta_3\Lambda(\eta_3), \label{eq:thm-mu-glm}
\end{align}
where $M_{3,1}, M_{3,2}$, and $M_{3,3}$ are positive constants, depending only on $(B_0, B_1, A_1, \xi_1, \nu_1, \eta_1)$ from Theorem \ref{thm:ps-error}, $(\sigma_0, A_2, \xi_2, \nu_2, \eta_2)$ from Remark \ref{rem:ORlm} and $(C_1, C_2, C_3, \eta_{2, 1}, \eta_{2, 2})$ from Theorem \ref{thm:or-error-glm}, $\eta_3 = (A_2 - 1)^{-1}M_3(|S_\gamma|\tilde{\lambda}_1 + |S_{\alpha_t}|\tilde{\lambda}_2)$, and $M_3$ is a constant such that the right-hand side of \eqref{eq:thm-or-error-glm} is upper-bounded by $\me^{\eta_{1, 3}}M_3(|S_\gamma|\tilde{\lambda}_1\tilde{\lambda}_2 + |S_{\alpha_t}|\tilde{\lambda}_2^2)$.
\end{thm}

Theorem \ref{thm:mu-glm} shows that $\hat{\mu}_t(\hat{m}^\#_{\text{RWL}}, \hat{\pi}_{\text{RCAL}})$ is doubly robust for $\mu_t$ provided $(|S_\gamma| + |S_{\alpha_t}|)\tilde{\lambda}_1 = o(1)$, that is, $(|S_\gamma| + |S_{\alpha_t}|)\{(K - 1) + \log(p+1)\}^{1/2} = o(n^{1/2})$. In addition, the error bounds imply that $\hat{\mu}_t(\hat{m}^\#_{\text{RWL}}, \hat{\pi}_{\text{RCAL}})$ admits the $n^{-1/2}$ asymptotic expansion (\ref{eq:desired-expan2}) provided $n^{1/2}(|S_\gamma| + |S_{\alpha_t}|)\tilde{\lambda}_1^2 = o(1)$, that is $(|S_\gamma| + |S_{\alpha_t}|)\{(K - 1) + \log(p+1)\} = o(n^{1/2})$ when PS model is correctly specified but OR model may be misspecified, because the term involving $\Lambda(\eta_3)$ vanishes when PS model \eqref{eq:PSmodel} is correctly specified. Unfortunately, asymptotic expansion may fail when PS model is misspecified. The following result gives the consistency of the variance estimator $\hat{V}_t$ to $V_t$ with a generalized linear OR model \eqref{eq:ORmodel} together with PS model (\ref{eq:PSmodel}).

\begin{thm} \label{thm:V-glm}
Under the conditions of Theorem \ref{thm:mu-glm}, if $\{(K - 1) + \log(p+1)/\epsilon\} / n \leq 1$, then we have with probability at least $1 - 9\epsilon$,
\begin{align}
|\tilde{\E}(\hat{\varphi}_{tc}^2 - \bar{\varphi}_{tc}^2)| \leq & 2M^{1/2}_{3,4}\{\tilde{\E}(\bar{\varphi}_{tc}^2)\}^{1/2}\{(K - 1) + \Lambda^2(\eta_3)\}^{1/2}(|S_\gamma|\tilde{\lambda}_1 + |S_{\alpha_t}|\tilde{\lambda}_2) \nonumber \\
& + M_{3,4}\{(K - 1) + \Lambda^2(\eta_3)\}(|S_\gamma|\tilde{\lambda}_1 + |S_{\alpha_t}|\tilde{\lambda}_2)^2,
\label{eq:thm-V-glm}
\end{align}
where $M_{3,4}$ is a positive constant depending only on $(B_0, B_1, A_1, \xi_1, \nu_1, \eta_1)$ from Theorem \ref{thm:ps-error}, $(\sigma_0, A_2, \allowbreak \xi_2, \nu_2, \eta_2)$ from Remark \ref{rem:ORlm} and $(C_1, C_2, C_3, \eta_{2, 1}, \eta_{2, 2})$ from Theorem \ref{thm:or-error-glm}.
\end{thm}

Inequality \eqref{eq:thm-V-glm} shows that $\hat{V}$ is a consistent estimator of $V$, that is, $\hat{V} - V = o(1)$, provided $(|S_\gamma| + |S_{\alpha_t}|)(K-1)^{1/2}\tilde{\lambda}_1 = o(1)$, that is, $(|S_\gamma| + |S_{\alpha_t}|)(K-1)^{1/2}\{(K - 1) + \log(p+1)\}^{1/2} = o(n^{1/2})$.
Combining Theorems \ref{thm:mu-lm} and \ref{thm:V-lm} gives the following model-assisted Wald confidence intervals for $\mu_t$.

\begin{pro} \label{pro:double-mu-glm}
Suppose that Assumptions \ref{ass1} and \ref{ass2} hold, and $(|S_\gamma| + |S_{\alpha_t}|)(K-1)^{1/2}\{(K - 1) + \log(p+1)\} = o(n^{1/2})$. For sufficiently large constants $A_1^\dag$ and $A_2^\dag$, if PS model \eqref{eq:PSmodel} is correctly specified but OR model \eqref{eq:ORmodel} may be misspecified, then (i)-(iii) in Proposition \ref{pro:double-mu-lm} hold. That is, a PS based, OR assisted confidence interval for $\mu_t$ is obtained.
\end{pro}

\section{ATT estimation} \label{sec:ATT}

Our method and theory can be extended for estimating ATTs using the same set of regularized calibrated estimators as in ATE estimation.
From Section~\ref{sec:setup}, the ATT for treatment $t$ vs $k$ in the $k$th treated group is
$ \E ( Y^{(t)} - Y^{(k)} | T=k)$.
The mean $\E ( Y^{(k)} | T= k )$ can be directly estimated as $ \tilde \E ( Y R^{(k)} ) / \tilde \E ( R^{(k)} )$, i.e.,
the sample average of $Y$ within the $k$th treated group.
In the following, we mainly discuss estimation of the mean $\nu_t^{(k)} =  \E ( Y^{(t)}  | T=k)$ for $t \not= k$.

By the relation $\nu_t^{(k)} = \E ( Y^{(t)}R^{(k)} ) / \E ( R^{(k)})$, our estimator for $\nu_t^{(k)} $ is
\begin{align*}
\hat{\nu}^{(k)}_{t,\text{RCAL}} = \frac{\tilde \E \{\varphi_t^{(k)} (Y,T,X; \hat\alpha_{t,\text{RWL}}^{(k)},\hat\gamma_{\text{RCAL}}) \}}{\tilde{\E}(R^{(k)} )} ,
\end{align*}
where $(\hat\alpha_{t,\text{RWL}}^{(k)},\hat\gamma_{\text{RCAL}})$ are defined as in $\hat \mu_{t,\text{RCAL}} = \hat \mu_t ( \hat m^\#_{\text{RWL}}, \hat \pi_{\text{RCAL}})$,
and $\varphi_t^{(k)} (Y,T,X; \alpha_t^{(k)},\gamma)$ is defined in (\ref{eq:phi3}) such that
$\tilde \E \{\varphi_t^{(k)} (Y,T,X; \hat\alpha_{t,\text{RWL}}^{(k)},\hat\gamma_{\text{RCAL}}) \}$ is an augmented IPW estimator for $\E ( Y^{(t)}R^{(k)} )$.
Hence our estimators for $\mu_t$ and $\nu_t^{(k)}$ satisfy the natural decomposition:
\begin{align*}
\hat \mu_{t,\text{RCAL}} =\tilde \E ( Y R^{(t)} )+ \sum_{k\not= t}  \hat{\nu}^{(k)}_{t,\text{RCAL}}  \tilde{\E}(R^{(k)} ) ,
\end{align*}
in accordance with our earlier discussion about the estimator $\mu_t (\hat m^\#, \hat\pi)$ in (\ref{eq:ATE-aipw2}).
In this sense, our approach handles estimation of ATEs and ATTs in a unified manner.

Statistical properties of $\hat{\nu}^{(k)}_{t,\text{RCAL}}$ can be deduced similarly as in Section~\ref{sec:theory-aipw}.
In particular, if PS model (\ref{eq:PSmodel}) is correctly specified but OR model (\ref{eq:ORmodel2})
may be specified, then $\hat{\nu}^{(k)}_{t,\text{RCAL}} $ can be shown to admit the asymptotic expansion under suitable sparsity condition,
\begin{align}
\hat{\nu}^{(k)}_{t,\text{RCAL}}  = \frac{\tilde \E \{\varphi_t^{(k)} (Y,T,X; \bar\alpha_{t,\text{WL}}^{(k)},\bar\gamma_{\text{CAL}}) \} }{\tilde{\E}(R^{(k)})} + o_p(n^{-1/2}), \label{eq:nu-expan}
\end{align}
where $(\bar\alpha_{t,\text{WL}}^{(k)},\bar\gamma_{\text{CAL}})$ are the limit values of $ (\hat\alpha_{t,\text{RWL}}^{(k)},\hat\gamma_{\text{RCAL}})$ as before.
Then an asymptotic $(1-c)$ confidence interval for $\nu_t^{(k)}$ can be obtained as
$ \hat{\nu}^{(k)}_{t,\text{RCAL}} \pm z_{c/2}\sqrt{\hat U_t^{(k)} / n}$,
where
\begin{align}
\hat U_t^{(k)} = \tilde{\E} \left[\{ \varphi_t^{(k)} (Y, T, X; \hat{\alpha}_{t,\text{RWL}}^{(k)}, \hat{\gamma}_{\text{RCAL}}) - R^{(k)} \hat{\nu}^{(k)}_{t,\text{RCAL}}  \}^2\right]  / \tilde{\E}^2(R^{(k)}) . \label{eq:estimatedU}
\end{align}
For a linear OR model, the asymptotic expansion (\ref{eq:nu-expan}) can be established, with possible misspecification of
both models (\ref{eq:PSmodel}) and (\ref{eq:ORmodel2}). In this case, the confidence intervals for $\nu_t^{(k)}$ are doubly robust,
being valid when either model (\ref{eq:PSmodel}) or linear model (\ref{eq:ORmodel2}) is correctly specified.

\section{Simulation study} \label{sec:simulation}

We conduct a simulation study to evaluate the performance of $\hat{\mu}_t(\hat{m}^\#_{\text{RWL}}, \hat{\pi}_{\text{RCAL}})$ compared with $\hat{\mu}_t(\hat{m}_{\text{RMLs}}, \hat{\pi}_{\text{RML}})$ and $\hat{\mu}_t(\hat{m}_{\text{RMLg}}, \hat{\pi}_{\text{RML}})$ for ATE estimation,
where the fitted value $\hat{m}_{\text{RMLs}}(t,X)$ is based on the separate Lasso estimator $\hat\alpha_{t,\text{RMLs}}$
and $\hat{m}_{\text{RMLg}}(t,X)$ is based on the group Lasso estimator $\{\hat\alpha_{t,\text{RMLg}}: t\in\mathcal T\}$
as described in Section \ref{sec:setup}.
In addition, we also compare the performance of $\hat{\nu}^{(k)}_{t, \text{RCAL}}$ with $\hat{\nu}^{(k)}_{t, \text{RMLs}}$ and $\hat{\nu}^{(k)}_{t, \text{RMLg}}$
for ATT estimation and include the corresponding results in the Supplementary Material, where
$\hat{\nu}^{(k)}_{t, \text{RMLs}} = \tilde{\E}\{\varphi_t^{(k)}(Y, T, X; \hat{\alpha}_{t, \text{RMLs}}, \hat{\gamma}_{\text{RML}})\} / \tilde{\E}(R^{(k)})$ and $\hat{\nu}^{(k)}_{t, \text{RMLg}} = \tilde{\E}\{\varphi_t^{(k)}(Y, T, X; \hat{\alpha}_{t, \text{RMLg}}, \hat{\gamma}_{\text{RML}})\} / \tilde{\E}(R^{(k)})$
for $k\neq t$,
similarly as $\hat{\nu}^{(k)}_{t, \text{RCAL}}$ in Section \ref{sec:ATT}.

The algorithms for computing the RCAL estimators
$\hat{\gamma}_{\text{RCAL}}$ and $\hat{\alpha}^\#_{t, \text{RWL}}$ are described in Section ~\ref{sec:computation},
combining Fisher scoring, the MM technique and the block coordinate descent.
The constraint $\gamma_t \equiv 0$ is used in $\hat\gamma_{\text{RCAL}}$ when estimating $\mu_t$.
We compute the RML estimators $\hat{\gamma}_{\text{RML}}$, $\hat\alpha_{t,\text{RMLs}}$, and $\{\hat\alpha_{t,\text{RMLg}}: t\in\mathcal T\}$
similarly as in the R package \texttt{glmnet} (Friedman et al.~2010; Simon et al.~2013),
except that the one-to-zero constraint $\gamma_0 \equiv 0$ is used in $\hat{\gamma}_{\text{RML}}$.
For completeness, we also conduct  simulations using the sum-to-zero constraint
$\sum_{k\in\mathcal T} \gamma_k \equiv 0$ in $\hat{\gamma}_{\text{RCAL}}$  and $\hat{\gamma}_{\text{RML}}$.
The results are presented in the Supplementary Material, and similar conclusions are obtained as below.
All the methods are implemented in the R package \texttt{mRCAL}, including RCAL and RMLs and RMLg.

For computing $\hat{\gamma}_{\text{RML}}$ or $\hat{\gamma}_{\text{RCAL}}$,
the group-Lasso tuning parameter is determined using 5-fold cross validation based on the corresponding loss function as follows. For $s = 1, \ldots, 5$, let $\mathcal I_s$ be a random subsample of size $n / 5$ form $\{1, 2, \ldots, n\}$. For a loss function $\ell(\gamma)$, either $\ell_{\text{ML}}(\gamma)$ in \eqref{eq:RML-gamma-obj} or $\ell_{\text{CAL}}(\gamma)$ in \eqref{eq:PSloss}, denote by $\ell(\gamma; \mathcal I)$ the loss function obtained when the sample average $\tilde{\E}()$ is computed over only the subsample $\mathcal I$. The 5-fold cross-validation criterion is defined as $\text{CV}_5(\lambda_1) = (1/5)\sum_{s=1}^5\ell(\hat{\gamma}^{(s)}_{\lambda_1}; \mathcal I_s)$, where $\hat{\gamma}^{(s)}_{\lambda_1}$ is a minimizer of the penalized loss $\ell(\gamma; \mathcal I^c_s) + \lambda_1 \sum_{j=1}^p \lVert \gamma_{j.} \rVert_2$ over the subsample $\mathcal I^c_s$ of size $4n / 5$, i.e., the complement to $\mathcal I_s$. Then $\lambda_1$ is selected by minimizing $\text{CV}_5(\lambda_1)$ over the discrete set $\{\lambda^*_1\times 0.01^{j/20}: j = 0, 1, \ldots, 20\}$, where for $\hat{\pi} = (\tilde{\E}(R^{(k)}): k\in\mathcal T\backslash\{t\})^\T$, the value $\lambda_1^*$ is computed as $\lambda_1^* = \max_{j=1}^p \lVert\tilde{\E}[g(T, \hat{\pi})f_j(X)]\rVert_2$ with
$g(T, \hat{\pi}) = (\hat{\pi}_k - R^{(k)}: k\in\mathcal T\backslash\{t\})^\T$ when the likelihood loss \eqref{eq:RML-gamma-obj} is used,
or $g(T, \hat{\pi}) = (g_k: k\in\mathcal T\backslash\{t\})^\T$ with $g_k = R^{(t)}\hat{\pi}_{k} / \hat{\pi}_{t} - R^{(k)}$ when the calibration loss \eqref{eq:PSloss} is used.

For selection of the tuning parameter in $\hat{\alpha}_{t, \text{RMLs}}$, $\{\hat{\alpha}_{t, \text{RMLg}}: t\in\mathcal T\}$, or $\hat{\alpha}^\#_{t, \text{RCAL}}$,
5-fold cross validation is conducted similarity as above using respectively the loss function $\ell_{\text{ML}}(\alpha_t)$ in \eqref{eq:RML-alpha-obj1}, $\sum_{t\in\mathcal T}\ell_{\text{ML}}(\alpha_t)$ in \eqref{eq:RML-alpha-obj2}, or $\ell_{\text{WL}}(\alpha^\#_t; \hat{\gamma}_{\text{RCAL}})$ in \eqref{eq:ORobj}. In the last case, $\hat{\gamma}_{\text{RCAL}}$ is determined separately and then fixed during cross validation for computing $\hat{\alpha}^\#_{t, \text{RWL}}$.

In our simulation, the number of treatments $K$ is 4, and sample size $n$ is 1000. Let $X = (X_1, \ldots, X_p)^\T$ be multivariate normal with means 0 and covariance $\cov(X_{j_1}, X_{j_2}) = 2^{-|j_1 - j_2|}$ for $1\leq j_1, j_2 \leq p$. In addition, let $W_j = X_j + \{(X_j + 1)_{+}\}^2$ for $j = 1, \ldots, 4$, where $c_{+} = \max(0, c)$, and $X^\dagger_j = \{W_j - \E(W_j)\} / \sqrt{\text{var}(W_j)}$. Consider the following the data-generating configurations.\vspace{-.1in}
\begin{itemize}\addtolength{\itemsep}{-.1in}
\item[(C1)] Generate $T$ given $X$ from a categorical distribution with $\log\{\P(T = 1 | X) / \P(T = 0 | X) \} = X_1 - 0.5X_2 - 0.25X_3 + 0.125X_4$, $\log\{\P(T = 2 | X) / \P(T = 0 | X) \} = -0.5X_1 - 0.25X_2 + 0.125X_3 + X_4$ and $\log\{\P(T = 3 | X) / \P(T = 0 | X) \} = -0.25X_1 + 0.125X_2 + X_3 - 0.5X_4$,
and, independently, $Y$ given $(T, X)$ is generated from a Normal distribution with variance 1 and mean $\E(Y | T = 0, X) = 0 + X_1 - X_5 - X_6 + X_7$, $\E(Y | T = 1, X) = 1 + X_2 - X_7 - X_8 + X_9$, $\E(Y | T = 2, X) = 2 + X_3 - X_9 - X_{10} + X_{11}$ and $\E(Y | T = 3, X) = 3 + X_4 - X_{11} - X_{12} + X_{13}$.

\item[(C2)] Generate $T$ given $X$ as in (C1). But $Y$ given $(T, X)$ is generated from a Normal distribution with variance 1 and mean $\E(Y | T = 0, X) = 0 + X^\dagger_1 - X^\dagger_5 - X^\dagger_6 + X^\dagger_7$, $\E(Y | T = 1, X) = 1 + X^\dagger_2 - X^\dagger_7 - X^\dagger_8 + X^\dagger_9$, $\E(Y | T = 2, X) = 2 + X^\dagger_3 - X^\dagger_9 - X^\dagger_{10} + X^\dagger_{11}$ and $\E(Y | T = 3, X) = 3 + X^\dagger_4 - X^\dagger_{11} - X^\dagger_{12} + X^\dagger_{13}$.

\item[(C3)] Generate $Y$ given $(T, X)$ as in (C1). But $T$ given $X$ is generated with $\log\{\P(T = 1 | X) / \P(T = 0 | X) \} = X^\dagger_1 - 0.5X^\dagger_2 - 0.25X^\dagger_3 + 0.125X^\dagger_4$, $ \log\{\P(T = 2 | X) / \P(T = 0 | X) \} = -0.5X^\dagger_1 - 0.25X^\dagger_2 + 0.125X^\dagger_3 + X^\dagger_4$ and $\log\{\P(T = 3 | X) / \P(T = 0 | X) \} = -0.25X^\dagger_1 + 0.125X^\dagger_2 + X^\dagger_3 - 0.5X^\dagger_4$.
\end{itemize}\vspace{-.1in}
Consider multi-class logistic propensity score model \eqref{eq:PSmodel} and linear outcome model \eqref{eq:ORmodel2}, both with $f_j(X) = X_j$ for $j = 1, \ldots, p$. Then the two models can be classified as follows, depending on the data configuration above:\vspace{-.1in}
\begin{itemize}\addtolength{\itemsep}{-.1in}
\item[(C1)] PS and OR models both correctly specified;
\item[(C2)] PS model correctly specified, but OR model misspecified;
\item[(C3)] PS model misspecified, but OR model correctly specified.
\end{itemize}\vspace{-.1in}
Partly because the regressor $X_j^\dag$ is a monotone nonlinear transformation of $X_j$ for $j=1,\ldots, 4$,
the misspecified OR model in (C2) or PS model in (C1) appears to be nearly correct by standard model diagnosis.
See the Supplement Material for boxplots of $X_j$ and scatterplots of $Y$ against $X_j$  within different treatment groups for $j=1,\ldots,4$.

\begin{table} 
\caption{\footnotesize Summary of $\hat{\mu}_t$ for $t = 0, 1, 2, 3$.} \label{tb:mu_cons} \vspace{-4ex}
\begin{center}
\resizebox{1.0\textwidth}{!}{\begin{tabular}{lccccccccccccccccccccccc}
\hline
& \multicolumn{3}{c}{(C1) cor PS, cor OR} &~~& \multicolumn{3}{c}{(C2) cor PS, mis OR} &~~& \multicolumn{3}{c}{\small (C3) mis PS, cor OR} &~~& \multicolumn{3}{c}{(C1) cor PS, cor OR} &~~& \multicolumn{3}{c}{(C2) cor PS, mis OR} &~~& \multicolumn{3}{c}{\small (C3) mis PS, cor OR}\\
& RCAL & RMLs & RMLg &~~& RCAL & RMLs & RMLg &~~& RCAL & RMLs & RMLg &~~& RCAL & RMLs & RMLg &~~& RCAL & RMLs & RMLg &~~& RCAL & RMLs & RMLg \\
\hline
& \multicolumn{23}{c}{n = 1000, p = 50} \\
& \multicolumn{11}{c}{\footnotesize $\hat{\mu}_0$} &~~& \multicolumn{11}{c}{\footnotesize $\hat{\mu}_1$} \\
\cline{2-12}\cline{14-24}
Bias & -0.001 & -0.021 & -0.005 &~~& -0.006 & -0.021 & -0.011 &~~& -0.004 & -0.027 & -0.008 &~~& -0.015 & -0.067 & -0.020 &~~& -0.019 & -0.058 & -0.025 &~~& -0.012 & -0.033 & -0.016 \\
$\sqrt{\text{Var}}$ & 0.093 & 0.098 & 0.095 &~~& 0.100 & 0.109 & 0.102 &~~& 0.093 & 0.096 & 0.094 &~~& 0.095 & 0.101 & 0.098 &~~& 0.101 & 0.107 & 0.104 &~~& 0.095 & 0.102 & 0.100 \\
$\sqrt{\text{EVar}}$ & 0.090 & 0.091 & 0.088 &~~& 0.092 & 0.098 & 0.093 &~~& 0.088 & 0.088 & 0.086 &~~& 0.088 & 0.092 & 0.087 &~~& 0.089 & 0.096 & 0.090 &~~& 0.087 & 0.095 & 0.090 \\
Cov90 & 0.885 & 0.865 & 0.875 &~~& 0.869 & 0.864 & 0.869 &~~& 0.868 & 0.860 & 0.850 &~~& 0.869 & 0.793 & 0.861 &~~& 0.853 & 0.810 & 0.838 &~~& 0.854 & 0.854 & 0.845 \\
Cov95 & 0.937 & 0.924 & 0.939 &~~& 0.931 & 0.922 & 0.925 &~~& 0.930 & 0.916 & 0.920 &~~& 0.922 & 0.877 & 0.918 &~~& 0.907 & 0.881 & 0.904 &~~& 0.914 & 0.911 & 0.907 \\
& \multicolumn{11}{c}{\footnotesize $\hat{\mu}_2$} &~~& \multicolumn{11}{c}{\footnotesize $\hat{\mu}_3$} \\
\cline{2-12}\cline{14-24}
Bias & 0.001 & 0.034 & 0.000 &~~& -0.006 & 0.025 & -0.010 &~~& -0.003 & 0.021 & -0.005 &~~& -0.016 & -0.065 & -0.026 &~~& -0.025 & -0.072 & -0.045 &~~& -0.017 & -0.044 & -0.026 \\
$\sqrt{\text{Var}}$ & 0.100 & 0.106 & 0.103 &~~& 0.107 & 0.114 & 0.110 &~~& 0.098 & 0.105 & 0.102 &~~& 0.100 & 0.105 & 0.103 &~~& 0.105 & 0.111 & 0.109 &~~& 0.096 & 0.100 & 0.098 \\
$\sqrt{\text{EVar}}$ & 0.088 & 0.092 & 0.089 &~~& 0.090 & 0.097 & 0.092 &~~& 0.088 & 0.093 & 0.089 &~~& 0.088 & 0.090 & 0.086 &~~& 0.089 & 0.096 & 0.089 &~~& 0.087 & 0.089 & 0.086 \\
Cov90 & 0.852 & 0.823 & 0.839 &~~& 0.834 & 0.822 & 0.838 &~~& 0.852 & 0.844 & 0.837 &~~& 0.860 & 0.769 & 0.821 &~~& 0.826 & 0.760 & 0.792 &~~& 0.860 & 0.822 & 0.847 \\
Cov95 & 0.910 & 0.895 & 0.902 &~~& 0.913 & 0.890 & 0.900 &~~& 0.916 & 0.904 & 0.915 &~~& 0.917 & 0.841 & 0.886 &~~& 0.902 & 0.836 & 0.861 &~~& 0.917 & 0.897 & 0.902 \\
& \multicolumn{23}{c}{n = 1000, p = 300} \\
& \multicolumn{11}{c}{\footnotesize $\hat{\mu}_0$} &~~& \multicolumn{11}{c}{\footnotesize $\hat{\mu}_1$} \\
\cline{2-12}\cline{14-24}
Bias & 0.005 & -0.015 & 0.000 &~~& -0.002 & -0.013 & -0.006 &~~& 0.000 & -0.024 & -0.009 &~~& -0.014 & -0.073 & -0.027 &~~& -0.023 & -0.058 & -0.032 &~~& -0.013 & -0.056 & -0.027 \\
$\sqrt{\text{Var}}$ & 0.096 & 0.102 & 0.099 &~~& 0.101 & 0.110 & 0.107 &~~& 0.091 & 0.097 & 0.092 &~~& 0.094 & 0.105 & 0.099 &~~& 0.099 & 0.107 & 0.105 &~~& 0.092 & 0.099 & 0.096 \\
$\sqrt{\text{EVar}}$ & 0.087 & 0.087 & 0.087 &~~& 0.089 & 0.091 & 0.092 &~~& 0.086 & 0.086 & 0.085 &~~& 0.083 & 0.085 & 0.082 &~~& 0.083 & 0.085 & 0.083 &~~& 0.083 & 0.086 & 0.083 \\
Cov90 & 0.869 & 0.847 & 0.866 &~~& 0.855 & 0.823 & 0.854 &~~& 0.884 & 0.850 & 0.871 &~~& 0.845 & 0.716 & 0.809 &~~& 0.825 & 0.742 & 0.796 &~~& 0.859 & 0.791 & 0.829 \\
Cov95 & 0.928 & 0.910 & 0.925 &~~& 0.912 & 0.881 & 0.911 &~~& 0.938 & 0.899 & 0.933 &~~& 0.912 & 0.804 & 0.888 &~~& 0.882 & 0.826 & 0.868 &~~& 0.925 & 0.871 & 0.893 \\
& \multicolumn{11}{c}{\footnotesize $\hat{\mu}_2$} &~~& \multicolumn{11}{c}{\footnotesize $\hat{\mu}_3$} \\
\cline{2-12}\cline{14-24}
Bias & 0.003 & 0.045 & -0.002 &~~& -0.007 & 0.028 & -0.011 &~~& 0.000 & 0.029 & -0.005 &~~& -0.025 & -0.084 & -0.052 &~~& -0.044 & -0.091 & -0.080 &~~& -0.023 & -0.064 & -0.047 \\
$\sqrt{\text{Var}}$ & 0.095 & 0.103 & 0.102 &~~& 0.099 & 0.107 & 0.107 &~~& 0.095 & 0.101 & 0.100 &~~& 0.097 & 0.107 & 0.102 &~~& 0.102 & 0.108 & 0.108 &~~& 0.089 & 0.098 & 0.094 \\
$\sqrt{\text{EVar}}$ & 0.083 & 0.085 & 0.082 &~~& 0.084 & 0.086 & 0.084 &~~& 0.084 & 0.086 & 0.083 &~~& 0.084 & 0.083 & 0.081 &~~& 0.084 & 0.084 & 0.083 &~~& 0.084 & 0.083 & 0.082 \\
Cov90 & 0.842 & 0.787 & 0.820 &~~& 0.828 & 0.795 & 0.792 &~~& 0.852 & 0.822 & 0.829 &~~& 0.819 & 0.675 & 0.757 &~~& 0.782 & 0.650 & 0.675 &~~& 0.866 & 0.739 & 0.789 \\
Cov95 & 0.914 & 0.854 & 0.891 &~~& 0.895 & 0.864 & 0.873 &~~& 0.916 & 0.889 & 0.901 &~~& 0.900 & 0.764 & 0.831 &~~& 0.856 & 0.740 & 0.757 &~~& 0.930 & 0.823 & 0.865 \\
\hline
\end{tabular}}
\end{center}
\setlength{\baselineskip}{0.5\baselineskip}
\vspace{-.15in}\noindent{\tiny
\textbf{Note}: RCAL denotes $\hat{\mu}_t(\hat{m}^\#_{\text{RWL}}, \hat{\pi}_{\text{RCAL}})$, RMLs denotes $\hat{\mu}_t(\hat{m}_{\text{RMLs}}, \hat{\pi}_{\text{RML}})$ and RMLg denotes $\hat{\mu}_t(\hat{m}_{\text{RMLg}}, \hat{\pi}_{\text{RML}})$. Bias and Var are the Monte Carlo bias and variance of the point estimates. EVar is the mean of the variance estimates, and hence $\sqrt{\text{EVar}}$ also measures the $L_2$-average of lengths of confidence intervals. Cov90 or Cov95 is the coverage proportion of the 90\% or 95\% confidence intervals.}
\end{table}

Table \ref{tb:mu_cons} summarizes estimates of $\mu_t$ for $t = 0, 1, 2, 3$, based on 1000 repeated simulations. Three interesting findings can be obtained.
First, in terms of bias, RCAL overall has the smallest absolute bias, followed by RMLg and RMLs. For example, the biases of RMLs, RMLg and RCAL are -0.064, -0.047, and -0.023 for estimation of $\mu_3$ under (C3) with $p = 300$. Second, in terms of variance, the three methods perform similarly to each other.
Third, in terms of coverage, RCAL achieves coverage proportions overall the closet to the nominal levels. For example, the 90\% coverage proportions of RMLs, RMLg and RCAL are 0.739, 0.789 and 0.866 for estimation of $\mu_3$ under (C3) with $p = 300$. These properties can also seen from QQ plots of the $t$-statistics in the Supplement.

\section{Empirical application}\label{sec:empirical}

We provide an application to study the effects of maternal smoking during pregnancy on birth weights for singleton births in Pennsylvania between 1989 and 1991,
based on a dataset from the US National Center of Health Statistics. The original dataset was also used in Almond et al.~(2005) and Cattaneo (2010) among others.
To assess multi-valued treatment effects, the treatment $T$ is taken to be the number of cigarettes smoked per day collapsed into 6 levels $\{0, 1–5, 6–10, 11–15, 16–20, 21+\}$, labeled respectively as $0,1,\ldots,5$, as in Cattaneo (2010). The outcome $Y$ is defined as the log of the birth weight, hence different from that in Almond et al.~(2005) and Cattaneo (2010).
After dropping about 18\% observations with missing data and converting categorical covariates into dummy variables, the total sample size is 411609
and the number of covariates in $X$ is 33.
The sample sizes of the six treatment groups are 339659, 13798, 29746, 4923, 19517, 3966.
See Supplement Section \ref{sec:add-empi} for more information about data preprocessing.

To control for possible confounding beyond main effects of the covariates, we consider a multi-class logistic propensity score model and linear outcome model, where the regressor vector $f(X)$ includes all main effects and two-way interactions of $X$ except those with the fractions of nonzero values less than 0.8\% of the sample size
$n=411609$. The dimension of $f(X)$ is $p=399$, excluding the constant. All variables in $f(X)$ are standardized with sample means 0 and variances 1.

Given the large sample size $n=411609$ relative to the regressor size $p=399$, we investigate two separate analyses.
The first, full-sample analysis, is to apply the existing and proposed methods to the full sample as in a typical application.
The second, sub-sample analysis, is to repeatedly draw 1000 sub-samples, each of about 1/25 size of the full sample, and apply different methods to the sub-samples.
The sample sizes from treatment groups 0 to 5 are fixed at 13587, 552, 1190, 197, 781 and 159 in each sub-sample.
The sub-sample analysis is mainly designed to compare different methods in more challenging settings where sample sizes are
comparable to regressor sizes.
For space limitation, we present estimation results for the treatment means $\{\mu_t: t=0,\ldots,5\}$ in this section,
and defer those for the ATEs and ATTs to Supplement Section \ref{sec:add-empi}.

\textbf{Full-sample analysis.}\;
We apply regularized calibrated estimation (RCAL) and regularized maximum likelihood estimation with separate Lasso (RMLs) or group Lasso (RMLg),
similarly as in the simulation study.
Each separate or group Lasso tuning parameter $\lambda$ is selected by minimizing a 5-fold cross validation error over a discrete set
$\{\lambda^* \times 0.01^{j/20}: j = 0, 1, \ldots, 20\}$, where $\lambda^*$ is the value leading to a zero solution.
For comparison, we also apply the non-regularized calibrated (CAL) and maximum likelihood (ML) estimation using PS and OR models with main effects only.
The estimates of the treatment means $\{\mu_t: t=0,\ldots,5\}$ are summarized in Table \ref{tb:mu_full_cross}.

The following findings can be obtained from Table \ref{tb:mu_full_cross}.
First, the adjusted point estimates of $\mu_t$ are decreasing as the treatment level $t$ increases, i.e., the number of smoked cigarettes increases.
There is a slight increase from $\hat{\mu}_2$ to $\hat{\mu}_3$ in the unadjusted estimates.
Second, the point estimates and confidence intervals from RCAL, RMLs, and RMLg are overall similar to each other.
Third, compared with those from non-regularized estimation with main effects only,
regularized estimation incorporating interactions in PS and OR models leads to comparable or smaller standard errors.
For example, the width of the 95\% confidence interval for $\mu_5$ from CAL and RCAL are 0.0224 and 0.0185.
This difference not only agrees with the theoretical property that augmented IPW estimation
has a smaller asymptotic variance when using a larger (correct) PS model with a fixed number of regressors,
but also demonstrates the effectiveness of Lasso regularization in
suppressing variation that would otherwise be inflated with a large number of regressors.

\textbf{Sub-sample analysis.}\;
To further compare different methods, we apply RCAL, RMLs, RMLg estimation to 1000 sub-samples as described earlier.
We conduct 5-fold cross validation in each subsample, similarly as in the full-sample analysis,
but take $\lambda.min$ as the tuning parameter for PS and OR models in RCAL when estimating $\mu_0$ and
take $\lambda.1se$ for PS and OR models in RCAL when estimating $\mu_1,\ldots,\mu_5$,
where $\lambda.min$ gives the minimal cross validation error, and $\lambda.1se$ gives the most regularized model such that the cross validation error
is within one standard error of the minimum (Hastie et al.~2009).
Selecting $\lambda.min$ when estimating $\mu_1,\ldots,\mu_5$ leads to large ratios of the number of nonzero estimated coefficients
over the corresponding treatment group size, which would be incompatible with the sparsity assumptions for our theory.
Selecting $\lambda.min$ for treatment 0 does not suffer this issue because the sample size for treatment 0 is much larger.
See Supplement Figures \ref{fig:box_nnz_or_min} and \ref{fig:box_nnz_or_1se}.
For similar reasons, we also take $\lambda.min$ for OR model in RMLs for treatment 0 and take $\lambda.1se$ for OR model in RMLs for the other treatments,
and take $\lambda.1se$ as the group Lasso tuning parameter for OR model in RMLg.
Because RMLs or RMLg treats PS estimation independently of which $\mu_t$ is estimated, we take $\lambda.1se$ for PS model in RMLs and RMLg.
Results from always selecting $\lambda.min$ or $\lambda.1se$ in all methods can be found in the Supplement.

Table \ref{tb:mu_sub_cross} summarizes the estimates of $\{\mu_t: t=0,\ldots,5\}$ from 1000 subsamples similarly as in the simulation study.
Coverage proportions are calculated by treating the mean of the 1000 estimates as the true value for each method.
We see that RCAL, RMLs and RMLg perform similarly to each other in terms of the repeated-sampling means and variances.
However, the estimated variances from RCAL are close to the repeated-sampling variances, whereas those from RMLs and RMLg
show underestimation especially for $\mu_3$ and $\mu_5$.
Consistently with this observation,
the coverage proportions from RCAL are comparable to or more aligned with the nominal probabilities than from RMLs and RMLg. For example,
from RCAL, RMLs, RMLg, the 90\% coverage proportions are 0.900, 0.847 and 0.849 for $\mu_3$ and are 0.887, 0.810 and 0.812 for $\mu_5$.
These differences in coverage can also be seen from the QQ plots of standardized estimates in the Supplement.

\begin{figure}[tb!]
\centering
\includegraphics[scale=0.45]{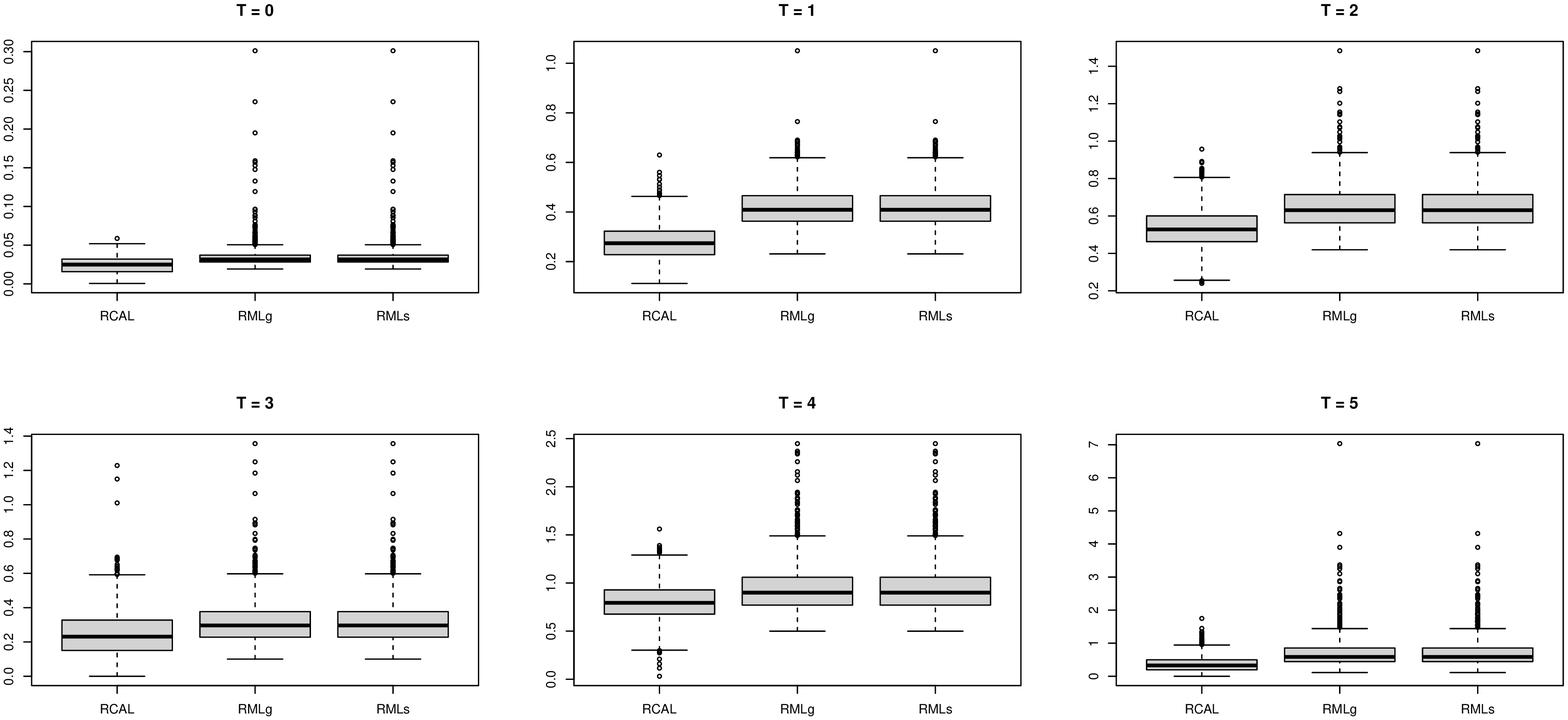}\vspace{-.1in}
\caption{Boxplots of relative variances of the inverse probability weights~(RV)}
\label{fig:box_rv_cross}
\end{figure}

Finally, to compare PS estimation from different methods,
we calculate the maximum absolute standardized calibration differences (MASCD) and the relative variances (RV) of the inverse probability weights
similarly as in Tan (2020a) by the following formulas:
\begin{align*}
\text{MASCD}_t & = \max_{j=1}^p \left\lvert \left[\frac{\tilde{\E}\left\{R^{(t)}f_j(X)\hat{\pi}^{-1}(t, X; \hat{\gamma})\right\}}{\tilde{\E}\left\{R^{(t)}\hat{\pi}^{-1}(t, X; \hat{\gamma})\right\}} - \tilde{\E}\{f_j(X)\}\right] \big/ \tilde{V}^{1/2}\{f_j(X)\} \right\rvert ,\\
\text{RV}_t & = \tilde{V}_t(1 / \{\hat{\pi}(t, X; \hat{\gamma})\}) \big/ \tilde{\E}_t^2[1 / \{\hat{\pi}(t, X; \hat{\gamma})\}],
\end{align*}
where $\tilde{V}$ denotes the sample variance, and $\tilde{\E}_t$ and $\tilde{V}_t$ denote the sample mean and variance over treatment $t$ respectively.
From Supplement Figures \ref{fig:box_nnz_ps_cross} and \ref{fig:box_masd_cross},
we see that RCAL, RMLs and RMLg are overall comparable to each other in terms of MASCD,
but RCAL is associated with a much smaller number of nonzero estimated coefficients.
From Figure \ref{fig:box_rv_cross}, we find that RCAL is associated with similar or smaller RVs compared with RMLs and RMLg,
while more outliers are produced by RMLs and RMLg. These differences indicate that
RCAL achieves greater sparsity in estimated coefficients and greater efficiency in weighting than RMLs and RMLg.

\begin{table} [tb!]
\caption{\footnotesize Summary of $\hat{\mu}_t$ on full sample for $t = 0, 1, \ldots, 5$.} \label{tb:mu_full_cross}\vspace{-4ex}
\begin{center}
\resizebox{1.0\textwidth}{!}{\begin{tabular}{lccccccccccc}
\hline
 & Est & SE & 95CI &~~& Est & SE & 95CI &~~ & Est & SE & 95CI  \\
\hline
  & \multicolumn{3}{c}{$\hat{\mu}_0~(n_0 = 13587)$} &~~& \multicolumn{3}{c}{$\hat{\mu}_1~(n_1 = 552)$} &~~& \multicolumn{3}{c}{$\hat{\mu}_2~(n_2 = 1190)$} \\
  \cline{2-4}\cline{6-8}\cline{10-12}
Unadj & 8.1272 & 0.0003 & (8.1266, 8.1279) &~~& 8.0576 & 0.0019 & (8.0538, 8.0614) &~~& 8.0398 & 0.0013 & (8.0372, 8.0424) \\
\rowcolor{lightgray}
CAL & 8.1247 & 0.0004 & (8.1240, 8.1254) &~~& 8.0763 & 0.0020 & (8.0723, 8.0803) &~~& 8.0570 & 0.0017 & (8.0537, 8.0602) \\
ML & 8.1252 & 0.0006 & (8.1240, 8.1263)  &~~& 8.0768 & 0.0019 & (8.0730, 8.0806)  &~~& 8.0573 & 0.0015 & (8.0542, 8.0603) \\
\rowcolor{lightgray}
RCAL & 8.1243 & 0.0004 & (8.1236, 8.1250) &~~& 8.0744 & 0.0022 & (8.0702, 8.0786) &~~& 8.0545 & 0.0016 & (8.0513, 8.0577 \\
RMLs & 8.1244 & 0.0004 & (8.1236, 8.1251) &~~& 8.0751 & 0.0020 & (8.0712, 8.0790) &~~& 8.0551 & 0.0017 & (8.0518, 8.0585) \\
RMLg & 8.1244 & 0.0004 & (8.1237, 8.1251) &~~& 8.0750 & 0.0020 & (8.0710, 8.0790) &~~& 8.0541 & 0.0016 & (8.0509, 8.0572) \\

  & \multicolumn{3}{c}{$\hat{\mu}_3~(n_3 = 197)$} &~~& \multicolumn{3}{c}{$\hat{\mu}_4~(n_4 = 781)$} &~~& \multicolumn{3}{c}{$\hat{\mu}_5~(n_5 = 159)$} \\
   \cline{2-4}\cline{6-8}\cline{10-12}
Unadj & 8.0474 & 0.0030 & (8.0416, 8.0533) &~~& 8.0365 & 0.0016 & (8.0334, 8.0396) &~~& 8.0318 & 0.0034 & (8.0251, 8.0384) \\
\rowcolor{lightgray}
CAL & 8.0517 & 0.0049 & (8.0421, 8.0612) &~~& 8.0460 & 0.0023 & (8.0415, 8.0505) &~~& 8.0442 & 0.0057 & (8.0330, 8.0554)  \\
ML & 8.0536 & 0.0045 & (8.0449, 8.0623)  &~~& 8.0460 & 0.0021 & (8.0418, 8.0502) &~~& 8.0413 & 0.0057 & (8.0301, 8.0525)  \\
\rowcolor{lightgray}
RCAL & 8.0494 & 0.0042 & (8.0411, 8.0576) &~~& 8.0439 & 0.0021 & (8.0399, 8.0480) &~~& 8.0417 & 0.0047 & (8.0324, 8.0509)  \\
RMLs & 8.0495 & 0.0036 & (8.0424, 8.0565) &~~& 8.0446 & 0.0020 & (8.0408, 8.0484) &~~& 8.0416 & 0.0044 & (8.0330, 8.0501)  \\
RMLg & 8.0484 & 0.0037 & (8.0412, 8.0556) &~~& 8.0448 & 0.0020 & (8.0408, 8.0488) &~~& 8.0410 & 0.0044 & (8.0323, 8.0497)  \\
\hline
\end{tabular}}
\end{center}
\setlength{\baselineskip}{0.5\baselineskip}
\vspace{-.15in}\noindent{\tiny
\textbf{Note}: Est, SE, or 95CI denotes point estimate, standard error, or 95\% confidence interval respectively.
Unadj denotes the unadjusted estimate, $\hat{\mu}_t = \tilde{\E}(Y^{(t)})$.
RCAL, RMLs, or RMLg denotes $\hat{\mu}_t(\hat{m}^\#_{\text{RWL}}, \hat{\pi}_{\text{RCAL}})$, $\hat{\mu}_t(\hat{m}_{\text{RMLs}}, \hat{\pi}_{\text{RML}})$,
or $\hat{\mu}_t(\hat{m}_{\text{RMLg}}, \hat{\pi}_{\text{RML}})$ respectively. CAL or ML denotes non-regularized estimation with main effects only in PS and OR models.}
\end{table}

\begin{table} [tb!]
\caption{\footnotesize Summary of $\hat{\mu}_t$ on sub-samples for $t = 0, 1, \ldots, 5$.} \label{tb:mu_sub_cross}\vspace{-4ex}
\begin{center}
\resizebox{1.0\textwidth}{!}{\begin{tabular}{lccccccccccccccccc}
\hline
 & Mean & $\sqrt{\text{Var}}$ & $\sqrt{\text{EVar}}$ & Cov90 & Cov95 &~~& Mean & $\sqrt{\text{Var}}$ & $\sqrt{\text{EVar}}$ & Cov90 & Cov95 &~~& Mean & $\sqrt{\text{Var}}$ & $\sqrt{\text{EVar}}$ & Cov90 & Cov95 \\
\hline
  & \multicolumn{5}{c}{$\hat{\mu}_0~(n_0 = 13587)$} &~~& \multicolumn{5}{c}{$\hat{\mu}_1~(n_1 = 552)$}  &~~& \multicolumn{5}{c}{$\hat{\mu}_2~(n_2 = 1190)$}\\
  \cline{2-6}\cline{8-12}\cline{14-18}
  \rowcolor{lightgray}
RCAL & 8.125 & 0.002 & 0.002 & 0.883 & 0.942 & ~~ & 8.069 & 0.009 & 0.010 & 0.913 & 0.950 & ~~ & 8.052 & 0.007 & 0.007 & 0.911 & 0.953 \\
RMLs & 8.125 & 0.002 & 0.002 & 0.888 & 0.946 & ~~ & 8.073 & 0.009 & 0.009 & 0.896 & 0.939 & ~~ & 8.052 & 0.007 & 0.007 & 0.906 & 0.944 \\
RMLg & 8.125 & 0.002 & 0.002 & 0.897 & 0.954 & ~~ & 8.073 & 0.009 & 0.009 & 0.895 & 0.939 & ~~ & 8.052 & 0.007 & 0.007 & 0.903 & 0.942 \\

& \multicolumn{5}{c}{$\hat{\mu}_3~(n_3 = 197)$} &~~& \multicolumn{5}{c}{$\hat{\mu}_4~(n_4 = 781)$} &~~& \multicolumn{5}{c}{$\hat{\mu}_5~(n_1 = 159)$} \\
 \cline{2-6}\cline{8-12}\cline{14-18}
 \rowcolor{lightgray}
RCAL & 8.050 & 0.018 & 0.017 & 0.900 & 0.947 & ~~ & 8.044 & 0.009 & 0.009 & 0.903 & 0.949 & ~~ & 8.038 & 0.018 & 0.018 & 0.887 & 0.940 \\
RMLs & 8.049 & 0.017 & 0.014 & 0.847 & 0.912 & ~~ & 8.041 & 0.009 & 0.009 & 0.881 & 0.929 & ~~ & 8.039 & 0.018 & 0.014 & 0.810 & 0.886 \\
RMLg & 8.049 & 0.017 & 0.014 & 0.849 & 0.912 & ~~ & 8.041 & 0.009 & 0.009 & 0.882 & 0.928 & ~~ & 8.039 & 0.018 & 0.015 & 0.812 & 0.889 \\

\hline
\end{tabular}}
\end{center}
\setlength{\baselineskip}{0.5\baselineskip}
\vspace{-.15in}\noindent{\tiny
\textbf{Note}: Mean, Var, EVar, Cov90, and Cov95 are calculated over the 1000 repeated subsamples, with
the mean treated as the true value. RCAL denotes $\hat{\mu}_t(\hat{m}^\#_{\text{RWL}}, \hat{\pi}_{\text{RCAL}})$. RMLs denotes $\hat{\mu}_t(\hat{m}_{\text{RMLs}}, \hat{\pi}_{\text{RML}})$. RMLg denotes $\hat{\mu}_t(\hat{m}_{\text{RMLg}}, \hat{\pi}_{\text{RML}})$.}
\end{table}

\section{Appendices}

\subsection{Appendix A: On multi-class calibration loss}

We establish an interesting relationship between the multi-class likelihood and calibration loss functions $\ell_{\text{CAL}}(\gamma)$ and $\ell_{\text{ML}}(\gamma)$,
which extends a similar result with binary treatments in Tan (2020a).
To allow for misspecification of model \eqref{eq:PSmodel}, we write $\ell_{\text{CAL}}(\gamma) = \kappa_{\text{CAL}}(\gamma^\T f(X))$ and $\ell_{\text{WL}}(\gamma) = \kappa_{\text{ML}}(\gamma^\T f(X))$, where for a $K\times 1$ function $h(x) = (h_k(x): k\in\mathcal T)^\T$,
\begin{align}
\kappa_{\text{CAL}}(h) & = \sum_{k\neq t}\tilde{\E}\left[R^{(t)} \me^{h_k(X) - h_t(X)} - R^{(k)}\{h_k(X) - h_t(X)\}\right] ,\\ 
\kappa_{\text{ML}}(h) & = \tilde{\E}\left[- \sum_{k\in\mathcal T} R^{(k)}h_k(X) + \log \sum_{k\in\mathcal T} \me^{h_k(X)}  \right]. 
\end{align}
Both $\kappa_{\text{CAL}}(h)$ and $\kappa_{\text{ML}}(h)$ are convex in $h$. For two $K\times 1$  functions $h(x)$ and $h^\prime(x)$, consider the Bregman divergences associated with $\kappa_{\text{CAL}}$ and $\kappa_{\text{ML}}$,
\begin{align*}
D_{\text{CAL}}(h, h^\prime) & = \kappa_{\text{CAL}}(h) - \kappa_{\text{CAL}}(h^\prime) - \sum_{k\in\mathcal T} \langle \nabla_k \kappa_{\text{CAL}}(h^\prime), h_k - h^\prime_k \rangle, \\
D_{\text{ML}}(h, h^\prime) & = \kappa_{\text{ML}}(h) - \kappa_{\text{ML}}(h^\prime) - \sum_{k\in\mathcal T} \langle \nabla_k \kappa_{\text{ML}}(h^\prime), h_k - h^\prime_k \rangle,
\end{align*}
where $h_k$ is identified as the vector $(h_{k1}, \dots, h_{kn})^\T $ with $h_{ki} = h_k(X_i)$,
$ \nabla_k \kappa_{\text{CAL}}(h^\prime)$ is identified as $(\partial \kappa_{\text{CAL}}(h^\prime) / \partial h^\prime_{k1}, \ldots,\partial \kappa_{\text{CAL}}(h^\prime) / \partial h^\prime_{kn})^\T $, and
$ \nabla_k \kappa_{\text{ML}}(h^\prime)$ is similarly defined.
For two probability vectors $\rho = (\rho_0, \rho_1, \ldots, \rho_{K-1})^\T $ and $\rho^\prime = (\rho^\prime_0, \rho^\prime_1, \ldots, \rho^\prime_{K-1})^\T $, the Kullback-Liebler divergence is $L(\rho, \rho^\prime) = \sum_{k=0}^{K-1}\rho^\prime_k\log(\rho^\prime_k / \rho_k)$. In addition, for $c, c^\prime \in (0, 1)$, let $K(c, c^\prime) = c^\prime / c - 1 - \log(c^\prime / c) \geq 0$, which is strictly convex in $c^\prime / c$, with a minimum of 0 at $c^\prime / c = 1$.

\begin{pro} \label{pro:diver}
(i) For any $K\times 1$ functions $h(x)$ and $h^\prime(x)$ and the corresponding functions
$\pi(x) = (\pi(k,x): k\in\mathcal T)^\T$ and $\pi^\prime(x) = (\pi^\prime(k,x) : k\in\mathcal T)^\T$,
where  $\pi(k, x) = \exp \{h_k(x)\} / \sum_{s\in\mathcal T} \exp\{h_s(x)\} $
and $\pi^\prime(k,x)$ is similarly defined from $h^\prime(x)$. Then
\begin{align*}
& D_{\text{CAL}}(h, h^\prime) = \tilde{\E}\left(\frac{R^{(t)}}{\pi^\prime(t, X)} \left[K\{\pi(t, X), \pi^\prime(t, X)\} + L\{\pi(X), \pi^\prime(X)\} \right]\right), \\
& D_{\text{ML}}(h, h^\prime) = \tilde{\E} \left[L\{\pi(X), \pi^\prime(X)\} \right].
\end{align*}
(ii) For any fixed $\gamma$, it holds that
\begin{align}
& \E\{\ell_{\text{CAL}}(\gamma) - \kappa_{\text{CAL}}(h^*)\} = \E \left[K\{\pi(t, X; \gamma), \pi^*(t, X)\} + L\{\pi(X; \gamma), \pi^*(X)\} \right], \label{eq:CALdiver} \\
& \E\{\ell_{\text{ML}}(\gamma) - \kappa_{\text{ML}}(h^*)\} = \E \left[L\{\pi(X; \gamma), \pi^*(X)\} \right], \label{eq:MLdiver}
\end{align}
where $\pi(x;\gamma) = ( \pi(k,x;\gamma): k\in\mathcal T)^\T$ and $\pi^*(x) = (\pi^*(k,x): k\in\mathcal T)^\T$ with the associated $K\times 1$  function
$h^*(x) = (h^*_k(x): k\in\mathcal T)^\T$.
\end{pro}

While the preceding result closely resembles Proposition 1 in Tan (2020a), the extension is nontrivial.
For the relations (\ref{eq:CALdiver})--(\ref{eq:MLdiver}), the Kullback-Liebler divergence is defined between
two multi-class probability vectors $\pi(X; \gamma)$ and $\pi^*(X)$,
but the divergence $K(\cdot,\cdot)$ involves only the scalar probabilities $\pi(t,X;\gamma)$ and $\pi^*(t,X)$ associated with class $t$.

From (\ref{eq:CALdiver})--(\ref{eq:MLdiver}),  the calibration and likelihood divergences can be shown to yield different bounds on
the mean squared relative error (MSRE) in propensity sores:
\begin{align*}
\text{MSRE}(\gamma) = \E[Q\{\pi(t,X; \gamma), \pi^*(t,X)\}] = \E\left[\left\{\frac{\pi^*(t, X)}{\pi(t, X; \gamma)} - 1\right\}^2\right],
\end{align*}
where $Q(c, c^\prime) = (c^\prime / c - 1)^2$ for two probabilities $c, c^\prime \in (0, 1)$.
As demonstrated in Tan (2020a), Proposition 2 (i),
this measure of relative errors in $\pi(t,X;\gamma)$ directly governs the mean squared error of an IPW estimator $\hat{\mu}_{t,\text{IPW}}(\gamma) = \tilde{\E}\{R^{(t)}Y / \pi(t, X; \gamma)\}$ for $\mu_t$.

\begin{cor} \label{cor:MSRE}
(i) If $\pi(t, X; \gamma) \geq a\pi^*(t, X)$ almost surely for some constant $a\in(0, 1/2]$, then
\begin{align}
\text{MSRE}(\gamma) \leq \frac{5}{3a}\E[K\{\pi(t, X; \gamma), \pi^*(t, X)\}] \leq \frac{5}{3a}\E\{\ell_{\text{CAL}}(\gamma) - \kappa_{\text{CAL}}(h^*)\}. \label{eq:MSRECAL}
\end{align}
The factor $5/(3a)$ in general connot be imcorved up to a constant, independent of $a$.

(ii) If $\pi(t, X; \gamma) \geq b$ almost surely for some constant $b\in(0, 1)$, then
\begin{align}
\text{MSRE}(\gamma) \leq \frac{1}{2b^2}\E[L\{\pi(X; \gamma), \pi^*(X)\}]  = \frac{1}{2b^2}\E\{\ell_{\text{ML}}(\gamma) - \kappa_{\text{ML}}(h^*)\}. \label{eq:MSREML}
\end{align}
The factor $1/(2b^2)$ in general connot be imcorved up to a divisor of order $\log(b^{-1})$.
\end{cor}

The two bounds (\ref{eq:MSRECAL})--(\ref{eq:MSREML}) are seemingly the same as those in Tan (2020a), Proposition 2, although the multi-class calibration and likelihood losses are involved.
The calibration divergence achieves strong control of the relative errors in propensity scores
through the additional term $\E[K\{\pi(t, X; \gamma), \pi^*(t, X)\}]$. By Lemma 1 in Tan (2020a), $K(c,c^\prime) \ge (3a/5) Q(c,c^\prime)$ for
any probabilities $c$ and $c^\prime$ satisfying $c/c^\prime \ge a$ with $a \in (0,1/2]$.
In contrast, the likelihood divergence mainly controls the absolute errors in propensity scores through Pinsker's inequality:
$L(\rho, \rho^\prime )  \geq \frac{1}{2} \| \rho - \rho^\prime  \|_1^2 \ge 2(\rho_t - \rho^\prime_t)^2$
for any probability vectors $\rho$ and $\rho^\prime$.
See Tan (2020a), Section 3.2, for further discussion which relates the minimum calibration and likelihood divergences to the mean squared errors of
$\hat{\mu}_{t,\text{IPW}}(\bar \gamma)$ at the corresponding target values $\bar\gamma$.

\subsection{Appendix B: Karush--Kuhn--Tucker conditions}

We discuss implications of the Karush--Kuhn--Tucker (KKT) conditions for regularized calibrated estimation in Section~\ref{sec:RCAL}.

First, by the KKT conditions for minimization of $\ell_{\text{RCAL}}(\gamma)$ in \eqref{eq:PSobj} with the one-to-zero constraint, the fitted propensity scores
$\hat{\pi}_{\text{RCAL}} (k, X) = \pi (k, X; \hat{\gamma}_{\text{RCAL}})$, $k\in \mathcal T$, satisfy
\begin{align*}
& \tilde{\E}\left\{ R^{(t)}\frac{\hat{\pi}_{\text{RCAL}}(k, X)}{\hat{\pi}_{\text{RCAL}}(t, X)} - R^{(k)}\right\} = 0, \quad k\neq t , \\ 
& \sum_{k\neq t}\tilde{\E}^2\left[\left\{R^{(t)}\frac{\hat{\pi}_{\text{RCAL}}(k, X)}{\hat{\pi}_{\text{RCAL}}(t, X)} - R^{(k)}\right\}f_j(X)\right] \leq \lambda_1^2, \quad j = 1, \ldots, p, 
\end{align*}
where equality holds in the second line for any $j$ such that the vector $\hat{\gamma}_{j., \text{RCAL}} = (\hat{\gamma}_{jk, \text{RCAL}}: k\in\mathcal T\backslash\{t\})^\T$ is nonzero. By summing the two sides of the first line over $k\neq t$ of the preceding equation and using inequality $|\sum_{k\neq t}a_k| \leq \sum_{k\neq t}|a_k| \leq \sqrt{K-1}\sqrt{\sum_{k\neq t}a_k^2}$ on the second line above, the fitted propensity score $\hat{\pi}_{\text{RCAL}}(t, X)$ for treatment $t$ satisfies
\begin{align}
& \frac{1}{n}\sum_{i=1}^n \frac{R_i^{(t)}}{\hat{\pi}_{\text{RCAL}}(t, X_i)} = 1, \label{eq:PS-KKT-t-1-cons} \\
& \frac{1}{n}\left\lvert \sum_{i=1}^n \frac{R_i^{(t)}f_j(X_i)}{\hat{\pi}_{\text{RCAL}}(t, X_i)} - \sum_{i=1}^n f_j(X_i) \right\rvert \leq \sqrt{K-1}\lambda_1,\quad j = 1, \ldots, p, \label{eq:PS-KKT-t-2-cons}
\end{align}
By \eqref{eq:PS-KKT-t-1-cons}, the inverse probability weights, $1 / \hat{\pi}_{\text{RCAL}}(t, X_i)$ with $R_i^{(t)} = 1$, sum to the sample size $n$.
By \eqref{eq:PS-KKT-t-2-cons}, the weighted average of each covariate function $f_j(X_i)$ in the $t$th treated group may differ from the overall sample average of $f_j(X_i)$ by no more than $\sqrt{K-1}\lambda_1$.
See the Supplement Section II.2 for a discussion on KKT conditions with the sum-to-zero constraint.

Second, by the KKT conditions for minimization of \eqref{eq:ORobj}, the fitted outcome regression functions $\hat m_{\text{RWL}}^{(k)} (t, X) = m(t, X; \hat{\alpha}_{t, \text{RWL}}^{(k)})$, $k\not=t$, satisfy
\begin{align}
& \tilde{\E}\left[R^{(t)}\frac{\hat{\pi}_{\text{RCAL}}(k, X)}{\hat{\pi}_{\text{RCAL}}(t, X)}\{Y - \hat m_{\text{RWL}}^{(k)} (t, X) \}\right] = 0, \quad k\neq t ,\label{eq:OR-KKT-1} \\
& \sum_{k\neq t}\tilde{\E}^2\left[R^{(t)}\frac{\hat{\pi}_{\text{RCAL}}(k, X)}{\hat{\pi}_{\text{RCAL}}(t, X)}\{Y - \hat m_{\text{RWL}}^{(k)} (t, X) \}f_j(X)\right] \leq \lambda^2_2, \quad j = 1, \ldots, p, \label{eq:OR-KKT-2}
\end{align}
where equality holds in \eqref{eq:OR-KKT-2} for any $j$ such that the vector $\hat{\alpha}^\#_{jt, \text{RWL}}$ is nonzero.
Equation \eqref{eq:OR-KKT-1} implies that by simple calculation, the estimator $\hat{\mu}_t(\hat{m}^\#_{\text{RWL}}, \hat{\pi}_{\text{RCAL}})$ can be recast as
\begin{align*}
\hat{\mu}_t(\hat{m}^\#_{\text{RWL}}, \hat{\pi}_{\text{RCAL}}) = \tilde{\E}\left\{R^{(t)}Y + \sum_{k\neq t}R^{(k)} \hat m_{\text{RWL}}^{(k)} (t, X) \right\}.
\end{align*}
As a consequence, $\hat{\mu}_t(\hat{m}^\#_{\text{RWL}}, \hat{\pi}_{\text{RCAL}} )$ always falls within the range of the observed outcomes
$\{Y_i: T_i = t, i = 1, \ldots, n\}$ and the predicted values $\{ \hat m_{\text{RWL}}^{(k)} (t, X) : T_i = k, i = 1, \ldots, n\}$ for $k\neq t$.
This boundedness property is not satisfied by the existing estimator $\hat\mu_t(\hat m_{\text{RMLs}}, \hat \pi_{\text{RML}} )$ or $\hat\mu_t(\hat m_{\text{RMLg}}, \hat \pi_{\text{RML}} )$.

\subsection{Appendix C: Control of gradient norms}

For a linear OR model, we discuss a subtle difference in the control of gradient norms, which may explain the difference between the error bound (\ref{eq:MT-error2})
and ours in Theorem \ref{thm:or-error-glm} as described in Remark \ref{rem:Lounici}.
Consider the fixed design where
each treatment subsample $\{i: T_i=k\}$ is of a fixed size $n_0 = n/K$ and taken as an individual task.
The least-square loss function  $\sum_{k=0}^{K-1} \tilde{\E} \{ R^{(k)} ( Y - \alpha_k^\T f(X) )^2 \}/ 2 $
can be written as $\sum_{k=0}^{K-1} \tilde{\E}_k \{ ( Y - \alpha_k^\T f(X) )^2 \}/ (2K) $,
where $\tilde{\E}_k(\cdot)$ denotes the average in the $k$th treatment subsample.
For the error bound (\ref{eq:MT-error2}), the gradient of the loss function at $\alpha^*$ is upper bounded in the sup-$L_2$ norm
with high probability as follows:
\begin{align}
 \sup_{j=0}^p\; \left\lVert \left[ K^{-1} \tilde{\E}_k \{ ( Y - \alpha_k^{*\T} f(X) ) f_j(X)\} : k=0,1,\ldots,K-1 \right] \right\rVert_2 \leq K^{-1/2} \dot\lambda . \label{eq:MT-grad-bound}
\end{align}
By comparison, the gradient of our loss function
$ \sum_{k\neq t}\tilde{\E}[R^{(t)}\omega(k, X; \bar{\gamma}_{\text{CAL}}) \{Y - \alpha_t^{(k)\T} f(X)\}^2] /2$
is upper bounded in the sup-$L_2$ norm with high probability in a random design as follows:
\begin{align}
 \sup_{j=0}^p\; \left\lVert \left[ \tilde{\E} \{R^{(t)}\omega(k, X; \bar{\gamma}_{\text{CAL}}) ( Y - \bar\alpha_{t, \text{WL}}^{(k)\T} f(X) ) f_j(X)\} : k \not= t \right] \right\rVert_2 \leq \tilde \lambda_2 . \label{eq:our-grad-bound}
\end{align}
See Supplement Lemma \ref{lem:or-score}.
While the two upper bounds in (\ref{eq:MT-grad-bound}) and (\ref{eq:our-grad-bound}) differ by a factor of $K^{-1/2}$,
this difference may be attributed to the fact that
the left-hand side of (\ref{eq:MT-grad-bound}) involves the $L_2$ norm of $K$ independent variables based on different treatment subsamples,
where that of (\ref{eq:our-grad-bound}) involves the $L_2$ norm of $K-1$ interdependent variables from the same treatment group $t$.

\vspace{.3in}
\centerline{\bf\Large References}

\begin{description}\addtolength{\itemsep}{-.1in}

\item Almond, D., Chay, K.Y., and Lee, D.S. (2005) The costs of low birth weight, {\em  Quarterly Journal of Economics}, 120, 1031–1083.

\item Avagyan, V. and Vansteelandt, S. (2017) Honest data-adaptive inference for the average treatment effect under model misspecification using penalised bias-reduced double-robust estimation, {\em arXiv preprint:1708.03787}.

\item Belloni, A., Chernozhukov, V., and Hansen, C. (2014) Inference on treatment effects after selection among high-dimensional controls, {\em Review of Economic Studies}, 81, 608–650.

\item Bohning, D. and Lindsay, B. G. (1988) Monotonicity of quadratic approximation algorithms, {\em Annals of the Institute of Statistical Mathematics}, 40, 641–663.

\item Bradic, J., Wager, S., and Zhu, Y. (2019) Sparsity double robust inference of average treatments effects, {\em arXiv preprint:1905.00744}.

\item Buhlmann, P. and van de Geer, S. (2011) {\em Statistics for High-Dimensional Data: Methods, Theory and Applications}, New York: Springer.

\item Cattaneo, M.D. (2010) Efficient semiparametric estimation of multi-valued treatment effects under ignorability, {\em Journal of Econometrics}, 155, 138–154.

\item Chernozhukov, V., Chetverikov, D., Demirer, M., Duflo, E., Hansen, C., Newey, W.K., and Robins, J.M. (2018) Double/debiased machine learning for treatment and structural parameters, {\em Econometrics Journal}, 21, C1–C68.

\item Farrell, M.H. (2015) Robust inference on average treatment effects with possibly more covariates than observations, {\em Journal of Econometrics}, 189, 1–23.

\item Folsom, R.E. (1991) Exponential and logistic weight adjustments for sampling and nonresponse error reduction,  {\em Proceedings of the American Statistical Association, Social Statistics Section}, 197–202.

\item Friedman, J., Hastie, T., and Tibshirani, R. (2010) Regularization paths for generalized linear models via coordinate descent, {\em Journal of statistical software}, 33, 1–22.

\item Ghosh, S. and Tan, Z. (2021) Doubly robust semiparametric inference using regularized calibrated estimation with high-dimensional Data, {\em  Bernoulli}, to appear.

\item Graham, B.S., de Xavier Pinto, C.C., and Egel, D. (2012) Inverse probability tilting for moment condition models with missing data, {\em Review of Economic Studies}, 79, 1053–1079.

\item Hainmueller,J. (2012) Entropy balancing for causal effects: A multivariate reweighting method to produce balanced samples in observational studies, {\em Political Analysis}, 20, 25–46.

\item Hastie, T., Tibshirani, R., and Friedman, J. (2009) {\em The elements of statistical learning}, New York: Springer.

\item Imai, K. and Ratkovic, M. (2014) Covariate balancing propensity score, {\em Journal of the Royal Statistical Society: Series B}, 76, 243–263.

\item Kang, J.D.Y. and Schafer, J.L. (2007) Demystifying double robustness: A comparison of alternative strategies for estimating a population mean from incomplete data (with discussion), {\em Statistical Science}, 22, 523–539.

\item Kim, J.K. and Haziza, D. (2014) Doubly robust inference with missing data in survey sampling, {\em Statistica Sinica}, 24, 375–394.

\item Lounici, K., Pontil, M.,  van de Geer, S., and Tsybakov, A.B. (2011) Oracle inequalities and optimal inference under group sparsity, {\em Annals of Statistics}, 39, 2164–2204.

\item McCullagh, P. and Nelder, J. (1989) {\em Generalized Linear Models}, New York: Chapman \& Hall, 2nd ed.

\item Neyman, J. (1923) On the application of probability theory to agricultural experiments. Essay on principles. Section 9, {\em Statistical. Science}, 5, 465–480.

\item Ning, Y., Peng, S., and Imai, K. (2020) Robust estimation of causal effects via a high-dimensional covariate balancing propensity score, {\em Biometrika}, 107, 533–554.

\item Robins, J. M., Rotnitzky, A. and Zhao, L. P. (1994) Estimation of regression coefficients when some regressors are not always observed, {\em Journal of the American Statistical Association}, 89, 846–866.

\item Rosenbaum, P. R. and Rubin, D. B. (1983) The central role of the propensity score in observational studies for causal effects, {\em Biometrika}, 70, 41–55.

\item Rubin, D.B. (1974) Estimating causal effects of treatments in randomized and non randomized studies, {\em Journal of educational Psychology}, 66, 688–701.

\item Rubin, D. B. (1976) Inference and missing data, {\em Biometrika}, 63, 581–590.

\item Rubin, D.B. (2001) Using propensity scores to help design observational studies: Application to the tobacco litigation, {\em Health Services \& Outcomes Research Methodology}, 2, 169–188.

\item Simon, N.,  Friedman, J., and Hastie, T. (2013) A blockwise descent algorithm for group-penalized multiresponse and multinomial regression, {\em arXiv preprint:1311.6529}.

\item Small, C. G., Wang, J., and Yang, Z. (2000) Eliminating multiple root problems in estimation (with discussion), {\em Statistical Science}, 15, 313–341.

\item Smucler, E., Rotnitzky, A., and Robins, J. M. (2019). A unifying approach for doubly-robust $\ell_1$ regularized estimation of causal contrasts, {\em arXiv preprint:1904.03737}.

\item Sun, B. and Tan, Z. (2021) High-dimensional model-assisted inference for local average treatment effects with instrumental variables, {\em Journal of Business and Economic Statistics}, to appear.

\item Tan, Z. (2007) Comment: Understanding OR, PS, and DR, {\em Statistical Science}, 22, 560–568.

\item Tan, Z. (2010) Bounded, efficient, and doubly robust estimation with inverse weighting, {\em Biometrika}, 97, 661–682.

\item Tan, Z. (2020a) Regularized calibrated estimation of propensity scores with model misspecification and high-dimensional data, {\em Biometrika}, 107, 137-158.

\item Tan, Z. (2020b) Model-assisted inference for treatment effects using regularized calibrated estimation with high-dimensional data, {\em Annals of Statistics}, 48, 811 - 837.

\item Tibshirani, R. (1996) Regression shrinkage and selection via the Lasso, {\em Journal of the Royal Statistical Society: Series B}, 58, 267–288.

\item Vermeulen. K. and Vansteelandt, S. (2015) Bias-reduced doubly robust estimation, {\em Journal of the American Statistical Association}, 110, 1024–1036.

\item Wu, T. T. and  Lang, K. (2010) The MM alternative to EM, {\em Statistical Science}, 25, 492–505.

\item Xu, W. and Tan, Z. (2022) {\em mRCAL: Regularized Calibrated Estimation with multi-valued treatments}, R package version 1.0, available at \url{http://www.stat.rutgers.edu/~ztan}.

\item Yuan, M. and Lin, Y. (2006) Model selection and estimation in regression with grouped variables,  {\em Journal of the Royal Statistical Society: Series B}, 68, 49–67.

\end{description}


\clearpage

\setcounter{page}{1}

\setcounter{section}{0}
\setcounter{equation}{0}

\setcounter{figure}{0}
\setcounter{table}{0}

\renewcommand{\theequation}{S\arabic{equation}}
\renewcommand{\thesection}{\Roman{section}}

\renewcommand\thefigure{S\arabic{figure}}
\renewcommand\thetable{S\arabic{table}}

\setcounter{lem}{0}
\renewcommand{\thelem}{S\arabic{lem}}

\setcounter{thm}{0}
\renewcommand{\thethm}{S\arabic{thm}}

\setcounter{pro}{0}
\renewcommand{\thepro}{S\arabic{pro}}

\setcounter{rem}{0}
\renewcommand{\therem}{S\arabic{rem}}

\begin{center}
{\Large Supplementary Material for}

{\Large ``High-dimensional model-assisted inference for treatment effects with multi-valued treatments''}

\vspace{.1in} {\large Wenfu Xu and Zhiqiang Tan}
\end{center}
The Supplementary Material contains Sections I--IV. 

Section I includes Sections I.1--I.6 which provide 
the proofs for Theorems \ref{thm:ps-error}--\ref{thm:V-glm} respectively, with the one-to-zero constraint used in PS estimation.

Section II provides various material when using the sum-to-zero constraint in PS estimation, including a correct proof 
of the sum-to-zero relationship for RML and RCAL estimates (Section II.1), KKT conditions (Section II.2), algorithm for computing $\hat{\gamma}_{\text{RCAL}}$ (Section II.3), and theoretical results with the one-to-zero constraint (Section II.4).

Section III provides additional results for the simulation study, including the average sample sizes of the treatment groups, boxplot of $X_j$ and scatterplots of $Y$ against $X_j$ within different treatment groups for $j = 1, \ldots, 4$, estimation results of $\hat{\nu}_t^{(k)}$, estimation results with sum-to-zero-constraint, and QQ plots of $t$-statistics.

Section IV provides additional material for empirical application, including data preprocessing, results of ATEs and ATTs, and results from always selecting $\lambda.min$ or $\lambda.1se$ in all methods.

\section{Proofs}

\subsection{Proof of Theorem \ref{thm:ps-error}}

Theorem \ref{thm:ps-error} is a direct result from Lemma \ref{lem:ps-error}, which is proved using Lemmas \ref{lem:ps-score}--\ref{lem:ps-empicc}.
Lemma \ref{lem:ps-score} gives a high-probability bound on the gradient of the loss $\ell_{\text{CAL}}(\gamma)$ at $\gamma=\bar\gamma_{\text{CAL}}$ in the dual of the  $\lVert \cdot \rVert_{2, 1}$ norm.
Lemma \ref{lem:ps-basic} derives a basic inequality by the definition of $\hat{\gamma}_{\text{RCAL}}$.
Lemma \ref{lem:ps-dual} is obtained by combining Lemmas \ref{lem:ps-score} and \ref{lem:ps-basic}.
Lemma \ref{lem:ps-dagger} gives a suitable lower bound on the symmetrized Bregman divergence for local analysis.
Lemma \ref{lem:ps-hessian} gives a high-probability bound which is used in Lemma \ref{lem:ps-empicc}
to derive the empirical compatibility condition.

The gradient of the loss $\ell_{\text{CAL}}(\gamma)$ with respect to $\gamma_{j\cdot}$ can be written as
 $Z_j(T, X; \gamma) = g(T, X; \gamma)f_j(X)$ for $j = 0, 1, \ldots, p$, where $g(T, X; \gamma) = (g_k: k\in\mathcal T\backslash \{t\})^\T$ with $g_k = R^{(t)}\omega(k,X; \gamma) - R^{(k)}$.

\begin{lem} \label{lem:ps-score}
Denote by $\Omega_{\gamma 1}$ the event that
\begin{equation}
\sup_{j=0, 1, \ldots, p} \left\lVert \tilde{\E}\left\{Z_j(T, X; \bar{\gamma}_{\text{CAL}})\right\}\right\rVert_2 \leq \tilde{\lambda}_1, \label{eq:ps-score}
\end{equation}
Under Assumptions \ref{ass1}(i)-(ii), if
\begin{align*}
\tilde{\lambda}_1 \geq \sqrt{3 /\log(2)} B_0B_1\sqrt{\{(2/3)(K - 1) + \log[(p+1)/\epsilon]\} / n},
\end{align*}
then $\P(\Omega_{\gamma 1}) \geq 1 - \epsilon$.
\end{lem}

\begin{prf}
Under Assumption \ref{ass1}(i) and $|a - b| \leq |a| + |b|$, we have
\begin{align*}
& \lVert Z_j(T, X; \bar{\gamma}_{\text{CAL}}) \rVert_2 \leq \lVert Z_j(T, X; \bar{\gamma}_{\text{CAL}}) \rVert_1
\leq B_0R^{(t)}\{1 / \pi(t, X; \bar{\gamma}_{\text{CAL}}) - 1\} + B_0\{1 - R^{(t)}\},
\end{align*}
which leads to $\lVert Z_j(T, X; \bar{\gamma}_{\text{CAL}}) \rVert_2 \leq B_0$ if $R^{(t)} = 0$ or $\lVert Z_j(T, X; \bar{\gamma}_{\text{CAL}}) \rVert_2 \leq B_0/\pi(t, X; \bar{\gamma}_{\text{CAL}}) - B_0$ if $R^{(t)} = 1$. Hence, $\lVert Z_j(T, X; \bar{\gamma}_{\text{CAL}}) \rVert_2 \leq B_0B_1$ under Assumption \ref{ass1}(ii). By the Cauchy--Schwartz inequality, we find
\begin{align*}
\E\exp\left\{\frac{(\beta^\T Z_j)^2}{B_0^2B_1^2 / \log2}\right\} \leq \E\exp\left\{\frac{\Vert \beta \rVert^2_2\lVert Z_j \rVert^2_2}{B_0^2B_1^2 / \log2}\right\} \leq 2
\end{align*}
for any $\beta\in\bbR^{K}$ such that $\lVert \beta \rVert_2 = 1$, which indicates $Z_j(T, X; \bar{\gamma}_{\text{CAL}})$ is a sub-gaussian random vector with parameter $B_0^2B_1^2 / \log2$.
By properties of sub-gaussian random vectors, it follows that $n\tilde{\E}\left\{Z_j(T, X; \bar{\gamma}_{\text{CAL}})\right\}$ is a sub-gaussian random vector with parameter $nB_0^2B_1^2 / \log2$.
By Hsu et al. (2012, Theorem 2.1), we have
\begin{align*}
& P\left( \lVert n\tilde{\E}\left\{Z_j(T, X; \bar{\gamma}_{\text{CAL}})\right\} \rVert^2_2 > n\sigma_1^2\{K - 1 + 2\sqrt{(K - 1)a} + 2a\} \right) \leq \me^{-a}
\end{align*}
for any constant $a > 0$, where $\sigma_1^2 = B_1^2B_0^2 / \log2$. Using the inequality $2\sqrt{(K - 1)a} \leq K - 1 + a$, we have
\begin{align*}
& P\left( \lVert \tilde{\E}\left\{Z_j(T, X; \bar{\gamma}_{\text{CAL}})\right\} \rVert_2 > \sqrt{3}\sigma_1\sqrt{\{(2 / 3)(K - 1) + a\} / n} \right) \leq \me^{-a}.
\end{align*}
Hence we have by the union bound
\begin{align*}
& P\left(\sup_{j = 0, 1, \ldots, p} \lVert \tilde{\E}\left\{Z_j(T, X; \bar{\gamma}_{\text{CAL}})\right\} \rVert_2 > \sqrt{3}\sigma_1\sqrt{\{(2 / 3)(K - 1) + a\} / n} \right) \leq (p + 1)\me^{-a}.
\end{align*}
Setting the right-hand side of the above inequality to be $\epsilon$, we complete the proof.
\end{prf}

Denote $\tilde{\Sigma}_{\gamma} = \tilde{\E}[\diag\{R^{(t)}\omega(k,X; \bar{\gamma}_{\text{CAL}}): k\neq t\}\otimes f(X)f^\T(X)]$, the sample version of $\Sigma_{\gamma}$ as defined in Assumption \ref{ass1}. Furthermore, denote $(\Sigma_{\gamma})_{j_1, j_2} = \E[\diag\{R^{(t)}\omega(k,X; \bar{\gamma}_{\text{CAL}}): k\neq t\}f_{j_1}(X)f_{j_2}(X)]$ and $(\tilde{\Sigma}_{\gamma})_{j_1, j_2}$ as the sample version of $(\Sigma_{\gamma})_{j_1, j_2}$.

\begin{lem}\label{lem:ps-hessian}
Denote by $\Omega_{\gamma 2}$ the event that
\begin{align}
\sup_{j_1, j_2 = 0, 1, \ldots, p}\left\lVert(\tilde{\Sigma}_{\gamma})_{j_1, j_2} - (\Sigma_{\gamma})_{j_1, j_2} \right\rVert_{\text{op}} \leq \tilde{\lambda}_1, \label{eq:ps-hessian}
\end{align}
Under Assumptions \ref{ass1}(i) and \ref{ass1}(ii), if
\begin{align*}
\tilde{\lambda}_1 \geq 4B_1B_0^2\sqrt{\{(1/2) \log(K-1) + \log[(p+1)/\epsilon]\}/n},
\end{align*}
then $\P(\Omega_{\gamma 2}) \geq 1 - 2\epsilon^2$.
\end{lem}

\begin{prf}
Denote $a_{j_1j_2} = (a_{j_1j_2, k}: k\in\mathcal T\backslash\{t\})^\T$ with $a_{j_1j_2,k} = R^{(t)}\omega(k,X; \bar{\gamma}_{\text{CAL}} )f_{j_1}(X)f_{j_2}(X)]$. Because
$(\tilde{\Sigma}_{\gamma})_{j_1, j_2} - (\Sigma_{\gamma})_{j_1, j_2}$ is a diagonal $(K-1)\times (K-1)$ matrix,  by the union bound, we have
\begin{align*}
& P\left(\sup_{j_1, j_2 = 0, 1, \ldots, p}\left\lVert(\tilde{\Sigma}_{\gamma})_{j_1, j_2} - (\Sigma_{\gamma})_{j_1, j_2} \right\rVert_{\text{op}} \geq \tilde{\lambda}_1\right) \\
& \leq (p+1)^2(K-1) \max_{j_1,j_2=0,1, \ldots, p, k\neq t}P\left( | \tilde{\E}(a_{j_1j_2,k}) - \E(a_{j_1j_2, k}) | \geq \tilde{\lambda}_1\right).
\end{align*}
Then the desired result follows from Lemma 7 of Tan (2020a) with $| \tilde{\E}a_{j_1j_2,k} - \E a_{j_1j_2, k} | \leq 2 B_1B_0^2$.
\end{prf}

\begin{lem} \label{lem:ps-basic}
For any coefficient matrix $\gamma$, we have
\begin{align}
& D_{\text{CAL}}(\hat{\gamma}_{\text{RCAL}}, \gamma) + D_{\text{CAL}}(\gamma, \hat{\gamma}_{\text{RCAL}}) + \langle \nabla\kappa_{\text{CAL}}(\gamma), \hat{\gamma}_{\text{RCAL}} - \gamma \rangle + \lambda_1 R(\hat{\gamma}_{\text{RCAL}}) \leq \lambda_1 R(\gamma), \label{eq:ps-basic1}
\end{align}
where $R(\gamma) = \sum_{j=1, \ldots, p} \lVert \gamma_{j\cdot} \rVert_2$ and $\gamma_{j\cdot} = (\gamma_{jk}: k\in\mathcal T\backslash\{t\})^\T$.
\end{lem}

\begin{prf}
For any $u\in(0, 1]$, the definition of $\hat{\gamma}_{\text{RCAL}}$ implies
\begin{align*}
\ell_{\text{CAL}}(\hat{\gamma}_{\text{RCAL}}) + \lambda_1 R(\hat{\gamma}_{\text{RCAL}}) \leq \ell_{\text{CAL}}\{u\gamma + (1-u)\hat{\gamma}_{\text{RCAL}}\} + \lambda_1 R(u\gamma + (1-u)\hat{\gamma}_{\text{RCAL}}),
\end{align*}
which by the convexity of $R(\cdot)$ gives
\begin{align*}
\ell_{\text{CAL}}(\hat{\gamma}_{\text{RCAL}}) - \ell_{\text{CAL}}\{u\gamma + (1-u)\hat{\gamma}_{\text{RCAL}}\} + \lambda_1 uR(\hat{\gamma}_{\text{RCAL}}) \leq \lambda_1 uR(\gamma).
\end{align*}
Dividing both sides of the preceding inequality by $u$ and letting $u\to 0^+$ yields
\begin{align*}
\langle \nabla\kappa_{\text{CAL}}(\hat{\gamma}_{\text{RCAL}}), \hat{\gamma}_{\text{RCAL}} - \gamma\rangle + \lambda_1 R(\hat{\gamma}) \leq \lambda_1 R(\gamma).
\end{align*}
Inequality \eqref{eq:ps-basic1} follows because, by direct calculation, $D_{\text{CAL}}(\hat{\gamma}_{\text{RCAL}}, \gamma) + D_{\text{CAL}}(\gamma, \hat{\gamma}_{\text{RCAL}})=$

\noindent
$\langle \nabla\kappa_{\text{CAL}}(\hat{\gamma}_{\text{RCAL}}) - \nabla\kappa_{\text{CAL}}(\gamma), \hat{\gamma}_{\text{RCAL}} - \gamma \rangle$.
\end{prf}

\begin{lem} \label{lem:ps-dual}
In the event $\Omega_{\gamma 1}$ from Lemma \ref{lem:ps-score}, we have
\begin{align}
& \lvert \langle\nabla\kappa_{\text{CAL}}(\bar{\gamma}_{\text{CAL}}), \hat{\gamma}_{\text{RCAL}} - \bar{\gamma}_{\text{CAL}} \rangle \rvert \leq \tilde{\lambda}_1\lVert \hat{\gamma}_{\text{RCAL}} - \bar{\gamma}_{\text{CAL}} \rVert_{2, 1}, \label{eq:ps-dual1}
\end{align}
and for any subset $S\subset \{0, 1, \ldots, p\}$ containing 0,
\begin{align}
& D_{\text{CAL}}(\hat{\gamma}_{\text{RCAL}}, \bar{\gamma}_{\text{CAL}}) + D_{\text{CAL}}(\bar{\gamma}_{\text{CAL}}, \hat{\gamma}_{\text{RCAL}}) + (A_1 - 1)\tilde{\lambda}_1\lVert \hat{\gamma}_{\text{RCAL}} - \bar{\gamma}_{\text{CAL}} \rVert_{2, 1} \nonumber \\
& \leq 2A_1\tilde{\lambda}_1 \left\{\sum_{j\in S}\lVert \hat{\gamma}_{j\cdot, \text{RCAL}} - \bar{\gamma}_{j\cdot, \text{CAL}} \rVert_2+ \sum_{j\notin S}\lVert \bar{\gamma}_{j\cdot, \text{CAL}} \rVert_2 \right\}. \label{eq:ps-dual2}
\end{align}
\end{lem}

\begin{prf}
By direct calculation from the definition of $\kappa_{\text{CAL}}()$, we find
\begin{align*}
\langle\nabla\kappa_{\text{CAL}}(\bar{\gamma}_{\text{CAL}}), \hat{\gamma}_{\text{RCAL}} - \bar{\gamma}_{\text{CAL}} \rangle  =  \sum_{j=0}^p\sum_{k\neq t}(\hat{\gamma}_{jk, \text{RCAL}} - \bar{\gamma}_{jk, \text{CAL}})\tilde{\E}\left\{Z_{jk}(T, X; \bar{\gamma}_{\text{CAL}})\right\},
\end{align*}
which by the Cauchy--Schwartz inequality gives
\begin{align*}
& \lvert \langle\nabla\kappa_{\text{CAL}}(\bar{\gamma}_{\text{CAL}}), \hat{\gamma}_{\text{RCAL}} - \bar{\gamma}_{\text{CAL}} \rangle \rvert
\leq \sum_{j=0}^p \lVert \hat{\gamma}_{j\cdot,\text{RCAL}} - \bar{\gamma}_{j\cdot, \text{CAL}} \rVert_2 \lVert \tilde{\E}\left\{Z_j(T, X; \bar{\gamma}_{\text{CAL}})\right\} \rVert_2 \\
& \leq \left\{\sum_{j=0}^p \lVert \hat{\gamma}_{j\cdot,\text{RCAL}} - \bar{\gamma}_{j\cdot, \text{CAL}} \rVert_2\right\}\left\{\sup_{j=0, 1, \ldots, p} \lVert \tilde{\E}\left\{Z_j(T, X; \bar{\gamma}_{\text{CAL}})\right\} \rVert_2 \right\} .
\end{align*}
Combining the preceding inequality and inequality \eqref{eq:ps-score} yields inequality \eqref{eq:ps-dual1}. Combining \eqref{eq:ps-basic1} and \eqref{eq:ps-dual1} and taking $\lambda_1 = A_1\tilde{\lambda}_1$ yields
\begin{align*}
& D_{\text{CAL}}(\hat{\gamma}_{\text{RCAL}}, \bar{\gamma}_{\text{CAL}}) + D_{\text{CAL}}(\bar{\gamma}_{\text{CAL}}, \hat{\gamma}_{\text{RCAL}}) + A_1\tilde{\lambda}_1R(\hat{\gamma}_{\text{RCAL}}) \\
& \leq \tilde{\lambda}_1\{\lVert \hat{\gamma}_{0\cdot,\text{RCAL}} - \bar{\gamma}_{0\cdot,\text{CAL}} \rVert _2+ R(\hat{\gamma}_{\text{RCAL}} - \bar{\gamma}_{\text{CAL}})\} + A_1\tilde{\lambda}_1R(\bar{\gamma}_{\text{CAL}}).
\end{align*}
Applying to the preceding inequality the triangle inequalities
\begin{align*}
& \lVert \hat{\gamma}_{j\cdot,\text{RCAL}} \rVert_2 = \lVert \hat{\gamma}_{j\cdot,\text{RCAL}}  - \bar{\gamma}_{j\cdot,\text{CAL}} + \bar{\gamma}_{j\cdot,\text{CAL}}\rVert_2  \\
& \geq
\left\{
\begin{array}{ll}
 \lVert \hat{\gamma}_{j\cdot,\text{RCAL}} - \bar{\gamma}_{j\cdot,\text{CAL}} \rVert_2 - \lVert \bar{\gamma}_{j\cdot,\text{CAL}} \rVert_2, & j\notin S, \\
\lVert \bar{\gamma}_{j\cdot,\text{CAL}} \rVert_2 - \lVert \hat{\gamma}_{j\cdot,\text{RCAL}} - \bar{\gamma}_{j\cdot,\text{CAL}} \rVert_2, & j\in S\backslash \{0\},
\end{array}
\right.
\end{align*}
and rearranging the result gives
\begin{align*}
& D_{\text{CAL}}(\hat{\gamma}_{\text{RCAL}}, \bar{\gamma}_{\text{CAL}}) + D_{\text{CAL}}(\bar{\gamma}_{\text{CAL}}, \hat{\gamma}_{\text{RCAL}}) + (A_1 - 1)\tilde{\lambda}_1R(\hat{\gamma}_{\text{RCAL}} - \bar{\gamma}_{\text{CAL}}) \\
& \leq \tilde{\lambda}_1\lVert \hat{\gamma}_{0\cdot,\text{RCAL}} - \bar{\gamma}_{0\cdot,\text{CAL}} \rVert_2 + 2A_1\tilde{\lambda}_1\left\{\sum_{j\in S\backslash\{0\}} \lVert \hat{\gamma}_{j\cdot,\text{RCAL}} - \bar{\gamma}_{j\cdot,\text{CAL}} \rVert_2 + \sum_{j\in S}\lVert \bar{\gamma}_{j\cdot,\text{CAL}} \rVert_2\right\}.
\end{align*}
The conclusion follows by adding $(A_1 - 1)\tilde{\lambda}_1\lVert \hat{\gamma}_{0\cdot,\text{RCAL}} - \bar{\gamma}_{0\cdot,\text{CAL}} \rVert_2$ to both sides above.
\end{prf}

\begin{lem} \label{lem:ps-dagger}
Suppose that Assumption \ref{ass1}(i) holds. Then for any two matrices $\gamma$ and $\gamma^\prime$,
\begin{align}
& D_{\text{CAL}}(\gamma, \gamma^\prime) + D_{\text{CAL}}(\gamma^\prime, \gamma) \geq \frac{1 - \me^{-B_0\lVert b \rVert_{2, 1}}}{B_0\lVert b \rVert_{2, 1}}\text{vec}^\T(b)\tilde{\Sigma}_{\gamma}\text{vec}(b), \label{eq:ps-dagger}
\end{align}
where $b = \gamma - \gamma^\prime$, and $vec(b) = (b_k^\T: k\in\mathcal T\backslash\{t\})^\T$.
\end{lem}

\begin{prf}
By direct calculation from the definition of $D_{\text{CAL}}()$, we find
\begin{align*}
& D_{\text{CAL}}(\gamma, \gamma^\prime) + D_{\text{CAL}}(\gamma^\prime, \gamma) = \tilde{\E}[(g(T, X; \gamma^\prime) - g(T, X; \gamma))^\T\{h^\prime(X) - h(X)\}] \\
& = \tilde{\E}\left(\int_0^1(h^\prime - h)^\T \diag\{R^{(t)}\omega(k, X; \gamma + u(\gamma^\prime - \gamma)): k\neq t\}(h^\prime - h)du\right) \\
& \geq \tilde{\E}\left(\int_0^1(h^\prime - h)^\T \diag\{R^{(t)}\omega(k, X; \gamma): k\neq t\}\me^{-u\lVert h^\prime - h \rVert_{\infty}}(h^\prime - h)du\right) \\
& = \tilde{\E}\left\{(h^\prime - h)^\T \diag\{R^{(t)}\omega(k, X; \gamma): k\neq t\} (h^\prime - h)\int_0^1\me^{-u\lVert h^\prime - h \rVert_{\infty}}du\right\},
\end{align*}
where $h(X) = \gamma^\T f(X)$, $h^\prime(X) = \gamma^{\prime\T} f(X)$ and the inequality follows from $\omega(k, X; \gamma + u(\gamma^\prime - \gamma)) = \me^{h_k(X) + u\{h_k^\prime(X) - h_k(X)\}} \geq \me^{h_k(X)}\me^{-u\lVert h^\prime - h \rVert_{\infty}}$.
By Assumption \ref{ass1}(i), we have
\begin{align*}
& \lVert h^\prime - h \rVert_{\infty} = \max_{k\neq t} \lvert \sum_{j=0}^p b_{jk}f_j(X) \rvert \leq B_0\max_{k\neq t} \sum_{j=0}^p |b_{jk}| \\
& \leq B_0\sum_{j=0}^p\max_{k\neq t}|b_{jk}| = B_0 \sum_{j=0}^p \lVert b_{j\cdot} \rVert_{\infty} \leq B_0 \sum_{j=0}^p \lVert b_{j\cdot} \rVert_2 = B_0 \lVert b \rVert_{2, 1},
\end{align*}
which gives
\begin{align}
& \lVert (\gamma^\prime - \gamma)^\T f(X) \rVert_{\infty} \leq B_0\lVert \gamma^\prime - \gamma \rVert_{2, 1}, \label{eq:ps-h-diff}
\end{align}
Collecting all the inequalities we have
\begin{align*}
D_{\text{CAL}}(\gamma, \gamma^\prime) + D_{\text{CAL}}(\gamma^\prime, \gamma) \geq \tilde{\E}\left\{(h^\prime - h)^\T \diag\{R^{(t)}\omega(k, X; \gamma): k\neq t\} (h^\prime - h)\right\}\left(\int_0^1\me^{-uB_0\lVert b \rVert_{2, 1}}du\right),
\end{align*}
which gives the desired result.
\end{prf}

\begin{lem}\label{lem:ps-empicc}
Suppose that Assumptions \ref{ass1}(iv)(a) holds. In the event $\Omega_{\gamma 2}$ from Lemma \ref{lem:ps-hessian}, Assumption \ref{ass1}(iii) (theoretical compatibility condition) implies an empirical compatibility condition: for any matrix $b = (b_0, b_1, \ldots, b_{K-1})$ satisfying \eqref{eq:ps-cc2},
\begin{align}
(1 - \eta_{1,1})\nu_1^2\left(\sum_{j\in S}\lVert b_{j\cdot} \rVert_2\right)^2 \leq |S|\text{vec}^\T(b)\tilde{\Sigma}_{\gamma}\text{vec}(b). \label{eq:ps-empicc}
\end{align}
\end{lem}

\begin{prf}
By direct calculation, we find
\begin{align*}
\text{vec}^\T(b)(\tilde{\Sigma}_{\gamma}- \Sigma_{\gamma})\text{vec}(b) = \sum_{j_1, j_2 = 0, 1, \ldots, p}b^\T_{j_1\cdot}\{(\tilde{\Sigma}_{\gamma})_{j_1j_2} - (\Sigma_{\gamma})_{j_1j_2}\}b_{j_2\cdot},
\end{align*}
which by the Cauchy--Schwartz inequality and \eqref{eq:ps-hessian} gives
\begin{align*}
\lvert \text{vec}^\T(b)(\tilde{\Sigma}_{\gamma}- \Sigma_{\gamma})\text{vec}(b) \rvert & \leq \sum_{j_1, j_2 = 0, 1, \ldots, p}\left\{ \lVert b_{j_1\cdot} \rVert_2 \times \lVert (\tilde{\Sigma}_{\gamma})_{j_1j_2} - (\Sigma_{\gamma})_{j_1j_2} \rVert_{\text{op}} \times \lVert b_{j_2\cdot} \rVert_2 \right\} \\
& \leq \sup_{j_1, j_2 = 0, 1, \ldots, p} \lVert (\tilde{\Sigma}_{\gamma})_{j_1j_2} - (\Sigma_{\gamma})_{j_1j_2} \rVert_{\text{op}} \sum_{j_1, j_2 = 0, 1, \ldots, p} \lVert b_{j_1\cdot}\rVert_2 \lVert b_{j_2\cdot} \rVert_2 \\
& \leq \tilde{\lambda}_1(\sum_{j=0}^p \lVert b_{j\cdot} \rVert_2)^2 .
\end{align*}
Then Assumption \ref{ass1}(iii) implies that for any matrix $b = (b_0, b_1, \ldots, b_{K-1})$ satisfying \eqref{eq:ps-cc2},
\begin{align*}
& \nu_1^2(\sum_{j\in S}\lVert b_{j\cdot} \rVert)^2\leq |S|\text{vec}^\T(b)\Sigma_{\gamma}\text{vec}(b) \leq |S|\left\{\text{vec}^\T(b)\tilde{\Sigma}_{\gamma}\text{vec}(b) + \tilde{\lambda}_1(\sum_{j=0}^p \lVert b_{j\cdot} \rVert_2)^2\right\} \\
& \leq |S|\text{vec}^\T(b)\tilde{\Sigma}_{\gamma}\text{vec}(b) + |S|\tilde{\lambda}_1(1 + \xi_1)^2(\sum_{j\in S} \lVert b_{j\cdot} \rVert)^2,
\end{align*}
where the last equality is due to $\sum_{j=0}^p \lVert b_{j\cdot} \rVert \leq (1 + \xi_1)\sum_{j\in S}\lVert b_{j\cdot} \rVert$ by \eqref{eq:ps-cc2}. Then \eqref{eq:ps-empicc} follows because $(1 + \xi_1)^2\nu_1^{-2}|S|\tilde{\lambda}_1\leq \eta_{1,1}(<1)$ by Assumption \ref{ass1}(iv)(a).
\end{prf}

\begin{lem} \label{lem:ps-error}
Suppose that Assumption \ref{ass1} holds, and $A_1 > (\xi_1 + 1) / (\xi_1 - 1)$. In the event $\Omega_{\gamma 1}\cap \Omega_{\gamma 2}$, we have
\begin{align}
D^{\dagger}_{\text{CAL}}(\hat{\gamma}_{\text{RCAL}}, \bar{\gamma}_{\text{CAL}}) + (A_1 - 1)\tilde{\lambda}_1\lVert \hat{\gamma}_{\text{RCAL}} - \bar{\gamma}_{\text{CAL}} \rVert_{2, 1} \leq  \xi_{1, 1}^2\nu_{1, 1}^{-2}|S|\tilde{\lambda}_1^2 + 2\xi_{1, 2}^{-1}A_1\tilde{\lambda}_1\sum_{j\notin S}\lVert \bar{\gamma}_{j\cdot,\text{CAL}} \rVert_2, \label{eq:ps-error}
\end{align}
where $\xi_{1, 1} = (\xi_1 + 1)(A_1 - 1)$, $\xi_{1, 2} = 1 - 2A_1 / \xi_{1, 1}\in(0, 1]$, and $\nu_{1, 1} = \nu_1(1 - \eta_{1,1})^{1/2}(1 - \eta_{1,2})^{1/2}$.
\end{lem}

\begin{prf}
Denote $b = \hat{\gamma}_{\text{RCAL}} - \bar{\gamma}_{\text{CAL}}$ and $D^{\ddagger}(\hat{\gamma}_{\text{RCAL}}, \bar{\gamma}_{\text{CAL}}) = D^{\dagger}(\hat{\gamma}_{\text{RCAL}}, \bar{\gamma}_{\text{CAL}}) + (A_1 - 1)\tilde{\lambda}_1\lVert \hat{\gamma}_{\text{RCAL}} - \bar{\gamma}_{\text{CAL}} \rVert_{2, 1}$, that is, the left-hand side of \eqref{eq:ps-error}. By Lemma \ref{lem:ps-dual} under \eqref{eq:ps-dual1}, inequality \eqref{eq:ps-dual2} with the subset $S_\gamma$ from Assumption \ref{ass1}(iii), we have
\begin{align*}
\xi_{1, 2}D^{\ddagger}(\hat{\gamma}_{\text{RCAL}}, \bar{\gamma}_{\text{CAL}}) + (1 - \xi_{1, 2})D^{\ddagger}(\hat{\gamma}_{\text{RCAL}}, \bar{\gamma}_{\text{CAL}}) \leq 2A_1\tilde{\lambda}_1\sum_{j\notin S}\lVert \bar{\gamma}_{j\cdot,\text{CAL}} \rVert_2 + 2A_1\tilde{\lambda}_1\sum_{j\in S}\lVert b_{j\cdot} \rVert_2,
\end{align*}
which leads to two possible cases: either
\begin{align}
\xi_{1, 2}D^{\ddagger}(\hat{\gamma}_{\text{RCAL}}, \bar{\gamma}_{\text{CAL}}) \leq 2A_1\tilde{\lambda}_1\sum_{j\notin S}\lVert \bar{\gamma}_{j\cdot,\text{CAL}} \rVert_2, \label{eq:ps-error-case1}
\end{align}
or
$(1 - \xi_{1, 2})D^{\ddagger}(\hat{\gamma}_{\text{RCAL}}, \bar{\gamma}_{\text{CAL}}) \leq 2A_1\tilde{\lambda}_1\sum_{j\in S}\lVert b_{j\cdot} \rVert_2$, that is,
\begin{align}
D^{\ddagger}(\hat{\gamma}_{\text{RCAL}}, \bar{\gamma}_{\text{CAL}}) \leq \xi_{1, 1}\tilde{\lambda}_1\sum_{j\in S}\lVert b_{.j} \rVert = (\xi_1 + 1)(A_1 - 1)\tilde{\lambda}_1\sum_{j\in S}\lVert b_{j\cdot} \rVert_2. \label{eq:ps-error-case2}
\end{align}
If \eqref{eq:ps-error-case2} holds, then $\sum_{j\notin S} \lVert b_{j\cdot} \rVert_2 \leq \xi_1\sum_{j\in S}\lVert b_{j\cdot} \rVert_2$, which by Lemma \ref{lem:ps-empicc} under \eqref{eq:ps-hessian} and Assumptions \ref{ass1}(iii) and \ref{ass1}(iv)(a), implies \eqref{eq:ps-empicc}, that is,
\begin{align}
\sum_{j\in S}\lVert b_{j\cdot} \rVert \leq (1 - \eta_{1,1})^{-1/2}\nu_1^{-1}|S|^{1/2}\{\text{vec}^\T(b)\tilde{\Sigma}_{\gamma}\text{vec}(b)\}^{1/2}.\label{eq:ps-empicc2}
\end{align}
By Lemma \ref{lem:ps-dagger} under Assumptions \ref{ass1}(i), we have
\begin{align*}
D^{\dagger}_{\text{CAL}}(\hat{\gamma}_{\text{RCAL}}, \bar{\gamma}_{\text{CAL}}) \geq \frac{1 - \me^{-B_0\lVert b \rVert_{2, 1}}}{B_0\lVert b \rVert_{2, 1}}\text{vec}^\T(b)\tilde{\Sigma}_{\gamma}\text{vec}(b), 
\end{align*}
which gives
\begin{align}
\text{vec}^\T(b)\tilde{\Sigma}_{\gamma}\text{vec}(b) \leq \frac{B_0\lVert b \rVert_{2, 1}}{1 - \me^{-B_0\lVert b \rVert_{2, 1}}} D^{\dagger}_{\text{CAL}}(\hat{\gamma}_{\text{RCAL}}, \bar{\gamma}_{\text{CAL}}) \leq \frac{B_0\lVert b \rVert_{2, 1}}{1 - \me^{-B_0\lVert b \rVert_{2, 1}}}D^{\ddagger}_{\text{CAL}}(\hat{\gamma}_{\text{RCAL}}, \bar{\gamma}_{\text{CAL}}). \label{eq:ps-quad}
\end{align}
Inequalities \eqref{eq:ps-quad} and \eqref{eq:ps-error-case2} together yield
\begin{align}
\text{vec}^\T(b)\tilde{\Sigma}_{\bar{\gamma}}\text{vec}(b) \leq \frac{B_0\lVert b \rVert_{2, 1}}{1 - \me^{-B_0\lVert b \rVert_{2, 1}}}\xi_{1, 1}\tilde{\lambda}_1\sum_{j\in S}\lVert b_{j\cdot} \rVert_2. \label{ps-quad2}
\end{align}
Combining \eqref{eq:ps-empicc2} and \eqref{ps-quad2}, we have
\begin{align*}
\{\text{vec}^\T(b)\tilde{\Sigma}_{\gamma}\text{vec}(b)\}^{1/2} \leq \frac{B_0\lVert b \rVert_{2, 1}}{1 - \me^{-B_0\lVert b \rVert_{2, 1}}}\xi_{1,1}\tilde{\lambda}_1(1 - \eta_{1,1})^{-1/2}\nu_1^{-1}|S|^{1/2}, 
\end{align*}
which, due to \eqref{eq:ps-empicc2} again, gives
\begin{align}
\sum_{j\in S_{\gamma}}\lVert b_{j\cdot} \rVert_2 \leq (1 - \eta_{1,1})^{-1}\nu_1^{-2}|S|\xi_{1,1}\tilde{\lambda}_1\frac{B_0\lVert b \rVert_{2, 1}}{1 - \me^{-B_0\lVert b \rVert_{2, 1}}}. \label{eq:ps-boundb}
\end{align}
Combining \eqref{eq:ps-error-case2} and \eqref{eq:ps-boundb}, we have
\begin{align}
D^{\ddagger}(\hat{\gamma}_{\text{RCAL}}, \bar{\gamma}_{\text{CAL}}) \leq \xi_{1, 1}\tilde{\lambda}_1\sum_{j\in S}\lVert b_{j\cdot} \rVert_2 \leq \xi_{1,1}^2(1 - \eta_{1,1})^{-1}\nu_1^{-2}|S|\tilde{\lambda}^2_1\frac{B_0\lVert b \rVert_{2, 1}}{1 - \me^{-B_0\lVert b \rVert_{2, 1}}}, \label{eq:ps-ddagger}
\end{align}
which gives
\begin{align*}
1 - \me^{-B_0\lVert b \rVert_{2, 1}} \leq B_0\xi^2_{1,1}(A_1 - 1)^{-1}(1 - \eta_{1,1})^{-1}\nu_1^{-2}|S|\tilde{\lambda}_1\leq \eta_{1,2}(<1)
\end{align*}
by the fact $(A_1 - 1)\tilde{\lambda}_1\lVert b \rVert_{2, 1}\leq D^{\ddagger}(\hat{\gamma}_{\text{RCAL}}, \bar{\gamma}_{\text{CAL}})$ and Assumption \ref{ass1}(iv)(b). As a result, $B_0\lVert b \rVert_{2, 1} \leq -\log(1 - \eta_{1,2})$ and hence
\begin{align}
\frac{1 - \me^{-B_0\lVert b \rVert_{2, 1}}}{B_0\lVert b \rVert_{2, 1}} = \int_{0}^1 \me^{-uB_0\lVert b \rVert_{2, 1}}du \geq \me^{-B_0 \lVert b \rVert_{2, 1}} \geq 1 - \eta_{1,2}. \label{eq:ps-boundfrac}
\end{align}
From \eqref{eq:ps-ddagger} and \eqref{eq:ps-boundfrac}, we see that
if \eqref{eq:ps-error-case2} holds, then $D^{\ddagger}(\hat{\gamma}_{\text{RCAL}}, \bar{\gamma}_{\text{CAL}}) \leq \xi_{1,1}^2\nu_1^{-2}(1 - \eta_{1,1})^{-1}(1 - \eta_{1,2})^{-1}|S|\tilde{\lambda}_1^2$. Therefore, \eqref{eq:ps-error} holds through \eqref{eq:ps-error-case1} and \eqref{eq:ps-error-case2} in the event $\Omega_{\gamma 1}\cap \Omega_{\gamma 2}$.
\end{prf}

\subsection{Proof of Theorem \ref{thm:or-error-glm}}
Theorem \ref{thm:or-error-glm} is a direct result from Lemma \ref{lem:or-error}, which is proved by combining Lemmas \ref{lem:or-score}--\ref{lem:or-empicc}.
Lemma \ref{lem:or-score} gives a high-probability bound on the gradient of the loss $\ell_{\text{WL}}(\cdot; \bar\gamma_{\text{CAL}})$ in the dual of the  $\lVert \cdot \rVert_{2, 1}$ norm.
Lemma \ref{lem:or-basic} derives a basic inequality by the definition of $\hat{\alpha}^\#_{t,\text{RWL}}$ depending on $\hat{\gamma}_{\text{RCAL}}$.
Lemmas \ref{lem:ps-Q} and \ref{lem:or-diff} are combined to obtain Lemma \ref{lem:or-pbasic}, which is used to handle the dependency on $\hat{\gamma}_{\text{RCAL}}$ in Lemma \ref{lem:or-dual}.
Lemma \ref{lem:or-dagger} gives a suitable lower bound on the symmetrized Bregman divergence for local analysis.
Lemma \ref{lem:or-empicc}  shows the empirical compatibility condition holds.
Lemma \ref{lem:psor-2} gives a high-probability bound which is used in Lemma \ref{lem:or-diff}.

\begin{lem}\label{lem:or-score}
Denote by $\Omega_{\alpha 1}$ the event that
\begin{equation}
\sup_{j=0, 1, \ldots, p}\left\lVert \tilde{\E}\left[ R^{(t)} \{\omega(k, X; \bar{\gamma}_{\text{CAL}}) (Y- m(t, X; \bar{\alpha}^{(k)}_{t,\text{WL}})): k\neq t\}^\T f_j(X)\right] \right\rVert_2 \leq \tilde{\lambda}_2, \label{eq:or-score}
\end{equation}
Under Assumptions \ref{ass1}(i), \ref{ass1}(ii) and \ref{ass2}(i), if
\begin{align*}
\tilde{\lambda}_2 \geq \sqrt{3}B_0(B_1 - 1)\sigma_0\sqrt{\{(2/3)(K-1) + \log[(p+1)/\epsilon]\} / n},
\end{align*}
then $\P(\Omega_{\alpha 1}) \geq 1 - \epsilon$.
\end{lem}

\begin{prf}
Denote $Z_j(T, X, Y) = (Z_{jk}: k\neq t)^\T$ with $Z_{jk} = R^{(t)}\omega(k, X; \bar{\gamma}_{\text{CAL}})\{Y - \bar{m}(t, X; \bar{\alpha}^{(k)}_{t, \text{WL}})\}f_j(X)$. We show that $Z_j(T, X, Y)$ is a sub-gaussian random vector for $j=0, 1, \ldots, p$ with parameter $B_0^2(B_1 - 1)^2\sigma_0^2$. Then Lemma \ref{lem:or-score} can be proved similarly to Lemma \ref{lem:ps-score}.

By the definition of a sub-gaussian random vector, we need to show that
$\beta^\T Z_j(T, X, Y)$ is a sub-gaussian random variable with parameter $B_0^2(B_1 - 1)^2\sigma_0^2$ for any $\beta = (\beta_k: k\neq t)^\T\in\bbR^{K-1}$ such that $\lVert \beta \rVert_2 = 1$. It suffices to show that conditionally on any $(T,X)$,
$\beta^\T Z_j(T, X, Y) $ is a sub-gaussian random variable with parameter $B_0^2(B_1 - 1)^2\sigma_0^2$. For notational simplicity,
we argue conditionally on $(T,X)$ and denote $Z_j(T, X, Y)$ as $Z_j(Y)$ below.

We need to show that $\E_{T, X}\exp[a\beta^\T\{Z_j(Y) - \E_{T,X} Z_j(Y)\}] \leq \exp\{ B_0^2(B_1 - 1)^2\sigma_0^2a^2 / 2\}$ holds for any $a\geq 0$, where $\E_{T, X}$ denotes the conditional expectation given $T$ and $X$.
By the definition of $Z_{jk}$, we find $Z_{jk} - \E_{T,X} (Z_{jk} )  = \omega(k, X; \bar{\gamma}_{\text{CAL}})f_j(X)R^{(t)}\{Y^{(t)} - \E_{T,X} Y^{(t)}\}$, and hence
\begin{align*}
\E_{T, X}\left[\me^{a\beta^\T\{Z_j(Y) - \E_{T,X} Z_j(Y)\}}\right] & = \E_{T, X}\left[\me^{\{Y^{(t)} - \E_{T,X} Y^{(t)}\}af_j(X)R^{(t)}\sum_{k\neq t}\beta_k\omega(k, X; \bar{\gamma}_{\text{CAL}})}\right] \\
& \leq \exp\left[\sigma_0^2a^2f^2_j(X)\{\sum_{k\neq t}\beta_k\omega(k, X; \bar{\gamma}_{\text{CAL}})\}^2 / 2\right],
\end{align*}
where the inequality follows from $Y^{(t)} - \bar{m}(t, X; \bar{\alpha}^{(k)}_{t,\text{WL}})$ being a sub-gaussian random variable with parameter $\sigma_0^2$
given $(T,X)$ under unconfoundedness and Assumption \ref{ass2}(i). Furthermore, by the Cauchy--Schwarz inequality, we have
\begin{align*}
\{\sum_{k\neq t}\beta_k\omega(k, X; \bar{\gamma}_{\text{CAL}})\}^2 \leq \lVert \beta \rVert_2^2\sum_{k\neq t}\omega^2(k, X; \bar{\gamma}_{\text{CAL}}) \leq (B_1 - 1)^2
\end{align*}
under Assumption \ref{ass1}(ii). Combing the preceding two inequalities yields
\begin{align*}
\E_{T, X}\left[\me^{a\beta^\T\{Z_j(Y) -\E_{T,X} Z_j(Y)\}} \right] \leq \me^{\frac{B_0^2(B_1 - 1)^2\sigma_0^2a^2}{2}},
\end{align*}
which implies that $\beta^\T Z_j(T, X, Y)$ is a sub-gaussian random variable with parameter $B_0^2(B_1 - 1)^2\sigma_0^2$ given any $(T,X)$.
This completes the proof as explained earlier.
\end{prf}

Denote $\Sigma_{\alpha 2} = \E[A\otimes f(X)f^\T(X)]$ with $A = \diag\{R^{(t)}\omega(k,X; \bar{\gamma}_{\text{CAL}})\{Y -  \bar{m}(t, X; \bar{\alpha}^{(k)}_{t,\text{WL}})\}^2: k\neq t\}$ and $\tilde{\Sigma}_{\alpha 2}$ as the sample version of $\Sigma_{\alpha 2}$. Furthermore, denote $(\Sigma_{\alpha 2})_{j_1, j_2} = \E[Af_{j_1}(X)f_{j_2}(X)]$ and $(\tilde{\Sigma}_{\alpha 2})_{j_1, j_2}$ as the sample version of $(\Sigma_{\alpha 2})_{j_1, j_2}$.

\begin{lem}\label{lem:psor-2}
Denote by $\Omega_{2}$ the event that
\begin{align}
& \sup_{j_1, j_2 = 0, 1, \ldots, p}\lVert (\tilde{\Sigma}_{\alpha 2})_{j_1, j_2}  - (\Sigma_{\alpha 2})_{j_1, j_2} \rVert_{\text{op}} \leq \sigma_0^2\tilde{\lambda}_1, \label{eq:psor-2}
\end{align}
Under Assumptions \ref{ass1}(i), \ref{ass1}(ii) and \ref{ass2}(i), if
\begin{align*}
\tilde{\lambda}_1 \geq 4B_0^2B_1\left[\{ (1/2) \log(K-1) + \log[(p+1)/\epsilon]\} / n + \sqrt{\{ (1/2) \log(K-1) + \log[(p+1)/\epsilon]\} / n}\right],
\end{align*}
then $\P(\Omega_2) \geq 1 - 2\epsilon^2$.
\end{lem}

\begin{prf}
Denote $a_k = R^{(t)}\omega(k,X, \bar{\gamma}_{\text{CAL}})\{Y -  \bar{m}(t, X; \bar{\alpha}^{(k)}_{t,\text{WL}})\}^2f_{j_1}(X)f_{j_2}(X)$, which is the product of $\omega(k,X; \bar{\gamma}_{\text{CAL}})f_{j_1}(X)f_{j_2}(X)$ and $R^{(t)}\{Y -  \bar{m}(t, X; \bar{\alpha}^{(k)}_{t,\text{WL}})\}^2$.
Note that $\omega(k,X; \bar{\gamma}_{\text{CAL}})  | f_{j_1}(X)f_{j_2}(X)|  \leq B_0^2B_1$ by Assumptions \ref{ass1}(i) and \ref{ass1}(ii) and $R^{(t)}\{Y -  \bar{m}(t, X; \bar{\alpha}^{(k)}_{t,\text{WL}})\}$ is a sub-gaussian random variable by Assumption \ref{ass2}(i). Applying Lemmas 16 and 18 of Tan (2020a) yields
$$
P\left\{|\tilde{\E}a_k - \E a_k| > 2B_0^2B_1\sigma_0^2(t + \sqrt{2t})\right\} \leq \frac{2\epsilon^2}{(K-1)(p+1)^2},
$$
where $t = \log[(K-1)(p+1)^2/\epsilon^2] / n$. By the union bound, we have
\begin{align*}
& P\left\{\sup_{j_1, j_2 = 0, 1, \ldots, p}\lVert (\tilde{\Sigma}_{\alpha 2})_{j_1, j_2}  - (\Sigma_{\alpha 2})_{j_1, j_2} \rVert_{\text{op}} > 2B_0^2B_1\sigma_0^2(t + \sqrt{2t})\right\} \\
& \leq (K-1)(p+1)^2P\left\{|\tilde{\E}a_k - \E a_k| > 2B_0^2B_1\sigma_0^2(t + \sqrt{2t})\right\} \leq 2\epsilon^2,
\end{align*}
which completes the proof.
\end{prf}

\begin{lem} \label{lem:or-basic}
For any coefficient matrix $\alpha^\#_t$, we have
\begin{align}
& D^\dag_{\text{WL}}(\hat{\alpha}^\#_{t, \text{RWL}}, \alpha^\#_t; \hat{\gamma}_{\text{RCAL}}) 
+ \langle \nabla\kappa_{\text{WL}}(\alpha^\#_t; \hat{\gamma}_{\text{RCAL}}), \hat{\alpha}^\#_{\text{RWL}} - \alpha^\#_t \rangle \nonumber \\
& \leq  \lambda_2 R(\alpha^\#_t) - \lambda_2 R(\hat{\alpha}^\#_{t,\text{RWL}}), \label{eq:or-basic}
\end{align}
where $R(\alpha^\#_t) = \sum_{j=1}^p \lVert \alpha^\#_{jt} \rVert_2$ and $\alpha^\#_{jt} = (\alpha_{jt}^{(k)}: k\neq t)^\T$.
\end{lem}

\begin{prf}
The result can be proved similarly to Lemma \ref{lem:ps-basic}.
\end{prf}

\begin{lem}\label{lem:ps-Q}
In the event $\Omega_{\gamma 1}\cap\Omega_{\gamma 2}$, we have
\begin{align}
\sum_{k\neq t}\tilde{\E}\left\{R^{(t)}\omega(k,X; \bar{\gamma}_{\text{CAL}})(\hat{h}_k - \bar{h}_k)^2\right\} \leq \me^{\eta_{1, 3}}M_1|S_\gamma|\tilde{\lambda}_1^2, \label{eq:ps-Q1}
\end{align}
where $\hat{h} = (\hat{h}_k: k\in\mathcal T\backslash\{t\})^\T$ with $\hat{h}_k = \hat{\gamma}^\T_{k,\text{RCAL}}f(X)$ and $\bar{h} = (\bar{h}_k: k\in\mathcal T\backslash\{t\})^\T$ with $\bar{h}_k = \bar{\gamma}^\T_{k,\text{CAL}}f(X)$ Moreover,  for any coefficient matrix $\alpha^\#_t$,
\begin{align}
D_{\text{WL}}^{\dagger}(\hat{\alpha}^\#_{t,\text{RWL}}, \alpha^\#_t; \hat{\gamma}_{\text{RCAL}}) \geq \me^{-\eta_{1, 3}}D_{\text{WL}}^{\dagger}(\hat{\alpha}^\#_{t,\text{RWL}}, \alpha^\#_t; \bar{\gamma}_{\text{CAL}}), \label{eq:ps-Q2}
\end{align}
where $\eta_{1, 3} = (A_1 - 1)^{-1}M_1\eta_1B_0$.
\end{lem}

\begin{prf}
By direct calculation from the definition of $D_{\text{CAL}}()$, we find
\begin{align*}
D_{\text{CAL}}^{\dagger}(\hat{\gamma}_{\text{RCAL}}, \bar{\gamma}_{\text{CAL}}) & = \sum_{k\neq t}\tilde{\E}\left\{R^{(t)}(\me^{\hat{h}_k} - \me^{\bar{h}_k})(\hat{h}_k - \bar{h}_k)\right\} \\
& = \sum_{k\neq t}\tilde{\E}\left\{R^{(t)}\me^{u (\hat{h}_k - \bar{h}_k) }\omega(k, X; \bar{\gamma}_{\text{CAL}})(\hat{h}_k - \bar{h}_k)^2\right\},
\end{align*}
where the second step uses the mean value theorem
\begin{align}
\me^{\hat{h}_k} - \me^{\bar{h}_k} = \me^{u \hat{h}_k + (1-u )\bar{h}_k}(\hat{\gamma}_{k,\text{RCAL}} - \bar{\gamma}_{k,\text{CAL}})^\T f(X) \label{eq:mean-value}
\end{align}
for some $u  \in(0, 1)$, with dependency on $\hat h$ and $\bar h$ suppressed.
In the event $\Omega_{\gamma 1}\cap\Omega_{\gamma 2}$ that \eqref{eq:thm-ps-error} holds, we have
\begin{align}
\lVert \hat{\gamma}_{\text{RCAL}} - \bar{\gamma}_{\text{CAL}} \rVert_{2, 1} \leq (A_1 - 1)^{-1}M_1|S_\gamma|\tilde{\lambda}_1\leq (A_1 - 1)^{-1}M_1\eta_1, \label{eq:gamma-bound}
\end{align}
where $\eta_1$ is a constant such that $|S_\gamma|\tilde{\lambda}_1\leq \eta_1$ under Assumption \ref{ass1}(iv) and $M_1 = \xi_{1, 1}^2\nu_{1, 1}^{-2}$. Combining \eqref{eq:ps-h-diff}, \eqref{eq:thm-ps-error} and \eqref{eq:gamma-bound} yields
\begin{align*}
M_1|S_\gamma|\tilde{\lambda}_1^2\geq D_{\text{CAL}}^{\dagger}(\hat{\gamma}_{\text{RCAL}}, \bar{\gamma}_{\text{CAL}}) \geq \me^{-\eta_{1, 3}}\sum_{k\neq t}\tilde{\E}\left\{R^{(t)}\omega(k,X; \bar{\gamma}_{\text{CAL}})(\hat{h}_k - \bar{h}_k)^2\right\},
\end{align*}
which gives the desired inequality \eqref{eq:ps-Q1}. In addition, we write
\begin{align*}
& D_{\text{WL}}^{\dagger}(\hat{\alpha}^\#_{t,\text{RWL}}, \alpha^\#_t; \hat{\gamma}_{\text{RCAL}}) \\
& = \sum_{k\neq t}\tilde{\E}\left(R^{(t)}\omega(k,X; \hat{\gamma}_{\text{RCAL}})[\psi\{\hat{\alpha}^{(k)\T}_{t, \text{RWL}}f(X)\} - \psi\{\alpha^{(k)\T}_tf(X)\}](\hat{\alpha}^{(k)}_{t,\text{RWL}} - \alpha^{(k)}_t)^\T f(X)\right) \\
& =  \sum_{k\neq t}\tilde{\E}\left(R^{(t)}\me^{\hat{h}_k - \bar{h}_k}\omega(k, X; \bar{\gamma}_{\text{CAL}})[\psi\{\hat{\alpha}^{(k)\T}_{t, \text{RWL}}f(X)\} - \psi\{\alpha^{(k)\T}_tf(X)\}](\hat{\alpha}^{(k)}_{t,\text{RWL}} - \alpha^{(k)}_t)^\T f(X)\right) \\
& \geq \sum_{k\neq t}\tilde{\E}\left(\me^{-|\hat{h}_k - \bar{h}_k|}R^{(t)}\omega(k, X; \bar{\gamma}_{\text{CAL}})[\psi\{\hat{\alpha}^{(k)\T}_{t, \text{RWL}}f(X)\} - \psi\{\alpha^{(k)\T}_tf(X)\}](\hat{\alpha}^{(k)}_{t,\text{RWL}} - \alpha^{(k)}_t)^\T f(X)\right) \\
& \geq \sum_{k\neq t}\tilde{\E}\left(\me^{-B_0\lVert \hat{\gamma}_{\text{RCAL}} - \bar{\gamma}_{\text{CAL}} \rVert_{2, 1}}R^{(t)}\omega(k, X; \bar{\gamma}_{\text{CAL}})[\psi\{\hat{\alpha}^{(k)\T}_{t, \text{RWL}}f(X)\} - \psi\{\alpha^{(k)\T}_tf(X)\}]\right. \\
&\hspace{.6in}~ \left. (\hat{\alpha}^{(k)}_{t,\text{RWL}} - \alpha^{(k)}_t)^\T f(X)\right), \\
\end{align*}
where the last inequality follows from inequality \eqref{eq:ps-h-diff}. Then, inequality \eqref{eq:ps-Q2} is obtained by inequality \eqref{eq:gamma-bound} in the event $\Omega_{\gamma 1}\cap\Omega_{\gamma 2}$.
\end{prf}

\begin{lem} \label{lem:or-diff}
In the event $\Omega_{\gamma 1}\cap\Omega_{\gamma 2}\cap\Omega_2$,  we have
\begin{align}
& | \langle\nabla\kappa_{\text{WL}}(\bar{\alpha}^\#_{t,\text{WL}}; \hat{\gamma}_{\text{RCAL}}) -  \nabla\kappa_{\text{WL}}(\bar{\alpha}^\#_{t,\text{WL}}; \bar{\gamma}_{\text{CAL}}), \hat{\alpha}^\#_{t,\text{RWL}} - \bar{\alpha}^\#_{t,\text{WL}}\rangle | \nonumber \\
& \leq \me^{\eta_{1,3}}\{M_{1,1}|S_\gamma|\tilde{\lambda}_1^2\}^{1/2}Q_{\text{WL}}^{1/2}(\hat{\alpha}^\#_{t,\text{RWL}}, \bar{\alpha}^\#_{t,\text{WL}}; \bar{\gamma}_{\text{CAL}}), \label{eq:or-diff}
\end{align}
where $M_{1, 1} = 3\sigma_0^2(A_1 - 1)^{-2}M_1^2\eta_1 + 2\sigma_0^2\me^{\eta_{1, 3}}M_1$.
\end{lem}

\begin{prf}
Let $\hat{h}_k = \hat{\gamma}^\T_{k, \text{RCAL}}f(X)$,  $\bar{h}_k = \bar{\gamma}^\T_{k, \text{CAL}}f(X)$, $\hat{\eta} = \hat{\alpha}^{\#\T}_{t,\text{RWL}}f(X)$,  $\bar{\eta}  = \bar{\alpha}^{\#\T}_{t,\text{WL}}f(X)$ and $\Delta = \langle\nabla\kappa_{\text{WL}}(\bar{\alpha}^\#_{t,\text{WL}}; \hat{\gamma}_{\text{RCAL}}) -  \nabla\kappa_{\text{WL}}(\bar{\alpha}^\#_{t,\text{WL}}; \bar{\gamma}_{\text{CAL}}), \hat{\alpha}^\#_{t,\text{RWL}} - \bar{\alpha}^\#_{t,\text{WL}}\rangle$
as in \eqref{eq:or-diff}, and $\bar{m}_t^{(k)} = m (t,X; \bar{\alpha}^{(k)}_{t,\text{WL}})$. By direct calculation from the definition of $\kappa_{\text{WL}}()$, we find
\begin{align*}
 \Delta & = -\tilde{\E}\left[\sum_{k\neq t}R^{(t)}(\me^{\hat{h}_k} - \me^{\bar{h}_k})\{Y - \bar{m}_t^{(k)}\}(\hat{\eta}_k - \bar{\eta}_k)\right] \\
 & = -\tilde{\E}\left[\sum_{k\neq t}\me^{u(\hat{h}_k - \bar{h}_k)}R^{(t)}\me^{\bar{h}_k}\{Y - \bar{m}_t^{(k)}\}(\hat{h}_k - \bar{h}_k)(\hat{\eta}_k - \bar{\eta}_k)\right],
\end{align*}
where the last step follows from \eqref{eq:mean-value}. Denote $d_k =\hat{h}_k - \bar{h}_k$.
Then we have
\begin{align}
& |\Delta|  \leq \tilde{\E}\left[\sum_{k\neq t}\me^{u  d_k} \sqrt{R^{(t)}\me^{\bar{h}_k}}\left\lvert (Y - \bar{m}_t^{(k)})d_k\right\rvert \sqrt{R^{(t)}\me^{\bar{h}_k}}\left\lvert \hat{\eta}_k - \bar{\eta}_k \right\rvert\right] \nonumber \\
& \leq \me^{B_0\lVert \hat{\gamma}_{\text{RCAL}} - \bar{\gamma}_{\text{CAL}} \rVert_{2, 1}} \sum_{k\neq t}\tilde{\E}\left[\sqrt{R^{(t)}\me^{\bar{h}_k}}\left\lvert (Y - \bar{m}_t^{(k)}) d_k\right\rvert \sqrt{R^{(t)}\me^{\bar{h}_k}}\left\lvert \hat{\eta}_k - \bar{\eta}_k \right\rvert\right] \nonumber \\
& \leq \me^{B_0\lVert \hat{\gamma}_{\text{RCAL}} - \bar{\gamma}_{\text{CAL}} \rVert_{2, 1}} \sum_{k\neq t}\left\{\tilde{\E}^{1/2}\left[R^{(t)}\me^{\bar{h}_k}(Y - \bar{m}_t^{(k)})^2d_k^2\right] \tilde{\E}^{1/2}\left[R^{(t)}\me^{\bar{h}_k}(\hat{\eta}_k - \bar{\eta}_k)^2\right]\right\} \nonumber \\
& \leq \me^{B_0\lVert \hat{\gamma}_{\text{RCAL}} - \bar{\gamma}_{\text{CAL}} \rVert_{2, 1}} \left\{\sum_{k\neq t}\tilde{\E}\left[R^{(t)}\me^{\bar{h}_k}(Y - \bar{m}_t^{(k)})^2d_k^2\right]\right\}^{1/2} \left\{\sum_{k\neq t}\tilde{\E}\left[R^{(t)}\me^{\bar{h}_k}(\hat{\eta}_k - \bar{\eta}_k)^2\right]\right\}^{1/2} \nonumber \\
& = \me^{B_0\lVert \hat{\gamma}_{\text{RCAL}} - \bar{\gamma}_{\text{CAL}} \rVert_{2, 1}} \left\{\sum_{k\neq t}\tilde{\E}\left[R^{(t)}\me^{\bar{h}_k}(Y - \bar{m}_t^{(k)})^2d_k^2\right]\right\}^{1/2} Q_{\text{WL}}^{1/2}(\hat{\alpha}^\#_{t,\text{RWL}}, \bar{\alpha}^\#_{t,\text{WL}}; \bar{\gamma}_{\text{CAL}}), \label{eq:or-diff-init}
\end{align}
where the second step uses inequality \eqref{eq:ps-h-diff}, and the third and fourth steps use the Cauchy--Schwartz inequality.
In the following, we upper bound the second term on \eqref{eq:or-diff-init} in several steps.

First, by direct calculation, we find
\begin{align*}
& (\tilde{\E} - \E)\left[\sum_{k\neq t}R^{(t)}\me^{\bar{h}_k} (Y - \bar{m}_t^{(k)})^2d_k^2\right] =  (\tilde{\E} - \E)\left[\sum_{j_1, j_2 = 0, 1, \ldots, p}b_{j_1}^\T Af_{j_1}(X)f_{j_2}(X)b_{j_2}\right] \\
& \leq \left\{\sup_{j_1, j_2 = 0, 1, \ldots, p}\left\lVert \tilde{\E}[Af_{j_1}(X)f_{j_2}(X)] - \E[Af_{j_1}(X)f_{j_2}(X)] \right\rVert_{\text{op}}\right\}(\sum_{j=0}^p\lVert b_{j\cdot} \rVert_2)^2,
\end{align*}
where $A = \diag\{R^{(t)}\me^{\bar{h}_k} (Y - \bar{m}_t^{(k)})^2: k\neq t\}$, $b_{j\cdot} = \hat{\gamma}_{j\cdot,\text{RCAL}} - \bar{\gamma}_{j\cdot,\text{CAL}}$, and the last step follows from the Cauchy--Schwartz inequality.
Then by Lemma \ref{lem:psor-2} we have in the event $\Omega_2$
\begin{align}
(\tilde{\E} - \E)\left[\sum_{k\neq t}R^{(t)}\me^{\bar{h}_k}(Y - \bar{m}_t^{(k)})^2d_k^2\right] \leq \sigma_0^2\tilde{\lambda}_1\lVert \hat{\gamma}_{\text{RCAL}} - \bar{\gamma}_{\text{CAL}} \rVert^2_{2, 1}. \label{eq:or-diff-first}
\end{align}
Second, by Assumption \ref{ass2}(i) and Lemma 17 of Tan (2020b), $\E[(Y^{(t)} - \bar{m}_t^{(k)})^2 | X] \leq 2\sigma_0^2$ and hence
\begin{align}
\E\left[\sum_{k\neq t}R^{(t)}\me^{\bar{h}_k}(Y - \bar{m}_t^{(k)})^2d_k^2\right] \leq 2\sigma_0^2\sum_{k\neq t}\E\left[R^{(t)}\me^{\bar{h}_k}d_k^2\right]. \label{eq:or-diff-second}
\end{align}
Third, we have by simple manipulation
\begin{align*}
& (\tilde{\E} - \E)\left[\sum_{k\neq t}R^{(t)}\me^{\bar{h}_k}d_k^2\right] = \sum_{j_1, j_2 = 0, 1, \ldots, p}b_{j_1}^\T[(\tilde{\Sigma}_{\gamma})_{j_1j_2} - (\Sigma_{\gamma})_{j_1j_2}]b_{j_2} \\
& \leq \left\{\sup_{j_1, j_2 = 0, 1, \ldots, p}\left\lVert  (\tilde{\Sigma}_{\gamma})_{j_1j_2} - (\Sigma_{\gamma})_{j_1j_2} \right\rVert\right\}(\sum_{j=0}^p\lVert b_{j\cdot} \rVert)^2.
\end{align*}
In the event $\Omega_{\gamma 2}$, we have by inequalities \eqref{eq:ps-hessian},
\begin{align}
(\tilde{\E} - \E)\left[\sum_{k\neq t}R^{(t)}\me^{\bar{h}_k}d_k^2\right] \leq \tilde{\lambda}_1\lVert \hat{\gamma} - \bar{\gamma} \rVert^2_{2, 1}. \label{eq:or-diff-third}
\end{align}
Combining inequalities \eqref{eq:or-diff-first}, \eqref{eq:or-diff-second} and \eqref{eq:or-diff-third}, we have in the event $\Omega_{\gamma 2}\cap\Omega_2$,
\begin{align}
& \sum_{k\neq t}\tilde{\E}\left[R^{(t)}\omega(k,X; \bar{\gamma}_{\text{CAL}})(Y - \bar{m}_t^{(k)})^2d_k^2\right] \nonumber \\
& \leq 3\sigma_0^2\tilde{\lambda}_1 \lVert \hat{\gamma}_{\text{RCAL}} - \bar{\gamma}_{\text{CAL}} \rVert^2_{2, 1} + 2\sigma_0^2\sum_{k\neq t}\tilde{\E}\left[R^{(t)}\omega(k,X ; \bar{\gamma}_{\text{CAL}})d_k^2\right]. \label{eq:or-diff-last}
\end{align}
Collecting inequalities \eqref{eq:or-diff-init} and \eqref{eq:or-diff-last} and applying \eqref{eq:ps-Q1} and \eqref{eq:gamma-bound} in the event $\Omega_{\gamma 1}\cap\Omega_{\gamma 2}$, we complete the proof.
\end{prf}

\begin{lem}\label{lem:or-pbasic}
Suppose that Assumption \ref{ass2}(i) holds. Then in the event $\Omega_{\gamma 1}\cap\Omega_{\gamma 2}\cap\Omega_2$, \eqref{eq:or-basic} implies
\begin{align}
& \me^{-\eta_{1, 3}}D_{\text{WL}}^{\dagger}(\hat{\alpha}^\#_{t,\text{RWL}}, \bar{\alpha}^\#_{t,\text{WL}}; \bar{\gamma}_{\text{CAL}}) + \lambda_2 \sum_{j=1}^p \lVert \hat{\alpha}^\#_{jt,\text{RWL}} \rVert_2 \nonumber \\
& \leq -\langle \nabla\kappa_{\text{WL}}(\bar{\alpha}^\#_{t,\text{WL}}; \bar{\gamma}_{\text{CAL}}), \hat{\alpha}^\#_{t,\text{RWL}} - \bar{\alpha}^\#_{t,\text{WL}}\rangle + \lambda_2 \sum_{j=1}^p \lVert \bar{\alpha}^\#_{jt,\text{WL}} \rVert_2 \nonumber \\
&~~ + \me^{\eta_{1,3}}\{M_{1,1}|S_\gamma|\tilde{\lambda}_1^2\}^{1/2}Q_{\text{WL}}^{1/2}(\hat{\alpha}^\#_{t,\text{RWL}}, \bar{\alpha}^\#_{t,\text{WL}}; \bar{\gamma}_{\text{CAL}}), \label{eq:or-pbasic}
\end{align}
\end{lem}

\begin{prf}
Inequality \eqref{eq:or-basic} can be rewritten as
\begin{align*}
& D_{\text{WL}}^{\dagger}(\hat{\alpha}^\#_{t,\text{RWL}}, \bar{\alpha}^\#_{t,\text{WL}}; \hat{\gamma}_{\text{RCAL}}) + \lambda_2 \sum_{j=1}^p \lVert \hat{\alpha}^\#_{jt,\text{RWL}} \rVert_2 \\
& \leq -\langle\nabla\kappa_{\text{WL}}(\bar{\alpha}^\#_{t,\text{WL}}; \bar{\gamma}_{\text{CAL}}), \hat{\alpha}^\#_{t,\text{RWL}} - \bar{\alpha}^\#_{t,\text{WL}} \rangle + \lambda_2 \sum_{j=1}^p \lVert \bar{\alpha}^\#_{jt,\text{WL}} \rVert_2 \\
& \quad - \langle\nabla\kappa_{\text{WL}}(\bar{\alpha}^\#_{t,\text{WL}}; \hat{\gamma}_{\text{RCAL}}) - \nabla\kappa_{\text{WL}}(\bar{\alpha}^\#_{t,\text{WL}}; \bar{\gamma}_{\text{CAL}}), \hat{\alpha}^\#_{t,\text{RWL}} - \bar{\alpha}^\#_{t,\text{WL}} \rangle .
\end{align*}
Then the desired result follows by applying inequalities \eqref{eq:ps-Q2} and \eqref{eq:or-diff}.
\end{prf}

\begin{lem}\label{lem:or-dual}
Suppose that Assumption \ref{ass2}(i) holds. In the event $\Omega_{\alpha  1}$, we have
\begin{align}
|\langle\nabla\kappa_{\text{WL}}(\bar{\alpha}^\#_{t,\text{WL}}; \bar{\gamma}_{\text{CAL}}), \hat{\alpha}^\#_{t,\text{RWL}} - \bar{\alpha}^\#_{t,\text{WL}}\rangle| \leq \tilde{\lambda}_2\lVert \hat{\alpha}^\#_{t,\text{RWL}} - \bar{\alpha}^\#_{t,\text{WL}} \rVert_{2, 1}. \label{eq:or-dual}
\end{align}
Moreover, in the event $\Omega_{\gamma 1}\cap\Omega_{\gamma 2}\cap\Omega_2\cap\Omega_{\alpha  1}$, we have
\begin{align}
& \me^{-\eta_{1, 3}}D_{\text{WL}}^{\dagger}(\hat{\alpha}^\#_{t,\text{RWL}}, \bar{\alpha}^\#_{t,\text{WL}}; \bar{\gamma}_{\text{CAL}}) + (A_2 - 1)\tilde{\lambda}_2\lVert \hat{\alpha}^\#_{t,\text{RWL}} - \bar{\alpha}^\#_{t,\text{WL}} \rVert_{2, 1} \nonumber \\
& \leq \me^{\eta_{1, 3}}\{M_{1, 1}|S_\gamma|\tilde{\lambda}_1^2\}^{1/2}Q^{1/2}_{\text{WL}}(\hat{\alpha}^\#_{t,\text{RWL}}, \bar{\alpha}^\#_{t,\text{WL}}; \bar{\gamma}_{\text{CAL}}) + 2A_2\tilde{\lambda}_2\sum_{j\in S_{\alpha_t}}\lVert \hat{\alpha}^\#_{jt,\text{RWL}} - \bar{\alpha}^\#_{jt,\text{WL}} \rVert_2, \label{eq:or-perror}
\end{align}
\end{lem}

\begin{prf}
By direct calculation from the definition of $\kappa_{\text{WL}}()$, we find
\begin{align*}
& \langle\nabla\kappa_{\text{WL}}(\bar{\alpha}^\#_{t,\text{WL}}; \bar{\gamma}_{\text{CAL}}), \hat{\alpha}^\#_{t,\text{RWL}} - \bar{\alpha}^\#_{t,\text{WL}}\rangle \\
& = -\sum_{j=0}^p\sum_{k\neq t}(\hat{\alpha}^{(k)}_{jt,\text{RWL}} - \bar{\alpha}^{(k)}_{jt,\text{WL}})\tilde{\E}\left[R^{(t)}\omega(k,X; \bar{\gamma}_{\text{CAL}})(Y - \bar{m}_t^{(k)})f_j(X)\right],
\end{align*}
which by the Cauchy--Schwartz inequality gives
\begin{align*}
& |\langle\nabla\kappa_{\text{WL}}(\bar{\alpha}^\#_{t,\text{WL}}; \bar{\gamma}_{\text{CAL}}), \hat{\alpha}^\#_{t,\text{RWL}} - \bar{\alpha}^\#_{t,\text{WL}}\rangle| \\
& \leq \sum_{j=0}^p \lVert \hat{\alpha}^\#_{jt,\text{RWL}} - \bar{\alpha}^\#_{jt,\text{WL}} \rVert_2 \times \lVert \tilde{\E}\left[R^{(t)} \{\omega(k, X; \bar{\gamma}_{\text{CAL}}) (Y- m(t, X; \bar{\alpha}^{(k)}_{t,\text{WL}})): k\neq t\}^\T f_j(X)\right] \rVert_2   \\
& \leq \lVert \hat{\alpha}^\#_{t,\text{RWL}} - \bar{\alpha}^\#_{t,\text{WL}} \rVert_{2, 1}\times\left\{\sup_{j=0, 1, \ldots, p} \lVert \tilde{\E}\left[R^{(t)} \{\omega(k, X; \bar{\gamma}_{\text{CAL}}) (Y- m(t, X; \bar{\alpha}^{(k)}_{t,\text{WL}})): k\neq t\}^\T f_j(X)\right] \rVert_2\right\}.
\end{align*}
Then inequality \eqref{eq:or-dual} holds in the event $\Omega_{\alpha  1}$.

From inequalities \eqref{eq:or-pbasic} and \eqref{eq:or-dual} and taking $\lambda_2 = A_2\tilde{\lambda}_2$, we have
\begin{align*}
& \me^{-\eta_{1, 3}}D_{\text{WL}}^{\dagger}(\hat{\alpha}^\#_{t,\text{RWL}}, \bar{\alpha}^\#_{t,\text{WL}}; \bar{\gamma}_{\text{CAL}}) + A_2\tilde{\lambda}_2 \sum_{j=1}^p \lVert \hat{\alpha}^\#_{jt,\text{RWL}} \rVert \nonumber \\
& \leq \tilde{\lambda}_2\lVert \hat{\alpha}^\#_{t,\text{RWL}} - \bar{\alpha}^\#_{t,\text{WL}} \rVert_{2, 1} + A_2\tilde{\lambda}_2 \sum_{j=1}^p \lVert \bar{\alpha}^\#_{jt,\text{WL}} \rVert_2 \\
&~~ + \me^{\eta_{1,3}}\{M_{1,1}|S_\gamma|\tilde{\lambda}_1^2\}^{1/2}Q_{\text{WL}}^{1/2}(\hat{\alpha}^\#_{t,\text{RWL}}, \bar{\alpha}^\#_{t,\text{WL}}; \bar{\gamma}_{\text{CAL}}).
\end{align*}
Applying to the preceding inequality the triangle inequalities
\begin{align*}
& \lVert\hat{\alpha}^\#_{jt,\text{RWL}}\rVert_2 = \lVert \hat{\alpha}^\#_{jt,\text{RWL}} - \bar{\alpha}^\#_{jt,\text{WL}} + \bar{\alpha}^\#_{jt,\text{WL}} \rVert_2 \\
& \geq \left\{
\begin{array}{ll}
\lVert \hat{\alpha}^\#_{jt,\text{RWL}} - \bar{\alpha}^\#_{jt,\text{WL}} \rVert_2 - \lVert \bar{\alpha}^\#_{jt,\text{WL}} \rVert_2, & j\notin S_{\alpha_t}, \\
\lVert \bar{\alpha}^\#_{jt,\text{WL}} \rVert_2 - \lVert \hat{\alpha}^\#_{jt,\text{RWL}} - \bar{\alpha}^\#_{jt,\text{WL}} \rVert_2, & j\in S_{\alpha_t}\backslash\{0\},
\end{array}
\right.
\end{align*}
and rearranging the results yield
\begin{align*}
& \me^{-\eta_{1, 3}}D_{\text{WL}}^{\dagger}(\hat{\alpha}^\#_{t,\text{RWL}}, \bar{\alpha}^\#_{t,\text{WL}}; \bar{\gamma}_{\text{CAL}}) + (A_2 - 1)\tilde{\lambda}_2\sum_{j=1}^p \lVert \hat{\alpha}^\#_{jt,\text{RWL}} - \bar{\alpha}^\#_{jt,\text{WL}} \rVert_2 \\
& \leq \tilde{\lambda}_2\lVert \hat{\alpha}^\#_{0t,\text{RWL}} - \bar{\alpha}^\#_{0t,\text{WL}} \rVert_2 + 2A_2\tilde{\lambda}_2\left\{\sum_{j\in S_{\alpha_t}\backslash\{0\}}\lVert \hat{\alpha}^\#_{jt,\text{RWL}} - \bar{\alpha}^\#_{jt,\text{WL}} \rVert_2 + \sum_{j\notin S_{\alpha_t}} \lVert \bar{\alpha}^\#_{jt,\text{WL}} \rVert_2\right\} \\
& \quad + \me^{\eta_{1,3}}\{M_{1,1}|S_\gamma|\tilde{\lambda}_1^2\}^{1/2}Q_{\text{WL}}^{1/2}(\hat{\alpha}^\#_{t,\text{RWL}}, \bar{\alpha}^\#_{t,\text{WL}}; \bar{\gamma}_{\text{CAL}}).
\end{align*}
Adding $(A_2 - 1)\tilde{\lambda}_2\lVert \hat{\alpha}^\#_{0t,\text{RWL}} - \bar{\alpha}^\#_{0t,\text{WL}} \rVert_2$ to both sides above, we have
\begin{align*}
& \me^{-\eta_{1, 3}}D_{\text{WL}}^{\dagger}(\hat{\alpha}^\#_{t,\text{RWL}}, \bar{\alpha}^\#_{t,\text{WL}}; \bar{\gamma}_{\text{CAL}}) + (A_2 - 1)\tilde{\lambda}_2\lVert \hat{\alpha}^\#_{t,\text{RWL}} - \bar{\alpha}^\#_{t,\text{WL}} \rVert_{2, 1} \\
& \leq 2A_2\tilde{\lambda}_2\left\{\sum_{j\in S_{\alpha_t}}\lVert \hat{\alpha}^\#_{jt,\text{RWL}} - \bar{\alpha}^\#_{jt,\text{WL}} \rVert_2 + \sum_{j\notin S_{\alpha_t}} \lVert \bar{\alpha}^\#_{jt,\text{WL}} \rVert_2\right\} \\
& \quad + \me^{\eta_{1,3}}\{M_{1,1}|S_\gamma|\tilde{\lambda}_1^2\}^{1/2}Q_{\text{WL}}^{1/2}(\hat{\alpha}^\#_{t,\text{RWL}}, \bar{\alpha}^\#_{t,\text{WL}}; \bar{\gamma}_{\text{CAL}}),
\end{align*}
which gives inequality \eqref{eq:or-perror}.
\end{prf}

Denote $\Sigma_\alpha = \E[\diag\{R^{(t)}\psi_2\{\alpha^{(k)\T}_tf(X)\}\omega(k,X; \bar{\gamma}_{\text{CAL}}):k\neq t\}\otimes f(X)f^\T(X)]$ and  the sample version of $\Sigma_\alpha$ as $\tilde{\Sigma}_\alpha$.

\begin{lem}\label{lem:or-dagger}
Suppose that Assumption \ref{ass2}(v) holds. Then for any two coefficient matrices $\alpha^\#_t$ and $\alpha^{\prime\#}_t$
\begin{align}
D_{\text{WL}}(\alpha^\#_t, \alpha^{\prime\#}_t; \bar{\gamma}_{\text{CAL}}) + D_{\text{WL}}(\alpha^{\prime\#}_t, \alpha^\#_t; \bar{\gamma}_{\text{CAL}}) \geq \frac{1 - \me^{-B_0C_3\lVert b \rVert_{2, 1}}}{B_0C_3\lVert b \rVert_{2, 1}}\text{vec}^\T(b)\tilde{\Sigma}_{\alpha}\text{vec}(b), \label{eq:or-dagger}
\end{align}
where $b = \alpha^\#_t - \alpha^{\prime\#}_t$.
\end{lem}

\begin{prf}
Denote $h(X) = \alpha^{\#\T}_tf(X)$ and $h^\prime(X) = \alpha^{\prime\#\T}_tf(X)$. Furthermore, denote $\psi(h) = (\psi\{h_k(X)\}: k\neq t)^\T$ and $\Psi_2\{h(X)\} = \diag\{\psi_2\{h_k(X)\}: k \neq t\}$. By direct calculation from the definition of $D_{\text{WL}}()$, we find
\begin{align*}
& D_{\text{WL}}(\alpha^\#_t, \alpha^{\prime\#}_t; \bar{\gamma}_{\text{CAL}}) + D_{\text{WL}}(\alpha^{\prime\#}_t, \alpha^\#_t; \bar{\gamma}_{\text{CAL}}) \\
& = \tilde{\E}\left(\left[R^{(t)}\diag\{\omega(k,X; \bar{\gamma}_{\text{CAL}}):k\neq t\}\{\psi(h^\prime) - \psi(h)\}\right]^\T\{h^\prime(X) - h(X)\}\right) \\
& = \tilde{\E}\left[\int_0^1 (h^\prime - h)^\T R^{(t)}\diag\{\omega(k,X; \bar{\gamma}_{\text{CAL}}):k\neq t\}\Psi_2\{h + u(h^\prime - h)\}(h^\prime - h)du\right],
\end{align*}
By Assumption \ref{ass2}(vi) and inequality \eqref{eq:ps-h-diff}, we have
\begin{align*}
& D_{\text{WL}}(\alpha^\#_t, \alpha^{\prime\#}_t; \bar{\gamma}_{\text{CAL}}) + D_{\text{WL}}(\alpha^{\prime\#}_t, \alpha^\#_t; \bar{\gamma}_{\text{CAL}}) \\
& \geq \tilde{\E}\left[\left(\int_0^1 \me^{-uC_3\lVert h - h^\prime \rVert_{\infty}}du\right)(h^\prime - h)^\T R^{(t)}\diag\{\omega(k,X; \bar{\gamma}_{\text{CAL}}):k\neq t\} \Psi_2\{h\}(h^\prime - h)\right] \\
& \geq \frac{1 - \me^{-B_0C_3\lVert b \rVert_{2, 1}}}{B_0C_3\lVert b \rVert_{2, 1}}\tilde{\E}\left[(h - h^\prime)^\T R^{(t)}\diag\{\omega(k,X; \bar{\gamma}_{\text{CAL}}):k\neq t\} \Psi_2\{h\}(h - h^\prime)\right],
\end{align*}
which completes the proof.
\end{prf}

\begin{lem}\label{lem:or-empicc}
Suppose that Assumption \ref{ass2}(ii) holds. In the event $\Omega_{\gamma 2}$, Assumption \ref{ass2}(ii) implies an empirical compatibility condition for $\tilde{\Sigma}_{\gamma}$: for any matrix $b = (b_k: k\neq t)$ such that $\sum_{j\notin S_{\alpha_t}} \lVert b_{j\cdot} \rVert_2 \leq \xi_2\sum_{j\in S_{\alpha_t}} \lVert b_{j\cdot} \rVert_2$, we have
\begin{align}
(1 - \eta_2)\nu_2^2\left(\sum_{j\in S_{\alpha_t}} \lVert b_{j\cdot} \rVert_2\right)^2 \leq |S_{\alpha_t}|\text{vec}^\T(b)\tilde{\Sigma}_{\gamma}\text{vec}(b). \label{eq:or-empicc}
\end{align}
\end{lem}

\begin{prf}
Similarly as in the proof of Lemma \ref{lem:ps-empicc}, we know
\begin{align*}
|\text{vec}^\T(b)(\tilde{\Sigma}_{\gamma} - \Sigma_{\gamma})\text{vec}(b)| \leq \tilde{\lambda}_1\lVert b \rVert^2_{2, 1}
\end{align*}
in the event $\Omega_{\gamma 2}$ by \eqref{lem:ps-hessian}. Then Assumption \ref{ass2}(ii) implies that for any matrix $b = (b_k: k\neq t)$ satisfying $\sum_{j\notin S_{\alpha_t}} \lVert b_{j\cdot} \rVert_2 \leq \xi_2\sum_{j\in S_{\alpha_t}} \lVert b_{j\cdot} \rVert_2$,
\begin{align*}
\nu_2^2\left(\sum_{j\in S_{\alpha_t}} \lVert b_{j\cdot} \rVert_2\right)^2 & \leq |S_{\alpha_t}|\text{vec}^\T(b)\Sigma_{\gamma}\text{vec}(b) \leq |S_{\alpha_t}|\text{vec}^\T(b)\tilde{\Sigma}_{\gamma}\text{vec}(b) + |S_{\alpha_t}|\tilde{\lambda}_1\left(\sum_{j\in S_{\alpha_t}} \lVert b_{j\cdot} \rVert_2 + \sum_{j\notin S_{\alpha_t}}\lVert b_{j\cdot} \rVert_2 \right)^2 \\
& \leq |S_{\alpha_t}|\text{vec}^\T(b)\tilde{\Sigma}_{\gamma}\text{vec}(b) + |S_{\alpha_t}|\tilde{\lambda}_2(1 + \xi_2)^2\left(\sum_{j\in S_{\alpha_t}} \lVert b_{j\cdot} \rVert_2\right)^2.
\end{align*}
Then \eqref{eq:or-empicc} follows because $(1 + \xi_2)^2\nu_2^{-2}|S_{\alpha_t}|\tilde{\lambda}_2\leq \eta_2(<1)$ by Assumption \ref{ass2}(vi)(a).
\end{prf}

\begin{lem}\label{lem:or-error}
Suppose that Assumption \ref{ass2} hold, and $A_2 \geq (\xi_2 + 1) / (\xi_2 - 1)$. In the event $\Omega_{\gamma 1}\cap\Omega_{\gamma 2}\cap\Omega_{\alpha  1}\cap\Omega_2$, we have
\begin{align}
& D^{\dagger}_{\text{WL}}(\hat{\alpha}^\#_{t,\text{RWL}}, \bar{\alpha}^\#_{t,\text{WL}}; \bar{\gamma}_{\text{CAL}}) + \me^{\eta_{1, 3}}(A_2 - 1)\tilde{\lambda}_2\lVert \hat{\alpha}^\#_{t,\text{RWL}} - \bar{\alpha}^\#_{t,\text{WL}} \rVert_{2, 1} \nonumber \\
& \leq \me^{4\eta_{1, 3}}\xi_{2, 3}^{-2}\{M_{1, 1}|S_{\gamma}|\tilde{\lambda}_1^2\} + \me^{2\eta_{1, 3}}\xi_{2, 2}^2\{\nu_{2, 2}^{-2}|S_{\alpha_t}|\tilde{\lambda}_2^2\}, \label{eq:or-error}
\end{align}
where $\xi_{2, 3} = \xi_{2, 1}C_2^{1/2}(1 - \eta_{2, 2})^{1/2}$, $\xi_{2, 2} = (\xi_2 + 1)(A_2 - 1)$, $\xi_{2, 1} = 1 - 2A_2 /\xi_{2, 2}\in(0, 1)$, $\nu_{2, 2} = \nu_{2, 1}C_2^{1/2}(1 - \eta_{2, 1})^{1/2}$ and $\nu_{2, 1} = \nu_2(1 - \eta_2)^{1/2}$.
\end{lem}

\begin{prf}
Denote $b = \hat{\alpha}^\#_{t,\text{RWL}} - \bar{\alpha}^\#_{t,\text{WL}}$, and
\begin{align*}
D^{\ddagger}_{\text{WL}}(\hat{\alpha}^\#_{t,\text{RWL}}, \bar{\alpha}^\#_{t,\text{WL}}; \bar{\gamma}_{\text{CAL}}) = \me^{-\eta_{1, 3}}D^{\dagger}_{\text{WL}}(\hat{\alpha}^\#_{t,\text{RWL}}, \bar{\alpha}^\#_{t,\text{WL}}; \bar{\gamma}_{\text{CAL}}) + (A_2 - 1)\tilde{\lambda}_2\lVert b \rVert_{2, 1}.
\end{align*}
In the event $\Omega_{\gamma 1}\cap\Omega_{\gamma 2}\cap\Omega_{\alpha  1}\cap\Omega_2$, inequality \eqref{eq:or-perror} from Lemma \ref{lem:or-dual} leads two possible case: either
\begin{align}
\xi_{2, 1}D^{\ddagger}_{\text{WL}}(\hat{\alpha}^\#_{t,\text{RWL}}, \bar{\alpha}^\#_{t,\text{WL}}; \bar{\gamma}_{\text{CAL}}) \leq \me^{\eta_{1, 3}}\{M_{1, 1}|S_{\gamma}|\tilde{\lambda}_1^2\}^{1/2}Q^{1/2}_{\text{WL}}(\hat{\alpha}^\#_{t,\text{RWL}}, \bar{\alpha}^\#_{t,\text{WL}}; \bar{\gamma}_{\text{CAL}}) \label{eq:or-error-case1}
\end{align}
or $(1 - \xi_{2, 1})D^{\ddagger}_{\text{WL}}(\hat{\alpha}^\#_{t,\text{RWL}}, \bar{\alpha}^\#_{t,\text{WL}}; \bar{\gamma}_{\text{CAL}}) \leq 2A_2\tilde{\lambda}_2\sum_{j\in S_{\alpha_t}} \lVert b_{j\cdot} \rVert_2$, that is
\begin{align}
D^{\ddagger}_{\text{WL}}(\hat{\alpha}^\#_{t,\text{RWL}}, \bar{\alpha}^\#_{t,\text{WL}}; \bar{\gamma}_{\text{CAL}}) \leq (1 - \xi_{2, 1})^{-1}2A_2\tilde{\lambda}_2\sum_{j\in S_{\alpha_t}} \lVert b_{j\cdot} \rVert_2 = \xi_{2, 2}\tilde{\lambda}_2\sum_{j\in S_{\alpha_t}} \lVert b_{j\cdot} \rVert_2. \label{eq:or-error-case2}
\end{align}
We deal with the above two cases separately as follows.

If \eqref{eq:or-error-case2} holds, then $\sum_{j\notin S_{\alpha_t}}\lVert b_{j\cdot} \rVert_2 \leq \xi_2\sum_{j\in S_{\alpha_t}} \lVert b_{j\cdot} \rVert_2$, which by Lemma \ref{lem:or-empicc} and Assumptions \ref{ass2}(ii) and \ref{ass2}(vi)(a), implies \eqref{eq:or-empicc}, that is,
\begin{align}
\sum_{j\in S_{\alpha_t}}\lVert b_{j\cdot} \rVert_2 \leq (1 - \eta_2)^{-1/2}\nu_2^{-1}|S_{\alpha_t}|^{1/2}\{\text{vec}^\T(b)\tilde{\Sigma}_{\gamma}\text{vec}(b)\}^{1/2}. \label{eq:or-empicc2}
\end{align}
By Assumption \ref{ass2}(iv), and Lemma \ref{lem:or-dagger} with Assumption \ref{ass2}(v), we have
\begin{align}
D^{\dagger}_{\text{WL}}(\alpha^\#_t, \alpha^{\prime\#}_t; \bar{\gamma}_{\text{CAL}}) \geq \frac{1 - \me^{-B_0C_3\lVert b \rVert_{2, 1}}}{B_0C_3\lVert b \rVert_{2, 1}}\text{vec}^\T(b)\tilde{\Sigma}_{\alpha}\text{vec}(b) \geq \frac{1 - \me^{-B_0C_3\lVert b \rVert_{2, 1}}}{B_0C_3\lVert b \rVert_{2, 1}}C_2\text{vec}^\T(b)\tilde{\Sigma}_{\gamma}\text{vec}(b). \label{eq:or-dagger2}
\end{align}
Because $D^{\dagger}(\alpha^\#_t, \alpha^{\prime\#}_t; \bar{\gamma}_{\text{CAL}}) \leq \me^{\eta_{1, 3}}D^{\ddagger}_{\text{WL}}(\alpha^\#_t, \alpha^{\prime\#}_t; \bar{\gamma}_{\text{CAL}})$, it follows from inequalities \eqref{eq:or-dagger2} and \eqref{eq:or-error-case2} that
\begin{align}
\text{vec}^\T(b)\tilde{\Sigma}_{\gamma}\text{vec}(b) \leq \frac{B_0C_3\lVert b \rVert_{2, 1}}{1 - \me^{-B_0C_3\lVert b \rVert_{2, 1}}}C_2^{-1}\me^{\eta_{1, 3}}\xi_{2, 2}\tilde{\lambda}_2\sum_{j\in S_{\alpha_t}} \lVert b_{j\cdot} \rVert_2. \label{eq:or-quad}
\end{align}
Combining inequalities \eqref{eq:or-quad} and \eqref{eq:or-empicc2} gives
\begin{align}
\{\text{vec}^\T(b)\tilde{\Sigma}_{\gamma}b\}^{1/2} \leq \frac{B_0C_3\lVert b \rVert_{2, 1}}{1 - \me^{-B_0C_3\lVert b \rVert_{2, 1}}}C_2^{-1}\me^{\eta_{1, 3}}\xi_{2, 2}\tilde{\lambda}_2(1 - \eta_2)^{-1/2}\nu_2^{-1}|S_{\alpha_t}|^{1/2}. \label{eq:or-quad2}
\end{align}
Inequalities \eqref{eq:or-quad2}, \eqref{eq:or-error-case2} and \eqref{eq:or-empicc2} yields
\begin{align}
D^{\ddagger}_{\text{WL}}(\alpha^\#_t, \alpha^{\prime\#}_t; \bar{\gamma}_{\text{CAL}}) \leq \me^{\eta_{1, 3}}\xi_{2, 2}^2(1 - \eta_2)^{-1}\nu_2^{-2}C_2^{-1}|S_{\alpha_t}|\tilde{\lambda}_2^2\frac{B_0C_3\lVert b \rVert_{2, 1}}{1 - \me^{-B_0C_3\lVert b \rVert_{2, 1}}}, \label{eq:or-ddagger}
\end{align}
which gives
\begin{align*}
1 - \me^{-B_0C_3\lVert b \rVert_{2, 1}} \leq B_0C_3(A_2 - 1)^{-1}\me^{\eta_{1, 3}}\xi_{2, 2}^2(1 - \eta_2)^{-1}\nu_2^{-2}C_2^{-1}|S_{\alpha_t}|\tilde{\lambda}_2 \leq \eta_{2, 1}
\end{align*}
by the fact $(A_2 - 1)\tilde{\lambda}_2\lVert b \rVert_{2, 1}\leq D^{\ddagger}_{\text{WL}}(\alpha^\#_t, \alpha^{\prime\#}_t; \bar{\gamma}_{\text{CAL}})$ and Assumption \ref{ass2}(vi)(b). As a result $B_0C_3\lVert b \rVert_{2, 1} \leq -\log(1 - \eta_{2, 1})$ and hence
\begin{align}
\frac{1 - \me^{-B_0C_3\lVert b \rVert_{2, 1}}}{B_0C_3\lVert b \rVert_{2, 1}} = \int_0^1 \me^{-uB_0C_3\lVert b \rVert_{2, 1}}du \geq 1 - \eta_{2, 1}. \label{eq:or-boundfrac}
\end{align}
Combining inequalities \eqref{eq:or-ddagger} and \eqref{eq:or-boundfrac} gives $D^{\ddagger}_{\text{WL}}(\alpha^\#_t, \alpha^{\prime\#}_t; \bar{\gamma}_{\text{CAL}}) \leq \me^{\eta_{1, 3}}\xi_{2, 2}^2\nu_{2, 2}^{-2}|S_{\alpha_t}|\tilde{\lambda}_2^2$, where $\nu_{2, 2} = (1 - \eta_{2, 1})^{1/2}C_2^{1/2}\nu_2(1 - \eta_2)^{1/2}$.

If \eqref{eq:or-error-case1} holds, we have
\begin{align}
& D^{\dagger}_{\text{WL}}(\alpha^\#_t, \alpha^{\prime\#}_t; \bar{\gamma}_{\text{CAL}}) \nonumber \\
& \leq \me^{\eta_{1,3}}D^{\ddagger}_{\text{WL}}(\alpha^\#_t, \alpha^{\prime\#}_t; \bar{\gamma}_{\text{CAL}}) \leq \xi_{2, 1}^{-1}\me^{2\eta_{1, 3}}\{M_{1, 1}|S_{\gamma}|\tilde{\lambda}_1^2\}^{1/2}\{\text{vec}^\T(b)\tilde{\Sigma}_{\gamma}\text{vec}(b)\}^{1/2} \label{eq:or-dagger3}
\end{align}
because $Q^{1/2}_{\text{WL}}(\alpha^\#_t, \alpha^{\prime\#}_t; \bar{\gamma}_{\text{CAL}}) = \text{vec}^\T(b)\tilde{\Sigma}_{\gamma}\text{vec}(b)$. Then, by inequality \eqref{eq:or-dagger2}, we have
\begin{align}
\{\text{vec}^\T(b)\tilde{\Sigma}_{\gamma}\text{vec}(b)\}^{1/2} \leq C_2^{-1}\xi_{2, 1}^{-1}\me^{2\eta_{1, 3}}\{M_{1, 1}|S_{\gamma}|\tilde{\lambda}_1^2\}^{1/2}\frac{B_0C_3\lVert b \rVert_{2, 1}}{1 - \me^{-B_0C_3\lVert b \rVert_{2, 1}}}. \label{eq:or-quad3}
\end{align}
Inequalities \eqref{eq:or-dagger3} and \eqref{eq:or-quad3} yield
\begin{align}
D^{\ddagger}_{\text{WL}}(\alpha^\#_t, \alpha^{\prime\#}_t; \bar{\gamma}_{\text{CAL}}) \leq C_2^{-1}\xi_{2, 1}^{-2}\me^{3\eta_{1, 3}}\{M_{1, 1}|S_{\gamma}|\tilde{\lambda}_1^2\}\frac{B_0C_3\lVert b \rVert_{2, 1}}{1 - \me^{-B_0C_3\lVert b \rVert_{2, 1}}}. \label{eq:or-ddagger2}
\end{align}
Using $(A_2 - 1)\tilde{\lambda}_2\lVert b \rVert_{2, 1} \leq D^{\ddagger}_{\text{WL}}(\alpha^\#_t, \alpha^{\prime\#}_t; \bar{\gamma}_{\text{CAL}})$, we find
\begin{align*}
1 - \me^{-B_0C_3\lVert b \rVert_{2, 1}} \leq \me^{3\eta_{1, 3}}\xi_{2, 1}^{-2}C_2^{-1}B_0C_3(A_2 - 1)^{-1}\{M_{1, 1}|S_{\gamma}|\tilde{\lambda}_1\} \leq \eta_{2, 2}
\end{align*}
with Assumption \ref{ass2}(vi)(c). As a result $B_0C_3\lVert b \rVert_{2, 1} \leq -\log(1 - \eta_{2, 2})$ and hence
\begin{align}
\frac{1 - \me^{-B_0C_3\lVert b \rVert_{2, 1}}}{B_0C_3\lVert b \rVert_{2, 1}} = \int_0^1 \me^{-uB_0C_3\lVert b \rVert_{2, 1}}du \geq 1 - \eta_{2, 2}. \label{eq:or-boundfrac2}
\end{align}
Combining inequalities \eqref{eq:or-ddagger2} and \eqref{eq:or-boundfrac2} gives $D^{\ddagger}_{\text{WL}}(\alpha^\#_t, \alpha^{\prime\#}_t; \bar{\gamma}_{\text{CAL}}) \leq \me^{3\eta_{1, 3}}\xi_{2, 3}^{-2}\{M_{1, 1}|S_{\gamma}|\tilde{\lambda}_1^2\}$, where $\xi_{2, 3} = C_2^{1/2}\xi_{2, 1}(1 - \eta_{2, 2})^{1/2}$.
Then inequality \eqref{eq:or-error} holds through \eqref{eq:or-error-case1} and \eqref{eq:or-error-case2}.
\end{prf}

\subsection{Proof of Theorem \ref{thm:mu-lm}}
First, we provide a high-probability bound in Lemma \ref{lem:psor-1}, which is used later in the proof.

Denote $\Sigma_{\alpha  1} = E[A\otimes f(X)f^T(X)]$, where $A = \diag\{R^{(t)}\omega(k,X; \bar{\gamma}_{\text{CAL}})|Y - \bar{m}(t, X; \bar{\alpha}^{(k)}_{t,\text{WL}})|: k\neq t\}$, and the sample version of $\Sigma_{\alpha  1}$ as $\tilde{\Sigma}_{\alpha  1}$. Furthermore, denote $(\Sigma_{\alpha  1})_{j_1, j_2} = \E[Af_{j_1}(X)f_{j_2}(X)]$ and  the sample version of $(\Sigma_{\alpha  1})_{j_1, j_2}$ as $(\tilde{\Sigma}_{\alpha  1})_{j_1, j_2}$.

\begin{lem}\label{lem:psor-1}
Denote by $\Omega_1$ the event that
\begin{align}
\sup_{j_1, j_2 = 0, 1, \ldots, p}\lVert (\tilde{\Sigma}_{\alpha  1})_{j_1, j_2} - (\Sigma_{\alpha  1})_{j_1, j_2} \rVert_{\text{op}} \leq \sigma_0\tilde{\lambda}_1, \label{eq:psor-1}
\end{align}
Under Assumptions \ref{ass1}(i), \ref{ass1}(ii) and \ref{ass2}(i), if
\begin{align*}
\tilde{\lambda}_1 \geq 4\sqrt{2}B_0^2B_1\sqrt{\{(1/2) \log(K - 1) + \log[(p+1)/\epsilon]\} / n}
\end{align*}
then $\P(\Omega_1) \geq 1 - 2\epsilon^2$.
\end{lem}

\begin{prf}
Denote $a_{j_1j_2,k} = R^{(t)}\omega(k,X; \bar{\gamma}_{\text{CAL}})|Y - \bar{m}(X; \bar{\alpha}^{(k)}_{t,\text{WL}})|f_{j_1}(X)f_{j_2}(X)$. By the union bound, we have
\begin{align*}
& P\left( \sup_{j_1, j_2 = 0, 1, \ldots, p}\lVert (\tilde{\Sigma}_{\alpha  1})_{j_1, j_2} - (\Sigma_{\alpha  1})_{j_1, j_2} \rVert_{\text{op}} > t \right) \\
& \leq (K - 1)(p + 1)^2 \max_{j_1,j_2 = 0,1,\ldots,p,k\neq t} P\left(|\tilde{\E}a_{j_1j_2,k} - \E a_{j_1j_2,k}| > t \right).
\end{align*}
Notice that $a_{j_1j_2,k}$ is the product of $\omega(k,X; \bar{\gamma}_{\text{CAL}})f_{j_1}(X)f_{j_2}(X)$ and $R^{(t)}|Y - \bar{m}(X; \bar{\alpha}^{(k)}_{t,\text{WL}})|$, where $\omega(k,X; \bar{\gamma}_{\text{CAL}}) | f_{j_1}(X)f_{j_2}(X)| \leq B_0^2B_1$ by Assumptions \ref{ass1}(i)-(ii), and $R^{(t)}|Y - \bar{m}(X; \bar{\alpha}^{(k)}_{t,\text{WL}})|$ is sub-gaussian by Assumption \ref{ass2}(i). Applying Lemma 15 of Tan (2020b)(with $c_1 = c_2 = B_0^2B_1\sigma_0$) yields
\begin{align*}
P\left(|\tilde{\E}a_k - \E a_k| > t \right) \leq 2\exp\{-\frac{nt^2}{8(B_0^4B_1^2\sigma_0^2 + B_0^4B_1^2\sigma_0^2})\}.
\end{align*}
Combining all the inequalities yields
\begin{align*}
P\left( \sup_{j_1, j_2 = 0, 1, \ldots, p}\lVert (\tilde{\Sigma}_{\alpha  1})_{j_1, j_2} - (\Sigma_{\alpha  1})_{j_1, j_2} \rVert_{\text{op}} > t \right) \leq 2(K - 1)(p + 1)^2\exp\{-\frac{nt^2}{8(B_0^4B_1^2\sigma_0^2 + B_0^4B_1^2\sigma_0^2)}\}.
\end{align*}
The proof is completed by setting the right-hand side of the preceding inequality to $2\epsilon^2$.
\end{prf}

In the following, we give the proof of Theorem \ref{thm:mu-lm}.

\begin{prf}
Denote $\hat{\varphi}_t^{(k)} = \varphi_t^{(k)}(Y, T, X; \hat{\alpha}^{(k)}_{t,\text{RWL}}, \hat{\gamma}_{\text{RCAL}})$ and $\bar{\varphi}_t^{(k)} = \varphi_t^{(k)}(Y, T, X; \bar{\alpha}^{(k)}_{t,\text{WL}}, \bar{\gamma}_{\text{CAL}})$. Then
\begin{align*}
\hat{\mu}_t(\hat{m}^\#_{\text{RWL}}, \hat{\pi}_{\text{RCAL}}) = \tilde{E}\left\{R^{(t)}Y + \sum_{k\neq t}\hat{\varphi}_t^{(k)}\right\}.
\end{align*}
Consider the following decomposition,
\begin{align}
\hat{\varphi}_t^{(k)} - \bar{\varphi}_t^{(k)} = & \{\hat{m}(t, X; \hat{\alpha}_t^{(k)}) - \bar{m}(t, X; \bar{\alpha}_t^{(k)})\}\left\{R^{(k)} - R^{(t)}\omega(k,X; \bar{\gamma}_{\text{CAL}})\right\} \nonumber \\
& + R^{(t)}\{Y - \bar{m}(t, X; \bar{\alpha}_t^{(k)})\}\left\{\omega(k,X; \hat{\gamma}_{\text{RCAL}}) - \omega(k,X; \bar{\gamma}_{\text{CAL}})\right\} \nonumber \\
& +  \{\hat{m}(t, X; \hat{\alpha}_t^{(k)}) - \bar{m}(t, X; \bar{\alpha}_t^{(k)})\}\left\{R^{(t)}\omega(k,X; \bar{\gamma}_{\text{CAL}}) - R^{(t)}\omega(k,X; \hat{\gamma}_{\text{RCAL}})\right\}, \label{eq:decom-varphi}
\end{align}
denoted as $\delta_{1k} + \delta_{2k} + \delta_{3k}$.

We show that in the event $\Omega_{\gamma 1}\cap\Omega_{\gamma 2}\cap\Omega_{\alpha  1}\cap\Omega_2\cap\Omega_1$, inequality \eqref{eq:thm-mu-lm} holds as in Theorem \ref{thm:mu-lm}.
The estimator $\hat{\mu}_t(\hat{m}^\#_{\text{RWL}}, \hat{\pi}_{\text{RCAL}})$ can be decomposed as
\begin{align*}
\hat{\mu}_t(\hat{m}^\#_{\text{RWL}}, \hat{\pi}_{\text{RCAL}}) = \bar{\mu}_t(\bar{m}^\#_{\text{WL}}, \bar{\pi}_{\text{CAL}}) + \Delta_1 + \Delta_2,
\end{align*}
where
\begin{align*}
& \Delta_1 = \sum_{k\neq t}\tilde{\E}(\delta_{1k} + \delta_{3k}) = \sum_{j=0}^p\sum_{k\neq t}(\hat{\alpha}^{(k)}_{jt} - \bar{\alpha}^{(k)}_{jt})\tilde{\E}\left[\{R^{(k)} - R^{(t)}\omega(k,X; \hat{\gamma}_{\text{CAL}})\}f_j(X)\right], \\
& \Delta_2 = \sum_{k\neq t}\tilde{\E}(\delta_{2k}) = \sum_{k\neq t}\tilde{\E}\left[R^{(t)}\{Y - \bar{m}(t, X; \bar{\alpha}^{(k)}_{t,\text{WL}})\}\left\{\omega(k,X; \hat{\gamma}_{\text{RCAL}}) - \omega(k,X; \bar{\gamma}_{\text{CAL}})\right\}\right].
\end{align*}
First, $\Delta_1$ can be bounded by Holder's inequality as
\begin{align*}
|\Delta_1| \leq \left\{\sup_{j=0,1,\ldots,p} \left \lVert \tilde{\E}\left[
\{ R^{(k)} - R^{(t)}\omega(k,X; \hat{\gamma}_{\text{RCAL}}) : k\neq t \}^\T f_j(X)
\right]\right\rVert_2 \right\}\times \lVert \hat{\alpha}^\#_{t,\text{RWL}} - \bar{\alpha}^\#_{t,\text{WL}} \rVert_{2, 1},
\end{align*}
Then, in the event  $\Omega_{\gamma 1}\cap\Omega_{\gamma 2}\cap\Omega_{\alpha  1}\cap\Omega_2$, we have
\begin{align}
|\Delta_1| \leq A_1\tilde{\lambda}_1\times (A_2 - 1)^{-1}M_2\{|S_{\gamma}|\tilde{\lambda}_1 + |S_{\alpha_t}|\tilde{\lambda}_2\} \label{eq:mu-linear-Delta1}
\end{align}
by Karush-Kuhn-Tucker conditions and inequality \eqref{eq:or-error} along with the definition of $M_2$. Furthermore, a second-order Taylor expansion of $\Delta_2$ yields
\begin{align*}
\Delta_2 & = \sum_{k\neq t}b_k^\T\tilde{\E}\left[R^{(t)}\{Y - \bar{m}(t, X; \bar{\alpha}^{(k)}_{t,\text{WL}})\}\omega(k,X; \bar{\gamma}_{\text{CAL}})f(X)\right] \\
& + \sum_{k\neq t}b_k^\T\tilde{\E}\left[R^{(t)}\{Y - \bar{m}(t, X; \bar{\alpha}^{(k)}_{t,\text{WL}})\}\me^{u\hat{h}_k + (1-u)\bar{h}_k}f(X)f^T(X)\right]b_k / 2,
\end{align*}
denoted as $\Delta_{21} + \Delta_{22}$, where $u\in(0, 1)$, $b_k = \hat{\gamma}_{k,\text{RCAL}} - \bar{\gamma}_{k,\text{CAL}}$ and $\hat{h}_k = \hat{\gamma}^\T_{k,\text{RCAL}}f(X)$ and $\bar{h}_k = \bar{\gamma}^\T_{k,\text{CAL}}f(X)$. By direct calculation, we have
\begin{align*}
|\Delta_{21}| & = |\sum_{j=0}^p\sum_{k\neq t}b_{jk}\tilde{\E}\left[R^{(t)}\{Y - \bar{m}(t, X; \bar{\alpha}^{(k)}_{t,\text{WL}})\}\omega(k,X; \bar{\gamma}_{\text{CAL}})f_j(X)\right]| \\
& \leq \sum_{j=0}^p \left\{\lVert b_{j\cdot} \rVert_2 \times \lVert \tilde{\E}\left[R^{(t)} \{(Y - \bar{m}(t, X; \bar{\alpha}^{(k)}_{t,\text{WL}}))\omega(k,X; \bar{\gamma}_{\text{CAL}}): k\not=t\}^\T f_j(X)\right] \rVert_2\right\} \\
& \leq \sum_{j=0}^p \lVert b_{j\cdot} \rVert_2 \times \left\{\sup_{j=0,1,\ldots,p} \lVert \tilde{\E}\left[R^{(t)}\{(Y - \bar{m}(t, X; \bar{\alpha}^{(k)}_{t,\text{WL}}))\omega(k,X; \bar{\gamma}_{\text{CAL}}): k\not=t\}^\T f_j(X)\right] \rVert_2\right\},
\end{align*}
where $b_{j\cdot} = (b_{jk}: k\neq t)^\T$ and the last two steps follow from the Cauchy--Schwartz inequality and Holder's inequality, respectively. Then, in the event $\Omega_{\gamma 1}\cap\Omega_{\gamma 2}\cap\Omega_{\alpha  1}$, we have
\begin{align}
|\Delta_{21}| \leq (A_1 - 1)^{-1}M_1|S_{\gamma}|\tilde{\lambda}_1 \times \tilde{\lambda}_2, \label{eq:mu-linear-Delta21}
\end{align}
by inequalities \eqref{eq:thm-ps-error} and \eqref{eq:or-score}. The term $\Delta_{22}$ can be bounded as
\begin{align}
|\Delta_{22}| \leq \me^{B_0\lVert \hat{\gamma}_{\text{RCAL}} - \bar{\gamma}_{\text{CAL}} \rVert_{2, 1}}\sum_{k\neq t}\tilde{\E}\left[R^{(t)}\omega(k,X; \bar{\gamma})_{\text{CAL}}|Y - \bar{m}(t, X; \bar{\alpha}^{(k)}_{t,\text{WL}})|d_k^2\right] / 2 \label{eq:mu-linear-Delta22-1}
\end{align}
by inequality \eqref{eq:ps-h-diff}, where $d_k = \hat{h}_k - \bar{h}_k$. By direct calculation, we find
\begin{align*}
& (\tilde{\E} - \E)\left[\sum_{k\neq t}R^{(t)}\omega(k,X; \bar{\gamma}_{\text{CAL}})|Y - \bar{m}(t, X; \bar{\alpha}^{(k)}_{t,\text{WL}})|d_k^2\right] \\
& = (\tilde{\E} - \E)\left[\sum_{j_1, j_2 = 0, 1, \ldots, p}b^\T_{j_1\cdot}Af_{j_1}(X)f_{j_2}(X)b_{j_2\cdot}\right] \\
& \leq \left\{\sup_{j_1, j_2 = 0, 1, \ldots, p}\lVert \tilde{\E}[Af_{j_1}(X)f_{j_2}(X)] - \E[Af_{j_1}(X)f_{j_2}(X)] \rVert_{\text{op}}\right\}\times (\sum_{j=0}^p \lVert b_{j\cdot} \rVert_2)^2,
\end{align*}
where $A = \diag\{R^{(t)}\omega(k,X; \bar{\gamma}_{\text{CAL}})|Y - m(t, X; \bar{\alpha}^{(k)}_{t,\text{WL}})|: k\neq t\}$ and the last step follows from the Cauchy--Schwartz inequality.  Then, in the event $\Omega_1$, we have
\begin{align}
(\tilde{\E} - \E)\left[\sum_{k\neq t}R^{(t)}\omega(k,X; \bar{\gamma}_{\text{CAL}})|Y - \bar{m}(t, X; \bar{\alpha}^{(k)}_{t,\text{WL}})|d_k^2\right] \leq \sigma_0\tilde{\lambda}_1\times \lVert \hat{\gamma}_{\text{RCAL}} - \bar{\gamma}_{\text{CAL}} \rVert_{2, 1}^2 \label{eq:mu-linear-Delta22-2}
\end{align}
by inequalities \eqref{eq:psor-1}. By Assumption \ref{ass2}(i) and Lemma 17 of Tan (2020b), we have $\E[\{Y - \bar{m}(t, X; \bar{\alpha}^{(k)}_{t,\text{WL}})\}^2 | X] \leq 2\sigma_0^2$, and hence $\E[|Y - \bar{m}(t, X; \bar{\alpha}^{(k)}_{t,\text{WL}}) | X ] \leq \sqrt{2}\sigma_0$, which leads to
\begin{align}
\E\left[\sum_{k\neq t}R^{(t)}\omega(k,X; \bar{\gamma}_{\text{CAL}})|Y - \bar{m}(t, X; \bar{\alpha}^{(k)}_{t,\text{WL}})|d_k^2\right] \leq \sqrt{2}\sigma_0\sum_{k\neq t}\E\left[R^{(t)}\omega(k,X; \bar{\gamma}_{\text{CAL}})d_k^2\right]. \label{eq:mu-linear-Delta22-3}
\end{align}
By direct calculation, we find
\begin{align*}
& (\tilde{\E} - \E)\left[R^{(t)}\omega(k,X; \bar{\gamma})d_k^2\right] = (\tilde{\E} - \E)\left[\sum_{j_1, j_2 = 0, 1, \ldots, p}b_{j_1\cdot}^\T\diag\{R^{(t)}\omega(X; \bar{\gamma}_{\text{CAL}}) \}b_{j_2\cdot} \right] \\
& \leq \left\{\sup_{j_1, j_2 = 0, 1, \ldots, p}\lVert (\tilde{\Sigma}_{\gamma})_{j_1, j_2} - (\Sigma_{\gamma})_{j_1, j_2} \rVert_{\text{op}}\right\}\times (\sum_{j=0}^p \lVert b_{j\cdot} \rVert_2)^2,
\end{align*}
where the last step follows from the Cauchy--Schwartz inequality. In the event $\Omega_{\gamma 2}$, we have
 \begin{align}
& (\tilde{\E} - \E)\left[R^{(t)}\omega(k,X; \bar{\gamma}_{\text{CAL}})d_k^2\right] \leq \tilde{\lambda}_1\times \lVert \hat{\gamma}_{\text{RCAL}} - \bar{\gamma}_{\text{CAL}} \rVert_{2, 1}^2 \label{eq:mu-linear-Delta22-4}
\end{align}
by inequalities \eqref{eq:ps-hessian}. Collecting inequalities \eqref{eq:mu-linear-Delta22-1}--\eqref{eq:mu-linear-Delta22-4} gives
\begin{align*}
|\Delta_{22}| & \leq (1/2)\me^{B_0\lVert \hat{\gamma}_{\text{RCAL}} - \bar{\gamma}_{\text{CAL}} \rVert_{2, 1}}\left\{(\sqrt{2} + 1)\sigma_0\tilde{\lambda}_1\lVert \hat{\gamma}_{\text{RCAL}} - \bar{\gamma}_{\text{CAL}} \rVert_{2, 1}^2 \right. \\
& \left. + \sqrt{2}\sigma_0\sum_{k\neq t}\tilde{\E}\left[R^{(t)}\omega(k,X; \bar{\gamma}_{\text{CAL}})d_k^2\right]\right\}.
\end{align*}
Then, inequalities \eqref{eq:thm-ps-error} and \eqref{eq:ps-Q1} yield
\begin{align}
|\Delta_{22}| \leq (1/2)\me^{\eta_{1, 3}}\left\{(\sqrt{2} + 1)\sigma_0\eta_{1, 4}|S_{\gamma}|\tilde{\lambda}_1^2 + \sqrt{2}\sigma_0\me^{\eta_{1, 3}}M_1|S_\gamma|\tilde{\lambda}_1^2\right\}, \label{eq:mu-linear-Delta22}
\end{align}
where $\eta_{1, 4} = (A_1 - 1)^{-2}M_1^2\eta_1$. Then \eqref{eq:thm-mu-lm} follows by collecting inequalities \eqref{eq:mu-linear-Delta1}, \eqref{eq:mu-linear-Delta21}, and \eqref{eq:mu-linear-Delta22}.
\end{prf}

\subsection{Proof of Theorem \ref{thm:V-lm}}

Using $a^2 - b^2 = 2(a-b)b + (a-b)^2$ and the Cauchy--Schwartz inequality, we find
\begin{align}
|\tilde{\E}(\hat{\varphi}_{tc}^2 - \bar{\varphi}_{tc}^2)|\leq 2\tilde{\E}^{1/2}(\bar{\varphi}_{tc}^2)\tilde{\E}^{1/2}\{(\hat{\varphi}_{tc} - \bar{\varphi}_{tc})^2\} + \tilde{\E}\{(\hat{\varphi}_{tc} - \bar{\varphi}_{tc})^2\}. \label{eq:V-linear-1}
\end{align}
Using $\hat{\varphi}_{tc} = \hat{\varphi}_t - \hat{\mu}_t(\hat{m}^\#_{\text{RWL}}, \hat{\pi}_{\text{RCAL}})$ and $\bar{\varphi}_{tc} = \bar{\varphi}_t - \bar{\mu}_t(\bar{m}^\#_{\text{WL}}, \bar{\pi}_{\text{CAL}})$, we find
\begin{align}
\tilde{\E}\{(\hat{\varphi}_{tc} - \bar{\varphi}_{tc})^2\} \leq 2\tilde{\E}\{(\hat{\varphi}_{t} - \bar{\varphi}_{t})^2\} + 2|\hat{\mu}_t(\hat{m}^\#_{\text{RWL}}, \hat{\pi}_{\text{RCAL}}) - \bar{\mu}_t(\bar{m}^\#_{\text{WL}}, \bar{\pi}_{\text{CAL}})|^2. \label{eq:V-linear-2}
\end{align}
To control $\tilde{\E}\{(\hat{\varphi}_{t} - \bar{\varphi}_{t})^2\}$, we use the decomposition \eqref{eq:decom-varphi}, denoted as $\delta_{1k} + \delta_{2k} + \delta_{3k}$.  Then, by the Cauchy--Schwartz inequality, we have
\begin{align}
& \tilde{\E}\{(\hat{\varphi}_{t} - \bar{\varphi}_{t})^2\} = \tilde{\E}\left\{\left[\sum_{k\neq t}\delta_{1k} + \sum_{k\neq t}\delta_{2k} + \sum_{k\neq t}\delta_{3k}\right]^2\right\} \nonumber \\
& \leq 3(K-1)\sum_{k\neq t}\tilde{\E}(\delta_{1k}^2) + 3(K-1)\sum_{k\neq t}\tilde{\E}(\delta_{2k}^2) + 3(K-1)\sum_{k\neq t}\tilde{\E}(\delta_{3k}^2). \label{eq:V-linear-3}
\end{align}
Denote $d_k = \hat{h}_k - \bar{h}_k$ with $\hat{h}_k = \hat{\gamma}^\T_{k,\text{RCAL}}f(X)$ and $\bar{h}_k = \bar{\gamma}^\T_{k,\text{CAL}}f(X)$ for $k \neq t$.
In the following, we upper bound the right-hand side of the preceding inequality in three steps.

First, by the mean value equation \eqref{eq:mean-value}, inequality \eqref{eq:ps-h-diff} and Assumptions \ref{ass1}(i) and \ref{ass1}(ii), we have
\begin{align}
& \sum_{k\neq t} \tilde{\E}(\delta_{2k}^2) = \sum_{k\neq t}\tilde{\E}\left[R^{(t)}\{Y - \bar{m}(t, X; \bar{\alpha}^{(k)}_{t,\text{WL}})\}^2\{\omega(k,X; \hat{\gamma}_{\text{RCAL}}) - \omega(k,X; \bar{\gamma}_{\text{CAL}})\}^2\right] \nonumber \\
& \leq B_1\me^{2B_0\lVert \hat{\gamma}_{\text{RCAL}} - \bar{\gamma}_{\text{CAL}} \rVert_{2, 1}}\sum_{k\neq t}\tilde{\E}\left[R^{(t)}\omega(k,X; \bar{\gamma}_{\text{CAL}})\{Y - \bar{m}(t, X; \bar{\alpha}^{(k)}_{t,\text{WL}})\}^2d_k^2\right]. \label{eq:V-linear-4} \\
& \leq B_1\me^{2B_0\lVert \hat{\gamma}_{\text{RCAL}} - \bar{\gamma}_{\text{CAL}} \rVert_{2, 1}}\left\{3\sigma_0^2\tilde{\lambda}_1\lVert \hat{\gamma}_{\text{RCAL}} - \bar{\gamma}_{\text{CAL}} \rVert^2_{2, 1} + 2\sigma_0^2\me^{\eta_{1, 3}}M_1|S_\gamma|\tilde{\lambda}_1^2\right\}, \nonumber
\end{align}
where the last inequality follows from inequality \eqref{eq:or-diff-last} and \eqref{eq:ps-Q1}.
Second, writing $\omega(k,X; \hat{\gamma}_{\text{RCAL}}) - \omega(k,X; \bar{\gamma}_{\text{CAL}}) = \omega(k,X; \bar{\gamma}_{\text{CAL}})(\me^{d_k} - 1)$ and using inequality \eqref{eq:ps-h-diff} and Assumptions \ref{ass1}(i) and \ref{ass1}(ii), we have
\begin{align}
& \sum_{k\neq t}\tilde{\E}(\delta_{3k}^2) = \sum_{k\neq t}\tilde{\E}\left[R^{(t)}\{\hat{m}(t, X; \hat{\alpha}^{(k)}_{t,\text{RWL}}) - \bar{m}(t, X; \bar{\alpha}^{(k)}_{t,\text{WL}})\}^2\{\omega(k, X; \hat{\gamma}_{\text{RCAL}}) - \omega(k,X; \bar{\gamma}_{\text{CAL}})\}^2\right] \nonumber \\
& \leq B_1(1 + \me^{B_0\lVert \hat{\gamma}_{\text{RCAL}} - \bar{\gamma}_{\text{CAL}} \rVert_{2, 1}})^2\sum_{k\neq t}\tilde{\E}\left[R^{(t)}\omega(k,X; \bar{\gamma}_{\text{CAL}})\{\hat{m}(t, X; \hat{\alpha}^{(k)}_{t,\text{RWL}}) - \bar{m}(t, X; \bar{\alpha}^{(k)}_{t,\text{WL}})\}^2\right] \label{eq:V-linear-5} \\
& \leq B_1(1 + \me^{B_0\lVert \hat{\gamma}_{\text{RCAL}} - \bar{\gamma}_{\text{CAL}} \rVert_{2, 1}})^2\me^{\eta_{1, 3}}M_2\{|S_\gamma|\tilde{\lambda}_1\tilde{\lambda}_2 + |S_{\alpha_t}|\tilde{\lambda}_2^2\}, \nonumber
\end{align}
where last inequality follows from inequality \eqref{eq:thm-or-error-lm} along with the definition of $M_2$.
Third, using Assumptions \ref{ass1}(i) and \ref{ass1}(ii), we also have
\begin{align}
& \sum_{k\neq t}\tilde{\E}(\delta_{1k}^2) = \sum_{k\neq t}\tilde{\E}\left[\{\hat{m}(t, X; \hat{\alpha}^{(k)}_{t,\text{RWL}}) - \bar{m}(t, X; \bar{\alpha}^{(k)}_{t,\text{WL}})\}^2\{R^{(k)} - R^{(t)}\omega(k,X; \bar{\gamma}_{\text{CAL}})\}^2\right]  \nonumber \\
& \le (1 + B_1)^2\sum_{j_1=0}^p\sum_{j_2=0}^p (\hat{\alpha}^\#_{j_1t,\text{RWL}} - \bar{\alpha}^\#_{j_1t,\text{WL}})^\T\tilde{\E}[f_{j_1}(X)f_{j_2}(X)](\hat{\alpha}^\#_{j_2t,\text{RWL}} - \bar{\alpha}^\#_{j_2t,\text{WL}}) \nonumber \\
& \leq (1 + B_1)^2B_0^2\lVert \hat{\alpha}^\#_{t,\text{RWL}} - \bar{\alpha}^\#_{t,\text{WL}} \rVert_{2, 1}^2 \label{eq:V-linear-7} \\
& \leq (1 + B_1)^2B_0^2(A_2 - 1)^{-2}M_2^2\{|S_\gamma|\tilde{\lambda}_1 + |S_{\alpha_t}|\tilde{\lambda}_2\}^2, \nonumber
\end{align}
where the last inequality follows from inequality \eqref{eq:thm-or-error-lm} along with the definition of $M_2$.
Inequality \eqref{eq:thm-V-lm} follows by collecting inequalities \eqref{eq:V-linear-1}-\eqref{eq:V-linear-7} and applying \eqref{eq:gamma-bound} for bounding $\lVert \hat{\gamma}_{\text{RCAL}} - \bar{\gamma}_{\text{CAL}} \rVert_{2, 1}$ and \eqref{eq:thm-mu-lm} for bounding $|\hat{\mu}_t(\hat{m}^\#_{\text{RWL}}, \hat{\pi}_{\text{RCAL}}) - \bar{\mu}_t(\bar{m}^\#_{\text{WL}}, \bar{\pi}_{\text{CAL}})|$.

\subsection{Proof of Theorem \ref{thm:mu-glm}}

We use the decomposition \eqref{eq:decom-varphi} and handle the three terms $\sum_{k\neq t}\tilde{\E}(\delta_{1k})$, $\sum_{k\neq t}\tilde{\E}(\delta_{2k})$, and

\noindent
$\sum_{k\neq t}\tilde{\E}(\delta_{3k})$ separately.

First, inequalities \eqref{eq:mu-linear-Delta21} and \eqref{eq:mu-linear-Delta22} in the proof of Theorem \ref{thm:mu-lm} gives
\begin{align}
|\sum_{k\neq t}\tilde{\E}(\delta_{2k})| \leq (A_2 - 1)^{-1}M_1|S_\gamma|\tilde{\lambda}_1\tilde{\lambda}_2 + (1/2)\me^{\eta_{1, 3}}\{(\sqrt{2} + 1)\sigma_0\eta_{1, 4} + \sqrt{2}\sigma_0\me^{\eta_{1, 3}}M_1\}|S_\gamma|\tilde{\lambda}_1^2. \label{eq:mu-Delta2}
\end{align}
By the mean value equation \eqref{eq:mean-value}, inequality \eqref{eq:ps-h-diff} and the Cauchy--Schwartz inequality, the term $\sum_{k\neq t}\tilde{\E}(\delta_{3k})$ can be bounded as
\begin{align}
|\sum_{k\neq t}\tilde{\E}(\delta_{3k})| & \leq \me^{B_0\lVert \hat{\gamma}_{\text{RCAL}} - \bar{\gamma}_{\text{CAL}} \rVert_{2, 1}}[\sum_{k\neq t}\tilde{\E}\{R^{(t)}\omega(k,X; \bar{\gamma}_{\text{CAL}})d_k^2\}]^{1/2} \nonumber \\
& \quad \times (\sum_{k\neq t}\tilde{\E}[R^{(t)}\omega(k,X; \bar{\gamma}_{\text{CAL}})\{\hat{m}(t, X; \hat{\alpha}^{(k)}_{t,\text{RWL}}) - \bar{m}(t, X; \bar{\alpha}^{(k)}_{t,\text{WL}})\}^2])^{1/2}, \label{eq:mu-Delta3-1}
\end{align}
where $d_k = \hat{h}_k - \bar{h}_k$ with $\hat{h}_k = \hat{\gamma}^\T_{k,\text{RCAL}}f(X)$ and $\bar{h}_k = \bar{\gamma}^\T_{k,\text{CAL}}f(X)$ for $k \neq t$.
Similarly as in Lemma \ref{lem:or-dagger} but arguing in the reverse direction by Assumptions \ref{ass1}(i) and \ref{ass2}(v), we find
\begin{align}
& \sum_{k\neq t}\tilde{\E}[R^{(t)}\omega(k,X; \bar{\gamma}_{\text{CAL}})\{\hat{m}(t, X; \hat{\alpha}^{(k)}_{t,\text{RWL}}) - \bar{m}(t, X; \bar{\alpha}^{(k)}_{t,\text{WL}})\}^2] \leq \me^{C_3B_0\lVert \hat{\alpha}^\#_{t,\text{RWL}} - \bar{\alpha}^\#_{t,\text{WL}} \rVert_{2, 1}} \nonumber \\
& \times \sum_{k\neq t}\tilde{\E}[R^{(t)}\omega(k,X; \bar{\gamma}_{\text{CAL}})\psi_2(\bar{\eta}_k)\{\hat{m}(t, X; \hat{\alpha}^{(k)}_{t,\text{RWL}}) - \bar{m}(t, X; \bar{\alpha}^{(k)}_{t,\text{WL}})\}\{\hat{\eta}_k - \bar{\eta}_k\}] \nonumber \\
& \leq \me^{C_3B_0\lVert  \hat{\alpha}^\#_{t,\text{RWL}} - \bar{\alpha}^\#_{t,\text{WL}} \rVert_{2, 1}}C_1D^{\dagger}_{\text{WL}}(\hat{\alpha}^\#_{t,\text{RWL}}, \bar{\alpha}^\#_{t,\text{WL}}; \bar{\gamma}_{\text{CAL}}), \label{eq:mu-Delta3-2}
\end{align}
where $\hat{\eta}_k = \hat{\alpha}^{(k)\T}_{t,\text{RWL}}f(X)$, $\bar{\eta}_k = \bar{\alpha}^{(k)\T}_{t,\text{WL}}f(X)$ and the second inequality follows from Assumption \ref{ass2}(iii). Combining \eqref{eq:mu-Delta3-1},  \eqref{eq:mu-Delta3-2}, \eqref{eq:ps-Q1}, \eqref{eq:gamma-bound} and \eqref{eq:thm-or-error-glm} yields
\begin{align}
|\sum_{k\neq t}\tilde{\E}(\delta_{3k})| & \leq \me^{2\eta_{1, 3}  + C_3B_0\eta_3}M_1^{1/2}C_1M_3^{1/2}(|S_\gamma|\tilde{\lambda}_1^2)^{1/2}(|S_\gamma|\tilde{\lambda}_1\tilde{\lambda}_2 + |S_{\alpha_t}|\tilde{\lambda}_2^2)^{1/2}, \label{eq:mu-Delta3}
\end{align}
where $M_3$ is a constant such that the right-hand side of \eqref{eq:thm-or-error-glm} is upper-bounded by $\me^{\eta_{1, 3}}M_3(S_\gamma|\tilde{\lambda}_1\tilde{\lambda}_2 + |S_{\alpha_t}|\tilde{\lambda}_2^2)$ and $\eta_3 = (A_2 - 1)^{-1}M_3(S_\gamma|\tilde{\lambda}_1 + |S_{\alpha_t}|\tilde{\lambda}_2)$.

In the following, we derive a bound on $\sum_{k\neq t}\tilde{\E}(\delta_{1k})$, which can be decomposed as
\begin{align*}
\sum_{k\neq t} \tilde{\E}(\delta_{1k}) = & \sum_{k\neq t}(\tilde{\E} - \E)\left[\{\hat{m}(t, X; \hat{\alpha}^{(k)}_{t,\text{RWL}}) - \bar{m}(t, X; \bar{\alpha}^{(k)}_{t,\text{WL}})\}\{R^{(k)} - R^{(t)}\omega(k,X; \bar{\gamma}_{\text{CAL}})\}\right] \\
& + \sum_{k\neq t}\E\left[\{\hat{m}(t, X; \hat{\alpha}^{(k)}_{t,\text{RWL}}) - \bar{m}(t, X; \bar{\alpha}^{(k)}_{t,\text{WL}})\}\{R^{(k)} - R^{(t)}\omega(k,X; \bar{\gamma}_{\text{CAL}})\}\right],
\end{align*}
denoted as $\Delta_{13} + \Delta_{14}$. By the mean value theorem, we have
\begin{align*}
\Delta_{14} & = \sum_{k\neq t}\E\left[\psi_2\{\tilde{\eta}_k\}\{\hat{\eta}_k - \bar{\eta}_k\}\{R^{(k)} - R^{(t)}\omega(k,X; \bar{\gamma}_{\text{CAL}})\}\right] \\
& = \sum_{j=0}^p\sum_{k\neq t}(\hat{\alpha}^{(k)}_{jt,\text{RWL}} - \bar{\alpha}^{(k)}_{jt,\text{WL}})\tilde{\E}\left[\psi_2\{\tilde{\eta}_k\}\{R^{(k)} - R^{(t)}\omega(k,X; \bar{\gamma}_{\text{CAL}})\}f_j(X)\right],
\end{align*}
where $\tilde{\eta}_k$ lies between $\hat{\eta}_k$ and $\bar{\eta}_k$. Then, we have from the Cauchy--Schwartz inequality and inequality \eqref{eq:thm-or-error-glm}
\begin{align}
|\Delta_{14}| \leq \lVert \hat{\alpha}^\#_{t,\text{RWL}} - \bar{\alpha}^\#_{t,\text{WL}} \rVert_{2, 1}\sup_{j=0, 1, \ldots, p}\lVert \E[\diag\{\psi_2\{\tilde{h}_k\}:k\neq t\}g(X; \bar{\gamma}_{\text{CAL}})f_j(X)] \rVert_2 \leq \eta_3\Lambda(\eta_3), \label{eq:mu-Delta14}
\end{align}
where $g(X; \bar{\gamma}_{\text{CAL}}) = (g_k: k\neq t)^\T$ with $g_k = R^{(t)}\omega(k,X; \bar{\gamma}_{\text{CAL}}) - R^{(k)}$.
Moreover, applying Lemma \ref{lem:last} shown below yields
\begin{align}
|\Delta_{13}| \leq 2C_1\me^{B_0C_3\eta_3}\eta_3\tilde{\lambda}_1. \label{eq:mu-Delta13}
\end{align}
Then \eqref{eq:thm-mu-glm} follows by combining  \eqref{eq:mu-Delta2}, \eqref{eq:mu-Delta3}, \eqref{eq:mu-Delta14} and \eqref{eq:mu-Delta13}.


\begin{lem}\label{lem:last}
For $r \geq 0$, denote by $\Omega_3$ the event that
\begin{align}
& \sup_{\lVert \alpha^\#_t - \bar{\alpha}^\#_{t,\text{WL}} \rVert_{2, 1} \leq r}\left\lvert (\tilde{\E} - \E)\sum_{k\neq t}\{m(t, X; \alpha_t^{(k)}) - \bar{m}(t, X; \bar{\alpha}^{(k)}_{t,\text{WL}})\}\{R^{(k)} - R^{(t)}\omega(k,X; \bar{\gamma}_{\text{CAL}})\}\right\rvert \nonumber \\
& \leq 2rC_1\me^{B_0C_3r}\tilde{\lambda}_1. \label{eq:last}
\end{align}
Under Assumptions \ref{ass1}(i), \ref{ass1}(ii), \ref{ass2}(iii) and \ref{ass2}(v), if
$$
\tilde{\lambda}_1 \geq \sqrt{3/\log(2)} B_0B_1\sqrt{\{(2/3)(K - 1) + \log[(p+1)/\epsilon]\}/n},
$$ then $\P(\Omega_3) > 1 - \epsilon$.
\end{lem}

\begin{prf}
Denote
\begin{align*}
q(T, X; \alpha^\#_t) = \sum_{k\neq t}\{m(t, X; \alpha_t^{(k)}) - \bar{m}(t, X; \bar{\alpha}^{(k)}_{t,\text{WL}})\}\{R^{(k)} - R^{(t)}\omega(k,X; \bar{\gamma}_{\text{CAL}})\}.
\end{align*}
By the mean value theorem, we have
\begin{align*}
q(T, X; \alpha^\#_t) & = \sum_{k\neq t}\psi_2\{\tilde{\eta}_k\}\{\eta_k - \bar{\eta}_k\}\{R^{(k)} - R^{(t)}\omega(k,X; \bar{\gamma}_{\text{CAL}})\} \\
& = \sum_{j=0}^p\sum_{k\neq t}(\alpha_{jt}^{(k)} - \bar{\alpha}^{(k)}_{jt,\text{WL}})\psi_2\{\tilde{\eta}_k\}\{R^{(k)} - R^{(t)}\omega(k, X; \bar{\gamma}_{\text{CAL}})\}f_j(X),
\end{align*}
where $\eta_k = \alpha^{(k)\T}_t f(X)$,  $\bar{\eta}_k = \bar{\alpha}^{(k)\T}_{t,\text{WL}}f(X)$, $\tilde{\eta}_k = \tilde{\alpha}^{(k)\T}_t f(X)$ with $\tilde{\alpha}^{(k)}_t$ lying between $\alpha_t^{(k)}$ and $\bar{\alpha}^{(k)}_{t,\text{WL}}$. Then we have from Holder's inequality
\begin{align*}
& |(\tilde{\E} - \E)\{q(T, X; \alpha^\#_t)\}| \\
& \leq \lVert \alpha^\#_t - \bar{\alpha}^\#_{t,\text{WL}} \rVert_{2, 1}\sup_{j=0, 1, \ldots, p}\lVert (\tilde{\E} - \E)[\diag\{\psi_2\{\tilde{\eta}_k\}: k\neq t\}g(X; \bar{\gamma}_{\text{CAL}})f_j(X)] \rVert_2 .
\end{align*}
Denote $\nu_j = \diag\{\psi_2\{\tilde{\eta}_k\}: k\neq t\}g(X; \bar{\gamma}_{\text{CAL}})f_j(X)$. By the definition of the operator norm, we find
\begin{align}
\lVert \nu_j \rVert_2 \leq \lVert \diag\{\psi_2\{\tilde{\eta}_k\}: k\neq t\} \rVert_{\text{op}} \times \lVert g(X; \bar{\gamma}_{\text{CAL}})f_j(X) \rVert_2 \leq \lVert \diag\{\psi_2\{\tilde{\eta}_k\}: k\neq t\} \rVert_{\text{op}} \times B_0B_1, \label{eq:last1}
\end{align}
where the second inequality follows from Assumption \ref{ass1}(i) and \ref{ass1}(ii).
By Assumption \ref{ass2}(v) and \ref{ass2}(iii), we have
\begin{align}
\psi_2\{\tilde{\eta}_k\} \leq \psi_2\{\bar{\eta}_k\}\me^{C_3|\tilde{\eta}_k - \bar{\eta}_k|} \leq  C_1\me^{C_3B_0\lVert \tilde{\alpha}^{(k)}_t - \bar{\alpha}^{(k)}_{t,\text{WL}} \rVert_1} \leq C_1\me^{C_3B_0\lVert \alpha^\#_t - \bar{\alpha}^\#_{t,\text{WL}} \rVert_{2,1}}. \label{eq:last2}
\end{align}
Collecting inequalities \eqref{eq:last1} and \eqref{eq:last2} yields $\lVert \nu_j \rVert_2 \leq B_0B_1C_1\me^{B_0C_3r}$, which implies that $\nu_j$ is a sub-gaussian random vector with parameter $\log^{-1}(2)B^2_0B_1^2C_1^2\me^{2B_0C_3r}$. Then, the proof is completed by similar steps as in the proof of Lemma \ref{lem:ps-score}.
\end{prf}

\subsection{Proof of Theorem \ref{thm:V-glm}}

The proof is similar to that of Theorem \ref{thm:V-lm}. First, \eqref{eq:V-linear-4} for $\sum_{k\neq t}\tilde{\E}(\delta_{2k}^2)$ remains valid. Applying \eqref{eq:gamma-bound} to inequality \eqref{eq:V-linear-4} gives
\begin{align}
\sum_{k\neq t}\tilde{\E}(\delta_{2k}^2) \leq B_1\me^{2\eta_{1, 3}}\{3\sigma_0^2(A_1 - 1)^{-2}M_1^2\eta_1 + 2\sigma_0^2\me^{\eta_{1, 3}}M_1\}|S_\gamma|\tilde{\lambda}_1^2. \label{eq:V-2}
\end{align}
Second, combining \eqref{eq:V-linear-5} and \eqref{eq:mu-Delta3-2} yields
\begin{align*}
\sum_{k\neq t}\tilde{\E}(\delta_{3k}^2) \leq B_1(1 + \me^{B_0\lVert \hat{\gamma}_{\text{RCAL}} - \bar{\gamma}_{\text{CAL}} \rVert_{2, 1}})^2\me^{C_3B_0\lVert \hat{\alpha}^\#_{t,\text{RWL}} - \bar{\alpha}^\#_{t,\text{WL}} \rVert_{2, 1}}C_1D^{\dagger}_{\text{WL}}(\hat{\alpha}^\#_{t,\text{RWL}}, \bar{\alpha}^\#_{t,\text{WL}}; \bar{\gamma}_{\text{CAL}}).
\end{align*}
Applying \eqref{eq:gamma-bound} and \eqref{eq:thm-or-error-glm} along with the definition of $\eta_3$ and $M_3$ to the preceding inequality yields
\begin{align}
\sum_{k\neq t}\tilde{\E}(\delta_{3k}^2) \leq B_1(1 + \me^{\eta_{1, 3}})^2\me^{\eta_{1, 3} + C_3B_0\eta_3}C_1M_3(|S_\gamma|\tilde{\lambda}_1\tilde{\lambda}_2 + |S_{\alpha_t}|\tilde{\lambda}_2^2). \label{eq:V-3}
\end{align}
Third, similarly as in \eqref{eq:V-linear-7} and \eqref{eq:mu-Delta3-2}, we have
\begin{align}
& \sum_{k\neq t}\tilde{\E}(\delta_{1k}^2) = \sum_{k\neq t}\tilde{\E}\left[\{\hat{m}(t, X; \hat{\alpha}^{(k)}_{t,\text{RWL}}) - \bar{m}(t, X; \bar{\alpha}^{(k)}_{t,\text{WL}})\}^2\{R^{(k)} - R^{(t)}\omega(k,X; \bar{\gamma}_{\text{CAL}})\}^2\right] \nonumber \\
& \leq (1 + B_1)^2\me^{2C_3B_0\lVert \hat{\alpha}^\#_{t,\text{RWL}} - \bar{\alpha}^\#_{t,\text{WL}} \rVert_{2, 1}}C_1^2\sum_{k\neq t}\tilde{E}\{(\hat{\eta}_k - \bar{\eta}_k)^2\} \nonumber \\
& \leq (1 + B_1)^2\me^{2C_3B_0\lVert \hat{\alpha}^\#_{t,\text{RWL}} - \bar{\alpha}^\#_{t,\text{WL}} \rVert_{2, 1}}B_0^2C_1^2 \lVert \hat{\alpha}^\#_{t,\text{RWL}} - \bar{\alpha}^\#_{t,\text{WL}} \rVert_{2, 1}^2 \nonumber \\
& \leq (1 + B_1)^2\me^{2C_3B_0\eta_3}B_0^2C_1^2(A_2 - 1)^{-2}M_3^2(|S_\gamma|\tilde{\lambda}_1 + |S_{\alpha_t}|\tilde{\lambda}_2)^2, \label{eq:V-1}
\end{align}
where $\hat{\eta}_k = \hat{\alpha}^{(k)\T}_{t,\text{RWL}}f(X)$ and $\bar{\eta}_k = \bar{\alpha}^{(k)\T}_{t,\text{WL}}f(X)$.
Inequality \eqref{eq:thm-V-glm} follows by collecting \eqref{eq:V-linear-1}--\eqref{eq:V-linear-3}, \eqref{eq:V-2}--\eqref{eq:V-1} and  applying \eqref{eq:thm-or-error-glm} for bounding $|\hat{\mu}_t(\hat{m}^\#_{\text{RWL}}, \hat{\pi}_{\text{RCAL}}) - \bar{\mu}_t(\bar{m}^\#_{\text{WL}}, \bar{\pi}_{\text{CAL}})|$.

\section{Sum-to-zero constraint in PS estimation} \label{sec:sum-to-zero}

In this section, we provide various material when using the sum-to-zero constraint in PS estimation.
First, we give a correct proof showing $\sum_{k\in\mathcal T}\hat{\gamma}_{jk, \text{RML}} = 0$, as mentioned under equation \eqref{eq:RML-gamma-obj}, and $\sum_{k\in\mathcal T}\hat{\gamma}_{jk, \text{RCAL}} = 0$ for $j = 1, \ldots, p$.
Second, we discuss the KKT conditions.
Third, we give the algorithm for computing $\hat{\gamma}_{\text{RCAL}}$. Fourth, we give the theoretical results.

\subsection{Sum-to-zero relationship} \label{sec:sum-to-zero-relationship}
\begin{lem}\label{lem:sum-to-zero}
Denote $\hat{\gamma}$ as a minimizer of the following group-Lasso penalized objective function
\begin{align*}
\ell(\gamma) + \lambda \sum_{j=1}^p \lVert \gamma_{j\cdot} \rVert_2,
\end{align*}
where $\gamma = (\gamma_k: k\in\mathcal T)$ is a $(p+1)\times K$ matrix, $\gamma_{j\cdot} = (\gamma_{jk}: k\in\mathcal T)^\T$ is the transpose of the row vector in $\gamma$, $\lambda > 0$ is a tuning parameter, and $\ell(\gamma)$ is a loss function of $\gamma$. If $\ell(\gamma)$ satisfies $\sum_{k\in\mathcal T}\partial \ell(\gamma) / \partial \gamma_k = 0$ for any $\gamma$, then $\sum_{k\in\mathcal T}\hat{\gamma}_{jk} = 0$ holds for $j = 1, \ldots, p$.
\end{lem}

\begin{prf}
If $\hat{\gamma}_{j\cdot} = 0$, the proof is completed. Otherwise, by the Karush--Kuhn--Tucker conditions, $\hat{\gamma}_{j\cdot}$ satisfies
\begin{align*}
\frac{\partial \ell(\gamma)}{\partial \gamma_{jk}} \big|_{\gamma = \hat{\gamma}} + \lambda \frac{\hat{\gamma}_{jk}}{\lVert \hat{\gamma}_{j\cdot} \rVert_2} = 0, \quad k\in\mathcal T
\end{align*}
for $ j = 1, \ldots, p$. Summing the two sides of the preceding equality over $k\in\mathcal T$ yields
\begin{align*}
\sum_{k\in\mathcal T}\frac{\partial \ell(\gamma)}{\partial \gamma_{jk}} \big|_{\gamma = \hat{\gamma}} + \lambda \frac{\sum_{k\in\mathcal T}\hat{\gamma}_{jk}}{\lVert \hat{\gamma}_{j\cdot} \rVert_2} = 0.
\end{align*}
Then, the proof is completed by the condition $\sum_{k\in\mathcal T}\partial \ell(\gamma) / \partial \gamma_k = 0$ for any $\gamma$.
\end{prf}

Notice that $\sum_{k\in\mathcal T}\partial \ell_{\text{ML}}(\gamma) / \partial \gamma_k = \sum_{k\in\mathcal T} \tilde{\E}[\{\pi(k, X; \gamma) - R^{(k)}\}f(X)]= 0$ and $\sum_{k\in\mathcal T}\partial \ell_{\text{CAL}}(\gamma) / \partial \gamma_k = \{1 - R^{(t)} / \pi(t, X; \gamma) + \sum_{k\neq t}R^{(t)}\pi(k, X; \gamma) / \pi(t, X; \gamma) - R^{(k)}\}f(X) = 0$. Then, by Lemma \ref{lem:sum-to-zero}, we have $\sum_{k\in\mathcal T}\hat{\gamma}_{jk, \text{RML}} = 0$ and $\sum_{k\in\mathcal T}\hat{\gamma}_{jk, \text{RCAL}} = 0$ for $j = 1, \ldots, p$.

\subsection{KKT conditions}

By the KKT conditions for minimization of $\ell_{\text{RCAL}}(\gamma)$ in \eqref{eq:PSobj} with the sum-to-zero constraint, the fitted propensity scores
$\hat{\pi}_{\text{RCAL}} (k, X) = \pi (k, X; \hat{\gamma}_{\text{RCAL}})$, $k\in \mathcal T$, satisfy
\begin{align*}
& \tilde{\E}\left\{ 1 - \frac{R^{(t)}}{\hat{\pi}_{\text{RCAL}}(t, X)}\right\} = 0,\quad
\tilde{\E}\left\{ R^{(t)}\frac{\hat{\pi}_{\text{RCAL}}(k, X)}{\hat{\pi}_{\text{RCAL}}(t, X)} - R^{(k)}\right\} = 0, \quad k\neq t , \\ 
& \tilde{\E}^2\left[\left\{1 - \frac{R^{(t)}}{\hat{\pi}_{\text{RCAL}}(t, X)}\right\}f_j(X)\right] + \sum_{k\neq t}\tilde{\E}^2\left[\left\{R^{(t)}\frac{\hat{\pi}_{\text{RCAL}}(k, X)}{\hat{\pi}_{\text{RCAL}}(t, X)} - R^{(k)}\right\}f_j(X)\right] \leq \lambda_1^2, \quad j = 1, \ldots, p, 
\end{align*}
where equality holds in the second line for any $j$ such that the vector $\hat{\gamma}_{j., \text{RCAL}} = (\hat{\gamma}_{jk, \text{RCAL}}: k\in\mathcal T)^\T$ is nonzero.
Then  the fitted propensity score $\hat{\pi}_{\text{RCAL}}(t, X)$ for
treatment $t$ satisfies equality \eqref{eq:PS-KKT-t-1-cons}, which indicates the inverse probability weights, $1 / \hat{\pi}_{\text{RCAL}}(t, X_i)$ with $R_i^{(t)} = 1$, sum to the sample size $n$, and
\begin{align}
& \frac{1}{n}\left\lvert \sum_{i=1}^n \frac{R_i^{(t)}f_j(X_i)}{\hat{\pi}_{\text{RCAL}}(t, X_i)} - \sum_{i=1}^n f_j(X_i) \right\rvert \leq \lambda_1,\quad j = 1, \ldots, p, \label{eq:PS-KKT-t-2}
\end{align}
which indicates the weighted average of each covariate function $f_j(X_i)$ in the $t$th treated group may differ from the overall sample average of $f_j(X_i)$ by no more than $\lambda_1$.

\subsection{Algorithm for computing $\hat{\gamma}_{\normalfont \text{RCAL}}$} \label{sec:algorithm}
Compared to the algorithm with one-to-zero constraint, the main difference is that the weight matrix $H(X; \gamma)$ in \eqref{eq:taylor-gamma} is no longer diagonal since the loss function $\ell_{\text{CAL}}(\gamma)$ is not separable in $(\gamma_k: k\in\mathcal T)$ as mentioned under equation \eqref{eq:PSobj}. The non-diagonal matrix $H(X; \gamma)$ makes application of the majorization-minimization (MM) technique (Wu and Lange 2010) less straightforward than in Section \ref{sec:computation}. To solve the aforementioned problem, we show that the weight matrix $H(X; \gamma)$ can be dominated by a diagonal matrix.

The second-order Taylor expansion \eqref{eq:taylor-gamma} still holds under the sum-to-zero constraint. But the definitions of $g(X; \gamma)$ and $H(X; \gamma)$ are changed to the following: $g(X; \gamma) = (g_k: k\in\mathcal T)^\T$ with $g_t = 1 - R^{(t)} / \pi(t, X; \gamma)$ and $g_k = R^{(t)}\pi(k, X; \gamma)  - R^{(k)}$ for $k\neq t$, and
\begin{align}
H(X; \gamma) =  \left(
\begin{array}{ccc}
\diag\{\pi_{0:(t-1)}\} & -\pi_{0:(t-1)} & 0 \\
 -\pi^\T_{0:(t-1)} & 1 - \pi(t, X; \gamma) & -\pi^\T_{(t+1):(K-1)} \\
 0 & -\pi_{(t+1):(K-1)} & \diag\{\pi_{(t+1):(K-1)}\} \\
\end{array}
\right) \label{eq:H-gamma}
\end{align}
with $\pi_{0:(t-1)} = (\pi(k, X; \gamma): k = 0, 1, \ldots, t-1)^\T$ and $\pi_{(t+1):(K-1)} = (\pi(k, X; \gamma): k = t+1, \ldots, K-1)^\T$.

\begin{lem} \label{lem:Hbound}
For any $X$ and $\gamma$, we have
\begin{align*}
H(X; \gamma) \preceq 2\, \diag\{ \pi^\T_{0:(K-1)}, 1 - \pi(t, X; \gamma), \pi^\T_{(t+1):(K-1)} \},
\end{align*}
where $\pi_{0:(K-1)}$ and $\pi_{(t+1):(K-1)}$ are defined as in $H(X;\gamma)$ and $H_1 \preceq H_2 $ indicates that $H_2- H_1$ is nonnegative definite for two matrices $H_1$ and $H_2$.
\end{lem}

\begin{prf}
Let $D(X; \gamma) = 2\, \diag\{ \pi^\T_{0:(K-1)}, 1 - \pi(t, X; \gamma), \pi^\T_{(t+1):(K-1)} \} - H(X; \gamma)$. It suffices to show that $D(X; \gamma)$ is diagonally dominant and thus positive-definite. By direct calculation, we have
\begin{align*}
D(X; \gamma) = \left(
\begin{array}{ccc}
\diag\{\pi_{0:(t-1)}\} & \pi_{0:(t-1)} & 0 \\
 \pi^\T_{0:(t-1)} & 1 - \pi(t, X; \gamma) & \pi^\T_{(t+1):(K-1)} \\
 0 & \pi_{(t+1):(K-1)} & \diag\{\pi_{(t+1):(K-1)}\} \\
\end{array}
\right) .
\end{align*}
For each row of $D(X; \gamma)$, the diagonal entry is equal to the sum of all the other (non-diagonal) entries in the row. Thus $D(X; \gamma)$ is positive semi-definite, and so $H(X; \gamma) \preceq 2\, \diag\{ \pi^\T_{0:(K-1)}, 1 - \pi(t, X; \gamma), \pi^\T_{(t+1):(K-1)} \}$.
\end{prf}

Now if we consider $b_1 = \max_{i=1}^n 2\{1 - \pi(t, X_i; \tilde{\gamma})\}$, then preceding lemma gives us that $H(X_i; \tilde{\gamma}) \preceq b_1I$, and $I$ is the $K\times K$ identity matrix. Then, by replacing $g(X; \gamma)$ and $b_1$ in the update formula \eqref{eq:PSupdate} with their new definitions here, we can still use Algorithm 1 for computing $\hat{\gamma}_{\text{RCAL}}$.

\subsection{Theoretical results}
First, we give the theoretical results of the estimation of regression coefficients $\gamma$. Second, we give the theoretical results of the estimation of regression coefficients $\alpha_t$. Third, we give the theoretical results of the estimation of treatment means.

{\bf Estimation of regression coefficients $\gamma$.} Compared to the analysis with one-to-zero constraint, the analysis with sum-to-zero constraint is more complex and this complexity mainly results from the Hessian matrix $R^{(t)} / \pi(t, X; \gamma)H(X; \gamma)$ with respect to the linear predictors is no longer diagonal, where
$H(X; \gamma)$ is defined in (\ref{eq:H-gamma}).
For one-to-zero constraint in Section \ref{sec:computation}, $H(X; \gamma) = \diag\{\pi(k, X; \gamma): k\neq t\}$ is a $(K-1)\times (K-1)$ diagonal matrix.

Because the definition of $H(X; \gamma)$ is different here, all the analyses involving $H(X; \gamma)$ with one-to-zero constraint need to be modified. First, as discussed under equation \eqref{eq:ps-cc1}, Assumption \ref{ass1}(iii) amounts to a compatibility condition on the Hessian matrix $\Sigma_\gamma = \E[R^{(t)} / \pi(t, X; \bar{\gamma}_{\text{CAL}})H(X; \bar{\gamma}_{\text{CAL}})\otimes f(X)f^\T(X)]$, where $H(X; \gamma) = \diag\{\pi(k, X; \gamma): k\neq t\}$.
For sum-to-zero constraint, the compatibility condition is defined as follows: for any $(p+1)\times K$ matrix $b = (b_0, b_1, \ldots, b_{K-1})$ satisfying
\begin{align*}
\sum_{k\in\mathcal T} b_k =0 \quad \text{and}\quad  \sum_{j\notin S_\gamma} \lVert b_{j\cdot} \rVert_2 \leq \xi_1\sum_{j\in S_\gamma} \lVert b_{j\cdot} \rVert_2,
\end{align*}
it holds that
\begin{align*}
\nu_1^2\left(\sum_{j\in S_\gamma} \lVert b_{j\cdot} \rVert_2\right)^2 \leq
|S_\gamma| \, \E \left\{f^\T(X)b\frac{R^{(t)}}{\pi(t, X; \bar{\gamma}_{\text{CAL}})} H(X; \bar{\gamma}_{\text{CAL}})b^\T f(X)\right\} ,
\end{align*}
or equivalently by the definition of $H(X; \gamma)$  in (\ref{eq:H-gamma}),
\begin{align}
\nu_1^2\left(\sum_{j\in S_\gamma }\lVert b_{j\cdot} \rVert_2\right)^2 \leq |S_\gamma|\, \sum_{k\neq t}(b_k - b_t)^\T\E\left\{R^{(t)} \omega(k,X; \bar{\gamma}_{\text{CAL}})f(X)f^\T(X)\right\}(b_k - b_t),
\label{eq:ps-cc}
\end{align}
where $b_k = (b_{jk}: j = 0, 1, \ldots, p)^\T$ for $k \in \mathcal T$, $b_{j\cdot} = (b_{jk}: k\in\mathcal T)^\T$ is the transpose of the $(j+1)$th row vector in $b$, and $\omega(k,X; \gamma) = \pi(k, X; \gamma) / \pi(t, X;\gamma)$ with dependency on $t$ suppressed.
The sum-to-zero constraint $\sum_{k\in\mathcal T} b_k =0$ is tied to the definition of $\hat{\gamma}_{\text{RCAL}}$ and $\bar{\gamma}_{\text{CAL}}$ satisfying the same constraint.
As seen from the Taylor expansion (\ref{eq:taylor-gamma}),
the right-hand side of (\ref{eq:ps-cc}) can be expressed as $|S_\gamma| \text{vec}^\T(b)\Sigma_{\gamma1}\text{vec}(b)$,
where $\text{vec}(b) = (b^\T_0, b^\T_1, \ldots, b^\T_{K-1})^\T$ and $\Sigma_{\gamma1} = \E[R^{(t)}/ \pi(t, X; \bar{\gamma}_{\text{CAL}})H(X; \bar{\gamma}_{\text{CAL}})\otimes f(X)f^\T(X)]$ is the Hessian matrix of $\E\{\ell_{\text{CAL}}(\gamma)\}$ at $\gamma = \bar{\gamma}_{\text{CAL}}$.

Second, Assumption \ref{ass1}(iv)(a) involved in Lemma \ref{lem:ps-empicc} also need to be modified. We assume $(\xi_1 + 1)^2\nu_1^{-2}|S_\gamma|(K-1)\tilde{\lambda}_1 \leq \eta_{1,1}$ for a constant $0 < \eta_{1, 1} < 1$. Compared to Assumption \ref{ass1}{(iv)(a)}, there is an additional factor of $(K - 1)$, which arises from Lemma \ref{lem:ps-hessian-cons} showing that the empirical Hessian is close to the theoretical Hessian. Furthermore, Assumption \ref{ass1}(iv)(b) is replaced by $2B_0(\xi_1 + 1)^2(A_1 - 1)(1 - \eta_{1, 1})^{-1}\nu_1^{-2}|S_\gamma|\tilde{\lambda}_1 \leq \eta_{1, 2}$. Compared to that with one-to-zero constraint, there is an additional factor of 2. This factor arises from Lemma \ref{lem:4ps-dagger}, which will be discussed later.

Third, the tuning parameter in the penalized objective function \eqref{eq:PSobj} is specified as $\lambda_1 = A_1\tilde{\lambda}_1$, with a constant $A_1 > 1$, and
\begin{align*}
\tilde{\lambda}_1 = 8B_0^2B_1\sqrt{K / n + \log\{(p + 1) / \epsilon\} / n},
\end{align*}
where $(B_0, B_1)$ comes from Assumption \ref{ass1}, and $0 < \epsilon < 1$ is a tail probability for the error bound.
For sum-to-zero constraint, the corresponding Bregman divergence is defined as
\begin{align*}
D_{\text{CAL}}(\gamma, \gamma^\prime) =  \ell_{\text{CAL}}(\gamma) - \ell_{\text{CAL}} (\gamma^\prime)
- \sum_{k\in\mathcal T} (\gamma_k -\gamma^\prime_k)^\T (\partial /\partial \gamma^\prime_k)  \ell_{\text{CAL}} (\gamma^\prime),
\end{align*}
where the third term of the right hand side takes summation over $\mathcal T$ not over $\mathcal T\backslash\{t\}$ as n the case of one-to-zero constraint. The symmetrized Bregman divergence is easily shown to be
\begin{align*}
D^\dagger_{\text{CAL}}(\gamma, \gamma^\prime) & =  D_{\text{CAL}}(\gamma, \gamma^\prime) + D_{\text{CAL}}(\gamma^\prime, \gamma) \\
& = \sum_{k\neq t}\tilde{\E}\left[R^{(t)}\left\{ \me^{( \gamma_k - \gamma_t)^\T f(X)} - \me^{ (\gamma^\prime_k - \gamma^\prime_t)^\T f(X) }\right\}
\{(\gamma_k - \gamma_t) - (\gamma^\prime_k - \gamma^\prime_t) \}^\T f(X) \right].
\end{align*}
Compared to the symmetrized Bregman divergence with one-to-zero constraint, $\gamma_k$ and $\gamma^\prime_k$ are replaced with $\gamma_k - \gamma_t$ and $\gamma^\prime_k - \gamma^\prime_t$, respectively.
Let Assumption \ref{ass1}$^\prime$ be Assumption \ref{ass1} with Assumption \ref{ass1}(iii) and \ref{ass1}(iv) being modified as described above. Then, we have the following results.

\begin{thm}\label{thm:ps-error-cons}
Suppose that Assumption \ref{ass1}$^\prime$ holds. Then we have probability at least $1 - 3\epsilon$,
\begin{align}
& D^\dagger_{\text{CAL}}(\hat{\gamma}_{\text{RCAL}}, \bar{\gamma}_{\text{CAL}}) + (A_1 - 1)\tilde{\lambda}_1\lVert \hat{\gamma}_{\text{RCAL}} - \bar{\gamma}_{\text{CAL}} \rVert_{2, 1} \leq \xi_{1, 1}^2\nu_{1, 1}^{-2}|S_\gamma|\tilde{\lambda}_1^2,
\label{eq:thm-ps-error-cons}
\end{align}
where $\xi_{1, 1} = (\xi_1 + 1)(A_1 - 1)$, and $\nu_{1, 1} = \nu_1(1 - \eta_{1, 1})^{1/2}(1 - \eta_{1, 2})^{1/2}$.
\end{thm}

Theorem \ref{thm:ps-error-cons} shows that the convergence of $\hat{\gamma}_{\text{RCAL}}$ to $\bar{\gamma}_{\text{CAL}}$ in the $L_{2, 1}$ norm at rate $|S_\gamma|\{K / n + \log(p) / n\}^{1/2}$ and the associated Bregman divergence at the rate $|S_\gamma|\{K / n + \log(p) / n\}$. These two rates are same as those in the case of one-to-zero constraint.

The proof of Theorem \ref{thm:ps-error-cons} is similar with that for Theorem \ref{thm:ps-error} except that we replace Lemma \ref{lem:ps-hessian} with Lemma \ref{lem:ps-hessian-cons} below. Moreover, the proof of Lemma \ref{lem:ps-empicc} (the third line of first equation), which gives a suitable lower bound on the symmetrized Bregman divergence for local analysis, requires $R^{(t)} / \pi(t, X; \gamma)H(X; \gamma) \preceq R^{(t)} / \pi(t, X; \gamma^\prime)H(X; \gamma^\prime)\me^{c\lVert (\gamma - \gamma^\prime)^\T f(X) \rVert_{\infty}}$ for any two matrices $\gamma$ and $\gamma^\prime$.
With one-to-zero constraint,  it is immediate that $R^{(t)} / \pi(t, X; \gamma)H(X; \gamma) = R^{(t)}\diag\{\me^{\gamma^\T_kf(X)}: k\neq t\}$ meets this requirement with $c = 1$. With sum-to-zero constraint, we give Lemma \ref{lem:4ps-dagger} to show that Hessian matrix with respect to the linear predictors meets the requirement with $c = 2$.

Denote $(\Sigma_{\gamma1})_{j_1, j_2} = \E[\{R^{(t)} / \pi(t, X; \bar{\gamma}_{\text{CAL}})\}H(X; \bar{\gamma}_{\text{CAL}})f_{j_1}(X)f_{j_2}(X)]$ and $(\tilde{\Sigma}_{\gamma1})_{j_1, j_2}$ as the sample version of $(\Sigma_{\gamma1})_{j_1, j_2}$.

\begin{lem} \label{lem:ps-hessian-cons}
Denote by $\Omega_{\gamma 2^\prime}$ the event that
\begin{align*}
\sup_{j_1, j_2 = 0, 1, \ldots, p} \lVert (\tilde{\Sigma}_{\gamma1})_{j_1, j_2} - (\Sigma_{\gamma1})_{j_1, j_2} \rVert_{\text{op}} \leq (K-1)\tilde{\lambda}_1.
\end{align*}
Under Assumptions \ref{ass1}(i) and \ref{ass1}(ii), if
\begin{align*}
\tilde{\lambda}_1 \geq 8B_1B_0^2\sqrt{\{(1/2)\log(K-1) + \log[(p+1)/\epsilon]\}/n},
\end{align*}
then $\P(\Omega_{\gamma 2^\prime}) \geq 1 - 2\epsilon^2$.
\end{lem}

\begin{prf}
Denote $a_{j_1j_2} = (a_{j_1j_2, k}: k\in\mathcal T\backslash\{t\})^\T$ with $a_{j_1j_2,k} = R^{(t)}\omega(k,X; \bar{\gamma})f_{j_1}(X)f_{j_2}(X)$ and $\omega(k,X; \bar{\gamma}_{\text{CAL}}) = \pi(k, X; \bar{\gamma}_{\text{CAL}}) / \pi(t, X; \bar{\gamma}_{\text{CAL}})$. Furthermore, denote $\bar{a}^c_{j_1j_2} = (\bar{a}^c_{j_1j_2, k}: k\in\mathcal T\backslash\{t\})^\T$ with $\bar{a}^c_{j_1j_2,k} = \tilde{\E}a_{j_1j_2,k} - \E a_{j_1j_2,k}$.
By direct calculation using the definition of $H(X; \bar{\gamma}_{\text{CAL}})$, we find
\begin{align*}
\lVert (\tilde{\Sigma}_{\gamma1})_{j_1, j_2} - (\Sigma_{\gamma1})_{j_1, j_2} \rVert^2_{\text{op}} & =
\sup_{\lVert \beta \rVert = 1}\left[\sum_{k\neq t}\bar{a}^{c^2}_{j_1j_2,k}(\beta_k - \beta_t)^2 + \left\{\sum_{k\neq t}\bar{a}^c_{j_1j_2,k}(\beta_k - \beta_t)\right\}^2\right] \\
& \leq \sup_{\lVert \beta \rVert = 1}\{\max_{k\neq t}(\beta_k - \beta_t)^2 \lVert \bar{a}^c_{j_1j_2} \rVert^2_2 + \max_{k\neq t}(\beta_k - \beta_t)^2  \lVert \bar{a}^c_{j_1j_2} \rVert_1^2\} .
\end{align*}
Applying $(\beta_k - \beta_t)^2 \leq 2(\beta_k^2 + \beta_t^2) \leq 2$ and $\lVert \bar{a}^c_{j_1j_2} \rVert_2 \leq \lVert \bar{a}^c_{j_1j_2} \rVert_1 \leq (K-1)\lVert \bar{a}^c_{j_1j_2} \rVert_{\infty}$ to the preceding inequality gives
\begin{align*}
\lVert (\tilde{\Sigma}_{\gamma1})_{j_1, j_2} - (\Sigma_{\gamma1})_{j_1, j_2} \rVert_{\text{op}} \leq 2(K-1)\lVert \bar{a}^c_{j_1j_2} \rVert_{\infty}.
\end{align*}
By the union bound, we have
\begin{align*}
& \quad P\left(\sup_{j_1, j_2 = 0, \ldots, p} \lVert (\tilde{\Sigma}_{\gamma1})_{j_1, j_2} - (\Sigma_{\gamma1})_{j_1, j_2} \rVert_{\text{op}} > \tilde{\lambda}_1\right) \\
& \leq (p + 1)^2(K - 1) \max_{j_1,j_2=0,\ldots,p, k\not=t} P\left(|\bar{a}^c_{j_1j_2,k}| \geq \tilde{\lambda}_1 / 2(K - 1)\right).
\end{align*}
Under Assumptions \ref{ass1}(i) and \ref{ass1}(ii), we have $|R^{(t)}\omega(k,X; \bar{\gamma}_{\text{CAL}})f_{j_1}(X)f_{j_2}(X)| \leq B_1B_0^2$. Applying to the preceding inequality Lemma 7 of Tan (2020a) gives
\begin{align*}
P\left(\sup_{j_1, j_2 = 0, \ldots, p} \lVert (\tilde{\Sigma}_{\gamma1})_{j_1, j_2} - (\Sigma_{\gamma1})_{j_1, j_2} \rVert_{\text{op}} \geq \tilde{\lambda}_1\right) \leq 2(K-1)(p+1)^2\exp\left\{-\frac{n\tilde{\lambda}_1^2}{32(K-1)^2B_1^2B_0^4}\right\}.
\end{align*}
Setting the right-hand side of the above inequality being $2\epsilon^2$, we complete the proof.
\end{prf}

\begin{lem}\label{lem:4ps-dagger}
For any two matrices $\gamma$ and $\gamma^\prime$,
\begin{align}
\frac{R^{(t)}}{\pi(t, X; \gamma)}H(X; \gamma) \preceq \frac{R^{(t)}}{\pi(t, X; \gamma^\prime)}H(X; \gamma^\prime)\me^{2\lVert (\gamma - \gamma^\prime)^\T f(X) \rVert_{\infty}}, \label{eq:4ps-dagger}
\end{align}
where $H_1 \preceq H_2$ indicates that $H_2 - H_1$ is nonnegative definite for two matrices $H_1$ and $H_2$.
\end{lem}

\begin{prf}
Denote $A = \me^{2\lVert h - h' \rVert_{\infty}}\frac{R^{(t)}}{\pi(t, X; \gamma^\prime)}H(X; \gamma^\prime) - \frac{R^{(t)}}{\pi(t, X; \gamma)}H(X; \gamma)$, where
$h(X) = \gamma^\T f(X)$ and $h^\prime(X) = \gamma^{\prime\T}f(X)$.
To show (\ref{eq:4ps-dagger}), it suffices to show that  $A$ is a diagonally dominant matrix with non-negative diagonal entries.
Denote as $A_{ij}$ the $(i, j)$ entry of $A$ for $i, j = 0, 1, \ldots, K-1$. By direct calculation from the definition of $H()$, the entries on the $i$th row $(i\neq t)$ of A are
\begin{align*}
& A_{ii} = R^{(t)}\{\me^{2\lVert h - h^\prime \rVert_{\infty}} - \me^{-(h_t - h_t^\prime)}\me^{h_i - h_i^\prime}\}\me^{h_i^\prime - h_t^\prime},~A_{it} = -A_{ii},~A_{ij} = 0~\text{for}~j\neq i, t,
\end{align*}
and the entries on the $t$th row of A are
\begin{align*}
& A_{tt} = R^{(t)}\{\me^{2\lVert h - h^\prime \rVert_{\infty}}\sum_{j\neq t}\me^{h_j^\prime - h_t^\prime} - \me^{-(h_t - h_t^\prime)}\sum_{j\neq t}\me^{h_j - h_j^\prime}\me^{h_j^\prime - h_t^\prime}\}, \\
& A_{tj} = -R^{(t)}\{\me^{2\lVert h - h^\prime \rVert_{\infty}}\me^{h_j^\prime - h_t^\prime} - \me^{-(h_t - h_t^\prime)}\me^{h_j - h_j^\prime}\me^{h_j^\prime - h_t^\prime}\}~\text{for}~j\neq t.
\end{align*}
The diagonal entries of $A$ are nonnegative, because $\me^{|h_k - h_k^\prime|} \leq \me^{\lVert h - h^\prime \rVert_{\infty}}$ for $k = 0, 1, \ldots, K-1$. Moreover, it can be easily verified that
$|A_{ii}| = \sum_{j\neq i}|A_{ij}|$ for $i = 0, 1, \ldots, K-1$, which indicates that $A$ is a diagonally dominant matrix.
\end{prf}

{\bf Estimation of regression coefficients $\alpha_t$.} The tuning parameter in the penalized objective function \eqref{eq:ORobj} is specified as $\lambda_2 = A_2\tilde{\lambda}_2$, with a constant $A_2 > 1$, and
\begin{align*}
\tilde{\lambda}_2 = \max[\tilde{\lambda}_1, \sqrt{3}B_0(B_1 - 1)\sigma_0\sqrt{(K - 1) / n + \log\{(p + 1) / \epsilon\} / n},
\end{align*}
where $(B_0, B_1)$ comes from Assumptions \ref{ass1}(i)-(ii), $\sigma_0$ is from Assumption \ref{ass2}(i), and $0 < \epsilon < 1$ is a tail probability for the error bound. Let Assumption \ref{ass2}$^\prime$ be Assumption \ref{ass2} with Assumption \ref{ass2}(vi)(b) and Assumption \ref{ass2}(vi)(c) being replaced by Assumption \ref{ass2}$^\prime$(vi)(b) and Assumption \ref{ass2}$^\prime$(vi)(c) respectively, where  Assumption \ref{ass2}$^\prime$(vi)(b) is $\me^{2\eta_{1, 3}}B_0C_3C_2^{-1}\nu_2^{-2}(\xi_2 + 1)^2(A_2 - 1)(1 - \eta_2)^{-1}|S_{\alpha_t}|\tilde{\lambda}_2 \leq \eta_{2, 1}$ and Assumption \ref{ass2}$^\prime$(vi)(c) is $\me^{6\eta_{1, 3}}B_0C_3C_2^{-1}\xi_{2, 3}^{-2}(A_2 - 1)^{-1}M_{1, 1}(K-1)|S_\gamma|\tilde{\lambda}_1 \leq \eta_{2, 2}$ and $(\eta_{1, 3}, \xi_{2, 3}, M_{1, 1})$ are as in Theorem \ref{thm:or-error-glm-cons}. Then we have the following results.

\begin{thm}\label{thm:or-error-glm-cons}
Suppose that Assumption \ref{ass1}$^\prime$ and Assumption \ref{ass2}$^\prime$ except \ref{ass2}$^\prime$(iii) hold. If $\log\{K + \log(p + 1)/\epsilon\} / n \leq 1$, then for $A_1 > (\xi_1 + 1) / (\xi_1 - 1)$ and $A_2 > (\xi_2 + 1) / (\xi_2 - 1)$, we have with probability at least $1 - 8\epsilon$
\begin{align}
& D^{\dagger}_{\text{WL}}(\hat{\alpha}^\#_{t, \text{RWL}}, \bar{\alpha}^\#_{t, \text{WL}}; \bar{\gamma}_{\text{CAL}}) + \me^{2\eta_{1, 3}}(A_2 - 1)\tilde{\lambda}_2\lVert \hat{\alpha}^\#_{t, \text{RWL}} - \bar{\alpha}^\#_{t, \text{WL}} \rVert_{2, 1} \nonumber \\
& \leq \me^{6\eta_{1, 3}}\xi_{2, 3}^{-2}\{M_{1, 1}|S_\gamma|(K-1)\tilde{\lambda}_1^2\} + \me^{2\eta_{1, 3}}\xi_{2, 2}^2\{\nu_{2, 2}^{-2}|S_{\alpha_t}|\tilde{\lambda}_2^2\},
\label{eq:thm-or-error-glm-cons}
\end{align}
where $\xi_{2, 2} = (\xi_2 + 1)(A_2 - 1)$, $\xi_{2, 3} = \xi_{2, 1}(1 - \eta_{2, 2})^{1/2}C_2^{1/2}$, $\xi_{2, 1} = 1 - 2A_2 / \{(\xi_2 + 1)(A_2 - 1)\}$, $\nu_{2, 2} = \nu_{2, 1}(1 - \eta_{2, 1})^{1/2}C_2^{1/2}$,  and $\nu_{2, 1} = \nu_2(1 - \eta_2)^{1/2}$,  depending only on $(A_2, \xi_2, \nu_2, \eta_2, \eta_{2, 1}, \eta_{2, 2})$, and
$M_{1, 1} = 12\sigma_0^2(A_1 - 1)^{-2}M_1^2\eta_1 + 2\sigma_0^2\me^{2\eta_{1, 3}}(K-1)^{-1}M_1$, $\eta_{1, 3} = (A_1 - 1)^{-1}M_1\eta_1B_0$ and $M_1 = \xi_{1,1}^2\nu_{1, 1}^{-2}$, depending only on $(B_0, B_1, A_1, \xi_1, \nu_1)$ and $\sigma_0$, and $\eta_1$ is a constant such that $|S_\gamma|\tilde{\lambda}_1 \leq \eta_1$ under Assumption \ref{ass1}$^\prime$(iv), and $(\xi_{1, 1}, \nu_{1, 1})$ are as in Theorem \ref{thm:ps-error-cons}.
\end{thm}

Theorem \ref{thm:or-error-glm-cons} shows that the convergence of $\hat{\alpha}^\#_{t, \text{RWL}}$ to $\bar{\alpha}^\#_{t, \text{WL}}$ in the $L_{2, 1}$ norm at rate $\{|S_\gamma|(K - 1) + |S_{\alpha_t}|\}\{K / n + \log(p) / n\}^{1/2}$ and in the Bregman divergence at the rate $\{|S_\gamma|(K - 1) + |S_{\alpha_t}|\}\{K / n + \log(p) / n\}$. Compared with the corresponding results with one-to-zero constraint, an additional factor of $K - 1$ is multiplied with $|S_\gamma|$.

\begin{rem}[Linear outcome model with sum-to-zero constraint on $\gamma$] \label{rem}
Similarly as in Remark \ref{rem:ORlm}, Assumptions \ref{ass2}$^\prime$(iii)-(v) hold with $C_1 = C_2 = 1$ and $C_3 = 0$ and Assumptions \ref{ass2}$^\prime$(vi)(b) and \ref{ass2}$^\prime$(vi)(c) hold with $\eta_{2, 1} = \eta_{2, 2} = 0$. Therefore, under Assumptions \ref{ass2}$^\prime$(i), \ref{ass2}$^\prime$(ii) and \ref{ass2}$^\prime$(vi)(a), we have from Theorem \ref{thm:or-error-glm-cons} that
\begin{align}
& Q_{\text{WL}}(\hat{\alpha}^\#_{t, \text{RWL}}, \bar{\alpha}^\#_{t, \text{WL}}; \bar{\gamma}_{\text{CAL}}) + \me^{2\eta_{1, 3}}(A_2 - 1)\tilde{\lambda}_2\lVert \hat{\alpha}^\#_{t, \text{RWL}} - \bar{\alpha}^\#_{t, \text{WL}} \rVert_{2, 1} \nonumber \\
& \leq \me^{6\eta_{1, 3}}\xi_{2, 1}^{-2}\{M_{1, 1}|S_\gamma|(K-1)\tilde{\lambda}_1^2\} + \me^{2\eta_{1, 3}}\xi_{2, 2}^2\{\nu_{2, 1}^{-2}|S_{\alpha_t}|\tilde{\lambda}_2^2\},
\label{eq:thm-or-error-lm-cons}
\end{align}
where $(\xi_{2, 1}, \xi_{2, 2}, \nu_{2, 1}, M_{1, 1})$, only depending on $(A_2, \xi_2, \nu_2, \eta_2, \sigma_0, A_1, \xi_1, \nu_1, \eta_1, B_0, B_1)$, are the same as in Theorem \ref{thm:or-error-glm}.
\end{rem}

Assumptions \ref{ass2}$^\prime$(vi)(b) and \ref{ass2}$^\prime$(vi)(c) are similar to Assumptions \ref{ass2}(vi)(b) and \ref{ass2}(vi)(c) except that $\me^{\eta_{1, 3}}$ in Assumption \ref{ass2}(vi)(b) is replaced by $\me^{2\eta_{1, 3}}$ in Assumptions \ref{ass2}$^\prime$(vi)(b) and $\me^{3\eta_{1, 3}}$ in Assumption \ref{ass2}(vi)(c) is replaced with $\me^{6\eta_{1, 3}}$ in Assumptions \ref{ass2}$^\prime$(vi)(c). The additional factor of 2 arises from Lemma \ref{lem:ps-Q-cons}, which is used to replace Lemma \ref{lem:ps-Q}. Moreover, an additional factor of $K-1$ appears in Assumptions \ref{ass2}$^\prime$(vi)(c) compared to Assumption \ref{ass2}(vi)(c) and it also appears in \eqref{eq:thm-or-error-glm-cons} compared to \eqref{eq:thm-or-error-glm}. The factor of $K-1$ arises from
Lemma \ref{lem:or-diff-cons}, which is used to replace Lemma \ref{lem:or-diff} for
handling the dependency on $\hat{\gamma}_{\text{RCAL}}$ as discussed in Remark \ref{rem:key-step}.
With these replacements, \eqref{eq:thm-or-error-glm-cons} can be obtained similarly as \eqref{eq:thm-or-error-glm} in Theorem \ref{thm:or-error-glm}.

\begin{lem}\label{lem:ps-Q-cons}
In the event $\Omega_{\gamma 1}\cap\Omega_{\gamma 2^\prime}$, we have
\begin{align}
\sum_{k\neq t}\tilde{\E}\left[R^{(t)}\omega(k,X; \bar{\gamma}_{\text{CAL}})\{(\hat{h}_k - \hat{h}_t) - (\bar{h}_k - \bar{h}_t)\}^2\right] \leq \me^{2\eta_{1, 3}}M_1|S_\gamma|\tilde{\lambda}_1^2, \label{eq:ps-Q1-cons}
\end{align}
where $\hat{h} = (\hat{h}_k: k\in\mathcal T)^\T$ with $\hat{h}_k = \hat{\gamma}^\T_{k,\text{RCAL}}f(X)$ and $\bar{h} = (\bar{h}_k: k\in\mathcal T)^\T$ with $\bar{h}_k = \bar{\gamma}^\T_{k,\text{CAL}}f(X)$ Moreover,  for any coefficient matrix $\alpha^\#_t$
\begin{align}
D_{\text{WL}}^{\dagger}(\hat{\alpha}^\#_{t,\text{RWL}}, \alpha^\#_t; \hat{\gamma}_{\text{RCAL}}) \geq \me^{-2\eta_{1, 3}}D_{\text{WL}}^{\dagger}(\hat{\alpha}^\#_{t,\text{RWL}}, \alpha^\#_t; \bar{\gamma}_{\text{CAL}}), \label{eq:ps-Q2-cons}
\end{align}
where $\eta_{1, 3} = (A_1 - 1)^{-1}M_1\eta_1B_0$.
\end{lem}
Compared with Lemma \ref{lem:ps-Q}, there are two differences. First, exponential parts in \eqref{eq:ps-Q1-cons} and  \eqref{eq:ps-Q2-cons} are $\me^{2\eta_{1, 3}}$ and $\me^{-2\eta_{1, 3}}$ respectively, and the corresponding parts in \eqref{eq:ps-Q1} and  \eqref{eq:ps-Q2} are $\me^{\eta_{1, 3}}$ and $\me^{-\eta_{1, 3}}$ respectively. Second,  $\hat{h}_k - \bar{h}_k$ in the left hand side of equation \eqref{eq:ps-Q1} is replaced by $(\hat{h}_k - \hat{h}_t) - (\bar{h}_k - \bar{h}_t)$ in \eqref{eq:ps-Q1-cons}.

\begin{prf}
The proof is similar to that of Lemma \ref{lem:ps-Q}. We replace $\hat{h}_k$ and $\bar{h}_k$ in the proof of Lemma \ref{lem:ps-Q} with $\hat{h}_k - \hat{h}_t$ and $\bar{h}_k - \bar{h}_t$ and replace $\hat{\gamma}_{k, \text{RCAL}}$ and $\bar{\gamma}_{k, \text{CAL}}$ with  $\hat{\gamma}_{k, \text{RCAL}} - \hat{\gamma}_{t, \text{RCAL}}$ and $\bar{\gamma}_{k, \text{CAL}} - \bar{\gamma}_{t, \text{CAL}}$. Furthermore, in the proof of Lemma \ref{lem:ps-Q}, $|\hat{h}_k - \bar{h}_k|$ can be bounded by $B_0\lVert \hat{\gamma}_{\text{RCAL}} - \bar{\gamma}_{\text{CAL}} \rVert_{2, 1}$. Here we bound $|(\hat{h}_k - \hat{h}_t) - (\bar{h}_k - \bar{h}_t)|$ by the following inequality:
\begin{align}
&|(\hat{h}_k - \hat{h}_t) - (\bar{h}_k - \bar{h}_t)| \leq |\hat{h}_k - \bar{h}_k| + |\hat{h}_t - \bar{h}_t| \leq 2B_0\lVert \hat{\gamma}_{\text{RCAL}} - \bar{\gamma}_{\text{CAL}} \rVert_{2, 1},
\label{eq:add-1}
\end{align}
thereby incurring an additional factor of 2.
\end{prf}

\begin{lem} \label{lem:or-diff-cons}
In the event $\Omega_{\gamma 1}\cap\Omega_{\gamma 2}\cap\Omega_{\gamma 2^\prime}\cap\Omega_2$,  we have
\begin{align}
& | \langle\nabla\kappa_{\text{WL}}(\bar{\alpha}^\#_{t,\text{WL}}; \hat{\gamma}_{\text{RCAL}}) -  \nabla\kappa_{\text{WL}}(\bar{\alpha}^\#_{t,\text{WL}}; \bar{\gamma}_{\text{CAL}}), \hat{\alpha}^\#_{t,\text{RWL}} - \bar{\alpha}^\#_{t,\text{WL}}\rangle | \nonumber \\
& \leq \me^{2\eta_{1,3}}\{M_{1,1}|S_\gamma|(K-1)\tilde{\lambda}_1^2\}^{1/2}Q_{\text{WL}}^{1/2}(\hat{\alpha}^\#_{t,\text{RWL}}, \bar{\alpha}^\#_{t,\text{WL}}; \bar{\gamma}_{\text{CAL}}), \label{eq:or-diff-cons}
\end{align}
where $M_{1, 1} = 12\sigma_0^2(A_1 - 1)^{-2}M_1^2\eta_1 + 2\sigma_0^2\me^{2\eta_{1, 3}}(K-1)^{-1}M_1$ and $M_1 = \xi^2_{1, 1}\nu^{-2}_{1, 1}$.
\end{lem}

Compared with Lemma \ref{lem:or-diff}, an additional factor of $K - 1$ is multiplied with $|S_\gamma|$.

\begin{prf}
The proof is similar to that of Lemma \ref{lem:or-diff}. We only describe the main differences. The first equation in the proof of Lemma \ref{lem:or-diff} uses the mean value theorem for weight difference $\omega(k, X; \hat{\gamma}_{\text{RCAL}}) - \omega(k, X; \bar{\gamma}_{\text{CAL}}) = \me^{\hat{h}_k - \hat{h}_t} - \me^{\bar{h}_k - \bar{h}_t}$, i.e., equation \eqref{eq:mean-value}, where $\hat{h}_k = \hat{\gamma}^\T_{k, \text{RCAL}}f(X)$ and $\bar{h}_k = \bar{\gamma}^\T_{k, \text{CAL}}f(X)$. With sum-to-zero constraint, equation \eqref{eq:mean-value} becomes $\me^{\hat{h}_k - \hat{h}_t} - \me^{\bar{h}_k - \bar{h}_t} = \me^{u(\hat{h}_k - \hat{h}_t) + (1- u)(\bar{h}_k - \bar{h}_t)}\{(\hat{h}_k - \hat{h}_t) - (\bar{h}_k - \bar{h}_t)\}$, so that we change $d_k$ defined in the proof of Lemma \ref{lem:or-diff} from $\hat{h}_k - \bar{h}_k$ to $(\hat{h}_k - \hat{h}_t) - (\bar{h}_k - \bar{h}_t)$ and also $b_{j\cdot}$
from $\hat{\gamma}_{j\cdot, \text{RCAL}} - \bar{\gamma}_{j\cdot, \text{CAL}}$ to $\hat{\gamma}^c_{j\cdot, \text{RCAL}} - \bar{\gamma}^c_{j\cdot, \text{CAL}}$, where $\hat{\gamma}^c_{j\cdot, \text{RCAL}} = (\hat{\gamma}_{jk, \text{RCAL}} - \hat{\gamma}_{jt, \text{RCAL}}: k\neq t)^\T$ and $\bar{\gamma}^c_{j\cdot, \text{CAL}} = (\bar{\gamma}_{jk, \text{CAL}} - \bar{\gamma}_{jt, \text{CAL}}: k\neq t)^\T$.
Furthermore, we need to bound $\sum_{j=0}^p \lVert b_{j\cdot} \rVert_2$ in the proof of Lemma \ref{lem:or-diff}. With one-to-zero constraint, $\sum_{j=0}^p \lVert b_{j\cdot} \rVert_2$ is  $\lVert \hat{\gamma}_{\text{RCAL}} - \bar{\gamma}_{\text{CAL}} \rVert_{2, 1}$ and can be easily bounded by Theorem \ref{thm:ps-error}. With sum-to-zero constraint, $\sum_{j=0}^p \lVert b_{j\cdot} \rVert_2$ becomes $\lVert \hat{\gamma}^c_{\text{RCAL}} - \bar{\gamma}^c_{\text{CAL}} \rVert_{2, 1}$, where $\hat{\gamma}^c_{\text{RCAL}} = (\hat{\gamma}_{k, \text{RCAL}} - \hat{\gamma}_{t, \text{RCAL}}: k\neq t)$ is centered $\hat{\gamma}_{\text{RCAL}}$ at $\hat{\gamma}_{t, \text{RCAL}}$ and $\bar{\gamma}^c_{\text{CAL}} = (\bar{\gamma}_{k, \text{CAL}} - \bar{\gamma}_{t, \text{CAL}}: k\neq t)$ is centered $\bar{\gamma}_{\text{CAL}}$ at $\bar{\gamma}_{t, \text{CAL}}$. We use the following inequality to bound $\lVert \hat{\gamma}^c_{\text{RCAL}} - \bar{\gamma}^c_{\text{CAL}} \rVert_{2, 1}$:
\begin{align}
& \lVert \hat{\gamma}^c_{\text{RCAL}} - \bar{\gamma}^c_{\text{CAL}} \rVert_{2, 1} = \sum_{j=0}^p \lVert \hat{\gamma}^c_{j\cdot, \text{RCAL}} - \bar{\gamma}^c_{j\cdot, \text{CAL}} \rVert_2 \leq \sqrt{K-1} \sum_{j=0}^p \lVert \hat{\gamma}^c_{j\cdot, \text{RCAL}} - \bar{\gamma}^c_{j\cdot, \text{CAL}} \rVert_{\infty} \nonumber \\
& \leq 2\sqrt{K-1}\sum_{j=0}^p \lVert \hat{\gamma}_{j\cdot,\text{RCAL}} - \bar{\gamma}_{j\cdot,\text{CAL}} \rVert_2 = 2\sqrt{K-1}\lVert \hat{\gamma}_{\text{RCAL}} - \bar{\gamma}_{\text{CAL}} \rVert_{2, 1},
\label{eq:add-2}
\end{align}
where the last step follows from inequality $|(\hat{\gamma}_{jk,\text{RCAL}} - \bar{\gamma}_{jk,\text{CAL}}) - (\hat{\gamma}_{jt,\text{RCAL}} - \bar{\gamma}_{jt,\text{CAL}})| \leq 2\lVert \hat{\gamma}_{j ,\text{RCAL}} - \bar{\gamma}_{j ,\text{CAL}} \rVert_{\infty}$.
\end{prf}

{\bf Estimation of treatment means.}  We can follow the line of proof for the theoretical results of estimation of treatment means with one-to-zero constraint to obtain similar theoretical results of estimation of treatment means with sum-to-zero constraint. Only two parts need to take care of. First, we replace $\hat{h}_k = \hat{\gamma}^\T_{k, \text{RCAL}}$ and $\bar{h}_k = \bar{\gamma}^\T_{k, \text{CAL}}$ during the proof for the theoretical results of estimation of treatment means with one-to-zero constraint with $\hat{h}^c_k = \hat{\gamma}^\T_{k, \text{RCAL}} - \hat{\gamma}^\T_{t, \text{RCAL}}$ and $\bar{h}^c_k = \bar{\gamma}^\T_{k, \text{CAL}} - \bar{\gamma}^\T_{t, \text{CAL}}$, respectively. Furthermore, we use inequality \eqref{eq:add-1} to bound $|\hat{h}^c_k - \bar{h}^c_k|$. Second, we replace $b_k = \hat{\gamma}_{k, \text{RCAL}} - \bar{\gamma}_{k, \text{CAL}}$ during the proof for the theoretical results of estimation of treatment means with one-to-zero constraint with $b^c_k = (\hat{\gamma}_{k, \text{RCAL}} - \hat{\gamma}_{t, \text{RCAL}}) - (\bar{\gamma}_{k, \text{CAL}} - \bar{\gamma}_{t, \text{CAL}})$ and use inequality \eqref{eq:add-1} to bound $\lVert \hat{\gamma}^c_{\text{RCAL}} - \bar{\gamma}^c_{\text{CAL}} \rVert_{2, 1}$.

First, we assume that linear OR model (\ref{eq:ORmodel2}) is used together with PS model (\ref{eq:PSmodel}), and develop theoretical analysis which leads to doubly robust Wald confidence intervals for $\mu_t$.
Similarly to Theorem \ref{thm:mu-lm}, the following result gives convergence of $\hat{\mu}_t(\hat{m}^\#_{\text{RWL}}, \hat{\pi}_{\text{RCAL}})$ to $\bar{\mu}_t(\bar{m}^\#_{\text{WL}}, \bar{\pi}_{\text{CAL}})$.

\begin{thm} \label{thm:mu-lm-cons}
Suppose that Assumption \ref{ass1}$^\prime$ and Assumptions \ref{ass2}$^\prime$(i)-(iii) hold. If $\log\{K + \log(p+1)/\epsilon\} / n\leq 1$, then for $A_1 > (\xi_1 + 1) / (\xi_1 - 1)$ and $A_2 > (\xi_2 + 1) / (\xi_2 - 1)$, we have with probability at least $1 - 10\epsilon$,
\begin{align}
& |\hat{\mu}_t(\hat{m}^\#_{\text{RWL}}, \hat{\pi}_{\text{RCAL}}) - \bar{\mu}_t(\bar{m}^\#_{\text{WL}}, \bar{\pi}_{\text{CAL}})| \nonumber \\
& \leq M_{2,1}|S_\gamma|(K-1)\tilde{\lambda}_1^2 + M_{2,2}|S_\gamma|(K-1)^{1/2}\tilde{\lambda}_1\tilde{\lambda}_2 + M_{2,3}|S_{\alpha_t}|\tilde{\lambda}_1\tilde{\lambda}_2,
\label{eq:thm-mu-lm-cons}
\end{align}
where $M_{2,1} = 2(\sqrt{2} + 1)\sigma_0\eta_{1, 4}\me^{2\eta_{1, 3}} + M_{2,3} + (\sqrt{2} / 2)\sigma_0M_1\me^{4\eta_{1, 3}} / (K-1)$, $M_{2,2} = 2(A_1 - 1)^{-1}M_1$, $M_{2,3} = A_1(A_2 - 1)^{-1}M_2$, $\eta_{1, 4} = (A_1 - 1)^{-2}M_1^2\eta_1$, $M_2$ is a constant such that the right-hand side of \eqref{eq:thm-or-error-lm-cons} in Remark \ref{rem} is upper bounded by $\me^{2\eta_{1, 3}}M_2\{|S_\gamma|(K-1)\tilde{\lambda}_1\tilde{\lambda}_2 + |S_{\alpha_t}|\tilde{\lambda}_2^2\}$ and $(M_1, \eta_1)$ are as in Theorem \ref{thm:or-error-glm-cons}.
\end{thm}

Theorem \ref{thm:mu-lm-cons} shows that $\hat{\mu}_t(\hat{m}^\#_{\text{RWL}}, \hat{\pi}_{\text{RCAL}})$ is doubly robust for $\bar{\mu}_t(\bar{m}^\#_{\text{WL}}, \bar{\pi}_{\text{CAL}})$ provided $\{|S_\gamma|(K-1) + |S_{\alpha_t}|\}\tilde{\lambda}_1^2 = o(1)$, that is, $\{|S_\gamma|(K-1) + |S_{\alpha_t}|\}\{K + \log(p+1)\} = o(n)$. In addition, Theorem \ref{thm:mu-lm-cons} gives the $n^{-1/2}$ asymptotic expansion (\ref{eq:desired-expan2}) provided $n^{1/2}\{|S_\gamma|(K-1) + |S_{\alpha_t}|\}\tilde{\lambda}_1^2 = o(1)$, that is $\{|S_\gamma|(K-1) + |S_{\alpha_t}|\}\{K + \log(p+1)\} = o(n^{1/2})$. Compared with the corresponding result with one-to-zero constraint, an additional factor of $K-1$ is multiplied with $|S_\gamma|$.

Similarly to Theorem \ref{thm:V-lm}, the following result gives convergence of the variance estimator $\hat{V}_t$ to $V_t$.
\begin{thm} \label{thm:V-lm-cons}
Under the conditions of Theorem \ref{thm:mu-lm-cons}, if $\{K + \log(p+1)/\epsilon\} / n \leq 1$, then we have with probability at least $1 - 10\epsilon$,
\begin{align}
|\tilde{\E}(\hat{\varphi}_{tc}^2 - \bar{\varphi}_{tc}^2)| \leq & 2M^{1/2}_{2, 4}\{\tilde{\E}(\bar{\varphi}_{tc}^2)\}^{1/2}(K-1)^{1/2}\{|S_\gamma|(K-1)\tilde{\lambda}_1 + |S_{\alpha_t}|\tilde{\lambda}_2\} \nonumber \\
& + M_{2, 4}(K-1)\{|S_\gamma|(K-1)\tilde{\lambda}_1 + |S_{\alpha_t}|\tilde{\lambda}_2\}^2,
\label{eq:thm-V-lm-cons}
\end{align}
where $M_{2, 4}$ is a positive constant depending only on $(B_0, B_1, A_1, \xi_1, \nu_1, \eta_1)$ in Theorem \ref{thm:ps-error-cons} and $(\sigma_0, A_2, \xi_2, \nu_2, \eta_2)$ in Theorem \ref{thm:or-error-glm-cons}.
\end{thm}

Inequality \eqref{eq:thm-V-lm-cons} shows that $\hat{V}_t$ is a consistent estimator of $V_t$, that is, $\hat{V} - V = o(1)$, provided $\{|S_\gamma|(K - 1) + |S_{\alpha_t}|\}(K-1)^{1/2}\tilde{\lambda}_1 = o(1)$, which means $\{|S_\gamma|(K-1) + |S_{\alpha_t}|\}(K-1)^{1/2}\{K + \log(p+1)\}^{1/2} = o(n^{1/2})$. Compared with the corresponding result with one-to-zero constraint, an additional factor of $K-1$ is multiplied with $|S_\gamma|$.

Combing Theorems \ref{thm:mu-lm-cons} and \ref{thm:V-lm-cons}, we have the following doubly robust Wald confidence intervals for $\hat{\mu}_t$. For simplicity, the group Lasso tuning parameters are denoted as $\lambda_1 = A_1^\dagger[\{K + \log(p+1)\} / n]^{1/2}$ for $\hat{\gamma}_{\text{RCAL}}$ and $\lambda_2 = A_2^\dagger[\{K + \log(p+1)\} / n]^{1/2}$ for $\hat{\alpha}^\#_{t,\text{RWL}}$.

\begin{pro} \label{pro:double-mu-lm-cons}
Suppose that Assumption \ref{ass1}$^\prime$ and Assumptions \ref{ass2}$^\prime$(i) , 2$^\prime$(ii), and 2$^\prime$(iv)(a) hold, and $\{|S_\gamma|(K - 1) + |S_{\alpha_t}|)(K-1)^{1/2}\{K + \log(p+1)\} = o(n^{1/2})$. Then for $\hat{\gamma}_{\text{RCAL}}$ and $\hat{\alpha}^\#_{t, \text{RWL}}$ with sufficiently large constants $A_1^{\dagger}$ and $A_2^{\dagger}$, asymptotic expansion (\ref{eq:desired-expan2}) is valid. Moreover, if either PS model (\ref{eq:PSmodel}) or linear OR model (\ref{eq:ORmodel2}) is correctly specified, the following results hold:\vspace{-.1in}
\begin{itemize}\addtolength{\itemsep}{-.1in}
\item[(i)] $n^{1/2}\{\hat{\mu}_t(\hat{m}^\#_{\text{RWL}}, \hat{\pi}_{\text{RCAL}}) - \mu_t\}\overset{D}{\to}N(0, V_t)$, where $V_t =\var\{\varphi_t(Y, T, X; \bar{\alpha}^\#_{t, \text{WL}}, \bar{\gamma}_{\text{CAL}})\}$;
\item[(ii)] a consistent estimator of $V$ is
\begin{align*}
\hat{V}_t = \tilde{\E}\left[\{\varphi_t(Y, T, X; \hat{\alpha}^\#_{t, \text{RWL}}, \hat{\gamma}_{\text{RCAL}}) - \hat{\mu}_t(\hat{m}^\#_{\text{RWL}}, \hat{\pi}_{\text{RCAL}})\}^2\right];
\end{align*}
\item[(iii)] an asymptotic $(1 - c)$ confidence interval for $\mu_t$ is $\hat{\mu}_t(\hat{m}^\#_{\text{RWL}}, \hat{\pi}_{\text{RCAL}})\pm z_{c/2}\sqrt{\hat{V}_t / n}$, where $z_{c/2}$ is the $(1 - c/2)$ quantile of $N(0, 1)$.
\end{itemize}\vspace{-.1in}
That is, a doubly robust confidence interval for $\mu_t$ is obtained.
\end{pro}
Proposition \ref{pro:double-mu-lm-cons} shows that doubly robust Wald confidence intervals for $\hat{\mu}_t$ can be obtained provided $\{|S_\gamma|(K - 1) + |S_{\alpha_t}|)(K-1)^{1/2}\{K + \log(p+1)\} = o(n^{1/2})$. Compared with the corresponding results with one-to-zero constraint, an additional factor of $K - 1$ is multiplied with $|S_\gamma|$.

Second, we assume that a generalized linear OR model (\ref{eq:ORmodel2}) is used together with PS model (\ref{eq:PSmodel}), and develop theoretical analysis which leads to valid Wald confidence intervals for $\mu_t$ if model (\ref{eq:PSmodel}) is correctly specified.
Similarly to Theorem \ref{thm:mu-glm}, the following result gives convergence of $\hat{\mu}_t(\hat{m}^\#_{\text{RWL}}, \hat{\pi}_{\text{RCAL}})$ to $\bar{\mu}_t(\bar{m}^\#_{\text{WL}}, \bar{\pi}_{\text{CAL}})$.
\begin{thm} \label{thm:mu-glm-cons}
Suppose that Assumptions \ref{ass1}$^\prime$ and \ref{ass2}$^\prime$ hold. If $\{K + \log(p+1) / \epsilon\} / n \leq 1$, then for $A_1 > (\xi_1 + 1) / (\xi_1 - 1)$ and $A_2 > (\xi_2 + 1) / (\xi_2 - 1)$, we have with probability at least $1 - 11\epsilon$
\begin{align}
& |\hat{\mu}_t(\hat{m}^\#_{\text{RWL}}, \hat{\pi}_{\text{RCAL}}) - \bar{\mu}_t(\bar{m}^\#_{\text{WL}}, \bar{\pi}_{\text{CAL}})| \nonumber \\
& \leq M_{3,1}|S_\gamma|(K-1)\tilde{\lambda}_1^2 + M_{3,2}|S_\gamma|(K-1)^{1/2}\tilde{\lambda}_1\tilde{\lambda}_2 + M_{3,3}|S_{\alpha_t}|\tilde{\lambda}_1\tilde{\lambda}_2 + \eta_3\Lambda(\eta_3), \label{eq:thm-mu-glm-cons}
\end{align}
where $M_{3,1}, M_{3,2}$, and $M_{3,3}$ are positive constants, depending only on $(B_0, B_1, A_1, \xi_1, \nu_1, \eta_1)$ from Theorem \ref{thm:ps-error-cons}, $(\sigma_0, A_2, \xi_2, \nu_2, \eta_2)$ from Remark \ref{rem} and $(C_1, C_2, C_3, \eta_{2, 1}, \eta_{2, 2})$ from Theorem \ref{thm:or-error-glm-cons}, $\eta_3 = (A_2 - 1)^{-1}M_3\{|S_\gamma|(K - 1)\tilde{\lambda}_1 + |S_{\alpha_t}|\tilde{\lambda}_2\}$, and $M_3$ is a constant such that the right-hand side of \eqref{eq:thm-or-error-glm-cons} is upper-bounded by $\me^{2\eta_{1, 3}}M_3\{|S_\gamma|(K - 1)\tilde{\lambda}_1\tilde{\lambda}_2 + |S_{\alpha_t}|\tilde{\lambda}_2^2\}$.
\end{thm}

Theorem \ref{thm:mu-glm-cons} shows that $\hat{\mu}_t(\hat{m}^\#_{\text{RWL}}, \hat{\pi}_{\text{RCAL}})$ is doubly robust for $\mu_t$ provided $\{|S_\gamma|(K - 1) + |S_{\alpha_t}|\}\tilde{\lambda}_1 = o(1)$, that is, $\{|S_\gamma|(K - 1) + |S_{\alpha_t}|\}\{K + \log(p + 1)\}^{1/2} = o(n^{1/2})$. In addition, the error bounds imply that $\hat{\mu}_t(\hat{m}^\#_{\text{RWL}}, \hat{\pi}_{\text{RCAL}})$ admits the $n^{-1/2}$ asymptotic expansion (\ref{eq:desired-expan2}) provided $n^{1/2}\{|S_\gamma|(K-1) + |S_{\alpha_t}|\}\tilde{\lambda}_1^2 = o(1)$, that is $\{|S_\gamma|(K-1) + |S_{\alpha_t}|\}\{K + \log(p+1)\} = o(n^{1/2})$ when PS model is correctly specified but OR model may be misspecified, because the term involving $\Lambda(\eta_3)$ vanishes when PS model  \eqref{eq:PSmodel} is correctly specified. Unfortunately, asymptotic expansion may fail when PS model is misspecified. Compared with the corresponding result with one-to-zero constraint, an additional factor of $K-1$ is multiplied with $|S_\gamma|$.

Similarly to Theorem \ref{thm:V-glm}, the following result giving convergence of the variance estimator $\hat{V}_t$ to $V_t$.
\begin{thm} \label{thm:V-glm-cons}
Under the conditions of Theorem \ref{thm:mu-glm-cons}, if $\{K + \log(p+1)/\epsilon\} / n \leq 1$, then we have with probability at least $1 - 11\epsilon$,
\begin{align}
|\tilde{\E}(\hat{\varphi}_{tc}^2 - \bar{\varphi}_{tc}^2)| \leq & 2M^{1/2}_{3,4}\{\tilde{\E}(\bar{\varphi}_{tc}^2)\}^{1/2}\{(K - 1) + \Lambda^2(\eta_3)\}^{1/2}\{|S_\gamma|(K-1)\tilde{\lambda}_1 + |S_{\alpha_t}|\tilde{\lambda}_2\} \nonumber \\
& + M_{3,4}\{(K - 1) + \Lambda^2(\eta_3)\}\{|S_\gamma|(K-1)\tilde{\lambda}_1 + |S_{\alpha_t}|\tilde{\lambda}_2\}^2,
\label{eq:thm-V-glm-cons}
\end{align}
where $M_{3,4}$ is a positive constant depending only on $(B_0, B_1, A_1, \xi_1, \nu_1, \eta_1)$ fromTheorem \ref{thm:ps-error-cons}, $(\sigma_0, A_2, \allowbreak \xi_2, \nu_2, \eta_2)$ from Remark \ref{rem} and $(C_1, C_2, C_3, \eta_{2, 1}, \eta_{2, 2})$ from Theorem \ref{thm:or-error-glm-cons}.
\end{thm}

Inequality \ref{eq:thm-V-glm-cons} shows that $\hat{V}$ is a consistent estimator of $V$, that is, $\hat{V} - V = o(1)$, provided $\{|S_\gamma|(K - 1) + |S_{\alpha_t}|\}(K - 1)^{1/2}\tilde{\lambda}_1 = o(1)$, that is, $\{|S_\gamma|(K - 1) + |S_{\alpha_t}|\}(K - 1)^{1/2}\{K + \log(p+1)\}^{1/2} = o(n^{1/2})$. Compared with related result with one-to-zero constraint, an additional factor of $K-1$ is attached with $|S_\gamma|$.

Similarly to Proposition \ref{pro:double-mu-glm}, the following result shows model-assisted Wald confidence intervals for $\mu_t$.

\begin{pro} \label{pro:double-mu-glm-cons}
Suppose that Assumption \ref{ass1}$^\prime$ and Assumption \ref{ass2}$^\prime$ hold, and $\{|S_\gamma|(K - 1) + |S_{\alpha_t}|\}(K - 1)^{1/2}\{K + \log(p+1)\} = o(n^{1/2})$. For sufficiently large constants $A_1^\dagger$ and $A_2^\dagger$, if logistic PS model \eqref{eq:PSmodel} is correctly specified but OR model \eqref{eq:ORmodel} may be misspecified, (i)-(iii) in Proposition \ref{pro:double-mu-lm-cons} hold. That is, a PS based ,OR assisted confidence interval for $\mu_t$ is obtained.
\end{pro}

\section{Additional results for simulation study}

Under (C1), we run 100 repeated simulations, each of sample size 100,000. The means of $T = t$ are $0.203, 0.269, 0.272$ and $0.256$ for $ t = 0, 1, 2$ and $3$ and the corresponding standard deviations are close to 0. The same experiment is done under (C3). The means of $T = t$ are $0.218, 0.263, 0.264$ and $0.255$ for $ t = 0, 1, 2$ and $3$ and the corresponding standard deviations are close to 0.

\begin{figure}[H]
\centering
\includegraphics[scale=0.47]{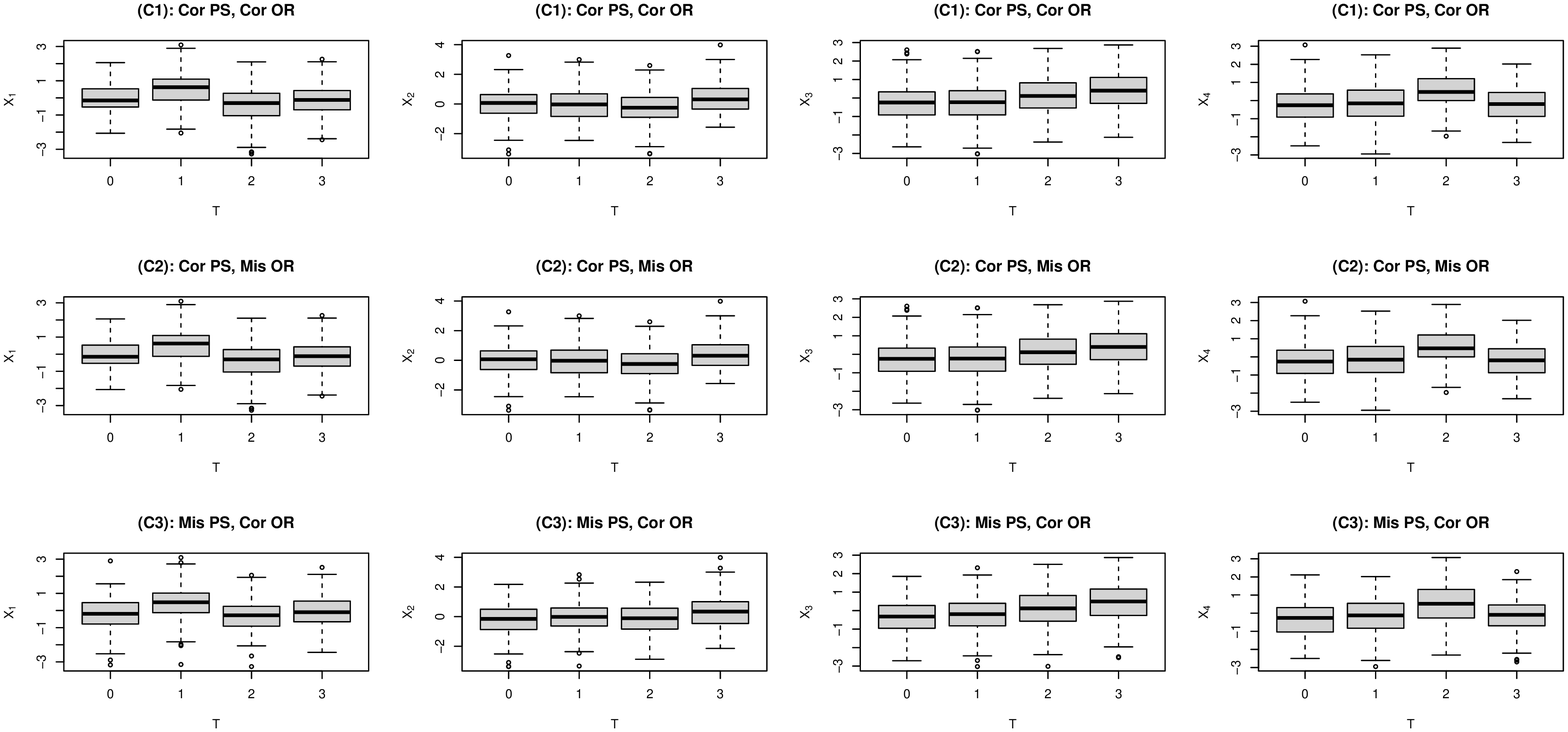}\vspace{-.1in}
\caption{Boxplots of $(X_1, \dots, X_4)$ within $\{T = 0, T = 1, T = 2, T = 3\}$ from a sample of size $n = 1000$.}
\label{fig:box_x}
\end{figure}

\begin{figure}[H]
\centering
\includegraphics[scale=0.47]{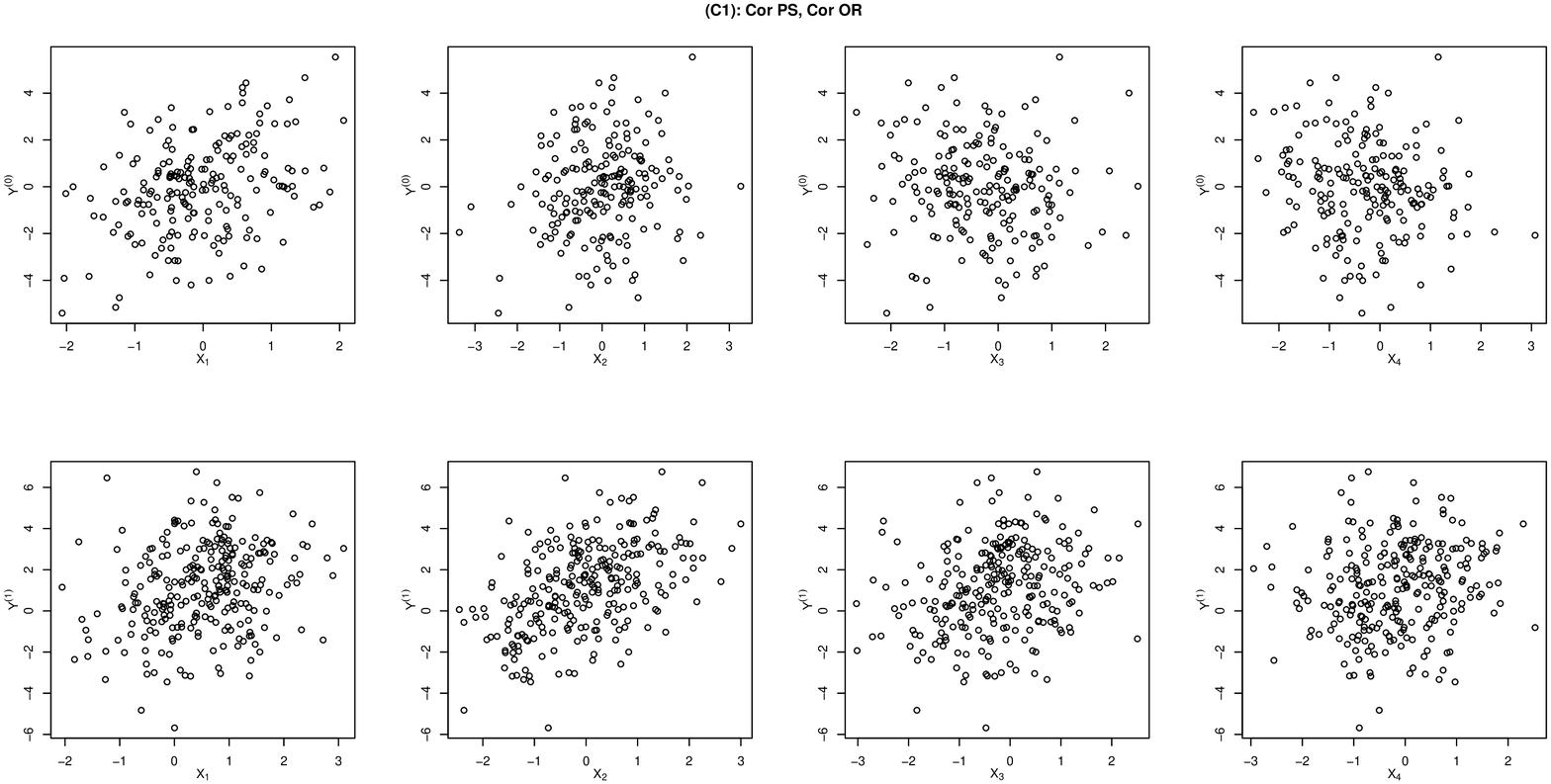}\vspace{-.1in}
\caption{Scatterplots of $Y$ against $(X_1, \dots, X_4)$ within $\{T = 0, T = 1\}$ from a sample of size $n = 1000$ in data configuration (C1).}
\label{fig:scatter_xy_c1_part1}
\end{figure}

\begin{figure}[H]
\centering
\includegraphics[scale=0.47]{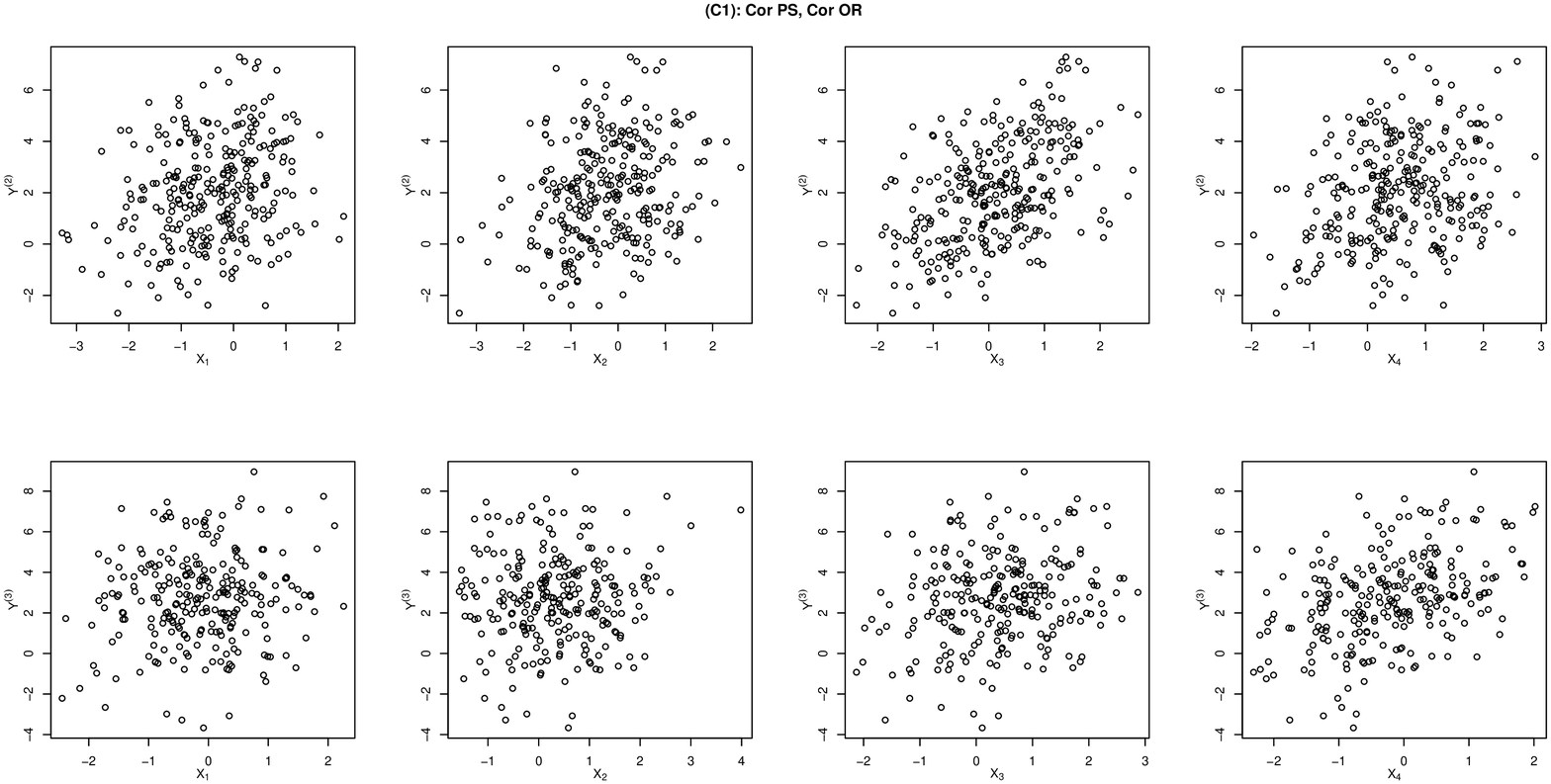}\vspace{-.1in}
\caption{Scatterplots of $Y$ against $(X_1, \dots, X_4)$ within $\{T = 2, T = 3\}$ from a sample of size $n = 1000$ in data configuration (C1).}
\label{fig:scatter_xy_c1_part2}
\end{figure}

\begin{figure}[H]
\centering
\includegraphics[scale=0.47]{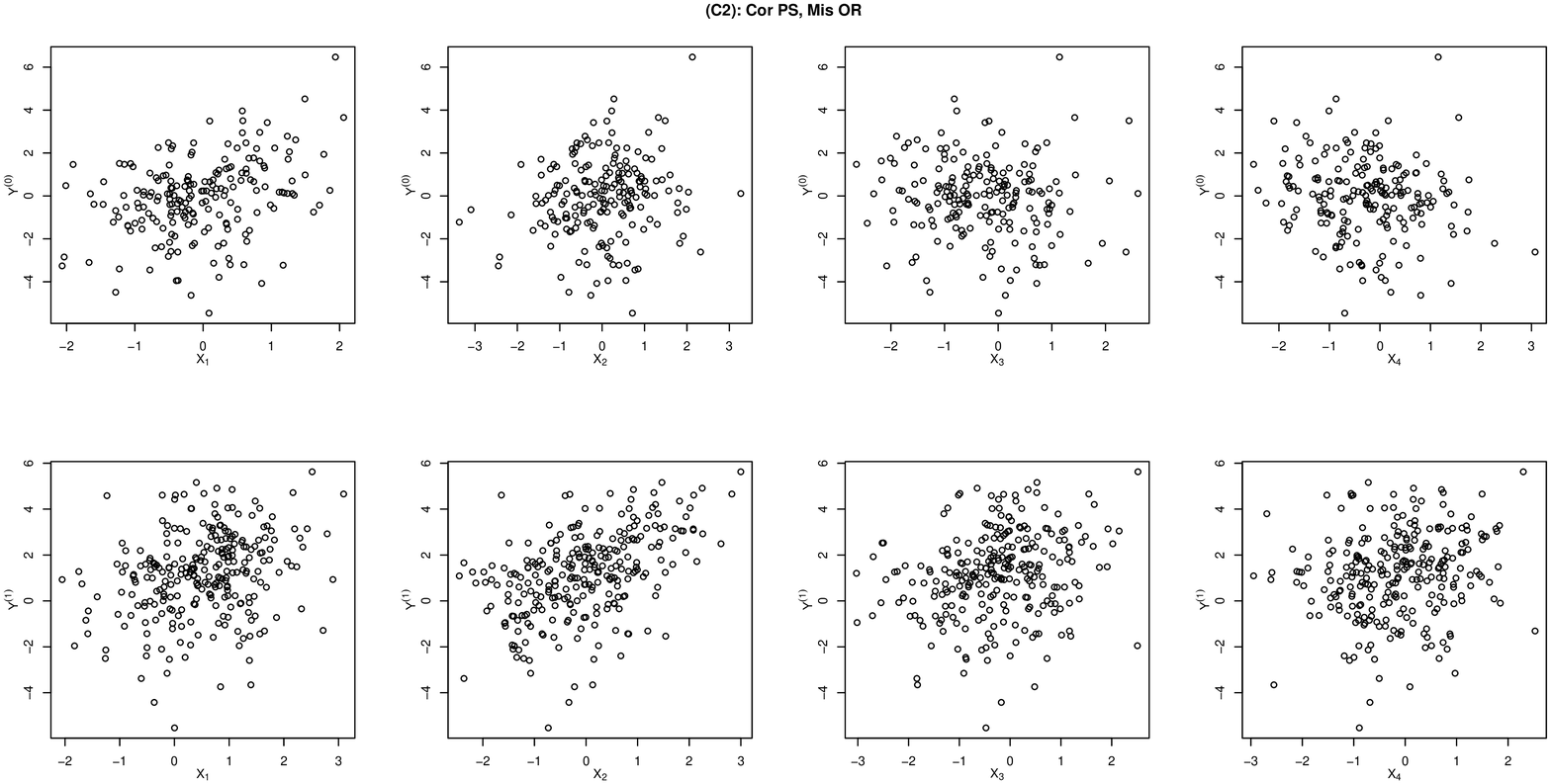}\vspace{-.1in}
\caption{Scatterplots of $Y$ against $(X_1, \dots, X_4)$ within $\{T = 0, T = 1\}$ from a sample of size $n = 1000$ in data configuration (C2).}
\label{fig:scatter_xy_c2_part1}
\end{figure}

\begin{figure}[H]
\centering
\includegraphics[scale=0.47]{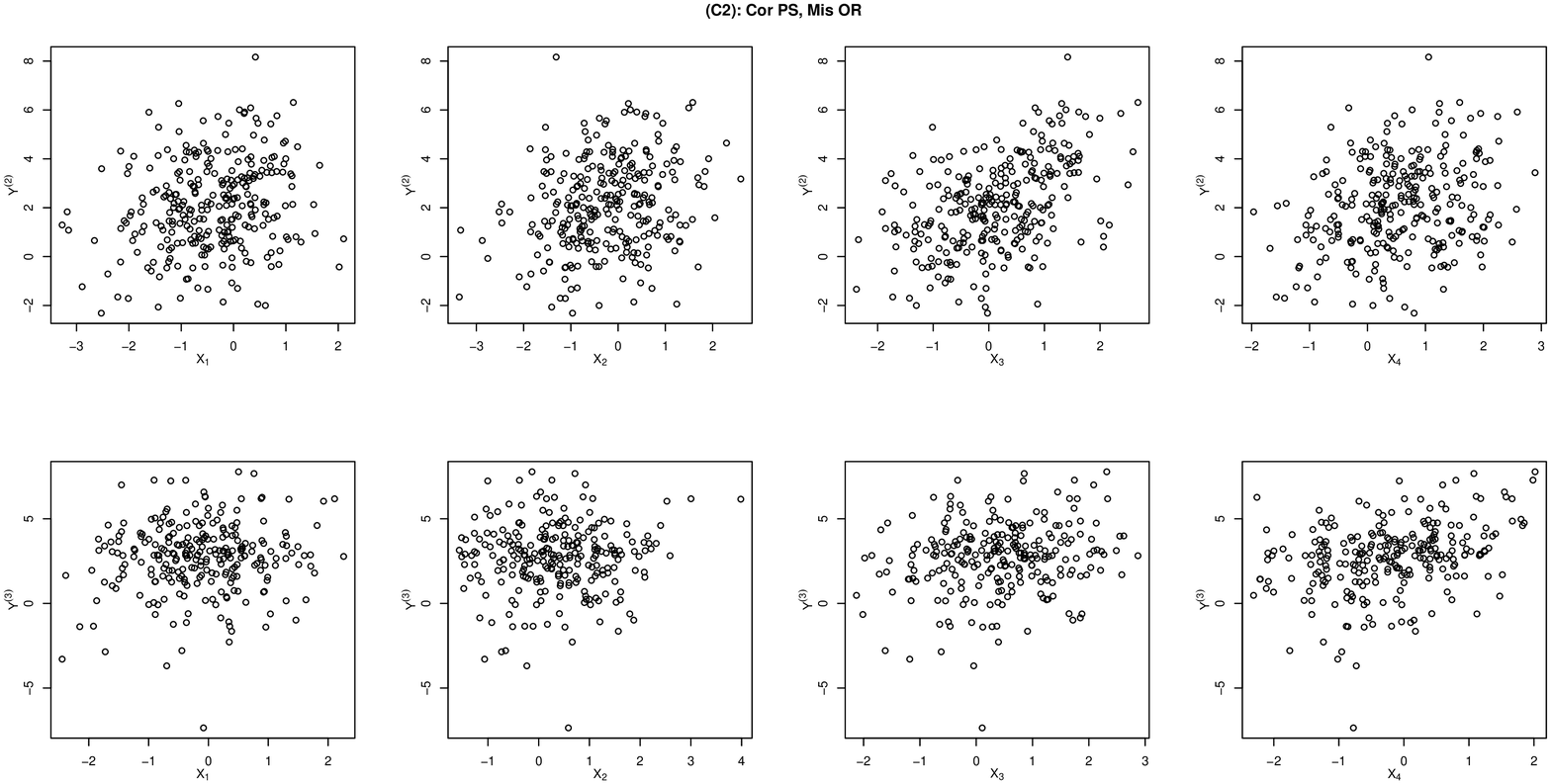}\vspace{-.1in}
\caption{Scatterplots of $Y$ against $(X_1, \dots, X_4)$ within \{T = 2, T = 3\} from a sample of size $n = 1000$ in data configuration (C2).}
\label{fig:scatter_xy_c2_part2}
\end{figure}

\begin{figure}[H]
\centering
\includegraphics[scale=0.47]{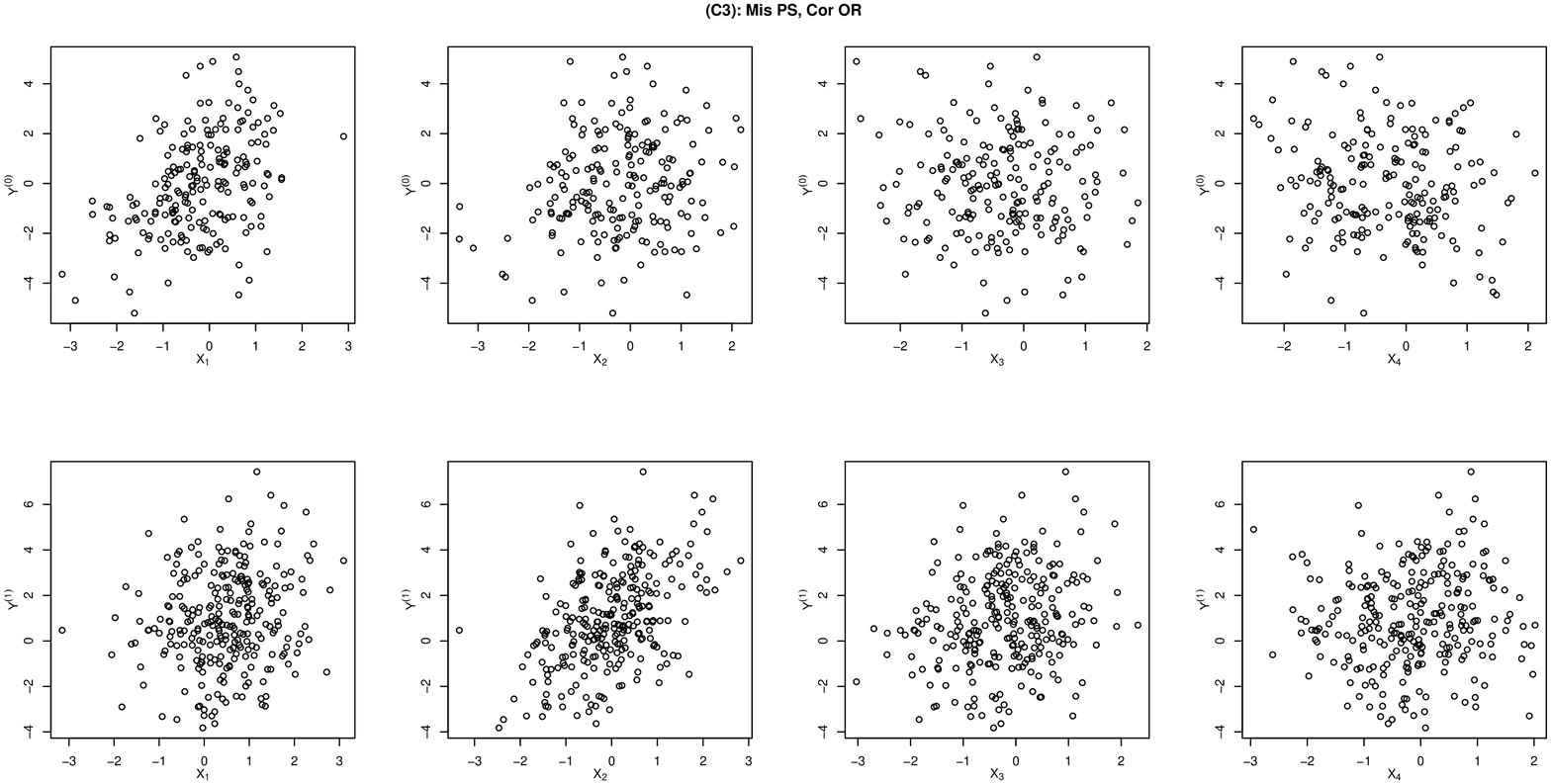}\vspace{-.1in}
\caption{Scatterplots of $Y$ against $(X_1, \dots, X_4)$ within \{T = 0, T = 1\} from a sample of size $n = 1000$ in data configuration (C3).}
\label{fig:scatter_xy_c3_part1}
\end{figure}

\begin{figure}[H]
\centering
\includegraphics[scale=0.47]{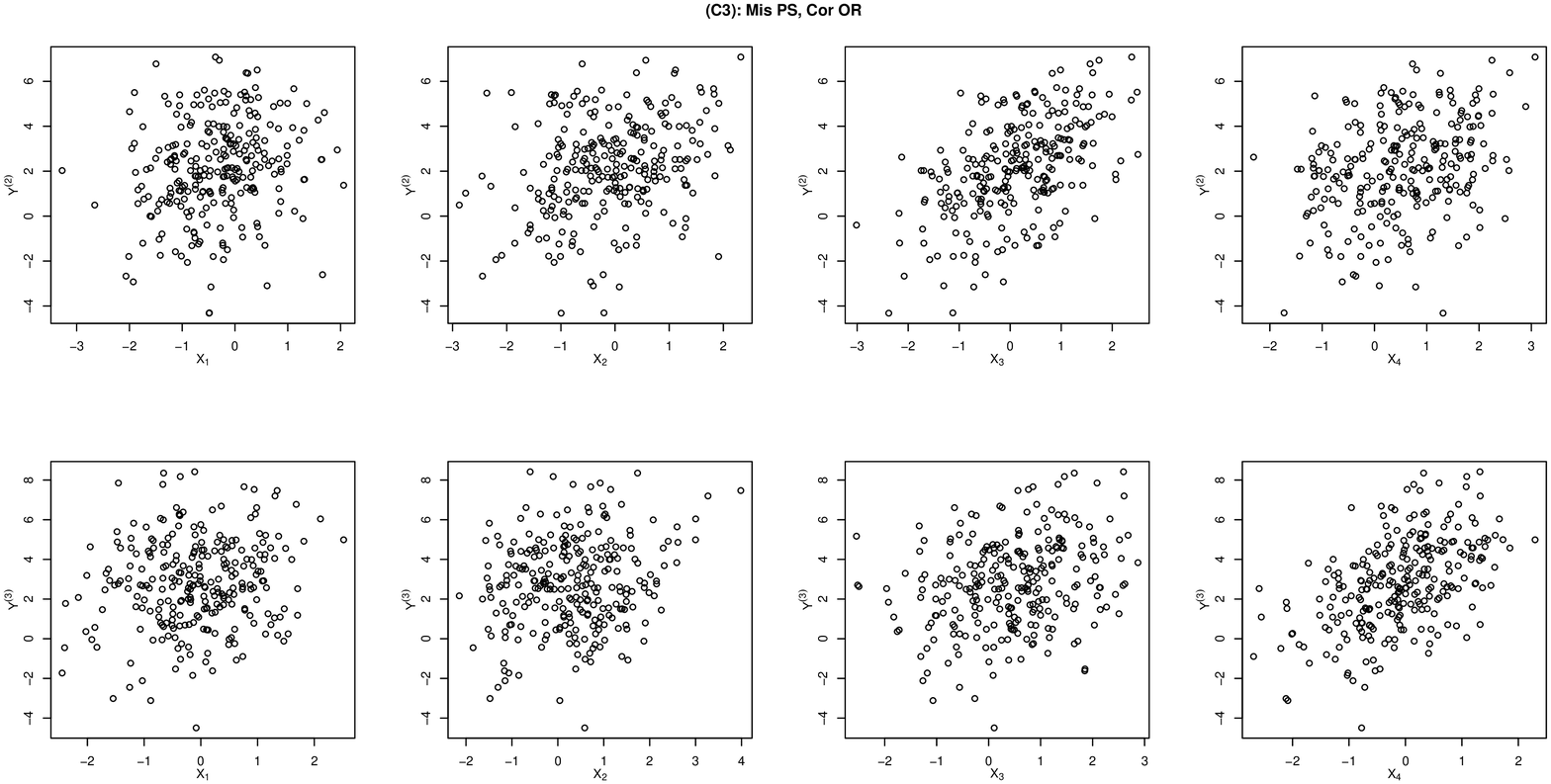}\vspace{-.1in}
\caption{Scatterplots of $Y$ against $(X_1, \dots, X_4)$ within \{T = 2, T = 3\} from a sample of size $n = 1000$ in data configuration (C3).}
\label{fig:scatter_xy_c3_part2}
\end{figure}

Figure \ref{fig:box_x} shows the boxplot of $(X_1, \ldots, X_4)$ by $T$. Figures \ref{fig:scatter_xy_c1_part1}--\ref{fig:scatter_xy_c3_part2} show  the scatterplots of $Y$ against $(X_1, \ldots, X_4)$ within $\{T = 0, T = 1, T = 2, T = 3\}$.

\begin{figure}[H]
\centering
\includegraphics[scale=0.47]{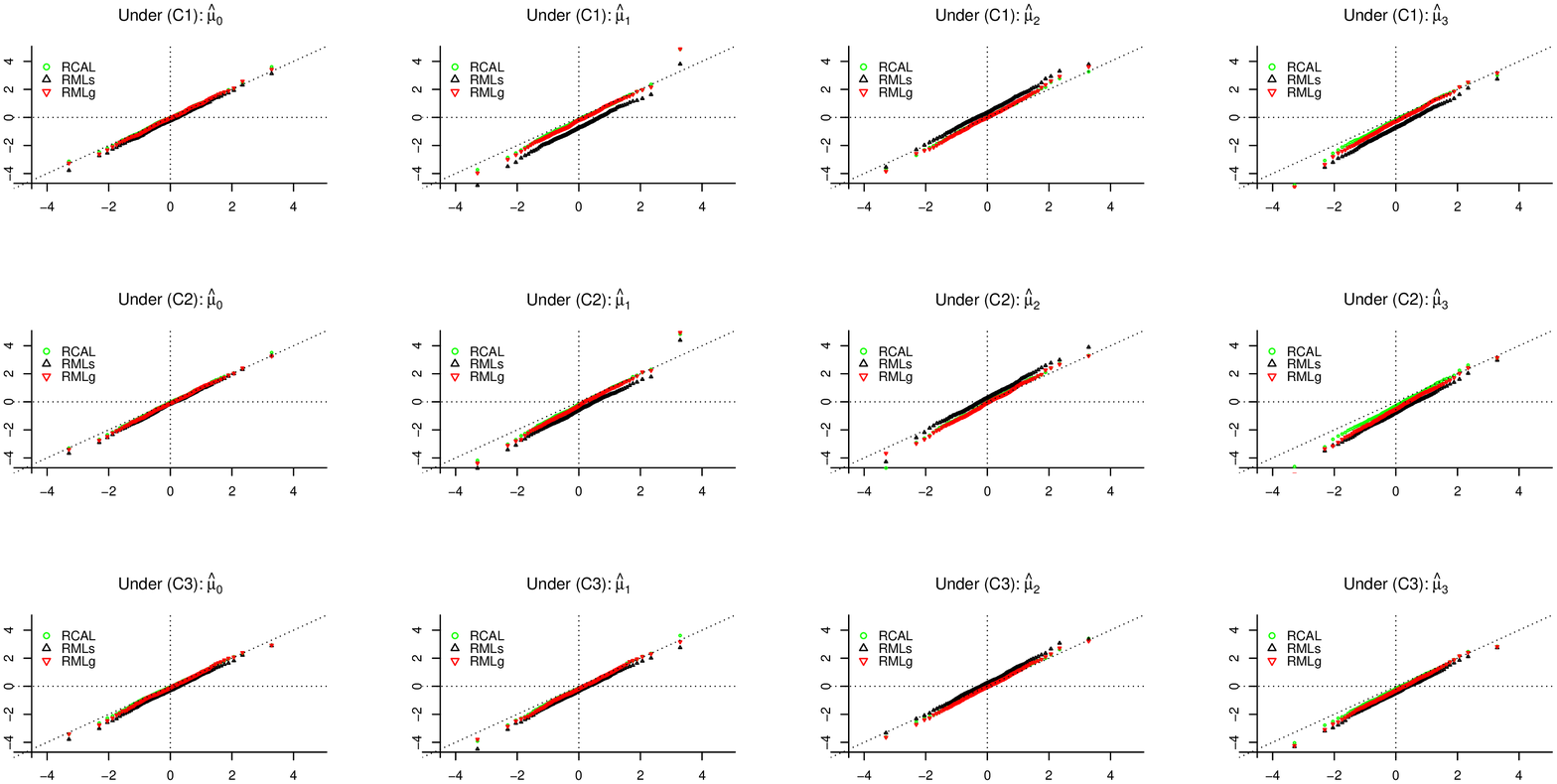}\vspace{-.1in}
\caption{QQ plots of the $t$-statistics against standard normal based on $\hat{\mu}_t$ with $n = 1000, p = 50$ with one-to-zero constraint.}
\label{fig:qq_mu_p50_cons}
\end{figure}

\begin{figure}[H]
\centering
\includegraphics[scale=0.47]{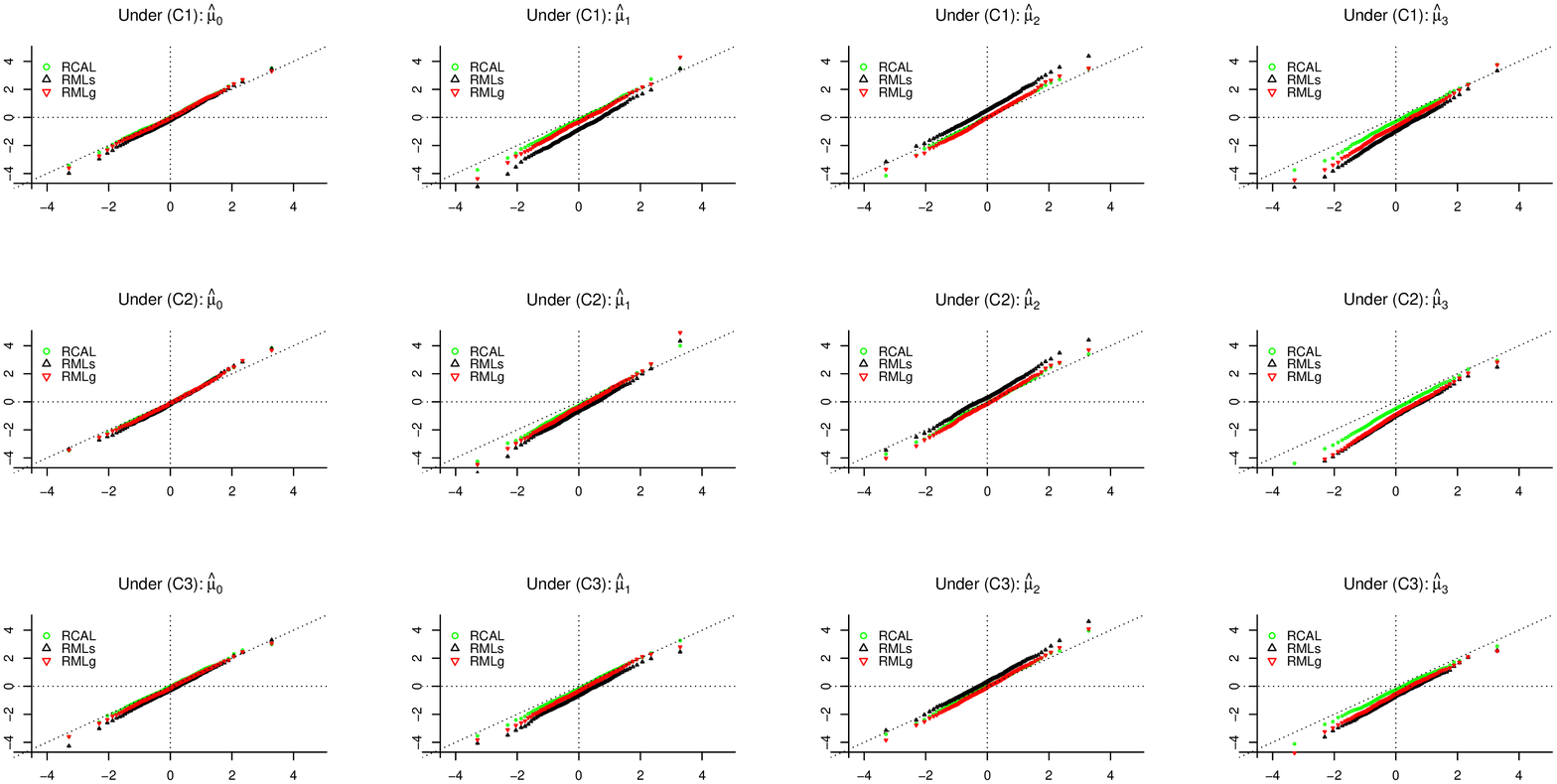}\vspace{-.1in}
\caption{QQ plots of the $t$-statistics against standard normal based on $\hat{\mu}_t$ with $n = 1000, p = 300$ with one-to-zero constraint.}
\label{fig:qq_mu_p300_cons}
\end{figure}

\begin{figure}[H]
\centering
\includegraphics[scale=0.47]{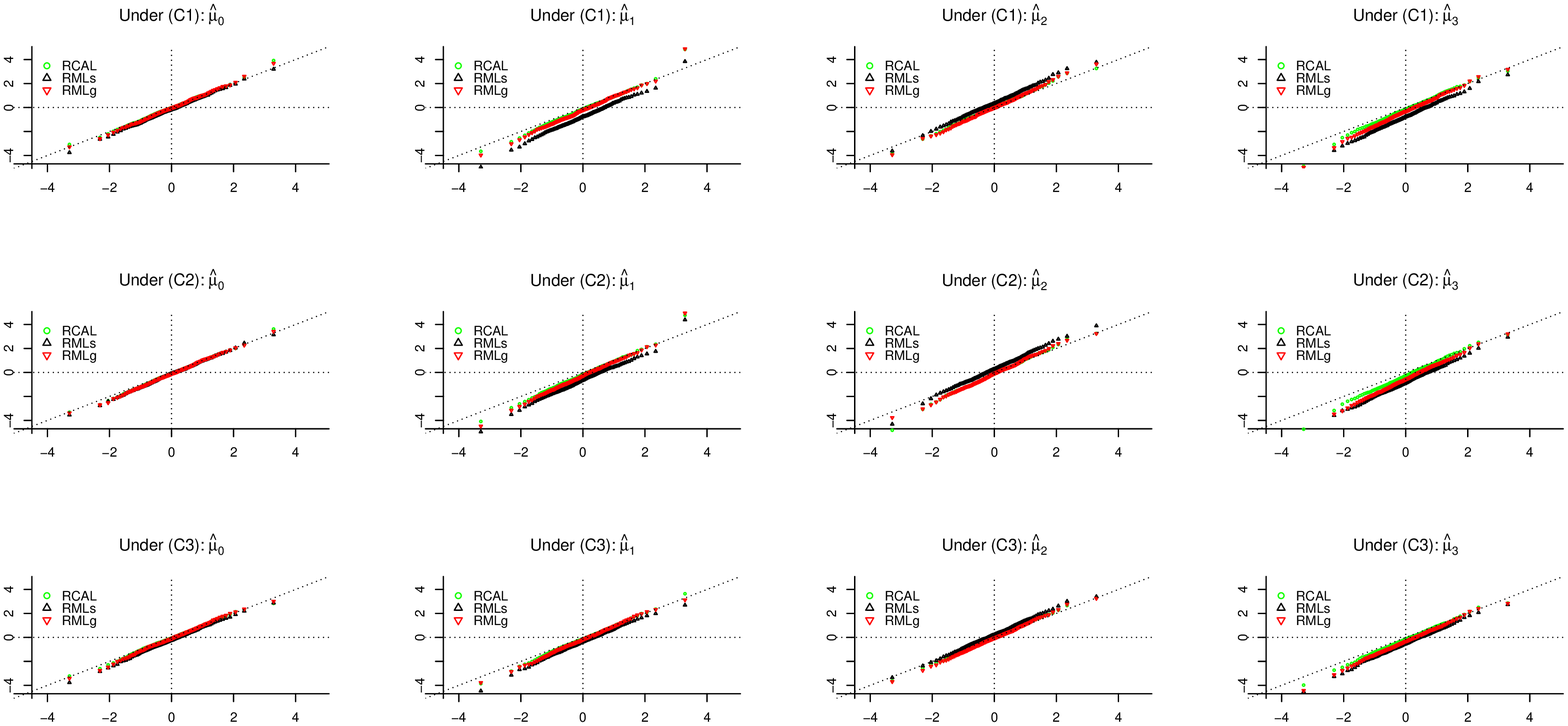}\vspace{-.1in}
\caption{QQ plots of the $t$-statistics against standard normal based on $\hat{\mu}_t$ with $n = 1000, p = 50$ with sum-to-zero constraint.}
\label{fig:qq_mu_p50}
\end{figure}

\begin{figure}[H]
\centering
\includegraphics[scale=0.47]{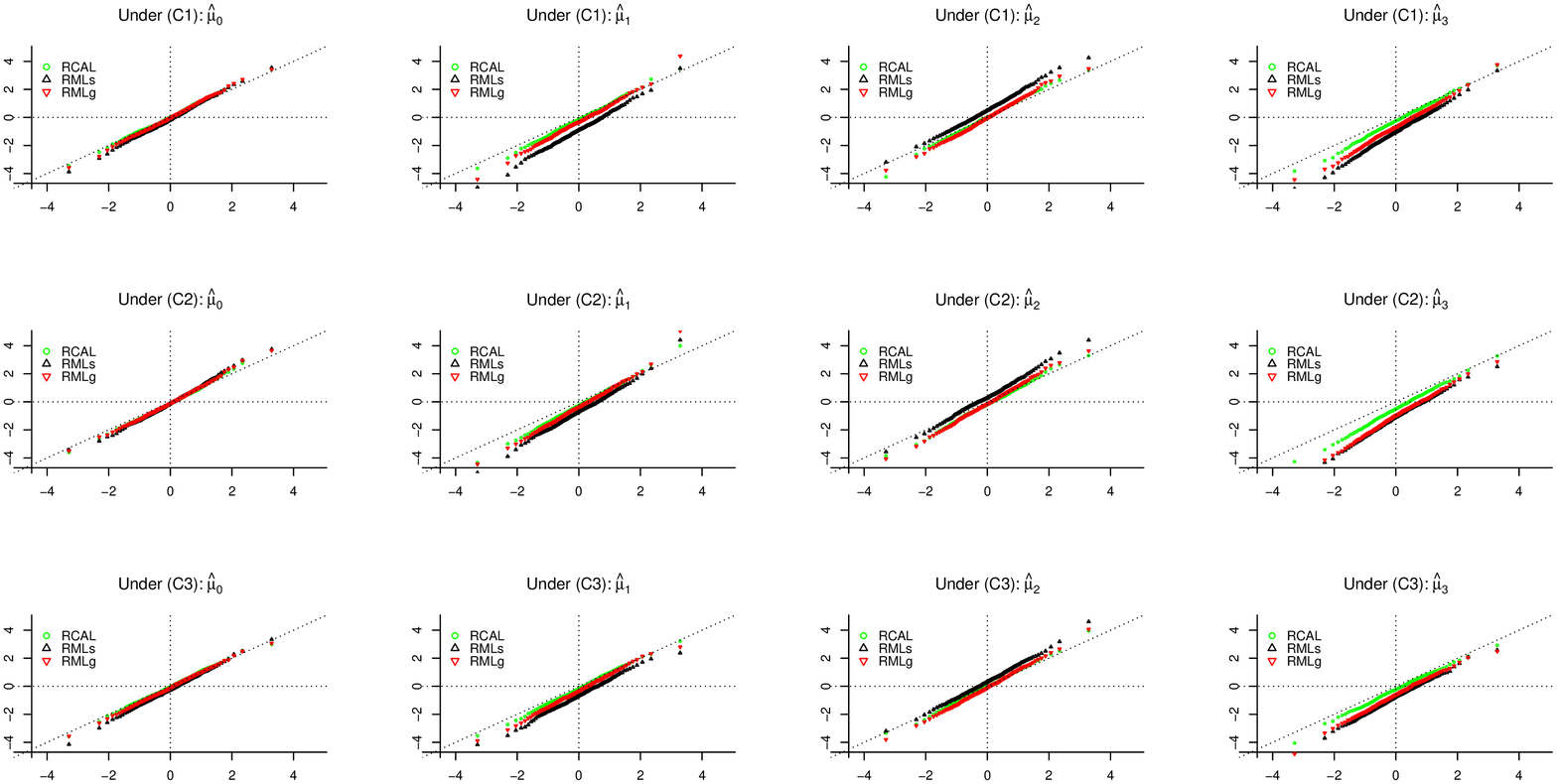}\vspace{-.1in}
\caption{QQ plots of the $t$-statistics against standard normal based on $\hat{\mu}_t$ with $n = 1000, p = 300$ with sum-to-zero constraint.}
\label{fig:qq_mu_p300}
\end{figure}

Table \ref{tb:mu} is summary of $\hat{\mu}_t$ with sum-to-zero constraint. Similarly to the results with one-to-zero constraint, RCAL has the smallest absolute bias and best coverage, and RMLs has the largest absolute bias, worst coverage. Figures \ref{fig:qq_mu_p50_cons}--\ref{fig:qq_mu_p300_cons} and  \ref{fig:qq_mu_p50}--\ref{fig:qq_mu_p300} are QQ plots of the $t$-statistics against standard normal based on $\hat{\mu}_t$ with one-to-zero constraint and sum-to-zero constraint, respectively. From these QQ plots, we easily see that RCAL has smaller bias, and the $t$-statistics of RCAL are more aligned with standard normal.

\begin{figure}[H]
\centering
\includegraphics[scale=0.47]{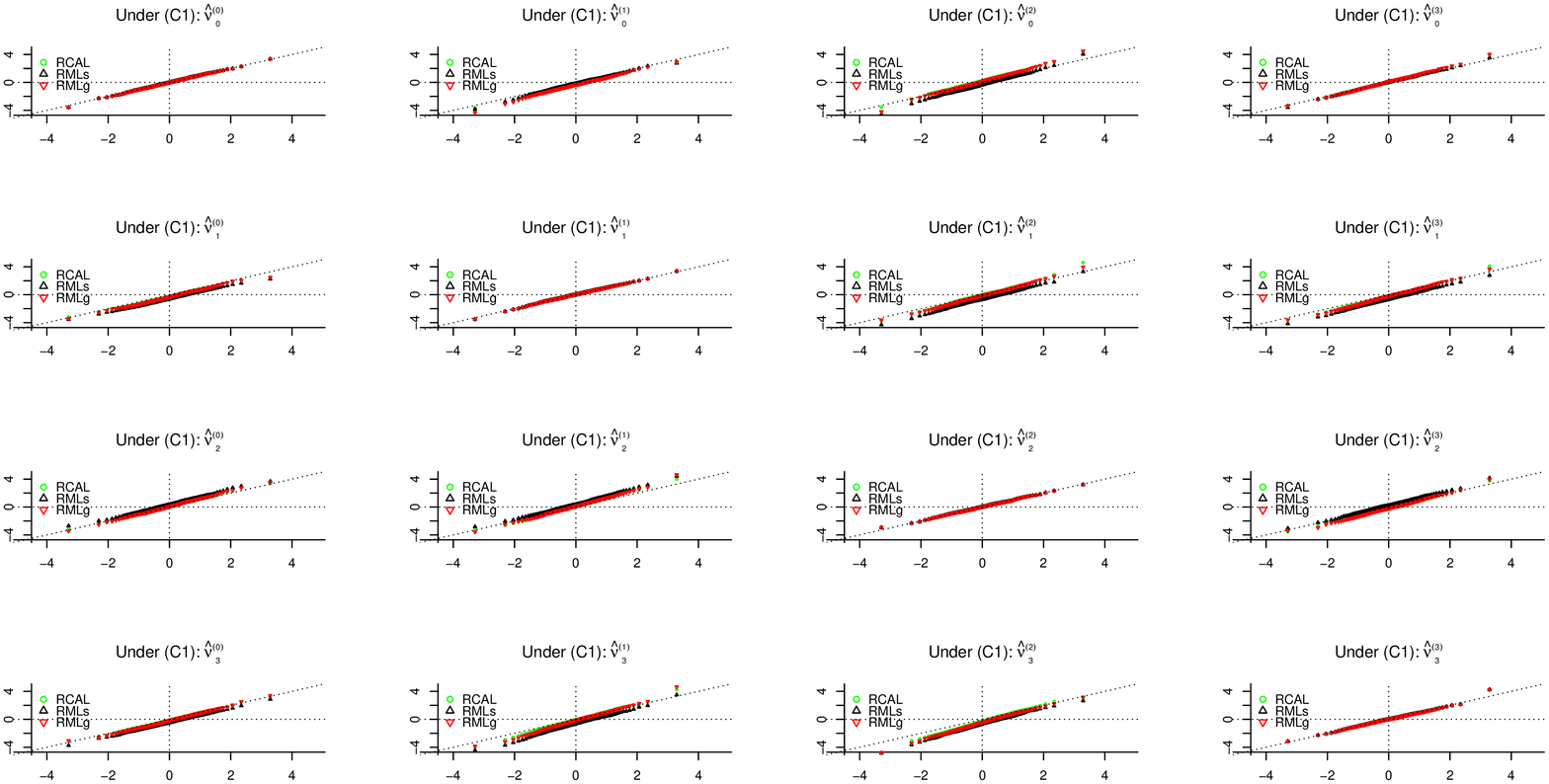}\vspace{-.1in}
\caption{QQ plots of the $t$-statistics against standard normal based on $\hat{\nu}^{(k)}_t$ with $n = 1000, p = 50$ and one-to-zero constraint under (C1).}
\label{fig:qq_nu_p50_c1_cons}
\end{figure}

\begin{figure}[H]
\centering
\includegraphics[scale=0.47]{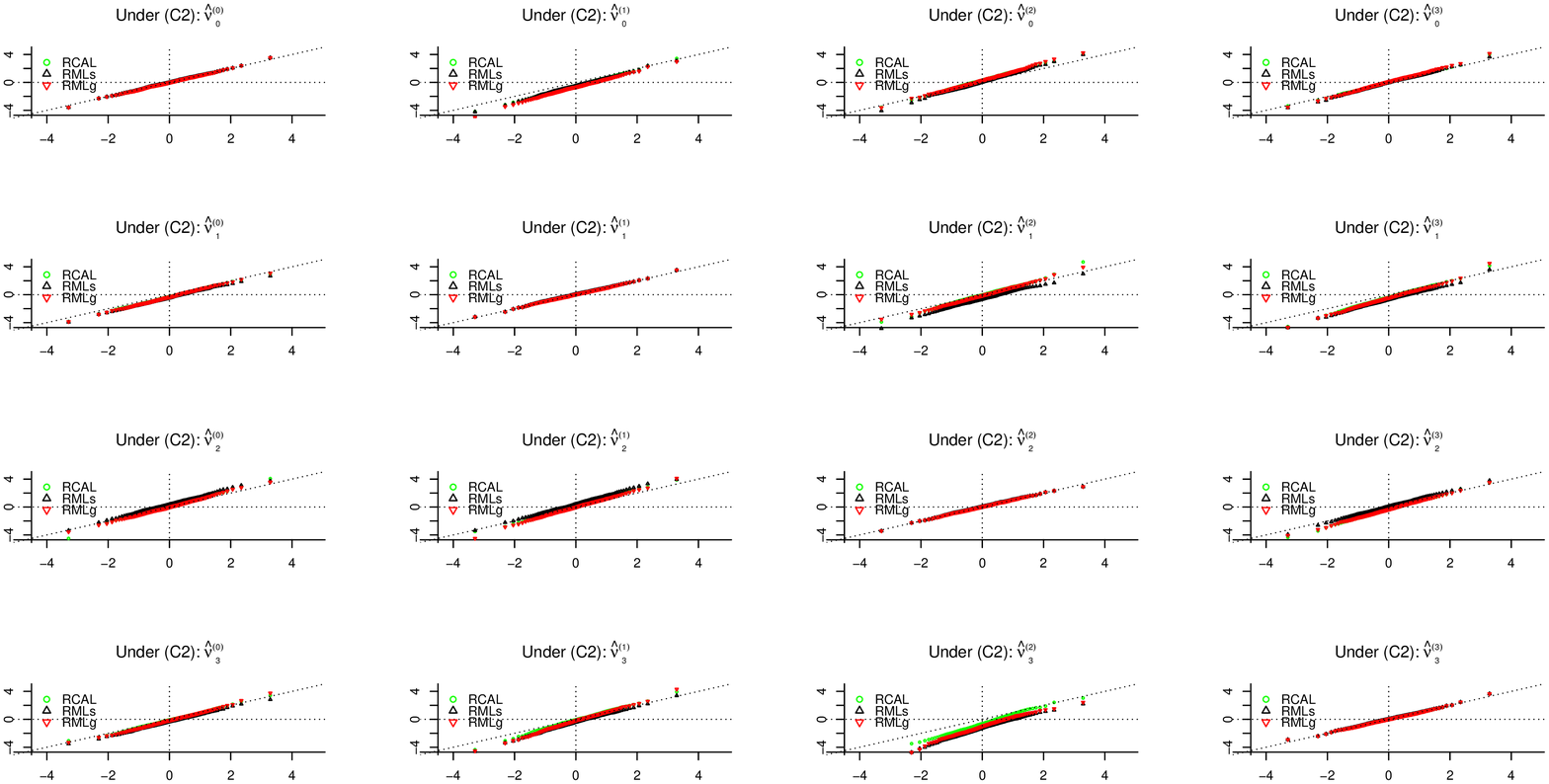}\vspace{-.1in}
\caption{QQ plots of the $t$-statistics against standard normal based on $\hat{\nu}^{(k)}_t$ with $n = 1000, p = 50$ and one-to-zero constraint under (C2).}
\label{fig:qq_nu_p50_c2_cons}
\end{figure}

\begin{figure}[H]
\centering
\includegraphics[scale=0.47]{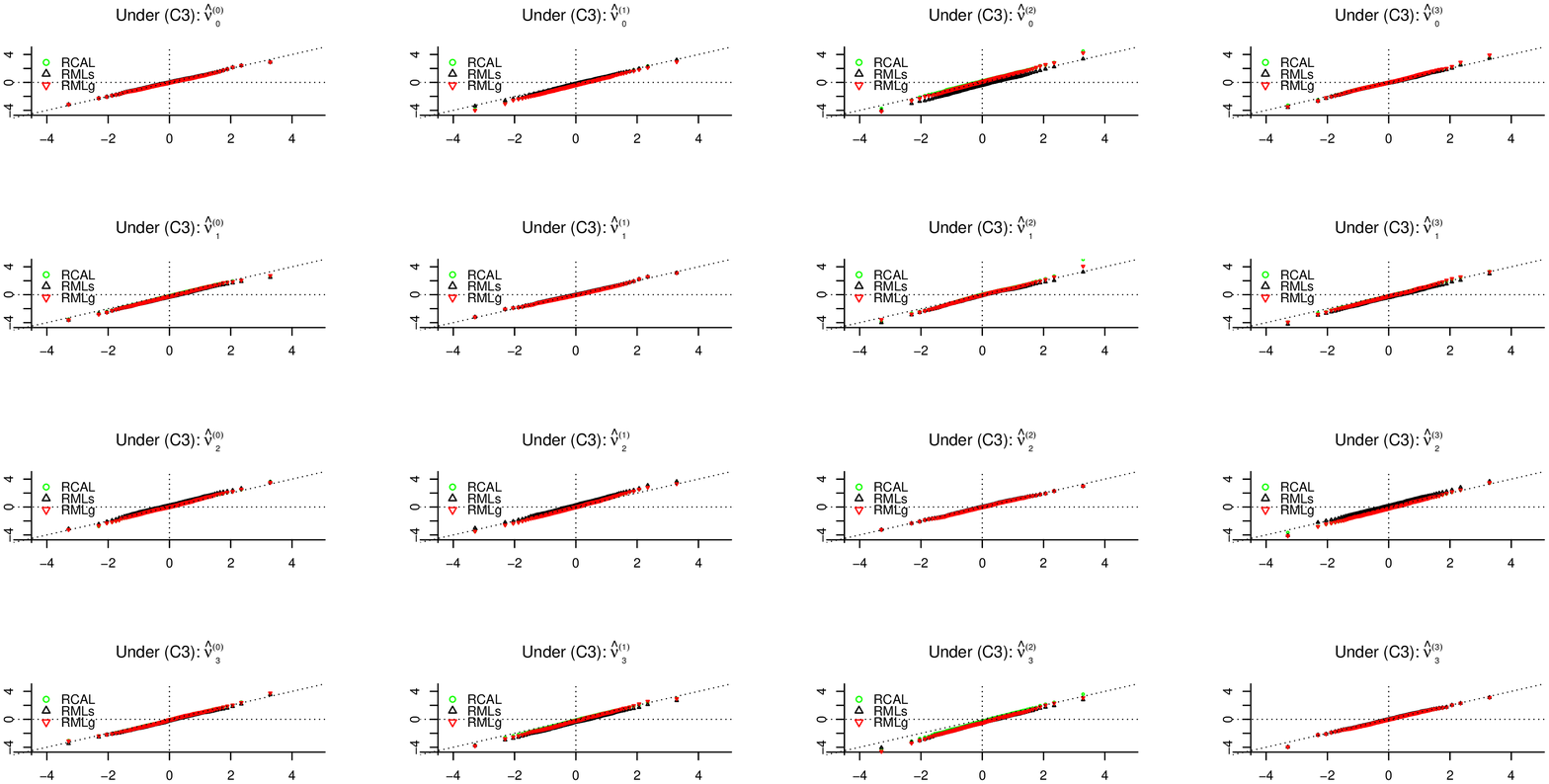}\vspace{-.1in}
\caption{QQ plots of the $t$-statistics against standard normal based on $\hat{\nu}^{(k)}_t$ with $n = 1000, p = 50$ and one-to-zero constraint under (C3).}
\label{fig:qq_nu_p50_c3_cons}
\end{figure}

\begin{figure}[H]
\centering
\includegraphics[scale=0.47]{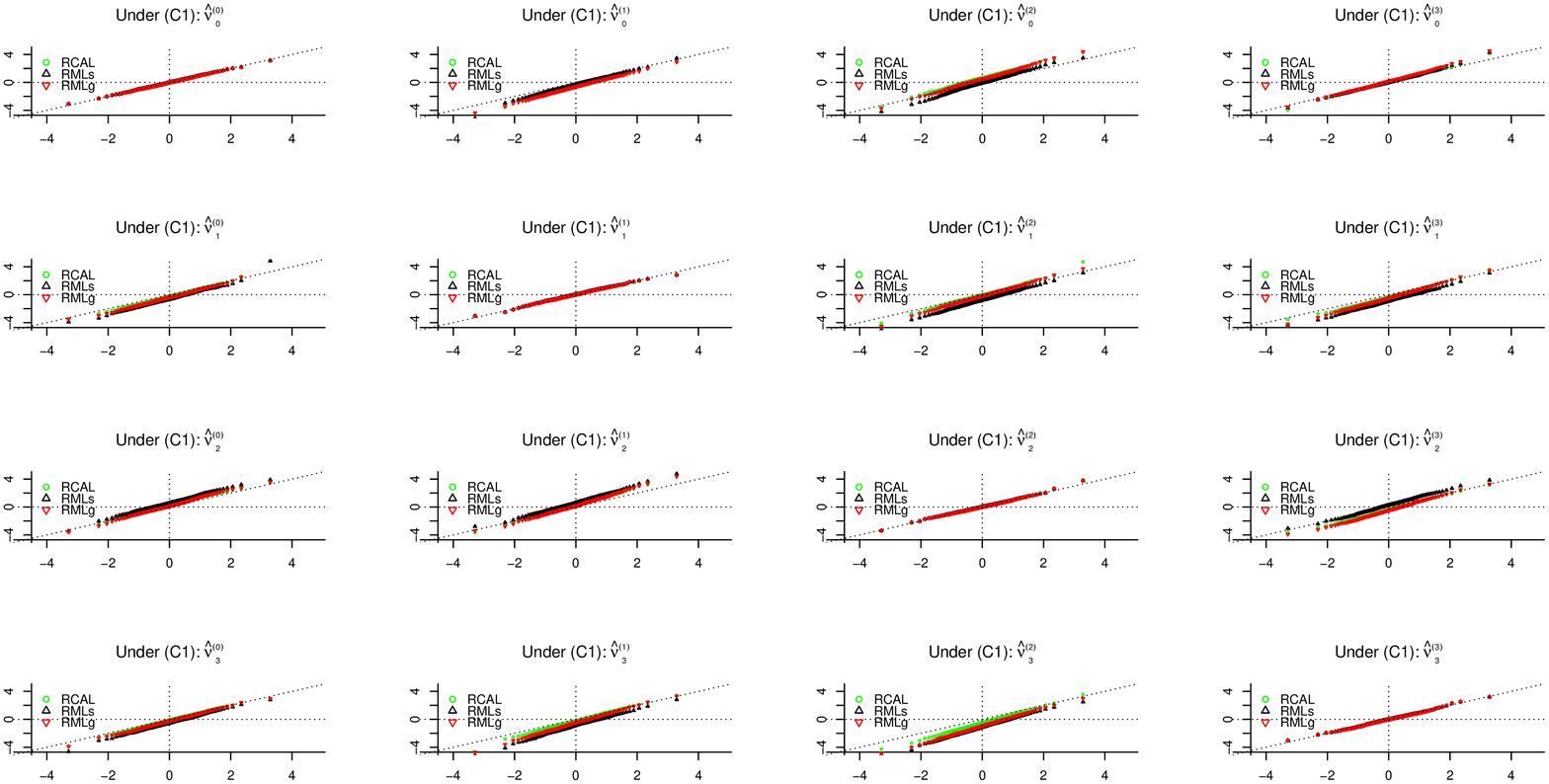}\vspace{-.1in}
\caption{QQ plots of the $t$-statistics against standard normal based on $\hat{\nu}^{(k)}_t$ with $n = 1000, p = 300$ and one-to-zero constraint under (C1).}
\label{fig:qq_nu_p300_c1_cons}
\end{figure}

\begin{figure}[H]
\centering
\includegraphics[scale=0.47]{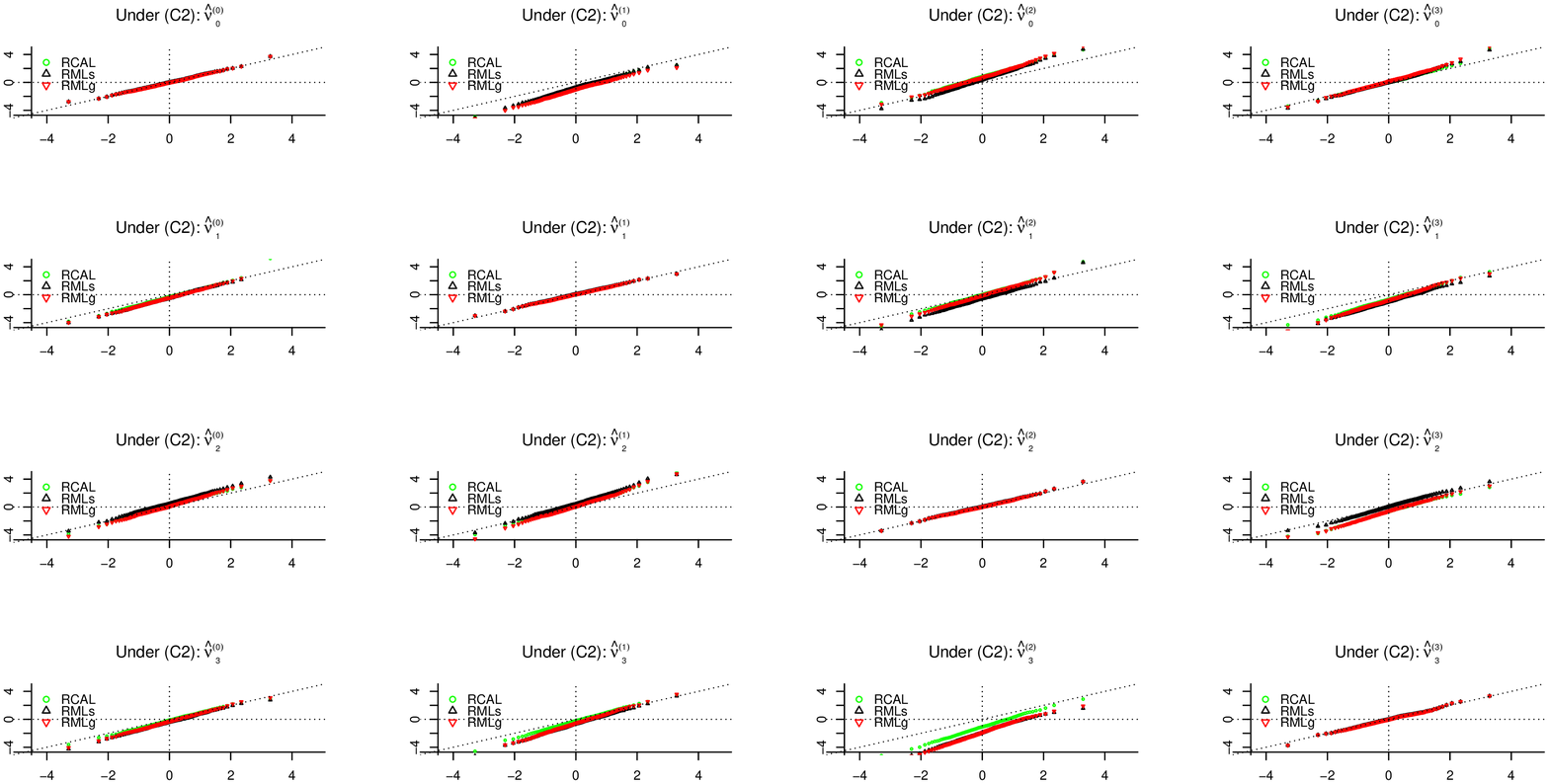}\vspace{-.1in}
\caption{QQ plots of the $t$-statistics against standard normal based on $\hat{\nu}^{(k)}_t$ with $n = 1000, p = 300$ and one-to-zero constraint under (C2).}
\label{fig:qq_nu_p300_c2_cons}
\end{figure}

\begin{figure}[H]
\centering
\includegraphics[scale=0.47]{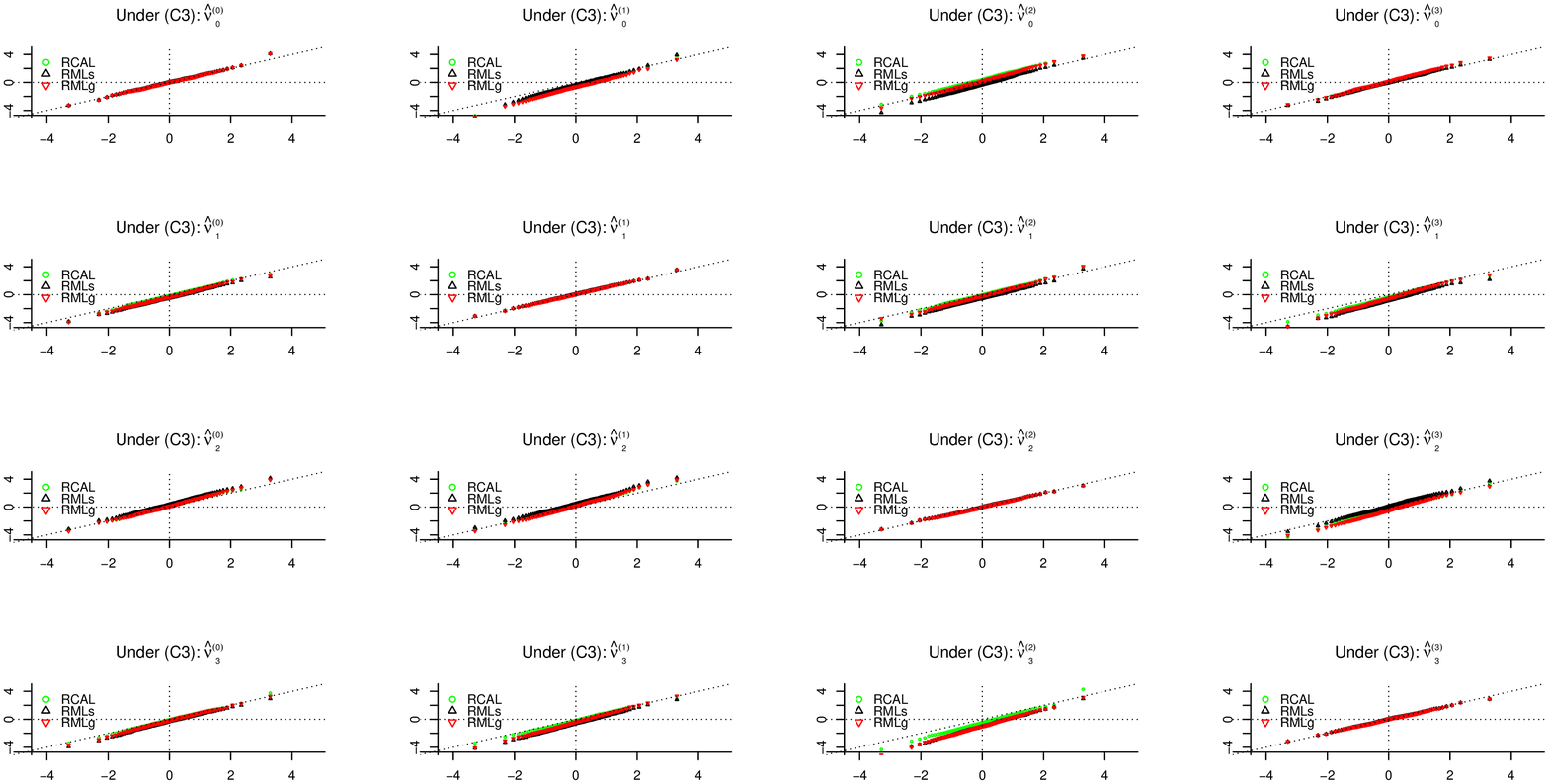}\vspace{-.1in}
\caption{QQ plots of the $t$-statistics against standard normal based on $\hat{\nu}^{(k)}_t$ with $n = 1000, p = 300$ and one-to-zero constraint under (C3).}
\label{fig:qq_nu_p300_c3_cons}
\end{figure}

\begin{figure}[H]
\centering
\includegraphics[scale=0.47]{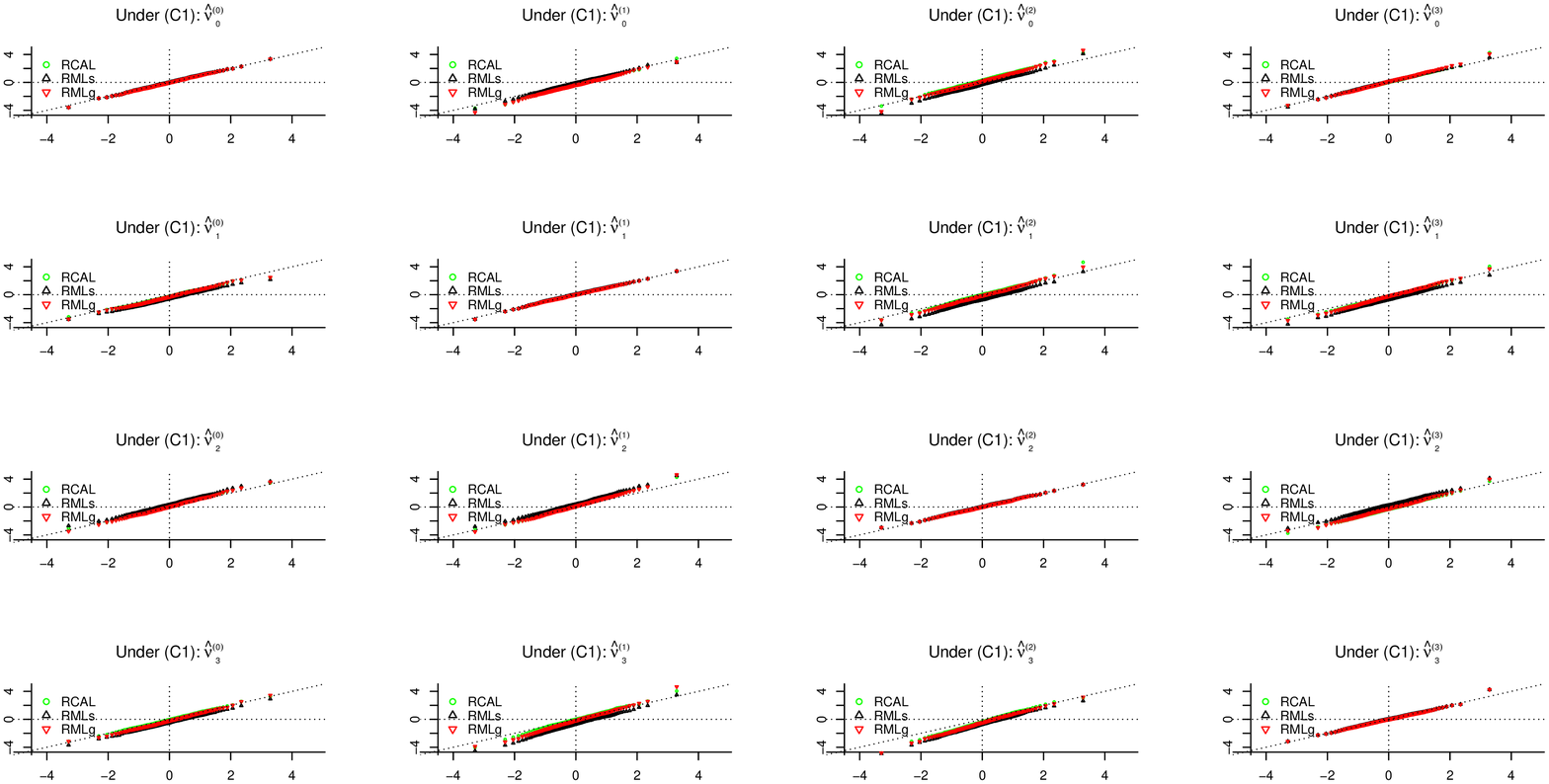}\vspace{-.1in}
\caption{QQ plots of the $t$-statistics against standard normal based on $\hat{\nu}^{(k)}_t$ with $n = 1000, p = 50$ and sum-to-zero constraint under (C1).}
\label{fig:qq_nu_p50_c1}
\end{figure}

\begin{figure}[H]
\centering
\includegraphics[scale=0.47]{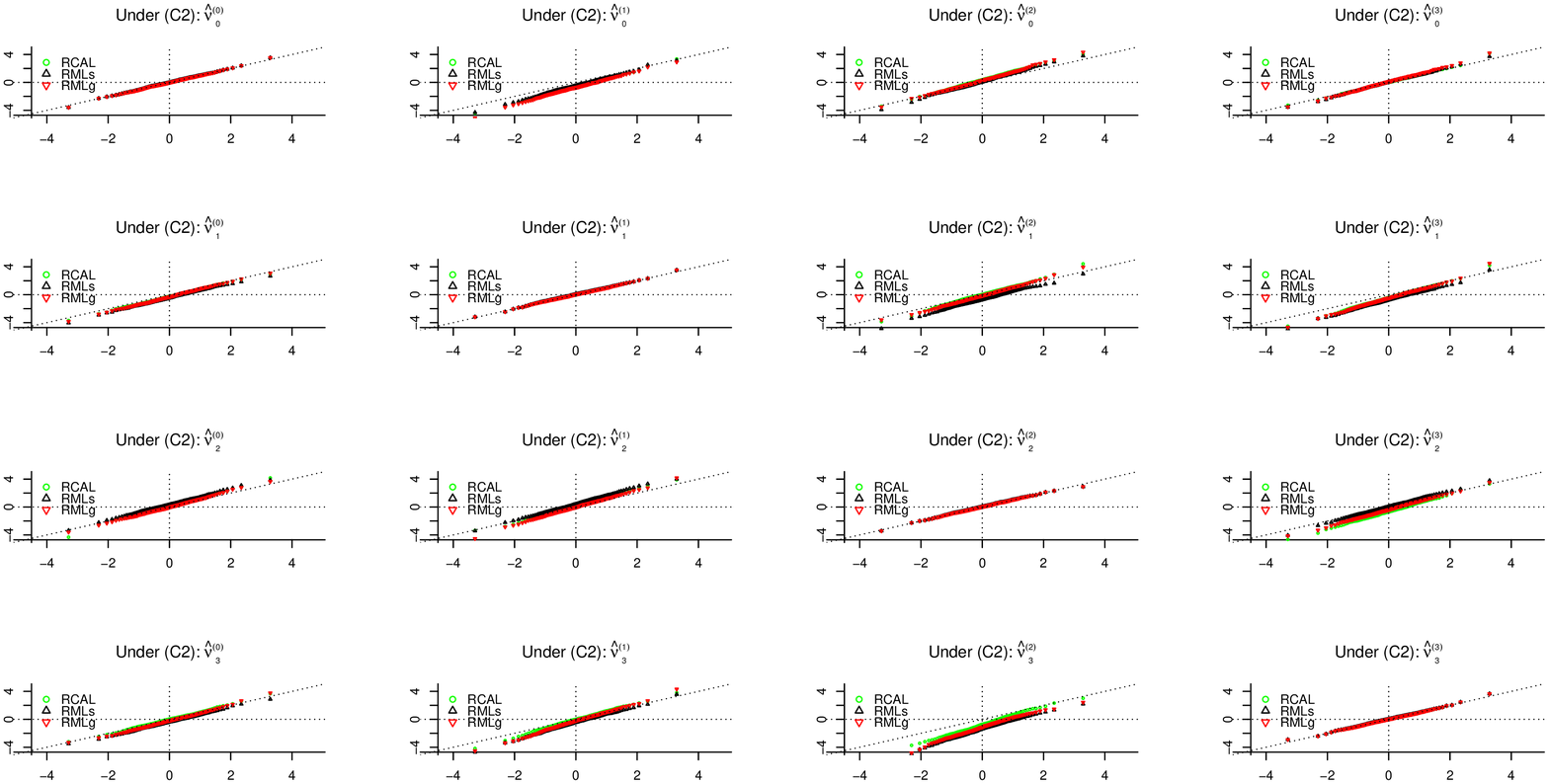}\vspace{-.1in}
\caption{QQ plots of the $t$-statistics against standard normal based on $\hat{\nu}^{(k)}_t$ with $n = 1000, p = 50$ and sum-to-zero constraint under (C2).}
\label{fig:qq_nu_p50_c2}
\end{figure}

\begin{figure}[H]
\centering
\includegraphics[scale=0.47]{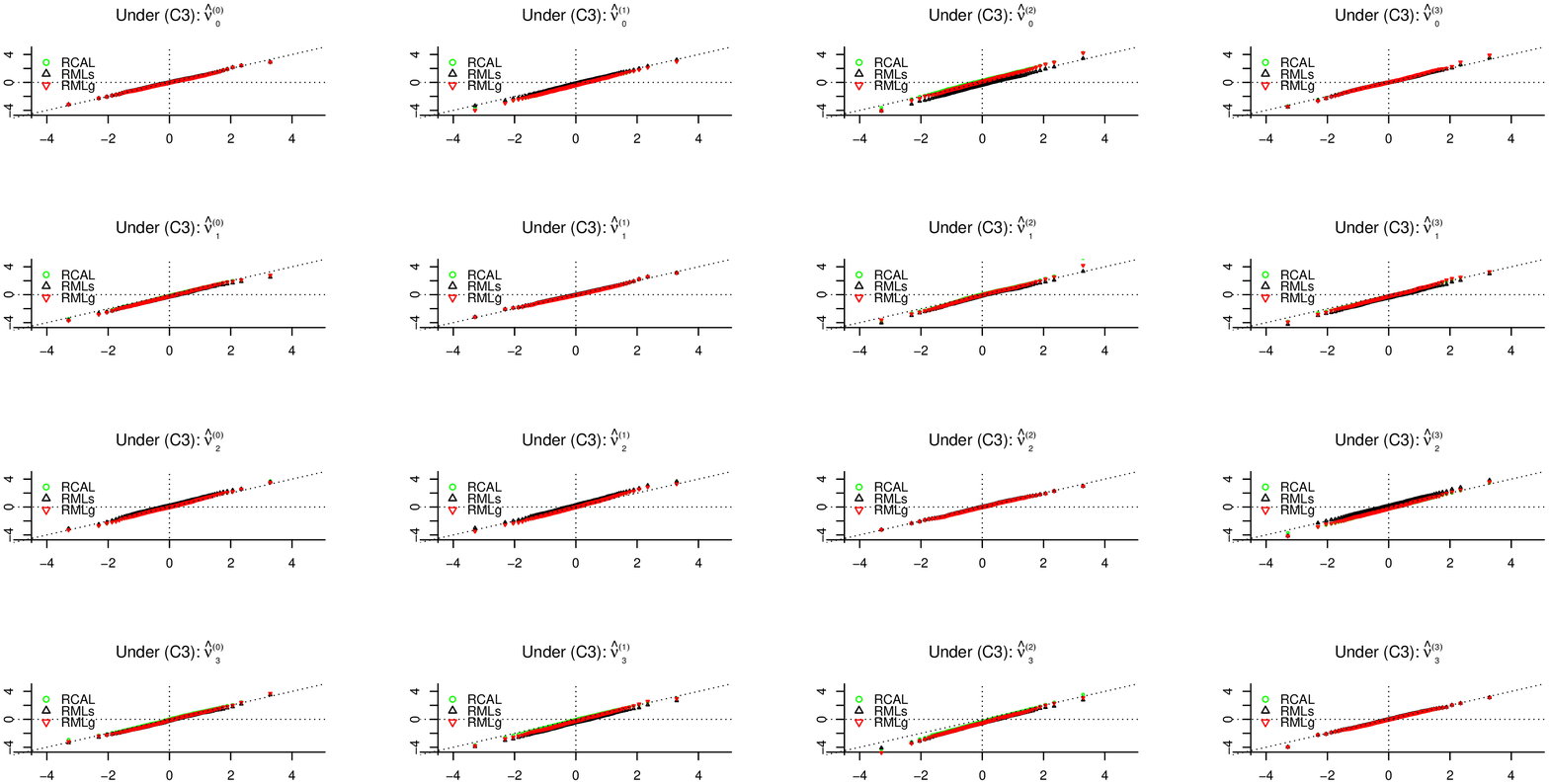}\vspace{-.1in}
\caption{QQ plots of the $t$-statistics against standard normal based on $\hat{\nu}^{(k)}_t$ with $n = 1000, p = 50$ and sum-to-zero constraint under (C3).}
\label{fig:qq_nu_p50_c3}
\end{figure}

\begin{figure}[H]
\centering
\includegraphics[scale=0.47]{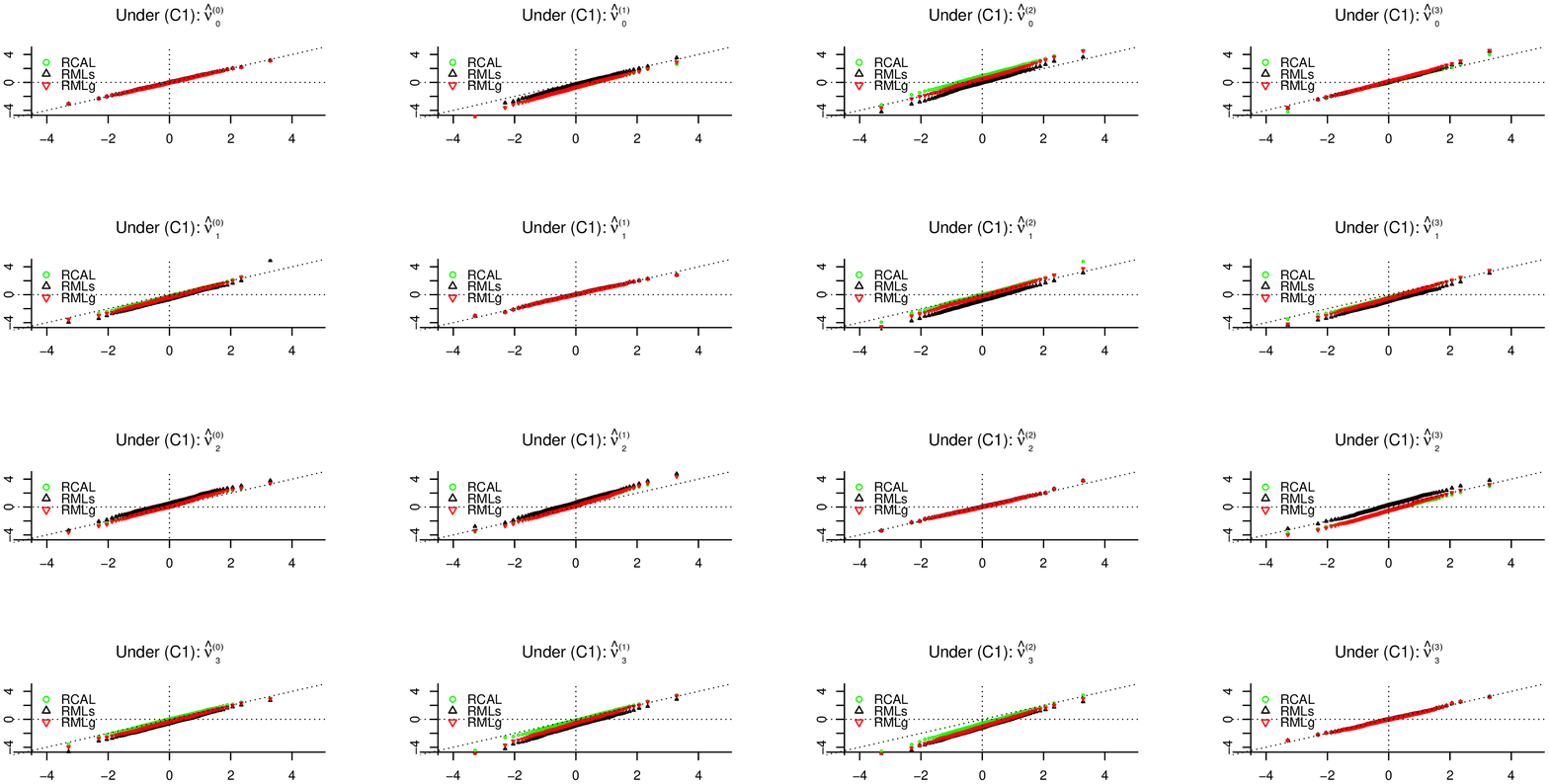}\vspace{-.1in}
\caption{QQ plots of the $t$-statistics against standard normal based on $\hat{\nu}^{(k)}_t$ with $n = 1000, p = 300$ and sum-to-zero constraint under (C1).}
\label{fig:qq_nu_p300_c1}
\end{figure}

\begin{figure}[H]
\centering
\includegraphics[scale=0.47]{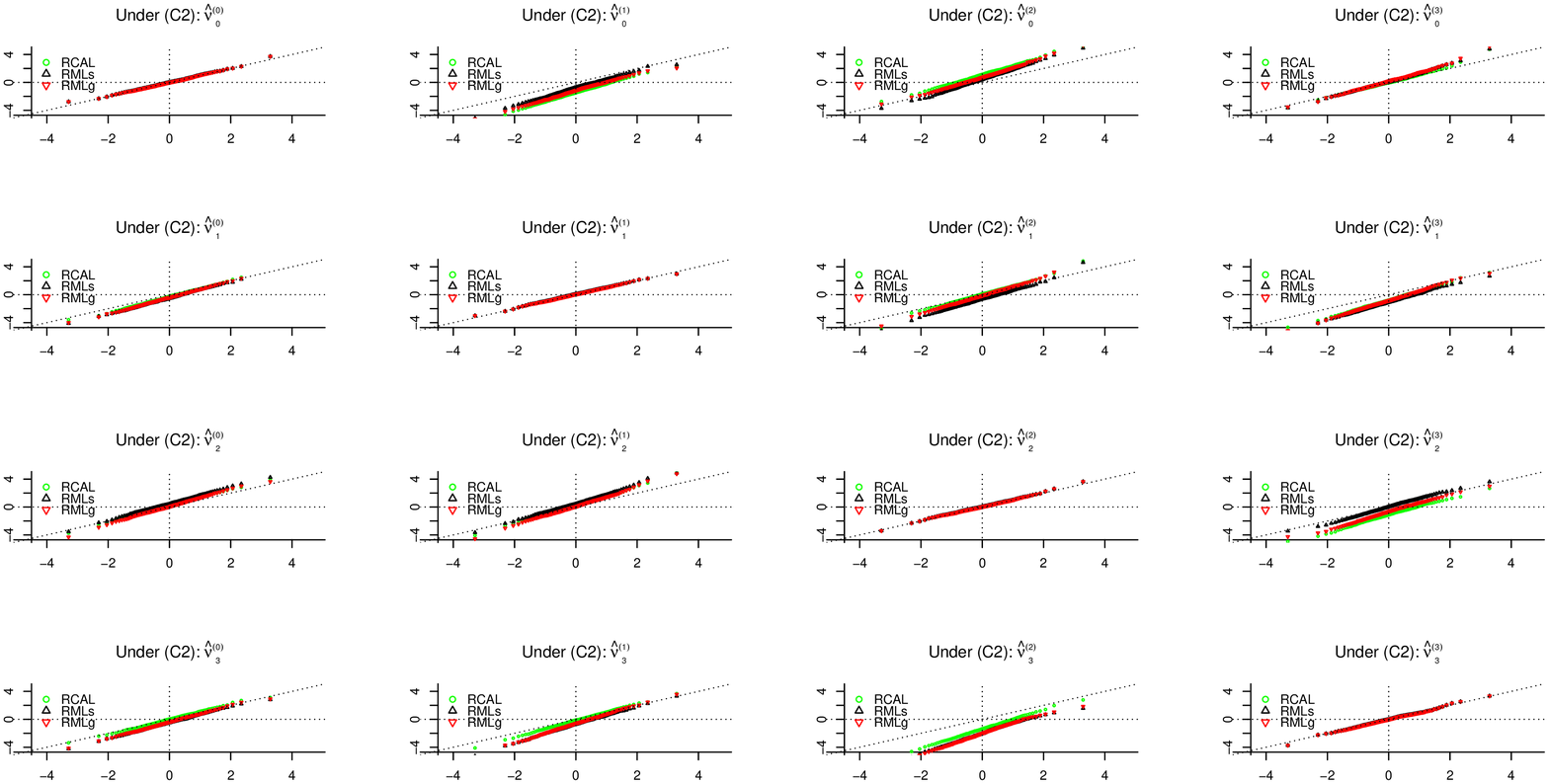}\vspace{-.1in}
\caption{QQ plots of the $t$-statistics against standard normal based on $\hat{\nu}^{(k)}_t$ with $n = 1000, p = 300$ and sum-to-zero constraint under (C2).}
\label{fig:qq_nu_p300_c2}
\end{figure}

\begin{figure}[H]
\centering
\includegraphics[scale=0.47]{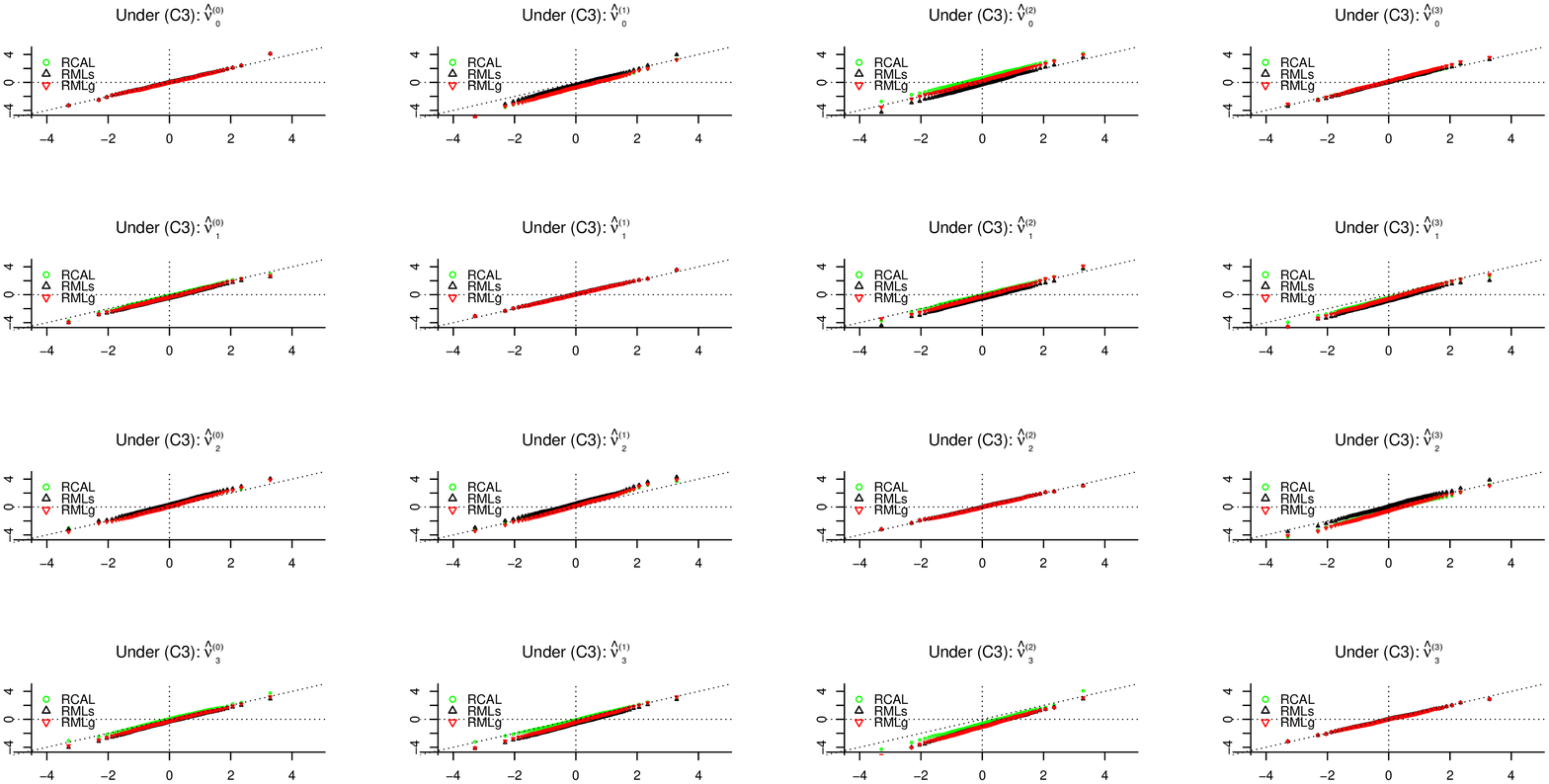}\vspace{-.1in}
\caption{QQ plots of the $t$-statistics against standard normal based on $\hat{\nu}^{(k)}_t$ with $n = 1000, p = 300$ and sum-to-zero constraint under (C3).}
\label{fig:qq_nu_p300_c3}
\end{figure}

Tables \ref{tb:nu_c1_cons}--\ref{tb:nu_c3_cons} and Tables \ref{tb:nu_c1}--\ref{tb:nu_c3} are summaries of $\hat{\nu}^{(k)}_t$ with one-to-zero constraint and sum-to-zero constraint, respectively. Figures \ref{fig:qq_nu_p50_c1_cons}--\ref{fig:qq_nu_p300_c3_cons} and Figures \ref{fig:qq_nu_p50_c1}--\ref{fig:qq_nu_p300_c3}  are QQ plots of the $t$-statistics against standard normal based on $\hat{\nu}^{(k)}_t$ with one-to-zero constraint and sum-to-zero constraint, respectively. The performances of RCAL, RMLs and RMLg on $\nu_t^{(k)}$ are similar to those on $\mu_t^{(k)}$ in general. RCAL has the smallest absolute biases and best coverage proportions and RMLs has the largest absolute biases and worst coverage proportions in most cases. There are some exceptions where RMLs has the smallest absolute biases and best coverage proportions. For example, in Table \ref{tb:nu_c1_cons}, bias of $\hat{\nu}_0^{(3)} $of RMLs, RCAL and RMLg are -0.017, 0.072 and 0.054, and the corresponding 90\% coverage proportions are 0.803, 0.774 and 0.778.

\begin{table}[H]
\caption{\footnotesize Summary of $\hat{\mu}_t$ with sum-to-zero constraint for $t = 0, 1, 2, 3$.} \label{tb:mu}\vspace{-4ex}
\begin{center}
\resizebox{1.0\textwidth}{!}{\begin{tabular}{lccccccccccccccccccccccc}
\hline
& \multicolumn{3}{c}{(C1) cor PS, cor OR} &~~& \multicolumn{3}{c}{(C2) cor PS, mis OR} &~~& \multicolumn{3}{c}{\small (C3) mis PS, cor OR} &~~& \multicolumn{3}{c}{(C1) cor PS, cor OR} &~~& \multicolumn{3}{c}{(C2) cor PS, mis OR} &~~& \multicolumn{3}{c}{\small (C3) mis PS, cor OR}\\
& RCAL & RMLs & RMLg &~~& RCAL & RMLs & RMLg &~~& RCAL & RMLs & RMLg &~~& RCAL & RMLs & RMLg &~~& RCAL & RMLs & RMLg &~~& RCAL & RMLs & RMLg \\
\hline
& \multicolumn{23}{c}{n = 1000, p = 50} \\
& \multicolumn{11}{c}{\footnotesize $\hat{\mu}_0$} &~~& \multicolumn{11}{c}{\footnotesize $\hat{\mu}_1$} \\
\cline{2-12}\cline{14-24}
Bias & -0.002 & -0.016 & -0.004 &~~& -0.010 & -0.015 & -0.013 &~~& -0.004 & -0.022 & -0.007 &~~& -0.014 & -0.070 & -0.020 &~~& -0.018 & -0.061 & -0.025 &~~& -0.012 & -0.036 & -0.016 \\
$\sqrt{\text{Var}}$ & 0.094 & 0.098 & 0.095 &~~& 0.100 & 0.108 & 0.103 &~~& 0.094 & 0.096 & 0.094 &~~& 0.096 & 0.101 & 0.098 &~~& 0.102 & 0.106 & 0.103 &~~& 0.096 & 0.101 & 0.099 \\
$\sqrt{\text{EVar}}$ & 0.090 & 0.091 & 0.089 &~~& 0.093 & 0.099 & 0.094 &~~& 0.089 & 0.089 & 0.086 &~~& 0.089 & 0.090 & 0.086 &~~& 0.091 & 0.094 & 0.088 &~~& 0.088 & 0.093 & 0.088 \\
Cov90 & 0.883 & 0.879 & 0.871 &~~& 0.869 & 0.872 & 0.865 &~~& 0.868 & 0.867 & 0.852 &~~& 0.871 & 0.776 & 0.859 &~~& 0.855 & 0.787 & 0.836 &~~& 0.861 & 0.842 & 0.841 \\
Cov95 & 0.941 & 0.931 & 0.936 &~~& 0.930 & 0.925 & 0.926 &~~& 0.929 & 0.922 & 0.925 &~~& 0.925 & 0.859 & 0.914 &~~& 0.905 & 0.873 & 0.898 &~~& 0.916 & 0.908 & 0.900 \\
& \multicolumn{11}{c}{\footnotesize $\hat{\mu}_2$} &~~& \multicolumn{11}{c}{\footnotesize $\hat{\mu}_3$} \\
\cline{2-12}\cline{14-24}
Bias & 0.001 & 0.033 & -0.001 &~~& -0.013 & 0.025 & -0.011 &~~& -0.003 & 0.021 & -0.006 &~~& -0.014 & -0.069 & -0.029 &~~& -0.024 & -0.078 & -0.048 &~~& -0.015 & -0.047 & -0.028 \\
$\sqrt{\text{Var}}$ & 0.101 & 0.105 & 0.103 &~~& 0.108 & 0.114 & 0.109 &~~& 0.099 & 0.104 & 0.102 &~~& 0.101 & 0.104 & 0.102 &~~& 0.107 & 0.110 & 0.108 &~~& 0.096 & 0.099 & 0.098 \\
$\sqrt{\text{EVar}}$ & 0.089 & 0.091 & 0.088 &~~& 0.091 & 0.096 & 0.091 &~~& 0.088 & 0.092 & 0.089 &~~& 0.089 & 0.089 & 0.085 &~~& 0.091 & 0.094 & 0.087 &~~& 0.088 & 0.088 & 0.085 \\
Cov90 & 0.847 & 0.824 & 0.836 &~~& 0.842 & 0.820 & 0.831 &~~& 0.847 & 0.848 & 0.837 &~~& 0.851 & 0.752 & 0.818 &~~& 0.829 & 0.733 & 0.779 &~~& 0.870 & 0.809 & 0.839 \\
Cov95 & 0.907 & 0.895 & 0.898 &~~& 0.908 & 0.888 & 0.898 &~~& 0.916 & 0.903 & 0.917 &~~& 0.919 & 0.828 & 0.881 &~~& 0.901 & 0.824 & 0.850 &~~& 0.916 & 0.892 & 0.898 \\
& \multicolumn{23}{c}{n = 1000, p = 300} \\
& \multicolumn{11}{c}{\footnotesize $\hat{\mu}_0$} &~~& \multicolumn{11}{c}{\footnotesize $\hat{\mu}_1$} \\
\cline{2-12}\cline{14-24}
Bias & 0.005 & -0.012 & -0.001 &~~& -0.010 & -0.010 & -0.008 &~~& -0.002 & -0.021 & -0.008 &~~& -0.014 & -0.075 & -0.027 &~~& -0.023 & -0.060 & -0.032 &~~& -0.013 & -0.058 & -0.027 \\
$\sqrt{\text{Var}}$ & 0.096 & 0.102 & 0.099 &~~& 0.101 & 0.111 & 0.107 &~~& 0.091 & 0.097 & 0.092 &~~& 0.094 & 0.104 & 0.098 &~~& 0.100 & 0.106 & 0.104 &~~& 0.092 & 0.098 & 0.096 \\
$\sqrt{\text{EVar}}$ & 0.087 & 0.086 & 0.087 &~~& 0.090 & 0.090 & 0.092 &~~& 0.086 & 0.085 & 0.085 &~~& 0.083 & 0.084 & 0.081 &~~& 0.084 & 0.084 & 0.082 &~~& 0.084 & 0.085 & 0.083 \\
Cov90 & 0.871 & 0.843 & 0.857 &~~& 0.854 & 0.827 & 0.849 &~~& 0.878 & 0.841 & 0.870 &~~& 0.839 & 0.706 & 0.809 &~~& 0.831 & 0.729 & 0.789 &~~& 0.863 & 0.778 & 0.820 \\
Cov95 & 0.932 & 0.906 & 0.920 &~~& 0.913 & 0.883 & 0.908 &~~& 0.938 & 0.906 & 0.928 &~~& 0.915 & 0.796 & 0.882 &~~& 0.886 & 0.812 & 0.865 &~~& 0.925 & 0.865 & 0.897 \\
& \multicolumn{11}{c}{\footnotesize $\hat{\mu}_2$} &~~& \multicolumn{11}{c}{\footnotesize $\hat{\mu}_3$} \\
\cline{2-12}\cline{14-24}
Bias & 0.001 & 0.043 & -0.004 &~~& -0.017 & 0.027 & -0.013 &~~& -0.002 & 0.028 & -0.008 &~~& -0.023 & -0.087 & -0.056 &~~& -0.045 & -0.096 & -0.085 &~~& -0.021 & -0.067 & -0.051 \\
$\sqrt{\text{Var}}$ & 0.095 & 0.103 & 0.102 &~~& 0.099 & 0.106 & 0.106 &~~& 0.095 & 0.101 & 0.100 &~~& 0.098 & 0.107 & 0.102 &~~& 0.102 & 0.108 & 0.107 &~~& 0.089 & 0.098 & 0.094 \\
$\sqrt{\text{EVar}}$ & 0.083 & 0.084 & 0.082 &~~& 0.084 & 0.085 & 0.083 &~~& 0.084 & 0.085 & 0.083 &~~& 0.085 & 0.082 & 0.081 &~~& 0.085 & 0.083 & 0.082 &~~& 0.084 & 0.083 & 0.081 \\
Cov90 & 0.843 & 0.785 & 0.818 &~~& 0.826 & 0.793 & 0.789 &~~& 0.852 & 0.818 & 0.829 &~~& 0.824 & 0.665 & 0.751 &~~& 0.786 & 0.640 & 0.661 &~~& 0.870 & 0.740 & 0.779 \\
Cov95 & 0.914 & 0.853 & 0.889 &~~& 0.891 & 0.862 & 0.870 &~~& 0.916 & 0.892 & 0.901 &~~& 0.897 & 0.752 & 0.823 &~~& 0.856 & 0.725 & 0.740 &~~& 0.931 & 0.819 & 0.856 \\
\hline
\end{tabular}}
\end{center}
\setlength{\baselineskip}{0.5\baselineskip}
\vspace{-.15in}
\noindent{\tiny
\textbf{Note}: RCAL denotes $\hat{\mu}_t(\hat{m}^\#_{\text{RWL}}, \hat{\pi}_{\text{RCAL}})$, RMLs denotes $\hat{\mu}_t(\hat{m}_{\text{RMLs}}, \hat{\pi}_{\text{RML}})$ and RMLg denotes $\hat{\mu}_t(\hat{m}_{\text{RMLg}}, \hat{\pi}_{\text{RML}})$. Bias and Var are the Monte Carlo bias and variance of the point estimates. EVar is the mean of the variance estimates, and hence $\sqrt{\text{EVar}}$ also measures the $L_2$-average of lengths of confidence intervals. Cov90 or Cov95 is the coverage proportion of the 90\% or 95\% confidence intervals.}
\end{table}

\begin{table}[H]
\caption{\footnotesize Under (C1), summary of $\hat{\nu}^{(k)}_t$ with one-to-zero constraint for $t, k = 0, 1, 2, 3$.} \label{tb:nu_c1_cons}\vspace{-4ex}
\begin{center}
\resizebox{\textwidth}{!}{\begin{tabular}{lccccccccccccccccccccccccccccccc}
\hline
& \multicolumn{3}{c}{$\hat{\nu}^{(0)}_t$} & $~~$ & \multicolumn{3}{c}{$\hat{\nu}^{(1)}_t$} & $~~$ & \multicolumn{3}{c}{$\hat{\nu}^{(2)}_t$} & $~~$ & \multicolumn{3}{c}{$\hat{\nu}^{(3)}_t$} &~~& \multicolumn{3}{c}{$\hat{\nu}^{(0)}_t$} & $~~$ & \multicolumn{3}{c}{$\hat{\nu}^{(1)}_t$} & $~~$ & \multicolumn{3}{c}{$\hat{\nu}^{(2)}_t$} & $~~$ & \multicolumn{3}{c}{$\hat{\nu}^{(3)}_t$} \\
& RCAL & RMLs & RMLg &~~& RCAL & RMLs & RMLg &~~& RCAL & RMLs & RMLg &~~& RCAL & RMLs & RMLg &~~& RCAL & RMLs & RMLg &~~& RCAL & RMLs & RMLg &~~& RCAL & RMLs & RMLg &~~& RCAL & RMLs & RMLg\\
\hline
& \multicolumn{31}{c}{p = 50} \\
& \multicolumn{15}{c}{\footnotesize t = 0} &~~& \multicolumn{15}{c}{\footnotesize t = 1} \\
\cline{2-16}\cline{18-32}
Bias & -0.002 & -0.002 & -0.002 &~~& -0.033 & -0.024 & -0.043 &~~& 0.029 & -0.048 & 0.019 &~~& 0.003 & -0.002 & 0.011 &~~& -0.032 & -0.085 & -0.038 &~~& 0.001 & 0.001 & 0.001 &~~& -0.007 & -0.094 & -0.018 &~~& -0.024 & -0.093 & -0.029 \\
$\sqrt{\text{Var}}$ & 0.144 & 0.144 & 0.144 &~~& 0.130 & 0.137 & 0.133 &~~& 0.141 & 0.156 & 0.145 &~~& 0.130 & 0.140 & 0.138 &~~& 0.133 & 0.142 & 0.138 &~~& 0.127 & 0.127 & 0.127 &~~& 0.149 & 0.167 & 0.161 &~~& 0.144 & 0.154 & 0.152 \\
$\sqrt{\text{EVar}}$ & 0.142 & 0.142 & 0.142 &~~& 0.122 & 0.132 & 0.124 &~~& 0.125 & 0.136 & 0.126 &~~& 0.124 & 0.134 & 0.125 &~~& 0.133 & 0.144 & 0.135 &~~& 0.128 & 0.128 & 0.128 &~~& 0.131 & 0.152 & 0.140 &~~& 0.129 & 0.143 & 0.133 \\
Cov90 & 0.891 & 0.891 & 0.891 &~~& 0.856 & 0.878 & 0.851 &~~& 0.853 & 0.842 & 0.851 &~~& 0.892 & 0.879 & 0.864 &~~& 0.878 & 0.830 & 0.865 &~~& 0.903 & 0.903 & 0.903 &~~& 0.852 & 0.796 & 0.839 &~~& 0.854 & 0.808 & 0.829 \\
Cov95 & 0.950 & 0.950 & 0.950 &~~& 0.918 & 0.930 & 0.909 &~~& 0.911 & 0.903 & 0.913 &~~& 0.939 & 0.940 & 0.924 &~~& 0.945 & 0.913 & 0.938 &~~& 0.948 & 0.948 & 0.948 &~~& 0.912 & 0.866 & 0.898 &~~& 0.914 & 0.877 & 0.905 \\
& \multicolumn{15}{c}{\footnotesize t = 2} &~~& \multicolumn{15}{c}{\footnotesize t = 3} \\
\cline{2-16}\cline{18-32}
Bias & 0.011 & 0.051 & 0.011 &~~& 0.019 & 0.055 & 0.015 &~~& 0.001 & 0.001 & 0.001 &~~& -0.027 & 0.031 & -0.026 &~~& -0.016 & -0.067 & -0.021 &~~& -0.009 & -0.092 & -0.023 &~~& -0.038 & -0.095 & -0.056 &~~& -0.003 & -0.003 & -0.003 \\
$\sqrt{\text{Var}}$ & 0.150 & 0.158 & 0.155 &~~& 0.157 & 0.170 & 0.166 &~~& 0.125 & 0.125 & 0.125 &~~& 0.143 & 0.153 & 0.151 &~~& 0.141 & 0.147 & 0.145 &~~& 0.151 & 0.165 & 0.161 &~~& 0.159 & 0.168 & 0.162 &~~& 0.128 & 0.128 & 0.128 \\
$\sqrt{\text{EVar}}$ & 0.136 & 0.145 & 0.138 &~~& 0.132 & 0.149 & 0.141 &~~& 0.127 & 0.127 & 0.127 &~~& 0.128 & 0.141 & 0.134 &~~& 0.134 & 0.143 & 0.135 &~~& 0.129 & 0.141 & 0.131 &~~& 0.129 & 0.141 & 0.132 &~~& 0.131 & 0.131 & 0.131 \\
Cov90 & 0.866 & 0.850 & 0.865 &~~& 0.822 & 0.830 & 0.828 &~~& 0.901 & 0.901 & 0.901 &~~& 0.855 & 0.850 & 0.838 &~~& 0.892 & 0.848 & 0.876 &~~& 0.841 & 0.779 & 0.808 &~~& 0.808 & 0.750 & 0.788 &~~& 0.902 & 0.902 & 0.902 \\
Cov95 & 0.922 & 0.921 & 0.915 &~~& 0.891 & 0.892 & 0.904 &~~& 0.949 & 0.949 & 0.949 &~~& 0.917 & 0.917 & 0.912 &~~& 0.940 & 0.919 & 0.931 &~~& 0.907 & 0.845 & 0.876 &~~& 0.877 & 0.846 & 0.862 &~~& 0.951 & 0.951 & 0.951 \\

\cline{2-32}
& \multicolumn{31}{c}{p = 300} \\
& \multicolumn{15}{c}{\footnotesize t = 0} &~~& \multicolumn{15}{c}{\footnotesize t = 1} \\
\cline{2-16}\cline{18-32}
Bias & -0.003 & -0.003 & -0.003 &~~& -0.052 & -0.036 & -0.071 &~~& 0.072 & -0.017 & 0.054 &~~& 0.002 & 0.000 & 0.019 &~~& -0.026 & -0.084 & -0.046 &~~& 0.002 & 0.002 & 0.002 &~~& -0.002 & -0.103 & -0.020 &~~& -0.038 & -0.116 & -0.054 \\
$\sqrt{\text{Var}}$ & 0.140 & 0.140 & 0.140 &~~& 0.127 & 0.133 & 0.130 &~~& 0.134 & 0.152 & 0.143 &~~& 0.124 & 0.133 & 0.133 &~~& 0.137 & 0.148 & 0.142 &~~& 0.126 & 0.126 & 0.126 &~~& 0.138 & 0.162 & 0.153 &~~& 0.136 & 0.146 & 0.146 \\
$\sqrt{\text{EVar}}$ & 0.141 & 0.141 & 0.141 &~~& 0.113 & 0.120 & 0.115 &~~& 0.115 & 0.120 & 0.116 &~~& 0.115 & 0.121 & 0.115 &~~& 0.126 & 0.131 & 0.126 &~~& 0.127 & 0.127 & 0.127 &~~& 0.119 & 0.133 & 0.125 &~~& 0.119 & 0.126 & 0.120 \\
Cov90 & 0.905 & 0.905 & 0.905 &~~& 0.821 & 0.842 & 0.795 &~~& 0.774 & 0.803 & 0.778 &~~& 0.882 & 0.866 & 0.849 &~~& 0.866 & 0.796 & 0.841 &~~& 0.905 & 0.905 & 0.905 &~~& 0.840 & 0.733 & 0.828 &~~& 0.827 & 0.698 & 0.799 \\
Cov95 & 0.958 & 0.958 & 0.958 &~~& 0.894 & 0.906 & 0.862 &~~& 0.865 & 0.871 & 0.861 &~~& 0.933 & 0.923 & 0.909 &~~& 0.929 & 0.868 & 0.901 &~~& 0.950 & 0.950 & 0.950 &~~& 0.896 & 0.813 & 0.883 &~~& 0.904 & 0.812 & 0.864 \\
& \multicolumn{15}{c}{\footnotesize t = 2} &~~& \multicolumn{15}{c}{\footnotesize t = 3} \\
\cline{2-16}\cline{18-32}
Bias & 0.019 & 0.074 & 0.021 &~~& 0.032 & 0.077 & 0.025 &~~& 0.003 & 0.003 & 0.003 &~~& -0.045 & 0.029 & -0.055 &~~& -0.015 & -0.073 & -0.031 &~~& -0.024 & -0.114 & -0.057 &~~& -0.056 & -0.140 & -0.110 &~~& -0.003 & -0.003 & -0.003 \\
$\sqrt{\text{Var}}$ & 0.141 & 0.149 & 0.149 &~~& 0.143 & 0.158 & 0.159 &~~& 0.126 & 0.126 & 0.126 &~~& 0.136 & 0.147 & 0.148 &~~& 0.136 & 0.147 & 0.138 &~~& 0.136 & 0.150 & 0.146 &~~& 0.145 & 0.155 & 0.151 &~~& 0.131 & 0.131 & 0.131 \\
$\sqrt{\text{EVar}}$ & 0.127 & 0.133 & 0.128 &~~& 0.118 & 0.130 & 0.124 &~~& 0.127 & 0.127 & 0.127 &~~& 0.118 & 0.125 & 0.120 &~~& 0.126 & 0.130 & 0.124 &~~& 0.119 & 0.122 & 0.116 &~~& 0.118 & 0.124 & 0.118 &~~& 0.130 & 0.130 & 0.130 \\
Cov90 & 0.857 & 0.800 & 0.824 &~~& 0.826 & 0.767 & 0.802 &~~& 0.909 & 0.909 & 0.909 &~~& 0.817 & 0.814 & 0.783 &~~& 0.860 & 0.803 & 0.835 &~~& 0.843 & 0.697 & 0.781 &~~& 0.793 & 0.624 & 0.670 &~~& 0.893 & 0.893 & 0.893 \\
Cov95 & 0.915 & 0.860 & 0.896 &~~& 0.882 & 0.847 & 0.864 &~~& 0.954 & 0.954 & 0.954 &~~& 0.883 & 0.902 & 0.856 &~~& 0.930 & 0.876 & 0.914 &~~& 0.905 & 0.792 & 0.851 &~~& 0.865 & 0.722 & 0.772 &~~& 0.952 & 0.952 & 0.952 \\
\hline
\end{tabular}}
\end{center}
\setlength{\baselineskip}{0.5\baselineskip}
\vspace{-.15in}\noindent{\tiny
\textbf{Note}: RCAL denotes $\hat{\nu}^{(k)}_{t,\text{RCAL}}$, RMLs denotes $\hat{\nu}^{(k)}_{t,\text{RMLs}}$ and RMLg denotes $\hat{\nu}^{(k)}_{t,\text{RMLg}}$. Bias and Var are the Monte Carlo bias and variance of the point estimates. EVar is the mean of the variance estimates, and hence $\sqrt{\text{EVar}}$ also measures the $L_2$-average of lengths of confidence intervals. Cov90 or Cov95 is the coverage proportion of the 90\% or 95\% confidence intervals.}
\end{table}

\begin{table}[H]
\caption{\footnotesize Under (C2), summary of $\hat{\nu}^{(k)}_t$ with one-to-zero constraint for $t, k = 0, 1, 2, 3$.} \label{tb:nu_c2_cons}\vspace{-4ex}
\begin{center}
\resizebox{\textwidth}{!}{\begin{tabular}{lccccccccccccccccccccccccccccccc}
\hline
& \multicolumn{3}{c}{$\hat{\nu}^{(0)}_t$} & $~~$ & \multicolumn{3}{c}{$\hat{\nu}^{(1)}_t$} & $~~$ & \multicolumn{3}{c}{$\hat{\nu}^{(2)}_t$} & $~~$ & \multicolumn{3}{c}{$\hat{\nu}^{(3)}_t$} &~~& \multicolumn{3}{c}{$\hat{\nu}^{(0)}_t$} & $~~$ & \multicolumn{3}{c}{$\hat{\nu}^{(1)}_t$} & $~~$ & \multicolumn{3}{c}{$\hat{\nu}^{(2)}_t$} & $~~$ & \multicolumn{3}{c}{$\hat{\nu}^{(3)}_t$} \\
& RCAL & RMLs & RMLg &~~& RCAL & RMLs & RMLg &~~& RCAL & RMLs & RMLg &~~& RCAL & RMLs & RMLg &~~& RCAL & RMLs & RMLg &~~& RCAL & RMLs & RMLg &~~& RCAL & RMLs & RMLg &~~& RCAL & RMLs & RMLg\\
\hline
& \multicolumn{31}{c}{p = 50} \\
& \multicolumn{15}{c}{\footnotesize t = 0} &~~& \multicolumn{15}{c}{\footnotesize t = 1} \\
\cline{2-16}\cline{18-32}
Bias & -0.003 & -0.003 & -0.003 &~~& -0.055 & -0.063 & -0.078 &~~& 0.035 & 0.001 & 0.032 &~~& 0.001 & -0.010 & 0.009 &~~& -0.034 & -0.055 & -0.037 &~~& 0.002 & 0.002 & 0.002 &~~& -0.005 & -0.085 & -0.015 &~~& -0.044 & -0.092 & -0.054 \\
$\sqrt{\text{Var}}$ & 0.127 & 0.127 & 0.127 &~~& 0.134 & 0.145 & 0.141 &~~& 0.144 & 0.158 & 0.149 &~~& 0.127 & 0.137 & 0.135 &~~& 0.124 & 0.129 & 0.128 &~~& 0.112 & 0.112 & 0.112 &~~& 0.147 & 0.163 & 0.161 &~~& 0.147 & 0.151 & 0.154 \\
$\sqrt{\text{EVar}}$ & 0.125 & 0.125 & 0.125 &~~& 0.116 & 0.129 & 0.124 &~~& 0.120 & 0.130 & 0.124 &~~& 0.115 & 0.123 & 0.117 &~~& 0.116 & 0.123 & 0.118 &~~& 0.114 & 0.114 & 0.114 &~~& 0.124 & 0.147 & 0.136 &~~& 0.121 & 0.133 & 0.125 \\
Cov90 & 0.895 & 0.895 & 0.895 &~~& 0.821 & 0.811 & 0.789 &~~& 0.817 & 0.838 & 0.817 &~~& 0.871 & 0.862 & 0.843 &~~& 0.857 & 0.860 & 0.850 &~~& 0.899 & 0.899 & 0.899 &~~& 0.847 & 0.810 & 0.842 &~~& 0.818 & 0.783 & 0.804 \\
Cov95 & 0.948 & 0.948 & 0.948 &~~& 0.879 & 0.882 & 0.856 &~~& 0.889 & 0.898 & 0.885 &~~& 0.927 & 0.912 & 0.907 &~~& 0.934 & 0.922 & 0.918 &~~& 0.950 & 0.950 & 0.950 &~~& 0.901 & 0.876 & 0.900 &~~& 0.880 & 0.857 & 0.868 \\
& \multicolumn{15}{c}{\footnotesize t = 2} &~~& \multicolumn{15}{c}{\footnotesize t = 3} \\
\cline{2-16}\cline{18-32}
Bias & 0.010 & 0.046 & 0.000 &~~& 0.011 & 0.053 & -0.002 &~~& 0.002 & 0.002 & 0.002 &~~& -0.045 & 0.004 & -0.041 &~~& -0.013 & -0.046 & -0.018 &~~& -0.011 & -0.057 & -0.026 &~~& -0.073 & -0.175 & -0.125 &~~& 0.000 & 0.000 & 0.000 \\
$\sqrt{\text{Var}}$ & 0.139 & 0.148 & 0.144 &~~& 0.157 & 0.173 & 0.166 &~~& 0.116 & 0.116 & 0.116 &~~& 0.149 & 0.158 & 0.157 &~~& 0.126 & 0.132 & 0.130 &~~& 0.146 & 0.158 & 0.158 &~~& 0.164 & 0.179 & 0.174 &~~& 0.115 & 0.115 & 0.115 \\
$\sqrt{\text{EVar}}$ & 0.121 & 0.132 & 0.126 &~~& 0.123 & 0.147 & 0.137 &~~& 0.116 & 0.116 & 0.116 &~~& 0.121 & 0.138 & 0.131 &~~& 0.117 & 0.125 & 0.117 &~~& 0.119 & 0.132 & 0.122 &~~& 0.124 & 0.150 & 0.135 &~~& 0.116 & 0.116 & 0.116 \\
Cov90 & 0.847 & 0.834 & 0.847 &~~& 0.794 & 0.808 & 0.824 &~~& 0.898 & 0.898 & 0.898 &~~& 0.810 & 0.857 & 0.816 &~~& 0.863 & 0.849 & 0.860 &~~& 0.827 & 0.806 & 0.789 &~~& 0.740 & 0.597 & 0.658 &~~& 0.910 & 0.910 & 0.910 \\
Cov95 & 0.912 & 0.901 & 0.909 &~~& 0.872 & 0.876 & 0.891 &~~& 0.953 & 0.953 & 0.953 &~~& 0.880 & 0.918 & 0.885 &~~& 0.921 & 0.910 & 0.915 &~~& 0.892 & 0.857 & 0.857 &~~& 0.825 & 0.686 & 0.746 &~~& 0.947 & 0.947 & 0.947 \\

\cline{2-32}
& \multicolumn{31}{c}{p = 300} \\
& \multicolumn{15}{c}{\footnotesize t = 0} &~~& \multicolumn{15}{c}{\footnotesize t = 1} \\
\cline{2-16}\cline{18-32}
Bias & -0.001 & -0.001 & -0.001 &~~& -0.092 & -0.087 & -0.117 &~~& 0.084 & 0.036 & 0.077 &~~& 0.001 & 0.001 & 0.018 &~~& -0.034 & -0.055 & -0.046 &~~& 0.005 & 0.005 & 0.005 &~~& 0.000 & -0.071 & -0.012 &~~& -0.070 & -0.115 & -0.082 \\
$\sqrt{\text{Var}}$ & 0.125 & 0.125 & 0.125 &~~& 0.131 & 0.140 & 0.139 &~~& 0.137 & 0.159 & 0.149 &~~& 0.118 & 0.130 & 0.130 &~~& 0.122 & 0.132 & 0.126 &~~& 0.112 & 0.112 & 0.112 &~~& 0.134 & 0.155 & 0.152 &~~& 0.136 & 0.142 & 0.147 \\
$\sqrt{\text{EVar}}$ & 0.125 & 0.125 & 0.125 &~~& 0.106 & 0.115 & 0.116 &~~& 0.108 & 0.114 & 0.114 &~~& 0.105 & 0.109 & 0.106 &~~& 0.107 & 0.111 & 0.107 &~~& 0.114 & 0.114 & 0.114 &~~& 0.108 & 0.123 & 0.117 &~~& 0.108 & 0.114 & 0.110 \\
Cov90 & 0.904 & 0.904 & 0.904 &~~& 0.700 & 0.730 & 0.674 &~~& 0.709 & 0.746 & 0.717 &~~& 0.867 & 0.831 & 0.823 &~~& 0.837 & 0.795 & 0.806 &~~& 0.902 & 0.902 & 0.902 &~~& 0.818 & 0.760 & 0.804 &~~& 0.765 & 0.675 & 0.706 \\
Cov95 & 0.949 & 0.949 & 0.949 &~~& 0.795 & 0.810 & 0.766 &~~& 0.801 & 0.818 & 0.801 &~~& 0.920 & 0.896 & 0.898 &~~& 0.899 & 0.876 & 0.873 &~~& 0.943 & 0.943 & 0.943 &~~& 0.890 & 0.840 & 0.874 &~~& 0.830 & 0.765 & 0.798 \\
& \multicolumn{15}{c}{\footnotesize t = 2} &~~& \multicolumn{15}{c}{\footnotesize t = 3} \\
\cline{2-16}\cline{18-32}
Bias & 0.018 & 0.055 & 0.012 &~~& 0.023 & 0.055 & 0.011 &~~& 0.004 & 0.004 & 0.004 &~~& -0.074 & 0.000 & -0.070 &~~& -0.017 & -0.050 & -0.033 &~~& -0.026 & -0.077 & -0.060 &~~& -0.119 & -0.219 & -0.208 &~~& -0.004 & -0.004 & -0.004 \\
$\sqrt{\text{Var}}$ & 0.130 & 0.139 & 0.139 &~~& 0.138 & 0.157 & 0.157 &~~& 0.117 & 0.117 & 0.117 &~~& 0.135 & 0.142 & 0.145 &~~& 0.121 & 0.129 & 0.125 &~~& 0.129 & 0.142 & 0.143 &~~& 0.149 & 0.157 & 0.157 &~~& 0.118 & 0.118 & 0.118 \\
$\sqrt{\text{EVar}}$ & 0.110 & 0.117 & 0.113 &~~& 0.108 & 0.123 & 0.117 &~~& 0.116 & 0.116 & 0.116 &~~& 0.108 & 0.116 & 0.113 &~~& 0.108 & 0.111 & 0.105 &~~& 0.106 & 0.109 & 0.105 &~~& 0.110 & 0.121 & 0.116 &~~& 0.115 & 0.115 & 0.115 \\
Cov90 & 0.829 & 0.789 & 0.806 &~~& 0.811 & 0.776 & 0.793 &~~& 0.906 & 0.906 & 0.906 &~~& 0.764 & 0.813 & 0.743 &~~& 0.859 & 0.806 & 0.825 &~~& 0.809 & 0.737 & 0.729 &~~& 0.641 & 0.433 & 0.446 &~~& 0.893 & 0.893 & 0.893 \\
Cov95 & 0.892 & 0.868 & 0.878 &~~& 0.871 & 0.851 & 0.866 &~~& 0.949 & 0.949 & 0.949 &~~& 0.837 & 0.887 & 0.825 &~~& 0.913 & 0.883 & 0.896 &~~& 0.888 & 0.818 & 0.815 &~~& 0.725 & 0.506 & 0.515 &~~& 0.936 & 0.936 & 0.936 \\
\hline
\end{tabular}}
\end{center}
\setlength{\baselineskip}{0.5\baselineskip}
\vspace{-.15in}\noindent{\tiny
\textbf{Note}: RCAL denotes $\hat{\nu}^{(k)}_{t,\text{RCAL}}$, RMLs denotes $\hat{\nu}^{(k)}_{t,\text{RMLs}}$ and RMLg denotes $\hat{\nu}^{(k)}_{t,\text{RMLg}}$. Bias and Var are the Monte Carlo bias and variance of the point estimates. EVar is the mean of the variance estimates, and hence $\sqrt{\text{EVar}}$ also measures the $L_2$-average of lengths of confidence intervals. Cov90 or Cov95 is the coverage proportion of the 90\% or 95\% confidence intervals.}
\end{table}

\begin{table}[H]
\caption{\footnotesize Under (C3), summary of $\hat{\nu}^{(k)}_t$ with one-to-zero constraint for $t, k = 0, 1, 2, 3$.} \label{tb:nu_c3_cons}\vspace{-4ex}
\begin{center}
\resizebox{\textwidth}{!}{\begin{tabular}{lccccccccccccccccccccccccccccccc}
\hline
& \multicolumn{3}{c}{$\hat{\nu}^{(0)}_t$} & $~~$ & \multicolumn{3}{c}{$\hat{\nu}^{(1)}_t$} & $~~$ & \multicolumn{3}{c}{$\hat{\nu}^{(2)}_t$} & $~~$ & \multicolumn{3}{c}{$\hat{\nu}^{(3)}_t$} &~~& \multicolumn{3}{c}{$\hat{\nu}^{(0)}_t$} & $~~$ & \multicolumn{3}{c}{$\hat{\nu}^{(1)}_t$} & $~~$ & \multicolumn{3}{c}{$\hat{\nu}^{(2)}_t$} & $~~$ & \multicolumn{3}{c}{$\hat{\nu}^{(3)}_t$} \\
& RCAL & RMLs & RMLg &~~& RCAL & RMLs & RMLg &~~& RCAL & RMLs & RMLg &~~& RCAL & RMLs & RMLg &~~& RCAL & RMLs & RMLg &~~& RCAL & RMLs & RMLg &~~& RCAL & RMLs & RMLg &~~& RCAL & RMLs & RMLg\\
\hline
& \multicolumn{31}{c}{p = 50} \\
& \multicolumn{15}{c}{\footnotesize t = 0} &~~& \multicolumn{15}{c}{\footnotesize t = 1} \\
\cline{2-16}\cline{18-32}
Bias & -0.004 & -0.004 & -0.004 &~~& -0.036 & -0.029 & -0.045 &~~& 0.022 & -0.057 & 0.012 &~~& 0.001 & -0.011 & 0.005 &~~& -0.023 & -0.050 & -0.030 &~~& -0.001 & -0.001 & -0.001 &~~& -0.004 & -0.026 & -0.010 &~~& -0.023 & -0.056 & -0.025 \\
$\sqrt{\text{Var}}$ & 0.134 & 0.134 & 0.134 &~~& 0.135 & 0.140 & 0.134 &~~& 0.138 & 0.149 & 0.141 &~~& 0.126 & 0.132 & 0.131 &~~& 0.140 & 0.147 & 0.144 &~~& 0.126 & 0.126 & 0.126 &~~& 0.146 & 0.165 & 0.160 &~~& 0.140 & 0.152 & 0.148 \\
$\sqrt{\text{EVar}}$ & 0.137 & 0.137 & 0.137 &~~& 0.124 & 0.132 & 0.124 &~~& 0.124 & 0.130 & 0.121 &~~& 0.122 & 0.128 & 0.121 &~~& 0.131 & 0.144 & 0.134 &~~& 0.129 & 0.129 & 0.129 &~~& 0.129 & 0.154 & 0.142 &~~& 0.129 & 0.142 & 0.132 \\
Cov90 & 0.902 & 0.902 & 0.902 &~~& 0.849 & 0.871 & 0.843 &~~& 0.857 & 0.819 & 0.842 &~~& 0.880 & 0.888 & 0.863 &~~& 0.870 & 0.869 & 0.864 &~~& 0.909 & 0.909 & 0.909 &~~& 0.860 & 0.863 & 0.851 &~~& 0.853 & 0.845 & 0.848 \\
Cov95 & 0.944 & 0.944 & 0.944 &~~& 0.917 & 0.926 & 0.911 &~~& 0.922 & 0.886 & 0.917 &~~& 0.938 & 0.941 & 0.928 &~~& 0.933 & 0.939 & 0.930 &~~& 0.956 & 0.956 & 0.956 &~~& 0.911 & 0.920 & 0.915 &~~& 0.925 & 0.910 & 0.911 \\
& \multicolumn{15}{c}{\footnotesize t = 2} &~~& \multicolumn{15}{c}{\footnotesize t = 3} \\
\cline{2-16}\cline{18-32}
Bias & 0.004 & 0.030 & 0.005 &~~& 0.007 & 0.033 & 0.002 &~~& -0.003 & -0.003 & -0.003 &~~& -0.024 & 0.024 & -0.025 &~~& -0.012 & -0.033 & -0.013 &~~& -0.016 & -0.061 & -0.025 &~~& -0.036 & -0.073 & -0.059 &~~& -0.004 & -0.004 & -0.004 \\
$\sqrt{\text{Var}}$ & 0.143 & 0.153 & 0.148 &~~& 0.148 & 0.162 & 0.159 &~~& 0.131 & 0.131 & 0.131 &~~& 0.144 & 0.154 & 0.150 &~~& 0.137 & 0.145 & 0.141 &~~& 0.140 & 0.152 & 0.148 &~~& 0.153 & 0.153 & 0.151 &~~& 0.131 & 0.131 & 0.131 \\
$\sqrt{\text{EVar}}$ & 0.131 & 0.143 & 0.136 &~~& 0.130 & 0.143 & 0.136 &~~& 0.128 & 0.128 & 0.128 &~~& 0.128 & 0.141 & 0.134 &~~& 0.130 & 0.139 & 0.132 &~~& 0.128 & 0.139 & 0.131 &~~& 0.130 & 0.136 & 0.128 &~~& 0.129 & 0.129 & 0.129 \\
Cov90 & 0.861 & 0.857 & 0.870 &~~& 0.850 & 0.838 & 0.843 &~~& 0.900 & 0.900 & 0.900 &~~& 0.852 & 0.863 & 0.859 &~~& 0.881 & 0.885 & 0.879 &~~& 0.871 & 0.827 & 0.854 &~~& 0.831 & 0.800 & 0.803 &~~& 0.905 & 0.905 & 0.905 \\
Cov95 & 0.915 & 0.910 & 0.913 &~~& 0.903 & 0.900 & 0.899 &~~& 0.955 & 0.955 & 0.955 &~~& 0.912 & 0.926 & 0.909 &~~& 0.942 & 0.933 & 0.937 &~~& 0.925 & 0.902 & 0.911 &~~& 0.894 & 0.881 & 0.876 &~~& 0.951 & 0.951 & 0.951 \\

\cline{2-32}
& \multicolumn{31}{c}{p = 300} \\
& \multicolumn{15}{c}{\footnotesize t = 0} &~~& \multicolumn{15}{c}{\footnotesize t = 1} \\
\cline{2-16}\cline{18-32}
Bias & -0.001 & -0.001 & -0.001 &~~& -0.057 & -0.047 & -0.078 &~~& 0.049 & -0.040 & 0.024 &~~& 0.007 & -0.005 & 0.019 &~~& -0.018 & -0.063 & -0.039 &~~& 0.007 & 0.007 & 0.007 &~~& -0.004 & -0.069 & -0.023 &~~& -0.044 & -0.105 & -0.060 \\
$\sqrt{\text{Var}}$ & 0.139 & 0.139 & 0.139 &~~& 0.131 & 0.137 & 0.132 &~~& 0.125 & 0.140 & 0.129 &~~& 0.122 & 0.129 & 0.127 &~~& 0.131 & 0.137 & 0.133 &~~& 0.129 & 0.129 & 0.129 &~~& 0.127 & 0.142 & 0.141 &~~& 0.136 & 0.145 & 0.144 \\
$\sqrt{\text{EVar}}$ & 0.137 & 0.137 & 0.137 &~~& 0.116 & 0.122 & 0.115 &~~& 0.115 & 0.118 & 0.112 &~~& 0.114 & 0.118 & 0.112 &~~& 0.123 & 0.130 & 0.124 &~~& 0.129 & 0.129 & 0.129 &~~& 0.118 & 0.131 & 0.125 &~~& 0.120 & 0.125 & 0.119 \\
Cov90 & 0.899 & 0.899 & 0.899 &~~& 0.817 & 0.837 & 0.780 &~~& 0.838 & 0.821 & 0.846 &~~& 0.855 & 0.861 & 0.840 &~~& 0.869 & 0.838 & 0.856 &~~& 0.905 & 0.905 & 0.905 &~~& 0.872 & 0.826 & 0.852 &~~& 0.824 & 0.735 & 0.774 \\
Cov95 & 0.948 & 0.948 & 0.948 &~~& 0.884 & 0.902 & 0.859 &~~& 0.902 & 0.887 & 0.900 &~~& 0.934 & 0.922 & 0.910 &~~& 0.921 & 0.907 & 0.922 &~~& 0.950 & 0.950 & 0.950 &~~& 0.924 & 0.892 & 0.910 &~~& 0.905 & 0.825 & 0.863 \\
& \multicolumn{15}{c}{\footnotesize t = 2} &~~& \multicolumn{15}{c}{\footnotesize t = 3} \\
\cline{2-16}\cline{18-32}
Bias & 0.018 & 0.051 & 0.018 &~~& 0.028 & 0.058 & 0.020 &~~& -0.003 & -0.003 & -0.003 &~~& -0.044 & 0.011 & -0.056 &~~& -0.011 & -0.047 & -0.022 &~~& -0.018 & -0.077 & -0.045 &~~& -0.056 & -0.124 & -0.113 &~~& -0.004 & -0.004 & -0.004 \\
$\sqrt{\text{Var}}$ & 0.136 & 0.144 & 0.141 &~~& 0.136 & 0.143 & 0.144 &~~& 0.129 & 0.129 & 0.129 &~~& 0.139 & 0.148 & 0.148 &~~& 0.129 & 0.137 & 0.131 &~~& 0.125 & 0.139 & 0.137 &~~& 0.139 & 0.149 & 0.143 &~~& 0.127 & 0.127 & 0.127 \\
$\sqrt{\text{EVar}}$ & 0.123 & 0.130 & 0.125 &~~& 0.119 & 0.126 & 0.121 &~~& 0.129 & 0.129 & 0.129 &~~& 0.120 & 0.127 & 0.122 &~~& 0.122 & 0.126 & 0.120 &~~& 0.118 & 0.121 & 0.116 &~~& 0.120 & 0.122 & 0.116 &~~& 0.130 & 0.130 & 0.130 \\
Cov90 & 0.854 & 0.823 & 0.837 &~~& 0.857 & 0.832 & 0.833 &~~& 0.903 & 0.903 & 0.903 &~~& 0.841 & 0.855 & 0.790 &~~& 0.888 & 0.853 & 0.867 &~~& 0.883 & 0.786 & 0.800 &~~& 0.804 & 0.690 & 0.692 &~~& 0.911 & 0.911 & 0.911 \\
Cov95 & 0.919 & 0.903 & 0.918 &~~& 0.911 & 0.902 & 0.897 &~~& 0.954 & 0.954 & 0.954 &~~& 0.901 & 0.910 & 0.875 &~~& 0.937 & 0.907 & 0.918 &~~& 0.935 & 0.858 & 0.887 &~~& 0.886 & 0.776 & 0.773 &~~& 0.952 & 0.952 & 0.952 \\
\hline
\end{tabular}}
\end{center}
\setlength{\baselineskip}{0.5\baselineskip}
\vspace{-.15in}\noindent{\tiny
\textbf{Note}: RCAL denotes $\hat{\nu}^{(k)}_{t,\text{RCAL}}$, RMLs denotes $\hat{\nu}^{(k)}_{t,\text{RMLs}}$ and RMLg denotes $\hat{\nu}^{(k)}_{t,\text{RMLg}}$. Bias and Var are the Monte Carlo bias and variance of the point estimates. EVar is the mean of the variance estimates, and hence $\sqrt{\text{EVar}}$ also measures the $L_2$-average of lengths of confidence intervals. Cov90 or Cov95 is the coverage proportion of the 90\% or 95\% confidence intervals.}
\end{table}

\begin{table}[H]
\caption{\footnotesize Under (C1), summary of $\hat{\nu}^{(k)}_t$ with sum-to-zero constraint for $t, k = 0, 1, 2, 3$.} \label{tb:nu_c1}\vspace{-4ex}
\begin{center}
\resizebox{\textwidth}{!}{\begin{tabular}{lccccccccccccccccccccccccccccccc}
\hline
& \multicolumn{3}{c}{$\hat{\nu}^{(0)}_t$} & $~~$ & \multicolumn{3}{c}{$\hat{\nu}^{(1)}_t$} & $~~$ & \multicolumn{3}{c}{$\hat{\nu}^{(2)}_t$} & $~~$ & \multicolumn{3}{c}{$\hat{\nu}^{(3)}_t$} &~~& \multicolumn{3}{c}{$\hat{\nu}^{(0)}_t$} & $~~$ & \multicolumn{3}{c}{$\hat{\nu}^{(1)}_t$} & $~~$ & \multicolumn{3}{c}{$\hat{\nu}^{(2)}_t$} & $~~$ & \multicolumn{3}{c}{$\hat{\nu}^{(3)}_t$} \\
& RCAL & RMLs & RMLg &~~& RCAL & RMLs & RMLg &~~& RCAL & RMLs & RMLg &~~& RCAL & RMLs & RMLg &~~& RCAL & RMLs & RMLg &~~& RCAL & RMLs & RMLg &~~& RCAL & RMLs & RMLg &~~& RCAL & RMLs & RMLg\\
\hline
& \multicolumn{31}{c}{p = 50} \\
& \multicolumn{15}{c}{\footnotesize t = 0} &~~& \multicolumn{15}{c}{\footnotesize t = 1} \\
\cline{2-16}\cline{18-32}
Bias & -0.002 & -0.002 & -0.002 &~~& -0.043 & -0.018 & -0.046 &~~& 0.038 & -0.043 & 0.022 &~~& 0.002 & 0.006 & 0.013 &~~& -0.030 & -0.083 & -0.036 &~~& 0.001 & 0.001 & 0.001 &~~& -0.001 & -0.100 & -0.019 &~~& -0.027 & -0.098 & -0.030 \\
$\sqrt{\text{Var}}$ & 0.144 & 0.144 & 0.144 &~~& 0.131 & 0.136 & 0.134 &~~& 0.140 & 0.157 & 0.146 &~~& 0.128 & 0.140 & 0.138 &~~& 0.135 & 0.142 & 0.138 &~~& 0.127 & 0.127 & 0.127 &~~& 0.148 & 0.165 & 0.159 &~~& 0.144 & 0.152 & 0.151 \\
$\sqrt{\text{EVar}}$ & 0.142 & 0.142 & 0.142 &~~& 0.121 & 0.131 & 0.123 &~~& 0.123 & 0.138 & 0.128 &~~& 0.123 & 0.134 & 0.125 &~~& 0.135 & 0.144 & 0.134 &~~& 0.128 & 0.128 & 0.128 &~~& 0.130 & 0.148 & 0.137 &~~& 0.129 & 0.140 & 0.130 \\
Cov90 & 0.891 & 0.891 & 0.891 &~~& 0.849 & 0.880 & 0.845 &~~& 0.843 & 0.850 & 0.854 &~~& 0.890 & 0.878 & 0.863 &~~& 0.881 & 0.838 & 0.867 &~~& 0.903 & 0.903 & 0.903 &~~& 0.847 & 0.781 & 0.836 &~~& 0.858 & 0.795 & 0.822 \\
Cov95 & 0.950 & 0.950 & 0.950 &~~& 0.911 & 0.930 & 0.908 &~~& 0.904 & 0.906 & 0.913 &~~& 0.937 & 0.939 & 0.920 &~~& 0.941 & 0.914 & 0.935 &~~& 0.948 & 0.948 & 0.948 &~~& 0.911 & 0.859 & 0.897 &~~& 0.912 & 0.868 & 0.899 \\
& \multicolumn{15}{c}{\footnotesize t = 2} &~~& \multicolumn{15}{c}{\footnotesize t = 3} \\
\cline{2-16}\cline{18-32}
Bias & 0.016 & 0.047 & 0.007 &~~& 0.028 & 0.056 & 0.015 &~~& 0.001 & 0.001 & 0.001 &~~& -0.042 & 0.031 & -0.029 &~~& -0.004 & -0.070 & -0.025 &~~& -0.002 & -0.098 & -0.025 &~~& -0.044 & -0.102 & -0.060 &~~& -0.003 & -0.003 & -0.003 \\
$\sqrt{\text{Var}}$ & 0.151 & 0.159 & 0.156 &~~& 0.156 & 0.167 & 0.164 &~~& 0.125 & 0.125 & 0.125 &~~& 0.144 & 0.151 & 0.150 &~~& 0.143 & 0.147 & 0.146 &~~& 0.150 & 0.164 & 0.160 &~~& 0.159 & 0.166 & 0.161 &~~& 0.128 & 0.128 & 0.128 \\
$\sqrt{\text{EVar}}$ & 0.137 & 0.146 & 0.139 &~~& 0.130 & 0.146 & 0.138 &~~& 0.127 & 0.127 & 0.127 &~~& 0.128 & 0.139 & 0.132 &~~& 0.136 & 0.143 & 0.135 &~~& 0.128 & 0.138 & 0.129 &~~& 0.127 & 0.138 & 0.129 &~~& 0.131 & 0.131 & 0.131 \\
Cov90 & 0.865 & 0.856 & 0.866 &~~& 0.820 & 0.830 & 0.821 &~~& 0.901 & 0.901 & 0.901 &~~& 0.833 & 0.849 & 0.836 &~~& 0.892 & 0.845 & 0.877 &~~& 0.835 & 0.760 & 0.807 &~~& 0.788 & 0.738 & 0.775 &~~& 0.902 & 0.902 & 0.902 \\
Cov95 & 0.922 & 0.923 & 0.914 &~~& 0.885 & 0.889 & 0.908 &~~& 0.949 & 0.949 & 0.949 &~~& 0.907 & 0.916 & 0.912 &~~& 0.939 & 0.921 & 0.927 &~~& 0.903 & 0.837 & 0.872 &~~& 0.871 & 0.830 & 0.856 &~~& 0.951 & 0.951 & 0.951 \\
\cline{2-32}
& \multicolumn{31}{c}{p = 300} \\
& \multicolumn{15}{c}{\footnotesize t = 0} &~~& \multicolumn{15}{c}{\footnotesize t = 1} \\
\cline{2-16}\cline{18-32}
Bias & -0.003 & -0.003 & -0.003 &~~& -0.085 & -0.035 & -0.079 &~~& 0.104 & -0.012 & 0.059 &~~& -0.001 & 0.006 & 0.021 &~~& -0.024 & -0.083 & -0.043 &~~& 0.002 & 0.002 & 0.002 &~~& 0.003 & -0.108 & -0.021 &~~& -0.044 & -0.120 & -0.056 \\
$\sqrt{\text{Var}}$ & 0.140 & 0.140 & 0.140 &~~& 0.127 & 0.134 & 0.131 &~~& 0.132 & 0.153 & 0.144 &~~& 0.122 & 0.134 & 0.133 &~~& 0.139 & 0.148 & 0.143 &~~& 0.126 & 0.126 & 0.126 &~~& 0.135 & 0.159 & 0.150 &~~& 0.136 & 0.145 & 0.144 \\
$\sqrt{\text{EVar}}$ & 0.141 & 0.141 & 0.141 &~~& 0.112 & 0.118 & 0.113 &~~& 0.112 & 0.121 & 0.116 &~~& 0.114 & 0.120 & 0.114 &~~& 0.126 & 0.131 & 0.125 &~~& 0.127 & 0.127 & 0.127 &~~& 0.117 & 0.130 & 0.122 &~~& 0.119 & 0.124 & 0.118 \\
Cov90 & 0.905 & 0.905 & 0.905 &~~& 0.769 & 0.838 & 0.772 &~~& 0.717 & 0.795 & 0.777 &~~& 0.888 & 0.855 & 0.847 &~~& 0.865 & 0.793 & 0.837 &~~& 0.905 & 0.905 & 0.905 &~~& 0.843 & 0.722 & 0.824 &~~& 0.819 & 0.690 & 0.788 \\
Cov95 & 0.958 & 0.958 & 0.958 &~~& 0.846 & 0.899 & 0.842 &~~& 0.806 & 0.874 & 0.855 &~~& 0.931 & 0.923 & 0.910 &~~& 0.926 & 0.869 & 0.900 &~~& 0.950 & 0.950 & 0.950 &~~& 0.898 & 0.806 & 0.881 &~~& 0.897 & 0.801 & 0.863 \\
& \multicolumn{15}{c}{\footnotesize t = 2} &~~& \multicolumn{15}{c}{\footnotesize t = 3} \\
\cline{2-16}\cline{18-32}
Bias & 0.026 & 0.068 & 0.014 &~~& 0.042 & 0.077 & 0.026 &~~& 0.003 & 0.003 & 0.003 &~~& -0.067 & 0.028 & -0.058 &~~& 0.001 & -0.078 & -0.039 &~~& -0.015 & -0.116 & -0.060 &~~& -0.067 & -0.145 & -0.115 &~~& -0.003 & -0.003 & -0.003 \\
$\sqrt{\text{Var}}$ & 0.141 & 0.150 & 0.151 &~~& 0.140 & 0.156 & 0.157 &~~& 0.126 & 0.126 & 0.126 &~~& 0.135 & 0.146 & 0.147 &~~& 0.136 & 0.148 & 0.140 &~~& 0.133 & 0.149 & 0.145 &~~& 0.144 & 0.155 & 0.150 &~~& 0.131 & 0.131 & 0.131 \\
$\sqrt{\text{EVar}}$ & 0.127 & 0.133 & 0.128 &~~& 0.117 & 0.128 & 0.122 &~~& 0.127 & 0.127 & 0.127 &~~& 0.118 & 0.123 & 0.119 &~~& 0.127 & 0.130 & 0.124 &~~& 0.117 & 0.121 & 0.115 &~~& 0.117 & 0.122 & 0.116 &~~& 0.130 & 0.130 & 0.130 \\
Cov90 & 0.845 & 0.801 & 0.822 &~~& 0.813 & 0.764 & 0.803 &~~& 0.909 & 0.909 & 0.909 &~~& 0.788 & 0.809 & 0.775 &~~& 0.870 & 0.791 & 0.830 &~~& 0.843 & 0.695 & 0.773 &~~& 0.778 & 0.610 & 0.659 &~~& 0.893 & 0.893 & 0.893 \\
Cov95 & 0.913 & 0.872 & 0.899 &~~& 0.877 & 0.841 & 0.859 &~~& 0.954 & 0.954 & 0.954 &~~& 0.865 & 0.900 & 0.846 &~~& 0.932 & 0.869 & 0.903 &~~& 0.910 & 0.788 & 0.847 &~~& 0.846 & 0.705 & 0.749 &~~& 0.952 & 0.952 & 0.952 \\
\hline
\end{tabular}}
\end{center}
\setlength{\baselineskip}{0.5\baselineskip}
\vspace{-.15in}\noindent{\tiny
\textbf{Note}: RCAL denotes $\hat{\nu}^{(k)}_{t,\text{RCAL}}$, RMLs denotes $\hat{\nu}^{(k)}_{t,\text{RMLs}}$ and RMLg denotes $\hat{\nu}^{(k)}_{t,\text{RMLg}}$. Bias and Var are the Monte Carlo bias and variance of the point estimates. EVar is the mean of the variance estimates, and hence $\sqrt{\text{EVar}}$ also measures the $L_2$-average of lengths of confidence intervals. Cov90 or Cov95 is the coverage proportion of the 90\% or 95\% confidence intervals.}
\end{table}

\begin{table}[H]
\caption{\footnotesize Under (C2), summary of $\hat{\nu}^{(k)}_t$ with sum-to-zero constraint for $t, k = 0, 1, 2, 3$.} \label{tb:nu_c2}\vspace{-4ex}
\begin{center}
\resizebox{\textwidth}{!}{\begin{tabular}{lccccccccccccccccccccccccccccccc}
\hline
& \multicolumn{3}{c}{$\hat{\nu}^{(0)}_t$} & $~~$ & \multicolumn{3}{c}{$\hat{\nu}^{(1)}_t$} & $~~$ & \multicolumn{3}{c}{$\hat{\nu}^{(2)}_t$} & $~~$ & \multicolumn{3}{c}{$\hat{\nu}^{(3)}_t$} &~~& \multicolumn{3}{c}{$\hat{\nu}^{(0)}_t$} & $~~$ & \multicolumn{3}{c}{$\hat{\nu}^{(1)}_t$} & $~~$ & \multicolumn{3}{c}{$\hat{\nu}^{(2)}_t$} & $~~$ & \multicolumn{3}{c}{$\hat{\nu}^{(3)}_t$} \\
& RCAL & RMLs & RMLg &~~& RCAL & RMLs & RMLg &~~& RCAL & RMLs & RMLg &~~& RCAL & RMLs & RMLg &~~& RCAL & RMLs & RMLg &~~& RCAL & RMLs & RMLg &~~& RCAL & RMLs & RMLg &~~& RCAL & RMLs & RMLg\\
\hline
& \multicolumn{31}{c}{p = 50} \\
& \multicolumn{15}{c}{\footnotesize t = 0} &~~& \multicolumn{15}{c}{\footnotesize t = 1} \\
\cline{2-16}\cline{18-32}
Bias & -0.003 & -0.003 & -0.003 &~~& -0.073 & -0.057 & -0.083 &~~& 0.041 & 0.005 & 0.031 &~~& 0.000 & -0.001 & 0.011 &~~& -0.030 & -0.057 & -0.033 &~~& 0.002 & 0.002 & 0.002 &~~& 0.002 & -0.091 & -0.016 &~~& -0.050 & -0.097 & -0.057 \\
$\sqrt{\text{Var}}$ & 0.127 & 0.127 & 0.127 &~~& 0.134 & 0.143 & 0.141 &~~& 0.141 & 0.159 & 0.152 &~~& 0.124 & 0.137 & 0.136 &~~& 0.127 & 0.129 & 0.129 &~~& 0.112 & 0.112 & 0.112 &~~& 0.145 & 0.160 & 0.158 &~~& 0.146 & 0.148 & 0.151 \\
$\sqrt{\text{EVar}}$ & 0.125 & 0.125 & 0.125 &~~& 0.114 & 0.127 & 0.122 &~~& 0.118 & 0.133 & 0.127 &~~& 0.115 & 0.123 & 0.117 &~~& 0.119 & 0.122 & 0.117 &~~& 0.114 & 0.114 & 0.114 &~~& 0.122 & 0.142 & 0.132 &~~& 0.120 & 0.129 & 0.122 \\
Cov90 & 0.895 & 0.895 & 0.895 &~~& 0.774 & 0.821 & 0.775 &~~& 0.797 & 0.844 & 0.824 &~~& 0.874 & 0.860 & 0.842 &~~& 0.865 & 0.857 & 0.852 &~~& 0.899 & 0.899 & 0.899 &~~& 0.845 & 0.801 & 0.836 &~~& 0.808 & 0.767 & 0.793 \\
Cov95 & 0.948 & 0.948 & 0.948 &~~& 0.851 & 0.876 & 0.841 &~~& 0.883 & 0.896 & 0.886 &~~& 0.927 & 0.916 & 0.910 &~~& 0.936 & 0.911 & 0.919 &~~& 0.950 & 0.950 & 0.950 &~~& 0.902 & 0.864 & 0.891 &~~& 0.877 & 0.836 & 0.862 \\
& \multicolumn{15}{c}{\footnotesize t = 2} &~~& \multicolumn{15}{c}{\footnotesize t = 3} \\
\cline{2-16}\cline{18-32}
Bias & 0.013 & 0.043 & 0.000 &~~& 0.016 & 0.054 & -0.003 &~~& 0.002 & 0.002 & 0.002 &~~& -0.082 & 0.003 & -0.045 &~~& -0.001 & -0.051 & -0.022 &~~& -0.004 & -0.062 & -0.029 &~~& -0.087 & -0.189 & -0.134 &~~& 0.000 & 0.000 & 0.000 \\
$\sqrt{\text{Var}}$ & 0.143 & 0.148 & 0.145 &~~& 0.156 & 0.171 & 0.163 &~~& 0.116 & 0.116 & 0.116 &~~& 0.150 & 0.154 & 0.154 &~~& 0.129 & 0.132 & 0.131 &~~& 0.145 & 0.155 & 0.156 &~~& 0.164 & 0.175 & 0.170 &~~& 0.115 & 0.115 & 0.115 \\
$\sqrt{\text{EVar}}$ & 0.123 & 0.134 & 0.127 &~~& 0.122 & 0.144 & 0.134 &~~& 0.116 & 0.116 & 0.116 &~~& 0.119 & 0.135 & 0.127 &~~& 0.120 & 0.125 & 0.117 &~~& 0.118 & 0.128 & 0.118 &~~& 0.122 & 0.144 & 0.130 &~~& 0.116 & 0.116 & 0.116 \\
Cov90 & 0.843 & 0.846 & 0.853 &~~& 0.791 & 0.804 & 0.816 &~~& 0.898 & 0.898 & 0.898 &~~& 0.742 & 0.854 & 0.812 &~~& 0.866 & 0.841 & 0.856 &~~& 0.818 & 0.792 & 0.787 &~~& 0.704 & 0.558 & 0.634 &~~& 0.910 & 0.910 & 0.910 \\
Cov95 & 0.908 & 0.907 & 0.911 &~~& 0.871 & 0.874 & 0.891 &~~& 0.953 & 0.953 & 0.953 &~~& 0.821 & 0.915 & 0.886 &~~& 0.925 & 0.903 & 0.911 &~~& 0.888 & 0.853 & 0.847 &~~& 0.800 & 0.651 & 0.719 &~~& 0.947 & 0.947 & 0.947 \\
\cline{2-32}
& \multicolumn{31}{c}{p = 300} \\
& \multicolumn{15}{c}{\footnotesize t = 0} &~~& \multicolumn{15}{c}{\footnotesize t = 1} \\
\cline{2-16}\cline{18-32}
Bias & -0.001 & -0.001 & -0.001 &~~& -0.152 & -0.086 & -0.128 &~~& 0.114 & 0.041 & 0.079 &~~& -0.002 & 0.008 & 0.020 &~~& -0.030 & -0.056 & -0.042 &~~& 0.005 & 0.005 & 0.005 &~~& 0.006 & -0.074 & -0.013 &~~& -0.081 & -0.119 & -0.085 \\
$\sqrt{\text{Var}}$ & 0.125 & 0.125 & 0.125 &~~& 0.130 & 0.139 & 0.139 &~~& 0.132 & 0.161 & 0.151 &~~& 0.115 & 0.130 & 0.130 &~~& 0.125 & 0.132 & 0.127 &~~& 0.112 & 0.112 & 0.112 &~~& 0.129 & 0.151 & 0.149 &~~& 0.136 & 0.140 & 0.145 \\
$\sqrt{\text{EVar}}$ & 0.125 & 0.125 & 0.125 &~~& 0.102 & 0.112 & 0.112 &~~& 0.104 & 0.115 & 0.116 &~~& 0.104 & 0.108 & 0.106 &~~& 0.109 & 0.111 & 0.106 &~~& 0.114 & 0.114 & 0.114 &~~& 0.105 & 0.119 & 0.113 &~~& 0.107 & 0.112 & 0.107 \\
Cov90 & 0.904 & 0.904 & 0.904 &~~& 0.545 & 0.716 & 0.633 &~~& 0.635 & 0.736 & 0.720 &~~& 0.869 & 0.826 & 0.820 &~~& 0.834 & 0.796 & 0.807 &~~& 0.902 & 0.902 & 0.902 &~~& 0.824 & 0.750 & 0.800 &~~& 0.738 & 0.660 & 0.701 \\
Cov95 & 0.949 & 0.949 & 0.949 &~~& 0.645 & 0.795 & 0.716 &~~& 0.743 & 0.804 & 0.799 &~~& 0.924 & 0.894 & 0.892 &~~& 0.901 & 0.869 & 0.877 &~~& 0.943 & 0.943 & 0.943 &~~& 0.905 & 0.831 & 0.873 &~~& 0.806 & 0.748 & 0.784 \\
& \multicolumn{15}{c}{\footnotesize t = 2} &~~& \multicolumn{15}{c}{\footnotesize t = 3} \\
\cline{2-16}\cline{18-32}
Bias & 0.024 & 0.053 & 0.009 &~~& 0.028 & 0.055 & 0.010 &~~& 0.004 & 0.004 & 0.004 &~~& -0.123 & -0.002 & -0.074 &~~& -0.001 & -0.056 & -0.041 &~~& -0.016 & -0.079 & -0.063 &~~& -0.145 & -0.228 & -0.217 &~~& -0.004 & -0.004 & -0.004 \\
$\sqrt{\text{Var}}$ & 0.131 & 0.140 & 0.141 &~~& 0.135 & 0.155 & 0.155 &~~& 0.117 & 0.117 & 0.117 &~~& 0.133 & 0.140 & 0.143 &~~& 0.122 & 0.130 & 0.126 &~~& 0.125 & 0.140 & 0.141 &~~& 0.147 & 0.155 & 0.155 &~~& 0.118 & 0.118 & 0.118 \\
$\sqrt{\text{EVar}}$ & 0.111 & 0.118 & 0.113 &~~& 0.107 & 0.120 & 0.115 &~~& 0.116 & 0.116 & 0.116 &~~& 0.105 & 0.114 & 0.111 &~~& 0.110 & 0.111 & 0.105 &~~& 0.104 & 0.107 & 0.103 &~~& 0.107 & 0.117 & 0.112 &~~& 0.115 & 0.115 & 0.115 \\
Cov90 & 0.826 & 0.790 & 0.804 &~~& 0.804 & 0.769 & 0.792 &~~& 0.906 & 0.906 & 0.906 &~~& 0.647 & 0.812 & 0.734 &~~& 0.863 & 0.794 & 0.802 &~~& 0.826 & 0.728 & 0.727 &~~& 0.568 & 0.412 & 0.416 &~~& 0.893 & 0.893 & 0.893 \\
Cov95 & 0.889 & 0.866 & 0.876 &~~& 0.876 & 0.846 & 0.868 &~~& 0.949 & 0.949 & 0.949 &~~& 0.742 & 0.888 & 0.820 &~~& 0.920 & 0.879 & 0.886 &~~& 0.886 & 0.807 & 0.813 &~~& 0.652 & 0.474 & 0.488 &~~& 0.936 & 0.936 & 0.936 \\
\hline
\end{tabular}}
\end{center}
\setlength{\baselineskip}{0.5\baselineskip}
\vspace{-.15in}\noindent{\tiny
\textbf{Note}: RCAL denotes $\hat{\nu}^{(k)}_{t,\text{RCAL}}$, RMLs denotes $\hat{\nu}^{(k)}_{t,\text{RMLs}}$ and RMLg denotes $\hat{\nu}^{(k)}_{t,\text{RMLg}}$. Bias and Var are the Monte Carlo bias and variance of the point estimates. EVar is the mean of the variance estimates, and hence $\sqrt{\text{EVar}}$ also measures the $L_2$-average of lengths of confidence intervals. Cov90 or Cov95 is the coverage proportion of the 90\% or 95\% confidence intervals.}
\end{table}

\begin{table}[H]
\caption{\footnotesize Under (C3), summary of $\hat{\nu}^{(k)}_t$ with sum-to-zero constraint for $t, k = 0, 1, 2, 3$.} \label{tb:nu_c3}\vspace{-4ex}
\begin{center}
\resizebox{\textwidth}{!}{\begin{tabular}{lccccccccccccccccccccccccccccccc}
\hline
& \multicolumn{3}{c}{$\hat{\nu}^{(0)}_t$} & $~~$ & \multicolumn{3}{c}{$\hat{\nu}^{(1)}_t$} & $~~$ & \multicolumn{3}{c}{$\hat{\nu}^{(2)}_t$} & $~~$ & \multicolumn{3}{c}{$\hat{\nu}^{(3)}_t$} &~~& \multicolumn{3}{c}{$\hat{\nu}^{(0)}_t$} & $~~$ & \multicolumn{3}{c}{$\hat{\nu}^{(1)}_t$} & $~~$ & \multicolumn{3}{c}{$\hat{\nu}^{(2)}_t$} & $~~$ & \multicolumn{3}{c}{$\hat{\nu}^{(3)}_t$} \\
& RCAL & RMLs & RMLg &~~& RCAL & RMLs & RMLg &~~& RCAL & RMLs & RMLg &~~& RCAL & RMLs & RMLg &~~& RCAL & RMLs & RMLg &~~& RCAL & RMLs & RMLg &~~& RCAL & RMLs & RMLg &~~& RCAL & RMLs & RMLg\\
\hline
& \multicolumn{31}{c}{p = 50} \\
& \multicolumn{15}{c}{\footnotesize t = 0} &~~& \multicolumn{15}{c}{\footnotesize t = 1} \\
\cline{2-16}\cline{18-32}
Bias & -0.004 & -0.004 & -0.004 &~~& -0.045 & -0.023 & -0.046 &~~& 0.031 & -0.054 & 0.016 &~~& 0.001 & -0.003 & 0.007 &~~& -0.021 & -0.047 & -0.029 &~~& -0.001 & -0.001 & -0.001 &~~& 0.000 & -0.034 & -0.011 &~~& -0.026 & -0.063 & -0.027 \\
$\sqrt{\text{Var}}$ & 0.134 & 0.134 & 0.134 &~~& 0.136 & 0.140 & 0.135 &~~& 0.137 & 0.148 & 0.141 &~~& 0.125 & 0.132 & 0.131 &~~& 0.141 & 0.147 & 0.145 &~~& 0.126 & 0.126 & 0.126 &~~& 0.144 & 0.162 & 0.157 &~~& 0.139 & 0.149 & 0.147 \\
$\sqrt{\text{EVar}}$ & 0.137 & 0.137 & 0.137 &~~& 0.123 & 0.132 & 0.123 &~~& 0.122 & 0.131 & 0.122 &~~& 0.121 & 0.128 & 0.121 &~~& 0.132 & 0.144 & 0.135 &~~& 0.129 & 0.129 & 0.129 &~~& 0.127 & 0.150 & 0.138 &~~& 0.129 & 0.140 & 0.130 \\
Cov90 & 0.902 & 0.902 & 0.902 &~~& 0.837 & 0.874 & 0.842 &~~& 0.852 & 0.831 & 0.846 &~~& 0.879 & 0.888 & 0.861 &~~& 0.878 & 0.875 & 0.863 &~~& 0.909 & 0.909 & 0.909 &~~& 0.856 & 0.858 & 0.852 &~~& 0.856 & 0.832 & 0.843 \\
Cov95 & 0.944 & 0.944 & 0.944 &~~& 0.898 & 0.926 & 0.907 &~~& 0.912 & 0.896 & 0.917 &~~& 0.934 & 0.941 & 0.925 &~~& 0.929 & 0.942 & 0.928 &~~& 0.956 & 0.956 & 0.956 &~~& 0.917 & 0.916 & 0.913 &~~& 0.926 & 0.900 & 0.911 \\
& \multicolumn{15}{c}{\footnotesize t = 2} &~~& \multicolumn{15}{c}{\footnotesize t = 3} \\
\cline{2-16}\cline{18-32}
Bias & 0.009 & 0.027 & 0.000 &~~& 0.014 & 0.034 & 0.002 &~~& -0.003 & -0.003 & -0.003 &~~& -0.035 & 0.024 & -0.027 &~~& -0.001 & -0.035 & -0.017 &~~& -0.009 & -0.067 & -0.027 &~~& -0.043 & -0.079 & -0.062 &~~& -0.004 & -0.004 & -0.004 \\
$\sqrt{\text{Var}}$ & 0.144 & 0.155 & 0.151 &~~& 0.146 & 0.160 & 0.157 &~~& 0.131 & 0.131 & 0.131 &~~& 0.143 & 0.152 & 0.149 &~~& 0.138 & 0.146 & 0.142 &~~& 0.138 & 0.150 & 0.147 &~~& 0.152 & 0.153 & 0.151 &~~& 0.131 & 0.131 & 0.131 \\
$\sqrt{\text{EVar}}$ & 0.132 & 0.145 & 0.138 &~~& 0.128 & 0.141 & 0.134 &~~& 0.128 & 0.128 & 0.128 &~~& 0.128 & 0.139 & 0.132 &~~& 0.131 & 0.140 & 0.132 &~~& 0.127 & 0.137 & 0.128 &~~& 0.129 & 0.134 & 0.127 &~~& 0.129 & 0.129 & 0.129 \\
Cov90 & 0.858 & 0.859 & 0.864 &~~& 0.854 & 0.835 & 0.839 &~~& 0.900 & 0.900 & 0.900 &~~& 0.840 & 0.862 & 0.856 &~~& 0.882 & 0.882 & 0.873 &~~& 0.873 & 0.820 & 0.846 &~~& 0.827 & 0.790 & 0.795 &~~& 0.905 & 0.905 & 0.905 \\
Cov95 & 0.916 & 0.915 & 0.916 &~~& 0.906 & 0.898 & 0.898 &~~& 0.955 & 0.955 & 0.955 &~~& 0.905 & 0.924 & 0.908 &~~& 0.945 & 0.935 & 0.935 &~~& 0.929 & 0.894 & 0.910 &~~& 0.885 & 0.871 & 0.869 &~~& 0.951 & 0.951 & 0.951 \\
\cline{2-32}
& \multicolumn{31}{c}{p = 300} \\
& \multicolumn{15}{c}{\footnotesize t = 0} &~~& \multicolumn{15}{c}{\footnotesize t = 1} \\
\cline{2-16}\cline{18-32}
Bias & -0.001 & -0.001 & -0.001 &~~& -0.086 & -0.044 & -0.082 &~~& 0.075 & -0.036 & 0.030 &~~& 0.005 & 0.001 & 0.023 &~~& -0.015 & -0.063 & -0.038 &~~& 0.007 & 0.007 & 0.007 &~~& 0.000 & -0.074 & -0.023 &~~& -0.050 & -0.109 & -0.062 \\
$\sqrt{\text{Var}}$ & 0.139 & 0.139 & 0.139 &~~& 0.131 & 0.137 & 0.132 &~~& 0.123 & 0.140 & 0.129 &~~& 0.120 & 0.128 & 0.126 &~~& 0.132 & 0.138 & 0.135 &~~& 0.129 & 0.129 & 0.129 &~~& 0.124 & 0.141 & 0.140 &~~& 0.135 & 0.144 & 0.143 \\
$\sqrt{\text{EVar}}$ & 0.137 & 0.137 & 0.137 &~~& 0.115 & 0.120 & 0.114 &~~& 0.113 & 0.119 & 0.113 &~~& 0.113 & 0.117 & 0.111 &~~& 0.124 & 0.130 & 0.124 &~~& 0.129 & 0.129 & 0.129 &~~& 0.117 & 0.128 & 0.122 &~~& 0.119 & 0.124 & 0.118 \\
Cov90 & 0.899 & 0.899 & 0.899 &~~& 0.767 & 0.837 & 0.765 &~~& 0.787 & 0.820 & 0.843 &~~& 0.860 & 0.861 & 0.839 &~~& 0.871 & 0.838 & 0.854 &~~& 0.905 & 0.905 & 0.905 &~~& 0.877 & 0.818 & 0.849 &~~& 0.820 & 0.725 & 0.772 \\
Cov95 & 0.948 & 0.948 & 0.948 &~~& 0.846 & 0.898 & 0.848 &~~& 0.875 & 0.885 & 0.898 &~~& 0.937 & 0.924 & 0.907 &~~& 0.923 & 0.908 & 0.919 &~~& 0.950 & 0.950 & 0.950 &~~& 0.928 & 0.889 & 0.907 &~~& 0.899 & 0.813 & 0.860 \\
& \multicolumn{15}{c}{\footnotesize t = 2} &~~& \multicolumn{15}{c}{\footnotesize t = 3} \\
\cline{2-16}\cline{18-32}
Bias & 0.025 & 0.045 & 0.011 &~~& 0.034 & 0.058 & 0.021 &~~& -0.003 & -0.003 & -0.003 &~~& -0.062 & 0.010 & -0.060 &~~& 0.003 & -0.052 & -0.031 &~~& -0.008 & -0.080 & -0.047 &~~& -0.071 & -0.128 & -0.118 &~~& -0.004 & -0.004 & -0.004 \\
$\sqrt{\text{Var}}$ & 0.136 & 0.146 & 0.142 &~~& 0.132 & 0.142 & 0.143 &~~& 0.129 & 0.129 & 0.129 &~~& 0.137 & 0.147 & 0.147 &~~& 0.128 & 0.138 & 0.133 &~~& 0.123 & 0.138 & 0.136 &~~& 0.138 & 0.148 & 0.143 &~~& 0.127 & 0.127 & 0.127 \\
$\sqrt{\text{EVar}}$ & 0.123 & 0.131 & 0.126 &~~& 0.117 & 0.125 & 0.119 &~~& 0.129 & 0.129 & 0.129 &~~& 0.120 & 0.126 & 0.121 &~~& 0.123 & 0.127 & 0.121 &~~& 0.117 & 0.120 & 0.115 &~~& 0.119 & 0.121 & 0.115 &~~& 0.130 & 0.130 & 0.130 \\
Cov90 & 0.841 & 0.827 & 0.845 &~~& 0.851 & 0.826 & 0.829 &~~& 0.903 & 0.903 & 0.903 &~~& 0.823 & 0.853 & 0.789 &~~& 0.892 & 0.845 & 0.859 &~~& 0.885 & 0.776 & 0.799 &~~& 0.791 & 0.677 & 0.677 &~~& 0.911 & 0.911 & 0.911 \\
Cov95 & 0.917 & 0.907 & 0.913 &~~& 0.906 & 0.900 & 0.896 &~~& 0.954 & 0.954 & 0.954 &~~& 0.887 & 0.910 & 0.870 &~~& 0.941 & 0.898 & 0.914 &~~& 0.942 & 0.854 & 0.881 &~~& 0.858 & 0.761 & 0.761 &~~& 0.952 & 0.952 & 0.952 \\
\hline
\end{tabular}}
\end{center}
\setlength{\baselineskip}{0.5\baselineskip}
\vspace{-.15in}\noindent{\tiny
\textbf{Note}: RCAL denotes $\hat{\nu}^{(k)}_{t,\text{RCAL}}$, RMLs denotes $\hat{\nu}^{(k)}_{t,\text{RMLs}}$ and RMLg denotes $\hat{\nu}^{(k)}_{t,\text{RMLg}}$. Bias and Var are the Monte Carlo bias and variance of the point estimates. EVar is the mean of the variance estimates, and hence $\sqrt{\text{EVar}}$ also measures the $L_2$-average of lengths of confidence intervals. Cov90 or Cov95 is the coverage proportion of the 90\% or 95\% confidence intervals.}
\end{table}

\section{Additional material for empirical application}\label{sec:add-empi}

\textbf{Data preprocessing.}\; The original dataset is downloaded from the US National Center of Health Statistics, and then preprocessed as follows for our analyses. The covariates used here are the same as in Almond et al. (2005) except for the covariate ``parity indicator'', which is used in Almond et al. (2005) but cannot be found in the original dataset. Based on the Manual of the dataset, we construct this covariate by setting its value to 0 if the value of ``interval since last live birth'' is 777, which means there is no previous live birth. Otherwise, we set its value to be 0. Furthermore, we replace ``777'' in the covariate ``interval since last live birth'' with $-1$. After dropping observations with missing data and converting categorical covariates into dummy variables, we have 411609 samples and 33 covariates.

\begin{figure}[H]
\centering
\includegraphics[scale=0.47]{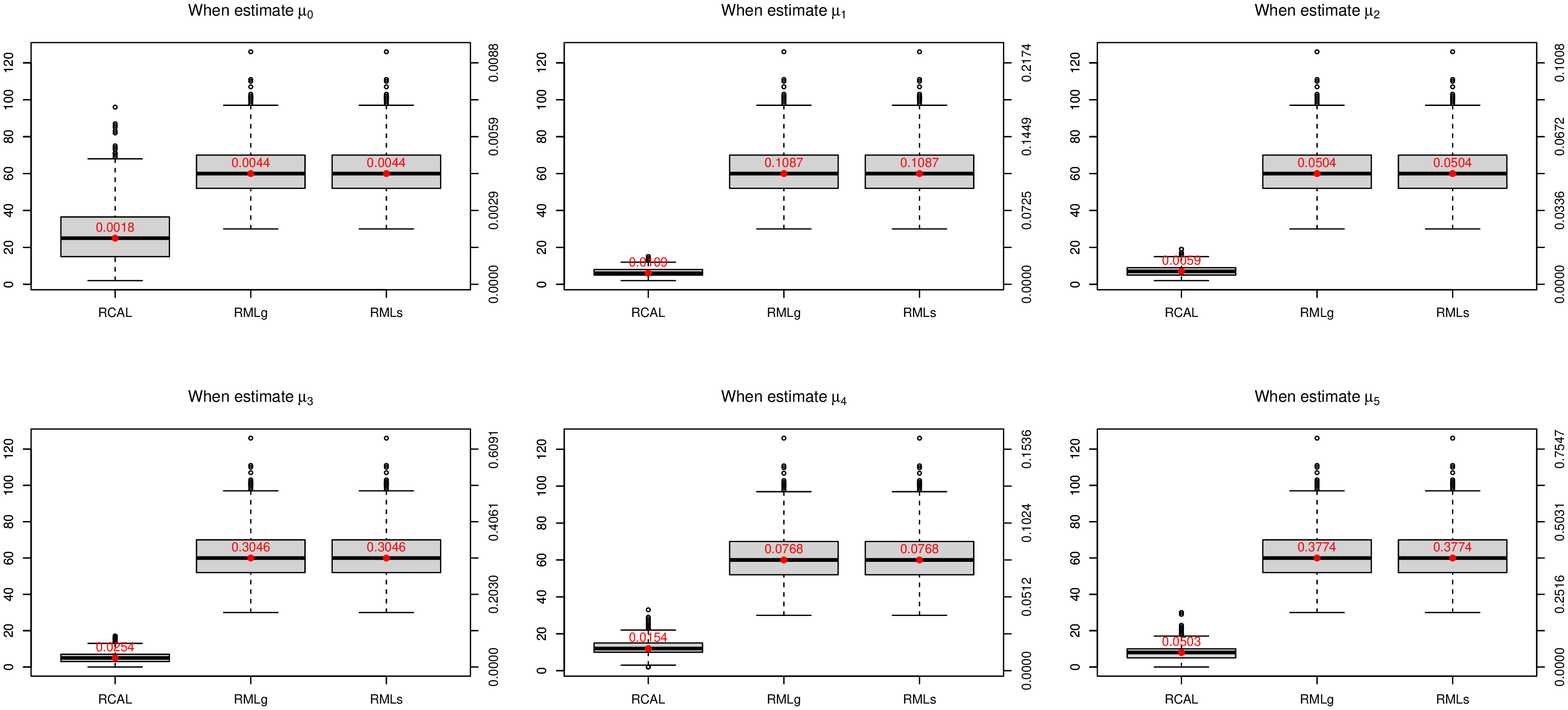}\vspace{-.1in}
\caption{Boxplot of number of nonzero estimated coefficients in PS model. Left y axis represents number of nonzero estimated coefficients. Right y axis represents ratio of number of nonzero estimated coefficients over corresponding treatment group size. Red number is the mean of ratios.}
\label{fig:box_nnz_ps_cross}
\end{figure}

\begin{figure}[H]
\centering
\includegraphics[scale=0.47]{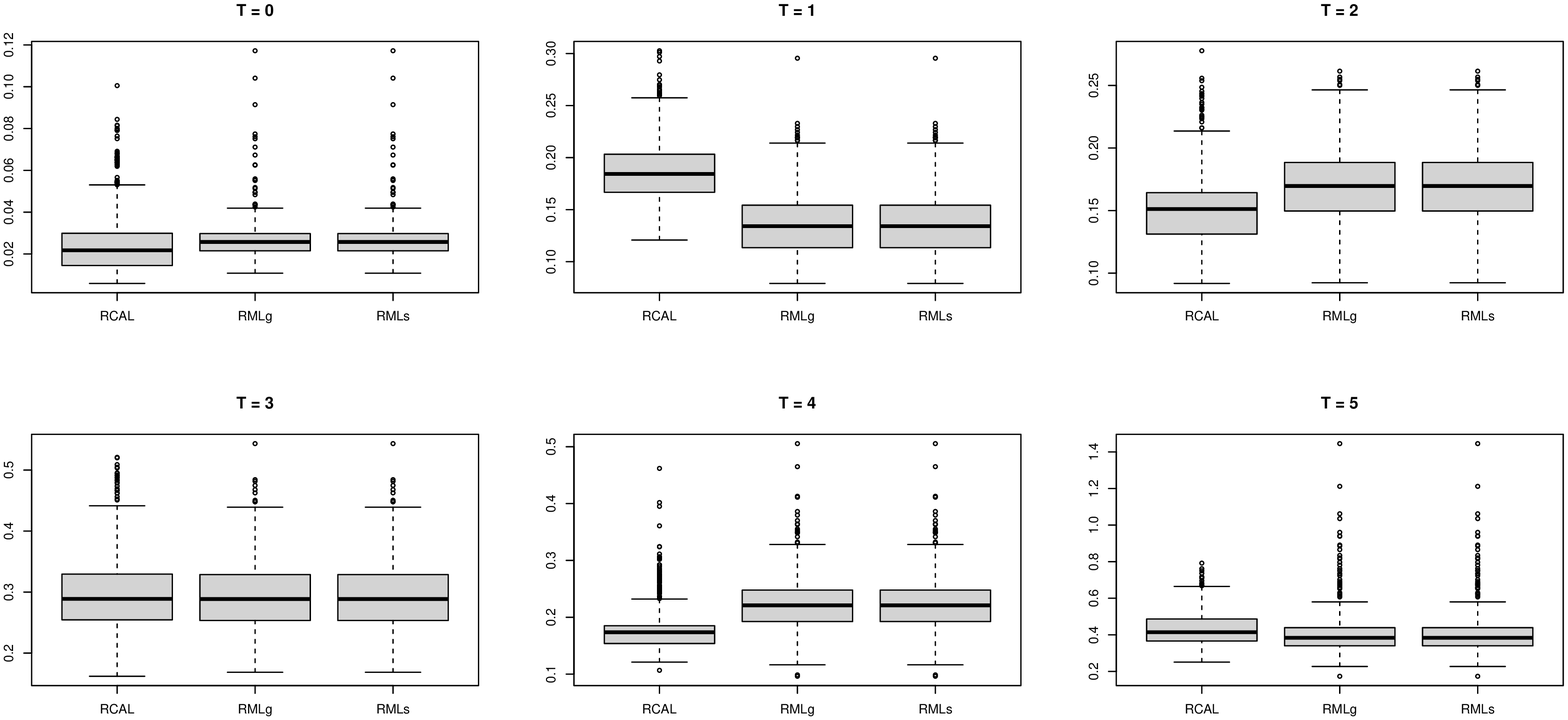}\vspace{-.1in}
\caption{Boxplots of maximum absolute standard calibration differences~(MASCD)}
\label{fig:box_masd_cross}
\end{figure}

\begin{figure}
\centering
\includegraphics[scale=0.47]{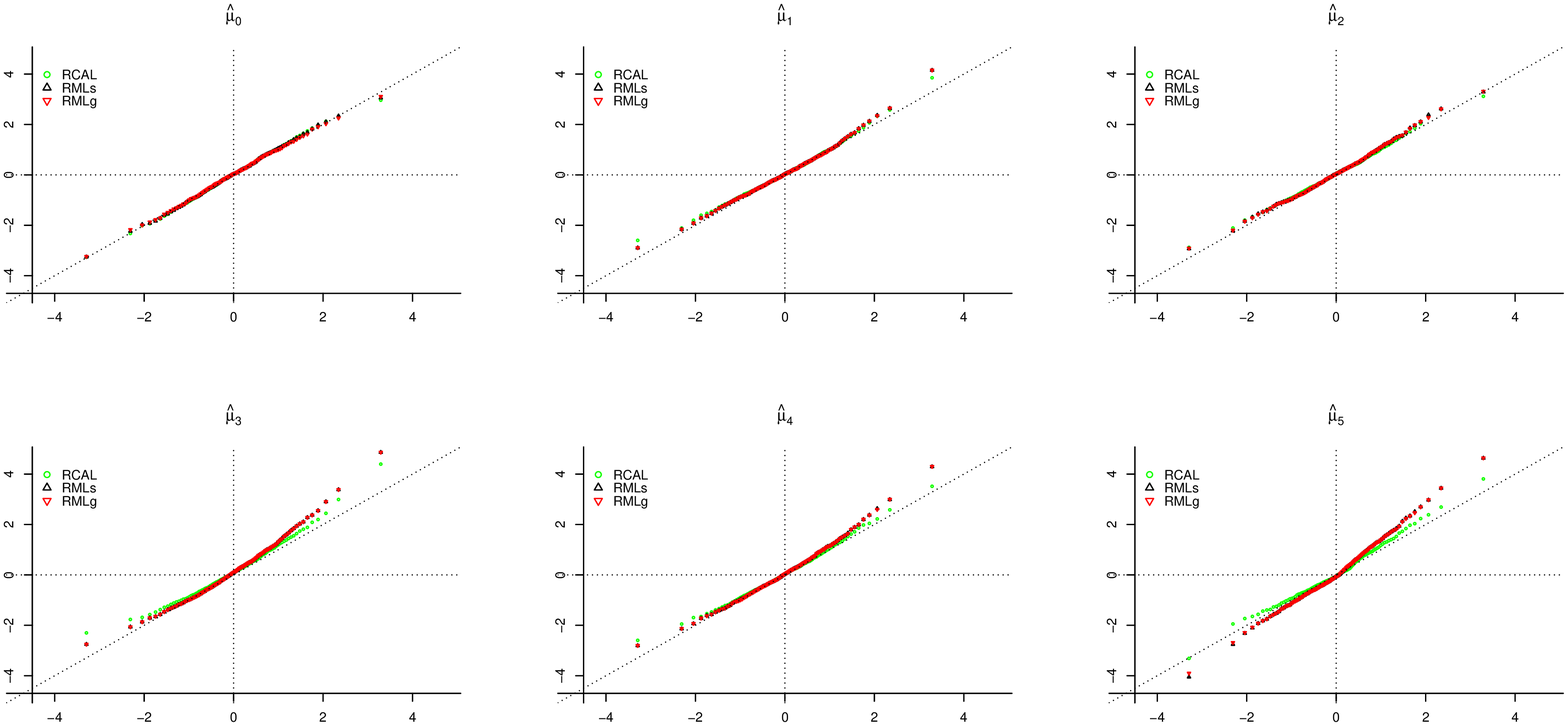}
\caption{QQ plots of the standardized $\hat{\mu}_t$ against standard normal.}
\label{fig:qq_mu_cross}
\end{figure}

\begin{figure}
\begin{subfigure}{0.8\textwidth}
\centering
\includegraphics[scale=0.47]{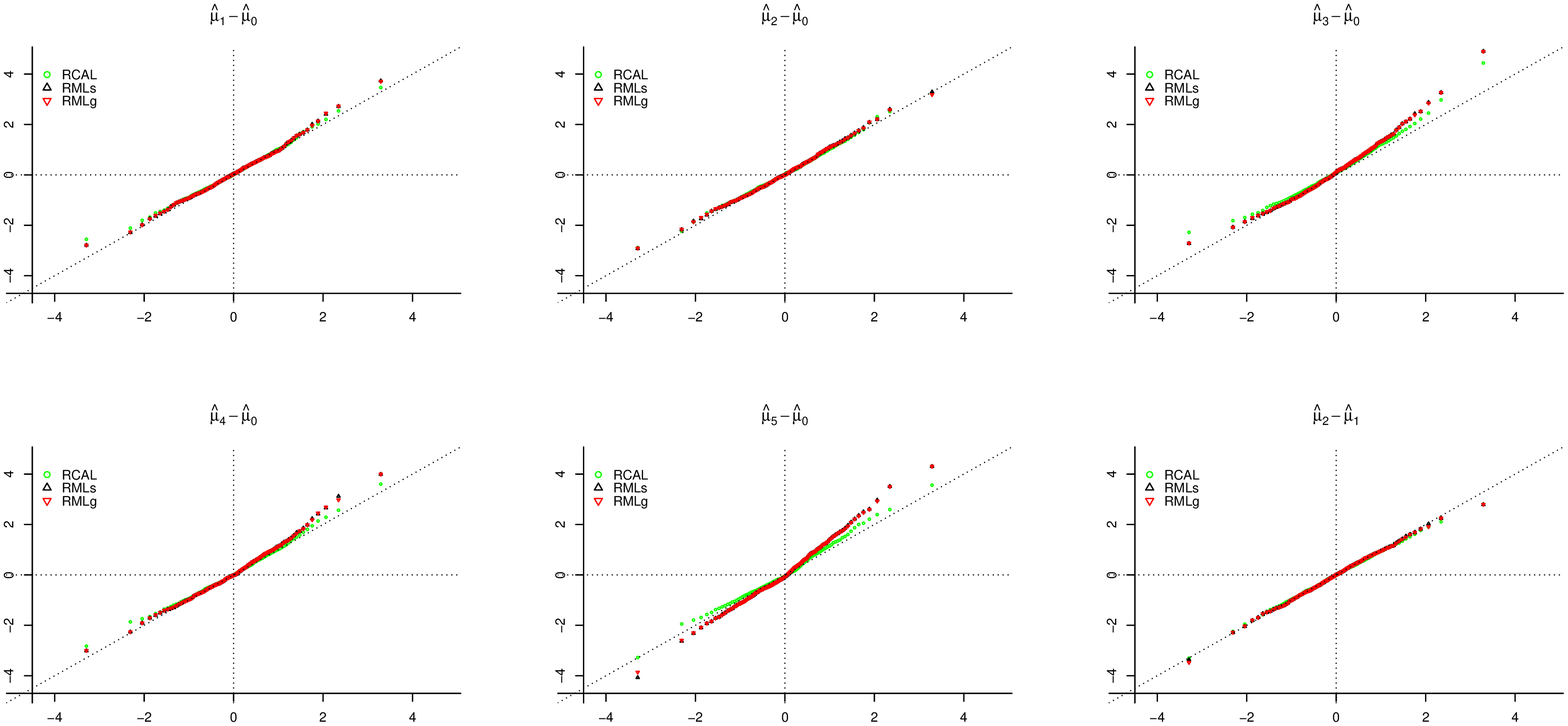}
\end{subfigure} %
\begin{subfigure}{0.8\textwidth}
\centering
\includegraphics[scale=0.47]{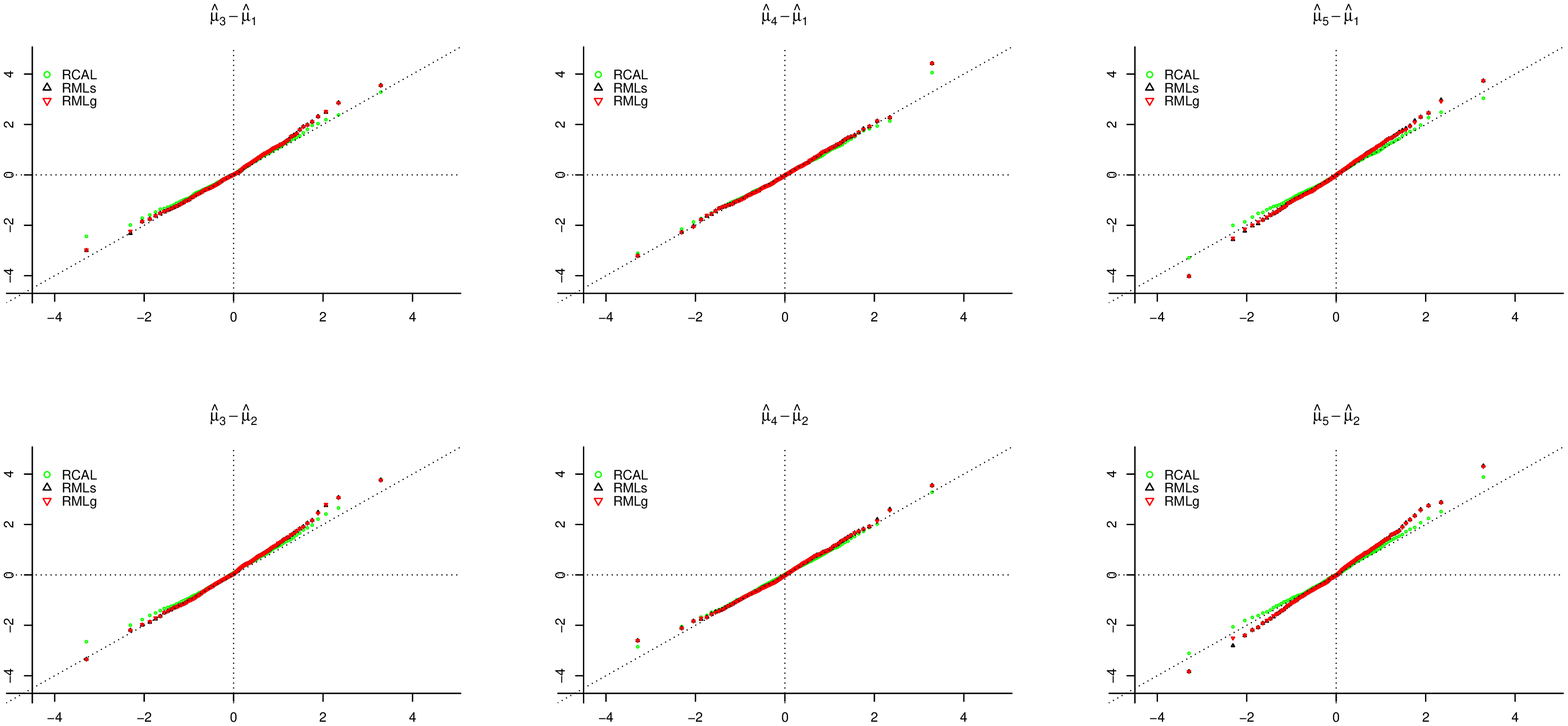}
\end{subfigure} %
\begin{subfigure}{0.8\textwidth}
\centering
\includegraphics[scale=0.47]{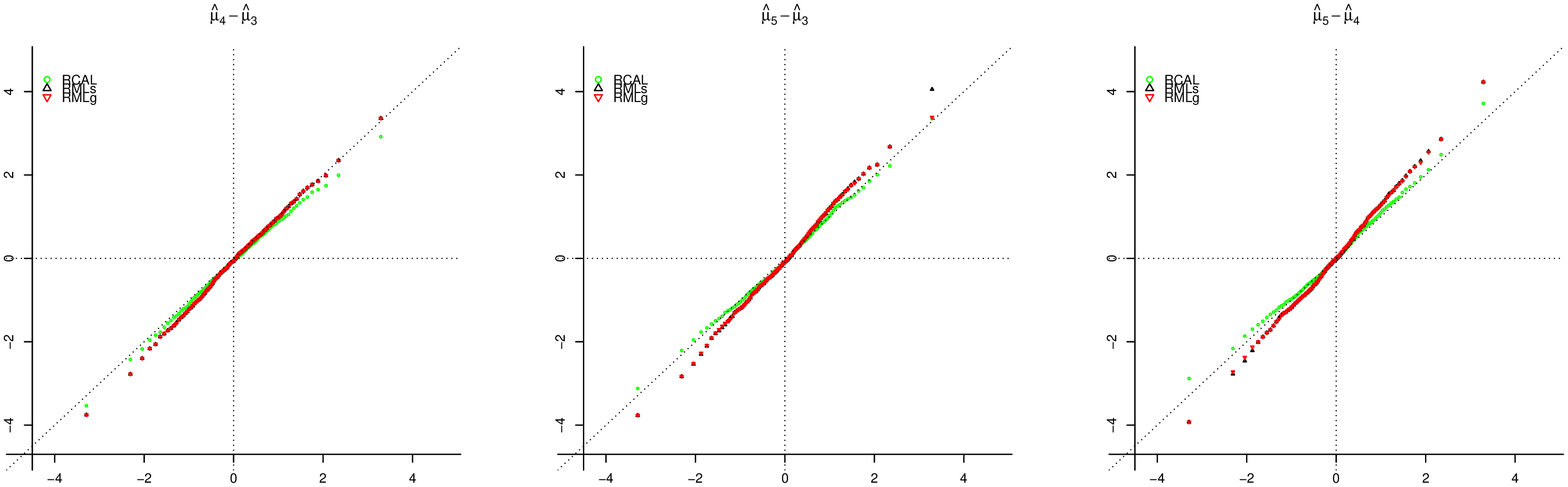}\vspace{-.1in}
\end{subfigure}
\caption{QQ plots of the standardized $\hat{\mu}_t - \hat{\mu}_k$ against standard normal.}
\label{fig:qq_ate_cross}
\end{figure}

\begin{figure}[H]
\centering
\includegraphics[scale=0.47]{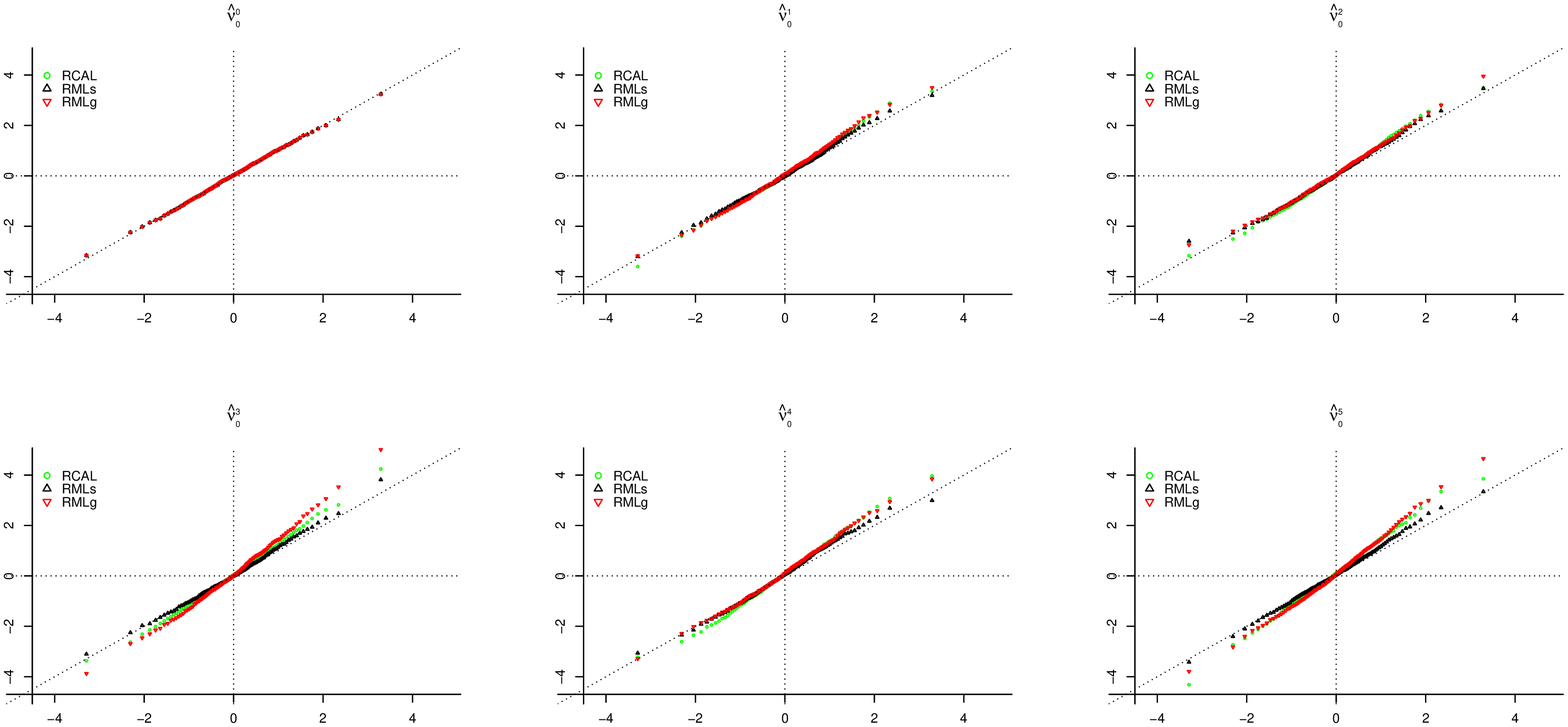}\vspace{-.1in}
\caption{QQ plots of the standardized $\hat{\nu}^{(k)}_t$against standard normal for $t = 0$}
\label{fig:qq_nu0_cross}
\end{figure}

\begin{figure}[H]
\centering
\includegraphics[scale=0.47]{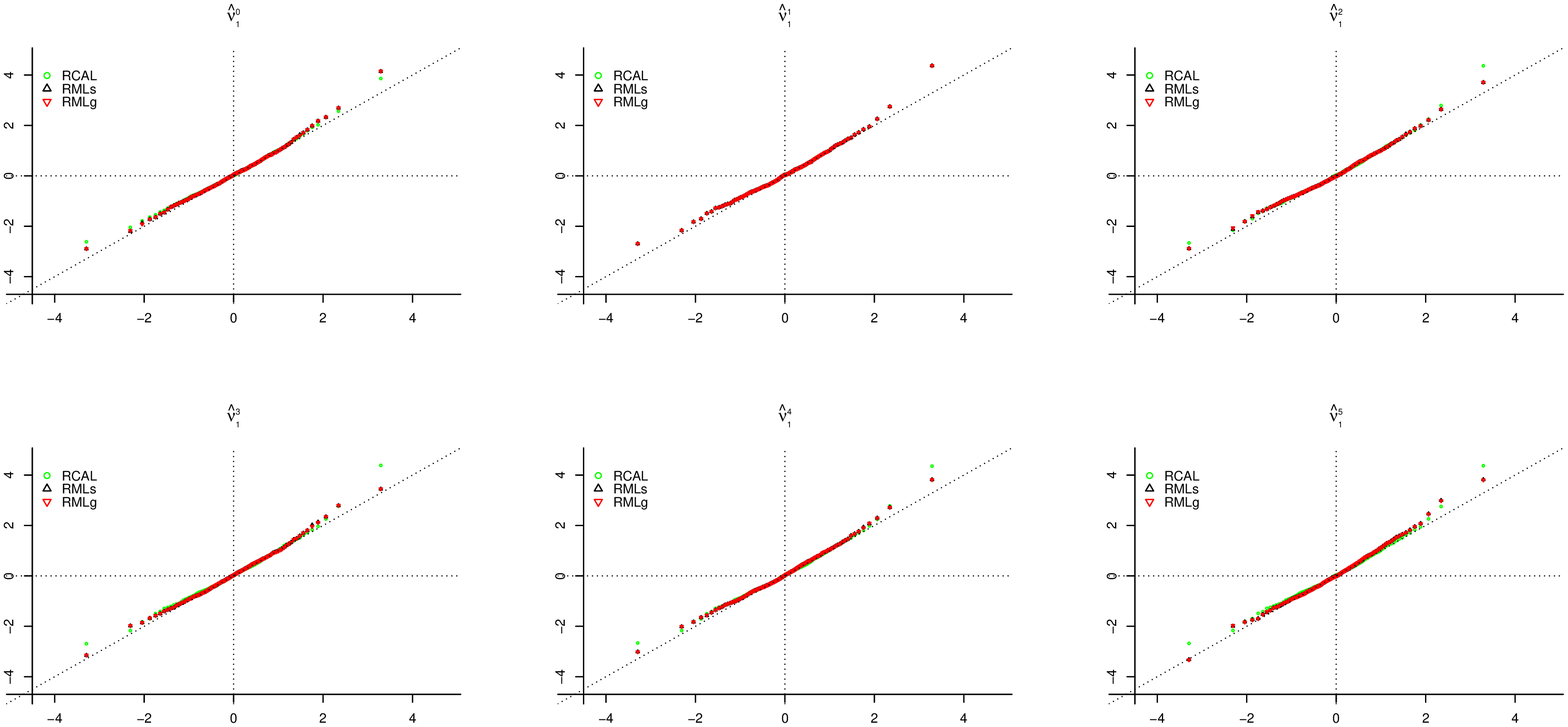}\vspace{-.1in}
\caption{QQ plots of the standardized $\hat{\nu}^{(k)}_t$against standard normal for $t = 1$}
\label{fig:qq_nu1_cross}
\end{figure}

\begin{figure}[H]
\centering
\includegraphics[scale=0.47]{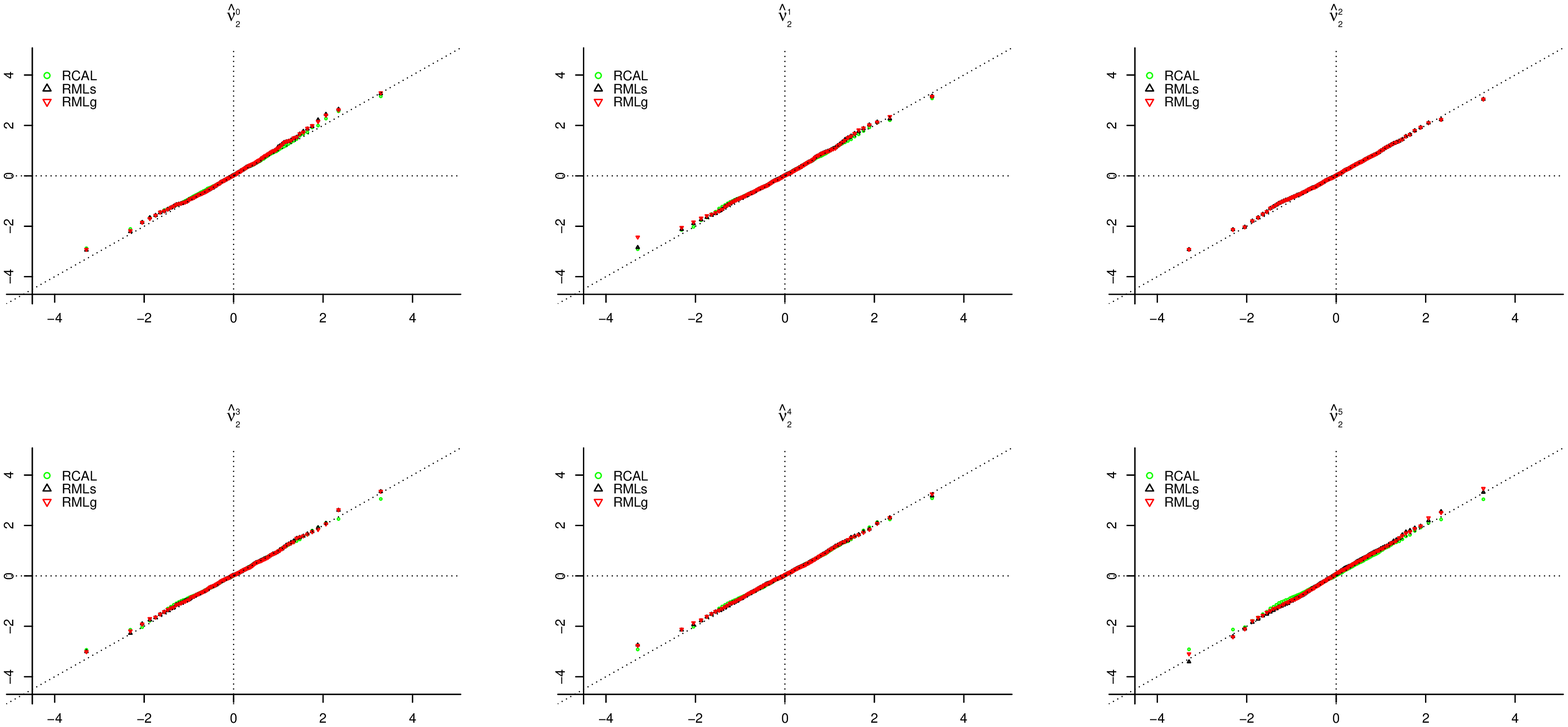}\vspace{-.1in}
\caption{QQ plots of the standardized $\hat{\nu}^{(k)}_t$against standard normal for $t = 2$}
\label{fig:qq_nu2_cross}
\end{figure}

\begin{figure}[H]
\centering
\includegraphics[scale=0.47]{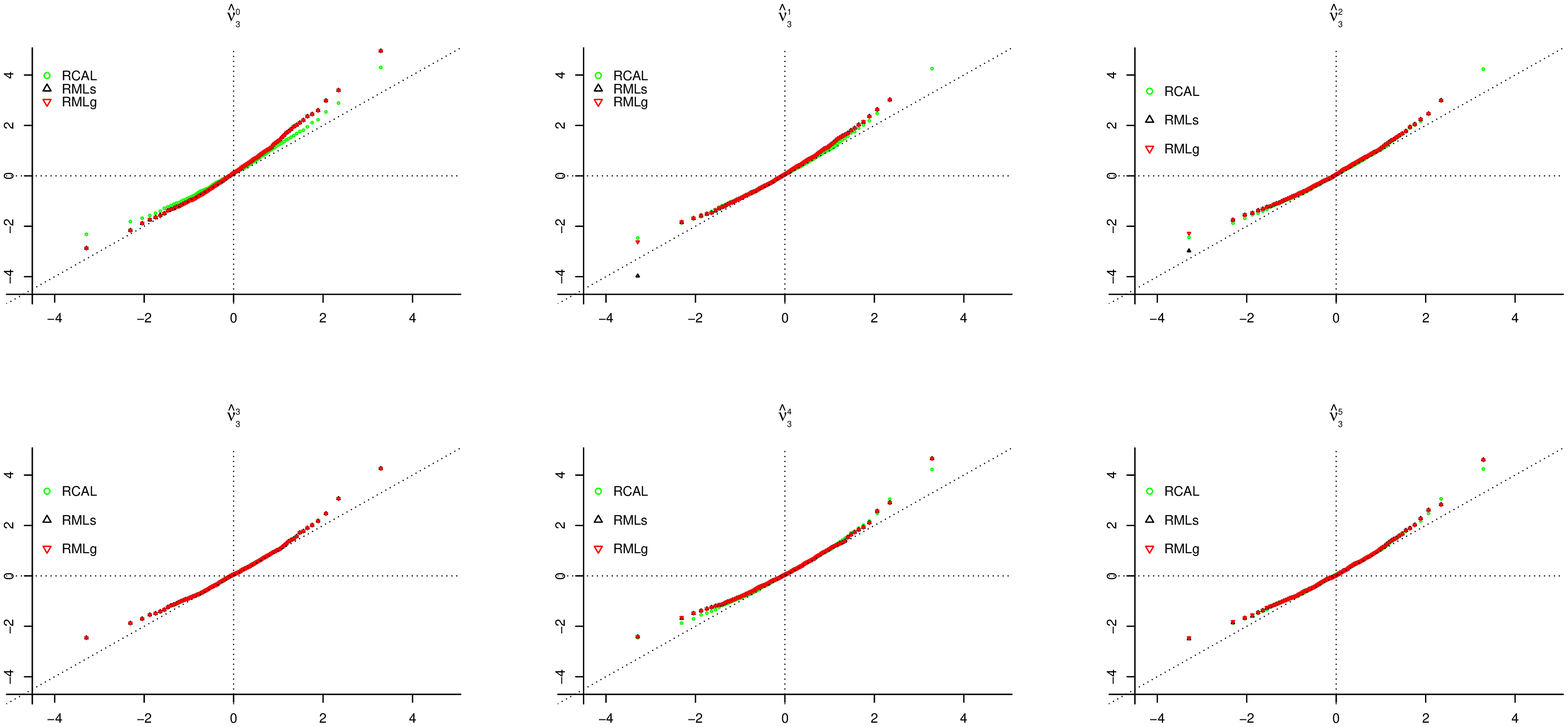}\vspace{-.1in}
\caption{QQ plots of the standardized $\hat{\nu}^{(k)}_t$against standard normal for $t = 3$}
\label{fig:qq_nu3_cross}
\end{figure}

\begin{figure}[H]
\centering
\includegraphics[scale=0.47]{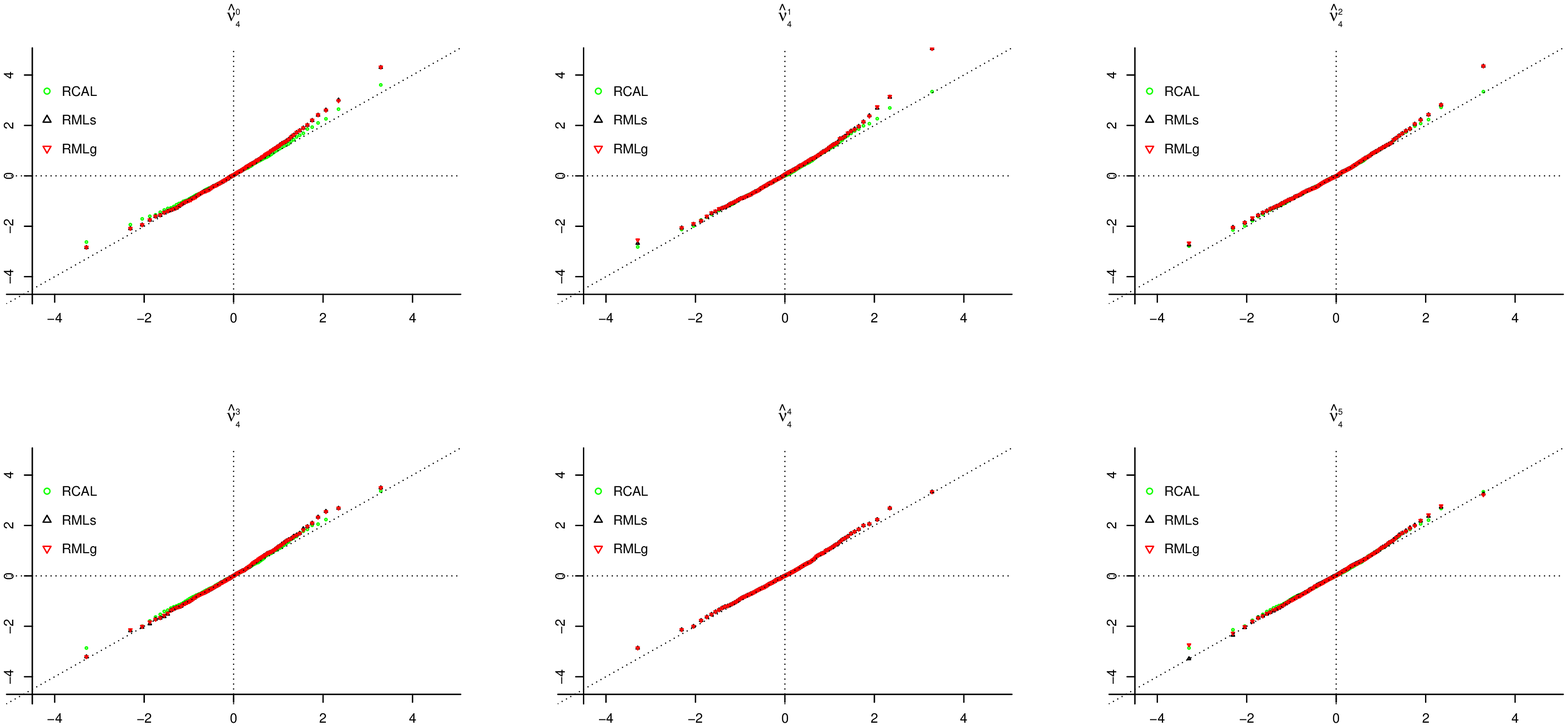}\vspace{-.1in}
\caption{QQ plots of the standardized $\hat{\nu}^{(k)}_t$against standard normal for $t = 4$}
\label{fig:qq_nu4_cross}
\end{figure}

\begin{figure}[H]
\centering
\includegraphics[scale=0.47]{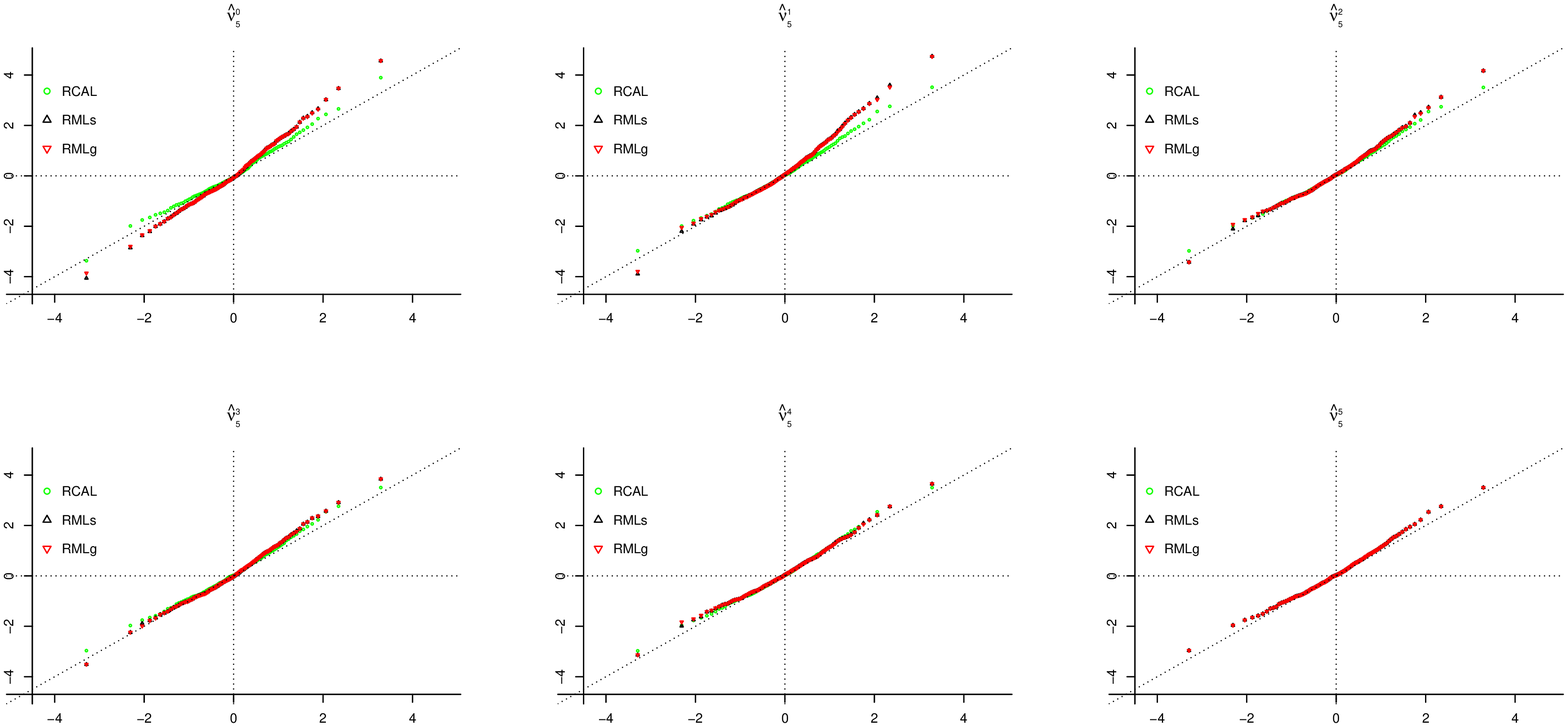}\vspace{-.1in}
\caption{QQ plots of the standardized $\hat{\nu}^{(k)}_t$against standard normal for $t = 5$}
\label{fig:qq_nu5_cross}
\end{figure}

\textbf{Additional results.}\;
Tables \ref{tb:ate_full_cross} and \ref{tb:nu_full_cross} are summaries of $\hat{\mu}_t - \hat{\mu}_k$ and $\hat{\nu}_t^{(k)}$ on full sample for $t, k = 0, 1, \ldots, 5$. Tables \ref{tb:ate_sub_cross} and \ref{tb:nu_sub_cross} are summarize of $\hat{\mu}_t - \hat{\mu}_k$ and $\hat{\nu}_t^{(k)}$ on sub samples for $t, k = 0, 1, \ldots, 5$.  Figures \ref{fig:box_nnz_ps_cross}  and \ref{fig:box_masd_cross} show boxplots of number of nonzero estimated coefficients in PS model and the boxplots of MASCD. Figures \ref{fig:qq_mu_cross}, \ref{fig:qq_ate_cross} and \ref{fig:qq_nu0_cross}--\ref{fig:qq_nu5_cross} are the QQ plots of the standardized $\hat{\mu}_t$, $\hat{\mu}_t - \hat{\mu}_k$ and $\hat{\nu}^{(k)}_t$ against standard normal.

From confidence intervals in Table \ref{tb:ate_full_cross}, we find that all methods CAL, RCAL, ML, RMLs and RMLg indicate that smoking could reduce birth weight and increasing the number of smoked cigarettes from 1-5 to 6-10 could further reduce birth weight at 95\% confidence level. Because 95\% confidence interval for $\mu_3 - \mu_2$ contains 0, there is less evidence that further increasing the number of smoked cigarettes from 6-10 could further reduce birth weight.

For estimation of $\hat{\mu}_t - \hat{\mu}_k$, we find (a) RCAL, RMLs and RMLg perform similarly to each other in terms of the repeated-sampling means and variances. However, the estimated variances from RCAL are close to the repeated-sampling variances, whereas those from RMLs and RMLg show underestimation; (b) in terms of coverage proportions, RCAL is either comparable or much better than RMLs and RMLg. For example, 90\% coverage proportions for $\mu_5 - \mu_0$ of RCAL, RMLs and RMLg are 0.890, 0.811 and 0.810, and 90\% coverage proportions for $\mu_5 - \mu_2$ of RCAL, RMLs and RMLg are 0.898, 0.822 and 0.822.

For estimation of $\hat{\nu}_t^{(k)}$, we find (a) RCAL, RMLs and RMLg perform similarly to each other in terms of variance; (b) in terms of coverage proportions, RCAL performs slightly better or much better than RMLs and RMLg in most cases. For example, 90\% coverage proportions for $\nu_3^{(0)}$ of RCAL, RMLs and RMLg are 0.894, 0.834 and 0.836, and 90\% coverage proportions for $\nu_5^{(0)}$ of RCAL, RMLs and RMLg are 0.889, 0.803 and 0.803. There are two cases that RMLs and RMLg are better than RCAL, they are $\nu^{(2)}_0$ and $\nu^{(4)}_0$. For $\nu^{(2)}_0$, 90\% coverage proportions of RMLs, RMLg and RCAL are 0.863, 0.857 and 0.836. For $\nu^{(4)}_0$, 90\% coverage proportions of RMLs, RMLg and RCAL are 0.851, 0.829 and 0.792.

From Figure \ref{fig:box_masd_cross} we see that RCAL is associated with comparable MASCD compared to RMLs and RMLg for treatments 0, 3, and 5, slightly smaller MASCD for treatments 2 and 4, and  slightly larger MASCD for treatment 1. But from Figure \ref{fig:box_nnz_ps_cross}, we see that RCAL is associated with a much smaller number of nonzero estimated coefficients, or greater sparsity, than RMLs and RMLg.

From the QQ plots, we easily see that standardized estimates of RCAL are more aligned with the standard normal in general.

\begin{figure}
\centering
\includegraphics[scale=0.47]{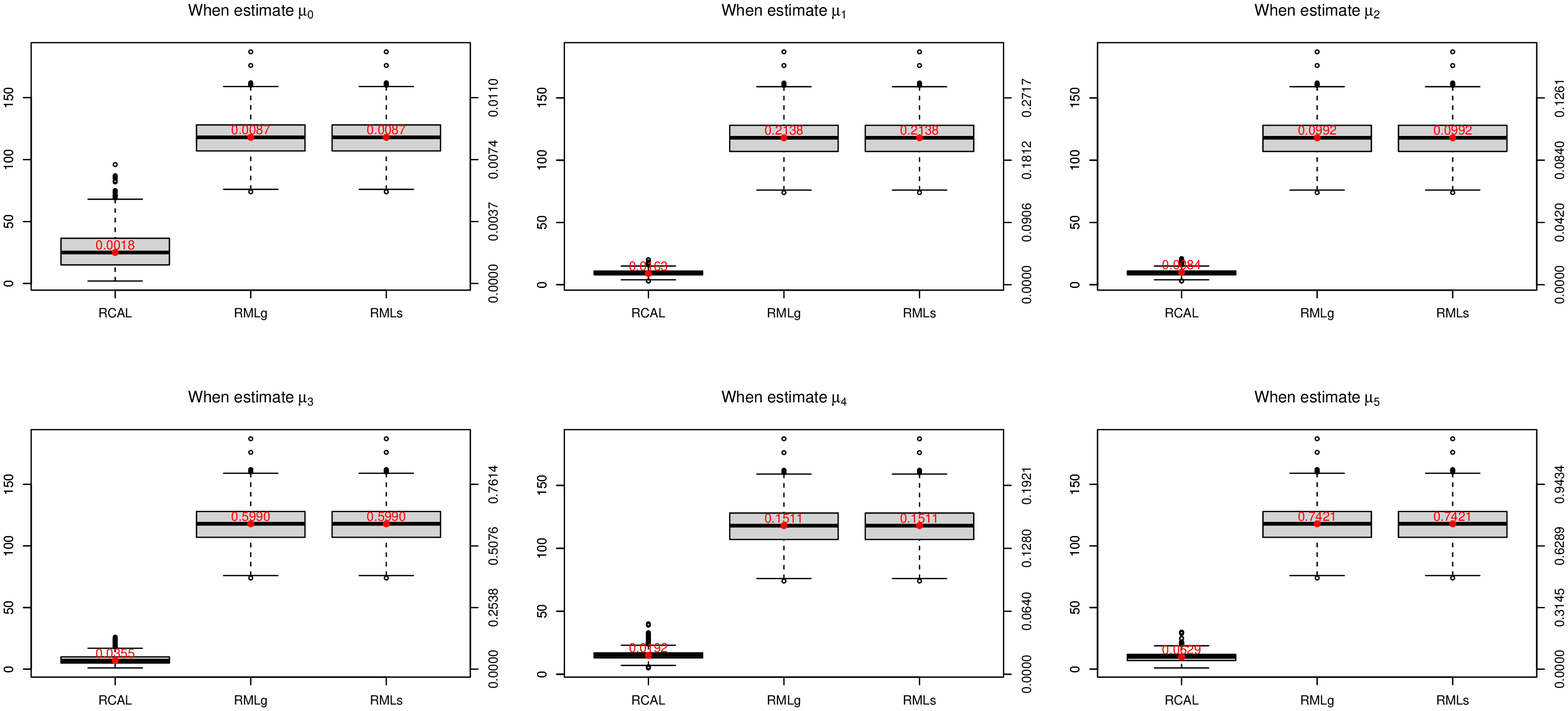}\vspace{-.1in}
\caption{Boxplot of number of nonzero estimated coefficients in PS model with tuning parameters selected as $\lambda.min$. Left y axis represents number of nonzero estimated coefficients. Right y axis represents ratio of number of nonzero estimated coefficients over corresponding treatment group size. Red number is the mean of ratios.}
\label{fig:box_nnz_ps_min}
\end{figure}

\begin{figure}
\centering
\includegraphics[scale=0.47]{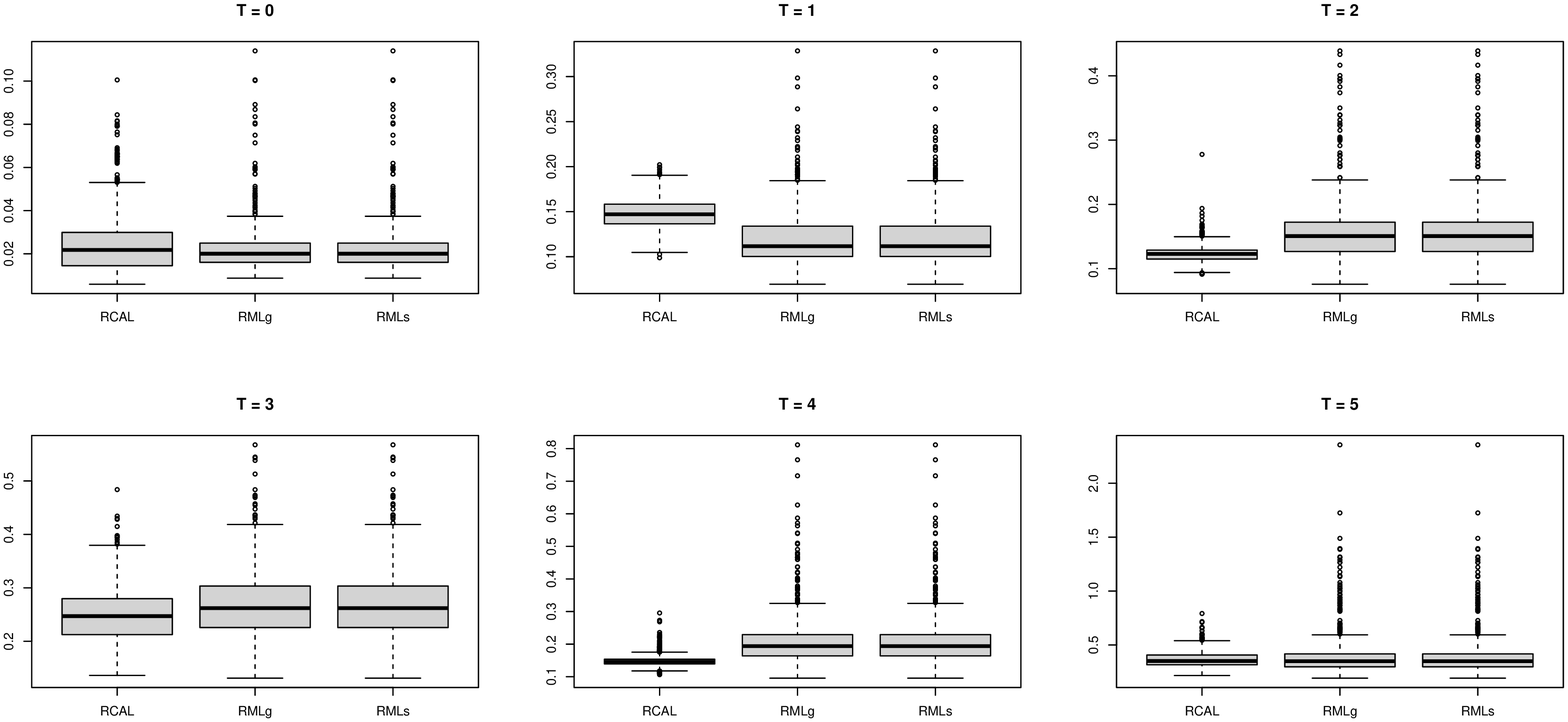}\vspace{-.1in}
\caption{Boxplot of MASCD with tuning parameters selected as $\lambda.min$.}
\label{fig:box_masd_min}
\end{figure}

\begin{figure}
\centering
\includegraphics[scale=0.47]{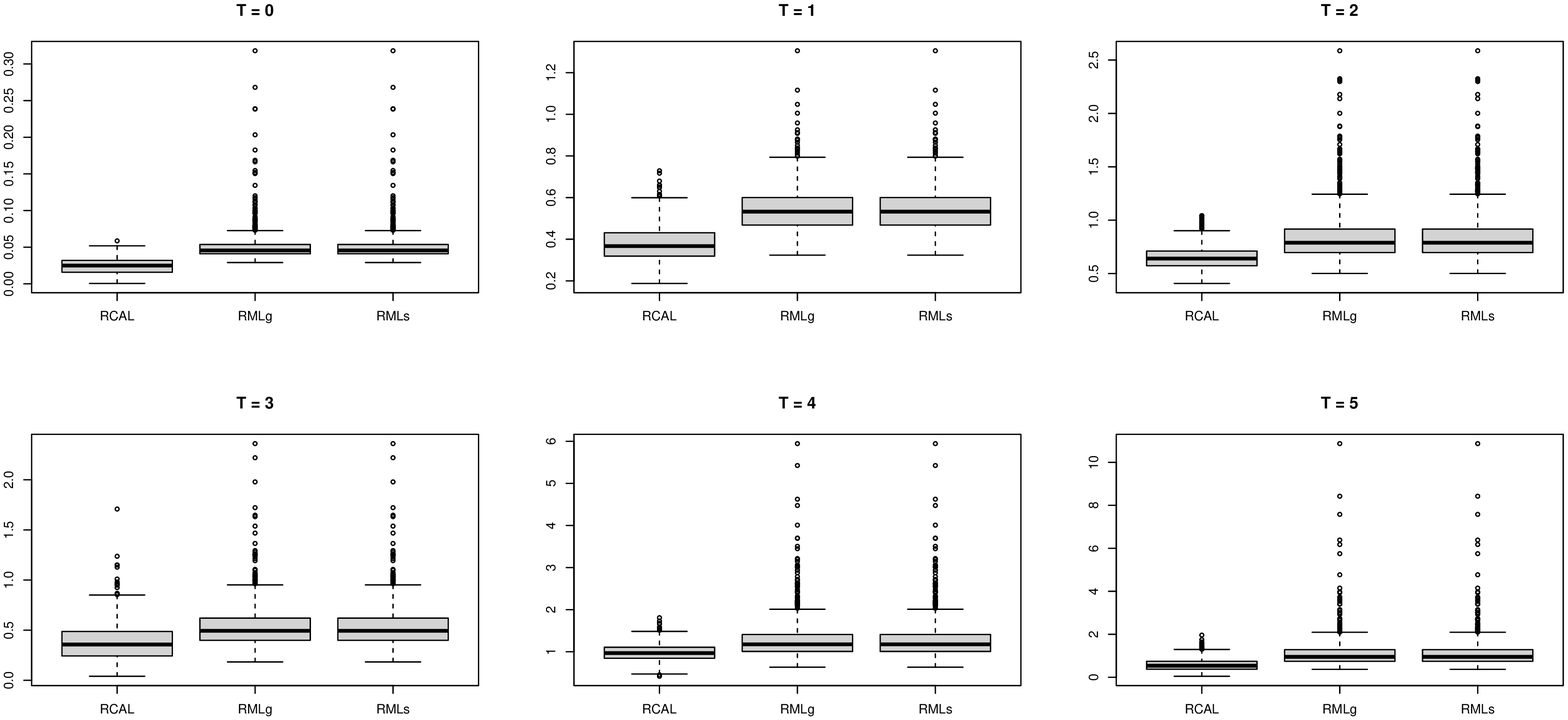}\vspace{-.1in}
\caption{Boxplot of RV with tuning parameters selected as $\lambda.min$.}
\label{fig:box_rv_min}
\end{figure}

\begin{figure}[H]
\centering
\includegraphics[scale=0.47]{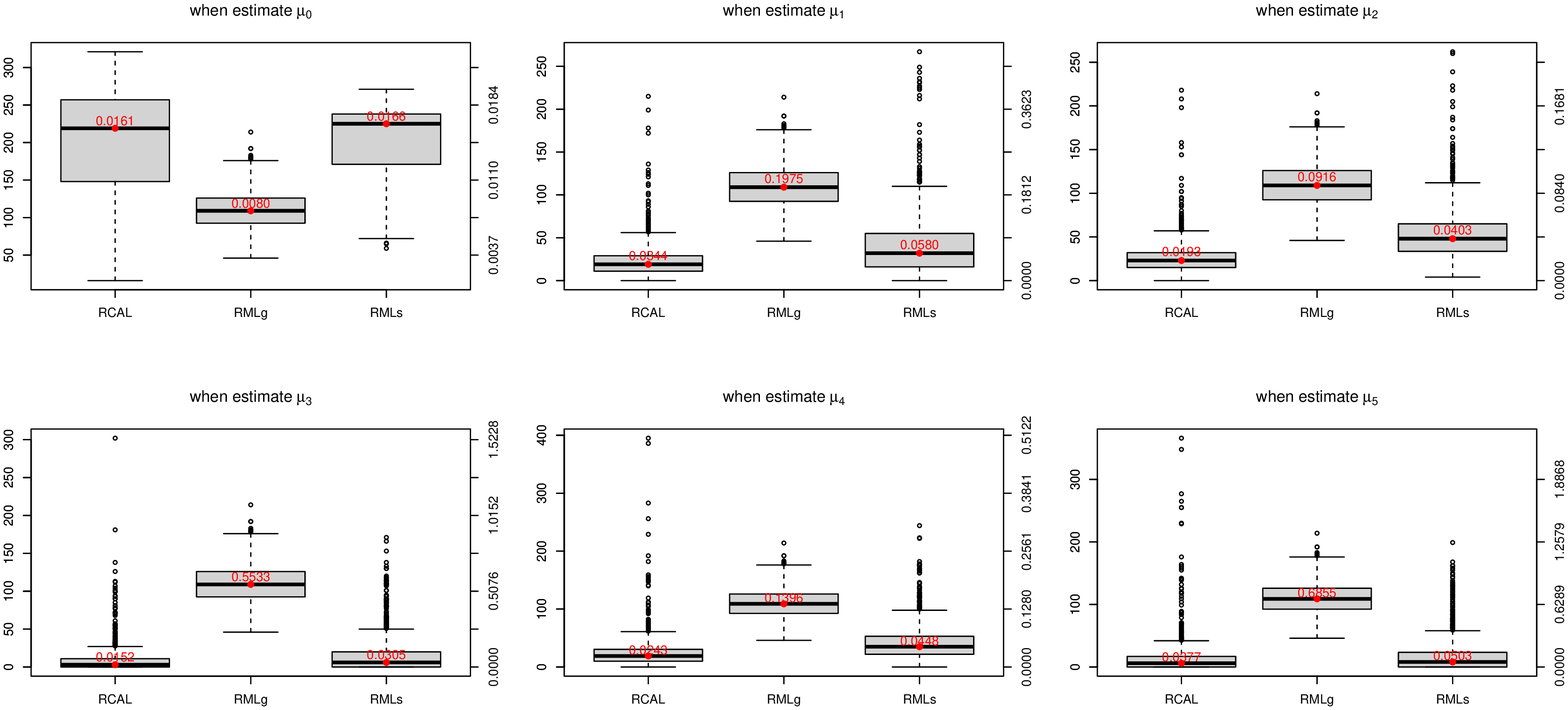}\vspace{-.1in}
\caption{Boxplot of number of nonzero estimated coefficients in OR model with tuning parameters selected as $\lambda.min$. Left y axis represents number of nonzero estimated coefficients. Right y axis represents ratio of number of nonzero estimated coefficients over corresponding treatment group size. Red number is the mean of ratios.}
\label{fig:box_nnz_or_min}
\end{figure}

\begin{figure}
\centering
\includegraphics[scale=0.47]{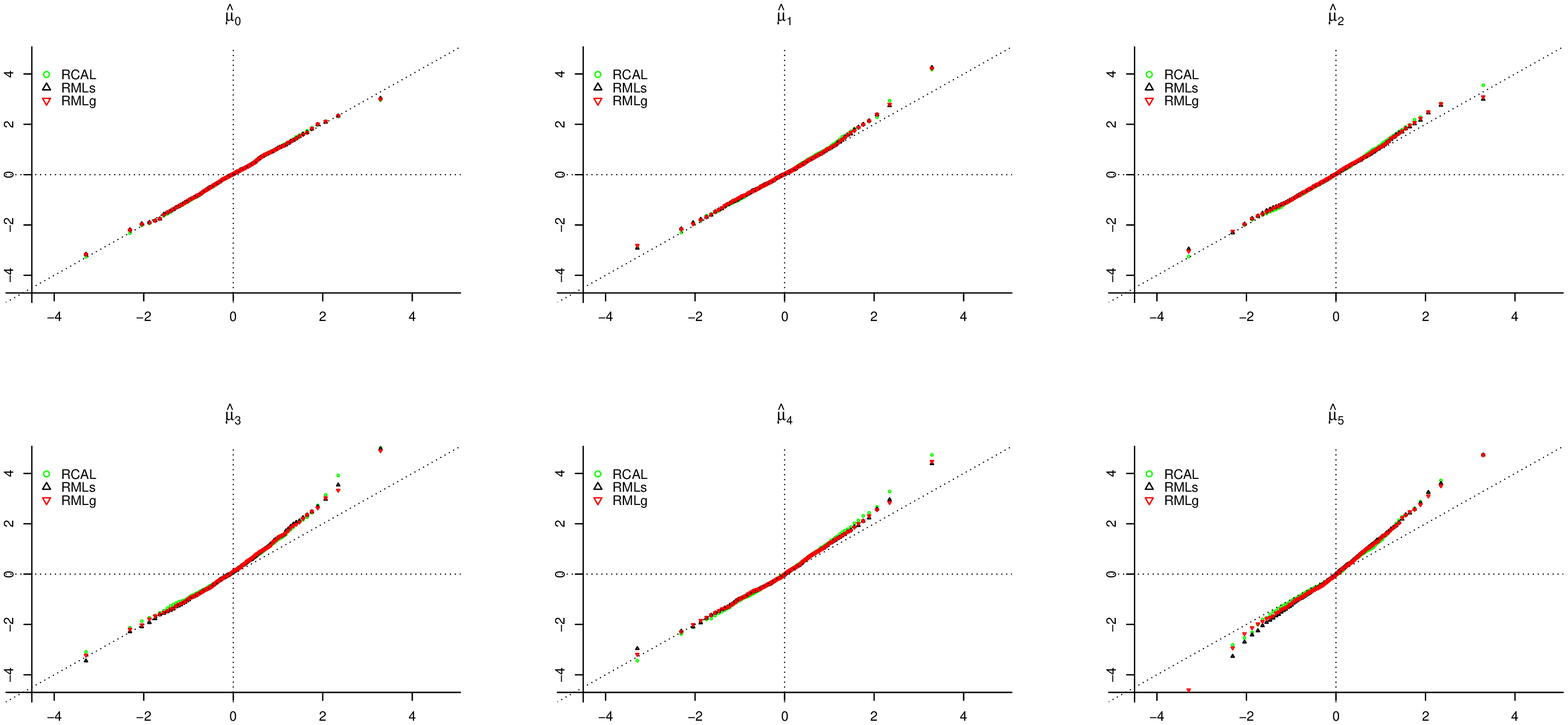}\vspace{-.1in}
\caption{QQ plots of the standardized $\hat{\mu}_t$ against standard normal with tuning parameters selected as $\lambda.min$.}
\label{fig:qq_mu_min}
\end{figure}

\begin{figure}
\begin{subfigure}{0.8\textwidth}
\centering
\includegraphics[scale=0.47]{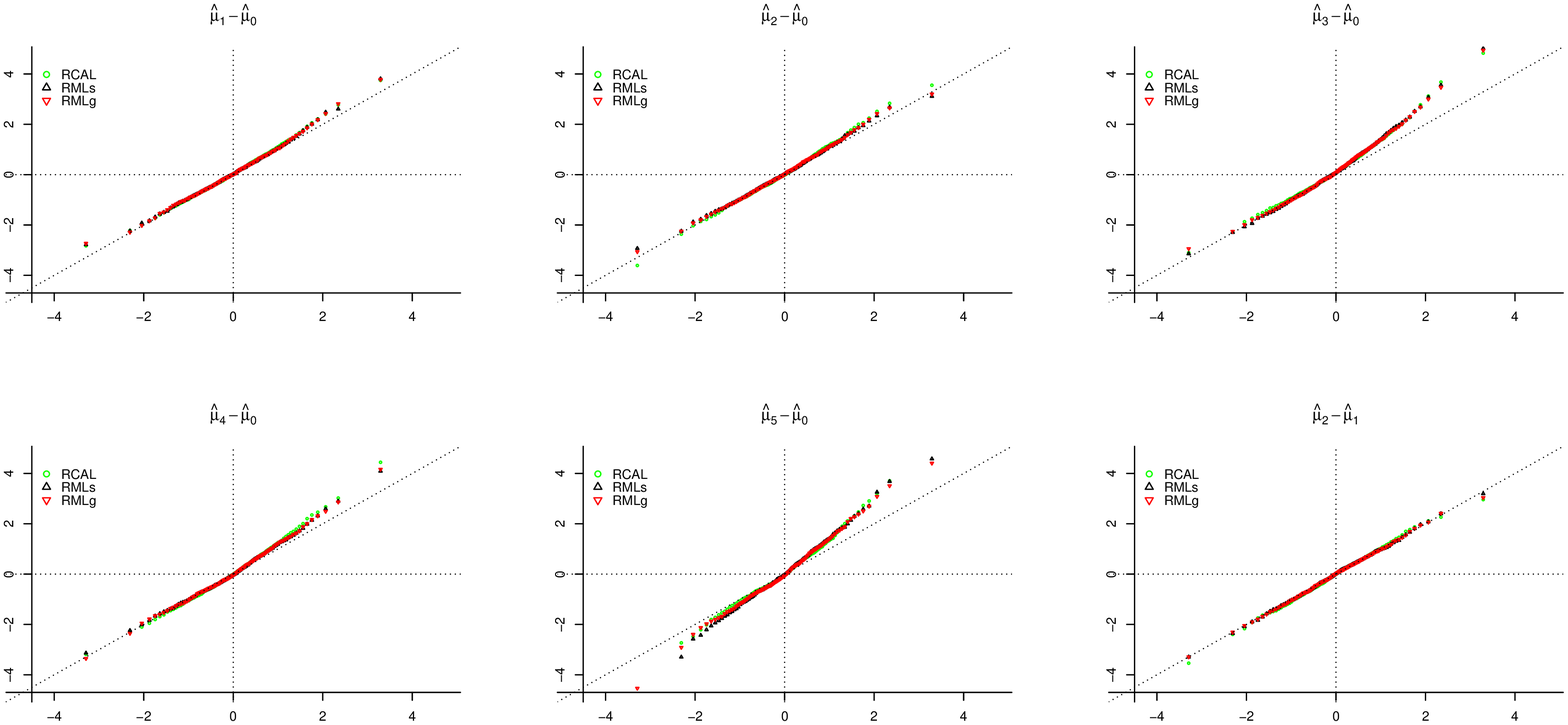}
\end{subfigure} %
\begin{subfigure}{0.8\textwidth}
\centering
\includegraphics[scale=0.47]{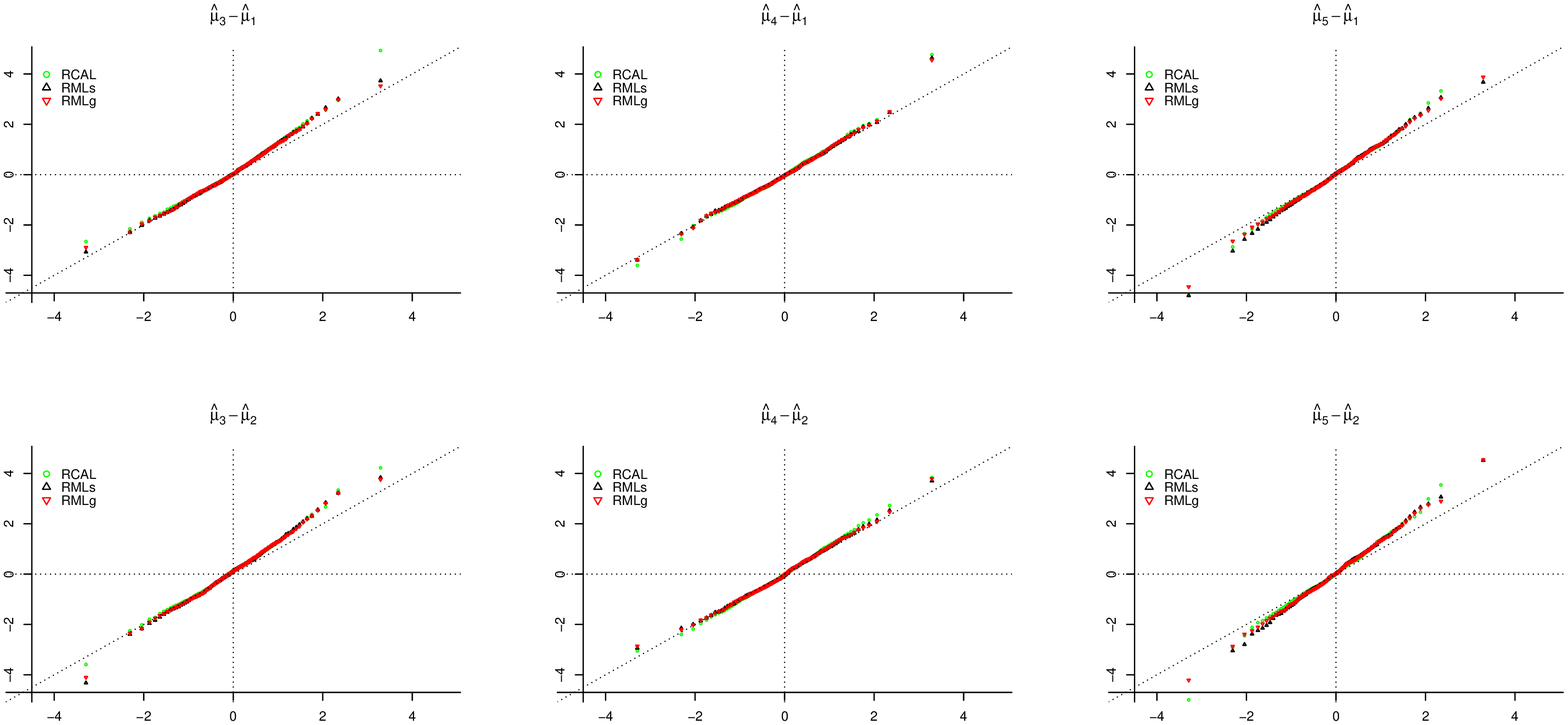}
\end{subfigure} %
\begin{subfigure}{0.8\textwidth}
\centering
\includegraphics[scale=0.47]{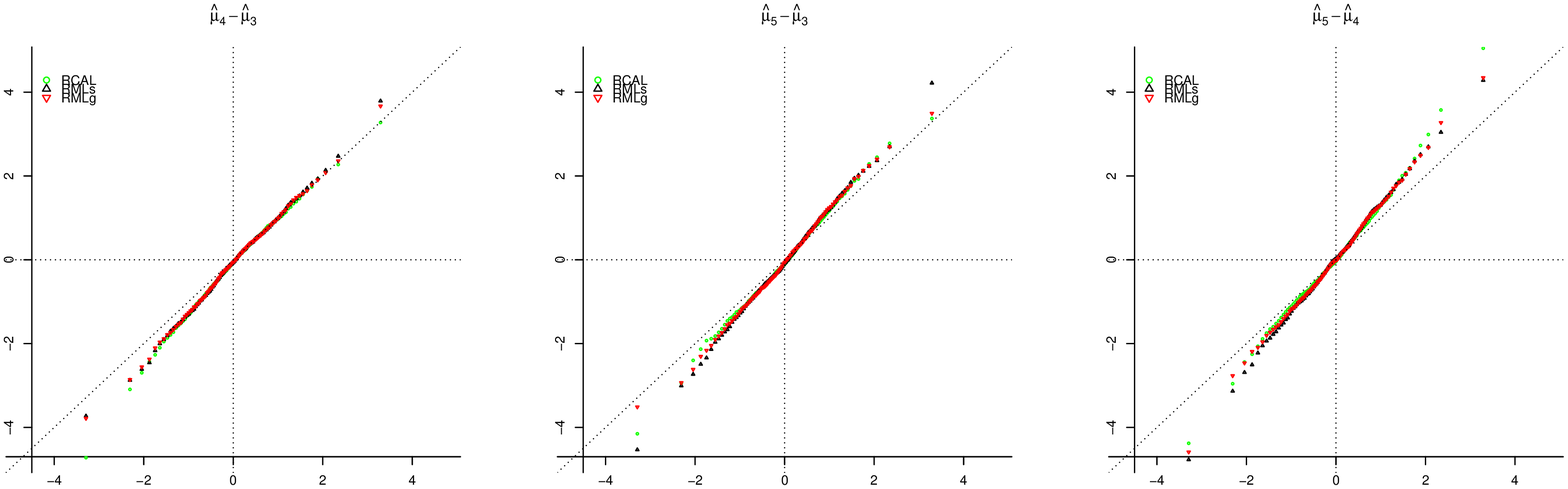}\vspace{-.1in}
\end{subfigure}
\caption{QQ plots of the standardized $\hat{\mu}_t - \hat{\mu}_k$ against standard normal with tuning parameters selected as $\lambda.min$.}
\label{fig:qq_ate_min}
\end{figure}

\begin{figure}[H]
\centering
\includegraphics[scale=0.47]{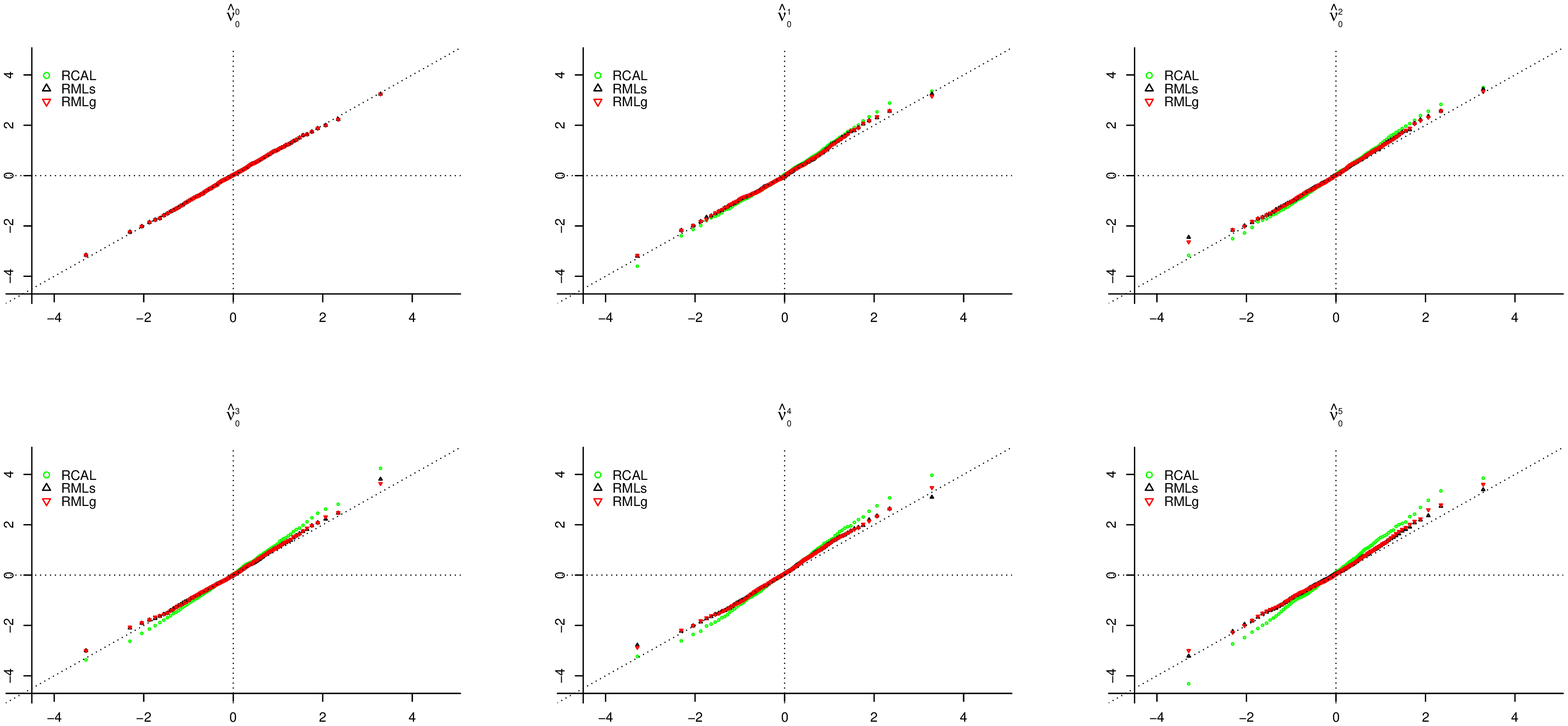}\vspace{-.1in}
\caption{QQ plots of the standardized $\hat{\nu}^{(k)}_t$against standard normal for $t = 0$ with tuning parameters selected as $\lambda.min$.}
\label{fig:qq_nu0_min}
\end{figure}

\begin{figure}[H]
\centering
\includegraphics[scale=0.47]{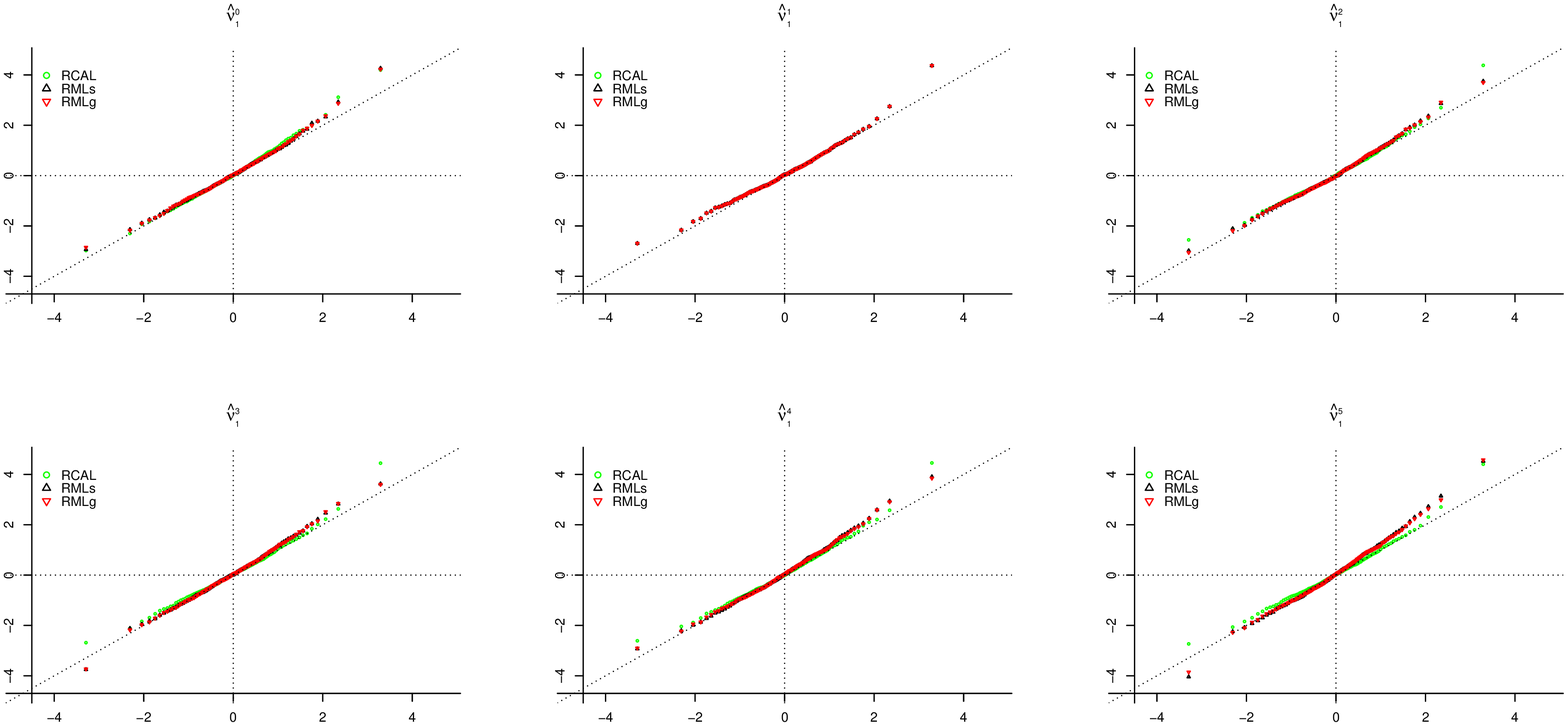}\vspace{-.1in}
\caption{QQ plots of the standardized $\hat{\nu}^{(k)}_t$against standard normal for $t = 1$ with tuning parameters selected as $\lambda.min$.}
\label{fig:qq_nu1_min}
\end{figure}

\begin{figure}[H]
\centering
\includegraphics[scale=0.47]{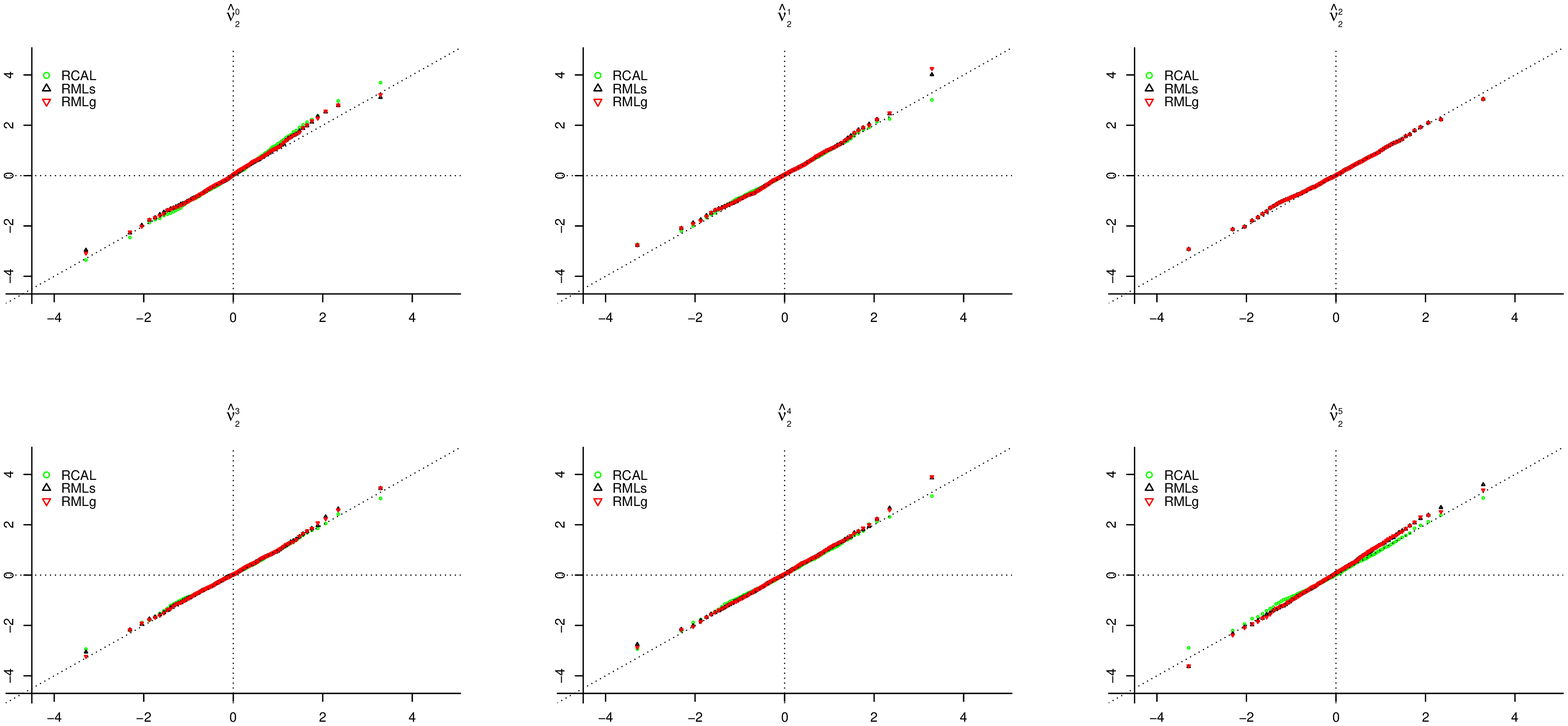}\vspace{-.1in}
\caption{QQ plots of the standardized $\hat{\nu}^{(k)}_t$against standard normal for $t = 2$ with tuning parameters selected as $\lambda.min$.}
\label{fig:qq_nu2_min}
\end{figure}

\begin{figure}[H]
\centering
\includegraphics[scale=0.47]{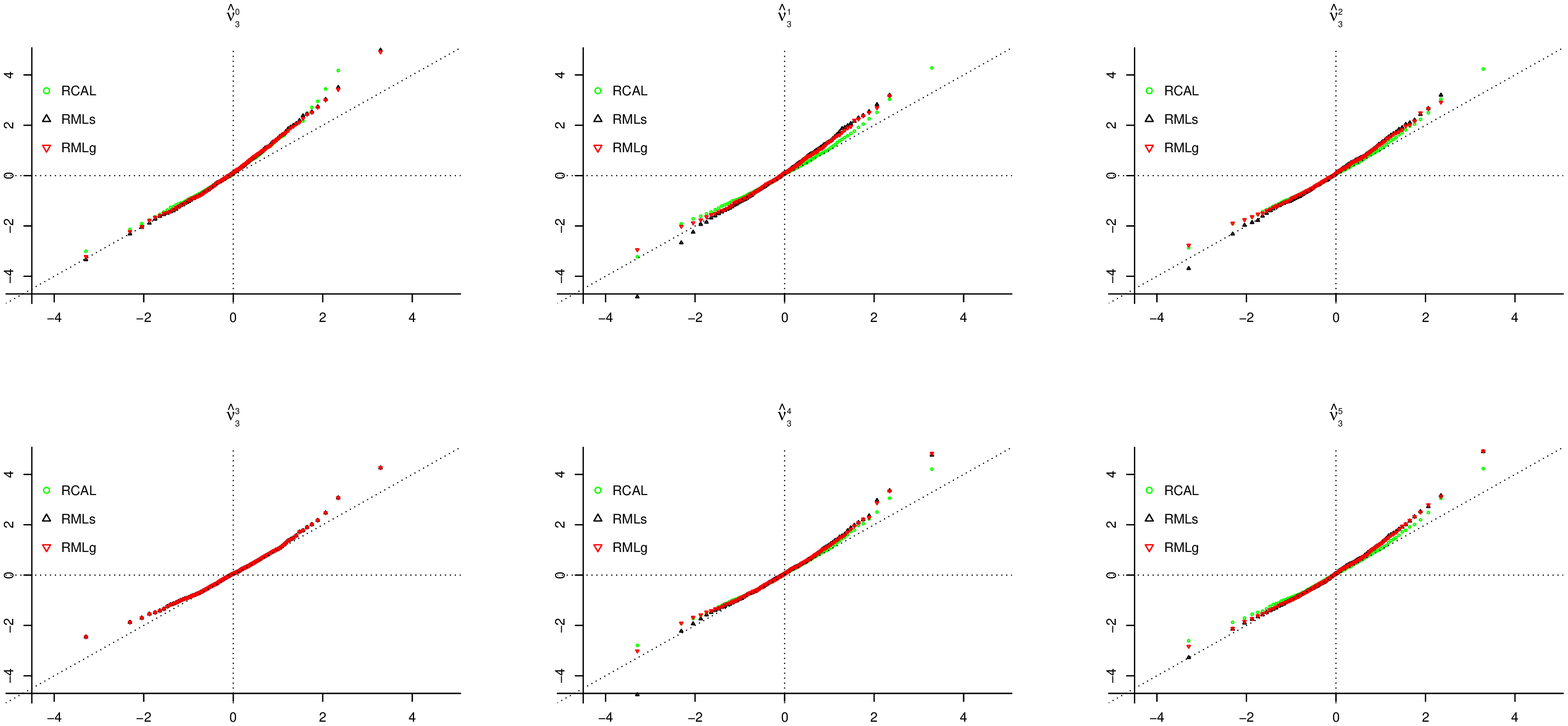}\vspace{-.1in}
\caption{QQ plots of the standardized $\hat{\nu}^{(k)}_t$against standard normal for $t = 3$ with tuning parameters selected as $\lambda.min$.}
\label{fig:qq_nu3_min}
\end{figure}

\begin{figure}[H]
\centering
\includegraphics[scale=0.47]{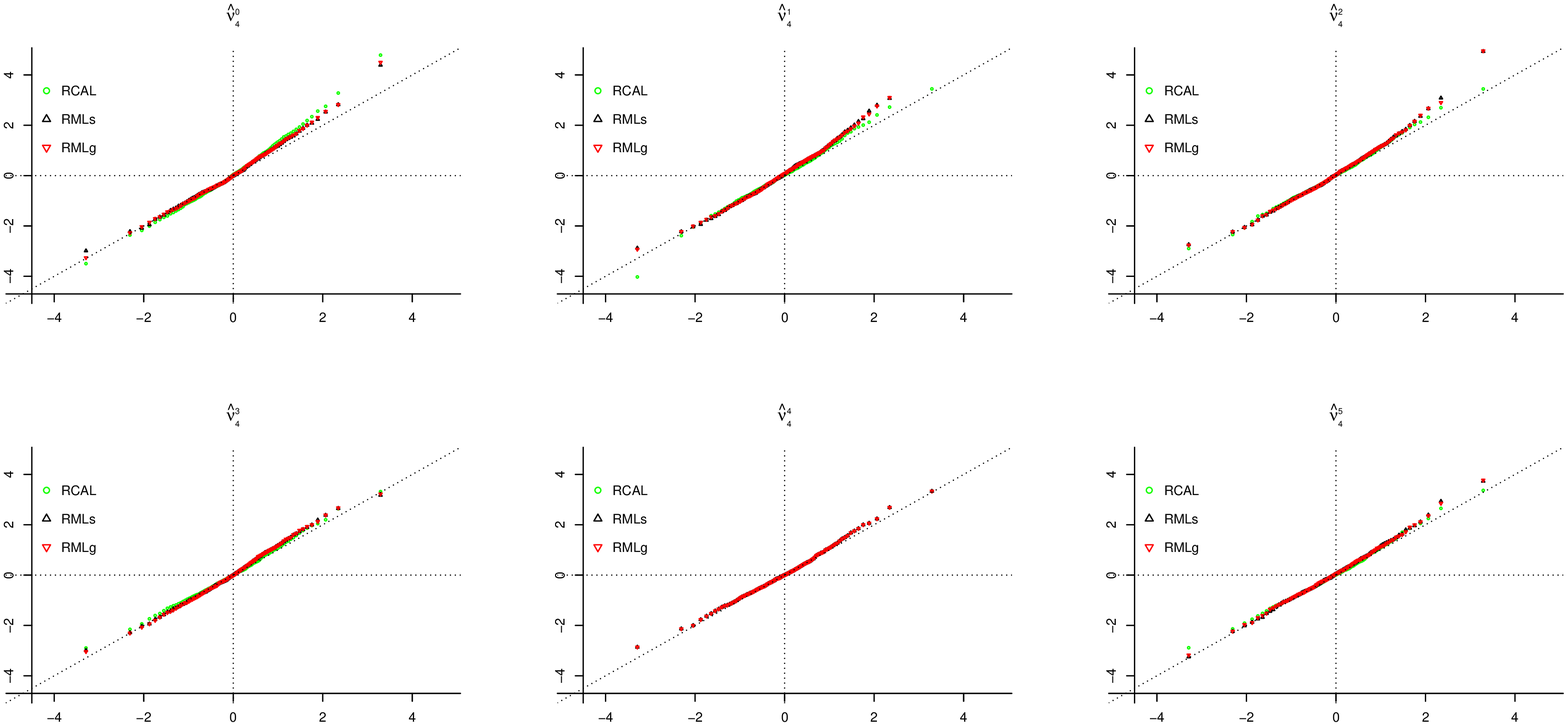}\vspace{-.1in}
\caption{QQ plots of the standardized $\hat{\nu}^{(k)}_t$against standard normal for $t = 4$ with tuning parameters selected as $\lambda.min$.}
\label{fig:qq_nu4_min}
\end{figure}

\begin{figure}[H]
\centering
\includegraphics[scale=0.47]{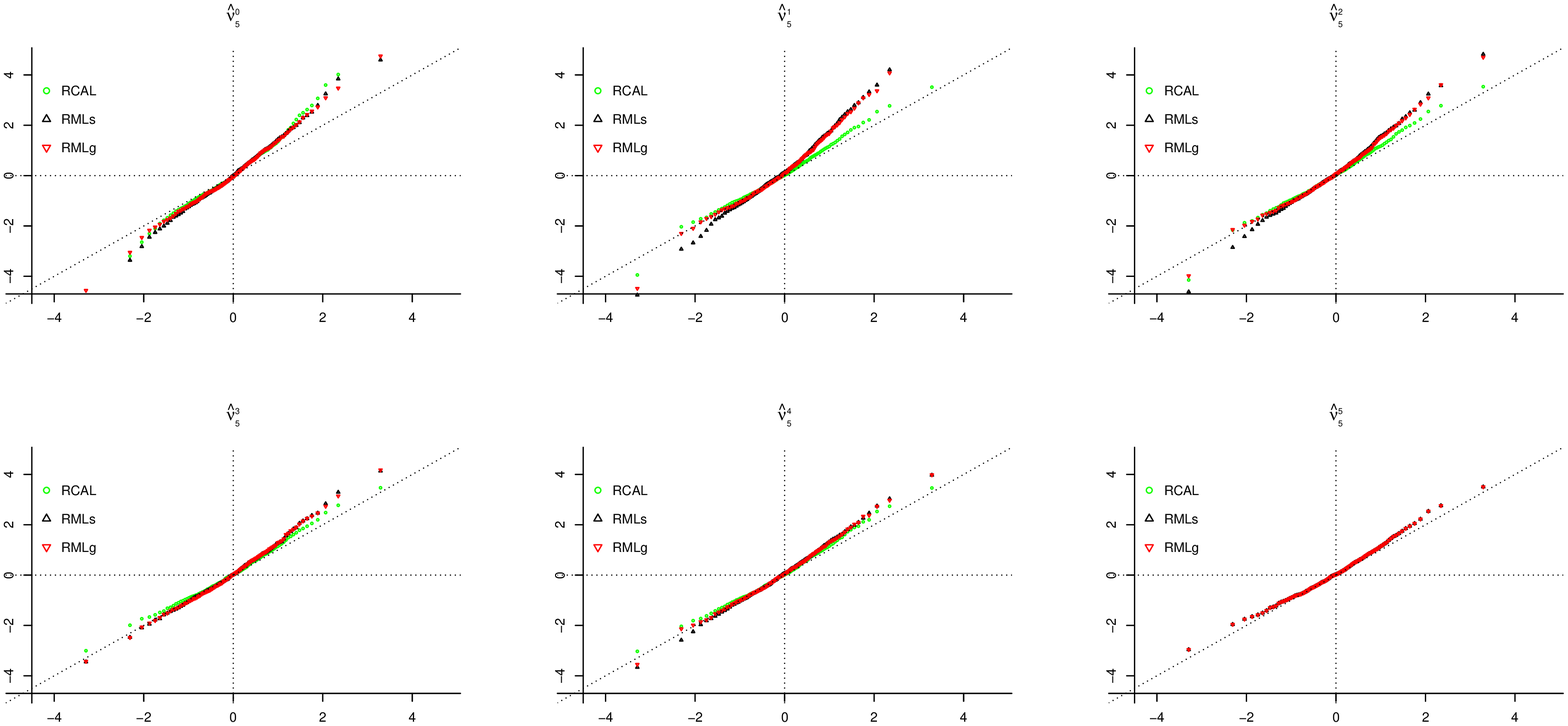}\vspace{-.1in}
\caption{QQ plots of the standardized $\hat{\nu}^{(k)}_t$against standard normal for $t = 5$ with tuning parameters selected as $\lambda.min$.}
\label{fig:qq_nu5_min}
\end{figure}

\textbf{Additional results with tuning parameters selected as $\lambda.min$.}\;
Tables \ref{tb:mu_full_min}, \ref{tb:ate_full_min} and \ref{tb:nu_full_min} are summarizes of $\hat{\mu}_t$, $\hat{\mu}_t - \hat{\mu}_k$ and $\hat{\nu}_t^{(k)}$ on full sample for $t, k = 0, 1, \ldots, 5$ with tuning parameters selected as $\lambda.min$ for PS and OR models in all methods.
Similarly, Tables \ref{tb:mu_sub_min}, \ref{tb:ate_sub_min} and \ref{tb:nu_sub_min} are summaries of  $\hat{\mu}_t$, $\hat{\mu}_t - \hat{\mu}_k$ and $\hat{\nu}_t^{(k)}$ on sub-samples for $t, k = 0, 1, \ldots, 5$.
Figures \ref{fig:box_nnz_ps_min}, \ref{fig:box_masd_min} and \ref{fig:box_rv_min} show boxplots of number of nonzero estimated coefficients in PS model, MASCD and RV.
Figure \ref{fig:box_nnz_or_min} shows  boxplots of number of nonzero estimated coefficients in OR model.
Figures \ref{fig:qq_mu_min}, \ref{fig:qq_ate_min} and \ref{fig:qq_nu0_min}--\ref{fig:qq_nu5_min} are the QQ plots of the standardized $\hat{\mu}_t$, $\hat{\mu}_t - \hat{\mu}_k$ and $\hat{\nu}^{(k)}_t$ against standard normal.
With $\lambda.min$ for all methods, similar conclusions are obtained as above although the advantage of RCAL over RMLs and RMLg is somewhat decreased.

\begin{figure}
\centering
\includegraphics[scale=0.47]{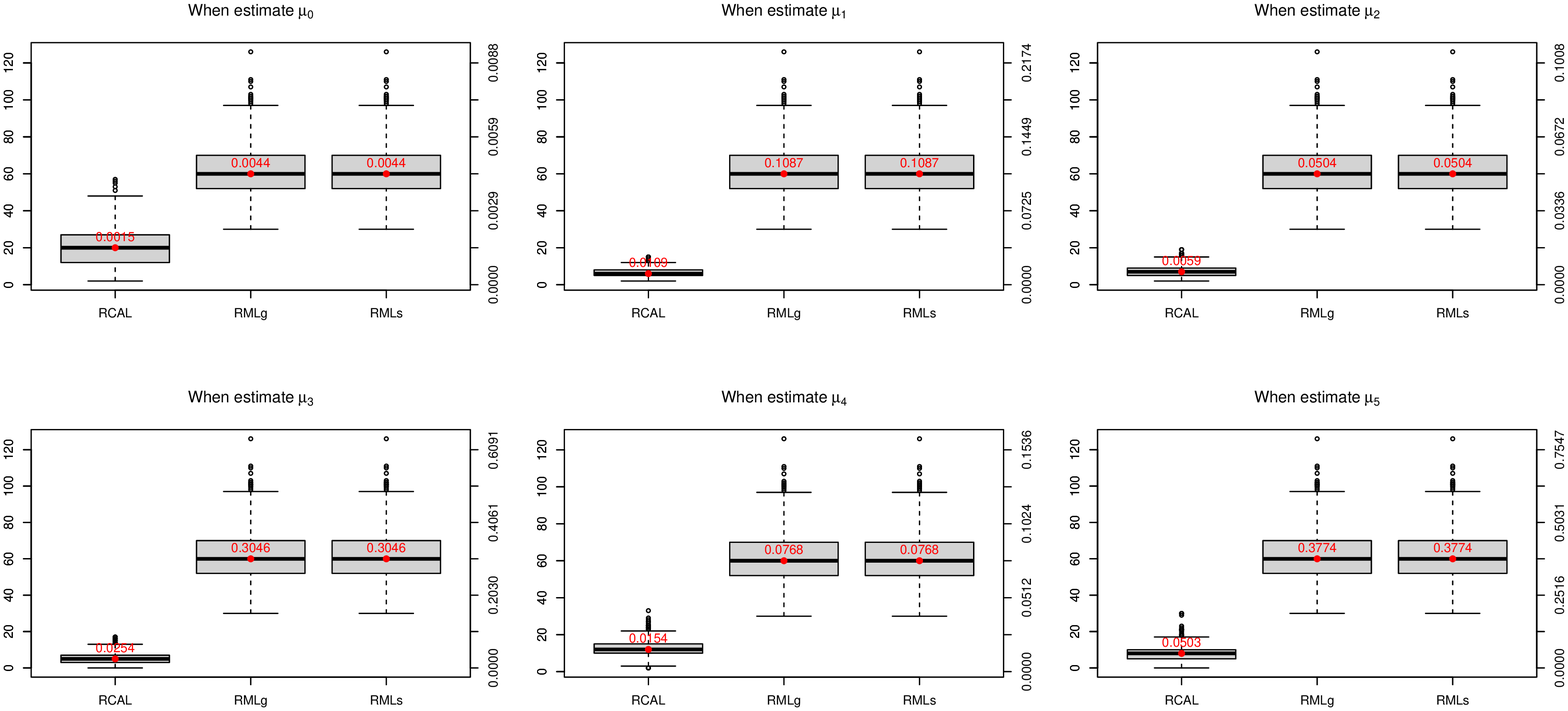}\vspace{-.1in}
\caption{Boxplot of number of nonzero estimated coefficients in PS model with tuning parameters selected as $\lambda.1se$. Left y axis represents number of nonzero estimated coefficients. Right y axis represents number of nonzero estimated coefficients over corresponding treatment group size. Red number is the mean of ratios.}
\label{fig:box_nnz_ps_1se}
\end{figure}

\begin{figure}
\centering
\includegraphics[scale=0.47]{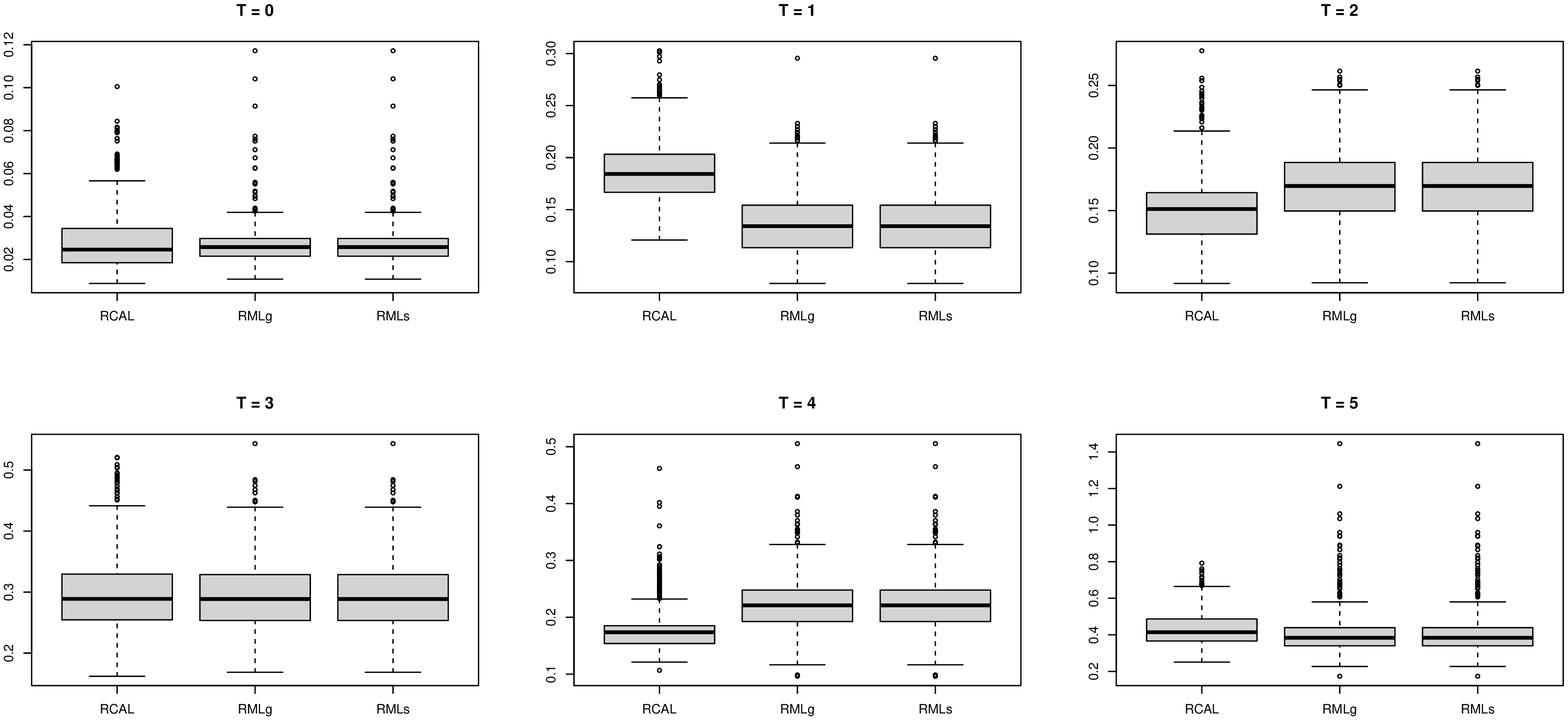}\vspace{-.1in}
\caption{Boxplot of MASCD with tuning parameters selected as $\lambda.1se$.}
\label{fig:box_masd_1se}
\end{figure}

\begin{figure}
\centering
\includegraphics[scale=0.47]{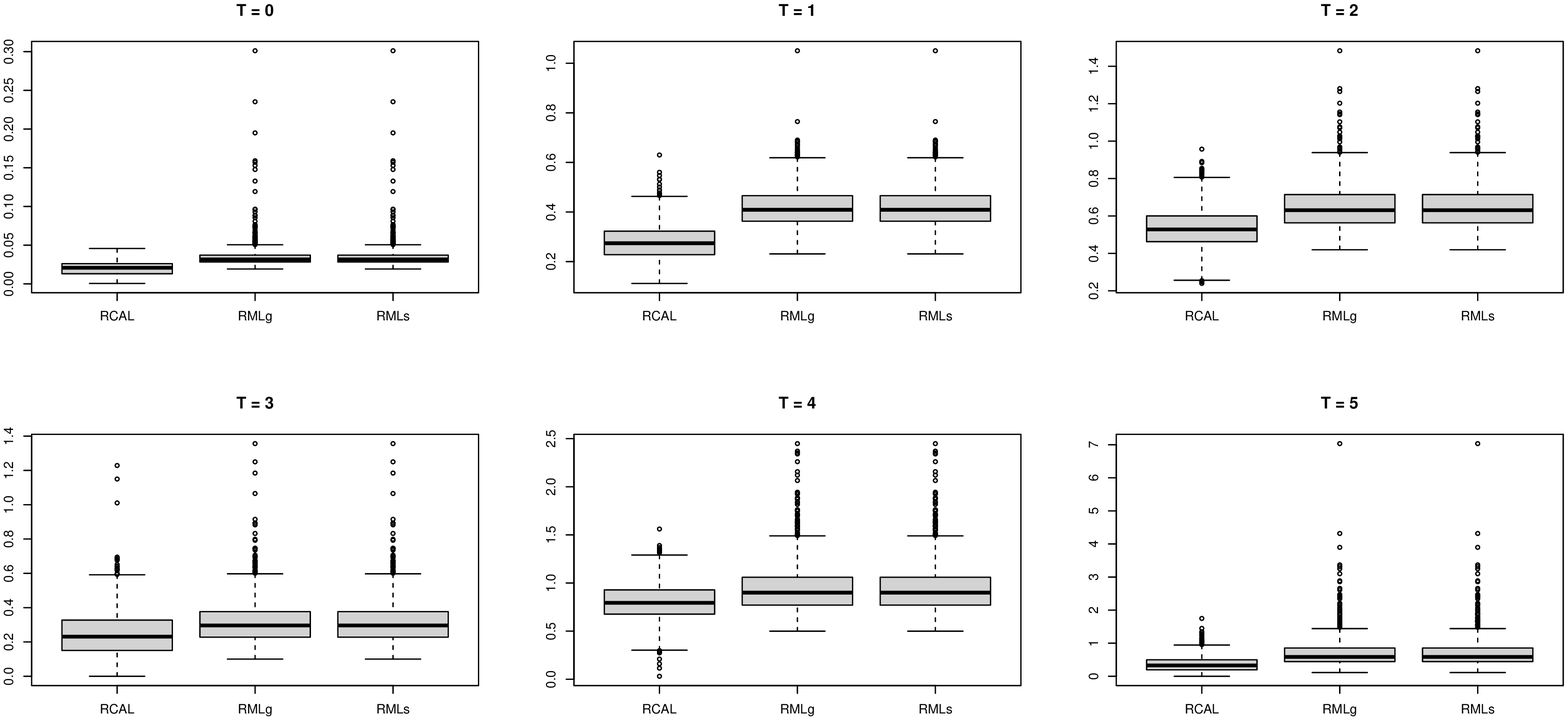}\vspace{-.1in}
\caption{Boxplot of RV with tuning parameters selected as $\lambda.1se$.}
\label{fig:box_rv_1se}
\end{figure}

\begin{figure}[H]
\centering
\includegraphics[scale=0.47]{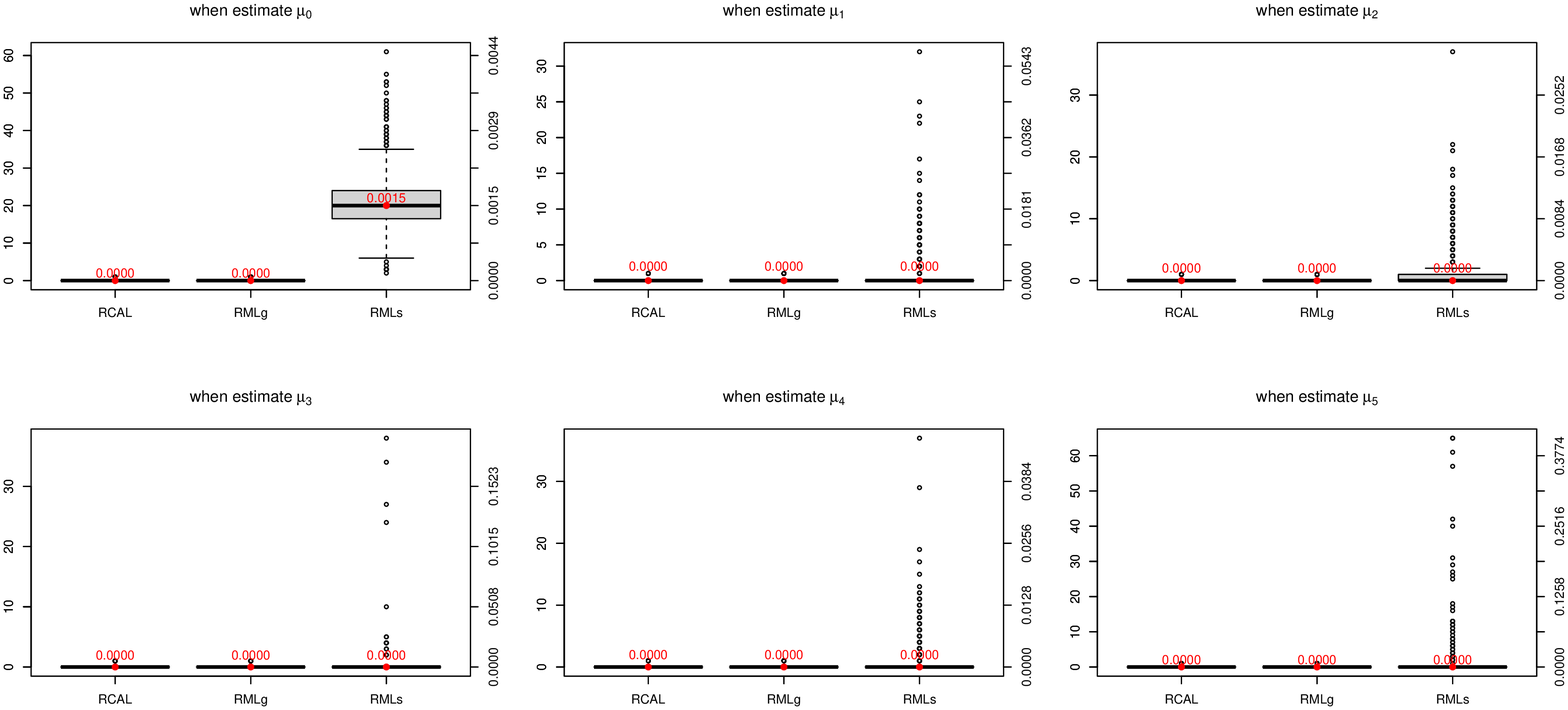}\vspace{-.1in}
\caption{Boxplot of number of nonzero estimated coefficients in OR model with tuning parameters selected as $\lambda.1se$. Left y axis represents number of nonzero estimated coefficients. Right y axis represents ratio of number of nonzero estimated coefficients over corresponding treatment group size. Red number is the mean of ratios.}
\label{fig:box_nnz_or_1se}
\end{figure}

\begin{figure}
\centering
\includegraphics[scale=0.47]{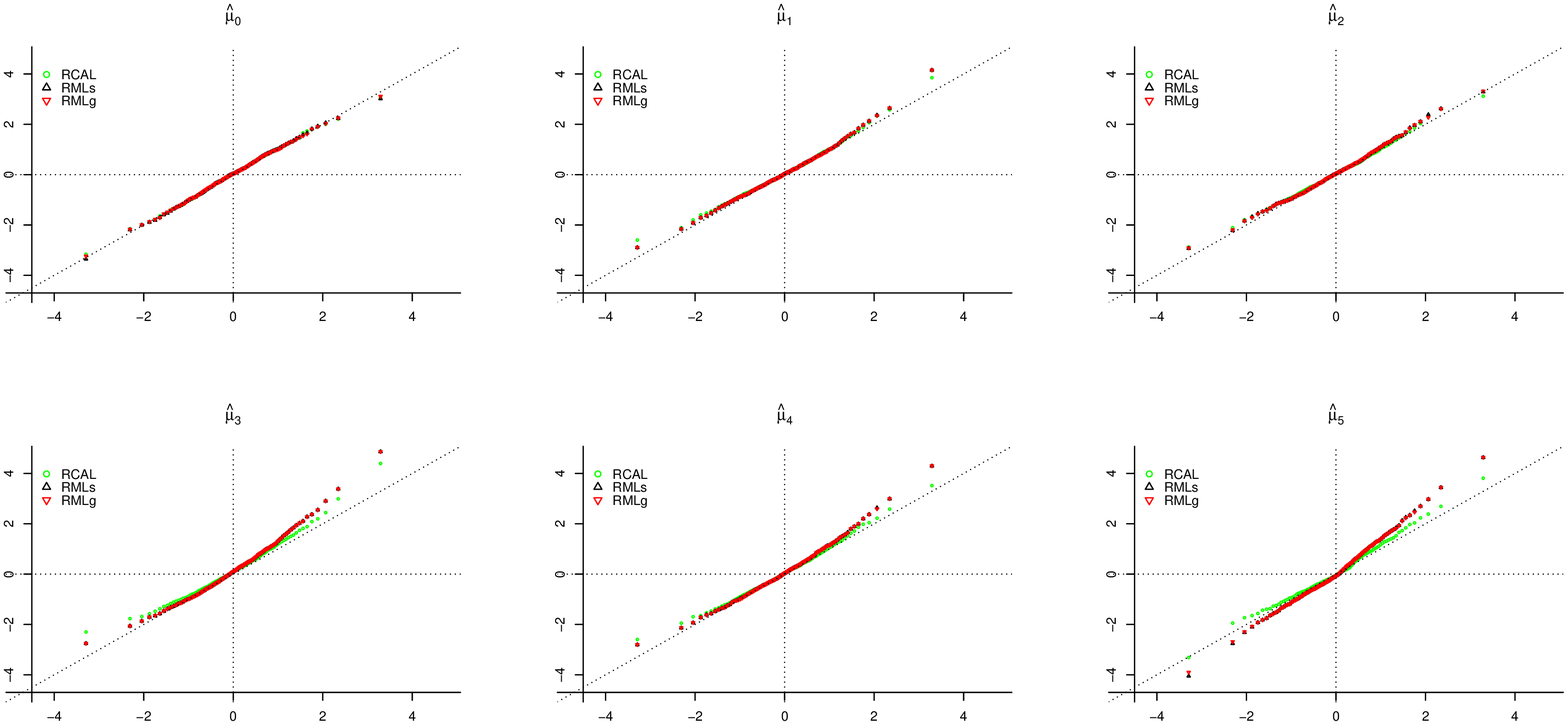}\vspace{-.1in}
\caption{QQ plots of the standardized $\hat{\mu}_t$ against standard normal with tuning parameters selected as $\lambda.1se$.}
\label{fig:qq_mu_1se}
\end{figure}

\begin{figure}
\begin{subfigure}{0.8\textwidth}
\centering
\includegraphics[scale=0.47]{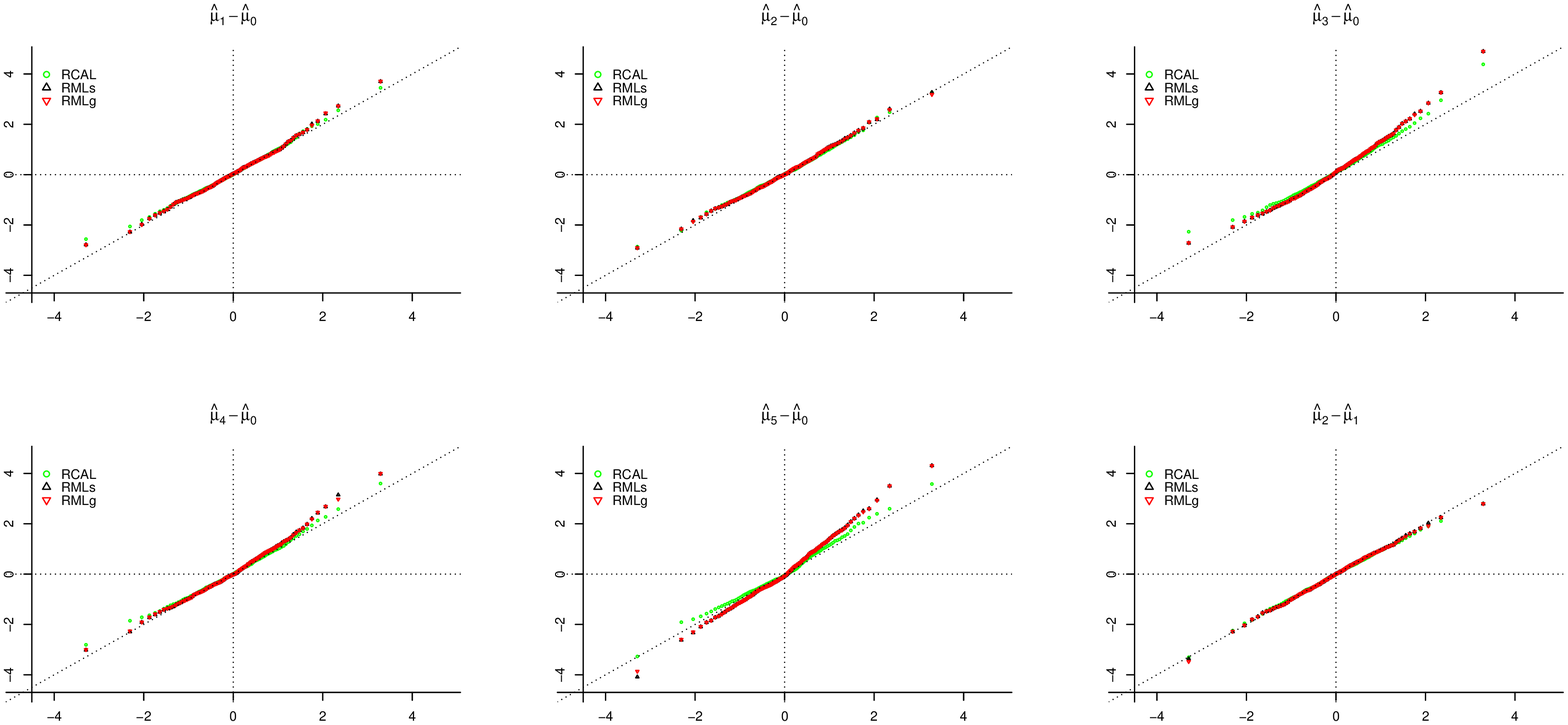}
\end{subfigure} %
\begin{subfigure}{0.8\textwidth}
\centering
\includegraphics[scale=0.47]{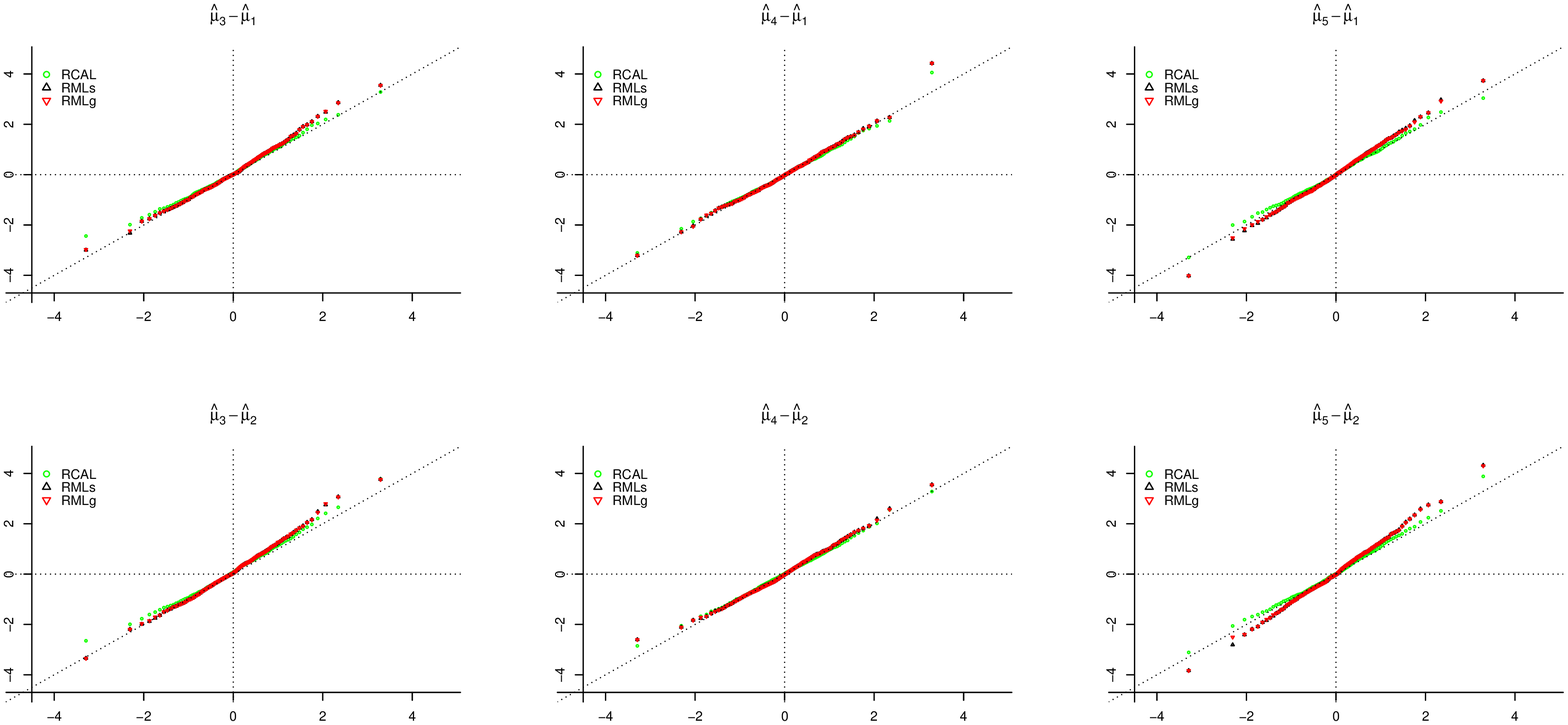}
\end{subfigure} %
\begin{subfigure}{0.8\textwidth}
\centering
\includegraphics[scale=0.47]{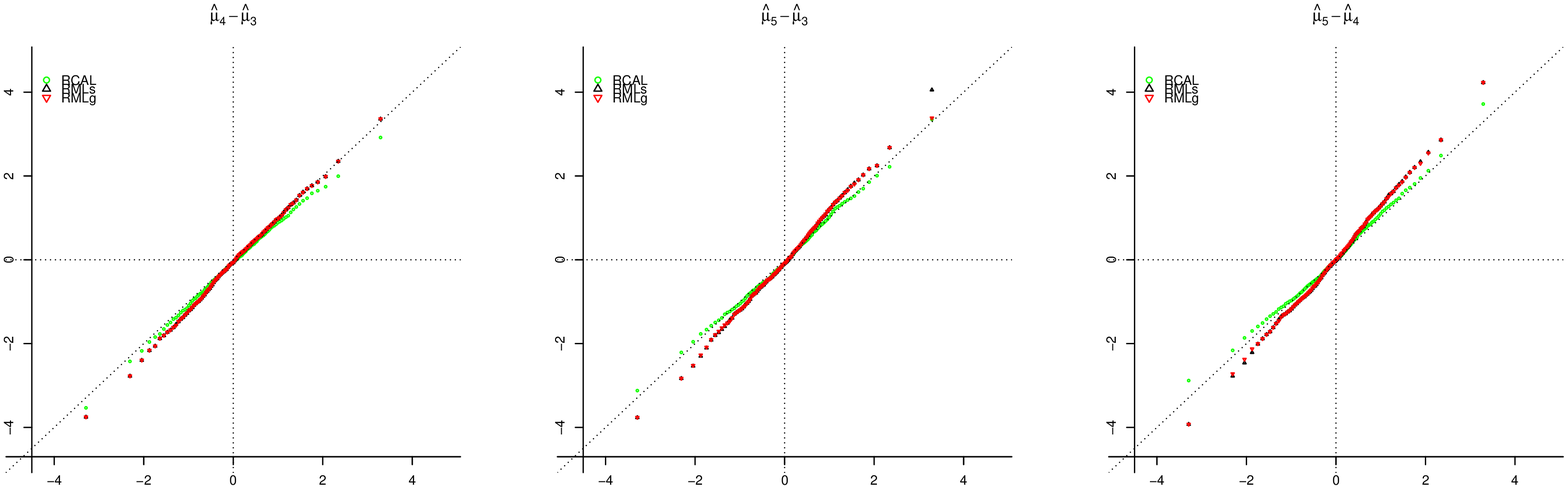}
\end{subfigure}
\caption{QQ plots of the standardized $\hat{\mu}_t - \hat{\mu}_k$ against standard normal with tuning parameters selected as $\lambda.1se$.}
\label{fig:qq_ate_1se}
\end{figure}

\begin{figure}[H]
\centering
\includegraphics[scale=0.47]{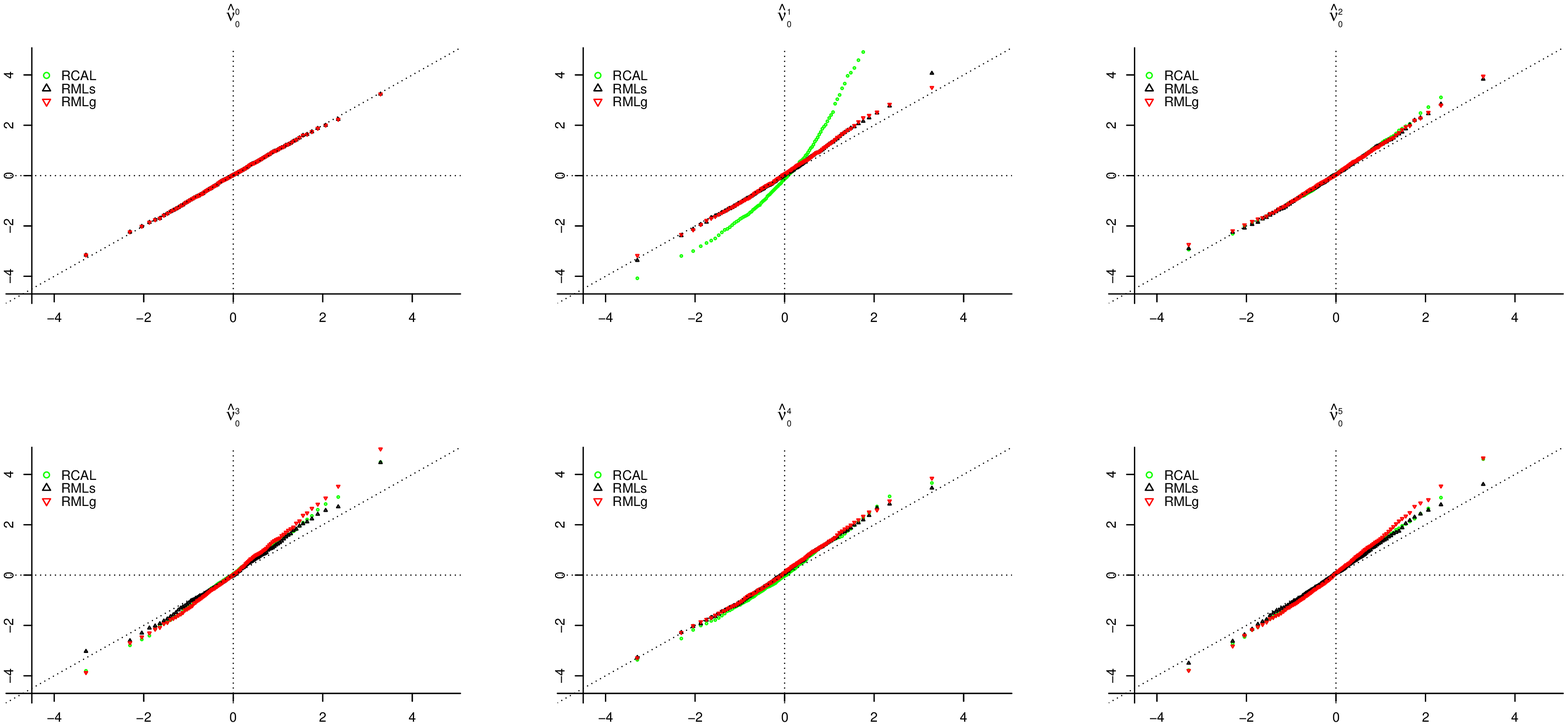}\vspace{-.1in}
\caption{QQ plots of the standardized $\hat{\nu}^{(k)}_t$against standard normal for $t = 0$ with tuning parameters selected as $\lambda.1se$.}
\label{fig:qq_nu0_1se}
\end{figure}

\begin{figure}[H]
\centering
\includegraphics[scale=0.47]{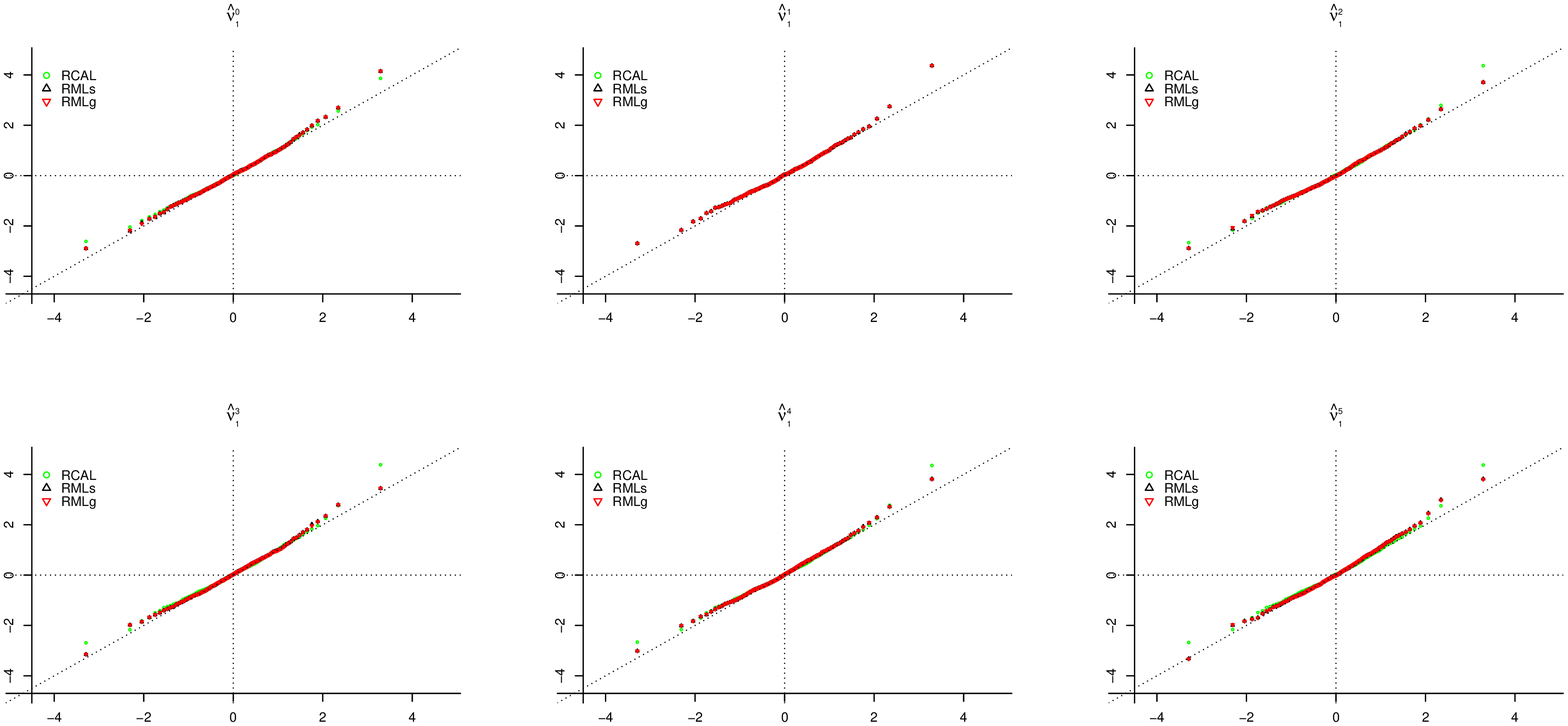}\vspace{-.1in}
\caption{QQ plots of the standardized $\hat{\nu}^{(k)}_t$against standard normal for $t = 1$ with tuning parameters selected as $\lambda.1se$.}
\label{fig:qq_nu1_1se}
\end{figure}

\begin{figure}[H]
\centering
\includegraphics[scale=0.47]{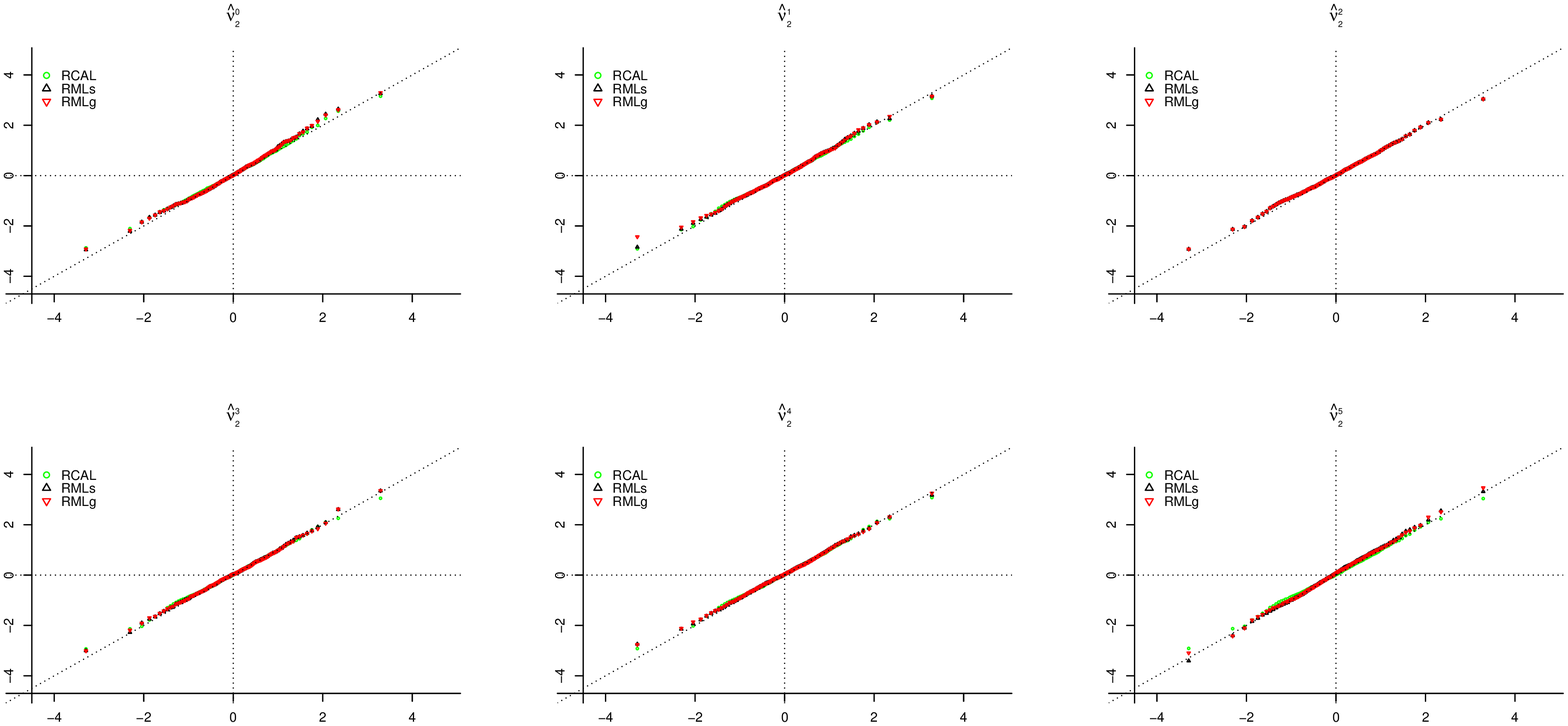}\vspace{-.1in}
\caption{QQ plots of the standardized $\hat{\nu}^{(k)}_t$against standard normal for $t = 2$ with tuning parameters selected as $\lambda.1se$.}
\label{fig:qq_nu2_1se}
\end{figure}

\begin{figure}[H]
\centering
\includegraphics[scale=0.47]{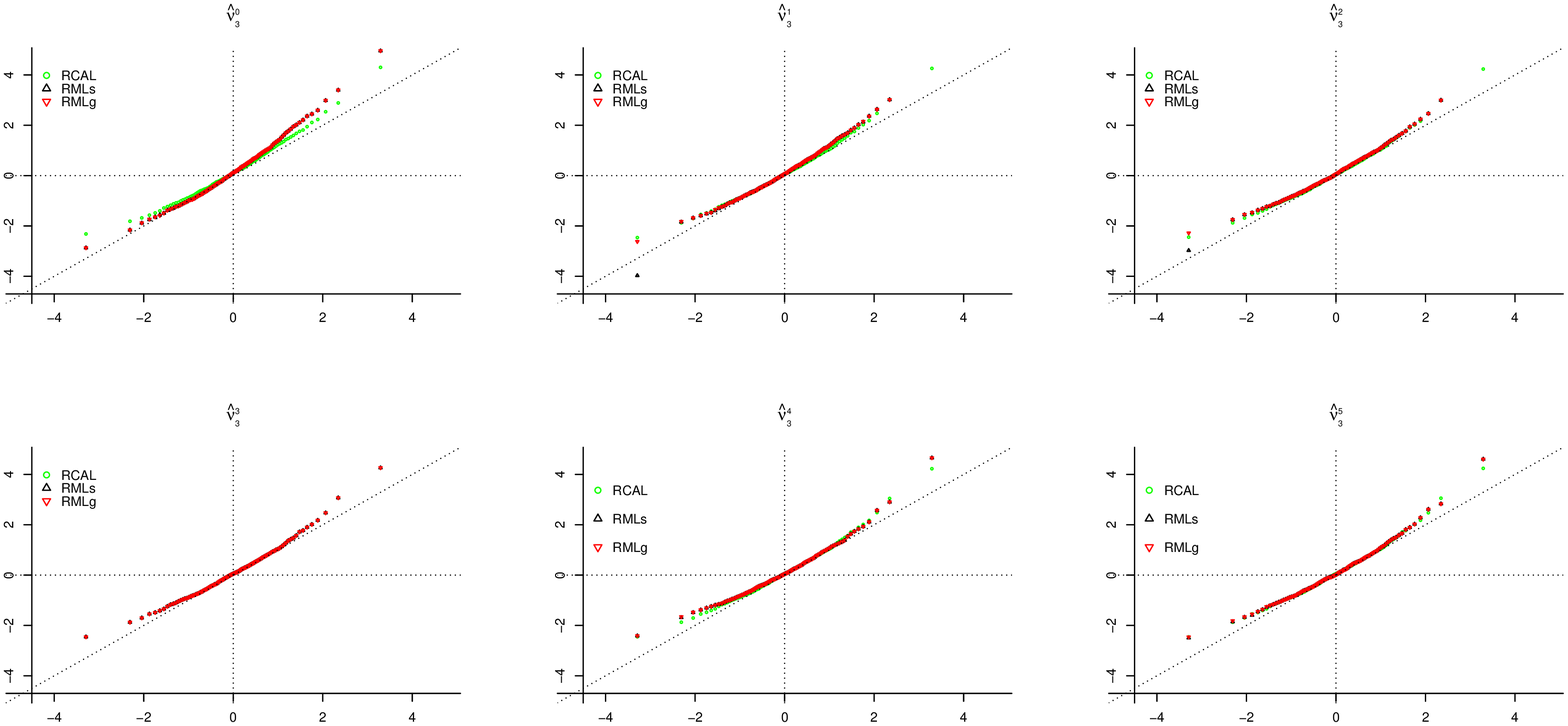}\vspace{-.1in}
\caption{QQ plots of the standardized $\hat{\nu}^{(k)}_t$against standard normal for $t = 3$ with tuning parameters selected as $\lambda.1se$.}
\label{fig:qq_nu3_1se}
\end{figure}

\begin{figure}[H]
\centering
\includegraphics[scale=0.47]{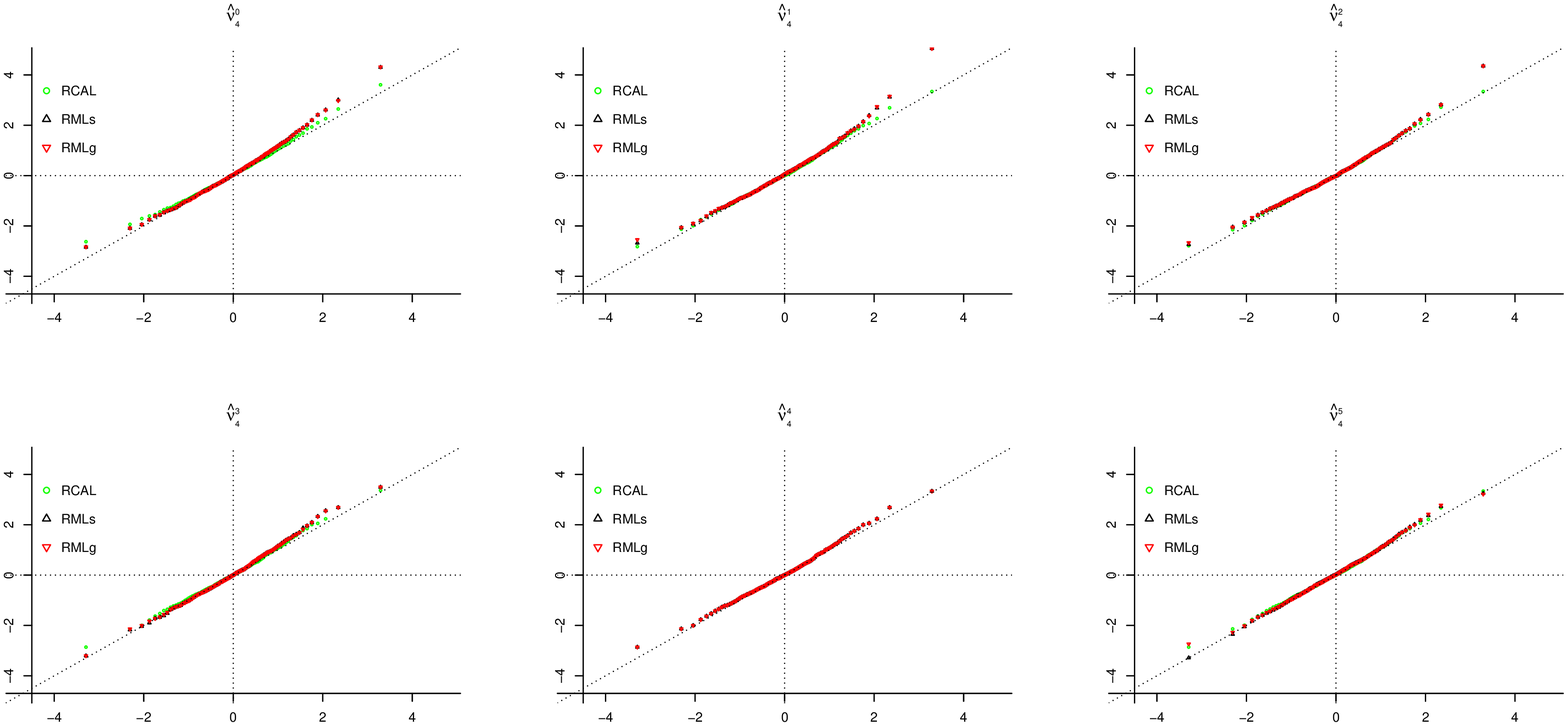}\vspace{-.1in}
\caption{QQ plots of the standardized $\hat{\nu}^{(k)}_t$against standard normal for $t = 4$ with tuning parameters selected as $\lambda.1se$.}
\label{fig:qq_nu4_1se}
\end{figure}

\begin{figure}[H]
\centering
\includegraphics[scale=0.47]{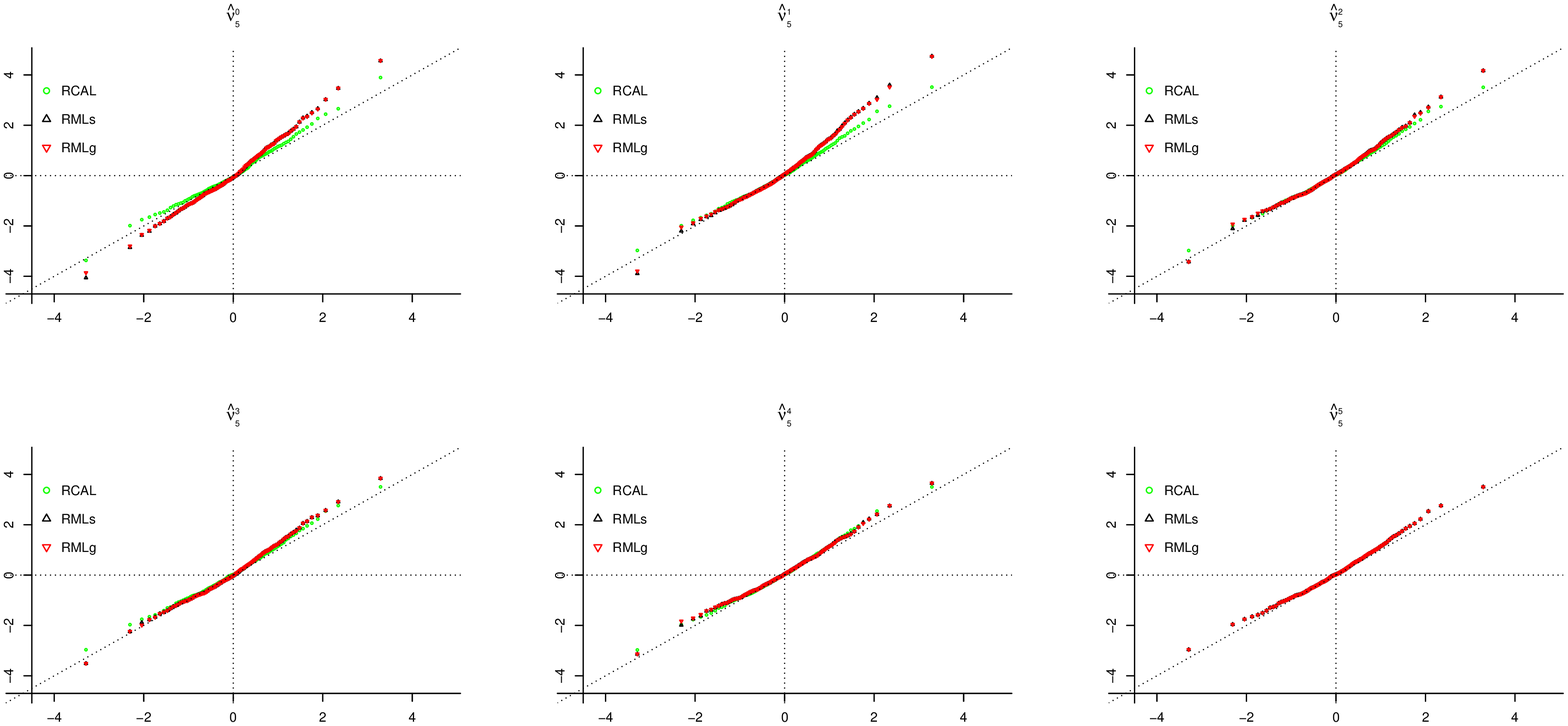}\vspace{-.1in}
\caption{QQ plots of the standardized $\hat{\nu}^{(k)}_t$against standard normal for $t = 5$ with tuning parameters selected as $\lambda.1se$.}
\label{fig:qq_nu5_1se}
\end{figure}

\textbf{Additional results with tuning parameters selected as $\lambda.1se$.}\;
Tables \ref{tb:mu_full_1se}, \ref{tb:ate_full_1se} and \ref{tb:nu_full_1se} are summarizes for $\hat{\mu}_t$, $\hat{\mu}_t - \hat{\mu}_k$ and $\hat{\nu}_t^{(k)}$ on full sample for $t, k = 0, 1, \ldots, 5$  with tuning parameters selected as $\lambda.1se$ for PS and OR models in all methods.
Tables \ref{tb:mu_sub_1se}, \ref{tb:ate_sub_1se} and \ref{tb:nu_sub_1se} are summaries of $\hat{\mu}_t$, $\hat{\mu}_t - \hat{\mu}_k$ and $\hat{\nu}_t^{(k)}$ on sub-samples for $t, k = 0, 1, \ldots, 5$.
Figures \ref{fig:box_nnz_ps_1se}, \ref{fig:box_masd_1se} and \ref{fig:box_rv_1se} show boxplots of number of nonzero estimated coefficients in PS model, MASCD and RV.
Figure \ref{fig:box_nnz_or_1se} shows boxplots of number of nonzero estimated coefficients in OR model.
Figures \ref{fig:qq_mu_1se}, \ref{fig:qq_ate_1se} and \ref{fig:qq_nu0_1se}--\ref{fig:qq_nu5_1se} are the QQ plots of the standardized $\hat{\mu}_t$, $\hat{\mu}_t - \hat{\mu}_k$ and $\hat{\nu}^{(k)}_t$ against standard normal.
The results with $\lambda.1se$ for all methods are the same as discussed in Section \ref{sec:empirical}
except for those related to treatment 0 such as $\hat{\mu}_0$, $\hat{\mu}_t - \hat{\mu}_0$  and $\hat{\nu}_0^{(t)}$ for $t = 1, 2, \ldots, 5$.
From Table \ref{tb:mu_sub_1se}, we see that RCAL, RMLs and RMLg performs similarly to each other in terms of variance and coverage for $\mu_0$. From Table \ref{tb:ate_sub_1se}, we see that RCAL has comparable or better coverage proportions than RMLs and RMLg for $\mu_t - \mu_0$. From Table \ref{tb:nu_sub_1se}, we see that RCAL performs similarly to RMLs and better than RMLg in terms of coverage for $\nu_0^{(t)}$ for $t = 2, 3, 4, 5$. The only case where RCAL has a worse coverage than RMLs and RMLg is $\hat{\nu}_0^{(2)}$. 

\begin{table}[H]
\caption{\footnotesize Summary of  $\hat{\mu}_t - \hat{\mu}_k$ on full sample for $t, k = 0, 1, \ldots, 5$.} \label{tb:ate_full_cross}\vspace{-4ex}
\begin{center}
\resizebox{1.0\textwidth}{!}{\begin{tabular}{lccccccccccc}
\hline
 & Est & SE & 95CI &~~& Est & SE & 95CI &~~& Est & SE & 95CI  \\
\hline
& \multicolumn{3}{c}{$\hat{\mu}_1 - \hat{\mu}_0$} &~~& \multicolumn{3}{c}{$\hat{\mu}_2 - \hat{\mu}_0$} &~~& \multicolumn{3}{c}{$\hat{\mu}_3 - \hat{\mu}_0$} \\
\cline{2-4}\cline{6-8}\cline{10-12}
\rowcolor{lightgray}
CAL & -0.0484 & 0.0021 & (-0.0524, -0.0443) &~~& -0.0677 & 0.0017 & (-0.0710, -0.0644) &~~& -0.0730 & 0.0049 & (-0.0826, -0.0634) \\
ML & -0.0484 & 0.0020 & (-0.0523, -0.0444) &~~& -0.0679 & 0.0016 & (-0.0711, -0.0647) &~~& -0.0716 & 0.0045 & (-0.0804, -0.0628) \\
\rowcolor{lightgray}
RCAL & -0.0499 & 0.0022 & (-0.0541, -0.0456) &~~& -0.0698 & 0.0017 & (-0.0730, -0.0666) &~~& -0.0749 & 0.0042 & (-0.0832, -0.0666) \\
RMLs & -0.0493 & 0.0020 & (-0.0532, -0.0453) &~~& -0.0692 & 0.0017 & (-0.0726, -0.0658) &~~& -0.0749 & 0.0036 & (-0.0819, -0.0678)  \\
RMLg & -0.0494 & 0.0021 & (-0.0535, -0.0454) &~~& -0.0703 & 0.0016 & (-0.0736, -0.0671) &~~& -0.0760 & 0.0037 & (-0.0832, -0.0688)  \\

& \multicolumn{3}{c}{$\hat{\mu}_4 - \hat{\mu}_0$} &~~& \multicolumn{3}{c}{$\hat{\mu}_5 - \hat{\mu}_0$} &~~& \multicolumn{3}{c}{$\hat{\mu}_2 - \hat{\mu}_1$}\\
\cline{2-4}\cline{6-8}\cline{10-12}
\rowcolor{lightgray}
CAL & -0.0787 & 0.0023 & (-0.0833, -0.0741)  &~~& -0.0805 & 0.0057 & (-0.0917, -0.0693) &~~& -0.0194 & 0.0026 & (-0.0245, -0.0142) \\
ML & -0.0792 & 0.0022 & (-0.0835, -0.0749) &~~& -0.0838 & 0.0057 & (-0.0951, -0.0726) &~~& -0.0195 & 0.0025 & (-0.0244, -0.0147) \\
\rowcolor{lightgray}
RCAL & -0.0804 & 0.0021 & (-0.0845, -0.0763) &~~& -0.0826 & 0.0047 & (-0.0919, -0.0734) &~~& -0.0199 & 0.0027 & (-0.0252, -0.0146)  \\
RMLs & -0.0797 & 0.0020 & (-0.0836, -0.0759) &~~& -0.0828 & 0.0044 & (-0.0913, -0.0742) &~~& -0.0199 & 0.0026 & (-0.0251, -0.0148)  \\
RMLg & -0.0796 & 0.0021 & (-0.0836, -0.0756) &~~& -0.0834 & 0.0045 & (-0.0922, -0.0747)  &~~& -0.0209 & 0.0026 & (-0.0260, -0.0158)  \\

& \multicolumn{3}{c}{$\hat{\mu}_3 - \hat{\mu}_1$} &~~& \multicolumn{3}{c}{$\hat{\mu}_4 - \hat{\mu}_1$} &~~& \multicolumn{3}{c}{$\hat{\mu}_5 - \hat{\mu}_1$}\\
\cline{2-4}\cline{6-8}\cline{10-12}
\rowcolor{lightgray}
CAL & -0.0247 & 0.0053 & (-0.0350, -0.0143)  &~~& -0.0303 & 0.0031 & (-0.0364, -0.0243) &~~& -0.0322 & 0.0061 & (-0.0440, -0.0203)  \\
ML & -0.0232 & 0.0048 & (-0.0327, -0.0137)  &~~& -0.0308 & 0.0029 & (-0.0365, -0.0252)  &~~& -0.0355 & 0.0060 & (-0.0473, -0.0237)  \\
\rowcolor{lightgray}
RCAL & -0.0250 & 0.0047 & (-0.0343, -0.0158) &~~& -0.0305 & 0.0030 & (-0.0364, -0.0246) &~~& -0.0328 & 0.0052 & (-0.0429, -0.0226)  \\
RMLs & -0.0256 & 0.0041 & (-0.0337, -0.0176) &~~& -0.0305 & 0.0028 & (-0.0359, -0.0250) &~~& -0.0335 & 0.0048 & (-0.0429, -0.0242)  \\
RMLg & -0.0266 & 0.0042 & (-0.0348, -0.0184) &~~& -0.0302 & 0.0029 & (-0.0358, -0.0245)  &~~& -0.0340 & 0.0049 & (-0.0436, -0.0244)  \\

& \multicolumn{3}{c}{$\hat{\mu}_3 - \hat{\mu}_2$} &~~& \multicolumn{3}{c}{$\hat{\mu}_4 - \hat{\mu}_2$} &~~& \multicolumn{3}{c}{$\hat{\mu}_5 - \hat{\mu}_2$}\\
\cline{2-4}\cline{6-8}\cline{10-12}
\rowcolor{lightgray}
CAL & -0.0053 & 0.0051 & (-0.0154, 0.0048)  &~~& -0.0110 & 0.0028 & (-0.0165, -0.0054) &~~& -0.0128 & 0.0059 & (-0.0245, -0.0011) \\
ML & -0.0037 & 0.0047 & (-0.0129, 0.0055) &~~& -0.0113 & 0.0026 & (-0.0164, -0.0061) &~~& -0.0160 & 0.0059 & (-0.0275, -0.0044) \\
\rowcolor{lightgray}
RCAL & -0.0051 & 0.0045 & (-0.0140, 0.0037) &~~& -0.0106 & 0.0026 & (-0.0157, -0.0054)  &~~& -0.0128 & 0.0050 & (-0.0226, -0.0031)  \\
RMLs & -0.0057 & 0.0040 & (-0.0135, 0.0021) &~~& -0.0105 & 0.0026 & (-0.0156, -0.0055)  &~~& -0.0136 & 0.0047 & (-0.0227, -0.0044)  \\
RMLg & -0.0057 & 0.0040 & (-0.0135, 0.0022) &~~& -0.0093 & 0.0026 & (-0.0143, -0.0042)  &~~& -0.0131 & 0.0047 & (-0.0223, -0.0039)  \\

& \multicolumn{3}{c}{$\hat{\mu}_4 - \hat{\mu}_3$} &~~& \multicolumn{3}{c}{$\hat{\mu}_5 - \hat{\mu}_3$} &~~& \multicolumn{3}{c}{$\hat{\mu}_5 - \hat{\mu}_4$}\\
\cline{2-4}\cline{6-8}\cline{10-12}
\rowcolor{lightgray}
CAL & -0.0057 & 0.0054 & (-0.0162, 0.0049)  &~~& -0.0075 & 0.0075 & (-0.0222, 0.0072)  &~~& -0.0018 & 0.0062 & (-0.0139, 0.0102)  \\
ML & -0.0076 & 0.0049 & (-0.0173, 0.0021)  &~~& -0.0123 & 0.0072 & (-0.0265, 0.0019)  &~~& -0.0047 & 0.0061 & (-0.0166, 0.0073)  \\
\rowcolor{lightgray}
RCAL & -0.0055 & 0.0047 & (-0.0147, 0.0037)  &~~& -0.0077 & 0.0063 & (-0.0201, 0.0047)  &~~& -0.0023 & 0.0051 & (-0.0123, 0.0078)  \\
RMLs & -0.0049 & 0.0041 & (-0.0129, 0.0031)  &~~& -0.0079 & 0.0056 & (-0.0189, 0.0032)  &~~& -0.0030 & 0.0048 & (-0.0124, 0.0063)  \\
RMLg & -0.0036 & 0.0042 & (-0.0118, 0.0046)  &~~& -0.0074 & 0.0057 & (-0.0187, 0.0038)  &~~& -0.0038 & 0.0049 & (-0.0134, 0.0057)  \\

\hline
\end{tabular}}
\end{center}
\setlength{\baselineskip}{0.5\baselineskip}
\vspace{-.15in}\noindent{\tiny
\textbf{Note}: Est, SE, or 95CI denotes point estimate, standard error, or 95\% confidence interval respectively.
RCAL, RMLs, or RMLg denotes
$\hat{\mu}_t(\hat{m}^\#_{\text{RWL}}, \hat{\pi}_{\text{RCAL}}) - \hat{\mu}_k(\hat{m}^\#_{\text{RWL}}, \hat{\pi}_{\text{RCAL}})$,
$\hat{\mu}_t(\hat{m}_{\text{RMLs}}, \hat{\pi}_{\text{RML}}) - \hat{\mu}_k(\hat{m}_{\text{RMLs}}, \hat{\pi}_{\text{RML}})$,
or $\hat{\mu}_t(\hat{m}_{\text{RMLg}}, \hat{\pi}_{\text{RML}}) - \hat{\mu}_k(\hat{m}_{\text{RMLg}}, \hat{\pi}_{\text{RML}})$
respectively. CAL or ML denotes non-regularized estimation with main effects only in PS and OR models.}
\end{table}

\begin{table}[H]
\caption{\footnotesize Summary of $\hat{\nu}^{(k)}_t$ on full sample for $t, k = 0, 1, \ldots, 5$.} \label{tb:nu_full_cross}\vspace{-4ex}
\begin{center}
\resizebox{1.0\textwidth}{0.5\textwidth}{\begin{tabular}{lccccccccccccccccccccccc}
\hline
& \multicolumn{3}{c}{$\hat{\nu}^{(0)}_t$} & $~~$ & \multicolumn{3}{c}{$\hat{\nu}^{(1)}_t$} & $~~$ & \multicolumn{3}{c}{$\hat{\nu}^{(2)}_t$} & $~~$ & \multicolumn{3}{c}{$\hat{\nu}^{(3)}_t$} & $~~$ & \multicolumn{3}{c}{$\hat{\nu}^{(4)}_t$} & $~~$ & \multicolumn{3}{c}{$\hat{\nu}^{(5)}_t$} \\
\cline{2-4}\cline{6-8}\cline{10-12}\cline{14-16}\cline{18-20}\cline{22-24}
& Est & SE & 95CI &~~& Est & SE & 95CI &~~& Est & SE & 95CI &~~& Est & SE & 95CI &~~& Est & SE & 95CI &~~& Est & SE & 95CI  \\
\hline
& \multicolumn{23}{c}{\footnotesize t = 0} \\
\rowcolor{lightgray}
CAL & 8.1272 & 0.0003 & (8.1266, 8.1279)  &~~& 8.1025 & 0.0009 & (8.1007, 8.1043) &~~& 8.1098 & 0.0008 & (8.1082, 8.1113) &~~& 8.1240 & 0.0011 & (8.1218, 8.1262) &~~& 8.1193 & 0.0009 & (8.1174, 8.1211) &~~& 8.1253 & 0.0015 & (8.1223, 8.1283) \\
ML & 8.1272 & 0.0003 & (8.1266, 8.1279) &~~& 8.1025 & 0.0009 & (8.1007, 8.1042) &~~& 8.1102 & 0.0010 & (8.1083, 8.1122) &~~& 8.1262 & 0.0024 & (8.1214, 8.1309) &~~& 8.1253 & 0.0061 & (8.1134, 8.1372) &~~& 8.1360 & 0.0095 & (8.1175, 8.1546) \\
\rowcolor{lightgray}
RCAL & 8.1272 & 0.0003 & (8.1266, 8.1279) &~~& 8.1011 & 0.0009 & (8.0994, 8.1028) &~~& 8.1072 & 0.0008 & (8.1057, 8.1088) &~~& 8.1227 & 0.0007 & (8.1213, 8.1242) &~~& 8.1162 & 0.0009 & (8.1145, 8.1180) &~~& 8.1225 & 0.0012 & (8.1203, 8.1248)  \\
RMLs & 8.1272 & 0.0003 & (8.1266, 8.1279) &~~& 8.1011 & 0.0009 & (8.0995, 8.1028) &~~& 8.1075 & 0.0008 & (8.1060, 8.1090) &~~& 8.1230 & 0.0011 & (8.1209, 8.1252) &~~& 8.1169 & 0.0010 & (8.1149, 8.1189) &~~& 8.1235 & 0.0018 & (8.1199, 8.1270) \\
RMLg & 8.1272 & 0.0003 & (8.1266, 8.1279) &~~& 8.1016 & 0.0007 & (8.1002, 8.1030) &~~& 8.1078 & 0.0007 & (8.1064, 8.1092) &~~& 8.1233 & 0.0007 & (8.1219, 8.1247) &~~& 8.1172 & 0.0009 & (8.1154, 8.1190) &~~& 8.1240 & 0.0013 & (8.1214, 8.1266) \\

& \multicolumn{23}{c}{\footnotesize t = 1} \\
\rowcolor{lightgray}
CAL & 8.0787 & 0.0022 & (8.0743, 8.0830) &~~& 8.0576 & 0.0019 & (8.0538, 8.0614) &~~& 8.0633 & 0.0019 & (8.0596, 8.0670) &~~& 8.0767 & 0.0021 & (8.0725, 8.0808) &~~& 8.0701 & 0.0022 & (8.0658, 8.0744) &~~& 8.0715 & 0.0031 & (8.0655, 8.0775) \\
ML & 8.0792 & 0.0021 & (8.0752, 8.0833) &~~& 8.0576 & 0.0019 & (8.0538, 8.0614) &~~& 8.0632 & 0.0018 & (8.0596, 8.0668) &~~& 8.0764 & 0.0021 & (8.0723, 8.0806) &~~& 8.0699 & 0.0020 & (8.0660, 8.0739) &~~& 8.0711 & 0.0026 & (8.0661, 8.0761) \\
\rowcolor{lightgray}
RCAL & 8.0772 & 0.0023 & (8.0727, 8.0817) &~~& 8.0576 & 0.0019 & (8.0538, 8.0614) &~~& 8.0609 & 0.0019 & (8.0572, 8.0646) &~~& 8.0612 & 0.0019 & (8.0575, 8.0648) &~~& 8.0645 & 0.0019 & (8.0609, 8.0682) &~~& 8.0609 & 0.0019 & (8.0572, 8.0646) \\
RMLs & 8.0773 & 0.0021 & (8.0731, 8.0814) &~~& 8.0576 & 0.0019 & (8.0538, 8.0614) &~~& 8.0622 & 0.0019 & (8.0584, 8.0660) &~~& 8.0766 & 0.0020 & (8.0727, 8.0805) &~~& 8.0693 & 0.0021 & (8.0652, 8.0734) &~~& 8.0716 & 0.0025 & (8.0666, 8.0765) \\
RMLg & 8.0771 & 0.0022 & (8.0729, 8.0814) &~~& 8.0576 & 0.0019 & (8.0538, 8.0614)  &~~& 8.0619 & 0.0020 & (8.0580, 8.0658) &~~& 8.0766 & 0.0019 & (8.0728, 8.0804) &~~& 8.0695 & 0.0021 & (8.0653, 8.0737) &~~& 8.0730 & 0.0024 & (8.0683, 8.0778)  \\

& \multicolumn{23}{c}{\footnotesize t = 2} \\
\rowcolor{lightgray}
CAL & 8.0603 & 0.0018 & (8.0567, 8.0639) &~~& 8.0358 & 0.0015 & (8.0329, 8.0387) &~~& 8.0398 & 0.0013 & (8.0372, 8.0424) &~~& 8.0511 & 0.0016 & (8.0480, 8.0542) &~~& 8.0441 & 0.0014 & (8.0412, 8.0469) &~~& 8.0434 & 0.0021 & (8.0392, 8.0476)  \\
ML & 8.0607 & 0.0017 & (8.0574, 8.0640) &~~& 8.0358 & 0.0015 & (8.0329, 8.0387) &~~& 8.0398 & 0.0013 & (8.0372, 8.0424) &~~& 8.0509 & 0.0016 & (8.0478, 8.0541) &~~& 8.0440 & 0.0014 & (8.0413, 8.0468) &~~& 8.0435 & 0.0019 & (8.0398, 8.0472) \\
\rowcolor{lightgray}
RCAL & 8.0576 & 0.0018 & (8.0541, 8.0610) &~~& 8.0393 & 0.0013 & (8.0367, 8.0419) &~~& 8.0398 & 0.0013 & (8.0372, 8.0424) &~~& 8.0406 & 0.0013 & (8.0380, 8.0432) &~~& 8.0408 & 0.0013 & (8.0382, 8.0433) &~~& 8.0400 & 0.0013 & (8.0374, 8.0426) \\
RMLs & 8.0581 & 0.0019 & (8.0543, 8.0618) &~~& 8.0363 & 0.0014 & (8.0335, 8.0390) &~~& 8.0398 & 0.0013 & (8.0372, 8.0424) &~~& 8.0516 & 0.0015 & (8.0487, 8.0546) &~~& 8.0444 & 0.0014 & (8.0416, 8.0471) &~~& 8.0441 & 0.0019 & (8.0404, 8.0479) \\
RMLg & 8.0567 & 0.0018 & (8.0532, 8.0601) &~~& 8.0367 & 0.0014 & (8.0340, 8.0394)  &~~& 8.0398 & 0.0013 & (8.0372, 8.0424) &~~& 8.0518 & 0.0013 & (8.0493, 8.0544) &~~& 8.0449 & 0.0014 & (8.0422, 8.0476) &~~& 8.0459 & 0.0017 & (8.0426, 8.0492) \\

& \multicolumn{23}{c}{\footnotesize t = 3} \\
\rowcolor{lightgray}
CAL & 8.0544 & 0.0055 & (8.0435, 8.0652) &~~& 8.0352 & 0.0040 & (8.0273, 8.0431) &~~& 8.0375 & 0.0034 & (8.0308, 8.0441) &~~& 8.0474 & 0.0030 & (8.0416, 8.0533) &~~& 8.0414 & 0.0032 & (8.0352, 8.0477) &~~& 8.0421 & 0.0038 & (8.0347, 8.0496)  \\
ML & 8.0567 & 0.0049 & (8.0471, 8.0663) &~~& 8.0345 & 0.0042 & (8.0263, 8.0428) &~~& 8.0373 & 0.0034 & (8.0307, 8.0439)  &~~& 8.0474 & 0.0030 & (8.0416, 8.0533) &~~& 8.0415 & 0.0032 & (8.0353, 8.0478) &~~& 8.0423 & 0.0035 & (8.0354, 8.0493) \\
\rowcolor{lightgray}
RCAL & 8.0499 & 0.0046 & (8.0408, 8.0590) &~~& 8.0467 & 0.0030 & (8.0408, 8.0526) &~~& 8.0463 & 0.0030 & (8.0404, 8.0523) &~~& 8.0474 & 0.0030 & (8.0416, 8.0533) &~~& 8.0470 & 0.0030 & (8.0412, 8.0529) &~~& 8.0474 & 0.0030 & (8.0415, 8.0532) \\
RMLs & 8.0520 & 0.0039 & (8.0444, 8.0595) &~~& 8.0334 & 0.0038 & (8.0259, 8.0408)  &~~& 8.0349 & 0.0037 & (8.0276, 8.0422) &~~& 8.0474 & 0.0030 & (8.0416, 8.0533) &~~& 8.0408 & 0.0034 & (8.0342, 8.0474) &~~& 8.0429 & 0.0034 & (8.0362, 8.0496)  \\
RMLg & 8.0507 & 0.0039 & (8.0431, 8.0584) &~~& 8.0337 & 0.0040 & (8.0258, 8.0416) &~~& 8.0346 & 0.0040 & (8.0268, 8.0425)  &~~& 8.0474 & 0.0030 & (8.0416, 8.0533)  &~~& 8.0405 & 0.0035 & (8.0336, 8.0474) &~~& 8.0434 & 0.0035 & (8.0365, 8.0503)  \\

& \multicolumn{23}{c}{\footnotesize t = 4} \\
\rowcolor{lightgray}
CAL & 8.0488 & 0.0026 & (8.0438, 8.0538)  &~~& 8.0261 & 0.0023 & (8.0217, 8.0306) &~~& 8.0312 & 0.0018 & (8.0278, 8.0347) &~~& 8.0426 & 0.0017 & (8.0392, 8.0459) &~~& 8.0365 & 0.0016 & (8.0334, 8.0396)  &~~& 8.0374 & 0.0020 & (8.0335, 8.0413)  \\
ML & 8.0488 & 0.0024 & (8.0441, 8.0534)  &~~& 8.0263 & 0.0021 & (8.0222, 8.0304)  &~~& 8.0312 & 0.0017 & (8.0279, 8.0345) &~~& 8.0425 & 0.0017 & (8.0391, 8.0459) &~~& 8.0365 & 0.0016 & (8.0334, 8.0396) &~~& 8.0373 & 0.0019 & (8.0336, 8.0411) \\
\rowcolor{lightgray}
RCAL & 8.0456 & 0.0023 & (8.0411, 8.0501) &~~& 8.0357 & 0.0016 & (8.0326, 8.0388) &~~& 8.0356 & 0.0016 & (8.0325, 8.0387) &~~& 8.0367 & 0.0016 & (8.0336, 8.0397) &~~& 8.0365 & 0.0016 & (8.0334, 8.0396) &~~& 8.0365 & 0.0016 & (8.0334, 8.0396) \\
RMLs & 8.0471 & 0.0021 & (8.0429, 8.0513) &~~& 8.0266 & 0.0019 & (8.0228, 8.0304) &~~& 8.0312 & 0.0017 & (8.0279, 8.0344) &~~& 8.0428 & 0.0016 & (8.0396, 8.0459) &~~& 8.0365 & 0.0016 & (8.0334, 8.0396) &~~& 8.0384 & 0.0018 & (8.0349, 8.0418) \\
RMLg & 8.0472 & 0.0022 & (8.0429, 8.0516)  &~~& 8.0282 & 0.0020 & (8.0243, 8.0321) &~~& 8.0315 & 0.0017 & (8.0281, 8.0348) &~~& 8.0428 & 0.0015 & (8.0399, 8.0457) &~~& 8.0365 & 0.0016 & (8.0334, 8.0396) &~~& 8.0386 & 0.0016 & (8.0354, 8.0417) \\

& \multicolumn{23}{c}{\footnotesize t = 5} \\
\rowcolor{lightgray}
CAL & 8.0482 & 0.0062 & (8.0361, 8.0603) &~~& 8.0160 & 0.0082 & (8.0000, 8.0320) &~~& 8.0231 & 0.0055 & (8.0123, 8.0339) &~~& 8.0377 & 0.0038 & (8.0303, 8.0451) &~~& 8.0303 & 0.0036 & (8.0232, 8.0375) &~~& 8.0318 & 0.0034 & (8.0251, 8.0384)  \\
ML & 8.0448 & 0.0061 & (8.0328, 8.0567) &~~& 8.0153 & 0.0081 & (7.9994, 8.0312) &~~& 8.0233 & 0.0050 & (8.0134, 8.0331) &~~& 8.0372 & 0.0037 & (8.0300, 8.0444) &~~& 8.0302 & 0.0035 & (8.0233, 8.0372) &~~& 8.0318 & 0.0034 & (8.0251, 8.0384) \\
\rowcolor{lightgray}
RCAL & 8.0440 & 0.0052 & (8.0337, 8.0542) &~~& 8.0304 & 0.0034 & (8.0237, 8.0372) &~~& 8.0302 & 0.0035 & (8.0233, 8.0370) &~~& 8.0319 & 0.0034 & (8.0253, 8.0385) &~~& 8.0315 & 0.0034 & (8.0249, 8.0382) &~~& 8.0318 & 0.0034 & (8.0251, 8.0384) \\
RMLs & 8.0446 & 0.0046 & (8.0355, 8.0536) &~~& 8.0224 & 0.0056 & (8.0114, 8.0335)  &~~& 8.0255 & 0.0044 & (8.0169, 8.0342) &~~& 8.0370 & 0.0033 & (8.0305, 8.0435) &~~& 8.0305 & 0.0035 & (8.0236, 8.0374) &~~& 8.0318 & 0.0034 & (8.0251, 8.0384) \\
RMLg & 8.0440 & 0.0047 & (8.0347, 8.0533)  &~~& 8.0218 & 0.0058 & (8.0105, 8.0331) &~~& 8.0246 & 0.0045 & (8.0157, 8.0335) &~~& 8.0366 & 0.0034 & (8.0300, 8.0432) &~~& 8.0296 & 0.0036 & (8.0225, 8.0367) &~~& 8.0318 & 0.0034 & (8.0251, 8.0384) \\

\hline
\end{tabular}}
\end{center}
\setlength{\baselineskip}{0.5\baselineskip}
\vspace{-.15in}\noindent{\tiny
\textbf{Note}: Est, SE, or 95CI denotes point estimate, standard error, or 95\% confidence interval respectively. CAL denotes $\hat{\nu}^{(k)}_{t, \text{CAL}}$. ML denotes $\hat{\nu}^{(k)}_{t, \text{ML}}$. RCAL denotes $\hat{\nu}^{(k)}_{t, \text{RCAL}}$. RMLs denotes $\hat{\nu}^{(k)}_{t, \text{RMLs}}$. RMLg denotes $\hat{\nu}^{(k)}_{t, \text{RMLg}}$.}
\end{table}

\begin{table}[H]
\caption{\footnotesize Summary of  $\hat{\mu}_t - \hat{\mu}_k$ on sub samples for $t, k = 0, 1, \ldots, 5$.} \label{tb:ate_sub_cross}\vspace{-4ex}
\begin{center}
\resizebox{1.0\textwidth}{!}{\begin{tabular}{lccccccccccccccccc}
\hline
 & Mean & $\sqrt{\text{Var}}$ & $\sqrt{\text{EVar}}$ & Cov90 & Cov95 &~~& Mean & $\sqrt{\text{Var}}$ & $\sqrt{\text{EVar}}$ & Cov90 & Cov95 &~~& Mean & $\sqrt{\text{Var}}$ & $\sqrt{\text{EVar}}$ & Cov90 & Cov95 \\
\hline
& \multicolumn{5}{c}{$\hat{\mu}_1 - \hat{\mu}_0$} &~~& \multicolumn{5}{c}{$\hat{\mu}_2 - \hat{\mu}_0$} &~~& \multicolumn{5}{c}{$\hat{\mu}_3 - \hat{\mu}_0$}\\
\cline{2-6}\cline{8-12}\cline{14-18}
\rowcolor{lightgray}
RCAL & -0.055 & 0.010 & 0.010 & 0.897 & 0.950 & ~~ & -0.073 & 0.007 & 0.008 & 0.914 & 0.949 & ~~ & -0.075 & 0.018 & 0.017 & 0.896 & 0.948 \\
RMLs & -0.051 & 0.009 & 0.009 & 0.898 & 0.938 & ~~ & -0.072 & 0.007 & 0.007 & 0.903 & 0.948 & ~~ & -0.076 & 0.017 & 0.014 & 0.865 & 0.910 \\
RMLg & -0.052 & 0.009 & 0.010 & 0.896 & 0.939 & ~~ & -0.073 & 0.007 & 0.007 & 0.905 & 0.950 & ~~ & -0.076 & 0.017 & 0.014 & 0.867 & 0.909 \\

& \multicolumn{5}{c}{$\hat{\mu}_4 - \hat{\mu}_0$} &~~& \multicolumn{5}{c}{$\hat{\mu}_5 - \hat{\mu}_0$} &~~& \multicolumn{5}{c}{$\hat{\mu}_2 - \hat{\mu}_1$} \\
\cline{2-6}\cline{8-12}\cline{14-18}
\rowcolor{lightgray}
RCAL & -0.080 & 0.009 & 0.010 & 0.903 & 0.954 & ~~ & -0.086 & 0.018 & 0.018 & 0.890 & 0.938 & ~~ & -0.018 & 0.012 & 0.012 & 0.919 & 0.966 \\
RMLs & -0.083 & 0.009 & 0.009 & 0.883 & 0.931 & ~~ & -0.085 & 0.018 & 0.015 & 0.811 & 0.883 & ~~ & -0.021 & 0.011 & 0.012 & 0.913 & 0.955 \\
RMLg & -0.084 & 0.009 & 0.009 & 0.883 & 0.932 & ~~ & -0.086 & 0.018 & 0.015 & 0.810 & 0.889 & ~~ & -0.021 & 0.011 & 0.012 & 0.912 & 0.956 \\

& \multicolumn{5}{c}{$\hat{\mu}_3 - \hat{\mu}_1$} &~~& \multicolumn{5}{c}{$\hat{\mu}_4 - \hat{\mu}_1$} &~~& \multicolumn{5}{c}{$\hat{\mu}_5 - \hat{\mu}_1$} \\
\cline{2-6}\cline{8-12}\cline{14-18}
\rowcolor{lightgray}
RCAL & -0.019 & 0.020 & 0.020 & 0.913 & 0.947 & ~~ & -0.025 & 0.013 & 0.014 & 0.911 & 0.969 & ~~ & -0.031 & 0.020 & 0.021 & 0.908 & 0.956 \\
RMLs & -0.024 & 0.019 & 0.017 & 0.880 & 0.930 & ~~ & -0.032 & 0.013 & 0.013 & 0.907 & 0.949 & ~~ & -0.034 & 0.020 & 0.017 & 0.847 & 0.915 \\
RMLg & -0.024 & 0.019 & 0.017 & 0.881 & 0.931 & ~~ & -0.032 & 0.013 & 0.013 & 0.909 & 0.949 & ~~ & -0.034 & 0.020 & 0.017 & 0.852 & 0.920 \\

& \multicolumn{5}{c}{$\hat{\mu}_3 - \hat{\mu}_2$} &~~& \multicolumn{5}{c}{$\hat{\mu}_4 - \hat{\mu}_2$} &~~& \multicolumn{5}{c}{$\hat{\mu}_5 - \hat{\mu}_2$} \\
\cline{2-6}\cline{8-12}\cline{14-18}
\rowcolor{lightgray}
RCAL & -0.002 & 0.019 & 0.018 & 0.900 & 0.947 & ~~ & -0.007 & 0.011 & 0.012 & 0.916 & 0.962 & ~~ & -0.013 & 0.020 & 0.019 & 0.898 & 0.952 \\
RMLs & -0.003 & 0.018 & 0.016 & 0.862 & 0.923 & ~~ & -0.011 & 0.011 & 0.011 & 0.889 & 0.957 & ~~ & -0.013 & 0.019 & 0.016 & 0.822 & 0.887 \\
RMLg & -0.003 & 0.018 & 0.016 & 0.859 & 0.924 & ~~ & -0.011 & 0.011 & 0.011 & 0.891 & 0.959 & ~~ & -0.013 & 0.019 & 0.016 & 0.822 & 0.889 \\

& \multicolumn{5}{c}{$\hat{\mu}_4 - \hat{\mu}_3$} &~~& \multicolumn{5}{c}{$\hat{\mu}_5 - \hat{\mu}_3$} &~~& \multicolumn{5}{c}{$\hat{\mu}_5 - \hat{\mu}_4$} \\
\cline{2-6}\cline{8-12}\cline{14-18}
\rowcolor{lightgray}
RCAL & -0.006 & 0.020 & 0.019 & 0.908 & 0.958 & ~~ & -0.012 & 0.025 & 0.025 & 0.915 & 0.958 & ~~ & -0.006 & 0.020 & 0.020 & 0.903 & 0.957 \\
RMLs & -0.008 & 0.019 & 0.017 & 0.860 & 0.931 & ~~ & -0.010 & 0.025 & 0.021 & 0.836 & 0.905 & ~~ & -0.002 & 0.021 & 0.017 & 0.825 & 0.893 \\
RMLg & -0.008 & 0.019 & 0.017 & 0.860 & 0.930 & ~~ & -0.010 & 0.025 & 0.021 & 0.838 & 0.907 & ~~ & -0.002 & 0.021 & 0.017 & 0.825 & 0.895 \\

\hline
\end{tabular}}
\end{center}
\setlength{\baselineskip}{0.5\baselineskip}
\vspace{-.15in}\noindent{\tiny
\textbf{Note}: Mean, Var, EVar, Cov90, and Cov95 are calculated over the 1000 repeated subsamples, with the mean treated as the true value. RCAL denotes $\hat{\mu}_t(\hat{m}^\#_{\text{RWL}}, \hat{\pi}_{\text{RCAL}}) - \hat{\mu}_k(\hat{m}^\#_{\text{RWL}}, \hat{\pi}_{\text{RCAL}})$. RMLs denotes $\hat{\mu}_t(\hat{m}_{\text{RMLs}}, \hat{\pi}_{\text{RML}}) - \hat{\mu}_k(\hat{m}_{\text{RMLs}}, \hat{\pi}_{\text{RML}})$. RMLg denotes $\hat{\mu}_t(\hat{m}_{\text{RMLg}}, \hat{\pi}_{\text{RML}}) - \hat{\mu}_k(\hat{m}_{\text{RMLg}}, \hat{\pi}_{\text{RML}})$.}
\end{table}

\begin{table}[H]
\caption{\footnotesize Summary of $\hat{\nu}^{(k)}_t$ on sub samples for $t, k = 0, 1, \ldots, 5$.} \label{tb:nu_sub_cross}\vspace{-4ex}
\begin{center}
\resizebox{\textwidth}{!}{\begin{tabular}{lccccccccccccccccccccccc}
\hline
& \multicolumn{3}{c}{$\hat{\nu}^{(0)}_t$} & $~~$ & \multicolumn{3}{c}{$\hat{\nu}^{(1)}_t$} & $~~$ & \multicolumn{3}{c}{$\hat{\nu}^{(2)}_t$} & $~~$ & \multicolumn{3}{c}{$\hat{\nu}^{(3)}_t$} & $~~$ & \multicolumn{3}{c}{$\hat{\nu}^{(4)}_t$} & $~~$ & \multicolumn{3}{c}{$\hat{\nu}^{(5)}_t$} \\
\cline{2-4}\cline{6-8}\cline{10-12}\cline{14-16}\cline{18-20}\cline{22-24}
& RCAL & RMLs & RMLg & $~~$ & RCAL & RMLs & RMLg & $~~$  & RCAL & RMLs & RMLg & $~~$ & RCAL & RMLs & RMLg & $~~$ & RCAL & RMLs & RMLg & $~~$ & RCAL & RML & RMLg \\
\hline
& \multicolumn{23}{c}{\footnotesize t = 0} \\
$\sqrt{\text{Var}}$ & 0.002& 0.002& 0.002& ~~& 0.004& 0.004& 0.003& ~~& 0.004& 0.004& 0.003& ~~& 0.004& 0.005& 0.003& ~~& 0.004& 0.004& 0.004& ~~& 0.005& 0.007& 0.004\\
$\sqrt{\text{EVar}}$ & 0.002& 0.002& 0.002& ~~& 0.004& 0.004& 0.003& ~~& 0.003& 0.003& 0.003& ~~& 0.003& 0.005& 0.002& ~~& 0.004& 0.004& 0.003& ~~& 0.003& 0.006& 0.003\\
Cov90 & 0.896& 0.896& 0.896& ~~& 0.853& 0.879& 0.842& ~~& 0.836& 0.863& 0.857& ~~& 0.827& 0.876& 0.770& ~~& 0.792& 0.851& 0.829& ~~& 0.789& 0.871& 0.781\\
Cov95 & 0.956& 0.956& 0.956& ~~& 0.914& 0.933& 0.907& ~~& 0.902& 0.926& 0.926& ~~& 0.891& 0.940& 0.854& ~~& 0.883& 0.929& 0.902& ~~& 0.859& 0.926& 0.851\\

& \multicolumn{23}{c}{\footnotesize t = 1} \\
$\sqrt{\text{Var}}$ & 0.010& 0.010& 0.010& ~~& 0.009& 0.009& 0.009& ~~& 0.009& 0.009& 0.009& ~~& 0.009& 0.009& 0.009& ~~& 0.009& 0.010& 0.010& ~~& 0.009& 0.010& 0.010\\
$\sqrt{\text{EVar}}$ & 0.010& 0.010& 0.010& ~~& 0.010& 0.010& 0.010& ~~& 0.010& 0.010& 0.010& ~~& 0.010& 0.009& 0.009& ~~& 0.010& 0.010& 0.010& ~~& 0.010& 0.010& 0.010\\
Cov90 & 0.907& 0.894& 0.893& ~~& 0.909& 0.909& 0.909& ~~& 0.911& 0.911& 0.913& ~~& 0.909& 0.899& 0.904& ~~& 0.912& 0.903& 0.906& ~~& 0.909& 0.887& 0.889\\
Cov95 & 0.951& 0.939& 0.940& ~~& 0.954& 0.954& 0.954& ~~& 0.953& 0.954& 0.954& ~~& 0.953& 0.947& 0.949& ~~& 0.954& 0.948& 0.950& ~~& 0.954& 0.950& 0.950\\

& \multicolumn{23}{c}{\footnotesize t = 2} \\
$\sqrt{\text{Var}}$ & 0.008& 0.007& 0.007& ~~& 0.006& 0.007& 0.007& ~~& 0.006& 0.006& 0.006& ~~& 0.006& 0.006& 0.006& ~~& 0.006& 0.006& 0.006& ~~& 0.006& 0.007& 0.007\\
$\sqrt{\text{EVar}}$ & 0.008& 0.007& 0.007& ~~& 0.007& 0.007& 0.007& ~~& 0.007& 0.007& 0.007& ~~& 0.007& 0.006& 0.006& ~~& 0.007& 0.007& 0.007& ~~& 0.007& 0.007& 0.007\\
Cov90 & 0.905& 0.897& 0.896& ~~& 0.908& 0.895& 0.901& ~~& 0.910& 0.910& 0.910& ~~& 0.910& 0.902& 0.910& ~~& 0.913& 0.913& 0.915& ~~& 0.911& 0.887& 0.891\\
Cov95 & 0.949& 0.945& 0.942& ~~& 0.950& 0.946& 0.950& ~~& 0.950& 0.950& 0.950& ~~& 0.952& 0.955& 0.955& ~~& 0.952& 0.958& 0.960& ~~& 0.951& 0.939& 0.946\\

& \multicolumn{23}{c}{\footnotesize t = 3} \\
$\sqrt{\text{Var}}$ & 0.019& 0.018& 0.018& ~~& 0.014& 0.017& 0.016& ~~& 0.014& 0.016& 0.016& ~~& 0.014& 0.014& 0.014& ~~& 0.014& 0.016& 0.016& ~~& 0.014& 0.015& 0.015\\
$\sqrt{\text{EVar}}$ & 0.018& 0.015& 0.015& ~~& 0.015& 0.016& 0.016& ~~& 0.015& 0.017& 0.017& ~~& 0.015& 0.015& 0.015& ~~& 0.015& 0.017& 0.017& ~~& 0.015& 0.016& 0.016\\
Cov90 & 0.894 & 0.834& 0.836& ~~& 0.903& 0.890& 0.890& ~~& 0.906& 0.913& 0.914& ~~& 0.903& 0.903& 0.903& ~~& 0.902& 0.919& 0.920& ~~& 0.902& 0.907& 0.908\\
Cov95 & 0.947& 0.894& 0.894& ~~& 0.945& 0.940& 0.941& ~~& 0.947& 0.948& 0.949& ~~& 0.946& 0.946& 0.946& ~~& 0.947& 0.957& 0.959& ~~& 0.946& 0.947& 0.948\\

& \multicolumn{23}{c}{\footnotesize t = 4} \\
$\sqrt{\text{Var}}$ & 0.010& 0.010& 0.010& ~~& 0.008& 0.009& 0.009& ~~& 0.008& 0.008& 0.008& ~~& 0.008& 0.008& 0.008& ~~& 0.008& 0.008& 0.008& ~~& 0.008& 0.008& 0.008\\
$\sqrt{\text{EVar}}$ & 0.010& 0.009& 0.009& ~~& 0.008& 0.009& 0.009& ~~& 0.008& 0.008& 0.008& ~~& 0.008& 0.007& 0.007& ~~& 0.008& 0.008& 0.008& ~~& 0.008& 0.008& 0.008\\
Cov90 & 0.909& 0.873& 0.876& ~~& 0.888& 0.880& 0.882& ~~& 0.893& 0.896& 0.898& ~~& 0.888& 0.867& 0.872& ~~& 0.888& 0.888& 0.888& ~~& 0.890& 0.879& 0.881\\
Cov95 & 0.952& 0.926& 0.927& ~~& 0.939& 0.930& 0.931& ~~& 0.937& 0.943& 0.942& ~~& 0.937& 0.926& 0.926& ~~& 0.937& 0.937& 0.937& ~~& 0.937& 0.933& 0.935\\

& \multicolumn{23}{c}{\footnotesize t = 5} \\
$\sqrt{\text{Var}}$ & 0.019& 0.019& 0.019& ~~& 0.017& 0.020& 0.020& ~~& 0.017& 0.019& 0.019& ~~& 0.017& 0.017& 0.017& ~~& 0.017& 0.018& 0.018& ~~& 0.017& 0.017& 0.017\\
$\sqrt{\text{EVar}}$ & 0.019& 0.015& 0.015& ~~& 0.017& 0.017& 0.017& ~~& 0.017& 0.018& 0.018& ~~& 0.017& 0.015& 0.015& ~~& 0.017& 0.018& 0.018& ~~& 0.017& 0.017& 0.017\\
Cov90 & 0.889 & 0.803& 0.803& ~~& 0.885& 0.831& 0.838& ~~& 0.887& 0.869& 0.872& ~~& 0.882& 0.860& 0.858& ~~& 0.885& 0.900& 0.906& ~~& 0.886& 0.886& 0.886\\
Cov95 & 0.942 & 0.873& 0.874& ~~& 0.939& 0.879& 0.886& ~~& 0.942& 0.927& 0.932& ~~& 0.941& 0.913& 0.913& ~~& 0.944& 0.942& 0.946& ~~& 0.943& 0.943& 0.943\\

\hline
\end{tabular}}
\end{center}
\setlength{\baselineskip}{0.5\baselineskip}
\vspace{-.15in}\noindent{\tiny
\textbf{Note}: Var, EVar, Cov90, and Cov95 are calculated over the 1000 repeated subsamples, with the mean treated as the true value. RCAL denotes $\hat{\nu}^{(k)}_{t, \text{RCAL}}$. RMLs denotes $\hat{\nu}^{(k)}_{t, \text{RMLs}}$. RMLg denotes $\hat{\nu}^{(k)}_{t, \text{RMLg}}$.}
\end{table}

\begin{table}[H]
\caption{\footnotesize Summary of $\hat{\mu}_t$ on full sample for $t = 0, 1, \ldots, 5$ with tuning parameters selected as $\lambda.min$.} \label{tb:mu_full_min}\vspace{-4ex}
\begin{center}
\resizebox{1.0\textwidth}{!}{\begin{tabular}{lccccccccccc}
\hline
 & Est & SE & 95CI &~~& Est & SE & 95CI &~~ & Est & SE & 95CI  \\
\hline
  & \multicolumn{3}{c}{$\hat{\mu}_0~(n_0 = 13587)$} &~~& \multicolumn{3}{c}{$\hat{\mu}_1~(n_1 = 552)$} &~~& \multicolumn{3}{c}{$\hat{\mu}_2~(n_2 = 1190)$} \\
  \cline{2-4}\cline{6-8}\cline{10-12}
Uuadj & 8.1272 & 0.0003 & (8.1266, 8.1279) &~~& 8.0576 & 0.0019 & (8.0538, 8.0614) &~~& 8.0398 & 0.0013 & (8.0372, 8.0424) \\
\rowcolor{lightgray}
CAL & 8.1247 & 0.0004 & (8.1240, 8.1254) &~~& 8.0763 & 0.0020 & (8.0723, 8.0803) &~~& 8.0570 & 0.0017 & (8.0537, 8.0602) \\
ML & 8.1252 & 0.0006 & (8.1240, 8.1263) &~~& 8.0768 & 0.0019 & (8.0730, 8.0806) &~~& 8.0573 & 0.0015 & (8.0542, 8.0603)  \\
\rowcolor{lightgray}
RCAL & 8.1243 & 0.0004 & (8.1236, 8.1250) &~~& 8.0763 & 0.0020 & (8.0723, 8.0803) &~~& 8.0570 & 0.0016 & (8.0540, 8.0601)  \\
RMLs & 8.1243 & 0.0004 & (8.1236, 8.1251) &~~& 8.0762 & 0.0019 & (8.0724, 8.0800) &~~& 8.0567 & 0.0017 & (8.0533, 8.0602) \\
RMLg & 8.1243 & 0.0004 & (8.1236, 8.1251) &~~& 8.0755 & 0.0019 & (8.0717, 8.0793) &~~& 8.0558 & 0.0017 & (8.0525, 8.0591) \\

  & \multicolumn{3}{c}{$\hat{\mu}_3~(n_3 = 197)$} &~~& \multicolumn{3}{c}{$\hat{\mu}_4~(n_4 = 781)$} &~~& \multicolumn{3}{c}{$\hat{\mu}_5~(n_5 = 159)$} \\
 \cline{2-4}\cline{6-8}\cline{10-12}
Uuadj & 8.0474 & 0.0030 & (8.0416, 8.0533) &~~& 8.0365 & 0.0016 & (8.0334, 8.0396) &~~& 8.0318 & 0.0034 & (8.0251, 8.0384) \\
\rowcolor{lightgray}
CAL & 8.0517 & 0.0049 & (8.0421, 8.0612) &~~& 8.0460 & 0.0023 & (8.0415, 8.0505) &~~& 8.0442 & 0.0057 & (8.0330, 8.0554) \\
ML & 8.0536 & 0.0045 & (8.0449, 8.0623) &~~& 8.0460 & 0.0021 & (8.0418, 8.0502) &~~& 8.0413 & 0.0057 & (8.0301, 8.0525) \\
\rowcolor{lightgray}
RCAL & 8.0520 & 0.0041 & (8.0440, 8.0599) &~~& 8.0474 & 0.0019 & (8.0436, 8.0512)  &~~& 8.0463 & 0.0044 & (8.0378, 8.0549) \\
RMLs & 8.0514 & 0.0035 & (8.0444, 8.0583) &~~& 8.0459 & 0.0019 & (8.0421, 8.0496) &~~& 8.0425 & 0.0042 & (8.0343, 8.0507)  \\
RMLg & 8.0491 & 0.0036 & (8.0421, 8.0562) &~~& 8.0454 & 0.0019 & (8.0416, 8.0491)  &~~& 8.0416 & 0.0043 & (8.0332, 8.0500) \\

\hline
\end{tabular}}
\end{center}
\setlength{\baselineskip}{0.5\baselineskip}
\vspace{-.15in}\noindent{\tiny
\textbf{Note}: Est, SE, or 95CI denotes point estimate, standard error, or 95\% confidence interval respectively.
Unadj denotes the unadjusted estimate, $\hat{\mu}_t = \tilde{\E}(Y^{(t)})$.
RCAL, RMLs, or RMLg denotes $\hat{\mu}_t(\hat{m}^\#_{\text{RWL}}, \hat{\pi}_{\text{RCAL}})$, $\hat{\mu}_t(\hat{m}_{\text{RMLs}}, \hat{\pi}_{\text{RML}})$,
or $\hat{\mu}_t(\hat{m}_{\text{RMLg}}, \hat{\pi}_{\text{RML}})$ respectively. CAL or ML denotes non-regularized estimation with main effects only in PS and OR models.}
\end{table}

\begin{table}[H]
\caption{\footnotesize Summary of  $\hat{\mu}_t - \hat{\mu}_k$ on full sample for $t, k = 0, 1, \ldots, 5$ with tuning parameters selected as $\lambda.min$.} \label{tb:ate_full_min}\vspace{-4ex}
\begin{center}
\resizebox{1.0\textwidth}{!}{\begin{tabular}{lccccccccccc}
\hline
 & Est & SE & 95CI &~~& Est & SE & 95CI &~~& Est & SE & 95CI  \\
\hline
& \multicolumn{3}{c}{$\hat{\mu}_1 - \hat{\mu}_0$} &~~& \multicolumn{3}{c}{$\hat{\mu}_2 - \hat{\mu}_0$} &~~& \multicolumn{3}{c}{$\hat{\mu}_3 - \hat{\mu}_0$} \\
\cline{2-4}\cline{6-8}\cline{10-12}
\rowcolor{lightgray}
CAL & -0.0484 & 0.0021 & (-0.0524, -0.0443) &~~& -0.0677 & 0.0017 & (-0.0710, -0.0644) &~~& -0.0730 & 0.0049 & (-0.0826, -0.0634) \\
ML & -0.0484 & 0.0020 & (-0.0523, -0.0444) &~~& -0.0679 & 0.0016 & (-0.0711, -0.0647) &~~& -0.0716 & 0.0045 & (-0.0804, -0.0628)  \\
\rowcolor{lightgray}
RCAL & -0.0480 & 0.0020 & (-0.0520, -0.0440) &~~& -0.0673 & 0.0016 & (-0.0704, -0.0642)  &~~& -0.0723 & 0.0041 & (-0.0803, -0.0643) \\
RMLs & -0.0482 & 0.0020 & (-0.0520, -0.0443) &~~& -0.0676 & 0.0018 & (-0.0711, -0.0641)  &~~& -0.0730 & 0.0036 & (-0.0799, -0.0660)  \\
RMLg & -0.0489 & 0.0020 & (-0.0527, -0.0450) &~~& -0.0685 & 0.0017 & (-0.0719, -0.0652) &~~& -0.0752 & 0.0036 & (-0.0823, -0.0682)  \\

& \multicolumn{3}{c}{$\hat{\mu}_4 - \hat{\mu}_0$} &~~& \multicolumn{3}{c}{$\hat{\mu}_5 - \hat{\mu}_0$} &~~& \multicolumn{3}{c}{$\hat{\mu}_2 - \hat{\mu}_1$}\\
\cline{2-4}\cline{6-8}\cline{10-12}
\rowcolor{lightgray}
CAL & -0.0787 & 0.0023 & (-0.0833, -0.0741) &~~& -0.0805 & 0.0057 & (-0.0917, -0.0693)  &~~& -0.0194 & 0.0026 & (-0.0245, -0.0142) \\
ML & -0.0792 & 0.0022 & (-0.0835, -0.0749)  &~~& -0.0838 & 0.0057 & (-0.0951, -0.0726)  &~~& -0.0195 & 0.0025 & (-0.0244, -0.0147)  \\
\rowcolor{lightgray}
RCAL & -0.0769 & 0.0020 & (-0.0808, -0.0731) &~~& -0.0780 & 0.0044 & (-0.0866, -0.0694) &~~& -0.0193 & 0.0025 & (-0.0243, -0.0143)  \\
RMLs & -0.0785 & 0.0020 & (-0.0823, -0.0747) &~~& -0.0819 & 0.0042 & (-0.0901, -0.0736) &~~& -0.0194 & 0.0026 & (-0.0245, -0.0144)  \\
RMLg & -0.0790 & 0.0020 & (-0.0828, -0.0752) &~~& -0.0828 & 0.0043 & (-0.0912, -0.0743)  &~~& -0.0196 & 0.0026 & (-0.0247, -0.0146)  \\

& \multicolumn{3}{c}{$\hat{\mu}_3 - \hat{\mu}_1$} &~~& \multicolumn{3}{c}{$\hat{\mu}_4 - \hat{\mu}_1$} &~~& \multicolumn{3}{c}{$\hat{\mu}_5 - \hat{\mu}_1$}\\
\cline{2-4}\cline{6-8}\cline{10-12}
\rowcolor{lightgray}
CAL & -0.0247 & 0.0053 & (-0.0350, -0.0143) &~~& -0.0303 & 0.0031 & (-0.0364, -0.0243)  &~~& -0.0322 & 0.0061 & (-0.0440, -0.0203) \\
ML & -0.0232 & 0.0048 & (-0.0327, -0.0137) &~~& -0.0308 & 0.0029 & (-0.0365, -0.0252)  &~~& -0.0355 & 0.0060 & (-0.0473, -0.0237) \\
\rowcolor{lightgray}
RCAL & -0.0243 & 0.0045 & (-0.0332, -0.0154) &~~& -0.0289 & 0.0028 & (-0.0344, -0.0235) &~~& -0.0300 & 0.0048 & (-0.0394, -0.0206)  \\
RMLs & -0.0248 & 0.0040 & (-0.0327, -0.0169) &~~& -0.0303 & 0.0027 & (-0.0356, -0.0250) &~~& -0.0337 & 0.0046 & (-0.0427, -0.0246) \\
RMLg & -0.0263 & 0.0041 & (-0.0343, -0.0184) &~~& -0.0301 & 0.0027 & (-0.0355, -0.0248) &~~& -0.0339 & 0.0047 & (-0.0431, -0.0246)  \\

& \multicolumn{3}{c}{$\hat{\mu}_3 - \hat{\mu}_2$} &~~& \multicolumn{3}{c}{$\hat{\mu}_4 - \hat{\mu}_2$} &~~& \multicolumn{3}{c}{$\hat{\mu}_5 - \hat{\mu}_2$}\\
\cline{2-4}\cline{6-8}\cline{10-12}
\rowcolor{lightgray}
CAL & -0.0053 & 0.0051 & (-0.0154, 0.0048)  &~~& -0.0110 & 0.0028 & (-0.0165, -0.0054)  &~~& -0.0128 & 0.0059 & (-0.0245, -0.0011) \\
ML & -0.0037 & 0.0047 & (-0.0129, 0.0055) &~~& -0.0113 & 0.0026 & (-0.0164, -0.0061) &~~& -0.0160 & 0.0059 & (-0.0275, -0.0044)  \\
\rowcolor{lightgray}
RCAL & -0.0050 & 0.0043 & (-0.0135, 0.0035)  &~~& -0.0097 & 0.0025 & (-0.0145, -0.0048) &~~& -0.0107 & 0.0046 & (-0.0198, -0.0016) \\
RMLs & -0.0054 & 0.0039 & (-0.0131, 0.0024) &~~& -0.0109 & 0.0026 & (-0.0159, -0.0058) &~~& -0.0143 & 0.0045 & (-0.0232, -0.0053) \\
RMLg & -0.0067 & 0.0040 & (-0.0145, 0.0011) &~~& -0.0105 & 0.0026 & (-0.0155, -0.0055)  &~~& -0.0142 & 0.0046 & (-0.0233, -0.0052) \\

& \multicolumn{3}{c}{$\hat{\mu}_4 - \hat{\mu}_3$} &~~& \multicolumn{3}{c}{$\hat{\mu}_5 - \hat{\mu}_3$} &~~& \multicolumn{3}{c}{$\hat{\mu}_5 - \hat{\mu}_4$}\\
\cline{2-4}\cline{6-8}\cline{10-12}
\rowcolor{lightgray}
CAL & -0.0057 & 0.0054 & (-0.0162, 0.0049)  &~~& -0.0075 & 0.0075 & (-0.0222, 0.0072) &~~& -0.0018 & 0.0062 & (-0.0139, 0.0102)  \\
ML & -0.0076 & 0.0049 & (-0.0173, 0.0021) &~~& -0.0123 & 0.0072 & (-0.0265, 0.0019) &~~& -0.0047 & 0.0061 & (-0.0166, 0.0073) \\
\rowcolor{lightgray}
RCAL & -0.0046 & 0.0045 & (-0.0134, 0.0042)  &~~& -0.0057 & 0.0060 & (-0.0173, 0.0060) &~~& -0.0011 & 0.0048 & (-0.0104, 0.0083)  \\
RMLs & -0.0055 & 0.0040 & (-0.0134, 0.0024) &~~& -0.0089 & 0.0055 & (-0.0196, 0.0019) &~~& -0.0034 & 0.0046 & (-0.0124, 0.0057) \\
RMLg & -0.0038 & 0.0041 & (-0.0118, 0.0042)  &~~& -0.0075 & 0.0056 & (-0.0185, 0.0034) &~~& -0.0038 & 0.0047 & (-0.0130, 0.0055)  \\

\hline
\end{tabular}}
\end{center}
\setlength{\baselineskip}{0.5\baselineskip}
\vspace{-.15in}\noindent{\tiny
\textbf{Note}: Est, SE, or 95CI denotes point estimate, standard error, or 95\% confidence interval respectively.
RCAL, RMLs, or RMLg denotes
$\hat{\mu}_t(\hat{m}^\#_{\text{RWL}}, \hat{\pi}_{\text{RCAL}}) - \hat{\mu}_k(\hat{m}^\#_{\text{RWL}}, \hat{\pi}_{\text{RCAL}})$,
$\hat{\mu}_t(\hat{m}_{\text{RMLs}}, \hat{\pi}_{\text{RML}}) - \hat{\mu}_k(\hat{m}_{\text{RMLs}}, \hat{\pi}_{\text{RML}})$,
or $\hat{\mu}_t(\hat{m}_{\text{RMLg}}, \hat{\pi}_{\text{RML}}) - \hat{\mu}_k(\hat{m}_{\text{RMLg}}, \hat{\pi}_{\text{RML}})$
respectively. CAL or ML denotes non-regularized estimation with main effects only in PS and OR models.}
\end{table}

\begin{table}[H]
\caption{\footnotesize Summary of $\hat{\nu}^{(k)}_t$ on full sample for $t, k = 0, 1, \ldots, 5$ with tuning parameters selected as $\lambda.min$.} \label{tb:nu_full_min}\vspace{-4ex}
\begin{center}
\resizebox{1.0\textwidth}{0.5\textwidth}{\begin{tabular}{lccccccccccccccccccccccc}
\hline
& \multicolumn{3}{c}{$\hat{\nu}^{(0)}_t$} & $~~$ & \multicolumn{3}{c}{$\hat{\nu}^{(1)}_t$} & $~~$ & \multicolumn{3}{c}{$\hat{\nu}^{(2)}_t$} & $~~$ & \multicolumn{3}{c}{$\hat{\nu}^{(3)}_t$} & $~~$ & \multicolumn{3}{c}{$\hat{\nu}^{(4)}_t$} & $~~$ & \multicolumn{3}{c}{$\hat{\nu}^{(5)}_t$} \\
\cline{2-4}\cline{6-8}\cline{10-12}\cline{14-16}\cline{18-20}\cline{22-24}
& Est & SE & 95CI &~~& Est & SE & 95CI &~~& Est & SE & 95CI &~~& Est & SE & 95CI &~~& Est & SE & 95CI &~~& Est & SE & 95CI  \\
\hline
& \multicolumn{23}{c}{\footnotesize t = 0} \\
\rowcolor{lightgray}
CAL & 8.1272 & 0.0003 & (8.1266, 8.1279) &~~& 8.1025 & 0.0009 & (8.1007, 8.1043) &~~& 8.1098 & 8e-04 & (8.1082, 8.1113) &~~& 8.1240 & 0.0011 & (8.1218, 8.1262) &~~& 8.1193 & 0.0009 & (8.1174, 8.1211) &~~& 8.1253 & 0.0015 & (8.1223, 8.1283)  \\
ML & 8.1272 & 0.0003 & (8.1266, 8.1279) &~~& 8.1025 & 0.0009 & (8.1007, 8.1042) &~~& 8.1102 & 0.0010 & (8.1083, 8.1122) &~~& 8.1262 & 0.0024 & (8.1214, 8.1309) &~~& 8.1253 & 0.0061 & (8.1134, 8.1372) &~~& 8.1360 & 0.0095 & (8.1175, 8.1546) \\
\rowcolor{lightgray}
RCAL & 8.1272 & 0.0003 & (8.1266, 8.1279) &~~& 8.1011 & 0.0009 & (8.0994, 8.1028) &~~& 8.1072 & 0.0008 & (8.1057, 8.1088) &~~& 8.1227 & 0.0007 & (8.1213, 8.1242) &~~& 8.1162 & 0.0009 & (8.1145, 8.1180) &~~& 8.1225 & 0.0012 & (8.1203, 8.1248) \\
RMLs & 8.1272 & 0.0003 & (8.1266, 8.1279) &~~& 8.1011 & 0.0009 & (8.0994, 8.1028) &~~& 8.1074 & 0.0008 & (8.1059, 8.1090) &~~& 8.1231 & 0.0011 & (8.1209, 8.1253) &~~& 8.1168 & 0.0011 & (8.1148, 8.1189) &~~& 8.1237 & 0.0020 & (8.1197, 8.1276) \\
RMLg & 8.1272 & 0.0003 & (8.1266, 8.1279) &~~& 8.1013 & 0.0009 & (8.0995, 8.1030) &~~& 8.1075 & 0.0008 & (8.1060, 8.1091) &~~& 8.1229 & 0.0011 & (8.1207, 8.1252) &~~& 8.1168 & 0.0011 & (8.1146, 8.1189) &~~& 8.1233 & 0.0022 & (8.1191, 8.1276) \\

& \multicolumn{23}{c}{\footnotesize t = 1} \\
\rowcolor{lightgray}
CAL & 8.0787 & 0.0022 & (8.0743, 8.0830) &~~& 8.0576 & 0.0019 & (8.0538, 8.0614) &~~& 8.0633 & 0.0019 & (8.0596, 8.0670) &~~& 8.0767 & 0.0021 & (8.0725, 8.0808) &~~& 8.0701 & 0.0022 & (8.0658, 8.0744) &~~& 8.0715 & 0.0031 & (8.0655, 8.0775) \\
ML & 8.0792 & 0.0021 & (8.0752, 8.0833) &~~& 8.0576 & 0.0019 & (8.0538, 8.0614) &~~& 8.0632 & 0.0018 & (8.0596, 8.0668) &~~& 8.0764 & 0.0021 & (8.0723, 8.0806) &~~& 8.0699 & 0.0020 & (8.0660, 8.0739) &~~& 8.0711 & 0.0026 & (8.0661, 8.0761) \\
\rowcolor{lightgray}
RCAL & 8.0787 & 0.0022 & (8.0744, 8.0829) &~~& 8.0576 & 0.0019 & (8.0538, 8.0614) &~~& 8.0632 & 0.0018 & (8.0597, 8.0668) &~~& 8.0723 & 0.0018 & (8.0687, 8.0759) &~~& 8.0708 & 0.0019 & (8.0671, 8.0744) &~~& 8.0692 & 0.0019 & (8.0655, 8.0729) \\
RMLs & 8.0786 & 0.0021 & (8.0746, 8.0827) &~~& 8.0576 & 0.0019 & (8.0538, 8.0614) &~~& 8.0622 & 0.0019 & (8.0585, 8.0660) &~~& 8.0768 & 0.0021 & (8.0727, 8.0809) &~~& 8.0691 & 0.0021 & (8.0650, 8.0733) &~~& 8.0720 & 0.0027 & (8.0668, 8.0772) \\
RMLg & 8.0777 & 0.0021 & (8.0736, 8.0818)  &~~& 8.0576 & 0.0019 & (8.0538, 8.0614) &~~& 8.0624 & 0.0019 & (8.0586, 8.0661) &~~& 8.0771 & 0.0020 & (8.0732, 8.0810) &~~& 8.0697 & 0.0021 & (8.0656, 8.0738) &~~& 8.0731 & 0.0025 & (8.0681, 8.0781) \\

& \multicolumn{23}{c}{\footnotesize t = 2} \\
\rowcolor{lightgray}
CAL & 8.0603 & 0.0018 & (8.0567, 8.0639) &~~& 8.0358 & 0.0015 & (8.0329, 8.0387) &~~& 8.0398 & 0.0013 & (8.0372, 8.0424) &~~& 8.0511 & 0.0016 & (8.0480, 8.0542) &~~& 8.0441 & 0.0014 & (8.0412, 8.0469) &~~& 8.0434 & 0.0021 & (8.0392, 8.0476) \\
ML & 8.0607 & 0.0017 & (8.0574, 8.0640) &~~& 8.0358 & 0.0015 & (8.0329, 8.0387) &~~& 8.0398 & 0.0013 & (8.0372, 8.0424) &~~& 8.0509 & 0.0016 & (8.0478, 8.0541) &~~& 8.0440 & 0.0014 & (8.0413, 8.0468) &~~& 8.0435 & 0.0019 & (8.0398, 8.0472) \\
\rowcolor{lightgray}
RCAL & 8.0604 & 0.0017 & (8.0571, 8.0638) &~~& 8.0361 & 0.0014 & (8.0335, 8.0388) &~~& 8.0398 & 0.0013 & (8.0372, 8.0424) &~~& 8.0465 & 0.0013 & (8.0438, 8.0491) &~~& 8.0446 & 0.0013 & (8.0420, 8.0472) &~~& 8.0416 & 0.0014 & (8.0389, 8.0444)  \\
RMLs & 8.0600 & 0.0020 & (8.0562, 8.0638) &~~& 8.0366 & 0.0014 & (8.0338, 8.0394) &~~& 8.0398 & 0.0013 & (8.0372, 8.0424) &~~& 8.0517 & 0.0016 & (8.0486, 8.0547) &~~& 8.0442 & 0.0014 & (8.0415, 8.0470) &~~& 8.0439 & 0.0020 & (8.0400, 8.0479) \\
RMLg & 8.0589 & 0.0019 & (8.0552, 8.0625) &~~& 8.0364 & 0.0014 & (8.0336, 8.0392) &~~& 8.0398 & 0.0013 & (8.0372, 8.0424) &~~& 8.0519 & 0.0015 & (8.0489, 8.0549) &~~& 8.0444 & 0.0014 & (8.0417, 8.0472) &~~& 8.0446 & 0.0020 & (8.0407, 8.0484) \\

& \multicolumn{23}{c}{\footnotesize t = 3} \\
\rowcolor{lightgray}
CAL & 8.0544 & 0.0055 & (8.0435, 8.0652) &~~& 8.0352 & 0.0040 & (8.0273, 8.0431) &~~& 8.0375 & 0.0034 & (8.0308, 8.0441) &~~& 8.0474 & 0.0030 & (8.0416, 8.0533) &~~& 8.0414 & 0.0032 & (8.0352, 8.0477) &~~& 8.0421 & 0.0038 & (8.0347, 8.0496) \\
ML & 8.0567 & 0.0049 & (8.0471, 8.0663) &~~& 8.0345 & 0.0042 & (8.0263, 8.0428) &~~& 8.0373 & 0.0034 & (8.0307, 8.0439) &~~& 8.0474 & 0.0030 & (8.0416, 8.0533) &~~& 8.0415 & 0.0032 & (8.0353, 8.0478) &~~& 8.0423 & 0.0035 & (8.0354, 8.0493) \\
\rowcolor{lightgray}
RCAL & 8.0536 & 0.0045 & (8.0449, 8.0624) &~~& 8.0443 & 0.0030 & (8.0385, 8.0501) &~~& 8.0426 & 0.0029 & (8.0369, 8.0484) &~~& 8.0474 & 0.0030 & (8.0416, 8.0533) &~~& 8.0452 & 0.0029 & (8.0394, 8.0510) &~~& 8.0467 & 0.0030 & (8.0409, 8.0526) \\
RMLs & 8.0542 & 0.0038 & (8.0468, 8.0617) &~~& 8.0340 & 0.0037 & (8.0266, 8.0413) &~~& 8.0355 & 0.0036 & (8.0284, 8.0426) &~~& 8.0474 & 0.0030 & (8.0416, 8.0533) &~~& 8.0407 & 0.0033 & (8.0342, 8.0472) &~~& 8.0431 & 0.0035 & (8.0363, 8.0499) \\
RMLg & 8.0516 & 0.0038 & (8.0441, 8.0591) &~~& 8.0335 & 0.0039 & (8.0259, 8.0411) &~~& 8.0348 & 0.0038 & (8.0273, 8.0423) &~~& 8.0474 & 0.0030 & (8.0416, 8.0533) &~~& 8.0406 & 0.0034 & (8.0339, 8.0473) &~~& 8.0432 & 0.0034 & (8.0365, 8.0499) \\

& \multicolumn{23}{c}{\footnotesize t = 4} \\
\rowcolor{lightgray}
CAL & 8.0488 & 0.0026 & (8.0438, 8.0538) &~~& 8.0261 & 0.0023 & (8.0217, 8.0306) &~~& 8.0312 & 0.0018 & (8.0278, 8.0347) &~~& 8.0426 & 0.0017 & (8.0392, 8.0459) &~~& 8.0365 & 0.0016 & (8.0334, 8.0396) &~~& 8.0374 & 0.0020 & (8.0335, 8.0413) \\
ML & 8.0488 & 0.0024 & (8.0441, 8.0534) &~~& 8.0263 & 0.0021 & (8.0222, 8.0304) &~~& 8.0312 & 0.0017 & (8.0279, 8.0345) &~~& 8.0425 & 0.0017 & (8.0391, 8.0459) &~~& 8.0365 & 0.0016 & (8.0334, 8.0396) &~~& 8.0373 & 0.0019 & (8.0336, 8.0411) \\
\rowcolor{lightgray}
RCAL & 8.0503 & 0.0021 & (8.0461, 8.0545) &~~& 8.0307 & 0.0016 & (8.0277, 8.0338) &~~& 8.0316 & 0.0015 & (8.0286, 8.0347) &~~& 8.0387 & 0.0016 & (8.0357, 8.0418) &~~& 8.0365 & 0.0016 & (8.0334, 8.0396) &~~& 8.0361 & 0.0016 & (8.0330, 8.0392) \\
RMLs & 8.0485 & 0.0021 & (8.0444, 8.0527) &~~& 8.0274 & 0.0020 & (8.0235, 8.0312) &~~& 8.0318 & 0.0017 & (8.0285, 8.0350) &~~& 8.0429 & 0.0017 & (8.0395, 8.0462) &~~& 8.0365 & 0.0016 & (8.0334, 8.0396) &~~& 8.0381 & 0.0019 & (8.0343, 8.0419) \\
RMLg & 8.0479 & 0.0021 & (8.0438, 8.0521)  &~~& 8.0272 & 0.0020 & (8.0234, 8.0310) &~~& 8.0316 & 0.0017 & (8.0283, 8.0348) &~~& 8.0429 & 0.0016 & (8.0397, 8.0461) &~~& 8.0365 & 0.0016 & (8.0334, 8.0396) &~~& 8.0384 & 0.0019 & (8.0348, 8.0421) \\

& \multicolumn{23}{c}{\footnotesize t = 5} \\
\rowcolor{lightgray}
CAL & 8.0482 & 0.0062 & (8.0361, 8.0603) &~~& 8.0160 & 0.0082 & (8.0000, 8.0320) &~~& 8.0231 & 0.0055 & (8.0123, 8.0339)  &~~& 8.0377 & 0.0038 & (8.0303, 8.0451) &~~& 8.0303 & 0.0036 & (8.0232, 8.0375) &~~& 8.0318 & 0.0034 & (8.0251, 8.0384) \\
ML & 8.0448 & 0.0061 & (8.0328, 8.0567) &~~& 8.0153 & 0.0081 & (7.9994, 8.0312) &~~& 8.0233 & 0.0050 & (8.0134, 8.0331) &~~& 8.0372 & 0.0037 & (8.0300, 8.0444) &~~& 8.0302 & 0.0035 & (8.0233, 8.0372) &~~& 8.0318 & 0.0034 & (8.0251, 8.0384) \\
\rowcolor{lightgray}
RCAL & 8.0494 & 0.0048 & (8.0400, 8.0589) &~~& 8.0312 & 0.0033 & (8.0246, 8.0377) &~~& 8.0306 & 0.0034 & (8.0240, 8.0371) &~~& 8.0333 & 0.0033 & (8.0268, 8.0398) &~~& 8.0326 & 0.0033 & (8.0261, 8.0391) &~~& 8.0318 & 0.0034 & (8.0251, 8.0384)  \\
RMLs & 8.0458 & 0.0045 & (8.0371, 8.0546) &~~& 8.0208 & 0.0054 & (8.0102, 8.0314) &~~& 8.0251 & 0.0042 & (8.0168, 8.0333) &~~& 8.0370 & 0.0033 & (8.0306, 8.0435) &~~& 8.0298 & 0.0034 & (8.0232, 8.0365) &~~& 8.0318 & 0.0034 & (8.0251, 8.0384) \\
RMLg & 8.0447 & 0.0046 & (8.0357, 8.0537) &~~& 8.0220 & 0.0056 & (8.0111, 8.0330) &~~& 8.0253 & 0.0044 & (8.0167, 8.0338) &~~& 8.0368 & 0.0033 & (8.0304, 8.0432) &~~& 8.0299 & 0.0035 & (8.0231, 8.0367) &~~& 8.0318 & 0.0034 & (8.0251, 8.0384) \\

\hline
\end{tabular}}
\end{center}
\setlength{\baselineskip}{0.5\baselineskip}
\vspace{-.15in}\noindent{\tiny
\textbf{Note}: Est, SE, or 95CI denotes point estimate, standard error, or 95\% confidence interval respectively. CAL denotes $\hat{\nu}^{(k)}_{t, \text{CAL}}$. ML denotes $\hat{\nu}^{(k)}_{t, \text{ML}}$. RCAL denotes $\hat{\nu}^{(k)}_{t, \text{RCAL}}$. RMLs denotes $\hat{\nu}^{(k)}_{t, \text{RMLs}}$. RMLg denotes $\hat{\nu}^{(k)}_{t, \text{RMLg}}$.}
\end{table}

\begin{table}[H]
\caption{\footnotesize Summary of $\hat{\mu}_t$ on sub samples for $t = 0, 1, \ldots, 5$ with tuning parameters selected as $\lambda.min$.} \label{tb:mu_sub_min}\vspace{-4ex}
\begin{center}
\resizebox{1.0\textwidth}{!}{\begin{tabular}{lccccccccccccccccc}
\hline
 & Mean & $\sqrt{\text{Var}}$ & $\sqrt{\text{EVar}}$ & Cov90 & Cov95 &~~& Mean & $\sqrt{\text{Var}}$ & $\sqrt{\text{EVar}}$ & Cov90 & Cov95 &~~& Mean & $\sqrt{\text{Var}}$ & $\sqrt{\text{EVar}}$ & Cov90 & Cov95 \\
\hline
  & \multicolumn{5}{c}{$\hat{\mu}_0~(n_0 = 13587)$} &~~& \multicolumn{5}{c}{$\hat{\mu}_1~(n_1 = 552)$}  &~~& \multicolumn{5}{c}{$\hat{\mu}_2~(n_2 = 1190)$}\\
  \cline{2-6}\cline{8-12}\cline{14-18}
  \rowcolor{lightgray}
RCAL & 8.125 & 0.002 & 0.002 & 0.883 & 0.942 & ~~ & 8.075 & 0.010 & 0.009 & 0.885 & 0.942 & ~~ & 8.056 & 0.008 & 0.007 & 0.871 & 0.926 \\
RMLs & 8.124 & 0.002 & 0.002 & 0.891 & 0.947 & ~~ & 8.075 & 0.009 & 0.009 & 0.888 & 0.937 & ~~ & 8.054 & 0.007 & 0.007 & 0.879 & 0.933 \\
RMLg & 8.124 & 0.002 & 0.002 & 0.887 & 0.946 & ~~ & 8.075 & 0.009 & 0.009 & 0.883 & 0.937 & ~~ & 8.055 & 0.007 & 0.007 & 0.885 & 0.929 \\

& \multicolumn{5}{c}{$\hat{\mu}_3~(n_3 = 197)$} &~~& \multicolumn{5}{c}{$\hat{\mu}_4~(n_4 = 781)$} &~~& \multicolumn{5}{c}{$\hat{\mu}_5~(n_5 = 159)$} \\
  \cline{2-6}\cline{8-12}\cline{14-18}
  \rowcolor{lightgray}
RCAL & 8.052 & 0.020 & 0.016 & 0.844 & 0.902 & ~~ & 8.046 & 0.010 & 0.009 & 0.838 & 0.907 & ~~ & 8.043 & 0.021 & 0.017 & 0.813 & 0.861 \\
RMLs & 8.050 & 0.018 & 0.014 & 0.826 & 0.879 & ~~ & 8.044 & 0.009 & 0.009 & 0.869 & 0.925 & ~~ & 8.040 & 0.019 & 0.014 & 0.775 & 0.854 \\
RMLg & 8.050 & 0.017 & 0.014 & 0.837 & 0.891 & ~~ & 8.044 & 0.009 & 0.009 & 0.870 & 0.921 & ~~ & 8.041 & 0.019 & 0.015 & 0.794 & 0.871 \\

\hline
\end{tabular}}
\end{center}
\setlength{\baselineskip}{0.5\baselineskip}
\vspace{-.15in}\noindent{\tiny
\textbf{Note}: Mean, Var, EVar, Cov90, and Cov95 are calculated over the 1000 repeated subsamples, with
the mean treated as the true value. RCAL denotes $\hat{\mu}_t(\hat{m}^\#_{\text{RWL}}, \hat{\pi}_{\text{RCAL}})$. RMLs denotes $\hat{\mu}_t(\hat{m}_{\text{RMLs}}, \hat{\pi}_{\text{RML}})$. RMLg denotes $\hat{\mu}_t(\hat{m}_{\text{RMLg}}, \hat{\pi}_{\text{RML}})$.}
\end{table}

\begin{table}[H]
\caption{\footnotesize Summary of  $\hat{\mu}_t - \hat{\mu}_k$ on sub samples for $t, k = 0, 1, \ldots, 5$ with tuning parameters selected as $\lambda.min$.} \label{tb:ate_sub_min}\vspace{-4ex}
\begin{center}
\resizebox{1.0\textwidth}{!}{\begin{tabular}{lccccccccccccccccc}
\hline
 & Mean & $\sqrt{\text{Var}}$ & $\sqrt{\text{EVar}}$ & Cov90 & Cov95 &~~& Mean & $\sqrt{\text{Var}}$ & $\sqrt{\text{EVar}}$ & Cov90 & Cov95 &~~& Mean & $\sqrt{\text{Var}}$ & $\sqrt{\text{EVar}}$ & Cov90 & Cov95 \\
\hline
& \multicolumn{5}{c}{$\hat{\mu}_1 - \hat{\mu}_0$} &~~& \multicolumn{5}{c}{$\hat{\mu}_2 - \hat{\mu}_0$} &~~& \multicolumn{5}{c}{$\hat{\mu}_3 - \hat{\mu}_0$}\\
\cline{2-6}\cline{8-12}\cline{14-18}
\rowcolor{lightgray}
RCAL & -0.050 & 0.010 & 0.009 & 0.879 & 0.934 & ~~ & -0.069 & 0.008 & 0.007 & 0.867 & 0.923 & ~~ & -0.073 & 0.020 & 0.017 & 0.848 & 0.904 \\
RMLs & -0.050 & 0.009 & 0.009 & 0.887 & 0.937 & ~~ & -0.070 & 0.007 & 0.007 & 0.890 & 0.944 & ~~ & -0.075 & 0.018 & 0.014 & 0.823 & 0.894 \\
RMLg & -0.049 & 0.009 & 0.009 & 0.889 & 0.933 & ~~ & -0.069 & 0.007 & 0.007 & 0.884 & 0.942 & ~~ & -0.075 & 0.018 & 0.014 & 0.831 & 0.904 \\

& \multicolumn{5}{c}{$\hat{\mu}_4 - \hat{\mu}_0$} &~~& \multicolumn{5}{c}{$\hat{\mu}_5 - \hat{\mu}_0$} &~~& \multicolumn{5}{c}{$\hat{\mu}_2 - \hat{\mu}_1$} \\
\cline{2-6}\cline{8-12}\cline{14-18}
\rowcolor{lightgray}
RCAL & -0.078 & 0.010 & 0.009 & 0.847 & 0.906 & ~~ & -0.081 & 0.021 & 0.017 & 0.813 & 0.864 & ~~ & -0.019 & 0.012 & 0.012 & 0.887 & 0.941 \\
RMLs & -0.081 & 0.009 & 0.009 & 0.881 & 0.925 & ~~ & -0.084 & 0.019 & 0.015 & 0.777 & 0.861 & ~~ & -0.021 & 0.012 & 0.011 & 0.892 & 0.943 \\
RMLg & -0.080 & 0.009 & 0.009 & 0.880 & 0.928 & ~~ & -0.083 & 0.019 & 0.015 & 0.793 & 0.875 & ~~ & -0.020 & 0.012 & 0.011 & 0.891 & 0.944 \\

& \multicolumn{5}{c}{$\hat{\mu}_3 - \hat{\mu}_1$} &~~& \multicolumn{5}{c}{$\hat{\mu}_4 - \hat{\mu}_1$} &~~& \multicolumn{5}{c}{$\hat{\mu}_5 - \hat{\mu}_1$} \\
\cline{2-6}\cline{8-12}\cline{14-18}
\rowcolor{lightgray}
RCAL & -0.023 & 0.022 & 0.019 & 0.864 & 0.920 & ~~ & -0.028 & 0.014 & 0.013 & 0.882 & 0.939 & ~~ & -0.032 & 0.023 & 0.019 & 0.845 & 0.898 \\
RMLs & -0.025 & 0.020 & 0.017 & 0.856 & 0.917 & ~~ & -0.031 & 0.013 & 0.013 & 0.893 & 0.944 & ~~ & -0.034 & 0.021 & 0.017 & 0.821 & 0.886 \\
RMLg & -0.025 & 0.019 & 0.017 & 0.861 & 0.926 & ~~ & -0.031 & 0.013 & 0.012 & 0.897 & 0.944 & ~~ & -0.034 & 0.020 & 0.017 & 0.832 & 0.905 \\

& \multicolumn{5}{c}{$\hat{\mu}_3 - \hat{\mu}_2$} &~~& \multicolumn{5}{c}{$\hat{\mu}_4 - \hat{\mu}_2$} &~~& \multicolumn{5}{c}{$\hat{\mu}_5 - \hat{\mu}_2$} \\
\cline{2-6}\cline{8-12}\cline{14-18}
\rowcolor{lightgray}
RCAL & -0.004 & 0.021 & 0.018 & 0.851 & 0.910 & ~~ & -0.009 & 0.012 & 0.011 & 0.872 & 0.926 & ~~ & -0.013 & 0.022 & 0.018 & 0.813 & 0.889 \\
RMLs & -0.004 & 0.019 & 0.016 & 0.841 & 0.898 & ~~ & -0.011 & 0.011 & 0.011 & 0.892 & 0.945 & ~~ & -0.014 & 0.020 & 0.016 & 0.808 & 0.860 \\
RMLg & -0.005 & 0.019 & 0.016 & 0.848 & 0.909 & ~~ & -0.011 & 0.011 & 0.011 & 0.891 & 0.951 & ~~ & -0.014 & 0.020 & 0.016 & 0.814 & 0.881 \\

& \multicolumn{5}{c}{$\hat{\mu}_4 - \hat{\mu}_3$} &~~& \multicolumn{5}{c}{$\hat{\mu}_5 - \hat{\mu}_3$} &~~& \multicolumn{5}{c}{$\hat{\mu}_5 - \hat{\mu}_4$} \\
\cline{2-6}\cline{8-12}\cline{14-18}
\rowcolor{lightgray}
RCAL & -0.005 & 0.022 & 0.019 & 0.852 & 0.913 & ~~ & -0.009 & 0.028 & 0.024 & 0.837 & 0.913 & ~~ & -0.003 & 0.023 & 0.019 & 0.827 & 0.882 \\
RMLs & -0.006 & 0.020 & 0.017 & 0.857 & 0.920 & ~~ & -0.009 & 0.026 & 0.021 & 0.808 & 0.879 & ~~ & -0.003 & 0.022 & 0.017 & 0.801 & 0.876 \\
RMLg & -0.006 & 0.020 & 0.017 & 0.855 & 0.923 & ~~ & -0.009 & 0.026 & 0.021 & 0.820 & 0.889 & ~~ & -0.003 & 0.021 & 0.017 & 0.811 & 0.884 \\

\hline
\end{tabular}}
\end{center}
\setlength{\baselineskip}{0.5\baselineskip}
\vspace{-.15in}\noindent{\tiny
\textbf{Note}: Mean, Var, EVar, Cov90, and Cov95 are calculated over the 1000 repeated subsamples, with the mean treated as the true value. RCAL denotes $\hat{\mu}_t(\hat{m}^\#_{\text{RWL}}, \hat{\pi}_{\text{RCAL}}) - \hat{\mu}_k(\hat{m}^\#_{\text{RWL}}, \hat{\pi}_{\text{RCAL}})$. RMLs denotes $\hat{\mu}_t(\hat{m}_{\text{RMLs}}, \hat{\pi}_{\text{RML}}) - \hat{\mu}_k(\hat{m}_{\text{RMLs}}, \hat{\pi}_{\text{RML}})$. RMLg denotes $\hat{\mu}_t(\hat{m}_{\text{RMLg}}, \hat{\pi}_{\text{RML}}) - \hat{\mu}_k(\hat{m}_{\text{RMLg}}, \hat{\pi}_{\text{RML}})$.}
\end{table}

\begin{table}[H]
\caption{\footnotesize Summary of $\hat{\nu}^{(k)}_t$ on sub samples for $t, k = 0, 1, \ldots, 5$ with tuning parameters selected as $\lambda.min$.} \label{tb:nu_sub_min}\vspace{-4ex}
\begin{center}
\resizebox{\textwidth}{!}{\begin{tabular}{lccccccccccccccccccccccc}
\hline
& \multicolumn{3}{c}{$\hat{\nu}^{(0)}_t$} & $~~$ & \multicolumn{3}{c}{$\hat{\nu}^{(1)}_t$} & $~~$ & \multicolumn{3}{c}{$\hat{\nu}^{(2)}_t$} & $~~$ & \multicolumn{3}{c}{$\hat{\nu}^{(3)}_t$} & $~~$ & \multicolumn{3}{c}{$\hat{\nu}^{(4)}_t$} & $~~$ & \multicolumn{3}{c}{$\hat{\nu}^{(5)}_t$} \\
\cline{2-4}\cline{6-8}\cline{10-12}\cline{14-16}\cline{18-20}\cline{22-24}
& RCAL & RMLs & RMLg & $~~$ & RCAL & RMLs & RMLg & $~~$  & RCAL & RMLs & RMLg & $~~$ & RCAL & RMLs & RMLg & $~~$ & RCAL & RMLs & RMLg & $~~$ & RCAL & RML & RMLg \\
\hline
& \multicolumn{23}{c}{\footnotesize t = 0} \\
$\sqrt{\text{Var}}$ & 0.002& 0.002& 0.002& ~~& 0.004& 0.004& 0.004& ~~& 0.004& 0.004& 0.004& ~~& 0.004& 0.005& 0.006& ~~& 0.004& 0.005& 0.005& ~~& 0.005& 0.008& 0.008\\
$\sqrt{\text{EVar}}$ & 0.002& 0.002& 0.002& ~~& 0.004& 0.004& 0.004& ~~& 0.003& 0.004& 0.004& ~~& 0.003& 0.005& 0.005& ~~& 0.004& 0.004& 0.004& ~~& 0.003& 0.007& 0.007\\
Cov90 & 0.896& 0.896& 0.896& ~~& 0.853& 0.879& 0.875& ~~& 0.836& 0.870& 0.872& ~~& 0.827& 0.883& 0.886& ~~& 0.792& 0.862& 0.870& ~~& 0.789& 0.880& 0.872\\
Cov95 & 0.956& 0.956& 0.956& ~~& 0.914& 0.931& 0.929& ~~& 0.902& 0.932& 0.932& ~~& 0.891& 0.941& 0.942& ~~& 0.883& 0.936& 0.932& ~~& 0.859& 0.933& 0.922\\

& \multicolumn{23}{c}{\footnotesize t = 1} \\
$\sqrt{\text{Var}}$ & 0.010& 0.010& 0.010& ~~& 0.009& 0.009& 0.009& ~~& 0.009& 0.010& 0.009& ~~& 0.009& 0.011& 0.010& ~~& 0.009& 0.011& 0.011& ~~& 0.009& 0.014& 0.013\\
$\sqrt{\text{EVar}}$ & 0.010& 0.010& 0.010& ~~& 0.010& 0.010& 0.010& ~~& 0.009& 0.009& 0.009& ~~& 0.009& 0.010& 0.010& ~~& 0.009& 0.010& 0.010& ~~& 0.010& 0.012& 0.011\\
Cov90 & 0.874& 0.882& 0.885& ~~& 0.909& 0.909& 0.909& ~~& 0.906& 0.889& 0.891& ~~& 0.918& 0.881& 0.876& ~~& 0.908& 0.868& 0.875& ~~& 0.909& 0.849& 0.861\\
Cov95 & 0.937& 0.938& 0.937& ~~& 0.954& 0.954& 0.954& ~~& 0.946& 0.933& 0.933& ~~& 0.954& 0.932& 0.932& ~~& 0.944& 0.930& 0.936& ~~& 0.955& 0.914& 0.915\\

& \multicolumn{23}{c}{\footnotesize t = 2} \\
$\sqrt{\text{Var}}$ & 0.008& 0.008& 0.008& ~~& 0.006& 0.007& 0.007& ~~& 0.006& 0.006& 0.006& ~~& 0.006& 0.008& 0.008& ~~& 0.006& 0.007& 0.007& ~~& 0.006& 0.010& 0.010\\
$\sqrt{\text{EVar}}$ & 0.007& 0.007& 0.007& ~~& 0.006& 0.007& 0.007& ~~& 0.007& 0.007& 0.007& ~~& 0.007& 0.008& 0.008& ~~& 0.006& 0.007& 0.007& ~~& 0.007& 0.009& 0.009\\
Cov90 & 0.847& 0.877& 0.874& ~~& 0.905& 0.897& 0.900& ~~& 0.910& 0.910& 0.910& ~~& 0.906& 0.899& 0.895& ~~& 0.908& 0.896& 0.897& ~~& 0.906& 0.860& 0.858\\
Cov95 & 0.913& 0.927& 0.923& ~~& 0.950& 0.947& 0.950& ~~& 0.950& 0.950& 0.950& ~~& 0.958& 0.950& 0.948& ~~& 0.953& 0.950& 0.944& ~~& 0.949& 0.918& 0.924\\

& \multicolumn{23}{c}{\footnotesize t = 3} \\
$\sqrt{\text{Var}}$ & 0.021& 0.019& 0.019& ~~& 0.015& 0.018& 0.018& ~~& 0.014& 0.017& 0.017& ~~& 0.014& 0.014& 0.014& ~~& 0.014& 0.016& 0.016& ~~& 0.014& 0.017& 0.017\\
$\sqrt{\text{EVar}}$ & 0.017& 0.015& 0.015& ~~& 0.015& 0.015& 0.016& ~~& 0.015& 0.015& 0.015& ~~& 0.015& 0.015& 0.015& ~~& 0.015& 0.015& 0.015& ~~& 0.015& 0.016& 0.015\\
Cov90 & 0.836& 0.821& 0.823& ~~& 0.900& 0.828& 0.854& ~~& 0.894& 0.847& 0.873& ~~& 0.903& 0.903& 0.903& ~~& 0.897& 0.879& 0.893& ~~& 0.902& 0.855& 0.870\\
Cov95 & 0.887& 0.876& 0.888& ~~& 0.943& 0.888& 0.917& ~~& 0.940& 0.915& 0.929& ~~& 0.946& 0.946& 0.946& ~~& 0.940& 0.918& 0.932& ~~& 0.945& 0.917& 0.916\\

& \multicolumn{23}{c}{\footnotesize t = 4} \\
$\sqrt{\text{Var}}$ & 0.011& 0.010& 0.010& ~~& 0.008& 0.010& 0.010& ~~& 0.008& 0.009& 0.009& ~~& 0.008& 0.009& 0.009& ~~& 0.008& 0.008& 0.008& ~~& 0.008& 0.010& 0.010\\
$\sqrt{\text{EVar}}$ & 0.009& 0.009& 0.009& ~~& 0.008& 0.009& 0.009& ~~& 0.008& 0.008& 0.008& ~~& 0.008& 0.008& 0.008& ~~& 0.008& 0.008& 0.008& ~~& 0.008& 0.009& 0.009\\
Cov90 & 0.829& 0.866& 0.867& ~~& 0.870& 0.847& 0.856& ~~& 0.885& 0.868& 0.868& ~~& 0.891& 0.866& 0.864& ~~& 0.888& 0.888& 0.888& ~~& 0.892& 0.878& 0.888\\
Cov95 & 0.901& 0.918& 0.923& ~~& 0.933& 0.908& 0.917& ~~& 0.929& 0.919& 0.921& ~~& 0.939& 0.925& 0.926& ~~& 0.937& 0.937& 0.937& ~~& 0.941& 0.936& 0.937\\

& \multicolumn{23}{c}{\footnotesize t = 5} \\
$\sqrt{\text{Var}}$ & 0.023& 0.020& 0.019& ~~& 0.017& 0.024& 0.022& ~~& 0.017& 0.020& 0.019& ~~& 0.017& 0.018& 0.017& ~~& 0.017& 0.018& 0.017& ~~& 0.017& 0.017& 0.017\\
$\sqrt{\text{EVar}}$ & 0.017& 0.015& 0.015& ~~& 0.016& 0.017& 0.017& ~~& 0.016& 0.016& 0.016& ~~& 0.016& 0.015& 0.015& ~~& 0.016& 0.015& 0.015& ~~& 0.017& 0.017& 0.017\\
Cov90 & 0.801& 0.784& 0.792& ~~& 0.870& 0.751& 0.785& ~~& 0.868& 0.795& 0.822& ~~& 0.882& 0.836& 0.834& ~~& 0.877& 0.844& 0.852& ~~& 0.886& 0.886& 0.886\\
Cov95 & 0.856& 0.852& 0.863& ~~& 0.936& 0.820& 0.848& ~~& 0.931& 0.870& 0.884& ~~& 0.942& 0.896& 0.896& ~~& 0.940& 0.904& 0.914& ~~& 0.943& 0.943& 0.943\\

\hline
\end{tabular}}
\end{center}
\setlength{\baselineskip}{0.5\baselineskip}
\vspace{-.15in}\noindent{\tiny
\textbf{Note}: Var, EVar, Cov90, and Cov95 are calculated over the 1000 repeated subsamples, with the mean treated as the true value. RCAL denotes $\hat{\nu}^{(k)}_{t, \text{RCAL}}$. RMLs denotes $\hat{\nu}^{(k)}_{t, \text{RMLs}}$. RMLg denotes $\hat{\nu}^{(k)}_{t, \text{RMLg}}$.}
\end{table}

\begin{table}[H]
\caption{\footnotesize Summary for $\hat{\mu}_t$ on full sample for $t = 0, 1, \ldots, 5$ with tuning parameters selected as $\lambda.1se$.} \label{tb:mu_full_1se}\vspace{-4ex}
\begin{center}
\resizebox{1.0\textwidth}{!}{\begin{tabular}{lccccccccccc}
\hline
 & Est & SE & 95CI &~~& Est & SE & 95CI &~~ & Est & SE & 95CI  \\
\hline
  & \multicolumn{3}{c}{$\hat{\mu}_0~(n_0 = 13587)$} &~~& \multicolumn{3}{c}{$\hat{\mu}_1~(n_1 = 552)$} &~~& \multicolumn{3}{c}{$\hat{\mu}_2~(n_2 = 1190)$} \\
  \cline{2-4}\cline{6-8}\cline{10-12}
Uuadj& 8.1272 & 0.0003 & (8.1266, 8.1279) &~~& 8.0576 & 0.0019 & (8.0538, 8.0614) &~~& 8.0398 & 0.0013 & (8.0372, 8.0424)  \\
\rowcolor{lightgray}
CAL & 8.1247 & 0.0004 & (8.1240, 8.1254) &~~& 8.0763 & 0.0020 & (8.0723, 8.0803) &~~& 8.0570 & 0.0017 & (8.0537, 8.0602) \\
ML & 8.1252 & 0.0006 & (8.1240, 8.1263) &~~& 8.0768 & 0.0019 & (8.0730, 8.0806) &~~& 8.0573 & 0.0015 & (8.0542, 8.0603) \\
\rowcolor{lightgray}
RCAL & 8.1244 & 0.0004 & (8.1237, 8.1251) &~~& 8.0744 & 0.0022 & (8.0702, 8.0786) &~~& 8.0545 & 0.0016 & (8.0513, 8.0577)  \\
RMLs & 8.1244 & 0.0004 & (8.1236, 8.1251) &~~& 8.0751 & 0.0020 & (8.0712, 8.0790) &~~& 8.0551 & 0.0017 & (8.0518, 8.0585) \\
RMLg & 8.1244 & 0.0004 & (8.1237, 8.1251) &~~& 8.0750 & 0.0020 & (8.0710, 8.0790) &~~& 8.0541 & 0.0016 & (8.0509, 8.0572) \\

  & \multicolumn{3}{c}{$\hat{\mu}_3~(n_3 = 197)$} &~~& \multicolumn{3}{c}{$\hat{\mu}_4~(n_4 = 781)$} &~~& \multicolumn{3}{c}{$\hat{\mu}_5~(n_5 = 159)$} \\
  \cline{2-4}\cline{6-8}\cline{10-12}
Uuadj & 8.0474 & 0.0030 & (8.0416, 8.0533) &~~& 8.0365 & 0.0016 & (8.0334, 8.0396) &~~& 8.0318 & 0.0034 & (8.0251, 8.0384)  \\
\rowcolor{lightgray}
CAL & 8.0517 & 0.0049 & (8.0421, 8.0612) &~~& 8.0460 & 0.0023 & (8.0415, 8.0505) &~~& 8.0442 & 0.0057 & (8.0330, 8.0554)  \\
ML & 8.0536 & 0.0045 & (8.0449, 8.0623) &~~& 8.0460 & 0.0021 & (8.0418, 8.0502) &~~& 8.0413 & 0.0057 & (8.0301, 8.0525) \\
\rowcolor{lightgray}
RCAL & 8.0494 & 0.0042 & (8.0411, 8.0576) &~~& 8.0439 & 0.0021 & (8.0399, 8.0480) &~~& 8.0417 & 0.0047 & (8.0324, 8.0509) \\
RMLs & 8.0495 & 0.0036 & (8.0424, 8.0565) &~~& 8.0446 & 0.0020 & (8.0408, 8.0484) &~~& 8.0416 & 0.0044 & (8.0330, 8.0501) \\
RMLg & 8.0484 & 0.0037 & (8.0412, 8.0556) &~~& 8.0448 & 0.0020 & (8.0408, 8.0488) &~~& 8.0410 & 0.0044 & (8.0323, 8.0497) \\

\hline
\end{tabular}}
\end{center}
\setlength{\baselineskip}{0.5\baselineskip}
\vspace{-.15in}\noindent{\tiny
\textbf{Note}: Est, SE, or 95CI denotes point estimate, standard error, or 95\% confidence interval respectively.
Unadj denotes the unadjusted estimate, $\hat{\mu}_t = \tilde{\E}(Y^{(t)})$.
RCAL, RMLs, or RMLg denotes $\hat{\mu}_t(\hat{m}^\#_{\text{RWL}}, \hat{\pi}_{\text{RCAL}})$, $\hat{\mu}_t(\hat{m}_{\text{RMLs}}, \hat{\pi}_{\text{RML}})$,
or $\hat{\mu}_t(\hat{m}_{\text{RMLg}}, \hat{\pi}_{\text{RML}})$ respectively. CAL or ML denotes non-regularized estimation with main effects only in PS and OR models.}
\end{table}

\begin{table}[H]
\caption{\footnotesize Summary for  $\hat{\mu}_t - \hat{\mu}_k$ on full sample for $t, k = 0, 1, \ldots, 5$ with tuning parameters selected as $\lambda.1se$.} \label{tb:ate_full_1se}\vspace{-4ex}
\begin{center}
\resizebox{1.0\textwidth}{!}{\begin{tabular}{lccccccccccc}
\hline
 & Est & SE & 95CI &~~& Est & SE & 95CI &~~& Est & SE & 95CI  \\
\hline
& \multicolumn{3}{c}{$\hat{\mu}_1 - \hat{\mu}_0$} &~~& \multicolumn{3}{c}{$\hat{\mu}_2 - \hat{\mu}_0$} &~~& \multicolumn{3}{c}{$\hat{\mu}_3 - \hat{\mu}_0$} \\
\cline{2-4}\cline{6-8}\cline{10-12}
\rowcolor{lightgray}
CAL & -0.0484 & 0.0021 & (-0.0524, -0.0443) &~~& -0.0677 & 0.0017 & (-0.0710, -0.0644) &~~& -0.0730 & 0.0049 & (-0.0826, -0.0634)  \\
ML & -0.0484 & 0.0020 & (-0.0523, -0.0444) &~~& -0.0679 & 0.0016 & (-0.0711, -0.0647) &~~& -0.0716 & 0.0045 & (-0.0804, -0.0628) \\
\rowcolor{lightgray}
RCAL & -0.0500 & 0.0022 & (-0.0542, -0.0457) &~~& -0.0699 & 0.0017 & (-0.0731, -0.0666) &~~& -0.0750 & 0.0042 & (-0.0833, -0.0667) \\
RMLs & -0.0493 & 0.0020 & (-0.0532, -0.0453) &~~& -0.0692 & 0.0017 & (-0.0726, -0.0658) &~~& -0.0749 & 0.0036 & (-0.0820, -0.0678)  \\
RMLg & -0.0494 & 0.0021 & (-0.0535, -0.0454) &~~& -0.0703 & 0.0016 & (-0.0736, -0.0671) &~~& -0.0760 & 0.0037 & (-0.0832, -0.0688) \\

& \multicolumn{3}{c}{$\hat{\mu}_4 - \hat{\mu}_0$} &~~& \multicolumn{3}{c}{$\hat{\mu}_5 - \hat{\mu}_0$} &~~& \multicolumn{3}{c}{$\hat{\mu}_2 - \hat{\mu}_1$}\\
\cline{2-4}\cline{6-8}\cline{10-12}
\rowcolor{lightgray}
CAL & -0.0787 & 0.0023 & (-0.0833, -0.0741) &~~& -0.0805 & 0.0057 & (-0.0917, -0.0693) &~~& -0.0194 & 0.0026 & (-0.0245, -0.0142) \\
ML & -0.0792 & 0.0022 & (-0.0835, -0.0749) &~~& -0.0838 & 0.0057 & (-0.0951, -0.0726) &~~& -0.0195 & 0.0025 & (-0.0244, -0.0147) \\
\rowcolor{lightgray}
RCAL & -0.0805 & 0.0021 & (-0.0846, -0.0763) &~~& -0.0827 & 0.0047 & (-0.0920, -0.0735) &~~& -0.0199 & 0.0027 & (-0.0252, -0.0146) \\
RMLs & -0.0798 & 0.0020 & (-0.0836, -0.0759) &~~& -0.0828 & 0.0044 & (-0.0913, -0.0742) &~~& -0.0199 & 0.0026 & (-0.0251, -0.0148)  \\
RMLg & -0.0796 & 0.0021 & (-0.0836, -0.0756) &~~& -0.0834 & 0.0045 & (-0.0922, -0.0747) &~~& -0.0209 & 0.0026 & (-0.0260, -0.0158) \\

& \multicolumn{3}{c}{$\hat{\mu}_3 - \hat{\mu}_1$} &~~& \multicolumn{3}{c}{$\hat{\mu}_4 - \hat{\mu}_1$} &~~& \multicolumn{3}{c}{$\hat{\mu}_5 - \hat{\mu}_1$}\\
\cline{2-4}\cline{6-8}\cline{10-12}
\rowcolor{lightgray}
CAL & -0.0247 & 0.0053 & (-0.0350, -0.0143) &~~& -0.0303 & 0.0031 & (-0.0364, -0.0243) &~~& -0.0322 & 0.0061 & (-0.0440, -0.0203)  \\
ML & -0.0232 & 0.0048 & (-0.0327, -0.0137) &~~& -0.0308 & 0.0029 & (-0.0365, -0.0252) &~~& -0.0355 & 0.0060 & (-0.0473, -0.0237) \\
\rowcolor{lightgray}
RCAL & -0.0250 & 0.0047 & (-0.0343, -0.0158) &~~& -0.0305 & 0.0030 & (-0.0364, -0.0246) &~~& -0.0328 & 0.0052 & (-0.0429, -0.0226) \\
RMLs & -0.0256 & 0.0041 & (-0.0337, -0.0176) &~~& -0.0305 & 0.0028 & (-0.0359, -0.0250) &~~& -0.0335 & 0.0048 & (-0.0429, -0.0242)  \\
RMLg & -0.0266 & 0.0042 & (-0.0348, -0.0184) &~~& -0.0302 & 0.0029 & (-0.0358, -0.0245) &~~& -0.0340 & 0.0049 & (-0.0436, -0.0244) \\

& \multicolumn{3}{c}{$\hat{\mu}_3 - \hat{\mu}_2$} &~~& \multicolumn{3}{c}{$\hat{\mu}_4 - \hat{\mu}_2$} &~~& \multicolumn{3}{c}{$\hat{\mu}_5 - \hat{\mu}_2$}\\
\cline{2-4}\cline{6-8}\cline{10-12}
\rowcolor{lightgray}
CAL & -0.0053 & 0.0051 & (-0.0154, 0.0048) &~~& -0.0110 & 0.0028 & (-0.0165, -0.0054) &~~& -0.0128 & 0.0059 & (-0.0245, -0.0011) \\
ML & -0.0037 & 0.0047 & (-0.0129, 0.0055) &~~& -0.0113 & 0.0026 & (-0.0164, -0.0061) &~~& -0.0160 & 0.0059 & (-0.0275, -0.0044) \\
\rowcolor{lightgray}
RCAL & -0.0051 & 0.0045 & (-0.0140, 0.0037) &~~& -0.0106 & 0.0026 & (-0.0157, -0.0054) &~~& -0.0128 & 0.0050 & (-0.0226, -0.0031)  \\
RMLs & -0.0057 & 0.0040 & (-0.0135, 0.0021) &~~& -0.0105 & 0.0026 & (-0.0156, -0.0055)  &~~& -0.0136 & 0.0047 & (-0.0227, -0.0044) \\
RMLg & -0.0057 & 0.0040 & (-0.0135, 0.0022) &~~& -0.0093 & 0.0026 & (-0.0143, -0.0042) &~~& -0.0131 & 0.0047 & (-0.0223, -0.0039) \\

& \multicolumn{3}{c}{$\hat{\mu}_4 - \hat{\mu}_3$} &~~& \multicolumn{3}{c}{$\hat{\mu}_5 - \hat{\mu}_3$} &~~& \multicolumn{3}{c}{$\hat{\mu}_5 - \hat{\mu}_4$}\\
\cline{2-4}\cline{6-8}\cline{10-12}
\rowcolor{lightgray}
CAL & -0.0057 & 0.0054 & (-0.0162, 0.0049) &~~& -0.0075 & 0.0075 & (-0.0222, 0.0072) &~~& -0.0018 & 0.0062 & (-0.0139, 0.0102)  \\
ML & -0.0076 & 0.0049 & (-0.0173, 0.0021) &~~& -0.0123 & 0.0072 & (-0.0265, 0.0019) &~~& -0.0047 & 0.0061 & (-0.0166, 0.0073) \\
\rowcolor{lightgray}
RCAL & -0.0055 & 0.0047 & (-0.0147, 0.0037) &~~& -0.0077 & 0.0063 & (-0.0201, 0.0047)  &~~& -0.0023 & 0.0051 & (-0.0123, 0.0078)  \\
RMLs & -0.0049 & 0.0041 & (-0.0129, 0.0031) &~~& -0.0079 & 0.0056 & (-0.0189, 0.0032)  &~~& -0.0030 & 0.0048 & (-0.0124, 0.0063) \\
RMLg & -0.0036 & 0.0042 & (-0.0118, 0.0046) &~~& -0.0074 & 0.0057 & (-0.0187, 0.0038) &~~& -0.0038 & 0.0049 & (-0.0134, 0.0057)  \\

\hline
\end{tabular}}
\end{center}
\setlength{\baselineskip}{0.5\baselineskip}
\vspace{-.15in}\noindent{\tiny
\textbf{Note}: Est, SE, or 95CI denotes point estimate, standard error, or 95\% confidence interval respectively.
RCAL, RMLs, or RMLg denotes
$\hat{\mu}_t(\hat{m}^\#_{\text{RWL}}, \hat{\pi}_{\text{RCAL}}) - \hat{\mu}_k(\hat{m}^\#_{\text{RWL}}, \hat{\pi}_{\text{RCAL}})$,
$\hat{\mu}_t(\hat{m}_{\text{RMLs}}, \hat{\pi}_{\text{RML}}) - \hat{\mu}_k(\hat{m}_{\text{RMLs}}, \hat{\pi}_{\text{RML}})$,
or $\hat{\mu}_t(\hat{m}_{\text{RMLg}}, \hat{\pi}_{\text{RML}}) - \hat{\mu}_k(\hat{m}_{\text{RMLg}}, \hat{\pi}_{\text{RML}})$
respectively. CAL or ML denotes non-regularized estimation with main effects only in PS and OR models.}
\end{table}

\begin{table}[H]
\caption{\footnotesize Summary for $\hat{\nu}^{(k)}_t$ on full sample for $t, k = 0, 1, \ldots, 5$ with tuning parameters selected as $\lambda.1se$.} \label{tb:nu_full_1se}\vspace{-4ex}
\begin{center}
\resizebox{1.0\textwidth}{0.5\textwidth}{\begin{tabular}{lccccccccccccccccccccccc}
\hline
& \multicolumn{3}{c}{$\hat{\nu}^{(0)}_t$} & $~~$ & \multicolumn{3}{c}{$\hat{\nu}^{(1)}_t$} & $~~$ & \multicolumn{3}{c}{$\hat{\nu}^{(2)}_t$} & $~~$ & \multicolumn{3}{c}{$\hat{\nu}^{(3)}_t$} & $~~$ & \multicolumn{3}{c}{$\hat{\nu}^{(4)}_t$} & $~~$ & \multicolumn{3}{c}{$\hat{\nu}^{(5)}_t$} \\
\cline{2-4}\cline{6-8}\cline{10-12}\cline{14-16}\cline{18-20}\cline{22-24}
& Est & SE & 95CI &~~& Est & SE & 95CI &~~& Est & SE & 95CI &~~& Est & SE & 95CI &~~& Est & SE & 95CI &~~& Est & SE & 95CI  \\
\hline
& \multicolumn{23}{c}{\footnotesize t = 0} \\
\rowcolor{lightgray}
CAL & 8.1272 & 0.0003 & (8.1266, 8.1279)  &~~& 8.1025 & 0.0009 & (8.1007, 8.1043) &~~& 8.1098 & 0.0008 & (8.1082, 8.1113) &~~& 8.1240 & 0.0011 & (8.1218, 8.1262) &~~& 8.1193 & 0.0009 & (8.1174, 8.1211) &~~& 8.1253 & 0.0015 & (8.1223, 8.1283)  \\
ML & 8.1272 & 0.0003 & (8.1266, 8.1279) &~~& 8.1025 & 0.0009 & (8.1007, 8.1042) &~~& 8.1102 & 0.0010 & (8.1083, 8.1122) &~~& 8.1262 & 0.0024 & (8.1214, 8.1309) &~~& 8.1253 & 0.0061 & (8.1134, 8.1372) &~~& 8.1360 & 0.0095 & (8.1175, 8.1546)  \\
\rowcolor{lightgray}
RCAL & 8.1272 & 0.0003 & (8.1266, 8.1279) &~~& 8.1015 & 0.0007 & (8.1002, 8.1029) &~~& 8.1075 & 0.0007 & (8.1061, 8.1089) &~~& 8.1231 & 0.0006 & (8.1219, 8.1243) &~~& 8.1170 & 0.0008 & (8.1153, 8.1186) &~~& 8.1238 & 0.0010 & (8.1218, 8.1258)  \\
RMLs & 8.1272 & 0.0003 & (8.1266, 8.1279) &~~& 8.1013 & 0.0008 & (8.0997, 8.1029) &~~& 8.1076 & 0.0008 & (8.1061, 8.1091) &~~& 8.1231 & 0.0010 & (8.1212, 8.1250) &~~& 8.1168 & 0.0009 & (8.1150, 8.1187) &~~& 8.1231 & 0.0015 & (8.1201, 8.1261) \\
RMLg & 8.1272 & 0.0003 & (8.1266, 8.1279) &~~& 8.1016 & 0.0007 & (8.1002, 8.1030) &~~& 8.1078 & 0.0007 & (8.1064, 8.1092) &~~& 8.1233 & 0.0007 & (8.1219, 8.1247) &~~& 8.1172 & 0.0009 & (8.1154, 8.1190) &~~& 8.1240 & 0.0013 & (8.1214, 8.1266) \\

& \multicolumn{23}{c}{\footnotesize t = 1} \\
\rowcolor{lightgray}
CAL & 8.0787 & 0.0022 & (8.0743, 8.0830) &~~& 8.0576 & 0.0019 & (8.0538, 8.0614) &~~& 8.0633 & 0.0019 & (8.0596, 8.0670) &~~& 8.0767 & 0.0021 & (8.0725, 8.0808) &~~& 8.0701 & 0.0022 & (8.0658, 8.0744) &~~& 8.0715 & 0.0031 & (8.0655, 8.0775) \\
ML & 8.0792 & 0.0021 & (8.0752, 8.0833) &~~& 8.0576 & 0.0019 & (8.0538, 8.0614) &~~& 8.0632 & 0.0018 & (8.0596, 8.0668) &~~& 8.0764 & 0.0021 & (8.0723, 8.0806) &~~& 8.0699 & 0.0020 & (8.0660, 8.0739) &~~& 8.0711 & 0.0026 & (8.0661, 8.0761)  \\
\rowcolor{lightgray}
RCAL & 8.0772 & 0.0023 & (8.0727, 8.0817) &~~& 8.0576 & 0.0019 & (8.0538, 8.0614) &~~& 8.0609 & 0.0019 & (8.0572, 8.0646) &~~& 8.0612 & 0.0019 & (8.0575, 8.0648) &~~& 8.0645 & 0.0019 & (8.0609, 8.0682) &~~& 8.0609 & 0.0019 & (8.0572, 8.0646) \\
RMLs & 8.0773 & 0.0021 & (8.0731, 8.0814) &~~& 8.0576 & 0.0019 & (8.0538, 8.0614) &~~& 8.0622 & 0.0019 & (8.0584, 8.0660) &~~& 8.0766 & 0.0020 & (8.0727, 8.0805) &~~& 8.0693 & 0.0021 & (8.0652, 8.0734) &~~& 8.0716 & 0.0025 & (8.0666, 8.0765) \\
RMLg & 8.0771 & 0.0022 & (8.0729, 8.0814) &~~& 8.0576 & 0.0019 & (8.0538, 8.0614) &~~& 8.0619 & 0.0020 & (8.0580, 8.0658) &~~& 8.0766 & 0.0019 & (8.0728, 8.0804) &~~& 8.0695 & 0.0021 & (8.0653, 8.0737) &~~& 8.0730 & 0.0024 & (8.0683, 8.0778)  \\

& \multicolumn{23}{c}{\footnotesize t = 2} \\
\rowcolor{lightgray}
CAL & 8.0603 & 0.0018 & (8.0567, 8.0639) &~~& 8.0358 & 0.0015 & (8.0329, 8.0387) &~~& 8.0398 & 0.0013 & (8.0372, 8.0424) &~~& 8.0511 & 0.0016 & (8.0480, 8.0542) &~~& 8.0441 & 0.0014 & (8.0412, 8.0469) &~~& 8.0434 & 0.0021 & (8.0392, 8.0476) \\
ML & 8.0607 & 0.0017 & (8.0574, 8.0640) &~~& 8.0358 & 0.0015 & (8.0329, 8.0387) &~~& 8.0398 & 0.0013 & (8.0372, 8.0424) &~~& 8.0509 & 0.0016 & (8.0478, 8.0541) &~~& 8.0440 & 0.0014 & (8.0413, 8.0468) &~~& 8.0435 & 0.0019 & (8.0398, 8.0472) \\
\rowcolor{lightgray}
RCAL & 8.0576 & 0.0018 & (8.0541, 8.0610) &~~& 8.0393 & 0.0013 & (8.0367, 8.0419) &~~& 8.0398 & 0.0013 & (8.0372, 8.0424) &~~& 8.0406 & 0.0013 & (8.0380, 8.0432) &~~& 8.0408 & 0.0013 & (8.0382, 8.0433) &~~& 8.0400 & 0.0013 & (8.0374, 8.0426) \\
RMLs & 8.0581 & 0.0019 & (8.0543, 8.0618)  &~~& 8.0363 & 0.0014 & (8.0335, 8.0390) &~~& 8.0398 & 0.0013 & (8.0372, 8.0424) &~~& 8.0516 & 0.0015 & (8.0487, 8.0546) &~~& 8.0444 & 0.0014 & (8.0416, 8.0471) &~~& 8.0441 & 0.0019 & (8.0404, 8.0479) \\
RMLg & 8.0567 & 0.0018 & (8.0532, 8.0601) &~~& 8.0367 & 0.0014 & (8.0340, 8.0394) &~~& 8.0398 & 0.0013 & (8.0372, 8.0424) &~~& 8.0518 & 0.0013 & (8.0493, 8.0544) &~~& 8.0449 & 0.0014 & (8.0422, 8.0476) &~~& 8.0459 & 0.0017 & (8.0426, 8.0492)  \\

& \multicolumn{23}{c}{\footnotesize t = 3} \\
\rowcolor{lightgray}
CAL & 8.0544 & 0.0055 & (8.0435, 8.0652) &~~& 8.0352 & 0.0040 & (8.0273, 8.0431) &~~& 8.0375 & 0.0034 & (8.0308, 8.0441) &~~& 8.0474 & 0.0030 & (8.0416, 8.0533) &~~& 8.0414 & 0.0032 & (8.0352, 8.0477)  &~~& 8.0421 & 0.0038 & (8.0347, 8.0496) \\
ML & 8.0567 & 0.0049 & (8.0471, 8.0663) &~~& 8.0345 & 0.0042 & (8.0263, 8.0428) &~~& 8.0373 & 0.0034 & (8.0307, 8.0439) &~~& 8.0474 & 0.0030 & (8.0416, 8.0533) &~~& 8.0415 & 0.0032 & (8.0353, 8.0478) &~~& 8.0423 & 0.0035 & (8.0354, 8.0493) \\
\rowcolor{lightgray}
RCAL & 8.0499 & 0.0046 & (8.0408, 8.0590) &~~& 8.0467 & 0.0030 & (8.0408, 8.0526) &~~& 8.0463 & 0.0030 & (8.0404, 8.0523) &~~& 8.0474 & 0.0030 & (8.0416, 8.0533) &~~& 8.0470 & 0.0030 & (8.0412, 8.0529) &~~& 8.0474 & 0.0030 & (8.0415, 8.0532) \\
RMLs & 8.0520 & 0.0039 & (8.0444, 8.0595) &~~& 8.0334 & 0.0038 & (8.0259, 8.0408) &~~& 8.0349 & 0.0037 & (8.0276, 8.0422) &~~& 8.0474 & 0.0030 & (8.0416, 8.0533) &~~& 8.0408 & 0.0034 & (8.0342, 8.0474) &~~& 8.0429 & 0.0034 & (8.0362, 8.0496) \\
RMLg & 8.0507 & 0.0039 & (8.0431, 8.0584) &~~& 8.0337 & 0.0040 & (8.0258, 8.0416) &~~& 8.0346 & 0.0040 & (8.0268, 8.0425) &~~& 8.0474 & 0.0030 & (8.0416, 8.0533) &~~& 8.0405 & 0.0035 & (8.0336, 8.0474) &~~& 8.0434 & 0.0035 & (8.0365, 8.0503) \\

& \multicolumn{23}{c}{\footnotesize t = 4} \\
\rowcolor{lightgray}
CAL & 8.0488 & 0.0026 & (8.0438, 8.0538) &~~& 8.0261 & 0.0023 & (8.0217, 8.0306) &~~& 8.0312 & 0.0018 & (8.0278, 8.0347) &~~& 8.0426 & 0.0017 & (8.0392, 8.0459) &~~& 8.0365 & 0.0016 & (8.0334, 8.0396) &~~& 8.0374 & 0.0020 & (8.0335, 8.0413) \\
ML & 8.0488 & 0.0024 & (8.0441, 8.0534) &~~& 8.0263 & 0.0021 & (8.0222, 8.0304) &~~& 8.0312 & 0.0017 & (8.0279, 8.0345) &~~& 8.0425 & 0.0017 & (8.0391, 8.0459) &~~& 8.0365 & 0.0016 & (8.0334, 8.0396) &~~& 8.0373 & 0.0019 & (8.0336, 8.0411) \\
\rowcolor{lightgray}
RCAL & 8.0456 & 0.0023 & (8.0411, 8.0501) &~~& 8.0357 & 0.0016 & (8.0326, 8.0388) &~~& 8.0356 & 0.0016 & (8.0325, 8.0387) &~~& 8.0367 & 0.0016 & (8.0336, 8.0397) &~~& 8.0365 & 0.0016 & (8.0334, 8.0396) &~~& 8.0365 & 0.0016 & (8.0334, 8.0396) \\
RMLs & 8.0471 & 0.0021 & (8.0429, 8.0513) &~~& 8.0266 & 0.0019 & (8.0228, 8.0304) &~~& 8.0312 & 0.0017 & (8.0279, 8.0344) &~~& 8.0428 & 0.0016 & (8.0396, 8.0459) &~~& 8.0365 & 0.0016 & (8.0334, 8.0396) &~~& 8.0384 & 0.0018 & (8.0349, 8.0418) \\
RMLg & 8.0472 & 0.0022 & (8.0429, 8.0516) &~~& 8.0282 & 0.0020 & (8.0243, 8.0321) &~~& 8.0315 & 0.0017 & (8.0281, 8.0348) &~~& 8.0428 & 0.0015 & (8.0399, 8.0457) &~~& 8.0365 & 0.0016 & (8.0334, 8.0396) &~~& 8.0386 & 0.0016 & (8.0354, 8.0417) \\

& \multicolumn{23}{c}{\footnotesize t = 5} \\
\rowcolor{lightgray}
CAL & 8.0482 & 0.0062 & (8.0361, 8.0603) &~~& 8.0160 & 0.0082 & (8.0000, 8.0320) &~~& 8.0231 & 0.0055 & (8.0123, 8.0339) &~~& 8.0377 & 0.0038 & (8.0303, 8.0451) &~~& 8.0303 & 0.0036 & (8.0232, 8.0375) &~~& 8.0318 & 0.0034 & (8.0251, 8.0384) \\
ML & 8.0448 & 0.0061 & (8.0328, 8.0567) &~~& 8.0153 & 0.0081 & (7.9994, 8.0312) &~~& 8.0233 & 0.0050 & (8.0134, 8.0331) &~~& 8.0372 & 0.0037 & (8.0300, 8.0444) &~~& 8.0302 & 0.0035 & (8.0233, 8.0372) &~~& 8.0318 & 0.0034 & (8.0251, 8.0384) \\
\rowcolor{lightgray}
RCAL & 8.0440 & 0.0052 & (8.0337, 8.0542) &~~& 8.0304 & 0.0034 & (8.0237, 8.0372) &~~& 8.0302 & 0.0035 & (8.0233, 8.0370) &~~& 8.0319 & 0.0034 & (8.0253, 8.0385) &~~& 8.0315 & 0.0034 & (8.0249, 8.0382) &~~& 8.0318 & 0.0034 & (8.0251, 8.0384) \\
RMLs & 8.0446 & 0.0046 & (8.0355, 8.0536) &~~& 8.0224 & 0.0056 & (8.0114, 8.0335) &~~& 8.0255 & 0.0044 & (8.0169, 8.0342) &~~& 8.0370 & 0.0033 & (8.0305, 8.0435) &~~& 8.0305 & 0.0035 & (8.0236, 8.0374) &~~& 8.0318 & 0.0034 & (8.0251, 8.0384) \\
RMLg & 8.0440 & 0.0047 & (8.0347, 8.0533) &~~& 8.0218 & 0.0058 & (8.0105, 8.0331) &~~& 8.0246 & 0.0045 & (8.0157, 8.0335) &~~& 8.0366 & 0.0034 & (8.0300, 8.0432) &~~& 8.0296 & 0.0036 & (8.0225, 8.0367) &~~& 8.0318 & 0.0034 & (8.0251, 8.0384) \\

\hline
\end{tabular}}
\end{center}
\setlength{\baselineskip}{0.5\baselineskip}
\vspace{-.15in}\noindent{\tiny
\textbf{Note}: Est, SE, or 95CI denotes point estimate, standard error, or 95\% confidence interval respectively. CAL denotes $\hat{\nu}^{(k)}_{t, \text{CAL}}$. ML denotes $\hat{\nu}^{(k)}_{t, \text{ML}}$. RCAL denotes $\hat{\nu}^{(k)}_{t, \text{RCAL}}$. RMLs denotes $\hat{\nu}^{(k)}_{t, \text{RMLs}}$. RMLg denotes $\hat{\nu}^{(k)}_{t, \text{RMLg}}$.}
\end{table}

\begin{table}[H]
\caption{\footnotesize Summary for $\hat{\mu}_t$ on sub samples for $t = 0, 1, \ldots, 5$ with tuning parameters selected as $\lambda.1se$.} \label{tb:mu_sub_1se}\vspace{-4ex}
\begin{center}
\resizebox{1.0\textwidth}{!}{\begin{tabular}{lccccccccccccccccc}
\hline
 & Mean & $\sqrt{\text{Var}}$ & $\sqrt{\text{EVar}}$ & Cov90 & Cov95 &~~& Mean & $\sqrt{\text{Var}}$ & $\sqrt{\text{EVar}}$ & Cov90 & Cov95 &~~& Mean & $\sqrt{\text{Var}}$ & $\sqrt{\text{EVar}}$ & Cov90 & Cov95 \\
\hline
  & \multicolumn{5}{c}{$\hat{\mu}_0~(n_0 = 13587)$} &~~& \multicolumn{5}{c}{$\hat{\mu}_1~(n_1 = 552)$}  &~~& \multicolumn{5}{c}{$\hat{\mu}_2~(n_2 = 1190)$}\\
  \cline{2-6}\cline{8-12}\cline{14-18}
  \rowcolor{lightgray}
RCAL & 8.125 & 0.002 & 0.002 & 0.889 & 0.954 & ~~ & 8.069 & 0.009 & 0.010 & 0.913 & 0.950 & ~~ & 8.052 & 0.007 & 0.007 & 0.911 & 0.953 \\
RMLs & 8.125 & 0.002 & 0.002 & 0.894 & 0.946 & ~~ & 8.073 & 0.009 & 0.009 & 0.896 & 0.939 & ~~ & 8.052 & 0.007 & 0.007 & 0.906 & 0.944 \\
RMLg & 8.125 & 0.002 & 0.002 & 0.897 & 0.954 & ~~ & 8.073 & 0.009 & 0.009 & 0.895 & 0.939 & ~~ & 8.052 & 0.007 & 0.007 & 0.903 & 0.942 \\

& \multicolumn{5}{c}{$\hat{\mu}_3~(n_3 = 197)$} &~~& \multicolumn{5}{c}{$\hat{\mu}_4~(n_4 = 781)$} &~~& \multicolumn{5}{c}{$\hat{\mu}_5(n_5 = 159)$} \\
 \cline{2-6}\cline{8-12}\cline{14-18}
  \rowcolor{lightgray}
RCAL & 8.050 & 0.018 & 0.017 & 0.900 & 0.947 & ~~ & 8.044 & 0.009 & 0.009 & 0.903 & 0.949 & ~~ & 8.038 & 0.018 & 0.018 & 0.887 & 0.940 \\
RMLs & 8.049 & 0.017 & 0.014 & 0.847 & 0.912 & ~~ & 8.041 & 0.009 & 0.009 & 0.881 & 0.929 & ~~ & 8.039 & 0.018 & 0.014 & 0.810 & 0.886 \\
RMLg & 8.049 & 0.017 & 0.014 & 0.849 & 0.912 & ~~ & 8.041 & 0.009 & 0.009 & 0.882 & 0.928 & ~~ & 8.039 & 0.018 & 0.015 & 0.812 & 0.889 \\

\hline
\end{tabular}}
\end{center}
\setlength{\baselineskip}{0.5\baselineskip}
\vspace{-.15in}\noindent{\tiny
\textbf{Note}: Mean, Var, EVar, Cov90, and Cov95 are calculated over the 1000 repeated subsamples, with
the mean treated as the true value. RCAL denotes $\hat{\mu}_t(\hat{m}^\#_{\text{RWL}}, \hat{\pi}_{\text{RCAL}})$. RMLs denotes $\hat{\mu}_t(\hat{m}_{\text{RMLs}}, \hat{\pi}_{\text{RML}})$. RMLg denotes $\hat{\mu}_t(\hat{m}_{\text{RMLg}}, \hat{\pi}_{\text{RML}})$.}
\end{table}

\begin{table}[H]
\caption{\footnotesize Summary for  $\hat{\mu}_t - \hat{\mu}_k$ on sub samples for $t, k = 0, 1, \ldots, 5$ with tuning parameters selected as $\lambda.1se$.} \label{tb:ate_sub_1se}\vspace{-4ex}
\begin{center}
\resizebox{1.0\textwidth}{!}{\begin{tabular}{lccccccccccccccccc}
\hline
 & Mean & $\sqrt{\text{Var}}$ & $\sqrt{\text{EVar}}$ & Cov90 & Cov95 &~~& Mean & $\sqrt{\text{Var}}$ & $\sqrt{\text{EVar}}$ & Cov90 & Cov95 &~~& Mean & $\sqrt{\text{Var}}$ & $\sqrt{\text{EVar}}$ & Cov90 & Cov95 \\
\hline
& \multicolumn{5}{c}{$\hat{\mu}_1 - \hat{\mu}_0$} &~~& \multicolumn{5}{c}{$\hat{\mu}_2 - \hat{\mu}_0$} &~~& \multicolumn{5}{c}{$\hat{\mu}_3 - \hat{\mu}_0$}\\
\cline{2-6}\cline{8-12}\cline{14-18}
RCAL & -0.056 & 0.010 & 0.010 & 0.901 & 0.953 & ~~ & -0.074 & 0.007 & 0.008 & 0.915 & 0.949 & ~~ & -0.075 & 0.018 & 0.017 & 0.896 & 0.949 \\
RMLs & -0.051 & 0.009 & 0.009 & 0.895 & 0.937 & ~~ & -0.073 & 0.007 & 0.007 & 0.903 & 0.950 & ~~ & -0.076 & 0.017 & 0.014 & 0.866 & 0.909 \\
RMLg & -0.052 & 0.009 & 0.010 & 0.896 & 0.939 & ~~ & -0.073 & 0.007 & 0.007 & 0.905 & 0.950 & ~~ & -0.076 & 0.017 & 0.014 & 0.867 & 0.909 \\

& \multicolumn{5}{c}{$\hat{\mu}_4 - \hat{\mu}_0$} &~~& \multicolumn{5}{c}{$\hat{\mu}_5 - \hat{\mu}_0$} &~~& \multicolumn{5}{c}{$\hat{\mu}_2 - \hat{\mu}_1$} \\
\cline{2-6}\cline{8-12}\cline{14-18}
RCAL & -0.081 & 0.009 & 0.010 & 0.907 & 0.954 & ~~ & -0.087 & 0.018 & 0.018 & 0.890 & 0.940 & ~~ & -0.018 & 0.012 & 0.012 & 0.919 & 0.966 \\
RMLs & -0.083 & 0.009 & 0.009 & 0.883 & 0.931 & ~~ & -0.086 & 0.018 & 0.015 & 0.811 & 0.884 & ~~ & -0.021 & 0.011 & 0.012 & 0.913 & 0.955 \\
RMLg & -0.084 & 0.009 & 0.009 & 0.883 & 0.932 & ~~ & -0.086 & 0.018 & 0.015 & 0.810 & 0.889 & ~~ & -0.021 & 0.011 & 0.012 & 0.912 & 0.956 \\

& \multicolumn{5}{c}{$\hat{\mu}_3 - \hat{\mu}_1$} &~~& \multicolumn{5}{c}{$\hat{\mu}_4 - \hat{\mu}_1$} &~~& \multicolumn{5}{c}{$\hat{\mu}_5 - \hat{\mu}_1$} \\
\cline{2-6}\cline{8-12}\cline{14-18}
RCAL & -0.019 & 0.020 & 0.020 & 0.913 & 0.947 & ~~ & -0.025 & 0.013 & 0.014 & 0.911 & 0.969 & ~~ & -0.031 & 0.020 & 0.021 & 0.908 & 0.956 \\
RMLs & -0.024 & 0.019 & 0.017 & 0.880 & 0.930 & ~~ & -0.032 & 0.013 & 0.013 & 0.907 & 0.949 & ~~ & -0.034 & 0.020 & 0.017 & 0.847 & 0.915 \\
RMLg & -0.024 & 0.019 & 0.017 & 0.881 & 0.931 & ~~ & -0.032 & 0.013 & 0.013 & 0.909 & 0.949 & ~~ & -0.034 & 0.020 & 0.017 & 0.852 & 0.920 \\

& \multicolumn{5}{c}{$\hat{\mu}_3 - \hat{\mu}_2$} &~~& \multicolumn{5}{c}{$\hat{\mu}_4 - \hat{\mu}_2$} &~~& \multicolumn{5}{c}{$\hat{\mu}_5 - \hat{\mu}_2$} \\
\cline{2-6}\cline{8-12}\cline{14-18}
RCAL & -0.002 & 0.019 & 0.018 & 0.900 & 0.947 & ~~ & -0.007 & 0.011 & 0.012 & 0.916 & 0.962 & ~~ & -0.013 & 0.020 & 0.019 & 0.898 & 0.952 \\
RMLs & -0.003 & 0.018 & 0.016 & 0.862 & 0.923 & ~~ & -0.011 & 0.011 & 0.011 & 0.889 & 0.957 & ~~ & -0.013 & 0.019 & 0.016 & 0.822 & 0.887 \\
RMLg & -0.003 & 0.018 & 0.016 & 0.859 & 0.924 & ~~ & -0.011 & 0.011 & 0.011 & 0.891 & 0.959 & ~~ & -0.013 & 0.019 & 0.016 & 0.822 & 0.889 \\

& \multicolumn{5}{c}{$\hat{\mu}_4 - \hat{\mu}_3$} &~~& \multicolumn{5}{c}{$\hat{\mu}_5 - \hat{\mu}_3$} &~~& \multicolumn{5}{c}{$\hat{\mu}_5 - \hat{\mu}_4$} \\
\cline{2-6}\cline{8-12}\cline{14-18}
RCAL & -0.006 & 0.020 & 0.019 & 0.908 & 0.958 & ~~ & -0.012 & 0.025 & 0.025 & 0.915 & 0.958 & ~~ & -0.006 & 0.020 & 0.020 & 0.903 & 0.957 \\
RMLs & -0.008 & 0.019 & 0.017 & 0.860 & 0.931 & ~~ & -0.010 & 0.025 & 0.021 & 0.836 & 0.905 & ~~ & -0.002 & 0.021 & 0.017 & 0.825 & 0.893 \\
RMLg & -0.008 & 0.019 & 0.017 & 0.860 & 0.930 & ~~ & -0.010 & 0.025 & 0.021 & 0.838 & 0.907 & ~~ & -0.002 & 0.021 & 0.017 & 0.825 & 0.895 \\

\hline
\end{tabular}}
\end{center}
\setlength{\baselineskip}{0.5\baselineskip}
\vspace{-.15in}\noindent{\tiny
\textbf{Note}: Mean, Var, EVar, Cov90, and Cov95 are calculated over the 1000 repeated subsamples, with the mean treated as the true value. RCAL denotes $\hat{\mu}_t(\hat{m}^\#_{\text{RWL}}, \hat{\pi}_{\text{RCAL}}) - \hat{\mu}_k(\hat{m}^\#_{\text{RWL}}, \hat{\pi}_{\text{RCAL}})$. RMLs denotes $\hat{\mu}_t(\hat{m}_{\text{RMLs}}, \hat{\pi}_{\text{RML}}) - \hat{\mu}_k(\hat{m}_{\text{RMLs}}, \hat{\pi}_{\text{RML}})$. RMLg denotes $\hat{\mu}_t(\hat{m}_{\text{RMLg}}, \hat{\pi}_{\text{RML}}) - \hat{\mu}_k(\hat{m}_{\text{RMLg}}, \hat{\pi}_{\text{RML}})$.}
\end{table}

\begin{table}[H]
\caption{\footnotesize Summary for $\hat{\nu}^{(k)}_t$ on sub samples for $t, k = 0, 1, \ldots, 5$ with tuning parameters selected as $\lambda.1se$.} \label{tb:nu_sub_1se}\vspace{-4ex}
\begin{center}
\resizebox{\textwidth}{!}{\begin{tabular}{lccccccccccccccccccccccc}
\hline
& \multicolumn{3}{c}{$\hat{\nu}^{(0)}_t$} & $~~$ & \multicolumn{3}{c}{$\hat{\nu}^{(1)}_t$} & $~~$ & \multicolumn{3}{c}{$\hat{\nu}^{(2)}_t$} & $~~$ & \multicolumn{3}{c}{$\hat{\nu}^{(3)}_t$} & $~~$ & \multicolumn{3}{c}{$\hat{\nu}^{(4)}_t$} & $~~$ & \multicolumn{3}{c}{$\hat{\nu}^{(5)}_t$} \\
\cline{2-4}\cline{6-8}\cline{10-12}\cline{14-16}\cline{18-20}\cline{22-24}
& RCAL & RMLs & RMLg & $~~$ & RCAL & RMLs & RMLg & $~~$  & RCAL & RMLs & RMLg & $~~$ & RCAL & RMLs & RMLg & $~~$ & RCAL & RMLs & RMLg & $~~$ & RCAL & RML & RMLg \\
\hline
& \multicolumn{23}{c}{\footnotesize t = 0} \\
$\sqrt{\text{Var}}$ & 0.002& 0.002& 0.002& ~~& 0.005& 0.004& 0.003& ~~& 0.003& 0.003& 0.003& ~~& 0.003& 0.004& 0.003& ~~& 0.003& 0.004& 0.004& ~~& 0.003& 0.005& 0.004\\
$\sqrt{\text{EVar}}$ & 0.002& 0.002& 0.002& ~~& 0.002& 0.003& 0.003& ~~& 0.003& 0.003& 0.003& ~~& 0.002& 0.003& 0.002& ~~& 0.003& 0.003& 0.003& ~~& 0.002& 0.004& 0.003\\
Cov90 & 0.896& 0.896& 0.896& ~~& 0.604& 0.847& 0.842& ~~& 0.850& 0.855& 0.857& ~~& 0.800& 0.819& 0.770& ~~& 0.835& 0.832& 0.829& ~~& 0.809& 0.834& 0.781\\
Cov95 & 0.956& 0.956& 0.956& ~~& 0.689& 0.915& 0.907& ~~& 0.917& 0.918& 0.926& ~~& 0.876& 0.884& 0.854& ~~& 0.903& 0.907& 0.902& ~~& 0.884& 0.896& 0.851\\

& \multicolumn{23}{c}{\footnotesize t = 1} \\
$\sqrt{\text{Var}}$ & 0.010& 0.010& 0.010& ~~& 0.009& 0.009& 0.009& ~~& 0.009& 0.009& 0.009& ~~& 0.009& 0.009& 0.009& ~~& 0.009& 0.010& 0.010& ~~& 0.009& 0.010& 0.010\\
$\sqrt{\text{EVar}}$ & 0.010& 0.010& 0.010& ~~& 0.010& 0.010& 0.010& ~~& 0.010& 0.010& 0.010& ~~& 0.010& 0.009& 0.009& ~~& 0.010& 0.010& 0.010& ~~& 0.010& 0.010& 0.010\\
Cov90 & 0.907& 0.894& 0.893& ~~& 0.909& 0.909& 0.909& ~~& 0.911& 0.911& 0.913& ~~& 0.909& 0.899& 0.904& ~~& 0.912& 0.903& 0.906& ~~& 0.909& 0.887& 0.889\\
Cov95 & 0.951& 0.939& 0.940& ~~& 0.954& 0.954& 0.954& ~~& 0.953& 0.954& 0.954& ~~& 0.953& 0.947& 0.949& ~~& 0.954& 0.948& 0.950& ~~& 0.954& 0.950& 0.950\\

& \multicolumn{23}{c}{\footnotesize t = 2} \\
$\sqrt{\text{Var}}$ & 0.008& 0.007& 0.007& ~~& 0.006& 0.007& 0.007& ~~& 0.006& 0.006& 0.006& ~~& 0.006& 0.006& 0.006& ~~& 0.006& 0.006& 0.006& ~~& 0.006& 0.007& 0.007\\
$\sqrt{\text{EVar}}$ & 0.008& 0.007& 0.007& ~~& 0.007& 0.007& 0.007& ~~& 0.007& 0.007& 0.007& ~~& 0.007& 0.006& 0.006& ~~& 0.007& 0.007& 0.007& ~~& 0.007& 0.007& 0.007\\
Cov90 & 0.905& 0.897& 0.896& ~~& 0.908& 0.895& 0.901& ~~& 0.910& 0.910& 0.910& ~~& 0.910& 0.902& 0.910& ~~& 0.913& 0.913& 0.915& ~~& 0.911& 0.887& 0.891\\
Cov95 & 0.949& 0.945& 0.942& ~~& 0.950& 0.946& 0.950& ~~& 0.950& 0.950& 0.950& ~~& 0.952& 0.955& 0.955& ~~& 0.952& 0.958& 0.960& ~~& 0.951& 0.939& 0.946\\

& \multicolumn{23}{c}{\footnotesize t = 3} \\
$\sqrt{\text{Var}}$ & 0.019& 0.018& 0.018& ~~& 0.014& 0.017& 0.016& ~~& 0.014& 0.016& 0.016& ~~& 0.014& 0.014& 0.014& ~~& 0.014& 0.016& 0.016& ~~& 0.014& 0.015& 0.015\\
$\sqrt{\text{EVar}}$ & 0.018& 0.015& 0.015& ~~& 0.015& 0.016& 0.016& ~~& 0.015& 0.017& 0.017& ~~& 0.015& 0.015& 0.015& ~~& 0.015& 0.017& 0.017& ~~& 0.015& 0.016& 0.016\\
Cov90 & 0.894& 0.834& 0.836& ~~& 0.903& 0.890& 0.890& ~~& 0.906& 0.913& 0.914& ~~& 0.903& 0.903& 0.903& ~~& 0.902& 0.919& 0.920& ~~& 0.902& 0.907& 0.908\\
Cov95 & 0.947& 0.894& 0.894& ~~& 0.945& 0.940& 0.941& ~~& 0.947& 0.948& 0.949& ~~& 0.946& 0.946& 0.946& ~~& 0.947& 0.957& 0.959& ~~& 0.946& 0.947& 0.948\\

& \multicolumn{23}{c}{\footnotesize t = 4} \\
$\sqrt{\text{Var}}$ & 0.010& 0.010& 0.010& ~~& 0.008& 0.009& 0.009& ~~& 0.008& 0.008& 0.008& ~~& 0.008& 0.008& 0.008& ~~& 0.008& 0.008& 0.008& ~~& 0.008& 0.008& 0.008\\
$\sqrt{\text{EVar}}$ & 0.010& 0.009& 0.009& ~~& 0.008& 0.009& 0.009& ~~& 0.008& 0.008& 0.008& ~~& 0.008& 0.007& 0.007& ~~& 0.008& 0.008& 0.008& ~~& 0.008& 0.008& 0.008\\
Cov90 & 0.909& 0.873& 0.876& ~~& 0.888& 0.880& 0.882& ~~& 0.893& 0.896& 0.898& ~~& 0.888& 0.867& 0.872& ~~& 0.888& 0.888& 0.888& ~~& 0.890& 0.879& 0.881\\
Cov95 & 0.952& 0.926& 0.927& ~~& 0.939& 0.930& 0.931& ~~& 0.937& 0.943& 0.942& ~~& 0.937& 0.926& 0.926& ~~& 0.937& 0.937& 0.937& ~~& 0.937& 0.933& 0.935\\

& \multicolumn{23}{c}{\footnotesize t = 5} \\
$\sqrt{\text{Var}}$ & 0.019& 0.019& 0.019& ~~& 0.017& 0.020& 0.020& ~~& 0.017& 0.019& 0.019& ~~& 0.017& 0.017& 0.017& ~~& 0.017& 0.018& 0.018& ~~& 0.017& 0.017& 0.017\\
$\sqrt{\text{EVar}}$ & 0.019& 0.015& 0.015& ~~& 0.017& 0.017& 0.017& ~~& 0.017& 0.018& 0.018& ~~& 0.017& 0.015& 0.015& ~~& 0.017& 0.018& 0.018& ~~& 0.017& 0.017& 0.017\\
Cov90 & 0.889& 0.803& 0.803& ~~& 0.885& 0.831& 0.838& ~~& 0.887& 0.869& 0.872& ~~& 0.882& 0.860& 0.858& ~~& 0.885& 0.900& 0.906& ~~& 0.886& 0.886& 0.886\\
Cov95 & 0.942& 0.873& 0.874& ~~& 0.939& 0.879& 0.886& ~~& 0.942& 0.927& 0.932& ~~& 0.941& 0.913& 0.913& ~~& 0.944& 0.942& 0.946& ~~& 0.943& 0.943& 0.943\\

\hline
\end{tabular}}
\end{center}
\setlength{\baselineskip}{0.5\baselineskip}
\vspace{-.15in}\noindent{\tiny
\textbf{Note}: Var, EVar, Cov90, and Cov95 are calculated over the 1000 repeated subsamples, with the mean treated as the true value. RCAL denotes $\hat{\nu}^{(k)}_{t, \text{RCAL}}$. RMLs denotes $\hat{\nu}^{(k)}_{t, \text{RMLs}}$. RMLg denotes $\hat{\nu}^{(k)}_{t, \text{RMLg}}$.}
\end{table}

\vspace{.3in}
\centerline{\bf\Large References}

\begin{description}\addtolength{\itemsep}{-.1in}

\item Hsu, D., Kakade, S.M., and Zhang, T. (2012) A tail inequality for quadratic forms of subgaussian random vectors, {\em Electronic Communications in Probability}, 17, 1–6.

\item Tan, Z. (2020a) Regularized calibrated estimation of propensity scores with model misspecification and high-dimensional data, {\em Biometrika}, 107, 137–158.

\item Tan, Z. (2020b) Model-assisted inference for treatment effects using regularized calibrated estimation with high-dimensional data, {\em Annals of Statistics}, 48, 811–837.

\end{description}

\end{document}